\documentclass[a4paper]{osudissert}

\usepackage{subfigure}
\usepackage{amsfonts}
\usepackage{color}
\usepackage{amsmath}
\usepackage{graphicx}  %
\usepackage{bm}  %
\usepackage{amssymb}
\usepackage{dcolumn}
\usepackage{multirow}
\usepackage{simplewick} 

\def\BibTeX{{\rm B\kern-.05em{\sc i\kern-.025em b}\kern-.08em
    T\kern-.1667em\lower.7ex\hbox{E}\kern-.125emX}}

\graphicspath{{./figures/}}

\newcommand{\beqn}{\begin{equation}}
\newcommand{\eeqn}{\end {equation}}
\newcommand{\bea}{\begin{eqnarray}}
\newcommand{\eea}{\end{eqnarray}}
\newcommand{\ben}{\begin{eqnarray*}}
\newcommand{\een}{\end{eqnarray*}}
\newcommand{\be}{\begin{enumerate}}
\newcommand{\ee}{\end{enumerate}}
\newcommand{\bi}{\begin{itemize}}
\newcommand{\ei}{\end{itemize}}
\newcommand{\bfig}{\begin{figure}}
\newcommand{\efig}{\end{figure}}
\newcommand{\bc}{\begin{center}}
\newcommand{\ec}{\end{center}}
\newcommand{\bt}{\begin{table} \begin{center}}
\newcommand{\et}{\end{center} \end{table}}

\newcommand{\kprime}{k^{\prime}}

\newcommand{\fm}{\, \rm \text{fm}}
\newcommand{\fmi}{\, \rm \text{fm}^{-1}}

\newcommand{\D}{\displaystyle}

\newcommand{\wt}{\widetilde}
\newcommand{\la}{\langle}
\newcommand{\ra}{\rangle}
\newcommand{\nexp}{n_{\rm \exp}}

\newcommand{\energy}[1]{E_{#1}}
\newcommand{\energyke}[1]{\epsilon_{#1}}
\newcommand{\adag}{a^\dagger}

\newcommand{\fet}[1]{\mbox{\boldmath $#1$}}

\newcommand{\vlowk}{V_{{\rm low}\,k}}

\newcommand{\Hbd}{H^{\rm bd}}
\newcommand{\flow}{s}
\newcommand{\Trel}{T_{\rm rel}}
\newcommand{\Hho}{H_{\rm ho}}

\newcommand{\abinit}{{\it ab initio }}

\newcommand{\nmax}{N_{\rm max}\,}
\newcommand{\mybar}[1]{\overline{{#1}}}

\newcommand{\vtwo}{V^{(2)}_s}
\newcommand{\vthree}{V^{(3)}_s}
\newcommand{\vbt}{\overline{V}^{(2)}_s}
\newcommand{\vbtr}{\overline{V}^{(3)}_s}
\newcommand{\noscr}{|n_1n_2\ra}
\newcommand{\hw}{\hbar\omega}
\newcommand{\kmax}{\rm k_{\rm max}}
\newcommand{\pmax}{p_{\rm max}}

\newcommand{\npi}{EFT($\pi$ \hspace{-.35 cm}/)}

\newcommand{\xeft}{$\chi$EFT}
\newcommand{\xefts}{$\chi$EFT's}

\newcommand{\singlpic}[1]{\includegraphics*[width=3in]{#1}}
\newcommand{\dblpic}[1]{\includegraphics*[width=2.8in]{#1}}
\newcommand{\dblpichgt}[1]{\includegraphics*[height=2.5in]{#1}}
\newcommand{\triplepic}[1]{\includegraphics*[height=1.8in]{#1}}
\newcommand{\strip}[1]{\includegraphics*[width=6in]{#1}}
\newcommand{\spuriousscale}[1]{\includegraphics*[height=1.1in]{#1}}
\newcommand{\captionspace}[1]{\vspace*{-.3in}\caption{#1}}
\renewcommand\typesetChapterTitle[1]{\uppercase{#1}}

\begin{document}
%
%
%
%
%
\author{Eric Jurgenson}
\title{Applications of the \\
Similarity Renormalization Group \\
to the Nuclear Interaction}
\authordegrees{B.A., M.S.}  
\unit{Graduate Program in Physics}
\advisorname{Prof. R. J. Furnstahl}
\member{Prof. R. Perry}
\member{Prof. E. Sugarbaker}
\member{Prof. T. Ho}

%
%

\maketitle


\disscopyright

%
\begin{abstract}
  The Similarity Renormalization Group (SRG) is investigated as a
powerful yet practical method to modify nuclear potentials so as to reduce
computational requirements for calculations of observables. The SRG
proves to be versatile and robust in its treatment of these
interactions and opens the door to a deeper understanding of the
renormalization process.

Chiral Effective Field Theory (\xeft) provides a consistent and
rigorous parametrization of the inter-nucleon interaction. While
already softer than other available potentials, transformation
via the Similarity Renormalization Group (SRG) brings numerous
computational benefits. The hierarchy of many-body forces inherent
in \xefts\ are also treated consistently by the SRG's simple
formalism. The SRG is a natural partner to this modern
program of formulating the nuclear interaction.

The key feature of SRG transformations that leads to computational
benefits is the decoupling of low-energy nuclear physics from
high-energy details of the inter-nucleon interaction. We 
examine decoupling quantitatively for two-body observables and
few-body binding energies. The universal nature of this decoupling is
illustrated and errors from suppressing high-momentum modes above the
decoupling scale are shown to be perturbatively small.

As implemented here, the SRG provides freedom to choose the form of
its transformations and can be tailored to a given application. We
explore the impacts of various choices and their decoupling
properties, specifically a block-type transformation inspired by
previous renormalization group techniques. Sharp and smooth block-diagonal
forms of phase-shift equivalent nucleon-nucleon potentials in momentum
space are generated as examples and compared to analogous low-momentum
interactions (``$\vlowk$'').

To explore the SRG evolution of many-body forces, we use as a
laboratory a one-dimensional system of bosons with short-range
repulsion and mid-range attraction, which emulates realistic nuclear
forces. The free-space SRG is implemented for few-body systems in a
symmetrized harmonic oscillator basis using a recursive construction
analogous to no-core shell model implementations. This approach is
fully unitary up to induced $A$-body forces when applied with an
$A$-particle basis (e.g., $A$-body bound-state energies are exactly
preserved). The oscillator matrix elements for a given $A$ can then be
used in larger systems. Errors from omitted induced many-body forces
show a hierarchy of decreasing contribution to binding energies. An
analysis of individual contributions to the growth of many-body forces
demonstrates such a hierarchy and provides an understanding of its
origins. Several other important sample calculations are explored in
this model for future use in realistic systems.

Building on one-dimensional results we performed the first practical
evolution of three-dimensional many-body forces within the No-Core
Shell Model basis. Results for the $^3$H binding energy are consistent
with previous calculations involving momentum-space evolution of only
two-body forces, and validate expectations from calculations in the
one-dimensional oscillator basis. When applied to $^4$He calculations,
the two- and three-body oscillator matrix elements yield rapid
convergence of the ground-state energy with a small net contribution
of the induced four-body force. The radius of light nuclei is also
explored in the three-dimensional basis.

\end{abstract}
%
%
%
\dedication{For Adrianne and Ethan}
%
%
\begin{acknowledgments}

I am deeply indebted to my adviser, Dr. Richard Furnstahl, for
his tireless efforts in my education. I am honored and privileged
to have worked with him. I can only hope I have absorbed some
measure of his educational, scientific, and collaborative
ethics. 

I am equally grateful to all the other members, past and present,
of the Nuclear Theory Group who have contributed to a vibrant
atmosphere of curiosity and helpfulness in research efforts.
Especially Robert Perry, my almost-co-adviser, Eric Anderson my office
mate, and the several postdocs, Scott Bogner, Lucas Platter, and
Joaqu{\'\i}n Drut.

Of course, without my collaborators much of this work would not
have happened. Among those not already named, I want to thank
especially Petr Navr\'atil and Achim Schwenk for their helpful
discussions and guidance.

I also want to thank my committee members for their time and effort in
evaluating and guiding me along this program.

This work would not be possible without the love and support of
my wife Adrianne and son Ethan. Their constant presence has given
me the emotional fuel for the day-to-day work.

I want to thank my parents and siblings, though far away, for
their quiet understanding and encouragement of my challenges and
successes.

Many friends have helped to make life fun here in Columbus - Jake
and Nichole Knepper, our graduate school peers and pinochle
partners, and their boys Joel and Zach, the Gethsemane Choir for
all their love and support especially our good friends, Dick and
Judy Reunning, Barb and Byron Ford, Bill Alsnauer for keeping me
on my toes, and John Jacobs for being the friend that he always
is.

I also must mention the numerous teachers who have instilled in
me the perseverance to take this career path: Julie Britton, Joe Hunt,
Barb Harken, Tom Stevens, Pat Mason, Uriel Nauenberg, and
many others.

\end{acknowledgments}

\begin{vita}

\dateitem{April 3, 1981}{Born - New Brunswick, NJ, U.S.A.}
\dateitem{May 2003}{B.A. Physics, University of Colorado,
Boulder, Colorado}
\dateitem{December 2008}{M.S. Physics, The Ohio State University,
Columbus, Ohio}
\dateitem{September 2003 - present}{Department of Physics, The
Ohio State University, Columbus, Ohio} 

\begin{publist}
\researchpubs
\pubitem{{``Decoupling in the Similarity Renormalization Group
for Nucleon-Nucleon Forces"}, 
E.D.~Jurgenson, S.K.~Bogner, R.J.~Furnstahl, R.J.~Perry, 
Phys.\ Rev.\ {\bf C} 78, 014003 (2008). arXiv:0711.4252 [nucl-th]}
\pubitem{{``Block Diagonalization using SRG Flow Equations"}, 
E.~Anderson, S.K.~Bogner, R.J.~Furnstahl, E.D.~Jurgenson,
R.J.~Perry, A.~Schwenk, 
Phys.\ Rev.\ {\bf C} 77, 037001 (2008). arXiv:0801.1098 [nucl-th]}
\pubitem{{``Similarity Renormalization Group Evolution of
Many-Body Forces in a One-Dimensional Model"}, 
E.D.~Jurgenson, R.J.~Furnstahl, 
Nucl.\ Phys.\ {\bf A} 818, 152 (2009). arXiv:0809.4199 [nucl-th]}
\pubitem{{``Evolution of Nuclear Many-Body Forces with the 
Similarity Renormalization Group"}, 
E.D.~Jurgenson, P.~Navr\'atil, R.J.~Furnstahl, 
Phys.\ Rev.\ Lett.\ {\bf 103}, 082501 (2009). arXiv:0905.1873 [nucl-th]}
\end{publist}

\begin{fieldsstudy} 
\majorfield{Physics} 
\onestudy{The Similarity Renormalization Group Approach to
the Nuclear Interaction}{Prof. Richard J. Furnstahl} 
\end{fieldsstudy}

\end{vita}

%
\tableofcontents
\listoftables
\listoffigures
%
\chapter{Introduction}
\label{chapt:introduction}

\section{History}


A complete understanding of the interaction between nucleons has been
sought by physicists for more than seventy years, beginning with the
pion exchange theory of Hideki Yukawa \cite{Yukawa:1935xg} in the
1930's, which set the paradigm of heavy boson exchange. Shortly after
the discovery of the pion and confirmation of Yukawa's conjecture,
nucleon-nucleon (NN) scattering experiments were interpreted as
meaning that the NN interaction had a strong repulsive core, an
intermediate ranged attraction, and strong tensor and spin-orbit
forces~\cite{machleidt_history}. Efforts to build an accurate NN
potential that incorporated these features and was based on Yukawa's
meson exchange idea were eventually quite successful by the late 
1990's~\cite{NN_experiment}. However, attempts to apply these meson-exchange
interactions to many-body problems proved difficult.

\begin{figure}[th]
\begin{center}
\strip{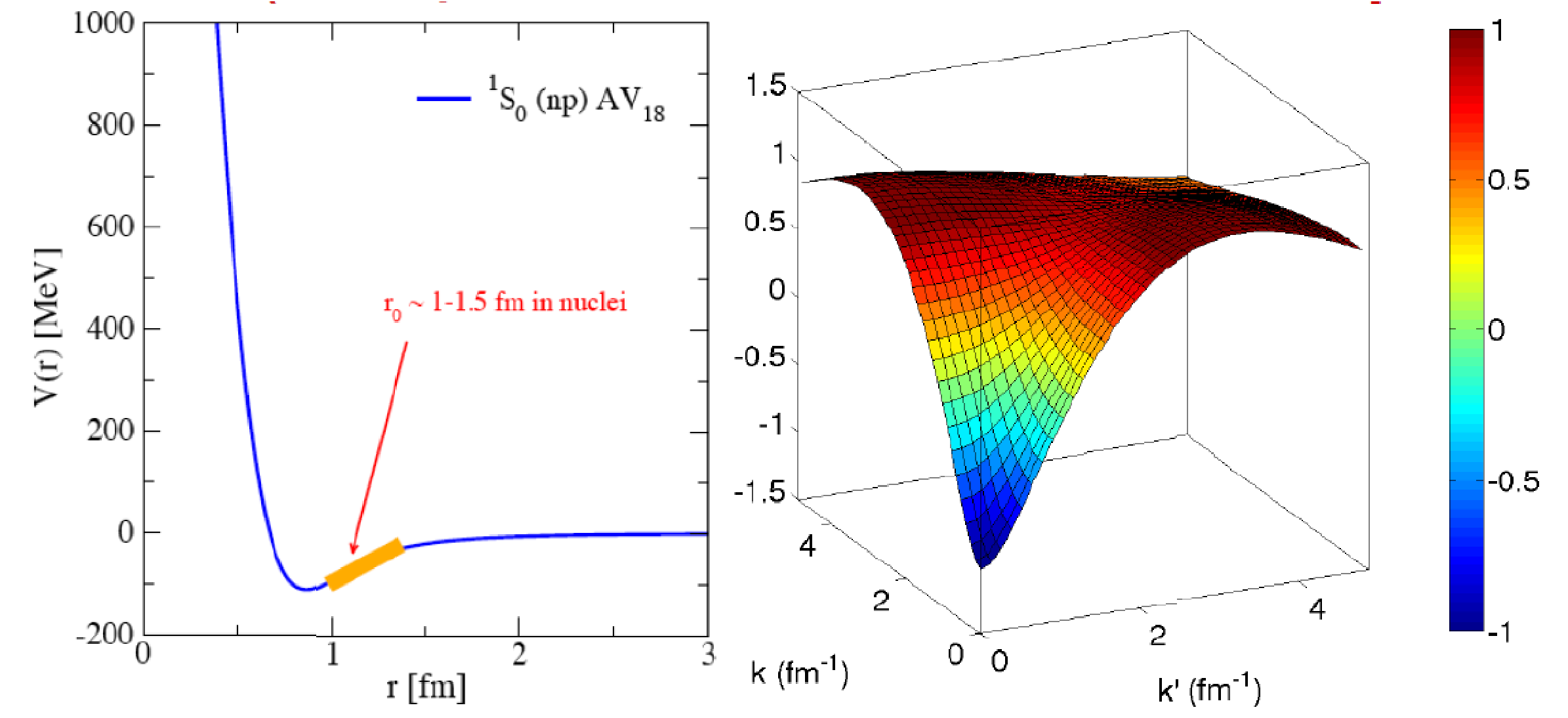}
\end{center}
\captionspace{The Argonne V18 nuclear potential in coordinate (left) and
momentum (right) representations. Notice the hard repulsive core at
short-distance/high-momentum and small mid-range attractive well.
The typical spacing within nuclei is indicated by the arrow.}
\label{fig:potential}
\end{figure}


Figure~\ref{fig:potential} shows an example of such an NN potential in
both coordinate and momentum representations. The hard repulsive core
in the left panel and the corresponding strong high-momentum matrix
elements in the right panel represent short-distance physics, the
details of which are irrelevant to low-energy physics. The specific
coupling of the low- and high-energy states in the high-energy matrix
elements shown here inflates the required basis size and hinders
attempts to compute larger nuclei ($A \ge 4$) using \abinit methods. A
softened potential, which has the low- and high-energy degrees of
freedom decoupled, would enable a truncation of the two-body basis and
ease the computational requirements of \abinit many-body problems.
Such soft potentials were tried long ago but deemed insufficient,
because they could not reproduce empirical nuclear matter saturation
properties~\cite{Bethe_review}. The problem was that many-body forces
were not correctly considered. Subsequently, the need for many-body
forces to accurately describe nuclear systems became apparent. For
instance, the highly successful NN interactions could not account for
the total binding energy of the triton~\cite{three_body_req}. However,
it proved to be difficult to systematically account for even
three-body forces.


Then in the early 1990's Steve Weinberg proposed generalizing chiral
perturbation theory~\cite{weinberg_eft} to develop an expansion
consistent with the symmetries of the known underlying theory of the
strong interaction, Quantum Chromodynamics (QCD).\footnote{Due to 
asymptotic freedom, QCD is weakly coupled at high energies and thus
easier to analyze. Unfortunately, it is strongly coupled at low
energies and not suitable for perturbative calculations in the regime
relevant to nuclear structure.} Such a program would provide a
systematic expansion of contributions, with controlled errors at every
order, known as Chiral Effective Field Theories (\xefts). The idea
sparked an industry in developing and properly parametrizing such a
theory of the nuclear interaction. Chiral EFTs provide a systematic
and controlled expansion of not only the NN interactions, but also
many-body forces with controlled errors, and in a natural framework for
including consistent external currents (electromagnetic) and
relativistic effects.


Chiral EFTs have provided softer interactions because there is a lower
cutoff of high-energy physics than in conventional potentials, but
\abinit methods are still computationally very difficult. Simply using
a lower cutoff to further soften \xeft\ potentials has resulted in
pathologies in the potentials and significantly diminished their
accuracy~\cite{N3LOW}. Recent work that softens these interactions
using a unitary block-diagonal transformation referred to as $\vlowk$
have had many successes, but many-body forces have only been included
approximately. The Similarity Renormalization Group (SRG), which
generates unitary transformations that can be used to soften
potentials, can also consistently treat many-body forces and is
particularly suited to the needs of the \xeft\  program.

Now, in an era of rapidly expanding computational capacity, nuclear
structure physics is experiencing a renaissance and helping to renew
collaborations across the larger physics community. Of paramount
importance to this program is the development of new techniques to
make nuclear many-body problems more tractable in a given
computational environment. The Similarity Renormalization Group
promises to be a cornerstone of these techniques for many years to
come. This thesis represents some of the earliest development of
applications of the SRG to problems in nuclear physics.


\section{Modern Many-Body Nuclear Physics}

The UNEDF (Universal Nuclear Energy Density Functional) collaboration
is a broad effort within the nuclear structure community to develop a
wide-reaching and self-consistent model of the structure and behavior
of large nuclei based on Density Functional Theory (DFT), which describes the
bulk properties of many-body systems in terms of functionals of the
nucleon densities. Because of its scaling properties, it is suited for
large many-body systems. The DFT strategy has been employed successfully
for atoms and molecules in quantum chemistry and also in nuclear
theory using empirical EDF's that are fit to nuclear properties. The UNEDF
collaboration is attempting to improve the nuclear DFT approach with
microscopic input such as from Chiral EFT.

\begin{figure}[th]
\begin{center}
\strip{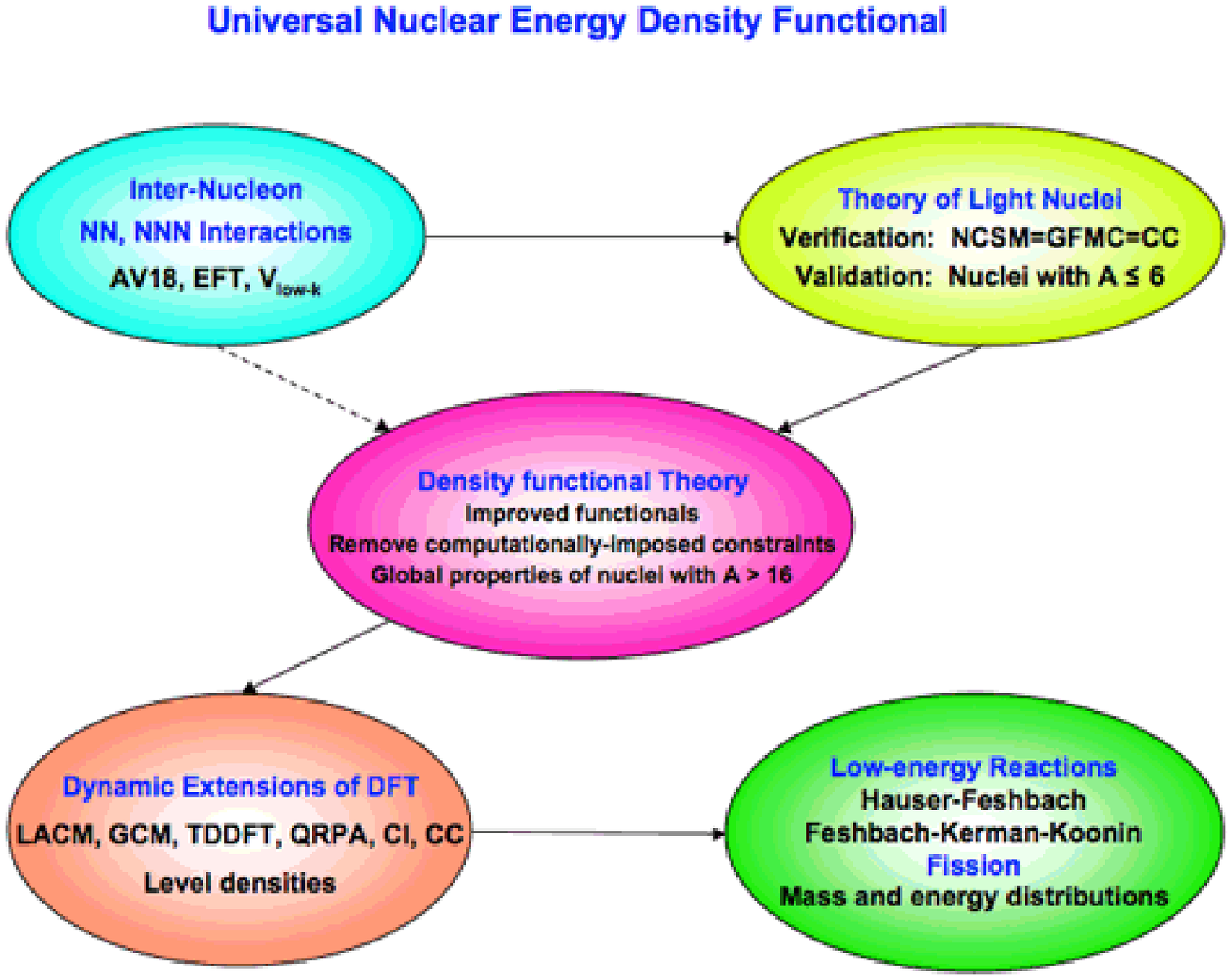}
\end{center}
\captionspace{Scientific interconnections within the UNEDF collaboration.}
\label{fig:UNEDF_connections}
\end{figure}

Figure \ref{fig:UNEDF_connections} shows a chart indicating the
relationships between the various scientific efforts involved in the
UNEDF project. The work in this thesis is directly connected to the
top two bubbles in the chart, and enables work on \abinit DFT's
represented in the center bubble~\cite{DFT_basic}. The upper left
corner represents work on the basic nuclear forces in few-body systems
and efforts to find optimal and practical parametrizations of the
nuclear interaction. Some examples listed are Argonne $v_{18}$ (also
denoted as AV18)~\cite{argonne_potential}, a phenomenological
potential developed in the mid-nineties and widely used today, EFT,
which refers to the Chiral Effective Field Theories described further
in section~\ref{sec:intro_eft} and appendix~\ref{chapt:app_eft}, and
$\vlowk$, which is a renormalization technique related to the SRG. The
SRG is applied in this thesis to treat the ``initial" interactions
(such as AV18 and \xefts) to ease computational requirements for
calculations elsewhere in the UNEDF program. Like $\vlowk$, it works
to soften potentials and is generally applicable to any interaction.
However, unlike $\vlowk$, the SRG handles many-body forces in a
straightforward way.

The upper right corner depicts the first step in practical
calculations with a given initial interaction. In this area of
research, \abinit calculations of the lightest nuclei are performed to
validate the properties of the initial interactions. This is a
computationally intensive program requiring large clusters of
processors numbering into the tens of thousands to accurately compute
the  nuclei above $^6$Li ($A=6$). The examples listed here are the
No-Core Shell Model\footnote{The moniker ``No-Core" simply means that
all nucleons are active instead of the more traditional shell model
which considers valence nucleons around an inert closed
shell~\cite{ncsm_basic,NCSM1a,NCSM1b,NCSM1c}.} (NCSM), which utilizes
a finite basis of harmonic oscillator wavefunctions, Green's Function
Monte Carlo (GFMC), which stochastically computes
coordinate-representation integrals numerically, and the
Coupled-Cluster (CC) method, which is a many-body technique based on a
wavefunction ansatz organized in subclusters of the total system. The
GFMC is currently limited to nuclei with $A \le 12$ due to its scaling
properties and is limited to using local potentials like the
phenomenological potential, AV18. However, work is proceeding to
address these challenges~\cite{Pieper:2002ne,Pieper:2001mp}.
Coupled-Cluster calculations~\cite{CC_orig,Hagen:2007hi} have the
farthest reach up the chart of nuclides of all \abinit techniques and
are an important source of validation of DFT efforts, but they are
limited to a small subset of nuclei near closed shells. 

The few-body calculations done in this thesis make use of the NCSM and
its antisymmetrized basis of harmonic oscillator (HO) wavefunctions.
The NCSM basis provides for a variational calculation in both the size
of the basis, $\nmax$, and the size of the oscillator parameter,
$\hw$. As $\nmax$ is increased the results converge, and at a given
$\nmax$ the oscillator parameter $\hw$ must be optimized, balancing
infrared and ultraviolet cutoffs. These features are explored in more
detail in appendix~\ref{chapt:app_osc_truncation}. In the literature,
the name NCSM usually implies use of the Lee-Suzuki (LS) procedure to
obtain an effective interaction that is not variational in the HO
basis parameters. Here, we will forgo this method in favor of the SRG,
which will provide a softened effective interaction and retain the
variational features of the basis.


\section{Resolution and Renormalization}

\begin{figure}[th]
\begin{center}
\dblpichgt{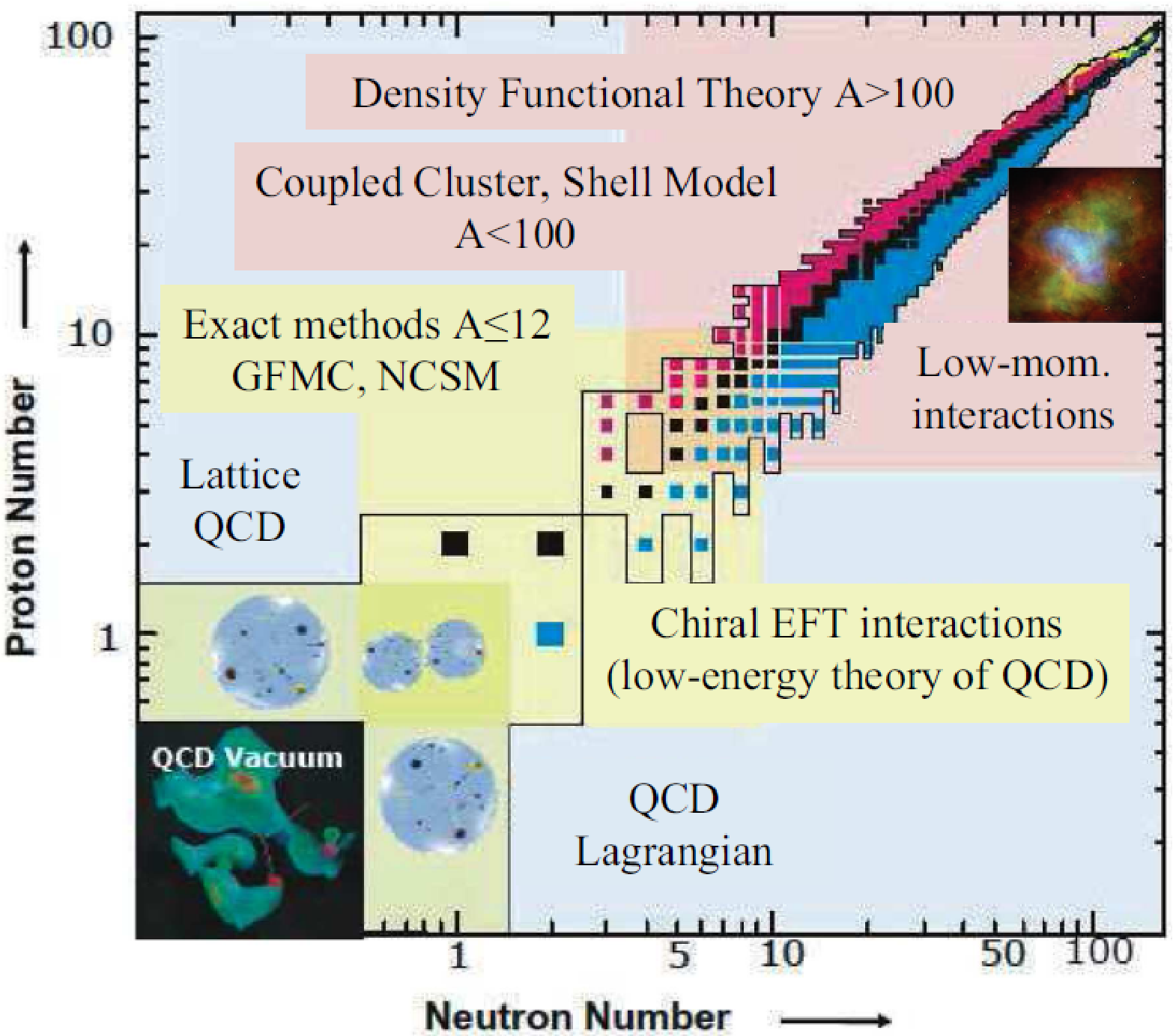}
\hspace{.1in}
\dblpichgt{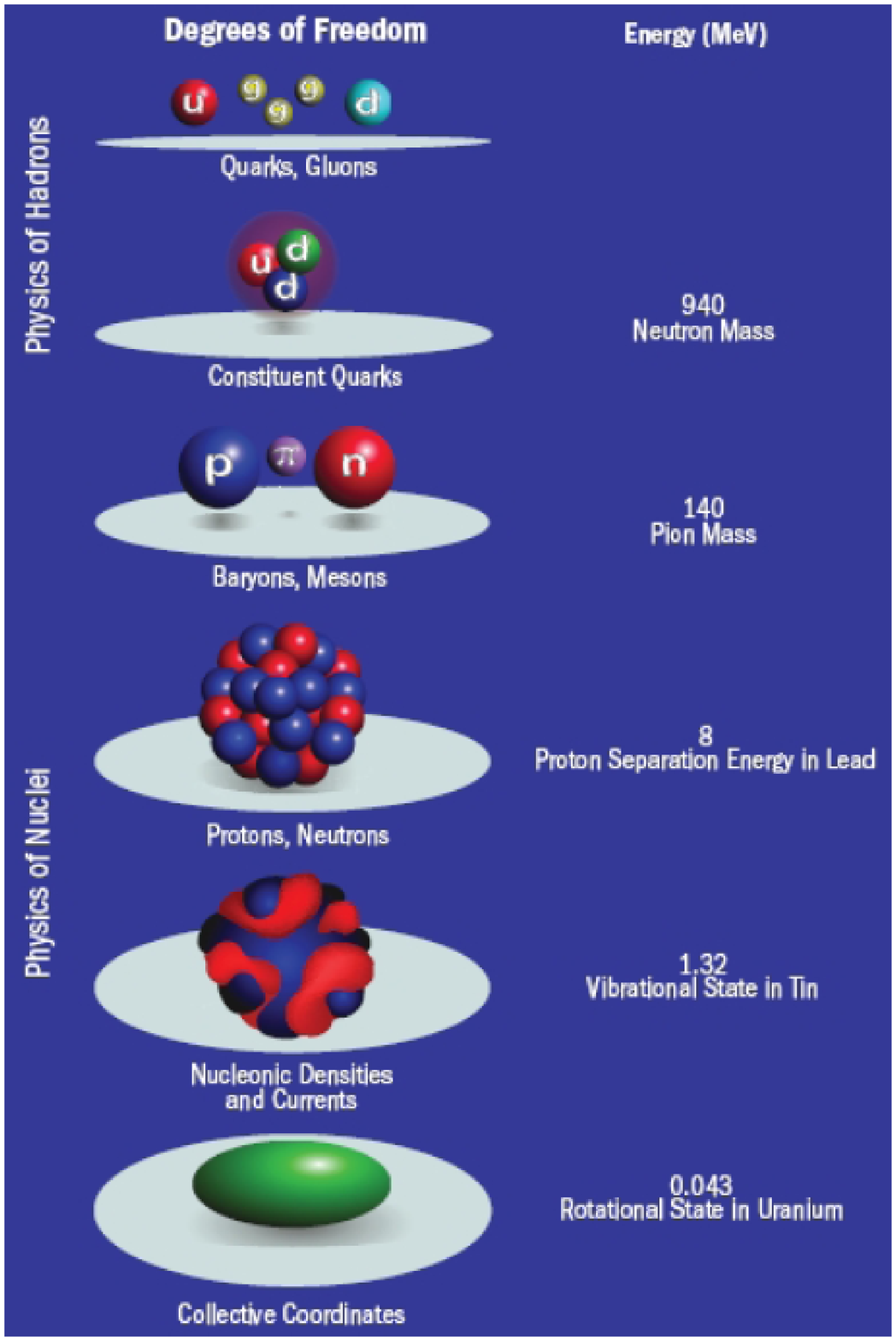}
\end{center}
\captionspace{Two pictures illustrating common degrees of freedom (dof),
interactions, and calculational techniques used in theoretical nuclear
physics.}
\label{fig:nuclear_scales}
\end{figure}

Figure \ref{fig:nuclear_scales} displays an overview of important
aspects of theoretical nuclear structure physics. The picture on the
left is the chart of nuclides on a log-log
plot~\cite{nuc_chart_loglog}. Above the diagonal of the chart are
indicated various calculational techniques positioned near their
relevant regimes, and below are listed interactions that can be used
in various parts of the chart. The right side of
Fig.~\ref{fig:nuclear_scales} is a picture from the UNEDF
collaboration showing some explicit degrees of freedom used in
studying nuclear phenomena~\cite{unedf_scales}. Any calculation can be
performed using any degree of freedom, but many choices may be
intractable. We should pick the degrees of freedom most useful to the
problem at hand. For instance, while Quantum Chromodynamics (QCD) is
the underlying theory to all nuclear interactions, trying to calculate
even the smallest nuclei with quark and gluon degrees of freedom is
computationally impossible with current techniques (though calculating
up to $A=4$ is a long-term goal for the next generation of
computers~\cite{savage_06}). Independent checks on calculations from
different methods is very important to the UNEDF program, and a
technique like the SRG is crucial to extending the overlap of
techniques like DFT, Coupled-Cluster, and NCSM.

In addition to the choice of gross degrees of freedom, the capacity to
adjust the parameters of an interaction within a given formulation is
desirable. The \abinit calculations in this thesis deal with explicit
nucleon degrees of freedom with interactions formulated in terms of
relative nucleon momenta. The computational difficulty of the few- and
many-body problem can be eased with an appropriate treatment of this
initial interaction. As the number of nucleons increases, the size of
the basis required for convergence increases very rapidly. Thus,
reducing the size of the basis required to encode NN physics will
reduce the required size of the $A$-body basis in which the NN physics
must be embedded. The SRG will provide a convenient way to accomplish
this adjustment in a manner that preserves initially chosen physics.
In addition it will also consistently treat many-body forces and other
observables.

\begin{figure}[th]
\begin{center}
\dblpichgt{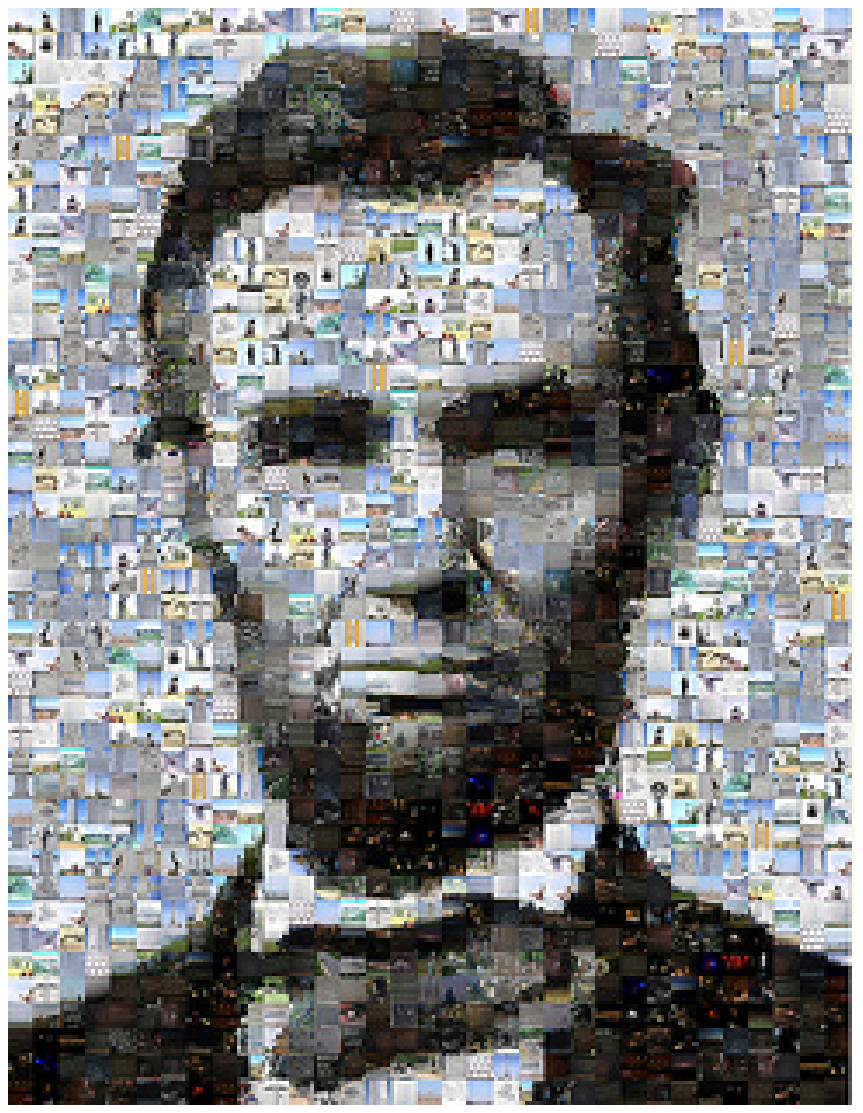}
\hspace{.1in}
\dblpichgt{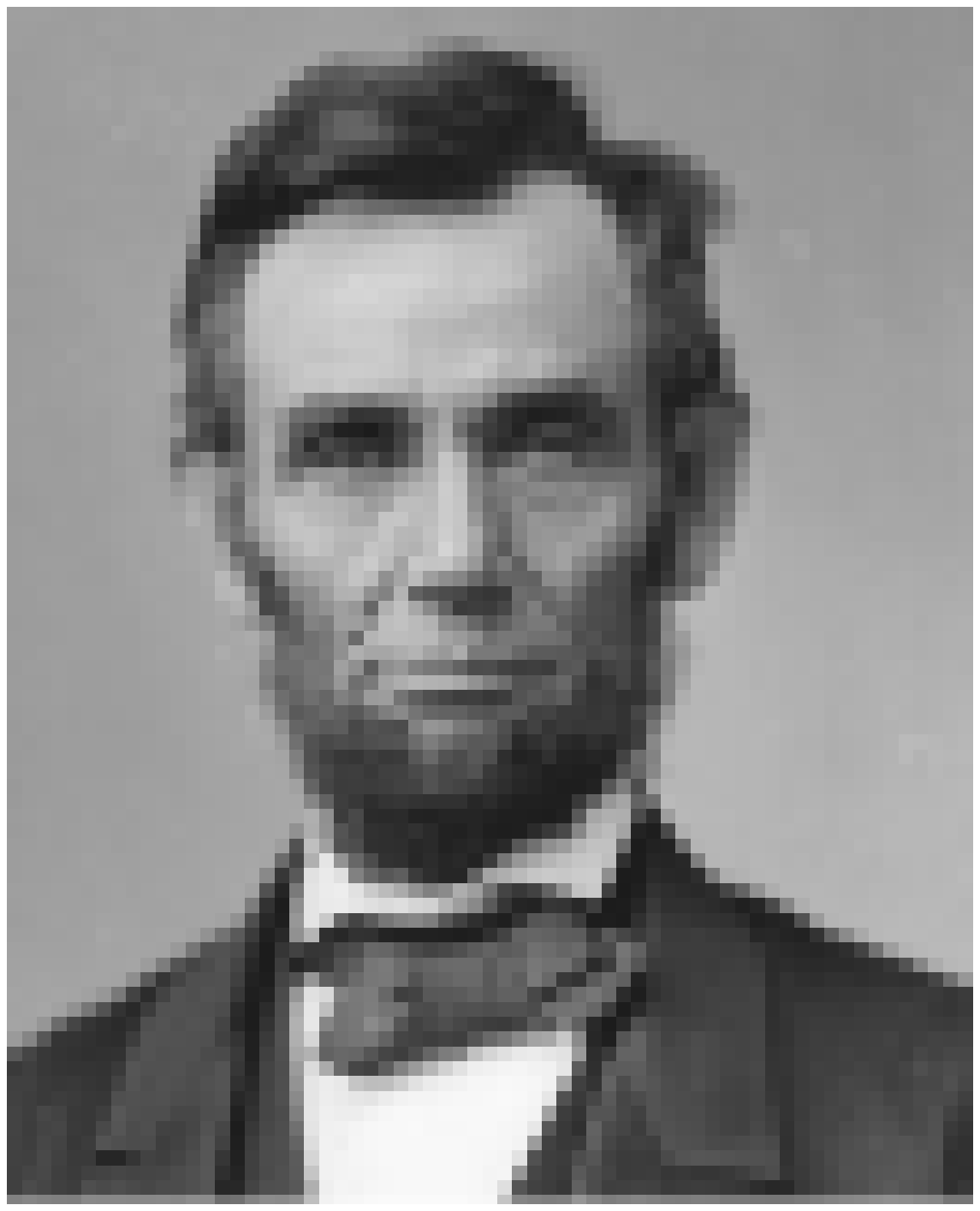}
\end{center}
\captionspace{Two pictures of our 16th president.}
\label{fig:lincoln}
\end{figure}

Figure \ref{fig:lincoln} presents a simple analogy of what the SRG can
do for us. The picture on the left is a photomosaic image where the
image of President Lincoln has been made up from hundreds of smaller
photos that have been fit to their particular location in the whole
image. The picture on the right is just the original photograph that
has been pixelated by standard photo software. Both of these images
get the basic long-distance information correct -- both show
Lincoln's face about equally well. However, the mosaic image of
Lincoln contains a lot of extra information that the viewer's brain
must work to filter and process, using lots of memory and power. Not
only is the image on the left harder to use but the extra information
it contains is even incorrect. All the details in the individual
photographs are not actually part of Lincoln's face and the viewer has
to filter them out. The image on the right has much less information
but still contains what is needed to recognize the face.  Ideally, a
calculation requiring only the long distance information of the right
picture would benefit from a technique which transforms the  mosaic
into a version of the pixelated right-hand side. This is analogous to
the renormalization group technique implemented here, in which
irrelevant details of the short distance information are removed in
favor of the dof's consistent with the relevant physical information,
simplifying the interaction for subsequent calculations. The AV18
potential is like the photomosaic on the left because it contains a
lot of short distance details irrelevant to low-energy nuclear
structure physics. The SRG is going to help filter out those details
and retain the long-distance, low-momentum, information like the
picture on the right.\footnote{This analogy breaks down in the sense
that a simple low-pass filter works well on the photomosaic but not on the
potential. The first is a classical problem, and the second is quantum
mechanical; the photo-mosaic {\it is} the observable, while the
potential is not. This will be described in more detail in
chapter~\ref{chapt:decoupling}.}

\section{Chiral Effective Field Theories}
\label{sec:intro_eft}

As mentioned above, Chiral Effective Field Theories (\xefts) have
evolved to play an important role in the development of a quantitative
understanding of the nuclear interaction. Here we give a brief
overview of their features, with various additional details involved
in constructing \xefts\ discussed in appendix~\ref{chapt:app_eft}.

All these theories are called ``effective'' because they encode the
details of short-distance (high-energy) physics in their coupling
constants, which are called Low Energy Constants (LECs). These are the
parameters that we must fit to make the \xefts\ consistent with the
data and predictive of low-energy physics. Ultimately, short-distance
behavior of the strong interaction is governed by Quantum
Chromodynamics (QCD). The degrees of freedom in a nuclear \xeft\ are
not the same as those in QCD because the quarks and gluons are
confined at the low energies we are considering. However, \xefts\ do
not need to incorporate explicitly the details of high-energy physics
in their Lagrangians because the high-energy behavior is parametrized
in a general way. Thus, \xefts\ exploit the fact that any of an
infinite number of possible parameterizations is sufficient to predict
the correct low-energy observables of nuclei.

Few-body forces are inevitable in any effective theory as a result of
elimination of degrees of freedom. In any process involving three
incoming nucleons there are a large number of possible intermediate
states, some of which will have high energy and occur at very small
distance and time scales. Because the details of this high-energy
physics are irrelevant to low-energy observables, we can use a cutoff
to eliminate explicit high-energy degrees of freedom. However to
account for their effects, we must introduce local interactions,
including many-body forces. The new LECs that come with such terms
must be fit to some experimental observables.  A major advantage of
\xefts\ is that they provide a consistent and systematic framework
with which to formulate such low-energy few-body interaction terms in
addition to the basic two-body (NN) interaction between nucleons. We
will consider in appendix~\ref{chapt:app_eft} the explicit form of
three- and four-body forces (3NF and 4NF). These terms become
important for reproducing the properties of larger nuclei (i.e.
$A\geq3$).

Building any good EFT depends on following three basic rules
\cite{crossing_border}. First, we must identify the degrees of freedom
we want the theory to encompass and write down the most general
Lagrangian that obeys the symmetries of the underlying theory (QCD).
In pionless EFT (\npi), we use only nucleons ($m_N \sim 1$ GeV) as the
degrees of freedom; any pion contributions will be parametrized in the
LEC's. With the pion mass ($m_\pi \sim 140$ MeV) as the cutoff scale
this theory will only be valid for nucleon momenta well below the pion
mass. In \xeft, we will take the cutoff scale to be much larger, such
as the nucleon or $\rho$ meson mass. Then we must include pions
explicitly and the spontaneous breaking of chiral symmetry found in
low energy QCD. Hence the name ``Chiral" in \xeft. The $\Delta$-isobar
($m_\Delta=1232$ MeV) might also be included in the theory due to its
mass being so close to the nucleon.  Second, we must declare a
regularization and renormalization procedure. Several different types
have been used with \xefts\ including  dimensional, cutoff, and
spectral regularization procedures combined with minimal and power
divergence (only used with dimensional regularization) subtraction
schemes. Due to its symmetry preserving properties, dimensional
regularization is often preferred but has not been successfully
applied within the non-perturbative resummations involved in
calculating nuclear bound states in \xeft. Instead, cutoff
regularization is used. Finally, we must choose a method of organizing
the contributions to the potential. The various methods of
organization are collectively referred to as \emph{power counting}
schemes. In any effective theory we integrate out higher momentum
contributions, and introduce an expansion of correction terms to
account for the removed degrees of freedom. Thus, we need a formal
scheme to keep track of the importance of these various contributions
and decide which terms are important at a desired level of accuracy.
The power counting one uses will lead to a particular expansion, or
hierarchy, of contributions to the total inter-nucleon interaction. 

\begin{figure}
\begin{center}
\strip{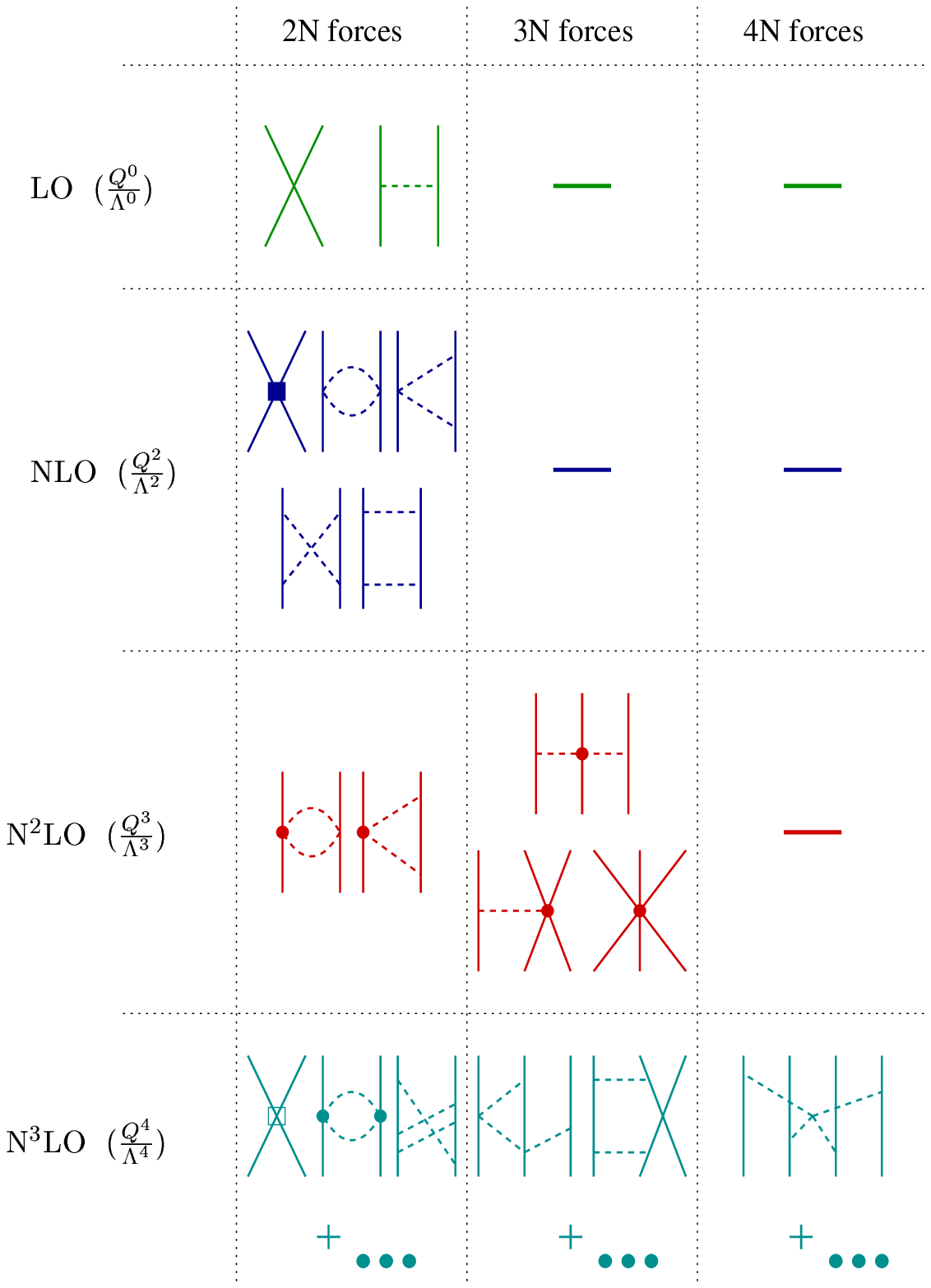}
\end{center}
\captionspace{A table showing the organization of EFT terms in
Weinberg power counting. The figure is from \cite{epelbaum_pic}.}
\label{fig:EFT_hierarchy}
\end{figure}

A picture showing a particular power counting scheme for \xeft\ terms
is given in Fig.~\ref{fig:EFT_hierarchy}. The vertical axis in the
table is the relative importance of terms entering into the amplitude
of a process. The order of the graph is expressed in powers of momenta
($Q/\Lambda$) where $Q$ is a momentum characteristic of the transfer
between interacting nucleons and $\Lambda$ is some large momentum
scale associated with the underlying physics and hence the scale at
which the EFT's predictive power fails.\footnote{Examples of $\Lambda$
are: for \npi\ $\Lambda\approx m_\pi=140$  MeV, for \xeft\ \ any one QCD
scale such as $m_\rho,m_N\simeq 1$  GeV, and for the Fermi theory of
weak interactions $\Lambda = M_W, M_Z \approx 80-90$  GeV
respectively.}  Each row is seen as a correction to the previous row.
Those higher-order terms require higher momenta in order to become
significant effects and therefore are suppressed at the low scale of
$Q$. The columns are separating basic types of terms for multi-body
processes. For instance, three body forces (3NF) first contribute at
the third-to-leading order although in a two-body problem they would
never be relevant.\footnote{3N ``tadpole'' graphs don't contribute to
NN amplitudes because $\overline{\rm N}$'s are not in the EFT. The pair
production that produces them requires a high energy and has been
integrated out. Thus, these effects are encoded in the NN contact
LECs.} Note that this represents a particular power counting choice
originally proposed by Weinberg~\cite{weinberg_eft}. The graphs
displayed in the figure are considered as a perturbative expansion of
the relevant contributions to the nuclear interaction, but at a
desired order in $Q/\Lambda$ all contributions must be summed
non-perturbatively to calculate bound states (in other schemes
only the leading order terms are summed).

The various LECs in a \xeft\ are currently obtained by fitting  to
experimental data. The fit theory can then be used to make predictions
about other observables. When more LECs enter the expansion, they must
be fit to additional experimental observables. At the outset, this
situation seems to limit the predictive power of \xefts\ in general,
but it is currently the most practical way to proceed with the \xeft\
program of many-body physics within large nuclei. The relevant
question is whether we can reach a desired accuracy with a number of
terms that we can handle. For example, hopefully nuclear matter will
only need a small set of $A$-body forces (hopefully no more than 4NFs)
for an accurate description and therefore a relatively small number of
LECs must be fit from experiment. Ultimately one would like to have an
\abinit calculation of nuclear LECs from the underlying theory, QCD.
Work is currently underway to develop techniques to calculate \xeft
LECs in this way, and recent progress makes this a plausible
future~\cite{savage_03, savage_06, green_lattice, kaiser}. In light of
these two approaches the \xeft\ program requires the development of
techniques both for controlling errors induced by fitting procedures,
and for self-consistent $A$-body renormalization capable of dealing
with each term in the EFT expansion, preserving the input physics. The
SRG is a prime example of a technique that satisfies both of these
needs.


\section{Similarity Renormalization Group}


While \xefts\ are already significantly softer than many
phenomenological potentials, further softening of these potentials
provides increased convergence in calculations of larger nuclei. A
potential like Argonne $v_{18}$ usually uses a momentum mesh out to
$\approx 30 \fmi$ while \xefts\ need around $\approx 7 \fmi$. Even
with the softer \xeft\ potentials the task of calculating properties
of $^4$He is computationally difficult, requiring large spaces in the
NCSM basis to achieve convergence (e.g., for $A=4$ at $\nmax=18$, the
basis dimension is 4750 states and the $^4$He binding energy is just
beginning to converge). However, typical momenta inside a nucleus are
of order $1 \fmi$ so we should be able to soften these potentials much
further and increase their convergence properties. 


\begin{figure}[th]
\begin{center}
\dblpic{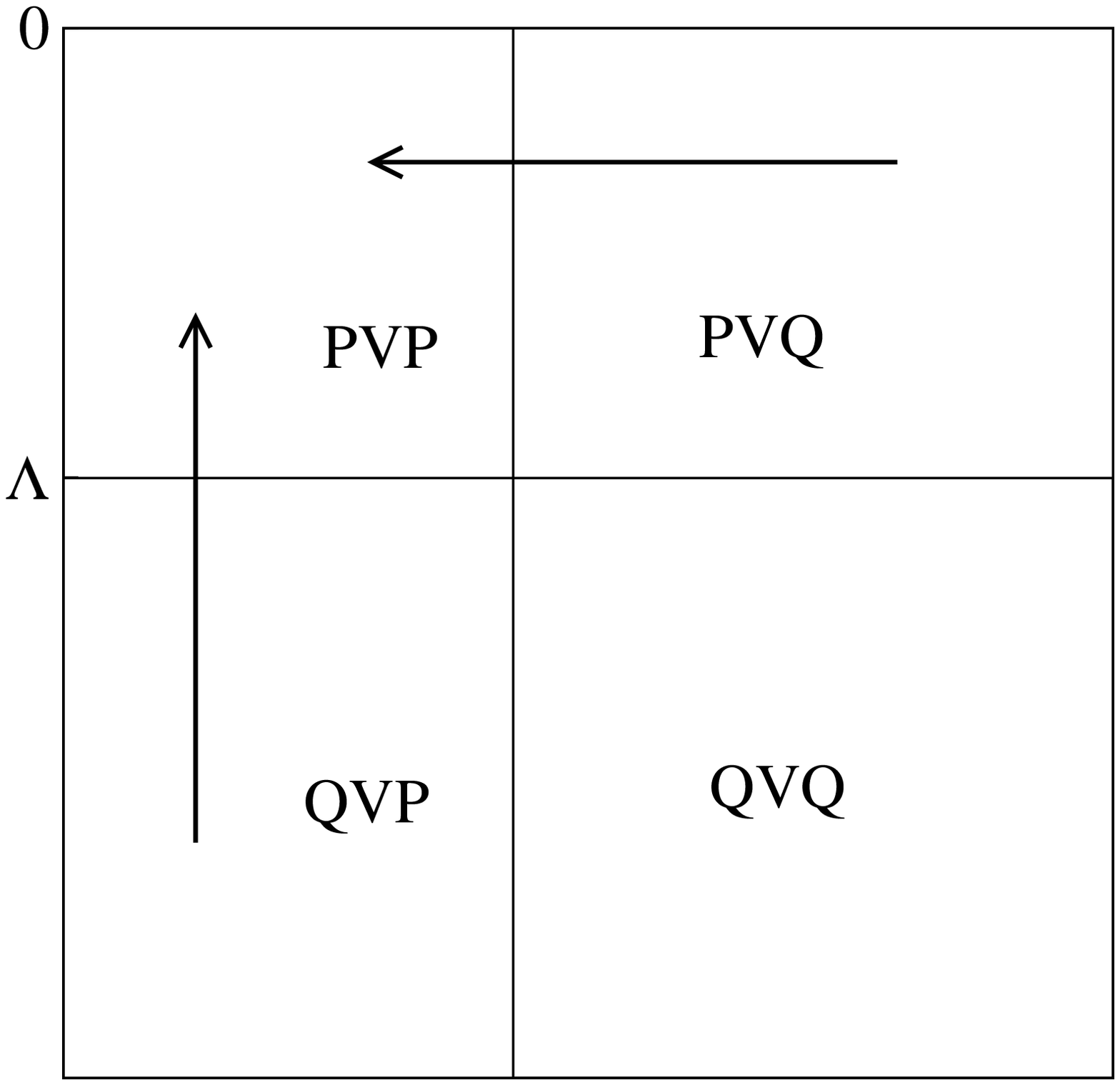}
\hfill
\dblpic{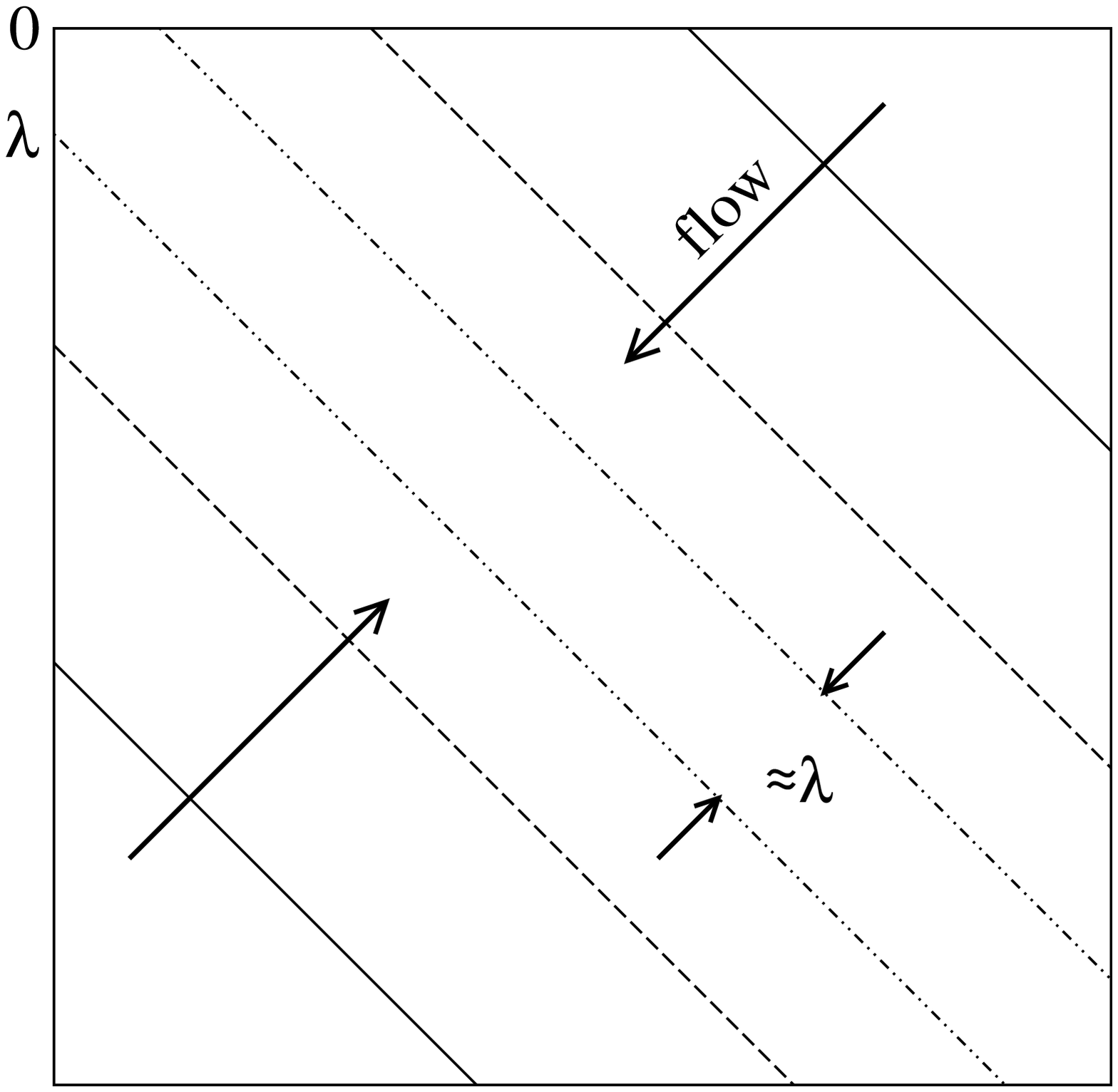}
\end{center}
\captionspace{A simple illustration of $\vlowk$'s block diagonalizing and SRG's band
diagonalizing forms.}
\label{fig:vlowk_vs_srg}
\end{figure}


A common established technique, called
$\vlowk$~\cite{vlowk_basic1,vlowk_basic2}, softens potentials and
improves convergence by transforming potentials to a block-diagonal
form.\footnote{$\vlowk$ is a momentum space implementation of the
block-diagonal transformation idea of  Lee and
Suzuki~\cite{LS_original}, though in the literature the term
Lee-Suzuki usually refers to a different implementation in the NCSM
basis.} In contrast the SRG, as typically implemented here, evolves
the potential into a band-diagonal form~\cite{Bogner:2006srg}. A
schematic is shown in Fig.~\ref{fig:vlowk_vs_srg} of the difference in
form between a potential renormalized by $\vlowk$ (left) and by the
SRG (right). Both of these renormalization strategies use a unitary
transformation to soften the initial potential and thus decouple high-
and low-energy degrees of freedom. On the left, $\vlowk$ transforms
the potential so that all interaction strength is contained in the two
diagonal blocks, $PVP$ and $QVQ$ (or simply $P$ and $Q$ blocks),
defined by the cutoff scale $\Lambda$\footnote{Usually the $\vlowk$
$Q$ block is set to zero in practice, but this is not required.}. In
contrast the SRG smoothly evolves the matrix element strength towards
a chosen form, here usually band diagonal in momentum with a width
$\sim\lambda$ which indicates the amount of coupling left between
neighboring states. However, the general form of SRG evolution is a
matter of choice and we will see that a particular choice reproduces
$\vlowk$ evolution. Note that the $\vlowk$ renormalization can be
achieved through both differential and integral procedures, while the
SRG currently only exists as a differential procedure (as fully
explained in chapter~\ref{chapt:srg}). In addition, a $\vlowk$
treatment of many-body forces is not straightforward and limits its
practical implementation with regard to many-body forces. A
quantitative investigation of the decoupling properties of the SRG
(and the behavior of residual coupling) is presented in
chapter~\ref{chapt:decoupling}.


In the NCSM's oscillator basis, a block-diagonal transformation
similar to that suggested by Lee and Suzuki~\cite{LS_original} has
successfully improved convergence properties for calculations of
many-body observables, it has significant drawbacks which the SRG is
able to address. This transformation is model-space dependent and not
variational and extrapolating its behavior in the basis size,
$\nmax$, is difficult. The SRG's improved convergence
properties are smoothly variational and therefore lend themselves to
extrapolation in basis size. This thesis shows SRG calculations to
have improved convergence over traditional Lee-Suzuki types of
treatments. As mentioned above, many-body forces are a natural
consequence of any renormalization program. Their hierarchy is
articulated naturally in the \xeft\ framework as opposed to the more
\emph{ad hoc} parametrizations using phenomenological potentials such
as IL-IX~\cite{illinois_nine}, which was designed for use
with AV18. The SRG provides for the consistent treatment of many-body
forces and is seen to work with this natural hierarchy as input and
complements \xefts\ as a general renormalization technique. 


In addition, the formal basis of the SRG allows a term-by-term
application to analyze its transformation of the initial \xeft\
hierarchy. This program was begun in this thesis by exploring how the
SRG preserves the hierarchy of input interaction terms in a
one-dimensional model and can be easily extended to the
three-dimensional case within the NCSM framework. A topic of future
investigations involves an analytic implementation  of the SRG to
study its preservation of the initial hierarchy and how it adjusts
LEC's in a realistic \xeft\ setting. Ultimately, this should provide a
deeper understanding of renormalization in the broader EFT hierarchy
and may lead to predictions about the quantitative importance and
relation of particular mechanisms and processes within nuclear
systems.


While this work focuses on the few-body \abinit program of nuclear
structure physics, the SRG is a general $A$-body technique that is
also being applied to infinite matter and density functional
calculations. These methods are currently under heavy development and
will also depend on the ability to soften the initial Hamiltonian
input. The SRG is a prime mover in bringing the few-body and many-body
regimes into quantitative contact with each other. Specifically, the
SRG will expand the useful range of wavefunction based methods like
the NCSM used in this thesis. This will be an important check on the
validity of the DFT calculations that develop.


\section{Thesis Organization}	  
	
The following topics in this thesis are original contributions to the
field: 

\be

\item A quantitative study of decoupling afforded by the SRG in
momentum representation NN forces and the calculational benefits it
provides.

\item The first studies using alternative forms of the SRG to test
different types of renormalization for nuclear interactions.

\item Development of a one-dimensional Jacobi coordinate harmonic
oscillator shell model as a laboratory to explore the behavior of the
SRG as it applies to the evolution of many-body forces. Within this
model we studied:

\bi

\item Vacuum expectation value decomposition and analysis of the
many-body force evolution.

\item Basis dependence issues and other SRG generators in the
oscillator representation.

\item Additional procedures such as fitting three-body forces to
evolved two-body interactions and evolving individual operators in
few-body systems.

\ei

\item First evolution of 3NFs for $^3$H and $^4$He using an
established No-Core Shell Model code, based on insight and experience
gained from the one-dimensional model.

\ee

Decoupling between high- and low-energy states is the feature of SRG
evolution that leads to increased convergence with basis size in
calculations of nuclear observables. While first demonstrated for the
NN interaction in Ref.~\cite{Bogner:2007srg}, in this work we achieved
a quantitative understanding of its roots and its behavior in the NN
partial-wave momentum representation. Residual coupling in the
interaction above momenta of interest is perturbatively small and
extends universally to many-body calculations like NCSM binding
energies. The form of the SRG's transformations can be tailored by
prudent choice of the operators entering the equations. We show that
such choices have little effect on the qualitative behavior of
decoupling in NN interactions.

The formal equations of the SRG indicate that decoupling should occur
for many-body forces in a similar manner as in the NN case. While the
SRG can be applied in any basis, momentum-basis calculations of
three-body potentials are less straightforward because of spectator
nucleons (those not participating in an interaction). Instead, we
choose to work in a discrete basis of harmonic oscillator
wavefunctions inspired by the no-core shell model. We first built a
model in one dimension (1D) for the purpose of rapid development and
intuition in SRG calculations involving many-body forces. Many
different types of calculations have been performed in one dimension
to gain intuition for their behavior in three-dimensional calculations
and provide a sanity check when working with the more complex
three-dimensional codes. It has also provided a simpler environment in
which to explore issues of basis dependence in calculations in general
and SRG calculations in particular.

Finally, we apply the insights gained in 1D to the full realistic
three-dimensional calculation of the NCSM. Here the 1D model has
proved to be highly predictive of the behavior of decoupling in
realistic calculations. The SRG improves convergence in calculations
of $^3$H and $^4$He, and the input hierarchy of many-body forces is
maintained as expected from the one-dimensional studies. These evolved
many-body interactions, which have much improved convergence
properties and well controlled errors, will be highly sought inputs to
various \abinit many-body calculations from across the UNEDF
collaboration. 


The general layout of this thesis is as follows: Chapter
\ref{chapt:srg} gives a brief introduction to the formalism employed
in making SRG calculations. In chapter \ref{chapt:decoupling}, a
quantitative study is made of the decoupling between high- and
low-energy states and its universal nature. With an eye towards
realistic many-body calculations, chapter \ref{chapt:OneD} develops a
one-dimensional analog to the No-Core Shell Model (NCSM) to explore
the behavior of SRG evolution in an $A$-body space. Many different
types of calculations are tried here to gain intuition for the real
three-dimensional case. Chapter \ref{chapt:ncsm} applies the insight
gained from the one-dimensional model to realistic few-body
calculations using the Jacobi coordinate NCSM. Finally, we conclude in
chapter \ref{chapt:conclusion} and discuss possible future
investigations. Several appendices explore technical details of the
basis and potentials involved in this work.

\chapter{Similarity Renormalization Group}
\label{chapt:srg}


The Similarity Renormalization Group
(SRG)~\cite{Glazek:1993rc,Wegner:1994,Kehrein:2006} provides a
compelling method for evolving internucleon forces to softer 
forms~\cite{Bogner:2006srg,Bogner:2007srg}. While observables are
unchanged by the SRG's unitary transformations, the contributions from
high-momentum intermediate states to low-energy observables is
modified by the running transformation. These transformations soften
initial interactions and can dramatically reduce the computational
requirements of low-energy many-body 
calculations~\cite{Bogner:2006srg,Bogner:2007srg,Bogner:2007rx}. At
the same time, the SRG induces many-body forces in response to its
transformation of the high-energy states in the Hamiltonian. In this
chapter I present the formalism used for analyzing SRG evolution and
the subsequently induced many-body forces.

\section{Derivation}

We apply the similarity renormalization group 
(SRG) transformations to inter-nucleon interactions based on the flow equation formalism 
of Wegner~\cite{Wegner:1994}. The evolution or flow of the Hamiltonian with a 
parameter $s$ is a series of unitary transformations
\beqn
	H_s = U_s H U^{\dagger}_s \equiv \Trel + V_s\;,
	\label{eq:srg1}
\eeqn
where $\Trel$ is the relative kinetic energy, $H = \Trel + V$ is the
initial Hamiltonian in the center-of-mass system, and $U_s$ is the
unitary transformation. Equation \eqref{eq:srg1} defines the evolved
potential $V_s$, with $\Trel$ here defined to be independent of $s$ in
all spaces\footnote{The center of mass kinetic energy, $T_{\rm cm}$,
will not contribute because we are working in a translationally
invariant system and it will commute with the other operators in
Eq.~\eqref{eq:srg1}}. Then $\frac{dH_s}{ds} = \frac{dV_s}{ds}$ evolves
according to
\bea
	\D\frac{dV_s}{ds}
    &=&\frac{dU_s}{ds} H U^{\dagger}_s + U_s H \frac{dU^{\dagger}}{ds} \nonumber \\
    &=&\frac{dU_s}{ds}U^{\dagger}_s U_s H U^{\dagger}_s + U_s H
    U^{\dagger}_s U_s \frac{dU^{\dagger}}{ds} \nonumber \\ 
    &=&\frac{dU_s}{ds}U^{\dagger}_s H_s + H_s U_s \frac{dU^{\dagger}}{ds}\;,
    \label{eq:srg2}
\eea		
which motivates us to focus on $\eta_s \equiv \D\frac{dU_s}{ds}
U^{\dagger}_s \Rightarrow \eta^{\dagger}_s \equiv U_s\D\frac{dU^{\dagger}_s}{ds}$. Therefore,
\beqn
	\D\frac{dV_s}{ds} = \eta_s H_s + H_s \eta^{\dagger}_s\;.
	\label{eq:srg3}
\eeqn
Using $U_s U^{\dagger}_s = 1$, we can establish the following
\beqn
	\D\frac{d}{ds}(U_s U^{\dagger}_s) = 0 
    \Rightarrow \left(\frac{d}{ds} U_s \right)  
    U^{\dagger}_s + U_s \left(\frac{d}{ds}U^{\dagger}_s\right) = 0
	\label{eq:srg4}
\eeqn
\beqn
	\Rightarrow \eta_s + \eta^{\dagger}_s = 0\;.
	\label{eq:srg5}
\eeqn
Therefore Eq.~\eqref{eq:srg3} can be written as
\beqn
	\D\frac{dV_s}{ds} = \eta_s H_s - H_s \eta_s = \left[\eta_s, H_s \right]\;.
	\label{eq:srg6}
\eeqn
The choice of $\eta_s$, as long as it is antihermitian, is
mathematically free and specifies the transformation. A natural choice
would involve parts of the Hamiltonian itself which, along with a
commutator structure, would satisfy the antihermiticity requirement.
Choosing one commutator argument to be the evolving Hamiltonian,
$H_s$, and leaving the other choice open, we can write
\beqn
	\eta_s = \left[G_s, H_s \right] \;.
	\label{eq:srg7}
\eeqn
The flow Eq.~\eqref{eq:srg6} now becomes
\beqn
	\D\frac{dV_s}{ds} = \left[\left[G_s, H_s \right], H_s \right] \;,
	\label{eq:srg8}
\eeqn		
where $G_s$ will be referred to as the generator, following
conventional usage, though
technically the generator of the group is $\eta_s$.  Various
possibilities for $G_s$ have been
proposed~\cite{Kehrein:2006,Bogner:2006srg,Anderson:2008mu},
though the most common in this thesis is $G_s = \Trel$, for which
the flow equation simplifies to
\beqn
	\D\frac{dV_s}{ds} = \left[\left[\Trel, V_s \right], H_s \right] \;.
	\label{eq:srg9}
\eeqn
Other choices that have been investigated to varying degrees are
listed in Table~\ref{tab:G_choices}.

\bt
\caption{Various choices for $G_s$ considered to date.}
\begin{tabular}{l|l}
 Name & Description \\
\hline
$\Trel$      & Relative kinetic energy between nucleons~\cite{Bogner:2006srg}  \\
$H_{\rm D}$  & Running diagonal part of the Hamiltonian advocated \\
                & by Wegner: $\Trel + V_{s,{\rm diag}}$ \cite{Wegner:1994} \\
$H_{\rm D,fixed}$  & Fixed, initial diagonal of the Hamiltonian:
$\Trel + V_{s=0,{\rm diag}}$ \\
$H_{\rm BD,sharp}$ & The running block diagonal part of the Hamiltonian. \\
                &  $P$ and $Q$ spaces defined by a sharp cutoff at $\Lambda$~\cite{Anderson:2008mu} \\
$H_{\rm BD,smooth}$ & Block diagonal as above but with a smooth
                    regulator of sharpness $n$ \\
$H_{\rm weird}$ & A strange choice with three different block diagonal regions \\
                & defined by two different cutoff $\Lambda$s~\cite{Anderson:2008mu} \\
$\Hho$         & The harmonic oscillator Hamiltonian in an
                    $A$-body space, which is \\ 
                    & diagonal in the HO basis \\
$\Trel + \alpha r^2$ & A hybrid $\Hho$ type choice for use in the oscillator
basis~\cite{Anderson:2009b} \\
$\Trel + V^{(2)}$ & The two-body Hamiltonian, suggested as a way to renormalize \\
                & only three-body forces
\end{tabular}
\label{tab:G_choices}
\et

Equations \eqref{eq:srg8} or \eqref{eq:srg9} are operator equations,
independent of any basis. They can be applied numerically by
projecting onto any convenient basis. For two-body systems, a
partial-wave momentum basis with states labeled by the initial and
final relative momenta, $k$ and $\kprime$, is convenient, accurate,
and straightforward. In higher-body systems we will resort to a
discrete basis\footnote{Work is underway to develop a three-body
momentum representation code which incorporates SRG
calculations~\cite{lplatter_pc}.} in which all the states can be
enumerated up to some basis limiting parameter. The bulk of this
thesis will employ a basis of harmonic oscillator functions known as
the No-Core Shell Model
(NCSM)~\cite{ncsm_orig1,ncsm_orig2,ncsm_orig3}. The NCSM usually uses
a renormalization procedure inspired by a Lee-Suzuki type
transformation~\cite{NCSM1a,NCSM1b,NCSM1c}. Calculations in this
thesis will not include any such transformations; the SRG is applied
to the ``bare" or initial potentials as given by the various
models/\xefts\ available. The SRG provides renormalization in addition
to that intrinsic to the formulation of input potentials (especially
those from \xeft).

Note that the dimension of $s$ is 1/energy$^2$. We can define a
momentum variable  $\lambda = 1/s^{1/4}$, with units of
fm$^{-1}$, that measures the spread of the off-diagonal strength
in units relevant to the basis. Using $\lambda$ as the flow
parameter Eq.~\eqref{eq:srg9} becomes
\beqn
\D\frac{dV_\lambda}{d\lambda} =
\frac{-4}{\lambda^5}\left[\left[\Trel,V_\lambda \right],H_\lambda
\right]\;,
\eeqn
so that the flow goes from $\lambda=\infty$ toward zero instead of
starting at $s=0$ and going toward infinity. Most of our formalism is
presented using $s$ for simplicity (since the flow is linear in
$s$) but most calculations are done using $\lambda$. Chapter
\ref{chapt:decoupling} will demonstrate how $\lambda$ is a
physically relevant measure of the extent of evolution
specifically as it applies to the decoupling of high- and
low-energy degrees of freedom. Matrix elements connecting states
with (kinetic) energies differing by more than $\lambda^2$ are
largely suppressed and therefore decoupled.

The earliest applications to nuclear physics have been in a
partial-wave momentum basis using $G_s = \Trel$~\cite{Bogner:2006srg}.
It is a straightforward matter to project Eq.~\eqref{eq:srg9} into
this basis. After some algebra we can show that, for each
partial-wave, Eq.~\eqref{eq:srg9} can be specified in the momentum
basis as
\beqn
	\frac{dV_s(k', k)}{ds} = -(k^2 - {k'}^2)^2 V_s(k', k) 
    + \frac{2}{\pi} \int_0^{\infty} q^2 dq\  
    (k^2 + {k'}^2 - 2q^2) V_s(k',q) V_s(q,k) \;.
	\label{eq:srg10}
\eeqn
In fact, it is simpler to implement the
commutator form of the flow equations on platforms like MATLAB or
Fortran which can easily handle matrix operations and coupled
differential equations. Once the Hamiltonian has been projected onto a
chosen mesh (usually Gaussian quadrature), the matrix multiplications
follow directly.

\begin{figure*}
\begin{center}
\strip{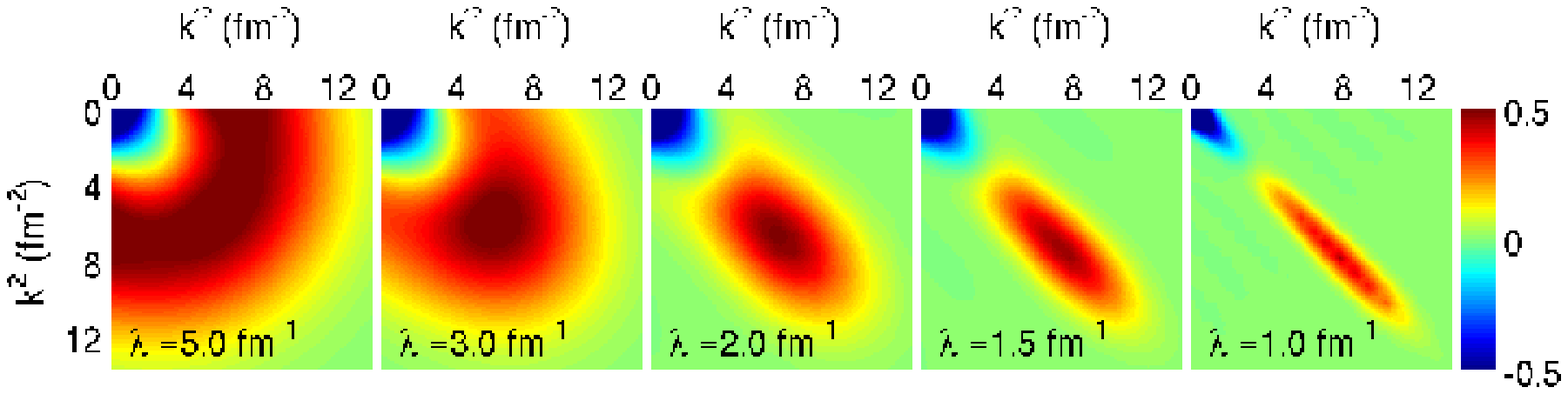} 

\strip{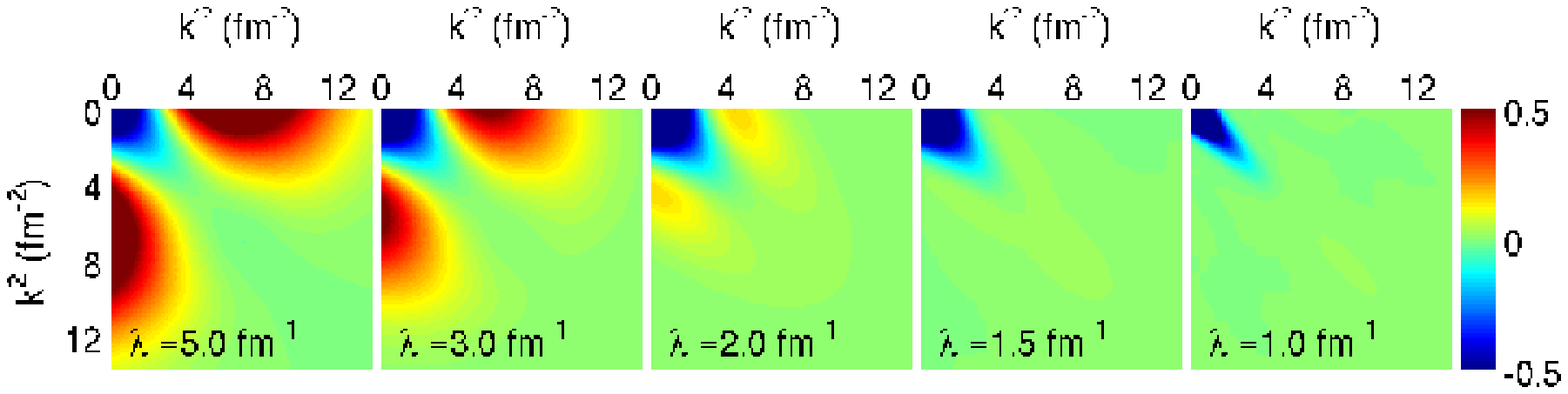}

\strip{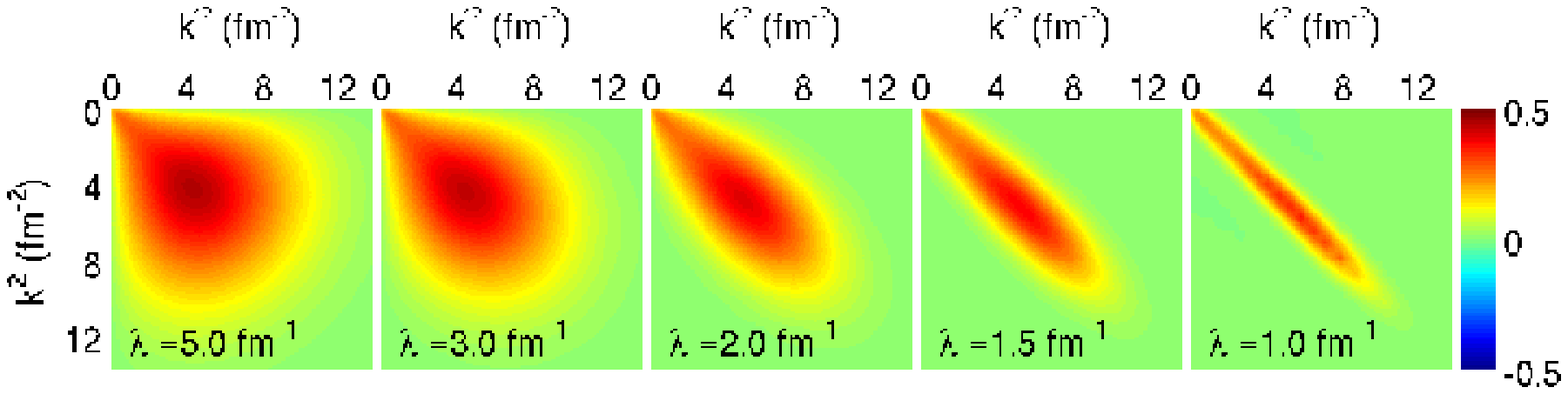} 

\strip{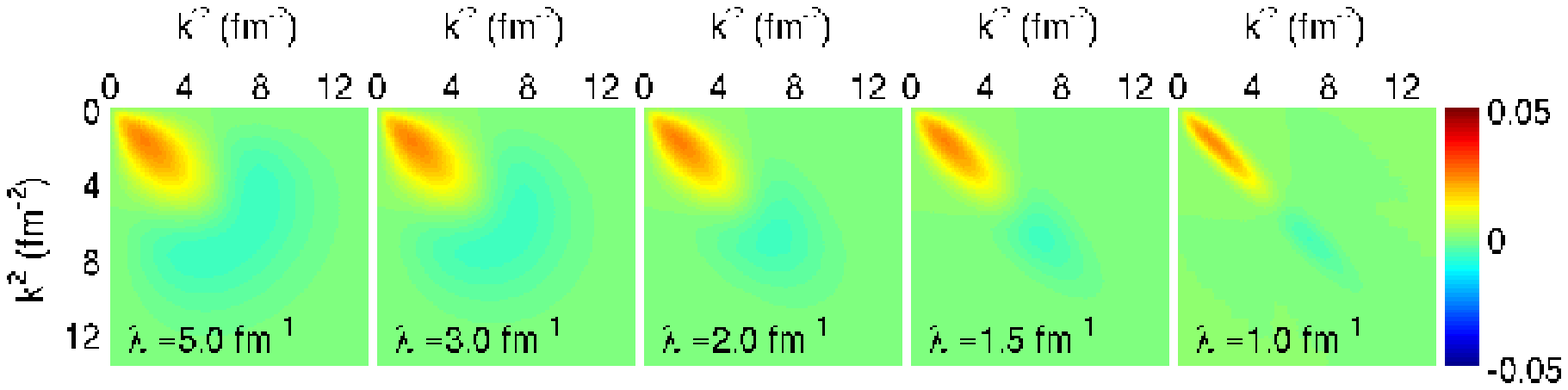} 
\end{center}
\captionspace{Film strips of evolution for representative partial waves in
the N$^3$LO (500 MeV) interaction of Ref.~\cite{N3LO} in the momentum
representation. Partial waves shown are $^1\!S_0$,
$^3\!S_1$, $^1\!P_1$, and $^3\!F_3$.}
\label{fig:ps_film_strips_500MeV}
\end{figure*}

\begin{figure*}
\begin{center}
\strip{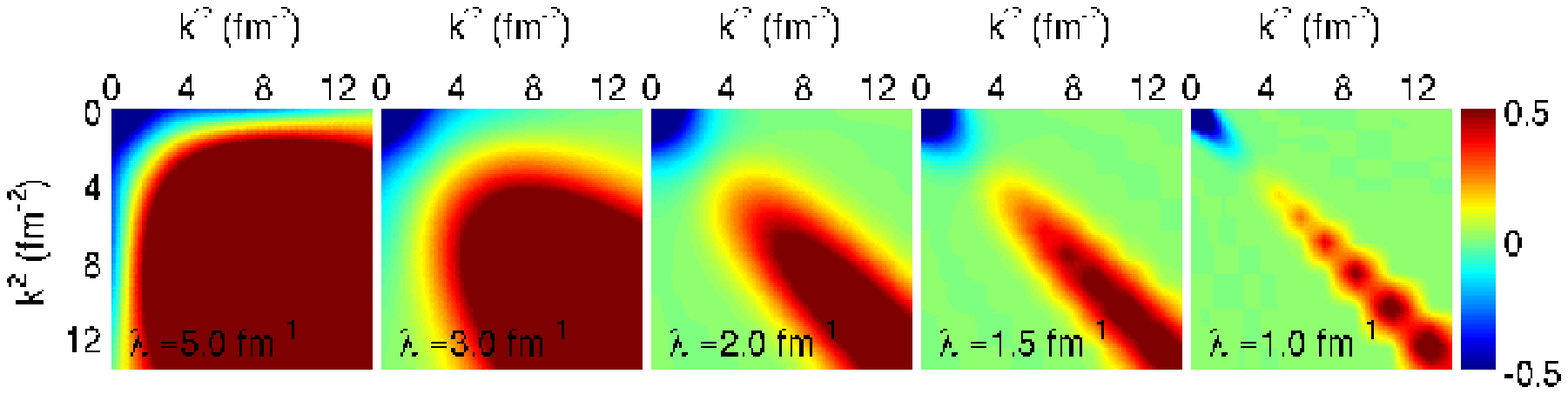} 

\strip{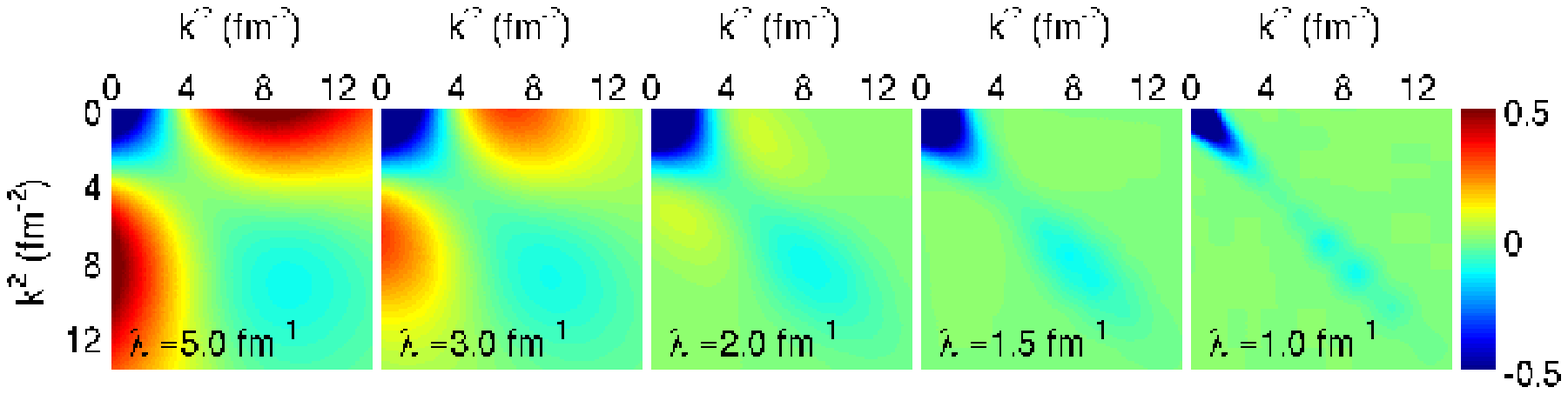}

\strip{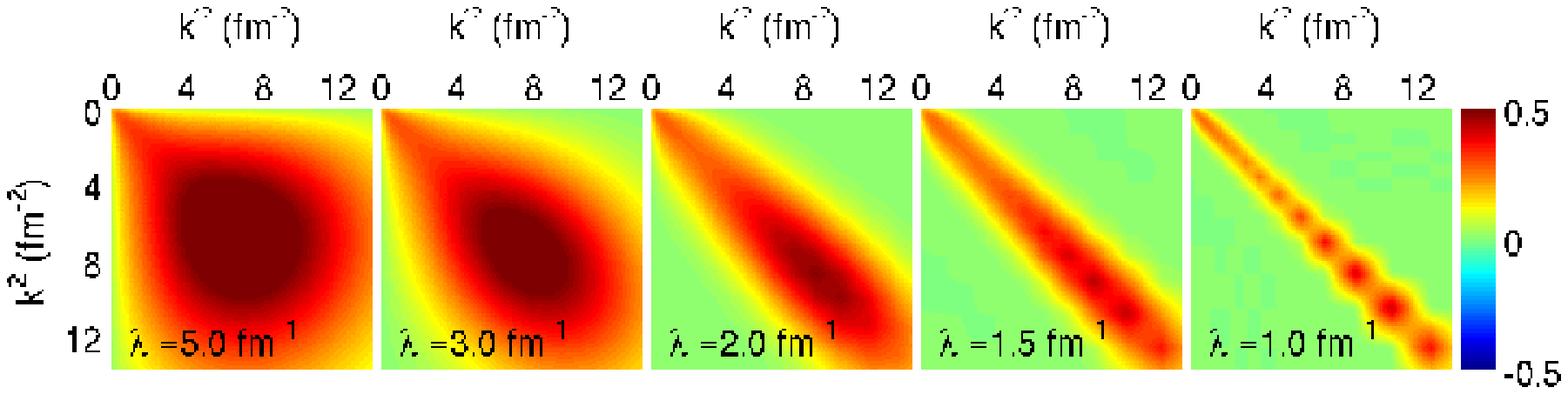} 

\strip{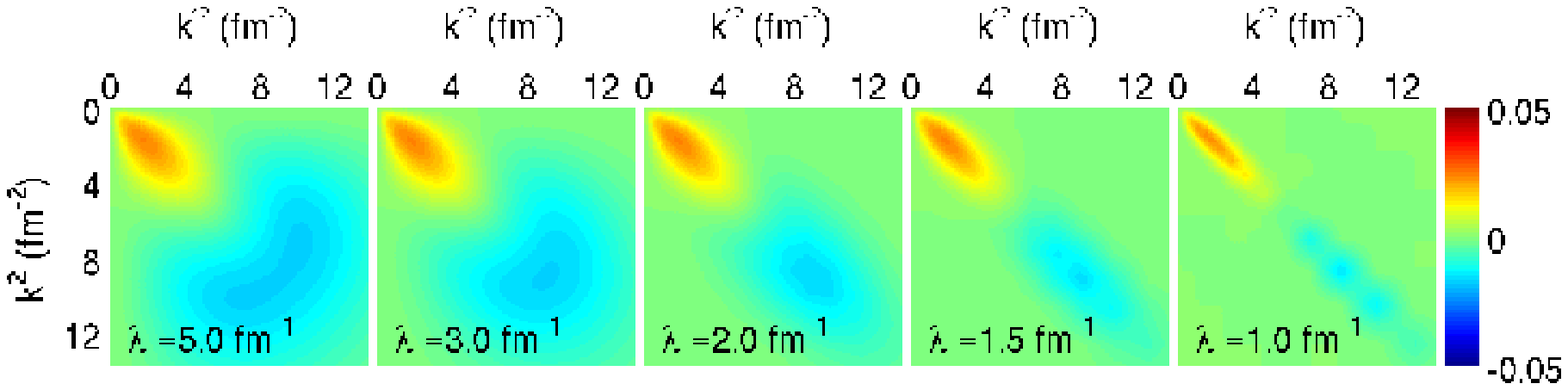} 
\end{center}
\captionspace{Film strips of evolution for representative partial waves in
the N$^3$LO (600 MeV) interaction of Ref.~\cite{N3LO} in the momentum
representation. Partial waves shown are $^1\!S_0$,
$^3\!S_1$, $^1\!P_1$, and $^3\!F_3$.}
\label{fig:ps_film_strips_600MeV}
\end{figure*}

\begin{figure*}
\begin{center}
\strip{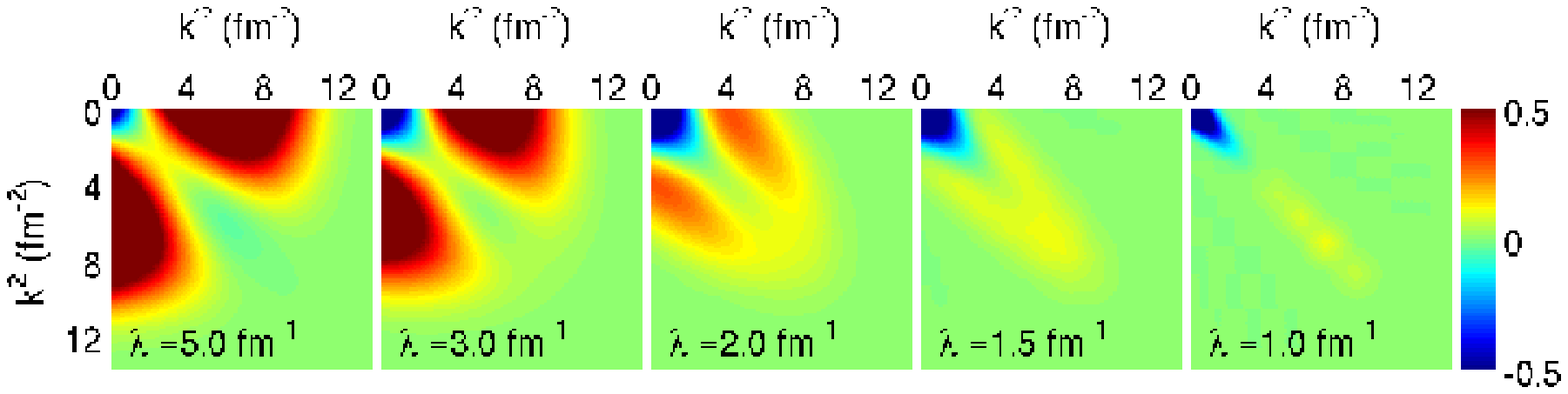} 

\strip{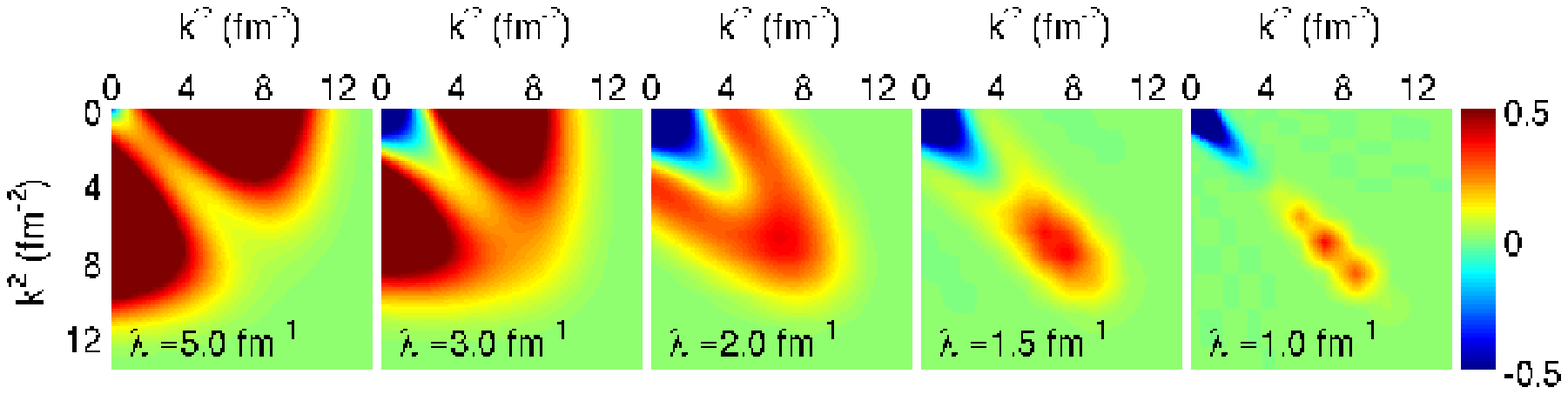}

\strip{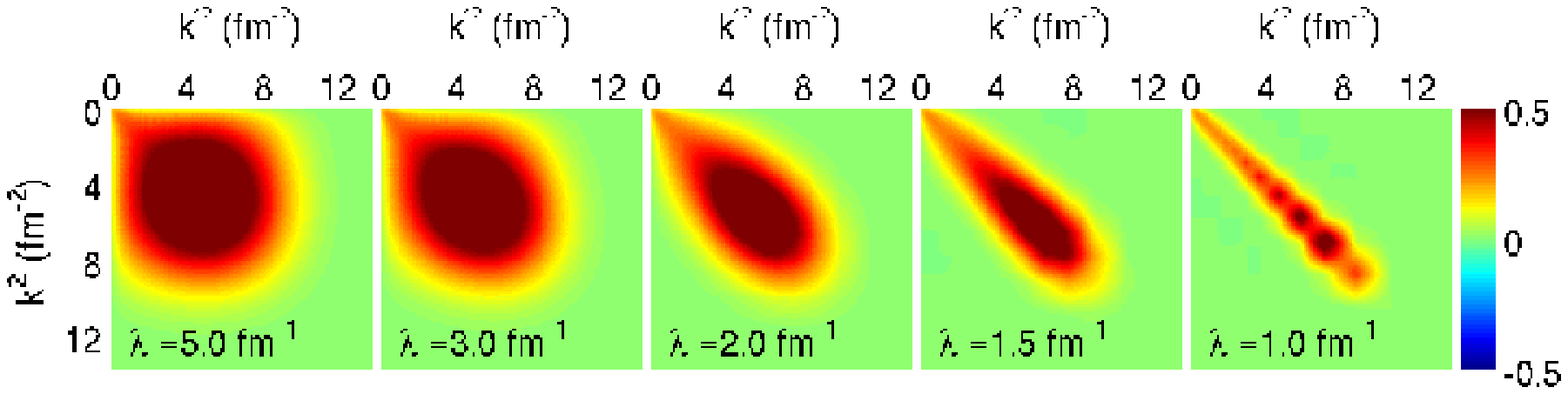} 

\strip{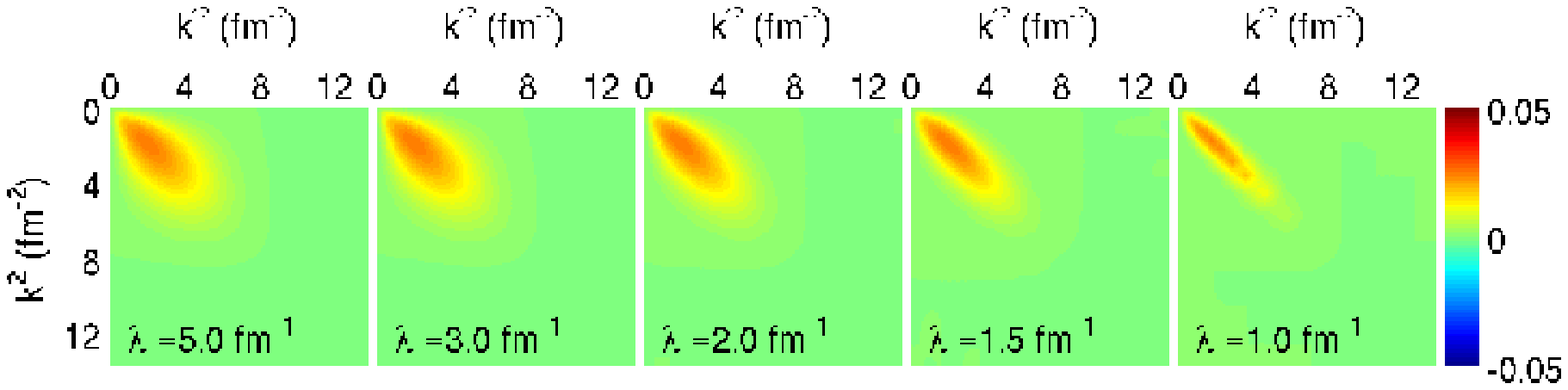} 
\end{center}
\captionspace{Film strips of evolution for representative partial waves in
the N$^3$LO (550/600 MeV) interaction of Ref.~\cite{N3LOEGM} in the
momentum representation. Partial waves shown are $^1\!S_0$, $^3\!S_1$,
$^1\!P_1$, and $^3\!F_3$.}
\label{fig:ps_film_strips_EGM}
\end{figure*}

\begin{figure*}
\begin{center}
\strip{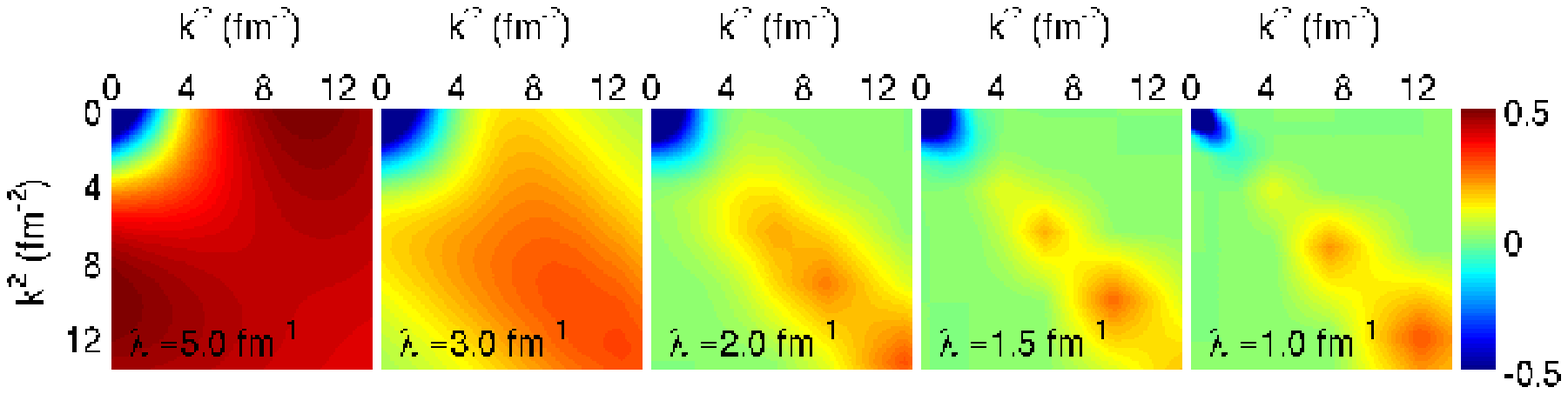} 

\strip{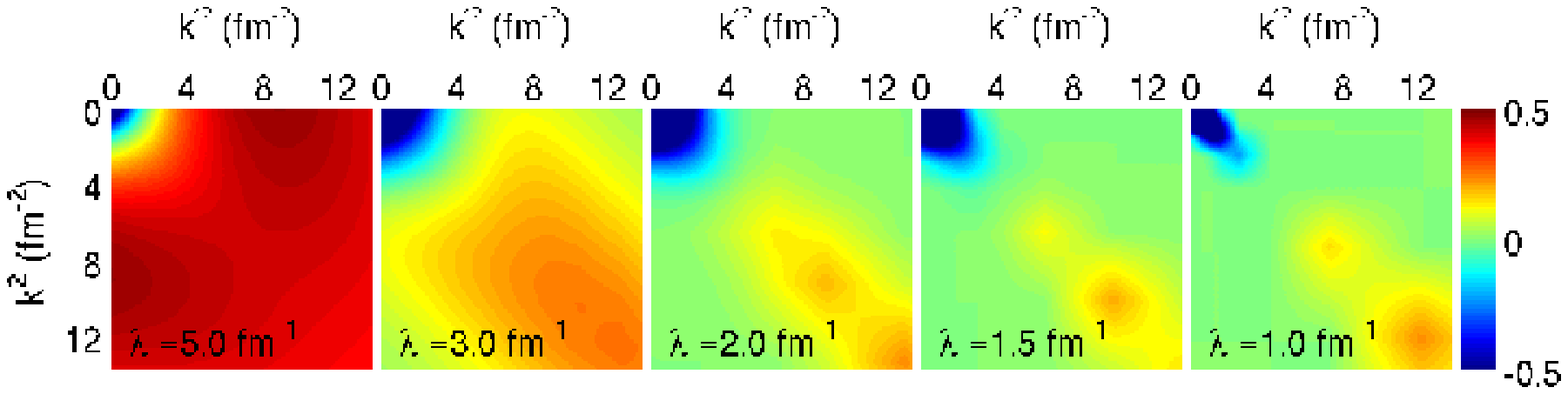}

\strip{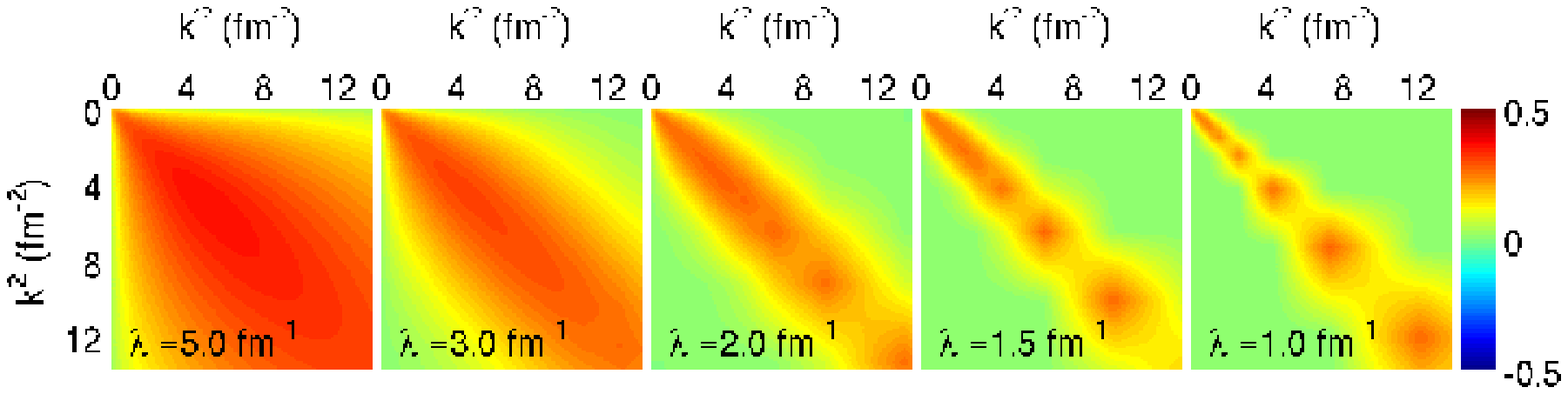} 

\strip{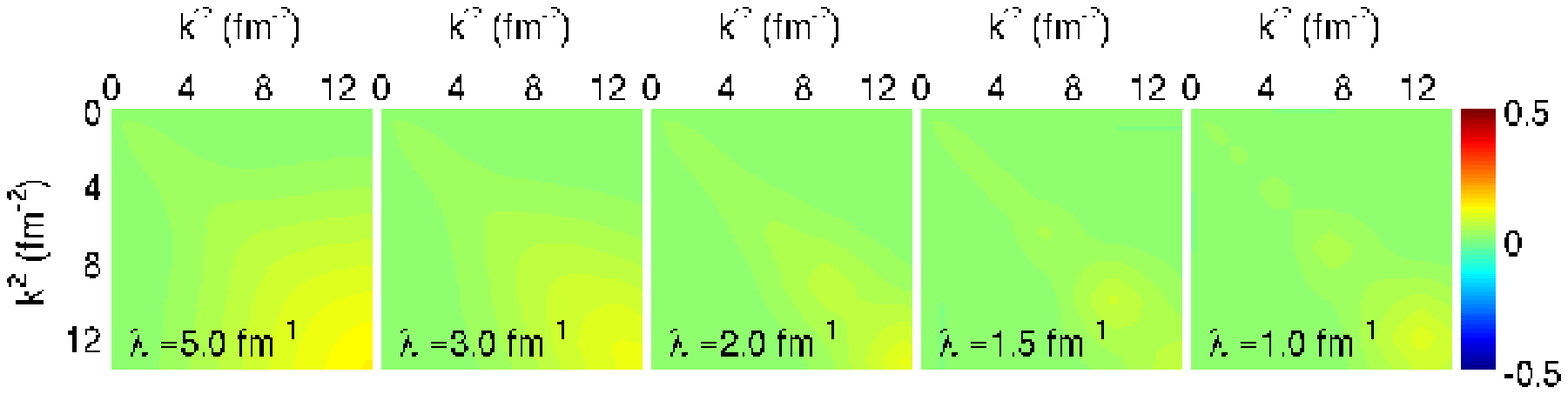} 
\end{center}
\captionspace{Film strips of evolution for representative partial waves in
the AV18 interaction of Ref.~\cite{argonne_potential} in the momentum
representation. Partial waves shown are $^1\!S_0$, $^3\!S_1$,
$^1\!P_1$, and $^3\!F_3$.}
\label{fig:ps_film_strips_argonne}
\end{figure*}


Figures
\ref{fig:ps_film_strips_500MeV}--\ref{fig:ps_film_strips_argonne}
document the evolution of various partial waves from several different
NN interactions in the momentum representation. Note the blue areas,
which indicate attraction and contribute to binding phenomena seen at
low energies, and the red areas which indicate repulsion and the hard
core probed by large momenta. However, the off-diagonal red matrix
elements which are coupling low- and high-momentum states are also
contributing to low-energy observables. The major feature to notice
looking across the film strips from, left to right, is that the
potential is driven to a band diagonal form with a width proportional
to the square of the momentum parameter $\lambda$. This behavior is a
major subject of chapter~\ref{chapt:decoupling}. In all of the cases,
the SRG is successful at transforming the potential to this
band-diagonal form, regardless of the specific input physics. Rather
the form of the SRG-evolved potentials is determined by the choice of
$G_s$, as shown formally in section~\ref{sec:decoupling_mechanics}. The
evolved potential approaches a universal soft form in a given partial
wave, regardless of the initial Hamiltonian. This suggests a natural
underlying hierarchy of renormalization particularly suitable to the
\xeft\ program, though it is equally successful at softening relatively
hard potentials such as AV18 (see
Fig.~\ref{fig:ps_film_strips_argonne})\footnote{Note however that SRG
evolution does in general introduce nonlocality to the initially local
AV18, spoiling its use in the GFMC program.}.


Once we have obtained wavefunctions of the evolved Hamiltonian through
the flow equations of Eq.~\eqref{eq:srg8}, we can directly construct
the unitary transformation, $U_s$, at each $s$
explicitly~\cite{Bogner:2007srg}. Given a complete set of eigenstates
of $H_s$, $\{|\psi_{\alpha}(s)\ra\}$, $U_s$ can be written as,
\beqn
U_s = \sum_{\alpha}|\psi_{\alpha}(s)\ra \la \psi_{\alpha}(0)|\;.
\eeqn
Because of the discretization, the summation over $\alpha$ is finite
and can be achieved through simple matrix multiplications using
platforms such as Fortran and MATLAB. With $U_s$, the wavefunctions
can be evolved independently,
\bea
U_s|\psi_{\beta}(0)\ra &=& \sum_{\alpha}|\psi_{\alpha}(s)\ra 
\la \psi_{\alpha}(0)|\psi_{\beta}(0)\ra \nonumber \\
&=& \sum_{\alpha}|\psi_{\alpha}(s)\ra \delta_{\alpha,\beta} \nonumber \\
&=& |\psi_{\beta}(s)\ra\;.
\eea
This form is particularly convenient for evolving individual operators
in parallel with the chosen initial interaction. In particular, any
operator can be evolved directly as 
\beqn
O_s = U_s O U_s^\dagger\;,
\label{eq:operator_transform}
\eeqn
as opposed to solving another flow equation,
\beqn
\D\frac{dO_s}{ds} = \left[\left[G_s, H_s \right], O_s \right]\;,
\label{eq:operator_flow}
\eeqn
just as for the Hamiltonian. These are identical strategies, but for
some operators evolving the potential and using $U_s$ as in
Eq.~\eqref{eq:operator_transform} is more numerically
robust~\cite{Anderson:2009a}. Ultimately, having the explicit unitary
transformation matrix facilitates rapid calculation of many different
operators with minimal coding.


To carry out the SRG evolution, we use a built-in MATLAB differential
equation solver, such as the MATLAB function {\tt ode23}~\cite{matlab}, 
which is an implementation of a Runge-Kutta
differential equation algorithm~\cite{rk_basic}.  We studied the
running time to evolve the various SRG schemes (i.e., the choice of
$G_s$) by plotting the time to run versus the evolution parameter,
$s$. We find a linear relationship to $s$, indicating no stiffness, in
every combination of potential and SRG scheme used to date. Details
regarding these timing studies and other computational scaling issues
are reviewed in Appendix \ref{chapt:app_scaling}.


\section{Second Quantization}
\label{sec:second_quantization}

To see how the two-, three-, and higher-body potentials are
identified, it is useful to decompose $H_s$ in second-quantized
form~\cite{fetter_walecka} which was first used for this purpose in
Ref.~\cite{Bogner:2007qb}. We can write an $A$-body Hamiltonian in
general as:
\beqn
  H = \sum_{ij} T_{ij} \adag_i a_j  + 
  \frac{1}{2!^2} \sum_{ijkl} V_{ijkl}^{(2)} \adag_i\adag_j a_l a_k
   + \frac{1}{3!^2} \sum_{ijklmn}  V_{ijklmn}^{(3)}
   \adag_i\adag_j\adag_k a_n a_m a_l + \cdots \;,
\label{eq:Ham_2ndquant}
\eeqn
where $\adag_i$ and $a_i$ represent creation and destruction operators
with respect to the vacuum in some single-particle basis. The
quantities $T_{ij}$, $V_{ijkl}^{(2)}$, and $V_{ijklmn}^{(3)}$
represent matrix elements of their respective operators.

As a straightforward way to see how the SRG induces many-body forces
in an $A$ body space, one can evaluate the flow equations using the
Hamiltonian in the form of Eq.~\eqref{eq:Ham_2ndquant}:
\bea
\frac{dH_s}{ds} = \left[ \left[\sum_{ij} T_{ij}\adag_i a_j, 
\sum_{ijkl} V_{ijkl,s}^{(2)} \adag_i\adag_j a_l a_k + \cdots \right],
 \sum_{ij} T_{ij}\adag_i a_j + \sum_{ijkl} V_{ijkl,s}^{(2)} 
 \adag_i\adag_j a_l a_k  + \cdots\right]\;.
\label{eq:flow_2ndquant}
\eea

For example consider the evolution of the potential, $V_s$, with the
Hamiltonian truncated to just the two-body interactions: 
\bea
\frac{dV_s}{ds} &=& \left[ \left[\sum_{ij} T_{ij}\adag_i a_j, 
\sum_{ijkl} V_{klmn,s}^{(2)} \adag_k\adag_l a_n a_m\right],
 \sum_{ijkl} V_{opqr,s}^{(2)} \adag_o\adag_p a_r a_q\right] \nonumber \\
 &=& T_{ij}V_{klmn}V_{opqr} \times\left[ (\adag_ia_j\adag_k\adag_la_na_m - \adag_k\adag_la_na_m\adag_ia_j),
 \adag_o\adag_pa_ra_q \right] \nonumber \\
&=& T_{ij}V_{klmn}V_{opqr} \times\left( \adag_ia_j\adag_k\adag_la_na_m\adag_o\adag_pa_ra_q 
- \adag_k\adag_la_na_m\adag_ia_j\adag_o\adag_pa_ra_q \right. \nonumber \\
&& \left. - \adag_o\adag_pa_ra_q\adag_ia_j\adag_k\adag_la_na_m 
+ \adag_o\adag_pa_ra_q\adag_k\adag_la_na_m\adag_ia_j \right)\;,
\label{eq:dVds_2ndquant}
\eea
resulting in the large strings of $a$'s and $\adag$'s.

These terms can be brought into normal order with respect to the
vacuum to simplify them. For this we will employ Wick's
theorem~\cite{wick_basic}, which states (using $A_i$ to represent
$a_i$ or $a^{\dagger}_i$)
\bea
A_iA_jA_k\cdots A_m &=& N(A_iA_jA_kA_l\cdots A_m) \nonumber \\
&& + N\left( (\contraction{}{A}{_i}{A} A_iA_jA_kA_l\cdots A_m) 
+ \mbox{all single contractions}\right) \nonumber \\
&& + N\left( (\contraction{}{A}{_i}{A} 
\contraction{A_iA_j}{A}{_k}{A} A_iA_jA_kA_l\cdots A_m) 
+ \mbox{all double contractions}\right) \nonumber \\
&& \vdots \nonumber \\
&& + N\left( \mbox{all full contractions}\right) \;,
\label{eq:wicks_theorem}
\eea
and its corollary
\bea
N(A_iA_j)N(A_kA_l) &=& N(A_iA_jA_kA_l) 
+ N(\contraction{}{A}{_iA_j}{A} A_iA_jA_kA_l) 
+ N(\contraction{}{A}{_iA_jA_k}{A} A_iA_jA_kA_l) \nonumber \\
&& + N(\contraction{A_i}{A}{_j}{A} A_iA_jA_kA_l) 
+ N(\contraction{A_i}{A}{_jA_k}{A} A_iA_jA_kA_l) 
+ N(\contraction[2ex]{}{A}{_iA_jA_k}{A}
\contraction{A_i}{A}{_j}{A} A_iA_jA_kA_l) \nonumber \\
&& + N(\contraction[2ex]{}{A}{_iA_j}{A} 
\contraction{A_i}{A}{_jA_k}{A} A_iA_jA_kA_l) \;,
\label{eq:wicks_corollary}
\eea
which helps to simplify products of multiple, individually
normal-ordered operator strings. There is one non-vanishing type of
contraction in the vacuum
\beqn
\contraction{}{a}{_i}{a} a_i\adag_j = \delta_{ij}\;,
\label{eq:contraction_def}
\eeqn
and the other three, $\contraction{}{a}{^{\dagger}_i}{a} \adag_i
a_j$,  $\contraction{}{a}{_i}{a} a_i a_j $, and 
$\contraction{}{a}{^{\dagger}_i}{a} \adag_i\adag_j$, give zero.

\begin{figure}[bt]
\begin{center}
\strip{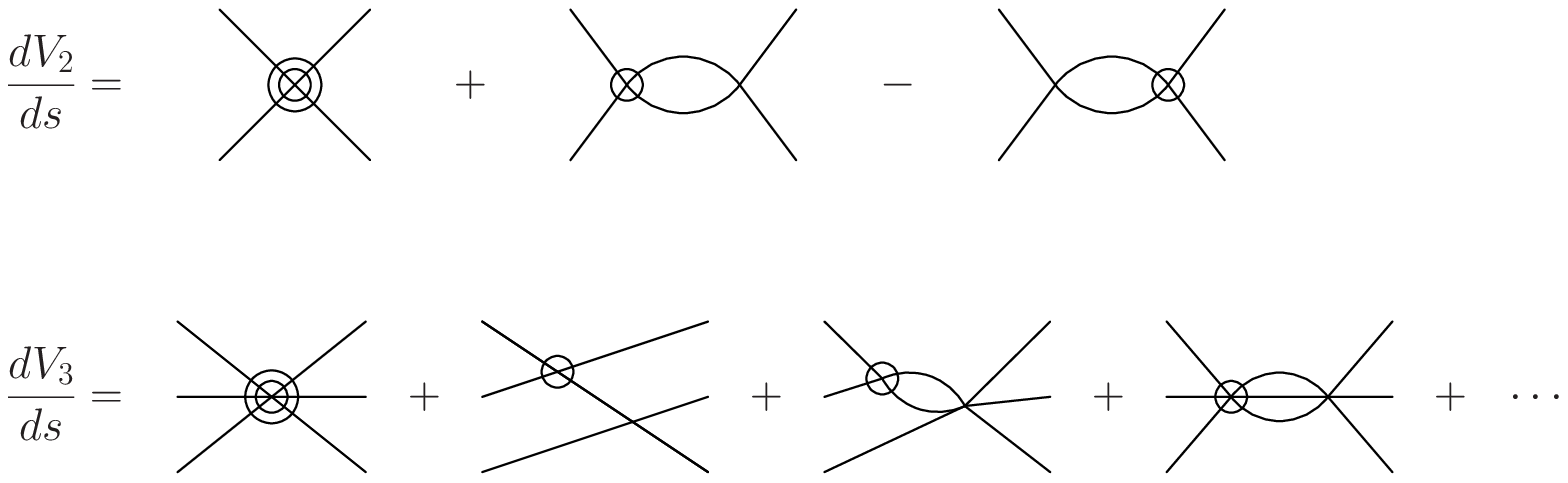}
\end{center}
\captionspace{A diagrammatic decomposition of the SRG induced
forces~\cite{Bogner:2007qb}. A circle at a vertex denotes a commutator
with $T_{\rm rel}$.}
\label{fig:srg_diagrams1}
\end{figure}

Returning to the result of Eq.~\eqref{eq:dVds_2ndquant}, we can now
see what many-body contributions it contains. The normal order of all
operators in each term is the same for all terms and thus cancel, so
no-five body operator survives at this order. All terms resulting from
a single contraction also cancel each other because of the double
commutator structure. For example N($\contraction{\adag_i}{a}{_j}{a}
\adag_ia_j\adag_k\adag_la_na_m\adag_o\adag_pa_ra_q$)  and N($-
\contraction{\adag_o\adag_pa_ra_q\adag_i}{a}{_j}{a}
\adag_o\adag_pa_ra_q\adag_ia_j\adag_k\adag_la_na_m$) are equal and
opposite. At the level of double contractions the commutator cannot
cancel all the terms and three-body terms survive. For example the
first term gives:
\bea
\adag_ia_j\adag_k\adag_la_na_m\adag_o\adag_pa_ra_q &=& 
   T_{ij}V_{klmn}V_{opqr} \times (\adag_i\adag_l\adag_pa_na_ra_q\delta_{jk}\delta_{mo}
  +\adag_i\adag_l\adag_pa_ma_ra_q\delta_{jk}\delta_{no}\nonumber \\
&&+\adag_i\adag_l\adag_oa_na_ra_q\delta_{jk}\delta_{mp}
  +\adag_i\adag_l\adag_oa_ma_ra_q\delta_{jk}\delta_{np} \nonumber \\
&&+\adag_i\adag_k\adag_pa_na_ra_q\delta_{jl}\delta_{mo}
  +\adag_i\adag_k\adag_pa_ma_ra_q\delta_{jl}\delta_{no}\nonumber \\
&&+\adag_i\adag_k\adag_oa_na_ra_q\delta_{jl}\delta_{mp}
  +\adag_i\adag_k\adag_oa_ma_ra_q\delta_{jl}\delta_{np})\;,
\label{eq:three_body_operators}
\eea
which can be simplified due to the symmetry of the potential matrix
elements. In fact all of these terms are permutations of a single
contribution in a diagrammatic representation shown in
Fig.~\ref{fig:srg_diagrams1}. Equation \eqref{eq:three_body_operators}
lists the explicit legs of the second diagram on the bottom, which is
the lone two-body only contribution to the induced three-body force.
Carrying out the same calculation for all triple contractions will
result in enumerating the legs associated with the two diagrams in the
top right of Fig.~\ref{fig:srg_diagrams1} which contribute to two-body
evolution. In an $A$-body system this means all disconnected
combinations of these two-body contributions.

As follows from these normal-ordering calculations, $A$-body forces
cannot be affected by the value of any initial higher-body terms; e.g.,
two-body evolution cannot be influenced by three-body forces. This
can be seen in second quantized form by writing out commutators
involving an $A$-body force. One can quickly see that it is impossible
to contract such terms to have less than $A$ pairs of $\adag$'s and
$a$'s. So for instance, an initial four-body force cannot contribute
to the evolution of two- or three-body forces. This feature is true
for calculations done in the vacuum. However, when the SRG is
implemented in the background of a nuclear medium, the rules for
contractions change. This has profound consequences for the use of the
SRG in nuclear matter calculations. Those calculations are outside the
scope of this thesis, so the ``in-medium SRG" will not
be considered further here.

The correspondence between equations like
Eq.~\eqref{eq:three_body_operators} and Fig.~\ref{fig:srg_diagrams1}
can be written more concisely by expanding the commutators of the SRG
flow equation, Eq.~\eqref{eq:srg8}. For instance the top row of
Fig.~\ref{fig:srg_diagrams1} corresponds to the expansion of
Eq.~\eqref{eq:srg8} in 2-body space:
\bea
\D\frac{d\vtwo}{ds} &=& \left[\left[\Trel, V_s \right], H_s \right] \nonumber \\
&=& \left[(\Trel V_s - V_s \Trel), (\Trel + V_s) \right] \nonumber \\
&=&  \left[\vbt, (\Trel + V_s) \right] \nonumber \\
&=&  \left[\vbt, \Trel \right] + \vbt\vtwo - \vtwo\vbt \nonumber \\
&=&  \overline{\overline{V}}^{(2)}_s + \vbt\vtwo - \vtwo\vbt\;,
\label{eq:diagrams_two_body}
\eea
where $\vbt$ represents a single commutator with $\Trel$ and
$\overline{\overline{V}}^{(2)}_s$ represents a double commutator with
$\Trel$, $[[\Trel,\vtwo],\Trel]$. These correspond to the vertices
with one and two circles in the top row of
Fig.~\ref{fig:srg_diagrams1}, so the last line of
Eq.~\eqref{eq:diagrams_two_body} is represented by those diagrams.
Similarly the bottom row represents the expansion in the 3-body space,
\beqn
\D\frac{d\vthree}{ds} = \overline{\overline{V}}^{(3)}_s + [\vbt,
\vtwo] + [\vbt,\vthree] + [\vbtr, \vtwo] + [\vbtr, \vthree]\;.
\label{eq:diagrams_three_body}
\eeqn
We stress that these diagrams are not Feynman diagrams, but merely a
mnemonic to summarize the contributions to the SRG transformation.
However, each vertex does represent a particular type of interaction
and the diagrammatic treatment can be used to explore the interplay
within the flow equations. An example of this kind of investigation
will be presented in chapter~\ref{chapt:OneD}.

Disconnected diagrams for each $A$-sector are canceled out by the
evolutions defined in lower sectors. Consider the flow equation for
the general three-body Hamiltonian~\cite{Bogner:2006srg} (in a
modified notation),
\beqn
\frac{dV_s}{ds} = \frac{dV_{12}}{ds} + \frac{dV_{13}}{ds} 
+ \frac{dV_{23}}{ds} + \frac{dV_{123}}{ds} = [[\Trel,V_s],H_s]\;,
\label{eq:three_body_flow}
\eeqn
where $V_{ij}$ is the two-body potential between the $i$th and $j$th
particles and $V_{123}$ is the three-body potential. The relative
kinetic energy $\Trel$ is the combination of the relative kinetic
energy between the $i$th and $j$th particles, $T_{ij} = (k_i -
k_j)^2/2m$, and the relative kinetic  between that pair's center of
mass and the third particle, $T_k = \frac{3}{2}((k_i+k_j)/2 -
k_k)^2/m$. Because $T_i$ and $V_{jk}$ commute the commutator of
$\Trel$ and $V_{ij}$ is 
\beqn
[\Trel,V_{ij}] = [T_{ij},V_{ij}]\;,
\label{eq:commutator_TV2}
\eeqn
so the flow equation for the two-body system is 
\beqn
\frac{dV_{ij}}{ds} = [[T_{ij},V_{ij}],(T_{ij}+V_{ij})]\;.
\label{eq:two_body_flow}
\eeqn
We can expand the three-body flow equation, shown in
Eq.~\eqref{eq:three_body_flow}, and using Eqs.~\ref{eq:commutator_TV2}
and \ref{eq:two_body_flow} see that the two-body evolution cancels out
precisely, leaving
\bea
\frac{dV_{123}}{ds} &=& [[T_{12},V_{12}],(T_3 + V_{13} + V_{23} + V_{123})] \nonumber \\
&& \quad + [[T_{13},V_{13}],(T_2 + V_{12} + V_{23} + V_{123})] \nonumber \\
&& \quad + [[T_{23},V_{23}],(T_1 + V_{12} + V_{13} + V_{123})] \nonumber \\
&& \quad + [[\Trel,V_{123}],H_s]\;.
\eea
So the three-body evolution contains no spectator graphs, but only
connected commutators of the form $[\overline{V_{ij}},V_{jk}]$,
$[\overline{V_{ij}},V_{ijk}]$, or $[\overline{V_{ijk}},V_{ijk}]$.  The
two-body evolution equations have completely cancelled out of the
three-body sector and are completely determined by the two-body
evolution. In other words, free-space $A$-body forces are {\it
defined} by evolution in the $A$-body space and are not altered or
evolved differently in higher sectors.

We can also see these cancellations in second-quantized form by
considering the one-loop two-body graphs on the top of
Fig.~\ref{fig:srg_diagrams1} coupled with a single line representing a
spectator particle in a three-particle system. Such graphs cannot be
generated from Eq.~\eqref{eq:dVds_2ndquant}, there not being enough
contraction opportunities to generate the necessary
creation/annihilation operators and momentum delta functions.

So, the SRG induces $A$-body forces using connected terms involving
lower-body forces; three-body forces can be initially zero but will
grow due to two-body-only contributions.  Eventually $A$-body forces
are induced as one can see by taking successive infinitesimal steps in
the evolution. In the first step, three-body forces appear which allow
induction of four-body forces in the next step. These in turn allow
higher-body forces ad infinitum. Including induced $A$-body forces is
a unitary calculation in an $A$-body space. In a two body-space,
three-body forces cannot be represented so evolution in the two-body
space is only approximately unitary when the evolved Hamiltonian is
used in the three-particle system. Because the induced three-body
forces are not included, we speak of ``missing induced
$A$-body forces". These induced forces appear in a hierarchy of
decreasing strength, which will be explored quantitatively in
chapter~\ref{chapt:OneD}.

\chapter{Decoupling in the momentum representation}
\label{chapt:decoupling}


\section{Benefits of Decoupling}

 While observables are unchanged by the SRG's unitary transformations,
the contributions from high-momentum intermediate eigenstate
components to low-energy
observables is modified by the running transformation. In particular,
the SRG as implemented in Refs.~\cite{Bogner:2006srg,Bogner:2007srg}
has the effect of partially diagonalizing the momentum-space potential
to a width of order the evolution parameter. Because of this partial
diagonalization, one anticipates a direct decoupling of low-energy
observables from high-energy degrees of freedom.

In Ref.~\cite{Bogner:2007srg}, evidence for decoupling at low momentum
was shown for the Argonne $v_{18}$ \cite{argonne_potential} potential in
calculations of phase shifts and the deuteron. In this chapter, we
extend the demonstration of decoupling to nucleon-nucleon (NN)
potentials from chiral effective field theory
(EFT)~\cite{N3LO,N3LOEGM} and to few-body nuclei up to $A = 6$ to
verify its universal nature and to show quantitatively that the
residual coupling is perturbative above the energy corresponding to
the SRG evolution parameter. 

The practical test for decoupling is whether changing high-momentum
matrix elements of the potential changes low-energy observables. The
strategy here is to first evolve the initial potential $V_{\rm NN}$ 
with the SRG equations to obtain the SRG potential $V_\flow$, where
$s$ denotes the flow parameter of the transformation.  Then we apply a
parametrized regulator to cut off the high-momentum part of the
evolved  potential in a controlled way. This cutoff potential is used
to calculate few-body  observables and their relative errors. By
varying the parameters of the regulator and correlating them with
errors in the calculated observables, we have a diagnostic tool to
quantitatively analyze the decoupling.

\begin{figure}[tbh!]
\begin{center}
\includegraphics*[height=2in]{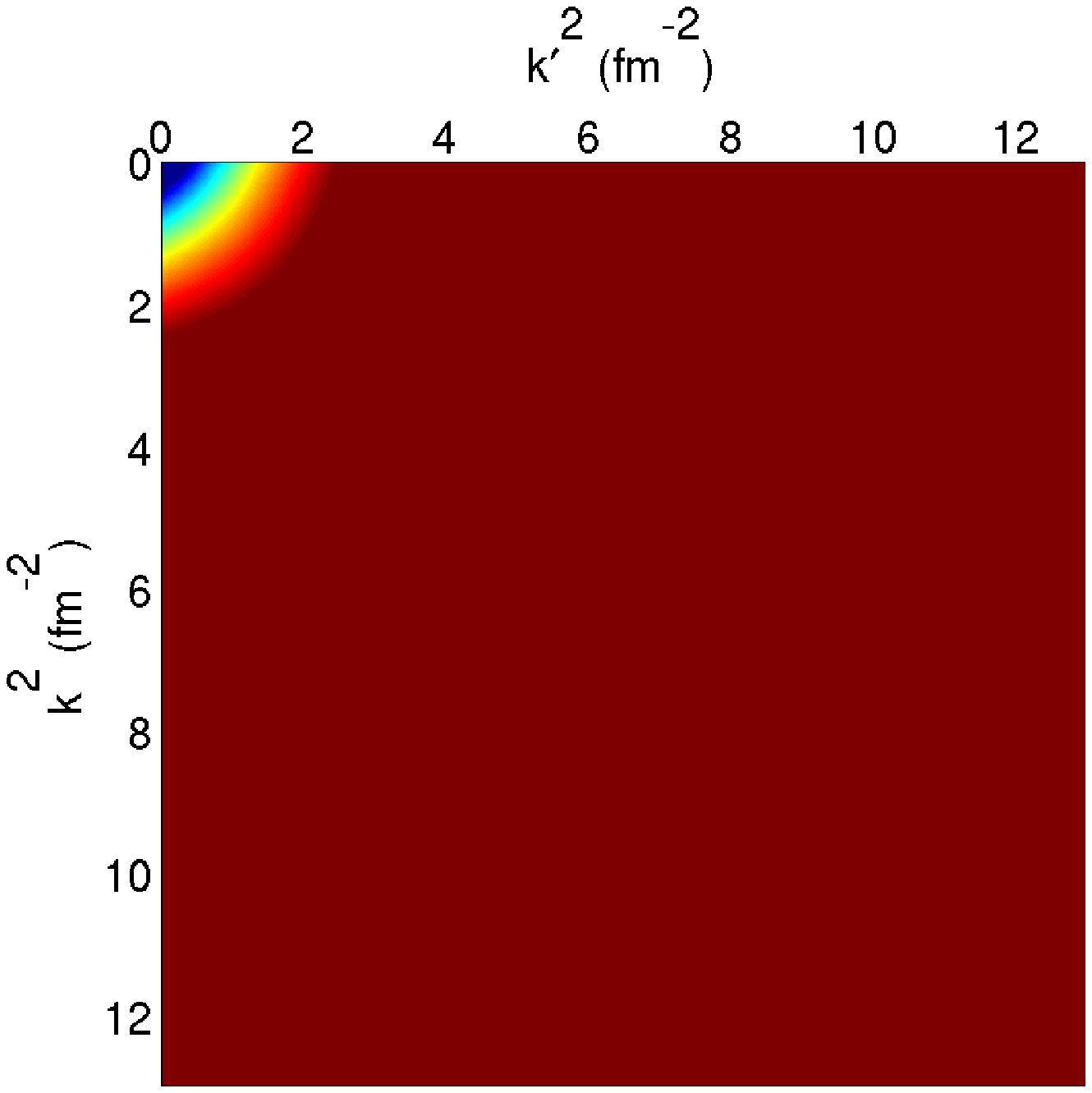}
\hfill
\includegraphics*[height=2in]{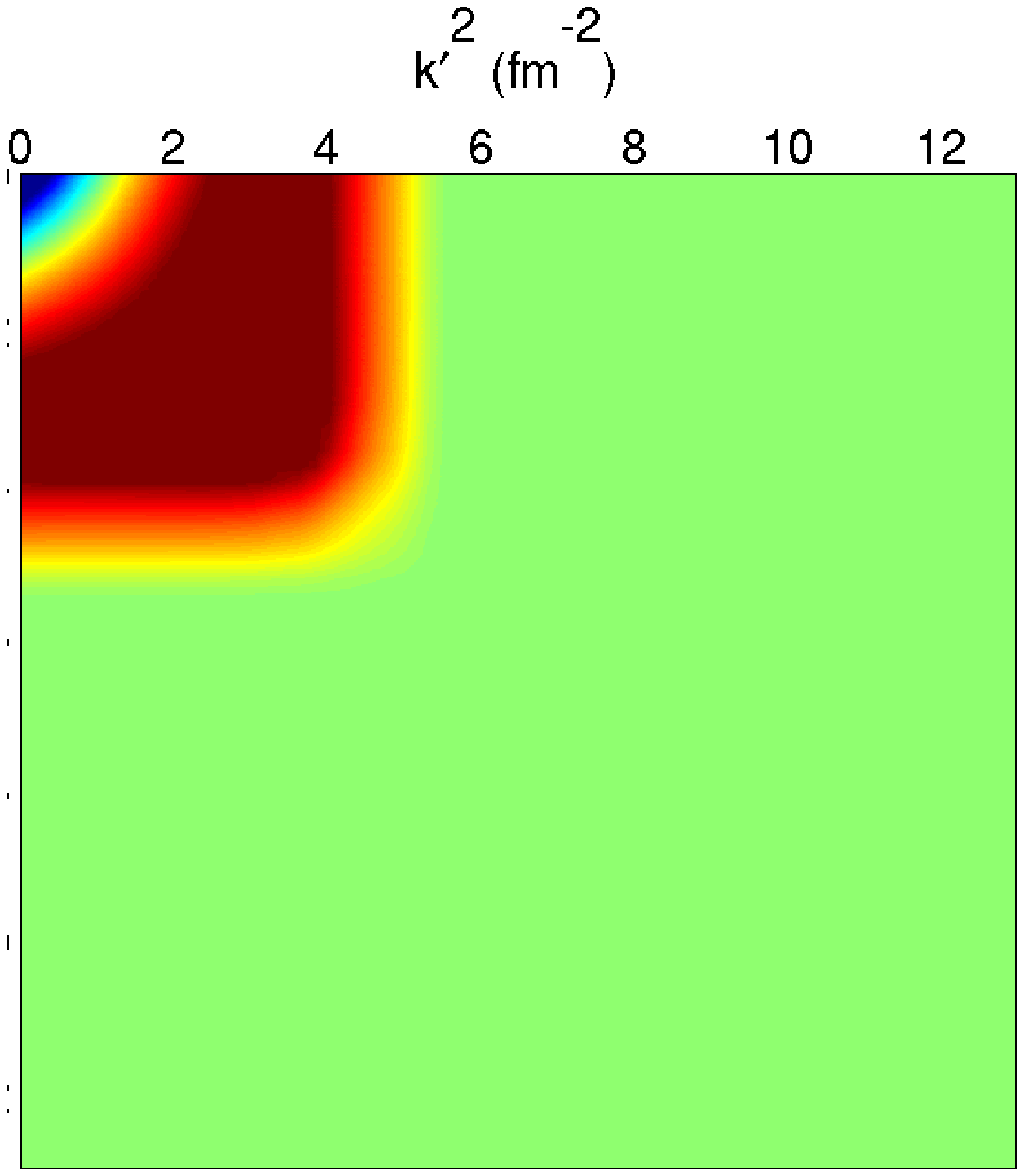}  
\hfill
\includegraphics*[height=1.7in]{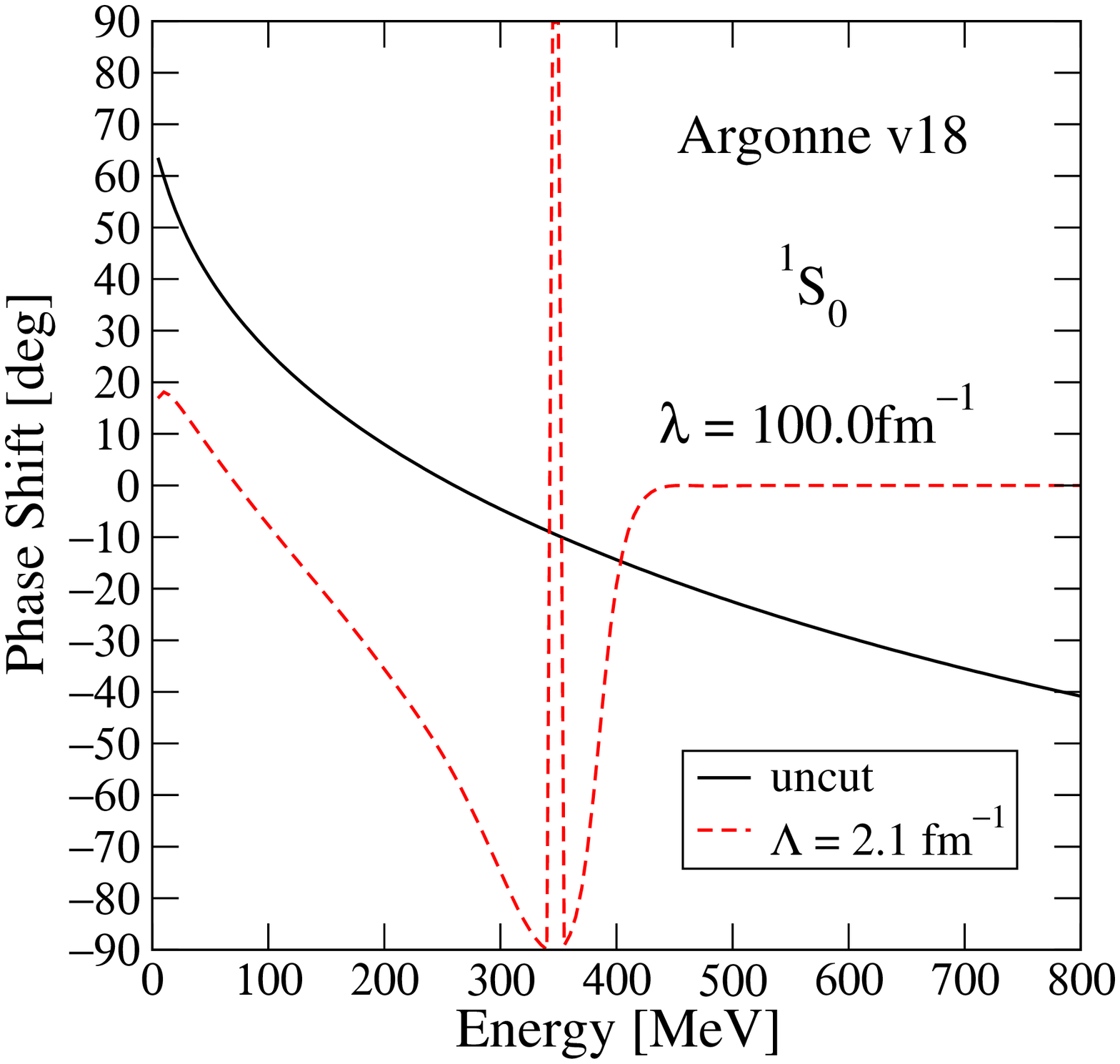}  

\includegraphics*[height=2in]{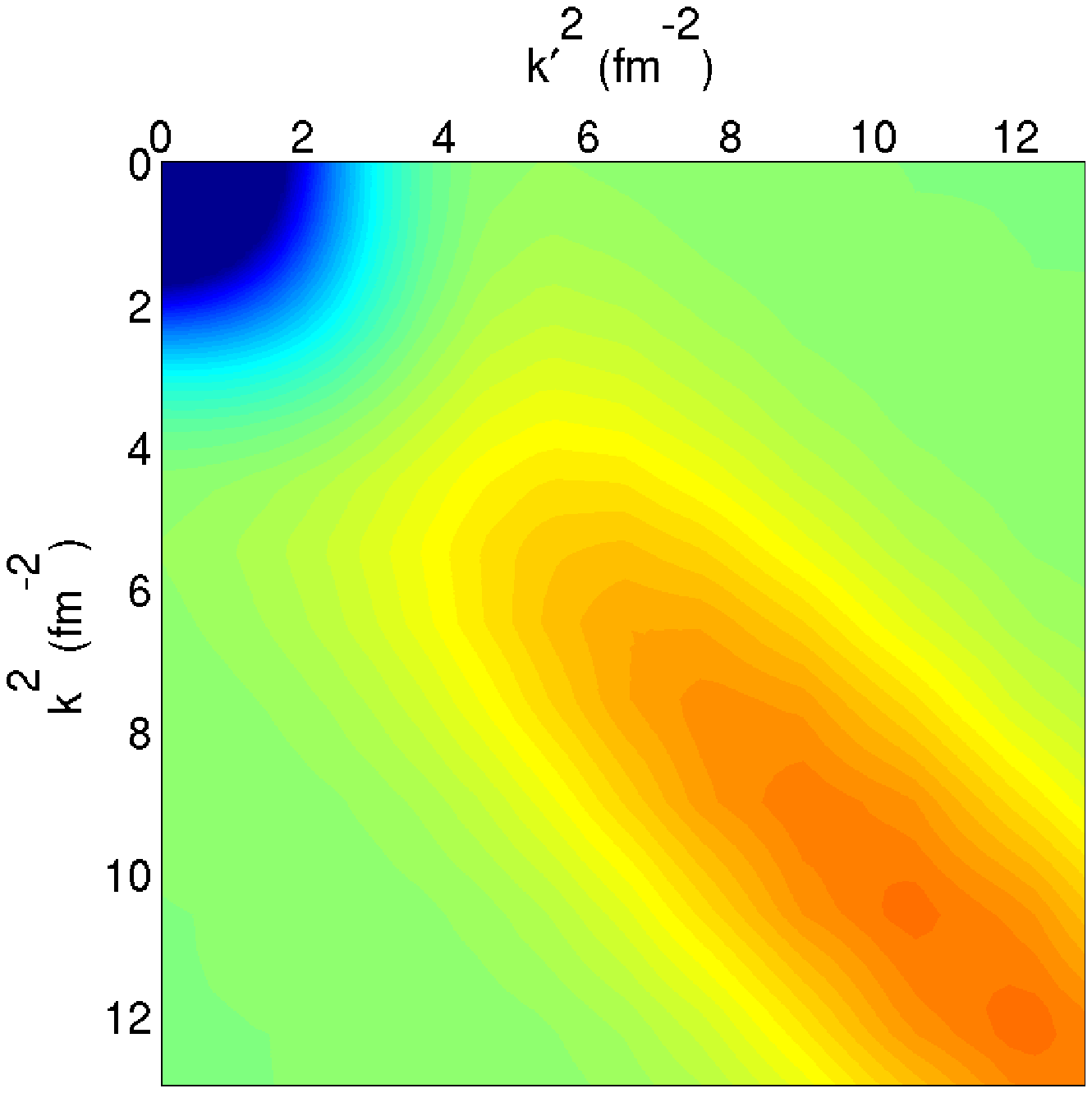}
\hfill
\includegraphics*[height=2in]{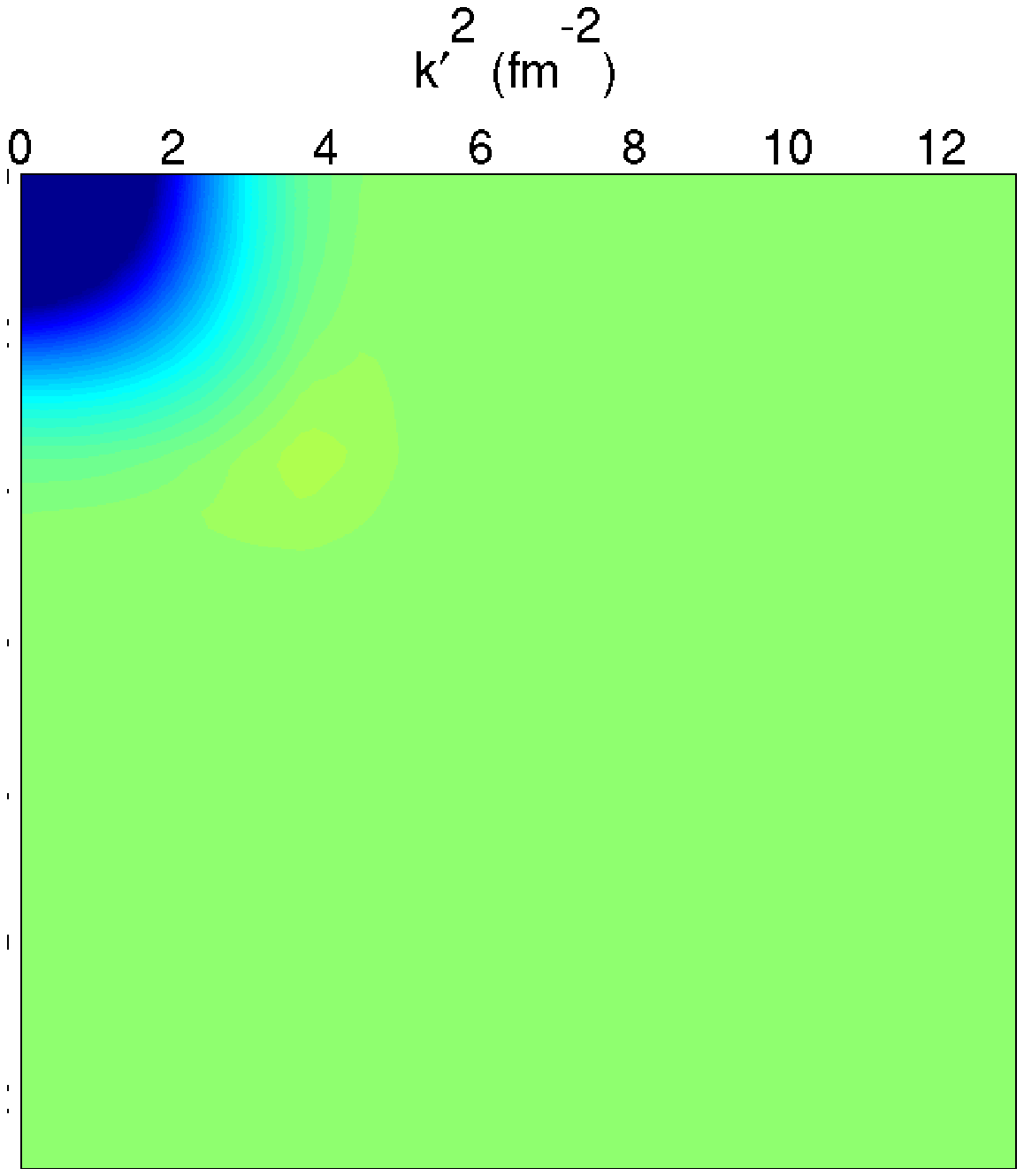}  
\hfill
\includegraphics*[height=1.7in]{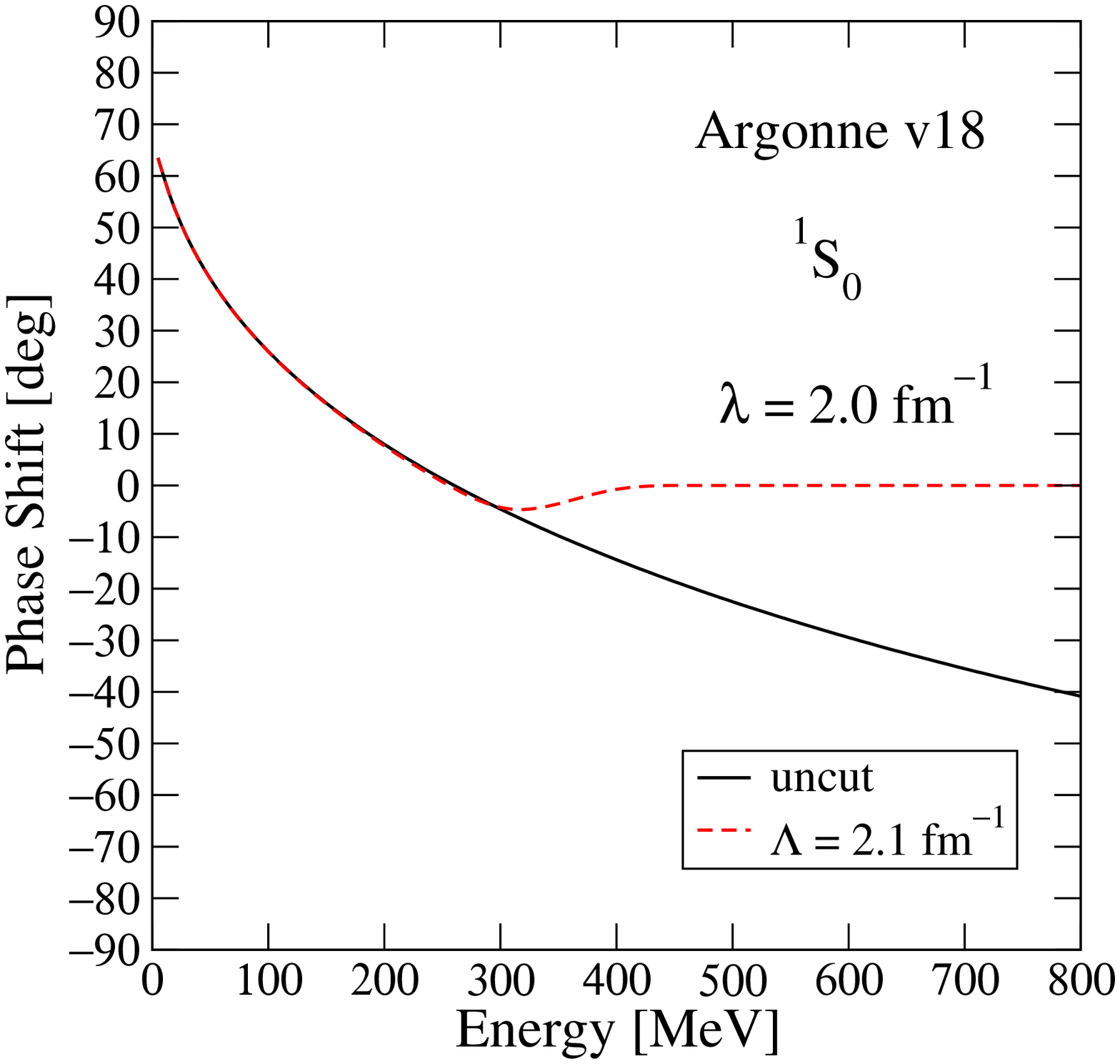}  
\end{center}
\captionspace{Phase shifts and relative errors in the $^1$S$_0$
channel for SRG potentials evolved from the Argonne $v_{18}$ potential
of Ref.~\cite{argonne_potential}. The upper row shows cut and uncut
potentials and phase shifts for the unevolved initial potential. The
lower row shows the same for the initial potential evolved to $\lambda
= 2.0 \fmi$.}
\label{fig:cutting_example}
\end{figure}

Shown in Figure~\ref{fig:cutting_example} is an example of the
strategy employed in this chapter to study the decoupling benefits of
the SRG. On the top row, the pictures from left to right are the uncut
Argonne $v_{18}$ potential in the $^1\!S_0$ partial wave, the same
potential cut at $\Lambda= 2.1 \fm^{-2}$ and the phase-shift
calculated from both potentials. The bottom row shows the same three
plots except now the Argonne potential is evolved with the SRG to
$\lambda = 2.0 \fmi$. In the top phase shift picture, the calculation
from the cut potential fails everywhere; the low-energy observable is
lost despite the low-energy matrix elements of the potential being
preserved. In the bottom phase shift picture the low-energy phase
shift is preserved because the SRG has renormalized the potential and
preserved all the information relevant to the calculation.

The phase shift, $\delta_\ell(E_{k_0})$, (where $\ell$ denotes a
particular partial wave) is a nuclear scattering observable calculated
non-perturbatively from the potential, via the relation
$R_\ell(k,k';E_{k_0}) = -\frac{tan
\delta_\ell(E_{k_0})}{m_Nk_0}$~\cite{landau_QM}, where $R_\ell$ is the
reaction matrix,
\bea
R_\ell(k,k',k_0) &=& V_\ell(k,k') + \frac{2}{\pi}P\int_0^\infty dp
\frac{p^2V_\ell(k,p)R_\ell(p,k',k_0)}{E_{k_0}-E_p} \nonumber \\
&=& V_\ell(k,k') + \frac{2}{\pi}P\int_0^\infty dp 
\frac{p^2V_\ell(k,p)V_\ell(p,k')}{E_{k_0}-E_p} \nonumber \\
&& \quad +\frac{4}{\pi^2}P\int_0^\infty dp dq 
\frac{p^2q^2V_\ell(k,p)V_\ell(p,q)V_\ell(q,k')}{(E_{k_0}-E_p)(E_{k_0}-E_q)} +
\ldots \nonumber \\
&=& \la k|V_\ell|k'\ra + \sum_{p} \frac{\la k|V_\ell|p\ra\la
p|V_\ell|k'\ra}{(k_0^2 - p^2)/m} + \ldots
\eea
Where $V_\ell$ represents the potential in a particular partial wave,
and $k_0$ is the momentum corresponding to the desired on-shell energy
of the calculation. The $P$ here indicates that we are using the
Cauchy principal value prescription which keeps matrix elements real,
a convenient property for numerical calculations. Cutting the
potential as shown in Fig.~\ref{fig:cutting_example} does not lose
information for the first order term (linear in $V$) since it
represents only one matrix element on the diagonal of the potential.
However, the second (and subsequent) terms involve sums over
intermediate states so that matrix elements along the whole row or
column involving the state $k_0$ are necessary to the calculation.
When these are arbitrarily cut off as in the top row, the phase shift
will fail at any energy.

However, in the bottom row of Fig.~\ref{fig:cutting_example} the
SRG has transformed, or renormalized, the potential so that all
the relevant information has been placed inside a low-energy
region of the potential. Now we can cut the high-energy matrix
elements, which are decoupled from the low-energy ones, leaving the
phase-shift intact for low energies. Thus we need less of the
total basis to represent the low-energy physics we are concerned
with reproducing.

The tool we will use to study decoupling is a smooth exponential 
regulator applied to the potential to cut off momenta above $\Lambda$:
\beqn
  V_{\lambda,\Lambda}(k,k') =  
  e^{-({k^2}/{\Lambda^2})^{\nexp}} V_{\lambda}(k,k') 
  e^{-({k'^2}/{\Lambda^2})^{\nexp}}  \;,
\label{eq:regulator}
\eeqn
where $\nexp$ takes on integer values.  From the cut potentials we
calculate observables such as phase shifts and  ground-state energies
and  compare to values calculated with the corresponding uncut
potential. If there is decoupling between matrix elements in a  given
potential (evolved or otherwise), we should be able to set those
elements to zero in this systematic way and use the relative error in
the observable as a metric of the degree of decoupling.  By varying
$\nexp$ we can identify quantitatively the residual coupling strength.

Here we are working with NN interactions only and therefore the SRG
transformations are truncated at the two-body level, which
means that they are only approximately unitary for $A \ge 3$ such that
those observables will vary with $\flow$. In these cases decoupling is
tested by comparing cut to uncut potentials at a fixed $\flow$. All
two-body observables calculated with the uncut $H_\flow$  are
independent of $\flow$ to within numerical precision.  The actual
numerical error depends on the details of the discretization (e.g.,
the number and distribution of mesh points, usually gaussian) and on
the accuracy and tolerances of the differential equation solver. While
in practice we can make such errors very small, to avoid mixing up
small errors we will also compare cut to uncut potentials rather than
to the unevolved ($s=0$) potential for two-body observables.


\section{Mechanics of Decoupling}
\label{sec:decoupling_mechanics}

The source of decoupling is the partial diagonalization of the
Hamiltonian by the SRG evolution~\cite{Jurgenson:2007td}. For the NN
interaction, the flow Eq.~\eqref{eq:srg8} can be simply evaluated in
the space of  discretized relative momentum NN
states~\cite{Bogner:2006srg}.   For a given partial wave, with units
where $\hbar^2/M = 1$, we define diagonal matrix elements of momentum
$k$ as\footnote{Note that the derivation here is only valid for a
discretized momentum basis such as the gaussian mesh used here.}
\beqn
   \langle k | H_\flow | k \rangle = \langle k | H_D | k \rangle
   \equiv e_k
   \;,  \label{eq:Hd_values}
\eeqn 
and
\beqn
   \langle k | T_{\rm rel} | k \rangle 
   \equiv \epsilon_k = k^2 \;. \label{eq:Trel_values}
\eeqn 
If we take $G_\flow = T_{\rm rel}$, the flow equation for each
matrix element is 
\bea
  \frac{d}{ds} \langle k|H_\flow|k'\rangle &=&
  \sum_q (\epsilon_k + \epsilon_{k'} - 2 \epsilon_q) \langle k|H_\flow|q\rangle
      \langle q|H_\flow|k'\rangle
      \nonumber \\
    &=& 
     -(\epsilon_k - \epsilon_{k'})^2 \langle k | V_s | k' \rangle
  + \sum_q (\epsilon_k +
  \epsilon_{k'} - 2 \epsilon_q) \langle k|V_\flow|q\rangle
  \langle q|V_\flow|k'\rangle \;.  
   \label{eq:Trel_flow}
\eea

\begin{figure}[tbh!]
\begin{center}
\strip{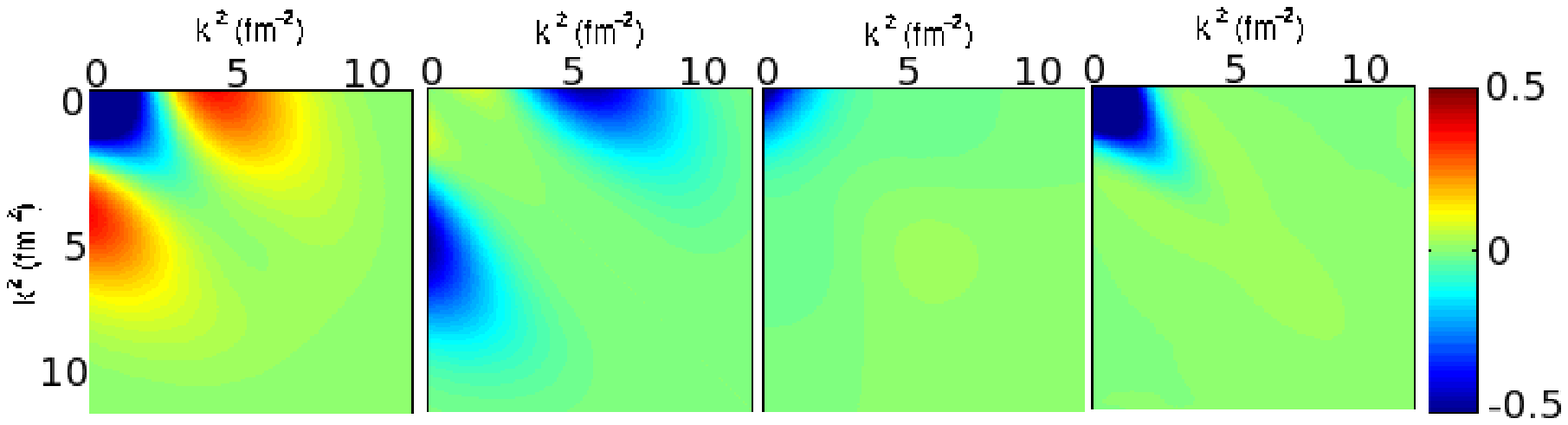}
\end{center}
\captionspace{A snapshot of the contributions to the flow of the
potential when using the $G_s = \Trel$ choice of SRG. From left to
right, the pictures are $V_{\lambda=2.5}$, 1st and 2nd rhs terms from
Eq.~\eqref{eq:Trel_flow}, and $V_{\lambda=1.5}$. The color scale for
the middle pictures is scaled up for visibility.}
\label{fig:srg_mechanics}
\end{figure}

A useful image of the mechanism by which the SRG diagonalizes the
potential is shown in Fig.~\ref{fig:srg_mechanics}. The pictures
represent a snap shot of the different parts of the flow equations
contributing to the evolution between two specific values of
$\lambda$. The plot on the left is the potential evolved down to
$\lambda = 2.5 \fmi$ and the plot on the far right is the potential
further evolved to $\lambda = 1.5 \fmi$. The two panels in the middle
are the first and second terms of the right-hand side from the flow
equations in Eq.~\eqref{eq:Trel_flow}. It is clear that the first term
is the dominant contributor to the suppression of off-diagonal matrix
elements outside the band of width $\lambda^2$. At the same time the
second term is providing for the flow of physics information to
low-energy states by increasing the magnitude of matrix elements
associated with those states. While the first term achieves decoupling
by suppressing off-diagonal matrix elements, the second term is
preserving unitarity by transforming the information of high-energy
matrix elements to low-momentum states.

\begin{figure}[tbh!]
\begin{center}
\dblpic{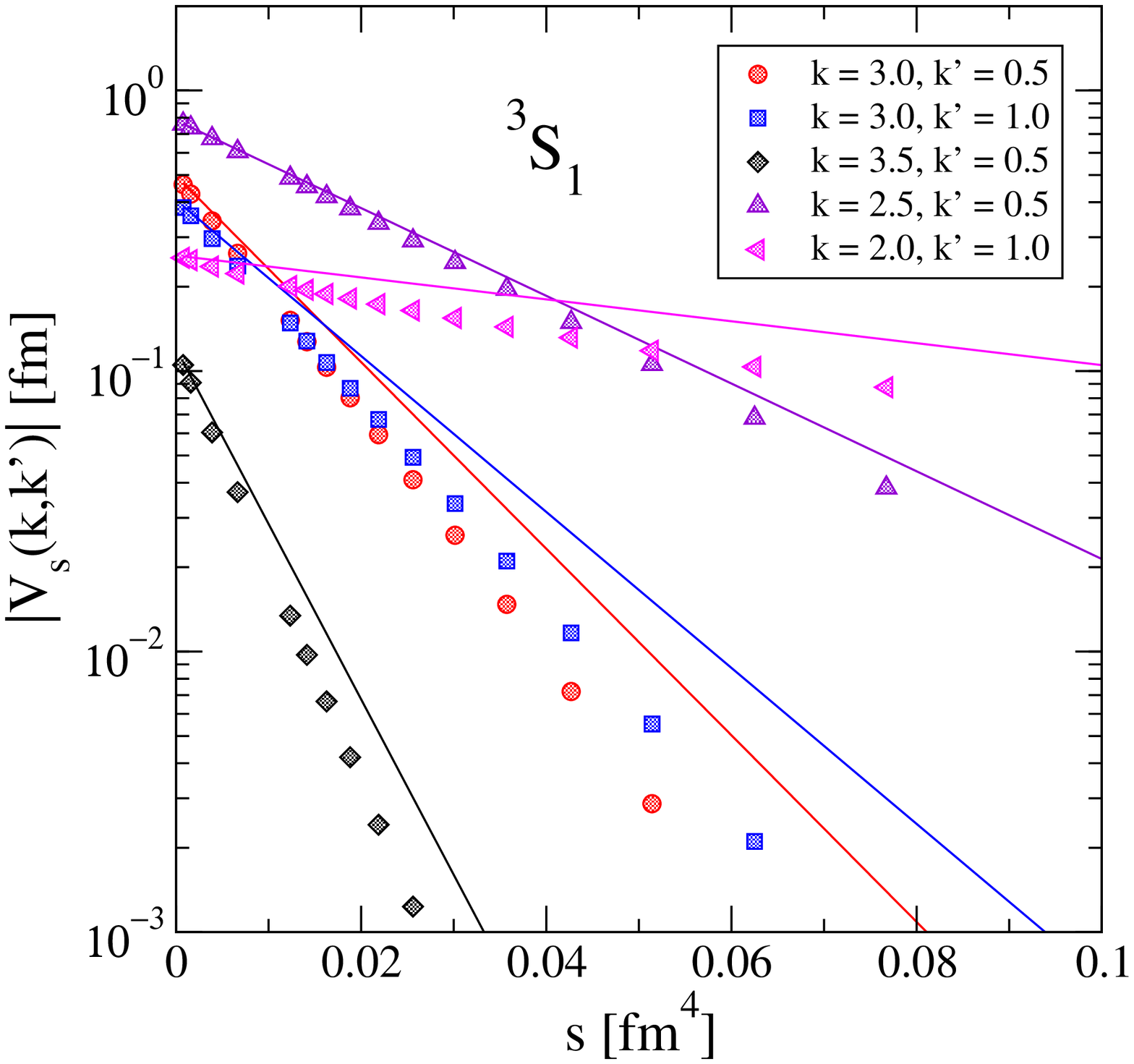}
\hfill
\dblpic{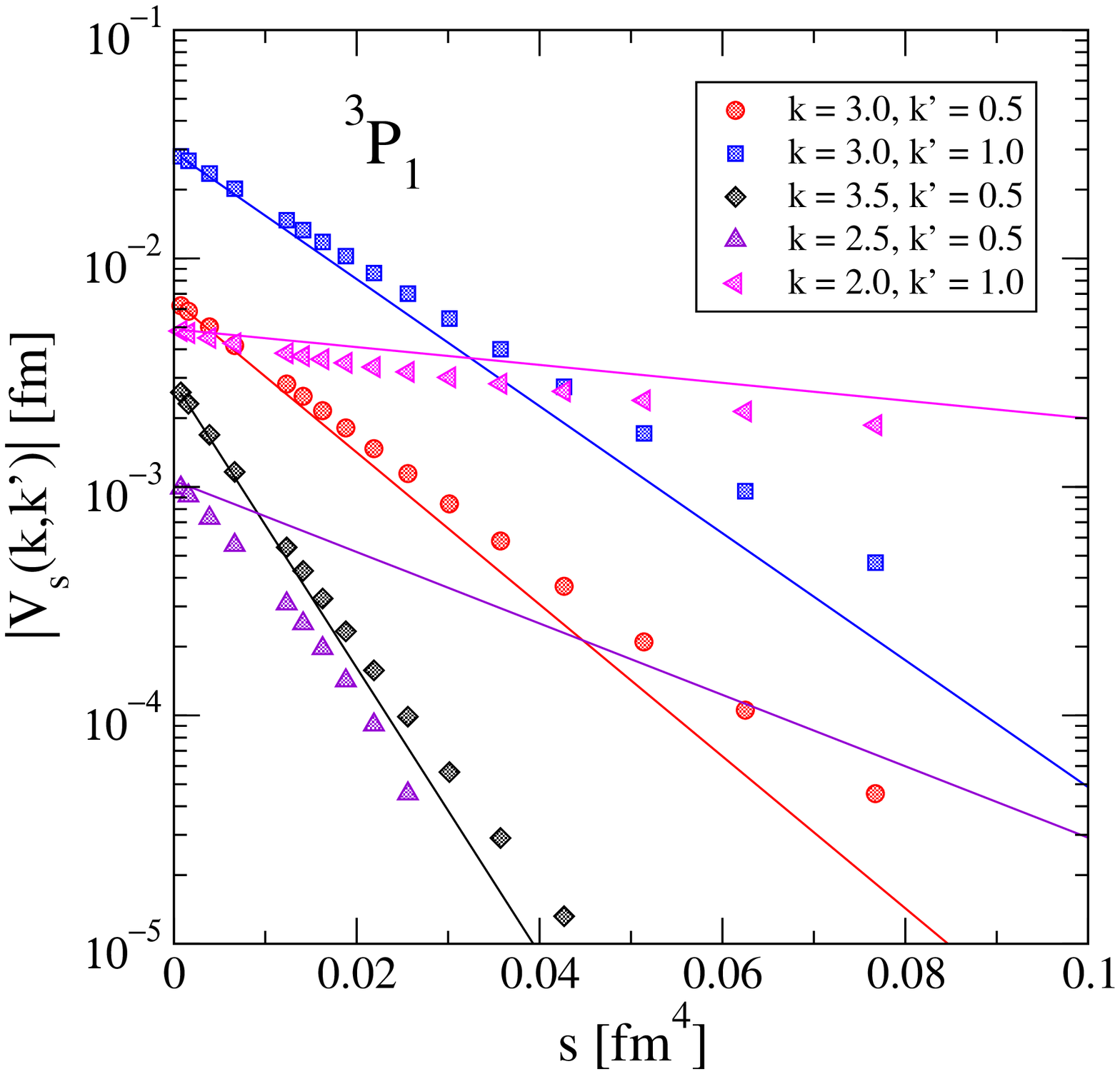}  
\end{center}
\captionspace{Absolute value of  the matrix element $\langle k | V_\flow |
k' \rangle $ for a representative sampling of off-diagonal $(k,k')$
pairs as a function of $s$, compared with the simple solutions from
Eq.~\eqref{eq:exponential}, which are straight lines (they agree at
$s=0$).  The partial waves $^3$S$_1$ and $^3$P$_1$ are shown.}
\label{fig:flowtest}
\end{figure}

We can find a semi-quantitative approximation for the flow of
off-diagonal matrix elements by keeping only the first term on the
right side of Eq.~\eqref{eq:Trel_flow}. Then the flow equation applied
to individual off-diagonal matrix elements simplifies to
\beqn 
  \frac{d}{ds} \langle k|H_\flow|k'\rangle =
  \frac{d}{d\flow} \langle k | V_\flow | k' \rangle 
  \approx - (\epsilon_k - \epsilon_{k'})^2 \, \langle k | V_\flow | k' \rangle
  \;,
\eeqn
which has the simple exponential solution
\beqn
\langle k | V_\flow | k' \rangle \approx \langle k | V_{s=0} | k' 
       \rangle e^{-\flow(\epsilon_k - \epsilon_{k'})^2}\;.
       \label{eq:exponential}
\eeqn
In Fig.~\ref{fig:flowtest}, we plot  $\langle k | V_\flow | k'
\rangle$ and the approximation from Eq.~\eqref{eq:exponential} versus
$s$ for some representative off-diagonal points in two partial waves.
In almost all cases the approximation gives a reasonable estimate of
the monotonic decrease to zero; in the one exception there is a
significantly more rapid decrease than predicted.

Equation~\eqref{eq:exponential} shows that it is convenient to switch
to the flow variable $\lambda = 1/s^{1/4}$, which has units of $\fmi$,
because it is a measure of the resulting diagonal width of $V_\flow$
in momentum space.  More precisely, the matrix $\langle k | V_\flow |
k' \rangle $  plotted as a function of kinetic energies $k^2$ and
$k'{}^2$ will rapidly go to zero outside of a diagonal band roughly of
width $\lambda^2$, which is verified by numerical
calculations~\cite{Jurgenson:2007td,srgwebsite}. For momenta within
$\lambda$ of the diagonal, the omitted quadratic part of the flow
equation is, of course, essential, and drives the  flow of physics
information necessary to preserve unitarity.


We can examine the flow of off-diagonal matrix elements as a whole by
considering the trace of $H_\flow^2$, Tr$[H_s^2] = $ 
Tr$[U_sHU_s^\dagger U_sHU_s^\dagger ] = $Tr$[H^2]$ using $UU^\dagger =
1$ and the cyclic property of traces~\cite{Wegner:1994}:
\bea
\frac{d}{ds}{\rm Tr}[H^2] = 0 &=& \frac{d}{ds}\sum_{i,j}H_{s,ij}H_{s,ji} \nonumber \\ 
&=& \frac{d}{ds}\sum_{i}|\la i|H_s|i\ra|^2 +
\frac{d}{ds}\sum_{i\neq j}|\la i|H_s|j\ra|^2 \;, 
\eea
and therefore we get,
\beqn
\frac{d}{ds}\sum_{i\neq j}|\la i|H_s|j\ra|^2 =
-\frac{d}{ds}\sum_i|\la i|H_s|i\ra|^2 =  
-2\sum_i \la i|H_s|i\ra \frac{d}{ds}\la i|H_s|i\ra \;.
\label{eq:trace_result}
\eeqn
In the case of Wegner's choice, $G_\flow = H_D$, the flow equation 
for each matrix element is
\beqn
  \frac{d}{ds} \langle k|H_\flow|k'\rangle =
  \sum_q (e_k + e_{k'} - 2 e_q) \langle k|H_\flow|q\rangle
      \langle q|H_\flow|k'\rangle
      \;.  
      \label{eq:Hd_flow}
\eeqn 
then \eqref{eq:trace_result} simplifies to
\bea \frac{d}{ds}\sum_{i\neq j}|\la i|H_s|j\ra|^2  &=&
-4\sum_{k\neq q} e_k(e_k - e_q)|\la k|H_s|q\ra|^2 \nonumber \\ 
&=& -2\sum_{k\neq q} e_k(e_k - e_q)|\la k|H_s|q\ra|^2 
-2\sum_{k\neq q} e_q(e_q - e_k)|\la q|H_s|k\ra|^2 \nonumber \\ 
&=& -2\sum_{k\neq q} (e_k - e_q)^2|\la k|H_s|q\ra|^2\;,
\label{eq:trace_Hd}
\eea
which shows a sum of positive definite terms whose derivative is
negative, so that the off-diagonal matrix elements decrease with $s$.
On the other hand using the choice $G_s = \Trel$ and
Eq.~\eqref{eq:Trel_flow} to simplify Eq.~\eqref{eq:trace_result} we get
\bea \frac{d}{ds}\sum_{i\neq j}|\la i|H_s|j\ra|^2  &=&
-4\sum_{k\neq q} \epsilon_k(e_k - e_q)|\la k|H_s|q\ra|^2 \nonumber \\ 
&=& -2\sum_{k\neq q} (e_k - e_q)(\epsilon_k - \epsilon_q)|\la
k|H_s|q\ra|^2\;,
\label{eq:trace_Trel}
\eea
so that off-diagonal elements are not guaranteed to decrease
monotonically if $e_k - e_q$ and $\epsilon_k - \epsilon_q$ have
opposite signs (see Ref.~\cite{Glazek:2008pg} for details). However,
this does not happen in the range of $s$ that has been considered in
the nuclear case because of the dominance of the kinetic energy. $T$ is
much larger than $V$ on the diagonal and therefore $T$ is effectively
$H_D$ in the nuclear case. We should see the difference when we evolve
down to a scale at which $T$ is no longer dominant. At such a scale new
bound states appear and the choice $G_s = \Trel$ cannot order them
properly~\cite{Glazek:2008pg}. We don't see any such pathologies in
the nuclear case because we have never needed to evolve so far in
$\lambda$ as the scale at which the deuteron appears, $\sim 0.5
\fmi$\footnote{In fact, we will not evolve this low also because we
expect the hierarchy of many-body forces to break down at such small
$\lambda$'s, as is discussed in detail in chapter's \ref{chapt:OneD}
and \ref{chapt:ncsm}.}.


\section{Phase Shift Errors}
\label{sec:phase_shifts}

\begin{figure}[tbh!]
\begin{center}
\dblpic{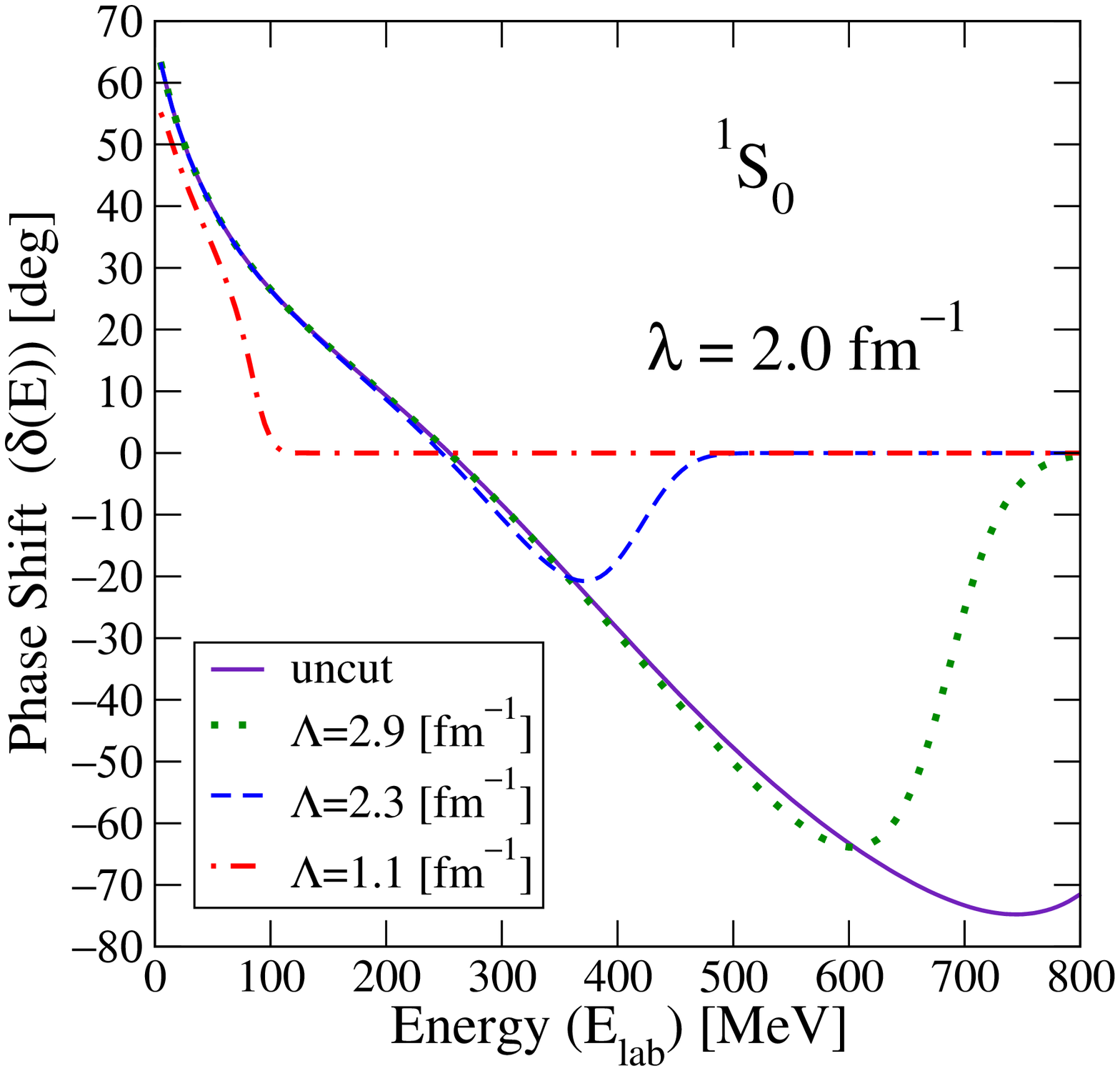}
\hfill
\dblpic{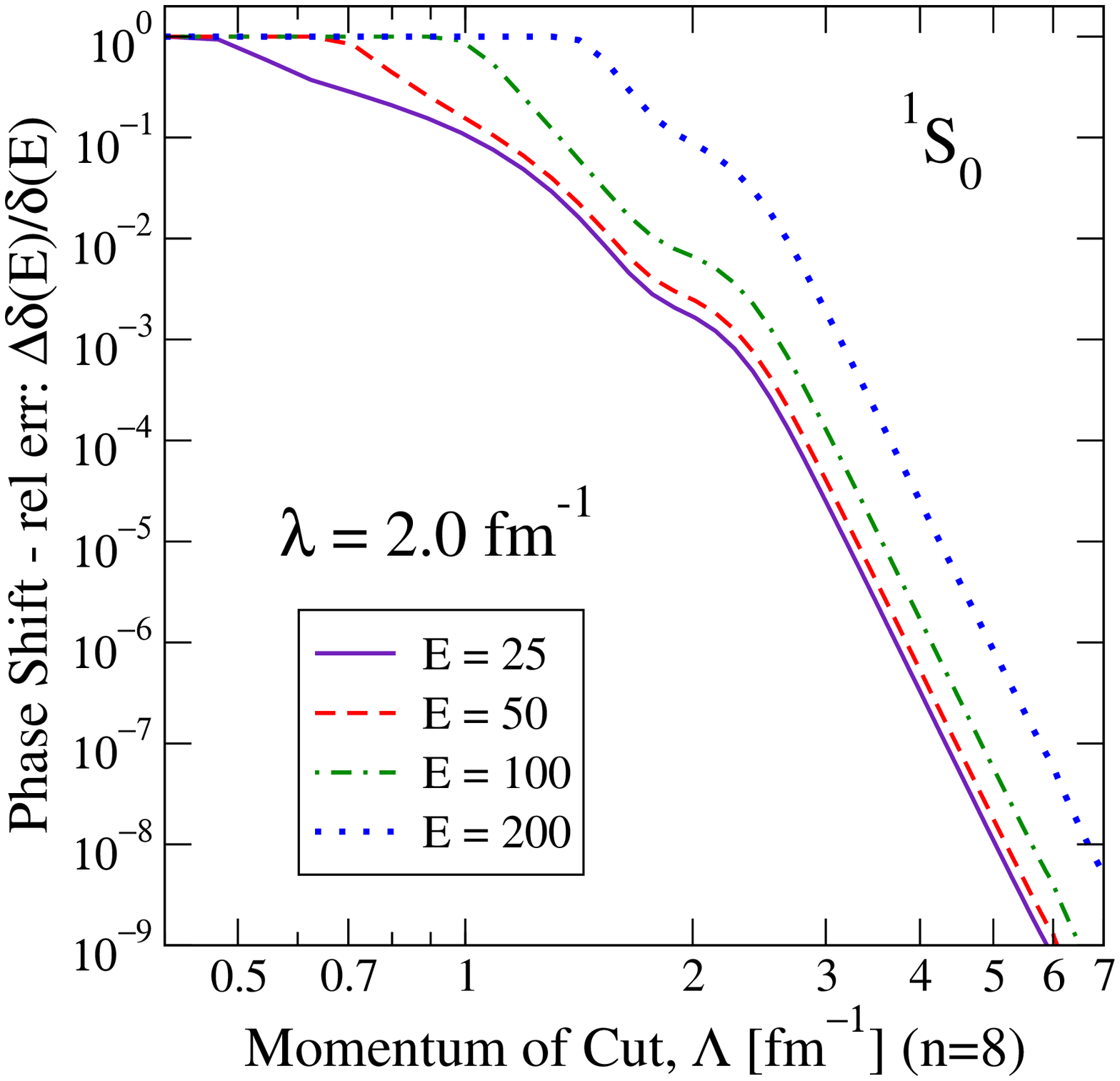}  

\vspace*{.1in}

\dblpic{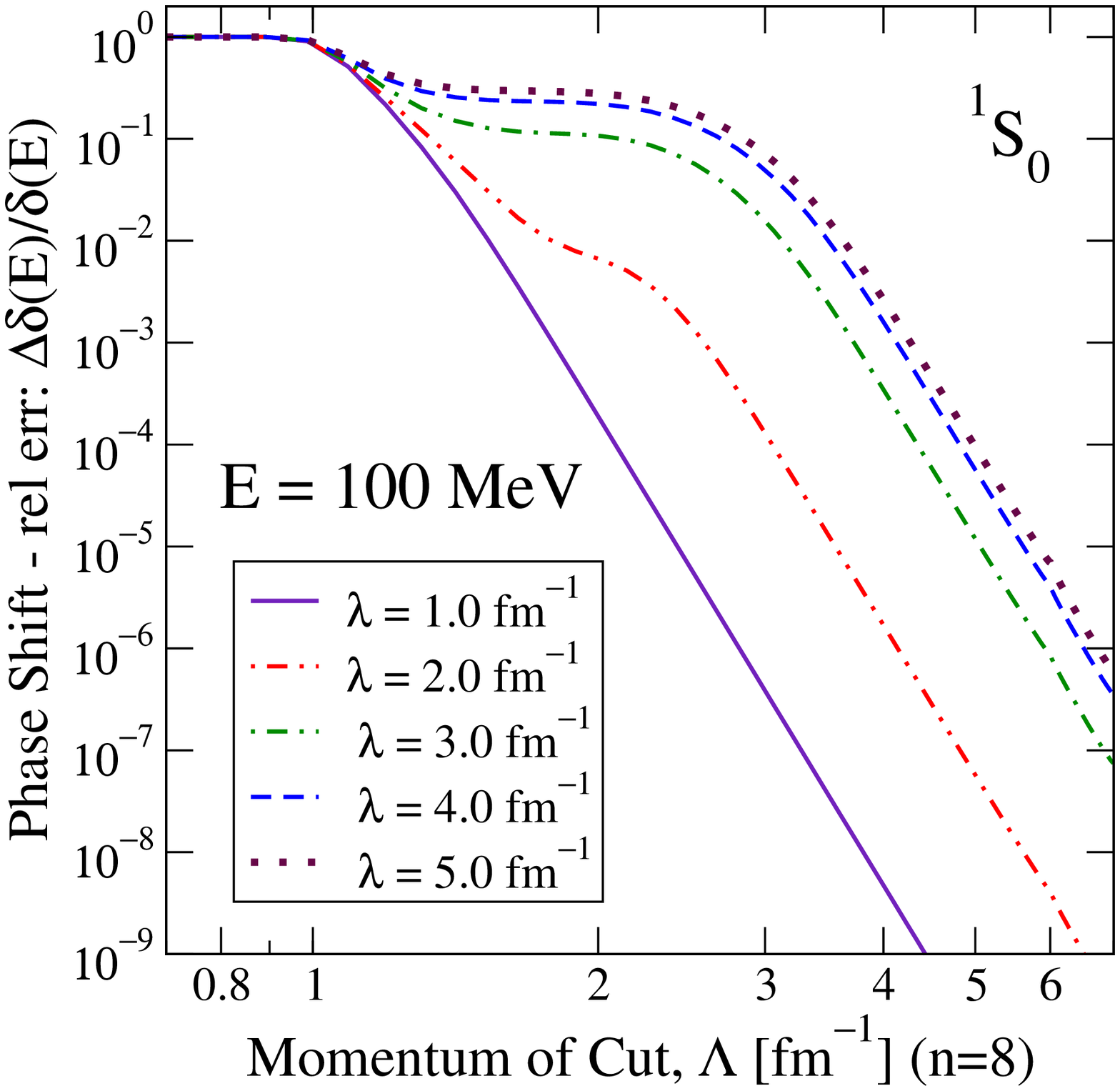}
\hfill
\dblpic{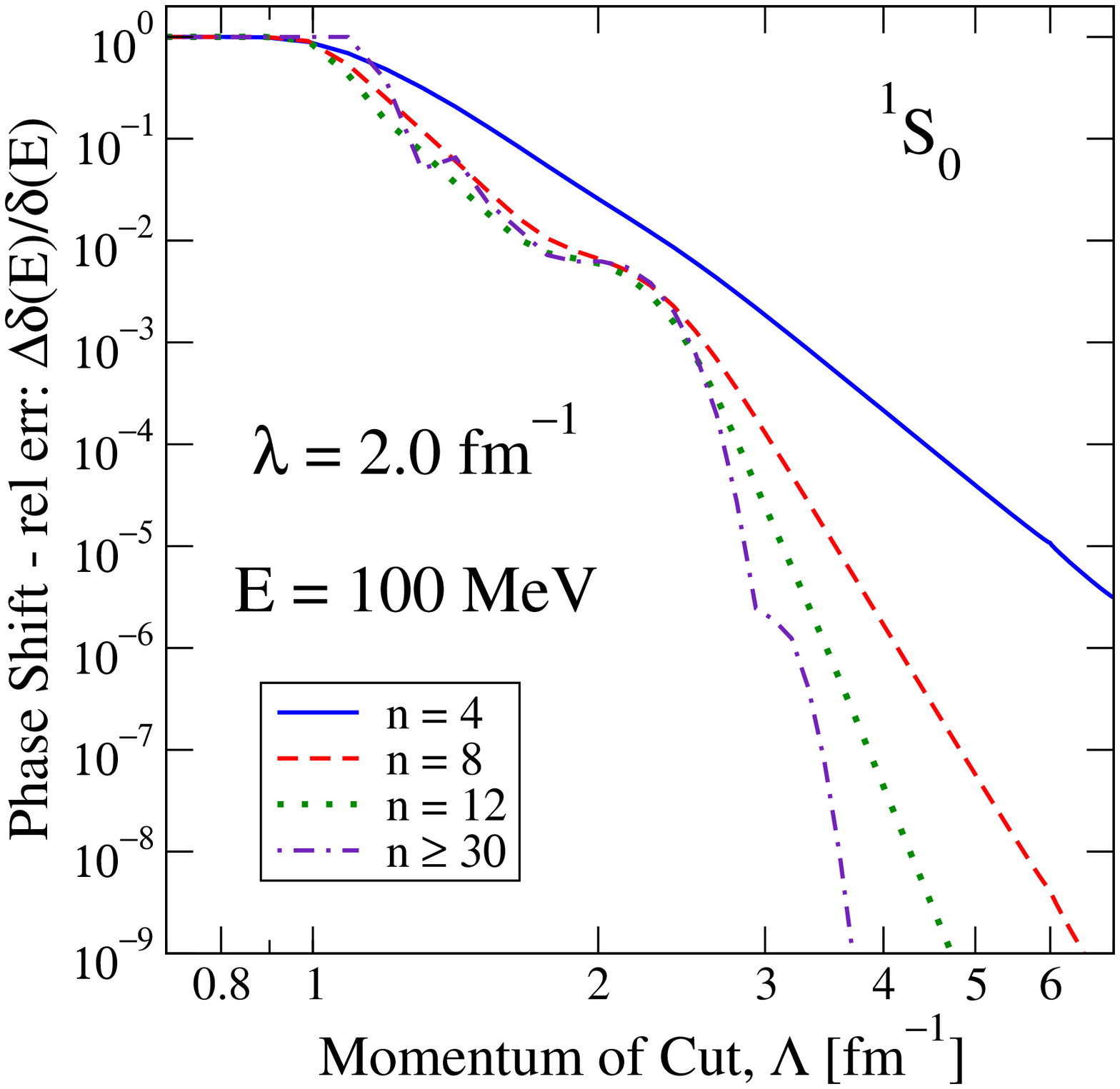}
\end{center}
\captionspace{Phase shifts and relative errors in the $^1$S$_0$ channel for
SRG potentials evolved from the N$^3$LO (500\,MeV) potential of
Ref.~\cite{N3LO}. The upper-left  graph shows the phase shifts vs.\
energy for the uncut $\lambda = 2\fmi$ potential and several   cut
versions with $n=8$.  The other panels show the relative error as a
function of the momentum  cut parameter $\Lambda$  at various energies
$E$, $\lambda$'s, and $n$'s, respectively.}
\label{fig:ps_err_vs_cut}
\end{figure}

In the upper-left panel of Fig.~\ref{fig:ps_err_vs_cut}, we show 
results for the $^1$S$_0$ phase shifts vs.\ energy calculated using 
the unevolved 500\,MeV N$^3$LO potential of Ref.~\cite{N3LO} and the 
corresponding SRG potential evolved to $\lambda = 2.0\fmi$ and then 
cut using the regulator of Eq.~\eqref{eq:regulator} with $n=8$. We  do
not explicitly show results from uncut SRG potentials, because  they
are indistinguishable from the unevolved results.

The qualitative pattern is that when the regulator parameter $\Lambda$
is greater than $\lambda$, there is good agreement of phase shifts
from uncut and cut potentials at small energies and reasonable
agreement up to  the energy corresponding to the momentum of the cut,
$E_{\rm lab} \approx 2\Lambda^2/m$ (with $\hbar = 1$). When $V_{\rm
srg}$ is cut below $\lambda$, there is poor agreement everywhere and
the phase shift is zero above this energy  (e.g., above $E_{\rm lab} =
100\,$MeV for $\Lambda = 1.1\fmi$).   Thus the decoupling of high and
low momentum means that we can explicitly cut out the high-momentum
part of the evolved potential without significantly distorting
low-energy phase shifts  as long as we don't cut below $\lambda$.
Cutting out the high-momentum part of conventional nuclear potentials
\textit{does} cause distortions, which has led to the misconception
that reproducing high-energy phase shifts is important for low-energy
nuclear structure observables~\cite{Bogner:2007srg}.

The quantitative systematics of SRG decoupling are documented in the
other panels of Fig.~\ref{fig:ps_err_vs_cut}, where we look at the
relative error as a function of the cutting momentum using log-log
plots. In these error plots, three main regions are evident.  In the
region below the $\Lambda$ corresponding to the fixed energy, the
predicted phase shift goes to zero since the potential has vanishing
matrix elements, so that the relative error goes to one. Starting at
$\Lambda$ slightly above the value of $\lambda$, there is a clear
power-law decrease in the error. In between is a transition region
without a definite pattern.

\begin{figure}[tbh!]
\begin{center}
\dblpic{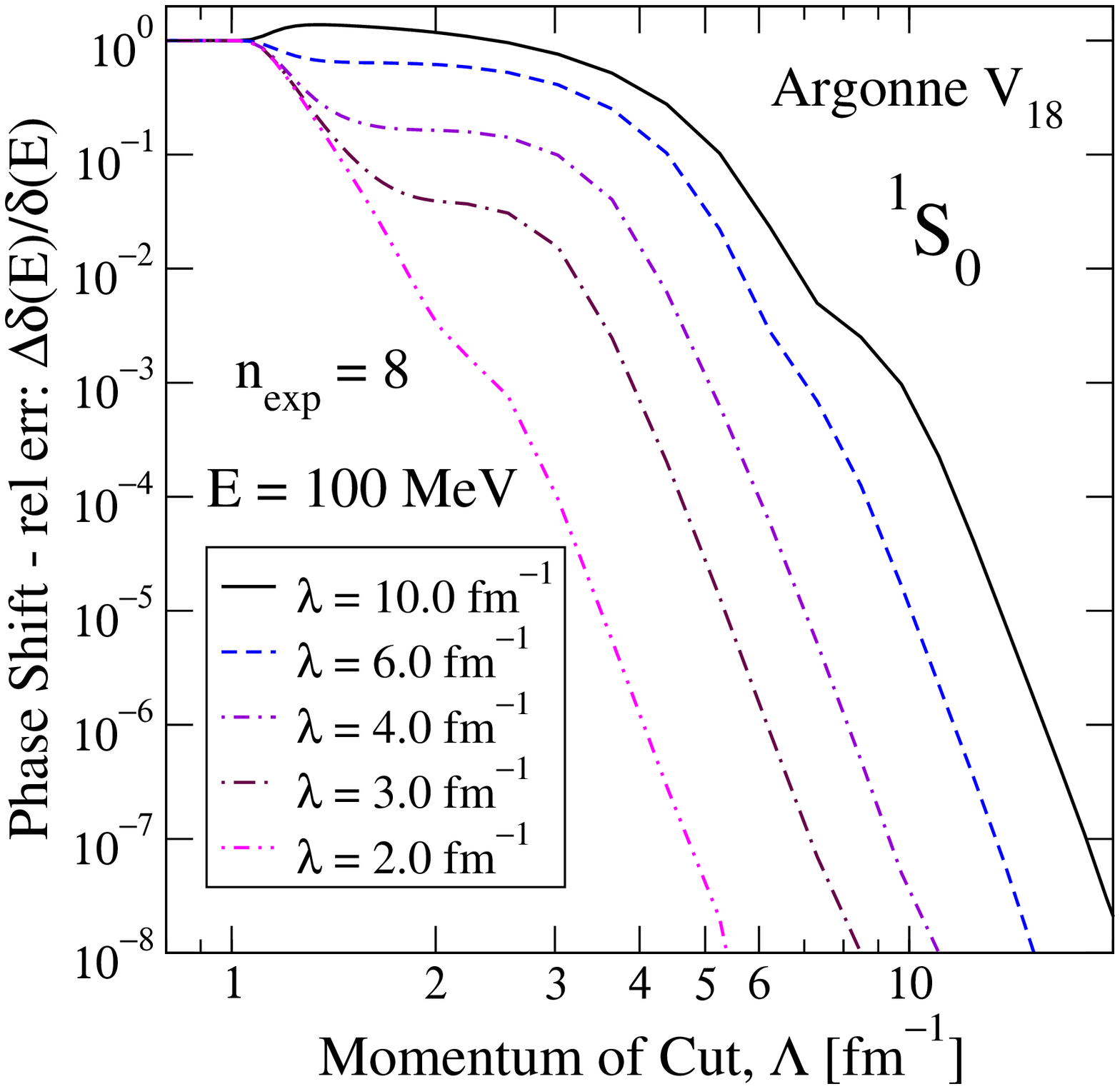} 
\end{center}
\captionspace{The phase shift errors computed in the $^1\!S_0$ channel of
the Argonne V$_{18}$ potential. Here the onset of the power-law slope
is much less saturated at higher $\lambda$'s, because this potential
has higher-momentum components to be renormalized.}
\label{fig:av18_saturation}
\end{figure}

We focus here on the power-law region. In the lower-left pane, we
find  that this decoupling starts with a shoulder at momenta slightly
above  $\lambda$. This effect saturates when $\lambda$ becomes
comparable to the underlying cutoff of the original potential (see
Fig.~\ref{fig:av18_saturation}). In the upper-right pane  we see that
the shoulder signaling the start of the power-law decrease is  not
affected by the energy, $E$. This holds for other values of $\lambda$ 
and $n$. In the lower-right pane we vary  the exponent of the
regulator,  $n$, which changes the smoothness of the regulator. The
smoothness affects the slope of the power law and the fine details in
the intermediate region,  but does not change the  position of the
shoulder near $\lambda$. As discussed  below, the power-law behavior
in the relative error signifies perturbative  decoupling with a
strength given by the sharpness of the regulator used to cut off the
potential.

\begin{figure}[ptbh!]
\begin{center}
\includegraphics*[width=5in]{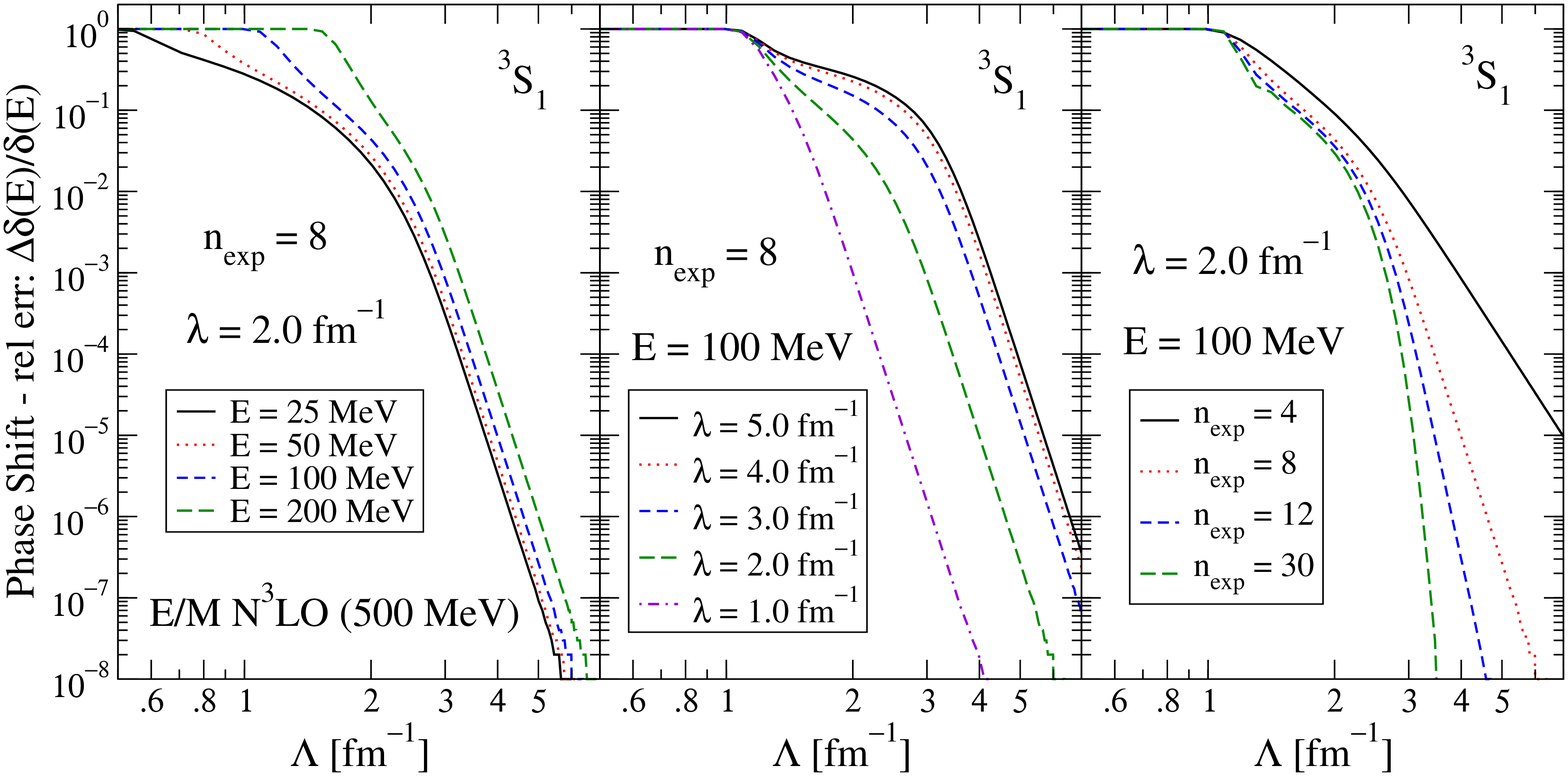} 

\vspace*{.1in}

\includegraphics*[width=5in]{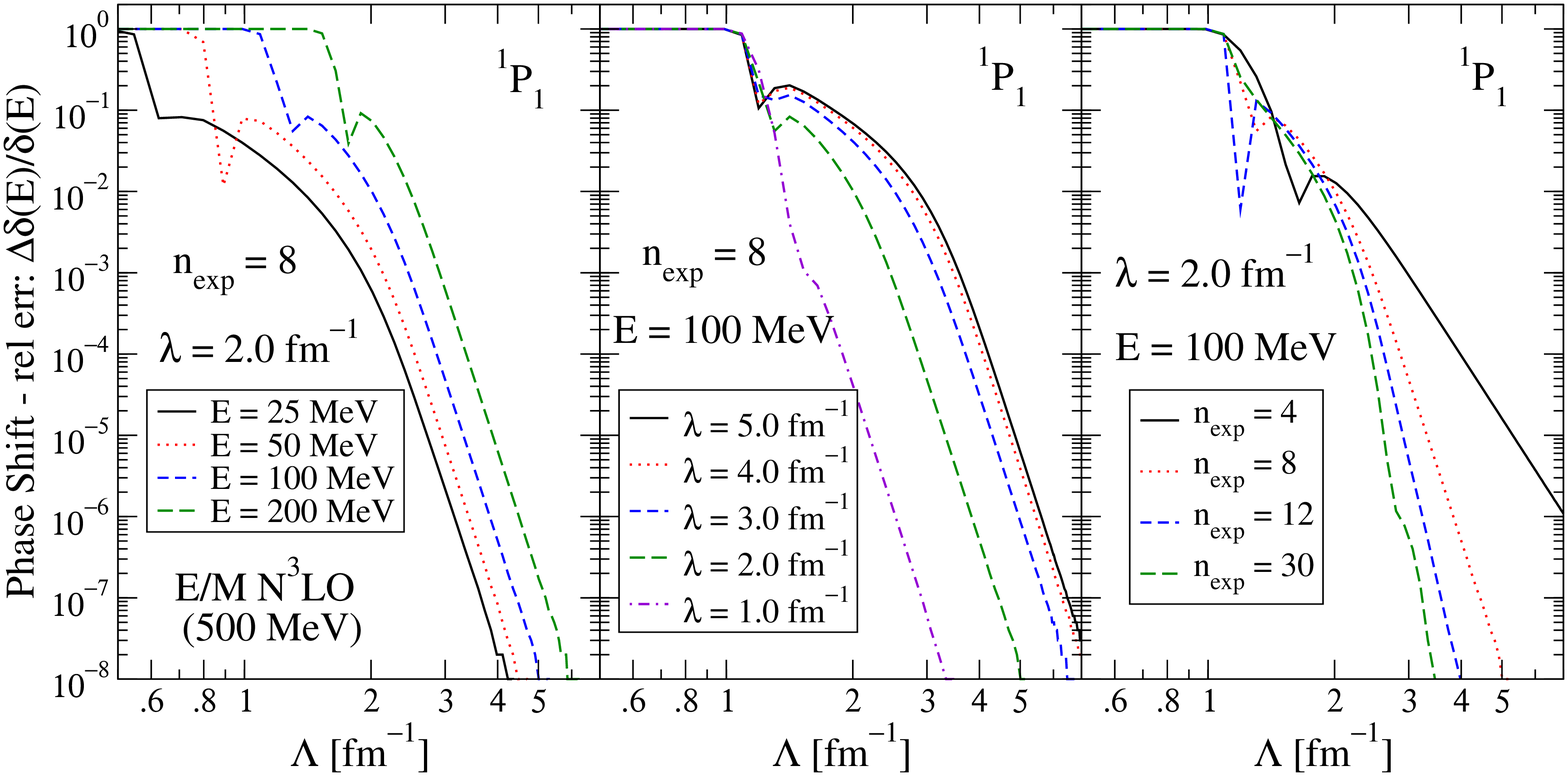}

\vspace*{.1in}

\includegraphics*[width=5in]{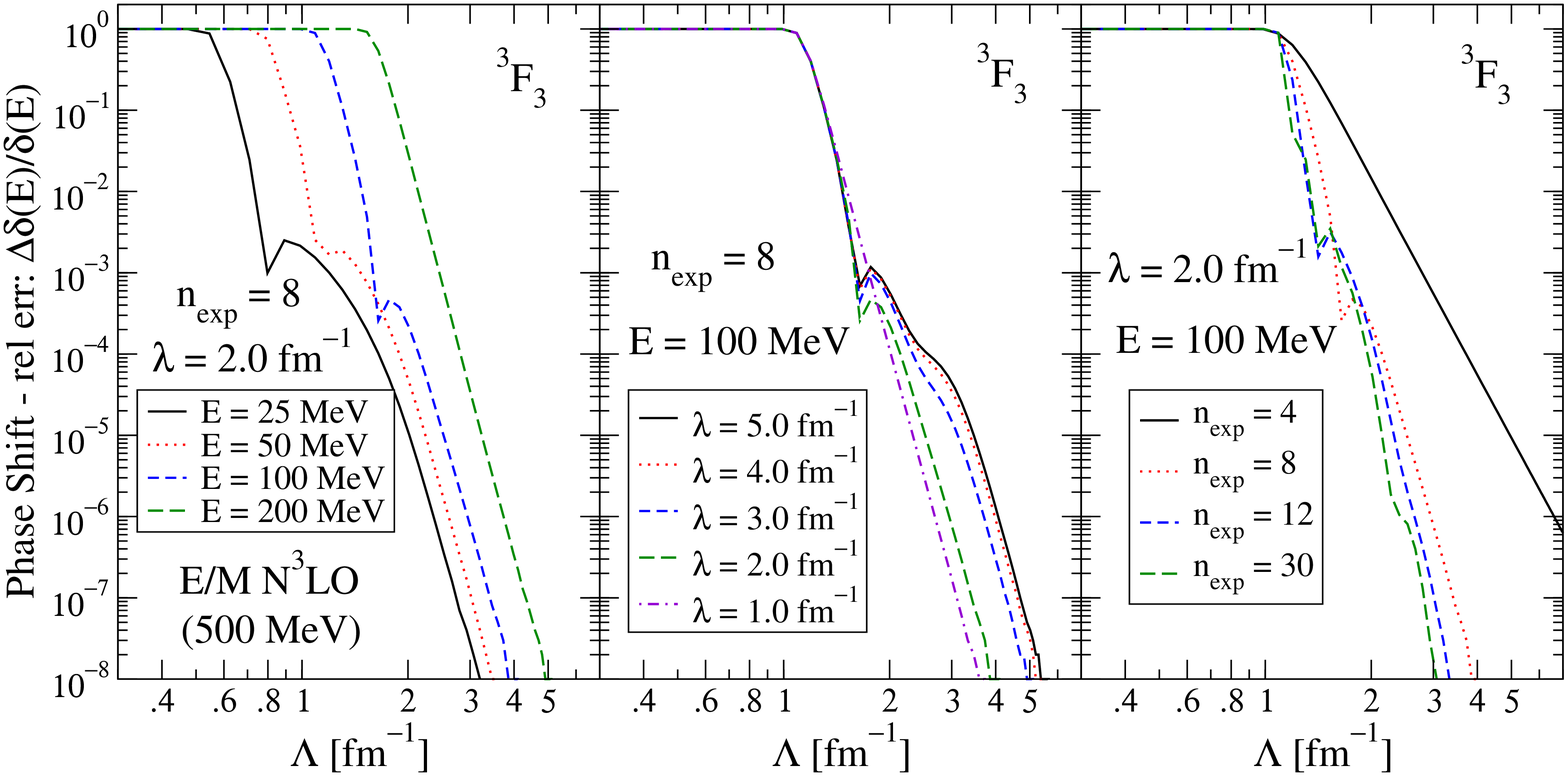}
\end{center}
\captionspace{The phase shift errors computed in select partial waves.
Other channels exhibit the same power-law dependence of the error
for $\Lambda > \lambda$.}
\label{fig:ps_err_other_channels}
\end{figure}

Indeed, the behavior of the errors in the decoupling region, where
$\Lambda > \lambda$, can be directly understood as a consequence of
the partial diagonalization of the evolved potential. The calculation
of the phase shift at a low-energy $k^2 \ll \lambda^2$ will involve an
integral over $p$ of $V_{\lambda,\Lambda}(k,p)$. But the potential
cuts off the integral at roughly  $p^2 \approx k^2 + \lambda^2 <
\Lambda^2$,  which means that we can expand the difference in the
uncut and cut potentials:
\beqn
 \delta V_{\lambda,\Lambda}(k,p) \equiv
    V_{\lambda}(k,p) - V_{\lambda,\Lambda}(k,p)
    \approx \left(\frac{k^{2n}}{\Lambda^{2n}}
                 + \frac{p^{2n}}{\Lambda^{2n}}  \right) V(k,p)
    \;. 
    \label{eq:deltaV}
\eeqn
Simple perturbation theory in $\delta V$ then predicts the dependence
of the phase shift error to be $1/\Lambda^{2n}$, which is the
power-law dependence seen in Figs.~\ref{fig:ps_err_vs_cut} and
\ref{fig:ps_err_other_channels}. The accuracy of first-order
perturbation theory is evidenced by the  constant slope of the error
curves, which translates into perturbatively small residual coupling.

The detailed dependence on the energy and $\lambda$ is not so
trivially extracted. However, the weak dependence on $E_{\rm lab} \leq
100\,\mbox{MeV}$ and strong dependence on $\lambda < 3 \fmi$ at fixed
$\Lambda$ seen in  Fig.~\ref{fig:ps_err_vs_cut} implies that the
integration picks up the scale $\lambda$, so that the dominant error
scales as $(\lambda/\Lambda)^{2n}$. This is, in fact, observed
numerically for intermediate values of $\lambda$ (e.g., for $1.8\fmi <
\lambda < 2.8\fmi$ when $\Lambda = 3\fmi$). 

We checked this decoupling behavior in different partial waves and for
other N$^3$LO potentials and found the same perturbative region in all
cases. The plots of Fig.~\ref{fig:ps_err_vs_cut} are reproduced in
Fig.~\ref{fig:ps_err_other_channels} for representative partial waves.
The potential in the S waves typically passes through zero for momenta
in the region of $\lambda = 2\fmi$ (as in
Fig.~\ref{fig:ps_film_strips_500MeV}), which might lead one to
associate decoupling with  this structure. The error plots for other
partial waves that lack this structure show that it is a more general
consequence of the SRG evolution.

Note that in higher partial waves, such as $^3\!F_3$, the phase shift
is already well decoupled and therefore doesn't benefit as much from
the SRG evolution. Errors shown for this partial wave in the bottom
strip of Fig.~\ref{fig:ps_err_other_channels} are below $10^{-4}$ at
the start of the decoupling region (note the shoulders at $\lambda$ in
the center panel). Such partial waves are the large angular momentum
components of the partial wave decomposition. Due to the centrifugal
barrier term they probe long-distance forces and have strength only at
low momenta. Therefore, low-momentum and high-momentum states are
automatically decoupled in these channels and little change is
provided by the SRG here.

As noted above, the shoulder in the relative error log-log plots
corresponding to the onset of the decoupling region saturates at
large  $\lambda$ due to the high-momentum cutoff introduced into the
initial potential during its initial formulation. In the plots of
Fig.~\ref{fig:ps_err_vs_cut} the shoulders for the $\lambda=4$ and 5
curves occurred at about $\lambda = 3$ since the \xeft\ potentials did
not have off-diagonal matrix elements at momenta larger than about
$k^2 \approx 10-11 \fm^{-2}$. As counterpoint to this effect,
Figure~\ref{fig:av18_saturation} shows the relative error plot, with a
range of $\lambda$'s at E = 100 MeV and $\nexp = 8$, using the Argonne
$v_{18}$ potential, which has matrix elements extending out to momenta
of 30 $\fmi$. Here one can see less saturation out to $\lambda = 10$
as there are plenty of off-diagonal matrix elements for the SRG to
suppress between $k^2 \ge 100 \fm^{-2}$ and $k^2 = 36 \fm^{-2}$, which
corresponds to $\lambda = 6 \fmi$. 

\begin{figure}[tbh!]
\begin{center}
\dblpic{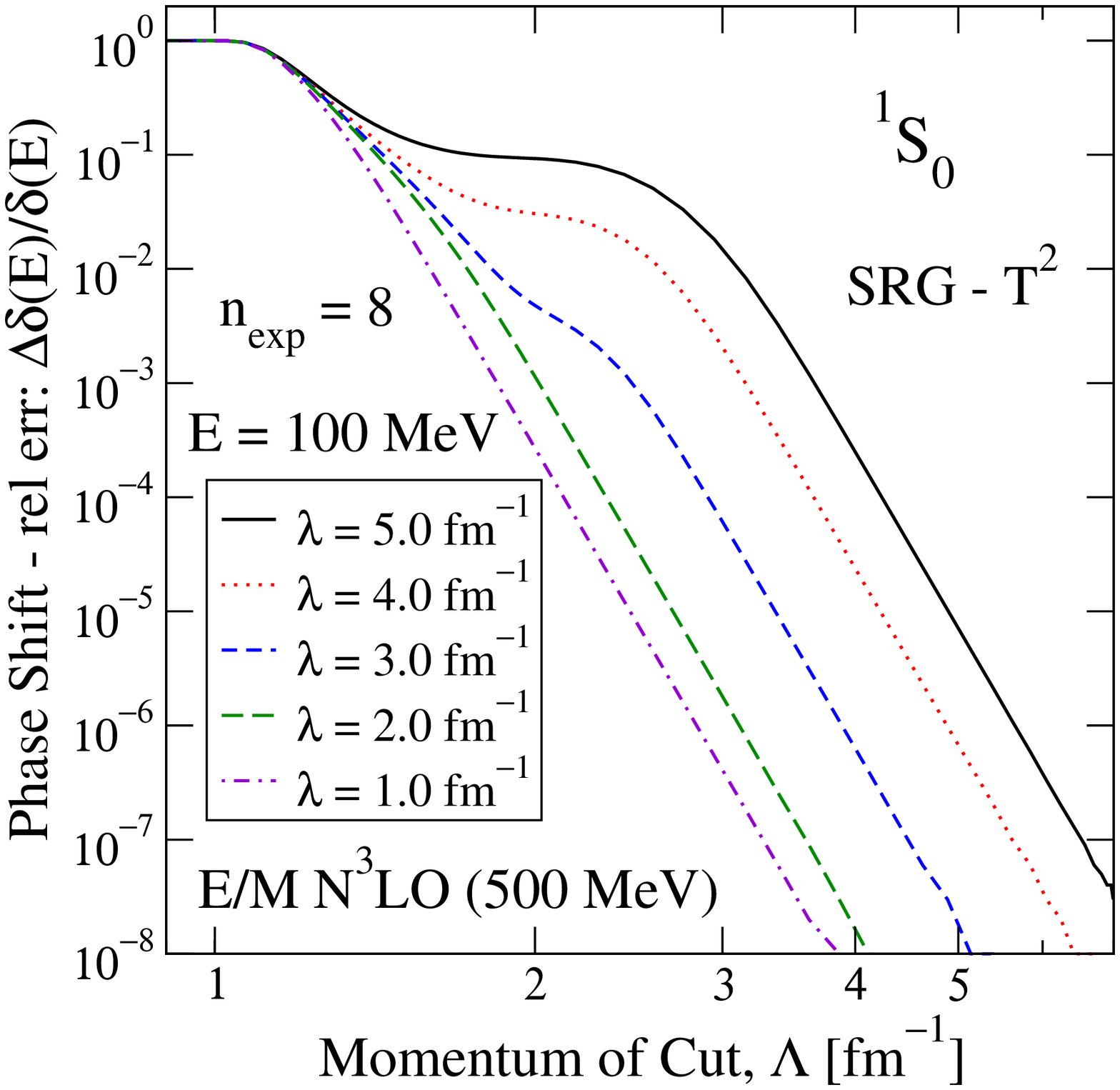} 
\hfill
\dblpic{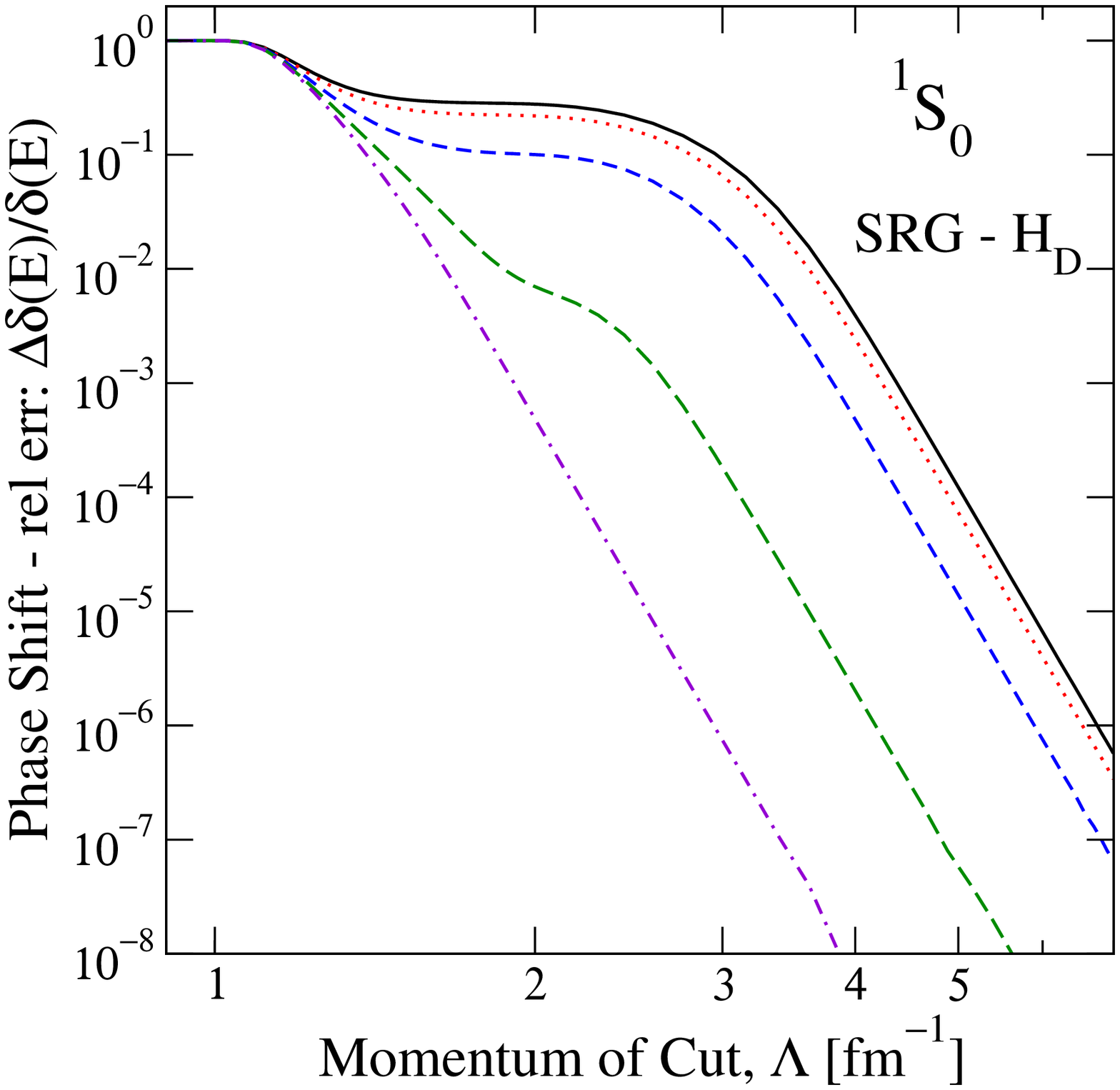} 

\dblpic{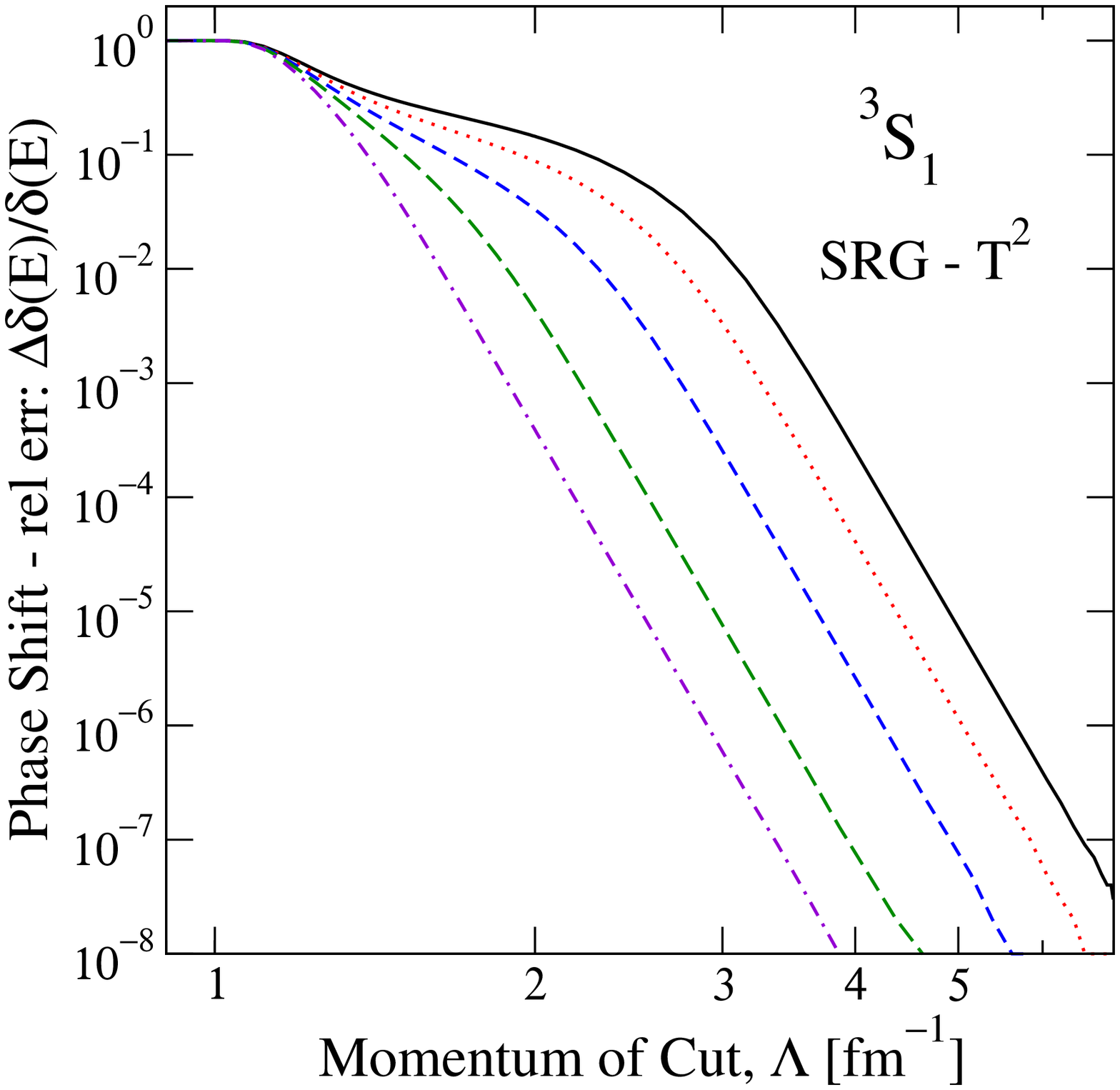} 
\hfill
\dblpic{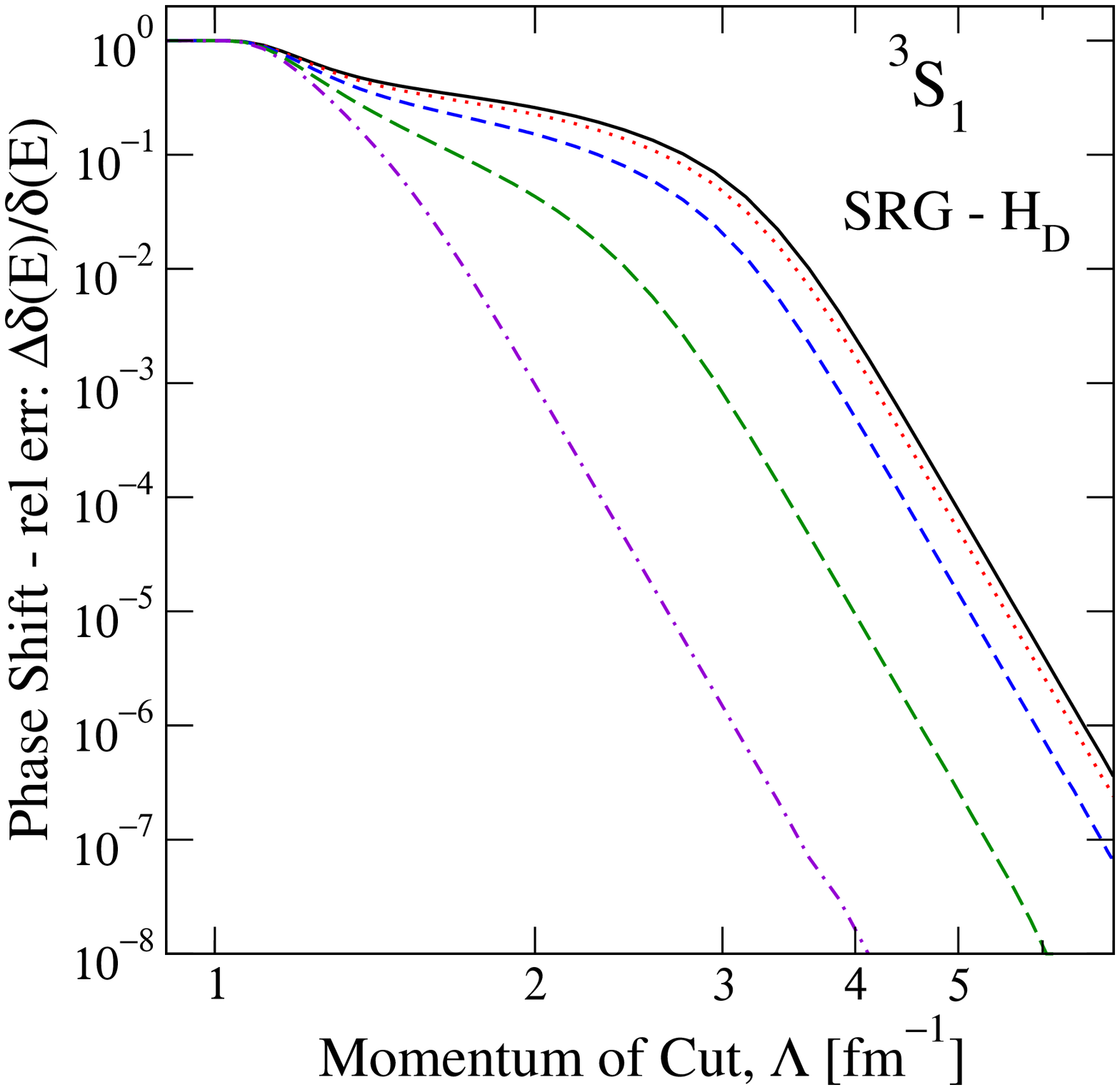} 
\end{center}
\captionspace{Sample phase shift error plots using other choices of $G_s$.
On the left is $G_s = \Trel^2$ and on the right is $G_s = H_D$ for
partial waves $^1S_0$ (top) and $^3S_1$ (bottom).}
\label{fig:other_Gs}
\end{figure}

We repeated the phase shift calculations with other choices of the SRG
generator, $G_s$, that are diagonal in momentum space, including
$\Trel^2$ and  $H_{\rm D} = \Trel + V_{\rm D}$, where $V_{\rm D}$ is
the (running) diagonal part of the bare potential. Samples of these
results are shown in Fig.~\ref{fig:other_Gs}. on the left (right) are
plots using $G_s = T^2$ ($H_D$) and on the top (bottom) are $^1S_0$
($^3S_1$). We found that these other choices for $G_s$ do not alter
the qualitative features of the power-law behavior region of the
previous error plots. This provides further evidence that the high-
and low-energy decoupling results primarily from the partially
diagonalized nature of the evolved potential.


\section{Decoupling and Deuteron Observables}
\label{sec:deuteron}

\begin{figure}[tbh!]
\begin{center}
\triplepic{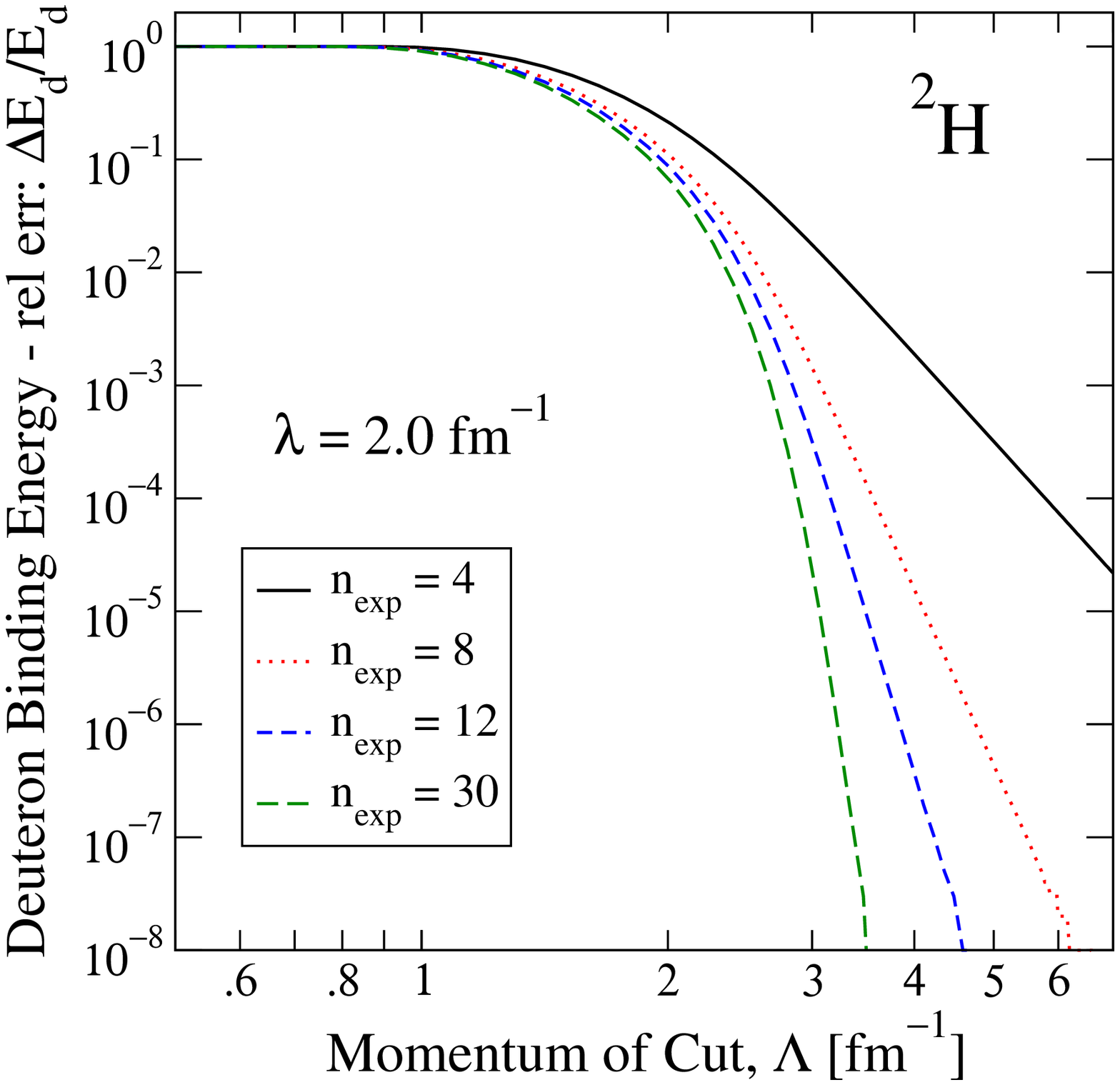} 
\hfill
\triplepic{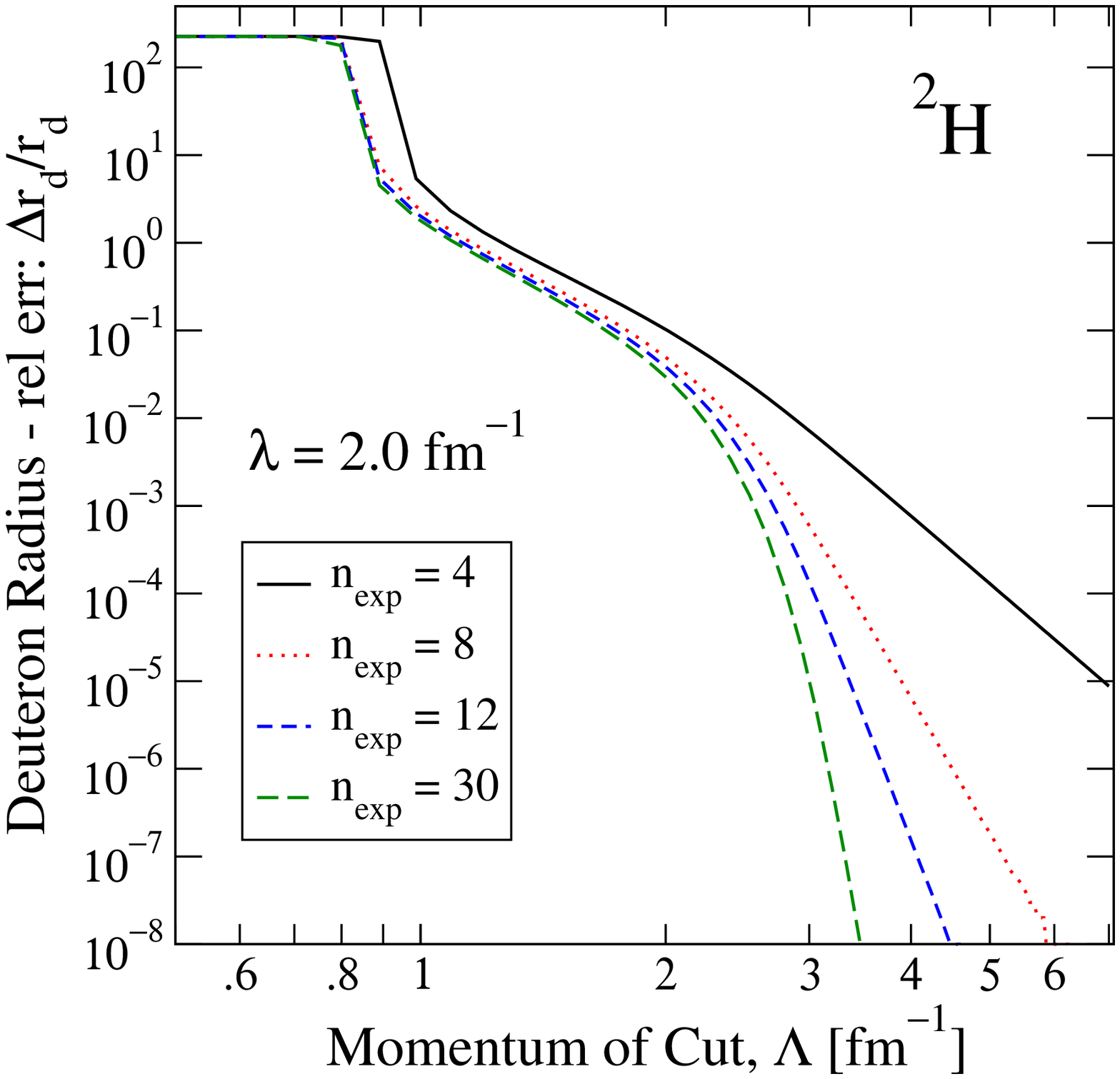} 
\hfill
\triplepic{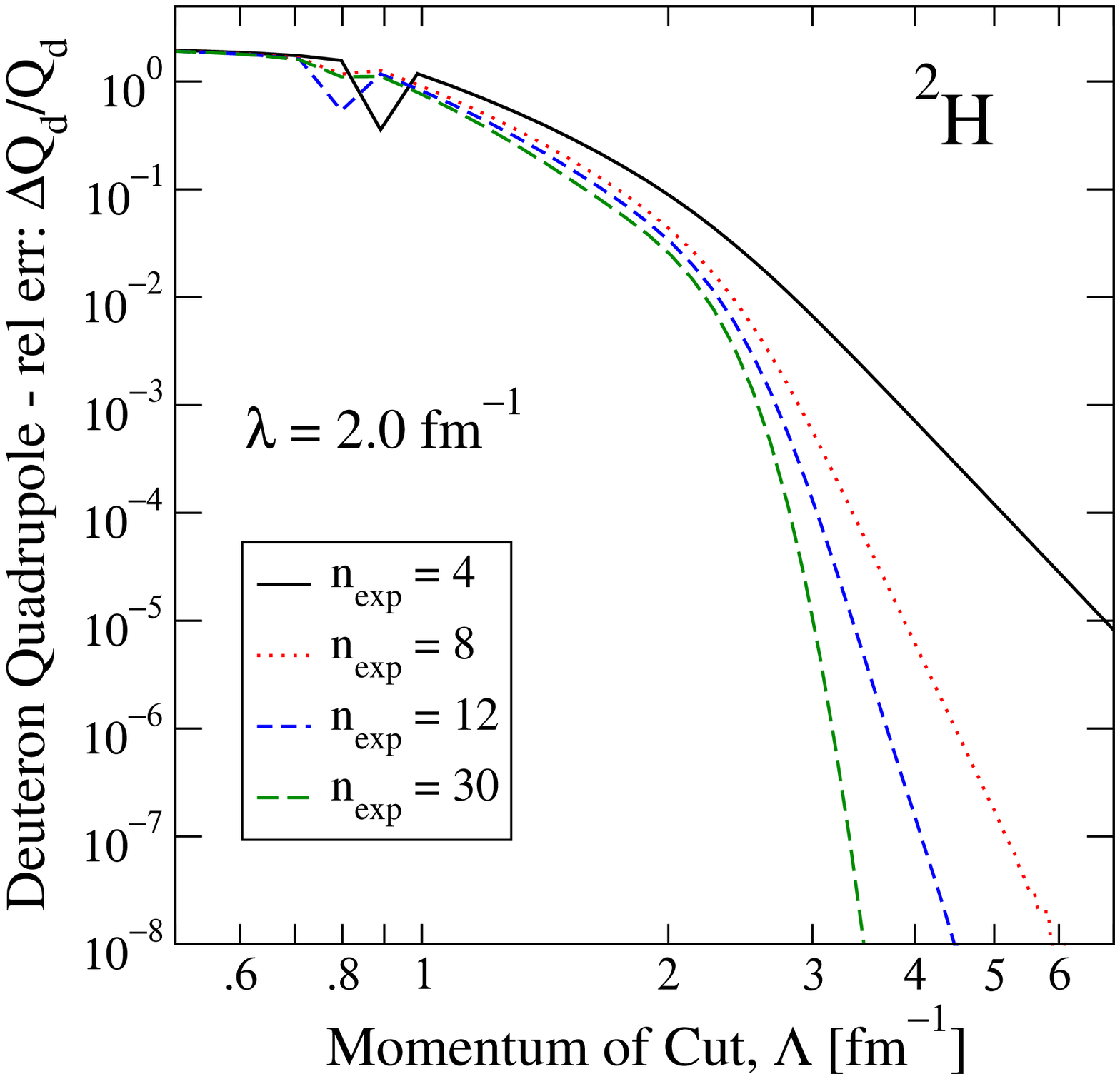} 
\end{center}
\captionspace{The relative error vs. cut parameter $\Lambda$ of  the
deuteron energy (left), rms radius (center), and quadrupole moment
(left) of the deuteron with several values
of the regulator  parameter $n$ indicated in the legends.  In
each case, the relative error is confirmed by the first-order
perturbation theory shown in Eq.~\eqref{eq:deltaV}.}
\label{fig:deut_err_vs_cut}
\end{figure}

To test the generality of the observations made for phase shifts,  the
same decoupling test techniques were applied to other low-energy observables such as
the deuteron binding energy, radius, and quadrupole moment.  The
binding energy and momentum-space wavefunction  were computed using
standard eigenvalue methods. The computation of $Q_d$ and $r_d$ from
the wavefunction uses~\cite{Bogner:2006vp},
\bea
Q_d = -\frac{1}{20} \int_0^\infty \! dk \, \biggl[
\sqrt{8} \, \biggl( k^2\, \frac{d\wt u(k)}{dk} \frac{d\wt w(k)}{dk}
+ 3 k\, \wt w(k) \frac{d\wt u(k)}{dk} \biggr) \nonumber \\
 + k^2 \biggl( \frac{d\wt w(k)}{dk}\biggr)^2 + 6 \, \wt w(k)^2 \biggr] \;, 
\label{eq:quad}
\eea
and
\beqn
r_d = \frac{1}{2} \biggl[ \int_0^\infty \! dk \, \biggl\{ \biggl(
k \, \frac{d\wt u(k)}{dk} \biggr)^2 + \biggl(k \, \frac{d\wt w(k)}{dk}\biggr)^2 
+ 6 \, \wt w(k)^2 \biggr\} \biggr]^{1/2} \;,
\label{eq:rad}
\eeqn
where $\wt u(k$) and $\wt w(k)$ correspond to the S and D components 
of the deuteron wavefunction respectively. We again computed relative 
errors in these observables and, as shown in
Fig.~\ref{fig:deut_err_vs_cut},  the errors show the same behavior as
observed for the phase shifts. That is, a power-law drop-off in the
error begins at $\Lambda$ just above  $\lambda$, with a slope
determined by the sharpness of the regulator as given by $n$.  

As with the phase shift, the analytic dependence of the error from
cutting the potential can be estimated directly in perturbation
theory.  In this case, partial diagonalization of the potential means
that the deuteron wave function has negligible momentum components
starting slightly above $\lambda$. This in turn validates the
expansion in Eq.~\eqref{eq:deltaV} and the dependence of the errors on
$1/\Lambda^{2n}$. The numerical calculation of the error in
perturbation theory is plotted in Fig.~\ref{fig:deut_err_vs_cut} and
shows close agreement in the decoupling region $\Lambda > \lambda$.


\section{Decoupling and Few-Body Energies with the NCSM}
\label{sec:ncsm_calcs}

The calculations described above have been only for two-particle
systems. We can test whether the high-energy decoupling behavior
extends to few-body systems by using No-Core Shell Model (NCSM)
calculations of ground-state energies with the Many-Fermion Dynamics
(MFD) code~\cite{MFD}.

As noted before, the NCSM is a harmonic oscillator basis calculation
of many-body systems where states can be explicitly enumerated. The
size of the basis is tracked by the parameter $\nmax$, the number of
oscillators functions used in the basis. Calculations are variational
in $\nmax$, improving accuracy with larger bases, and in $\hw$ with a
minimum occurring at the optimal value of $\hw$ for a given space and
$\nmax$. Here calculations are made for selected $\nmax$ and $\hw$ values.
Also, note here that the NCSM is a black box calculational tool, which
takes as input the harmonic oscillator matrix elements of potentials
evolved and cut in momentum-space and provides NN-only calculations of
few-body observables like the binding energies of the light nuclei.
This is opposed to subsequent chapters where the NCSM machinery will
be developed in order to access three-body interaction matrix
elements.

Conversion to the oscillator basis introduces a unique truncation of
the interaction based on the values of $\nmax$ and $\hw$ (as detailed in
Appendix~\ref{chapt:app_osc_truncation}). However, here the two-body
Hamiltonian is being evolved in the momentum basis so that the
evolved, and therefore decoupled, potential is unitarily equivalent to
the initial $A=2$ potential. We will take advantage of that decoupling
to avoid the oscillator basis truncations at a given basis size,
$\nmax$, and we will use the same systematic cutting as in the previous
sections to study the behavior of decoupling in few-body systems.

In this section, only NN interactions have been considered, with the
testing of decoupling for many-body forces deferred to later chapters.
Because induced three-body (and higher) forces are not included, the
converged values will differ significantly with $\lambda$. In spite of
this, we will still be able to see the evidence for NN decoupling in
these systems. Also, note that the general features of the SRG
exhibited in Section~\ref{sec:decoupling_mechanics} implies that
off-diagonal matrix elements of the three-body force will be
suppressed as well, with decoupling as an expected consequence. The
issue of three-body force evolution will be covered explicitly in
chapters~\ref{chapt:OneD} and \ref{chapt:ncsm}.

\begin{figure}[tbh!]
\begin{center}
\dblpic{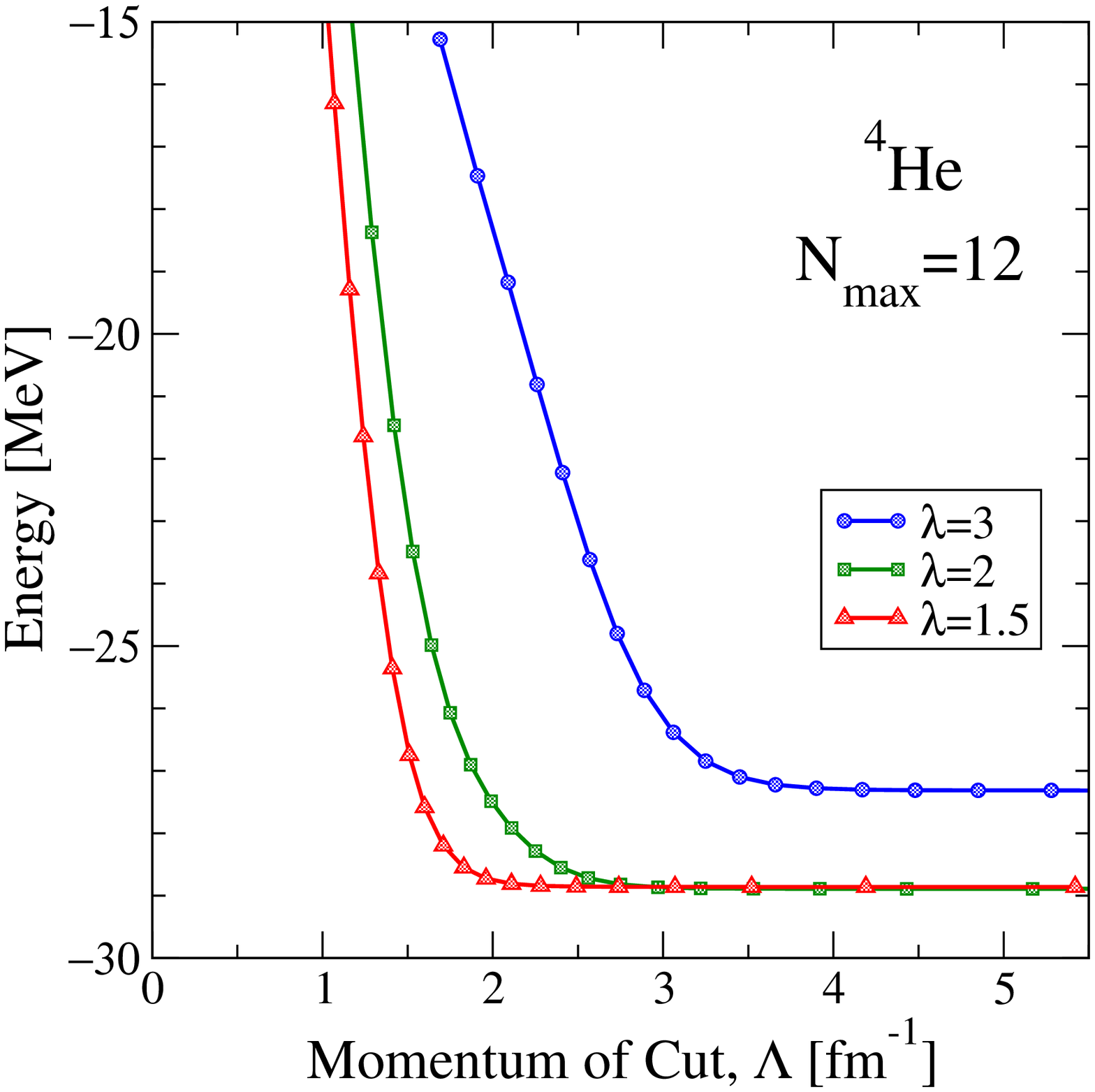} 
\hfill
\dblpic{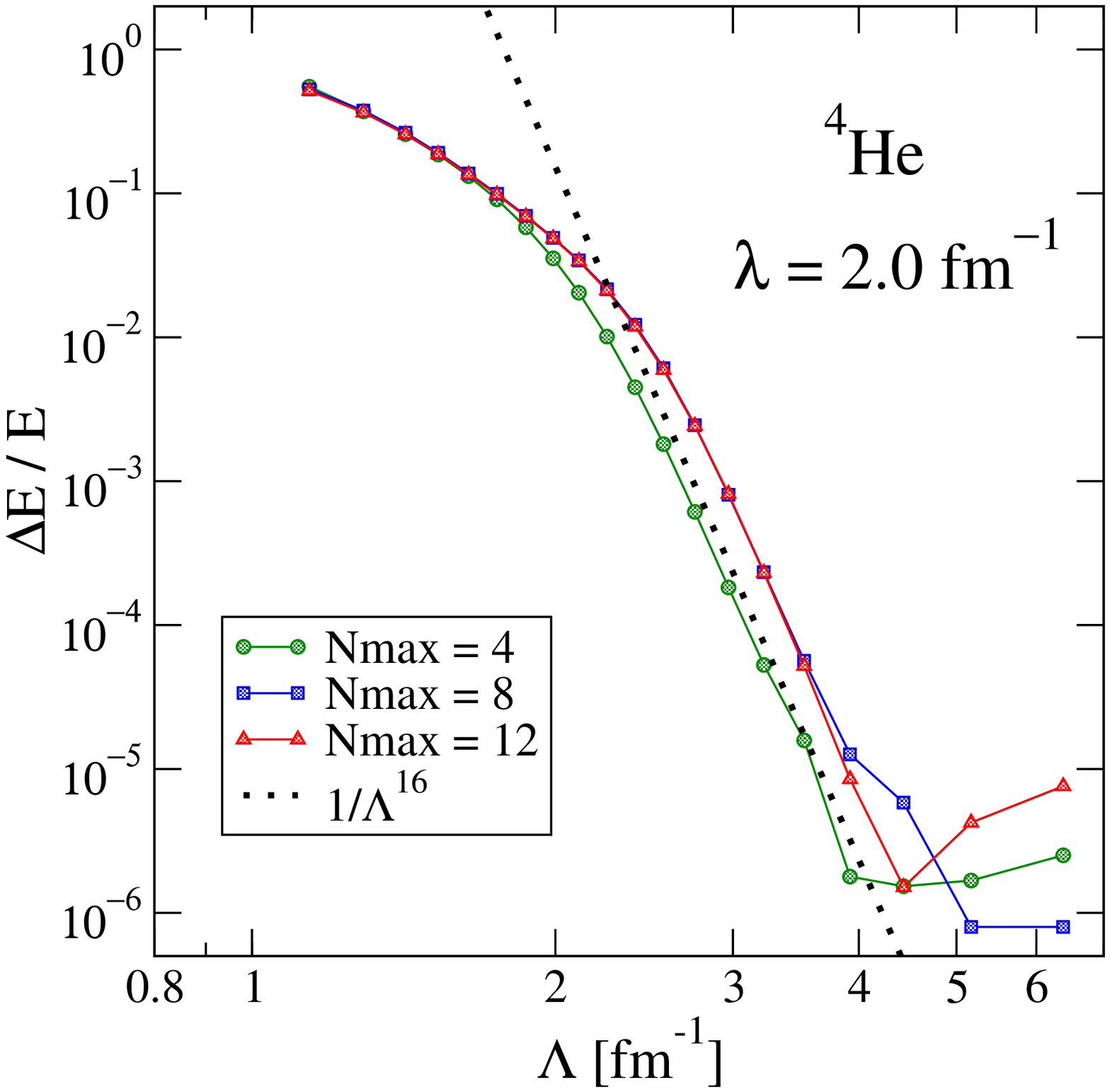}
\end{center}
\captionspace{Calculations of the $^4\rm{He}$ ground-state energy using the
NCSM. On the left is  the energy obtained from the NCSM for potentials
evolved to several different $\lambda$ values as a function of the cut
(regulator) momentum $\Lambda$ with $n=8$.  On the right is the
relative  error of the energy for the $\lambda = 2\fmi$ case as a
function of the cut momentum (with $n=8$) for several different
harmonic oscillator basis sizes. Also shown is the slope of the error
in the decoupling region predicted from perturbation theory (dotted
line).}
\label{fig:ncsm_nuclei_err_vs_cut_He4}
\end{figure}

\begin{figure}[tbh!]
\begin{center}
\dblpic{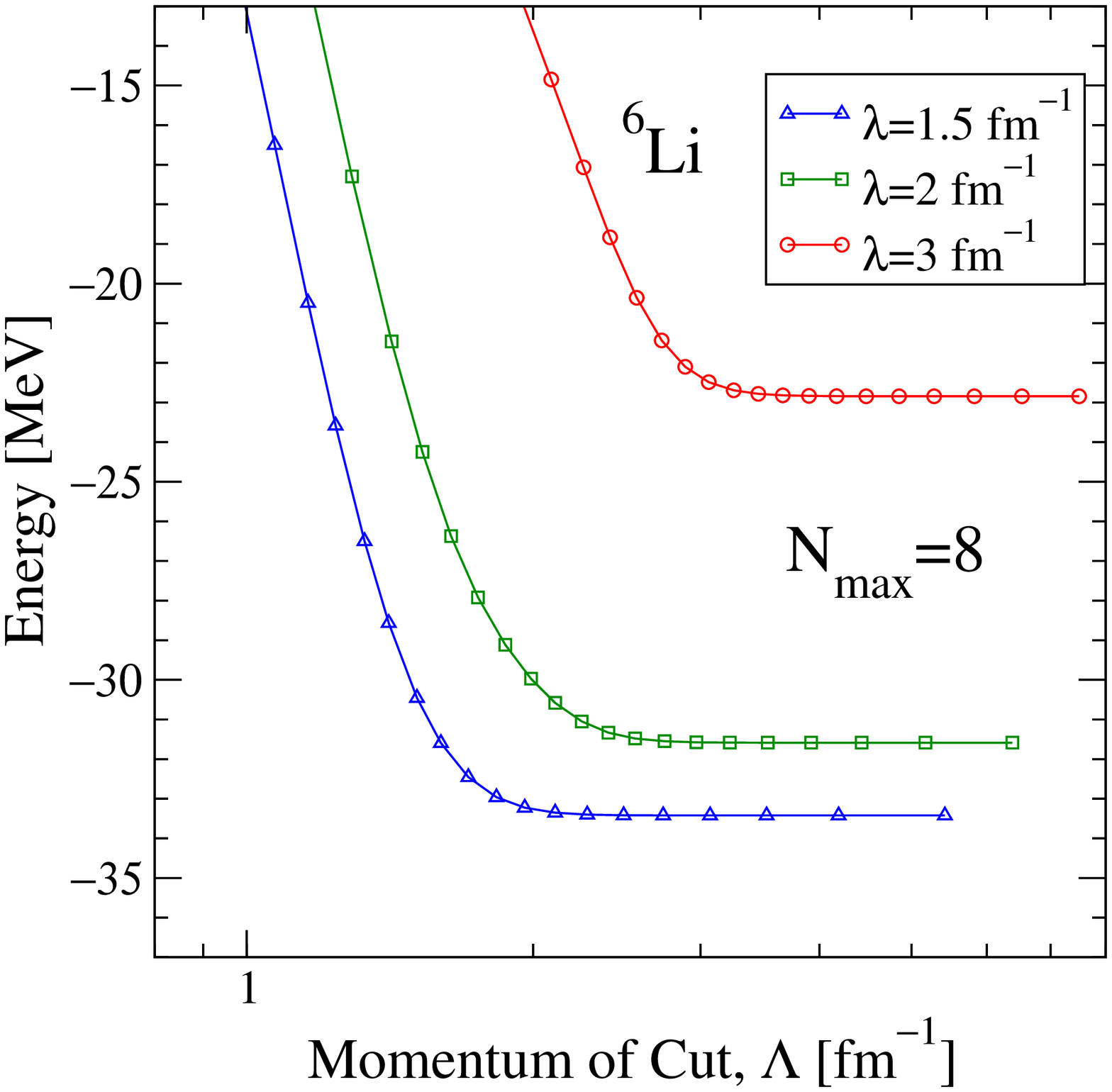} 
\hfill
\dblpic{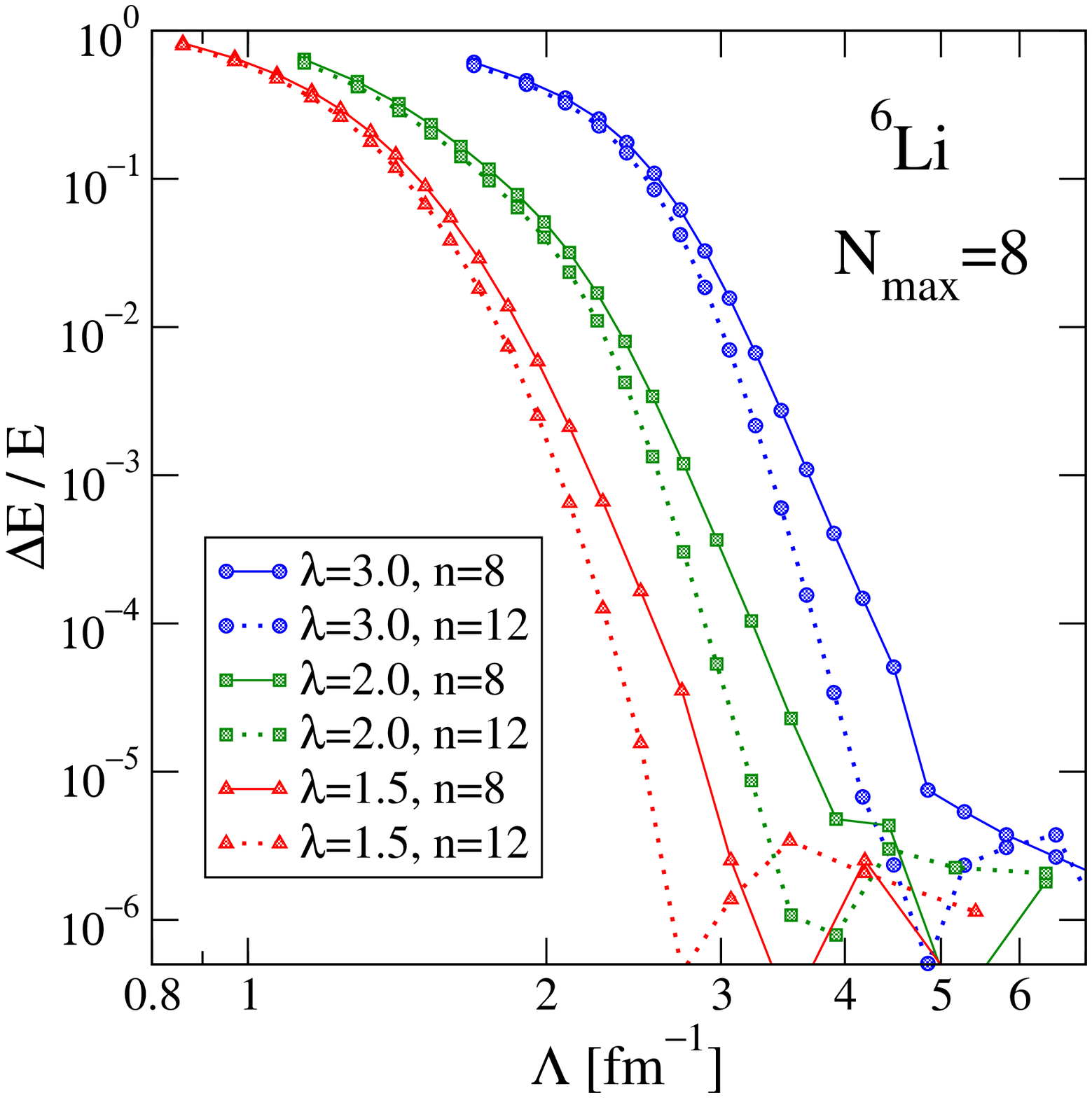}
\end{center}
\captionspace{Calculations of the $^6\rm{Li}$ ground-state energy using the
NCSM. On the left is  the energy obtained from the NCSM for potentials
evolved to several different $\lambda$ values as a function of the cut
(regulator) momentum $\Lambda$ with $n=8$.  On the right is the
relative  error of the energy for the same $\lambda$'s as a function
of the cut momentum for the same $\lambda$ values but with two values
of $n$.}
\label{fig:ncsm_nuclei_err_vs_cut}
\end{figure}

We first verified that the decoupling behavior already observed using
a direct calculation of the deuteron wavefunction is reproduced using
the MFD.  We then calculated a series of  larger nuclei, including
$^3$H, $^4$He, and $^6$Li, comparing results from uncut and a range of
cut potentials evolved to different values of $\lambda$. On the left
panel of Fig.~\ref{fig:ncsm_nuclei_err_vs_cut_He4}, the $^4$He
ground-state energy is plotted versus  the regulator parameter
$\Lambda$  for several different values of the SRG flow parameter
$\lambda$.  Each of the plotted points is at a basis size $\nmax = 12$
which is within several hundred keV of the energy from extrapolating
to $\nmax$$ = \infty$. A similar plot for $^6$Li is given in the left
panel of  Fig.~\ref{fig:ncsm_nuclei_err_vs_cut} using  a basis size
$\nmax = 8$, also within several hundred keV of the extrapolated
energy for $\lambda = 1.5\fmi$ but still several MeV off for  $\lambda
= 3.0\fmi$.

Both examples show that when the potential is cut with $\Lambda$
comparable to $\lambda$ or lower, the converged energy is
significantly different from the asymptotic uncut value, while it
approaches that value rapidly as $\Lambda$ moves above $\lambda$. 
This means that, for smaller $\lambda$, more high  momentum matrix
elements can be discarded without a loss of accuracy.  This decoupling
explains  the greatly improved convergence with basis size seen in
the  NCSM for corresponding $\lambda$ values~\cite{Bogner:2007rx}. As
noted above, the uncut ($\Lambda \rightarrow \infty$) energies vary
for each $\lambda$ because the SRG evolution includes the NN
interaction only; the closeness of the results for $\lambda = 2\fmi$
and $3\fmi$ for $^4$He is coincidental (see Ref.~\cite{Bogner:2007rx}
for further discussion about the running of the energies).

The quantitative behavior of the relative error parallels that
observed for two-body observables, as seen  on the right panels of
Figs.~\ref{fig:ncsm_nuclei_err_vs_cut_He4} and
\ref{fig:ncsm_nuclei_err_vs_cut}. In all cases, for a fixed value of
$\lambda$ the power decrease in the error starting with $\Lambda$
slightly above $\lambda$ is clearly seen, even though there are fewer
digits of precision in the NCSM results (so the relative error is in
the range $10^{-6}$--$10^{-5}$ at best). The same perturbative
residual coupling is seen for different basis sizes, with the slope
given by the dependence $1/\Lambda^{2n}$, although the onset of the
decoupling region shifts to higher $\Lambda$ until the calculation is
near convergence (see Fig.~\ref{fig:ncsm_nuclei_err_vs_cut_He4}). 
Similar results are found for other nuclei and for other values of
$\lambda$.


\section{Block Diagonalization}
\label{sec:decoupling_block_diagonal}

\begin{figure}[htb!]
\bc
\dblpichgt{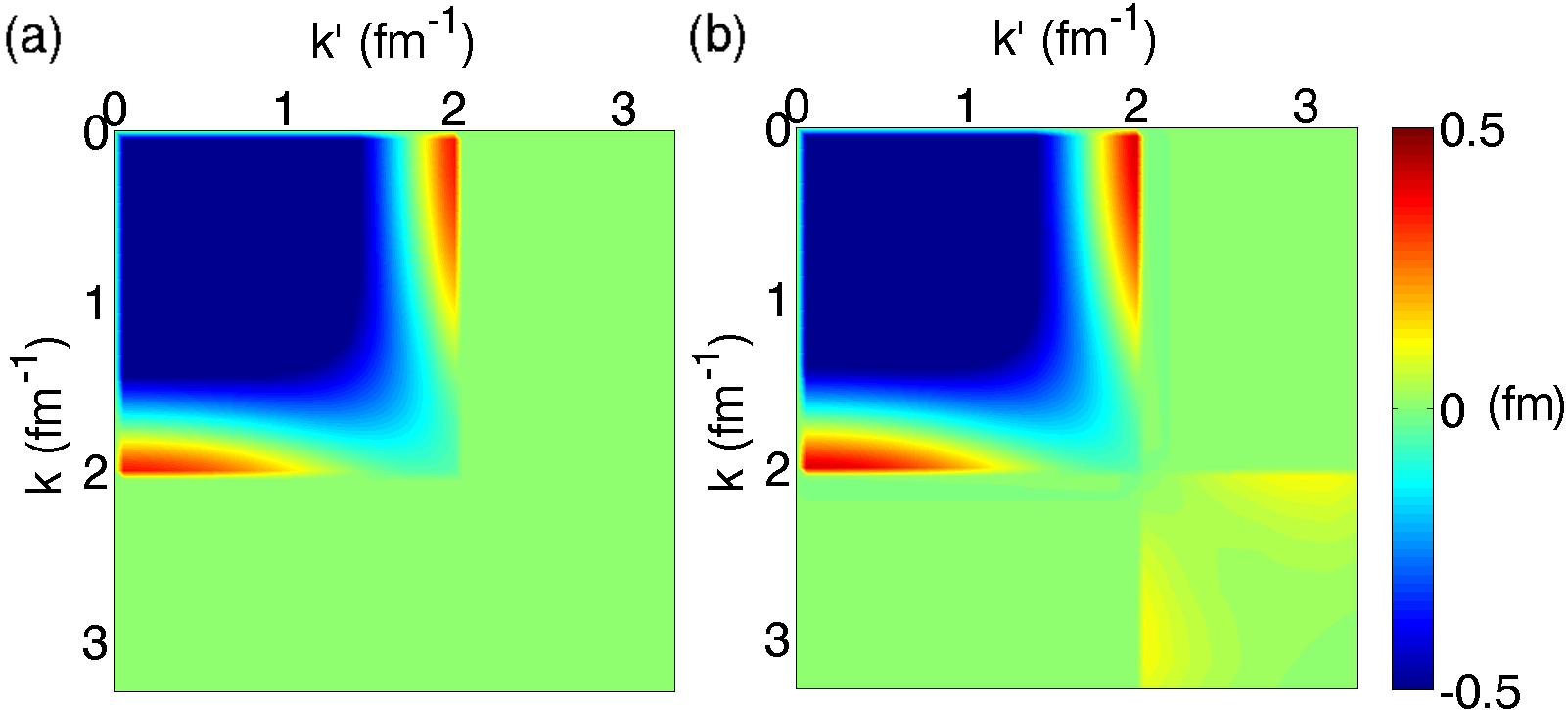}
\ec
\captionspace{Comparison of momentum-space $\vlowk$ (a)
and SRG (b) block-diagonal potentials with $\Lambda_{\rm BD}= 2\fmi$ 
evolved from an N$^3$LO $^3$S$_1$ potential~\cite{N3LO}.}
\label{fig:vlowk}
\end{figure}

\begin{figure}[htb!]
\bc
\dblpichgt{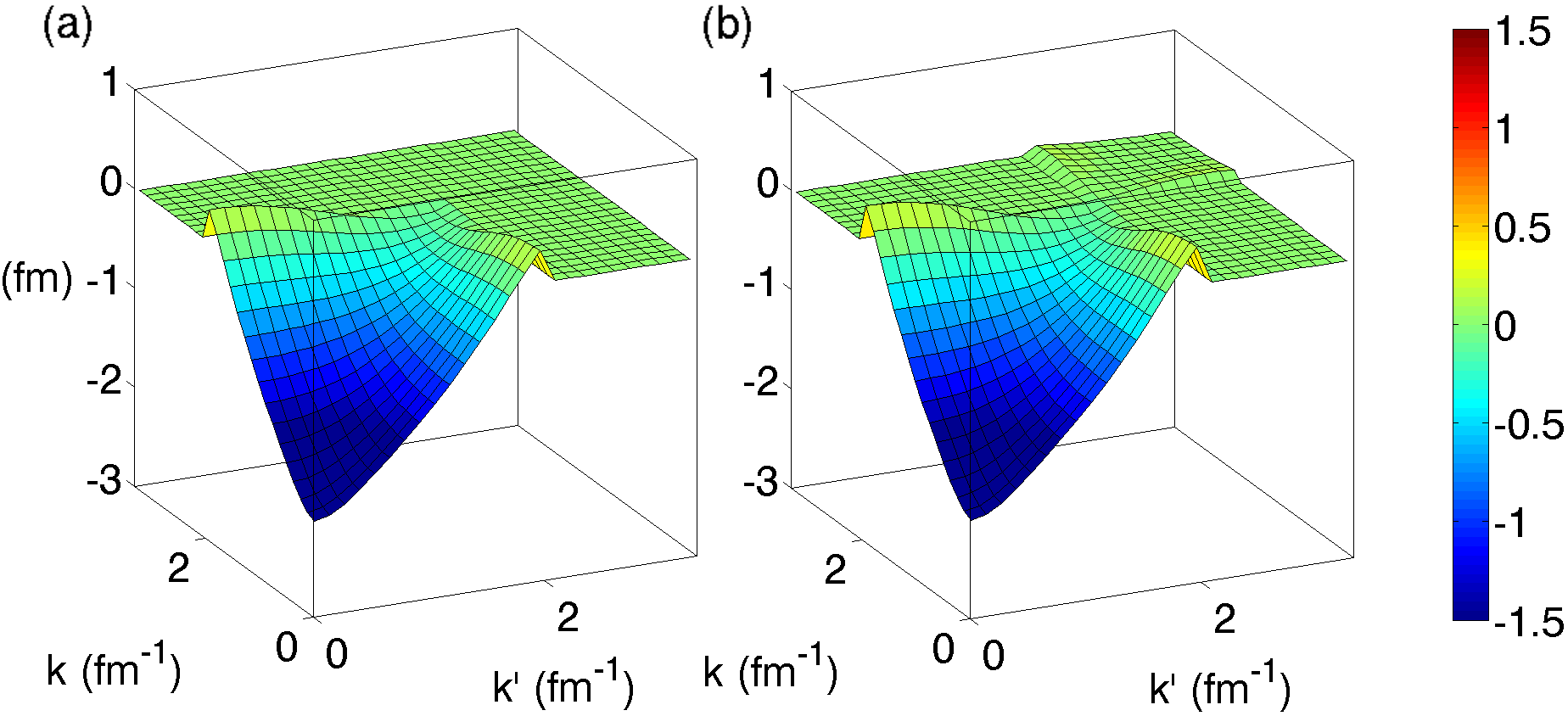}
\ec
\captionspace{Comparison of momentum-space $\vlowk$ (a)
and SRG (b)  block-diagonal potentials with $\Lambda_{\rm BD}= 2\fmi$
evolved from an N$^3$LO $^3$S$_1$ potential~\cite{N3LO}.}
\label{fig:vlowksurface}
\end{figure}

As demonstrated in sections~\ref{sec:phase_shifts} --
\ref{sec:ncsm_calcs}, decoupling between low-energy and high-energy
matrix elements is naturally achieved in a momentum basis by choosing
a momentum-diagonal flow operator such as the kinetic energy $\Trel$
or the diagonal of $H_s$; either drives the  Hamiltonian toward
\emph{band-diagonal} form. Renormalization Group (RG) methods that
evolve NN interactions with a sharp or smooth cutoff in relative
momentum, known generically as $\vlowk$, usually rely on equations
based on the half-on-shell invariance of the
two-nucleon T matrix~\cite{vlowk_diff_int,Bogner:2006vp}. These approaches
achieve a \emph{block-diagonal} form characterized by a cutoff
$\Lambda_{\rm BD}$ (see left plots in Figs.~\ref{fig:vlowk} and
\ref{fig:vlowksurface}) using a Lee-Suzuki type transformation which
unitarily transforms matrix elements into a low-momentum $P$ space and
a high-momentum $Q$ space. As usually implemented they set the
high-momentum matrix elements to zero but this is not required.

The SRG can also produce block-diagonal decoupling similar to the
sharp $\vlowk$ form by choosing a block-diagonal flow
operator~\cite{Elena99,Gubankova:2000cia}, 
\beqn
  G_s =  \left( 
        \begin{array}{cc}
          PH_{s}P & 0   \\
          0     & QH_{s}Q
        \end{array}
        \right) 
       \equiv \Hbd_s 
        \;,
     \label{eq:Hbd}   
\eeqn
with projection operators $P$ and $Q = 1 - P$.  In a partial-wave
momentum representation, $P$ and $Q$ are step functions defined by a
sharp cutoff $\Lambda_{\rm BD}$ on relative momenta. This choice for
$G_s$, which means that $\eta_s$ is non-zero only where $G_s$ is zero,
suppresses off-diagonal matrix elements such that the Hamiltonian
approaches a block-diagonal form as $s$ increases (or $\lambda$
decreases).

The plots in Figs.~\ref{fig:vlowk} and \ref{fig:vlowksurface} show a
comparison between the potentials renormalized via the $\vlowk$
procedure (left) and those evolved with the block-diagonal SRG of
Eq.~\eqref{eq:Hbd} (right). The initial potential used here is again
the N$^3$LO potential from Ref.~\cite{N3LO}. Figure \ref{fig:vlowk}
uses two-dimensional contour plots as usually presented and
Fig.~\ref{fig:vlowksurface} uses a three-dimensional surface plot for
comparison and reference. Both plots use a value $\Lambda_{\rm BD} = 2
\fmi$ with the SRG potential being evolved to $\lambda = 0.5\fmi$. The
agreement between $\vlowk$ and SRG potentials for momenta below
$\Lambda_{\rm BD}$ is striking. While the color scale is a bit saturated
in the former (because we tried to keep the $Q$-space visible), the
latter helps give perspective. A similar degree of universality is
found in the other partial waves. Deriving an explicit connection
between these approaches is the topic of an ongoing investigation.

One difference between these procedures is that $\vlowk$ can be
formulated as an integral transformation\footnote{$\vlowk$ has been
expressed in both integral and differential
forms~\cite{vlowk_diff_int}.} while the SRG is currently only
differential; here $\vlowk$ is implemented as a single transformation
on the initial matrix, while the SRG flows to the desired amount of
transformation. An integrated form of the SRG would be useful in
understanding the exact correspondence between the two procedures, and
to what extent $\vlowk$ is a subset of these block-diagonal SRG forms.

If one considers a measure of the off-diagonal coupling of the Hamiltonian,
\beqn
  {\rm Tr}[(Q H_s P)^{\dagger} (Q H_s P)]
    = {\rm Tr}[P H_s Q H_s P] \geqslant 0
    \;,
    \label{eq:QHPmeasure}
\eeqn
then its derivative is easily evaluated by applying the SRG 
equation, Eq.~\eqref{eq:srg8}:
\bea
  &&  \frac{d}{ds} {\rm Tr}[P H_s Q H_s P]
   \nonumber \\
    && \qquad = 
    {\rm Tr}[P\eta_s  Q(Q H_s Q H_s P - Q H_s P H_s P)]
    \nonumber \\
    & & \qquad\qquad\null +
        {\rm Tr}[(P H_s P H_s Q - P H_s Q H_s Q) Q\eta_s  P]
   \nonumber \\
   && \qquad 
     = -2 {\rm Tr} [(Q \eta_s P)^{\dagger} (Q \eta_s  P)]
     \leqslant 0
     \;.
\eea
Thus, the off-diagonal $Q H_s P$ block will decrease in general
as $s$ increases~\cite{Elena99,Gubankova:2000cia}.

\begin{figure}[tbh!]
\bc
\strip{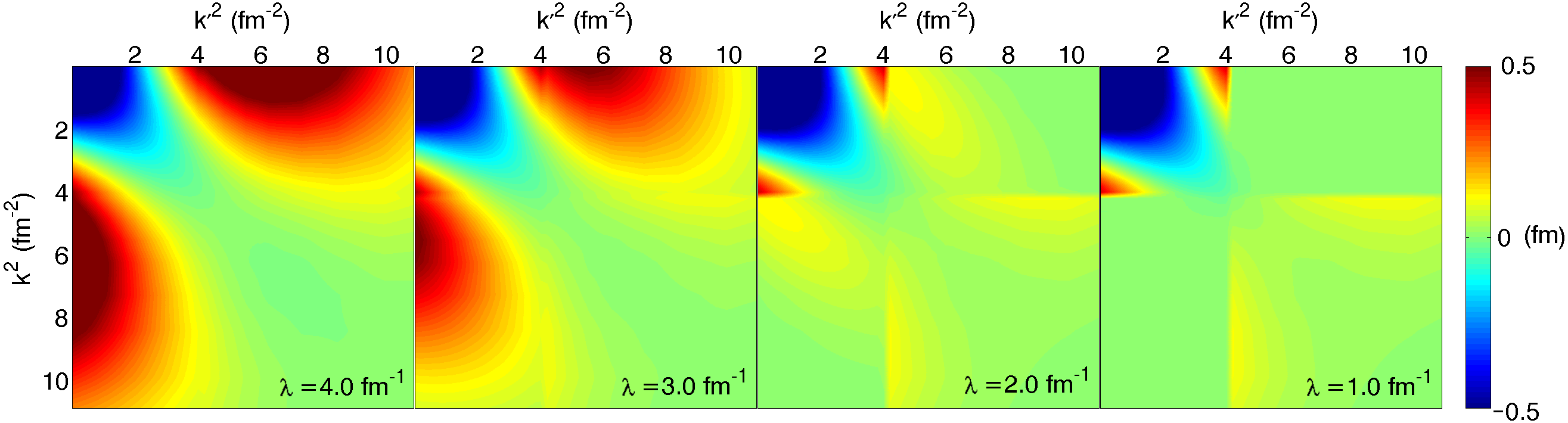}
\ec
\captionspace{Evolution of the $^3$S$_1$ partial  wave with a sharp
block-diagonal flow equation with $\Lambda_{\rm BD}=2\fmi$ at $\lambda =
4$, 3, 2, and $1\fmi$.  The initial N$^3$LO potential is from
Ref.~\cite{N3LO}. The axes are in units of $k^2$ from 0--11
fm$^{-2}$.  The color scale ranges from  $-0.5$ to $+0.5\fm$ as in
Fig.~\ref{fig:vlowk}. }
\label{fig:bd_srg_sharp_3s1}
\end{figure}

\begin{figure}[tbh!]
\bc
\strip{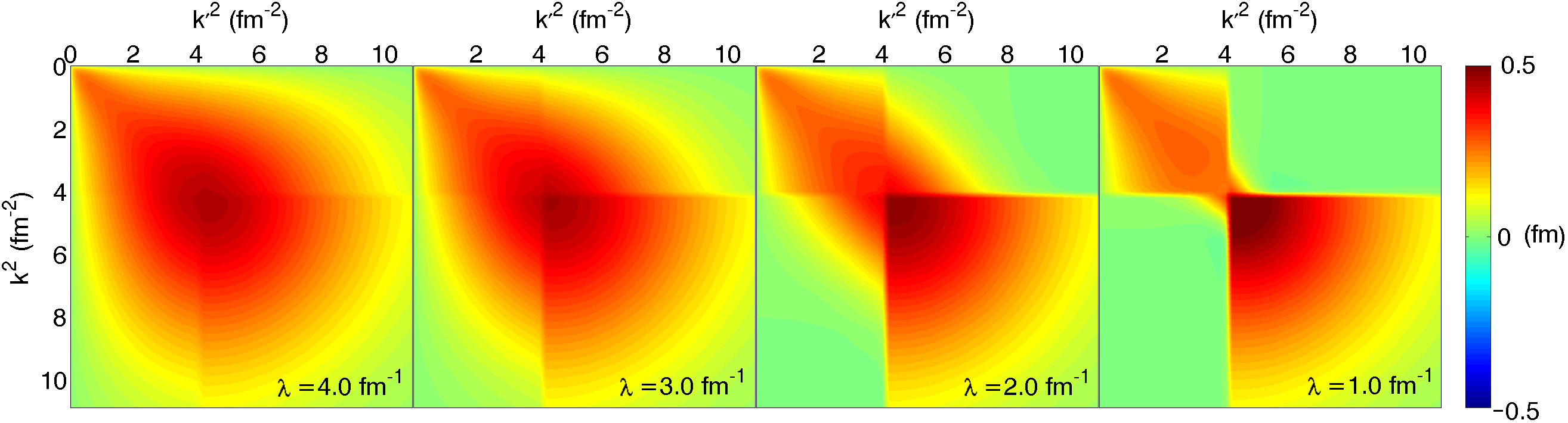}
\ec
\captionspace{Same as Fig.~\ref{fig:bd_srg_sharp_3s1} but for the $^1$P$_1$
partial wave.}
\label{fig:bd_srg_sharp_1p1}
\end{figure}

The evolution of the ``off-diagonal''  matrix elements (meaning those
outside the $PH_sP$ and $QH_sQ$ blocks) can be roughly understood from
the dominance of the kinetic energy on the diagonal. Let the indices
$p$ and $q$ run over indices of the momentum states in the $P$ and $Q$
spaces, respectively. Analogous to the approximation in
sec.~\ref{sec:decoupling_mechanics}, we can replace $P H_s P$ and $Q
H_s Q$ by their eigenvalues $E_p$ and $E_q$ in the SRG equations,
yielding~\cite{Elena99,Gubankova:2000cia}
\beqn
  \frac{d}{ds}h_{pq} \approx \eta_{pq} \energy{q} - \energy{p}\eta_{pq}
    = -(\energy{p}-\energy{q})\, \eta_{pq}
    \label{eq:hpq} 
\eeqn
and   
\beqn
  \eta_{pq} \approx \energy{p} h_{pq} - h_{pq} \energy{q} 
    = (\energy{p} - \energy{q})\, h_{pq} \;.
   \label{eq:etapq}
\eeqn
Combining these two results, we have the evolution of any
off-diagonal matrix element:
\beqn
  \frac{d}{ds}h_{pq} \approx - (\energy{p} - \energy{q})^2 \, h_{pq}
  \;. \label{eq:hpqapprox}
\eeqn
In the NN case we can approximate the difference of eigenvalues by
that for the relative 
kinetic energies,
giving an explicit solution
\beqn
   h_{pq}(s) \approx h_{pq}(0)\, e^{-s(\energyke{p} - \energyke{q})^2}
   \label{eq:explicit}
\eeqn
with $\energyke{p} \equiv p^2/M$. Thus the off-diagonal elements go to
zero with the energy differences just like with the SRG with $T_{\rm
rel}$; one can see the width of order $1/\sqrt{s} = \lambda^2$ in the
$k^2$ plots of the evolving potential in
Figs.~\ref{fig:bd_srg_sharp_3s1} and \ref{fig:bd_srg_sharp_1p1}. While
in principle the evolution to a sharp block-diagonal form means going
to $s = \infty$ ($\lambda = 0$), in practice we need only take $s$ as
large as needed to quantitatively achieve the decoupling implied by
Eq.~\eqref{eq:explicit}.

This is illustrated in Figs.~\ref{fig:bd_srg_sharp_3s1} and
\ref{fig:bd_srg_sharp_1p1} which show the evolution with $\lambda$ of
two representative partial waves ($^3$S$_1$ and $^1$P$_1$) starting
with the N$^3$LO potential from Ref.~\cite{N3LO}. One can clearly see
the transformation to a block-diagonal form. Notice the ``neck"
between the $P$ and $Q$ spaces. Its width, especially visible in the
$^1P_1$ channel, is proportional to $\lambda^2$. Presumably, though
needing further investigation, this finite and smooth interface between
the two spaces helps control errors, due to the mixing of $P$ and $Q$
space wavefunctions, that have plagued traditional Lee-Suzuki
transformations~\cite{haxton_LSerrors}.

\begin{figure}[tbh!]
\bc
\dblpic{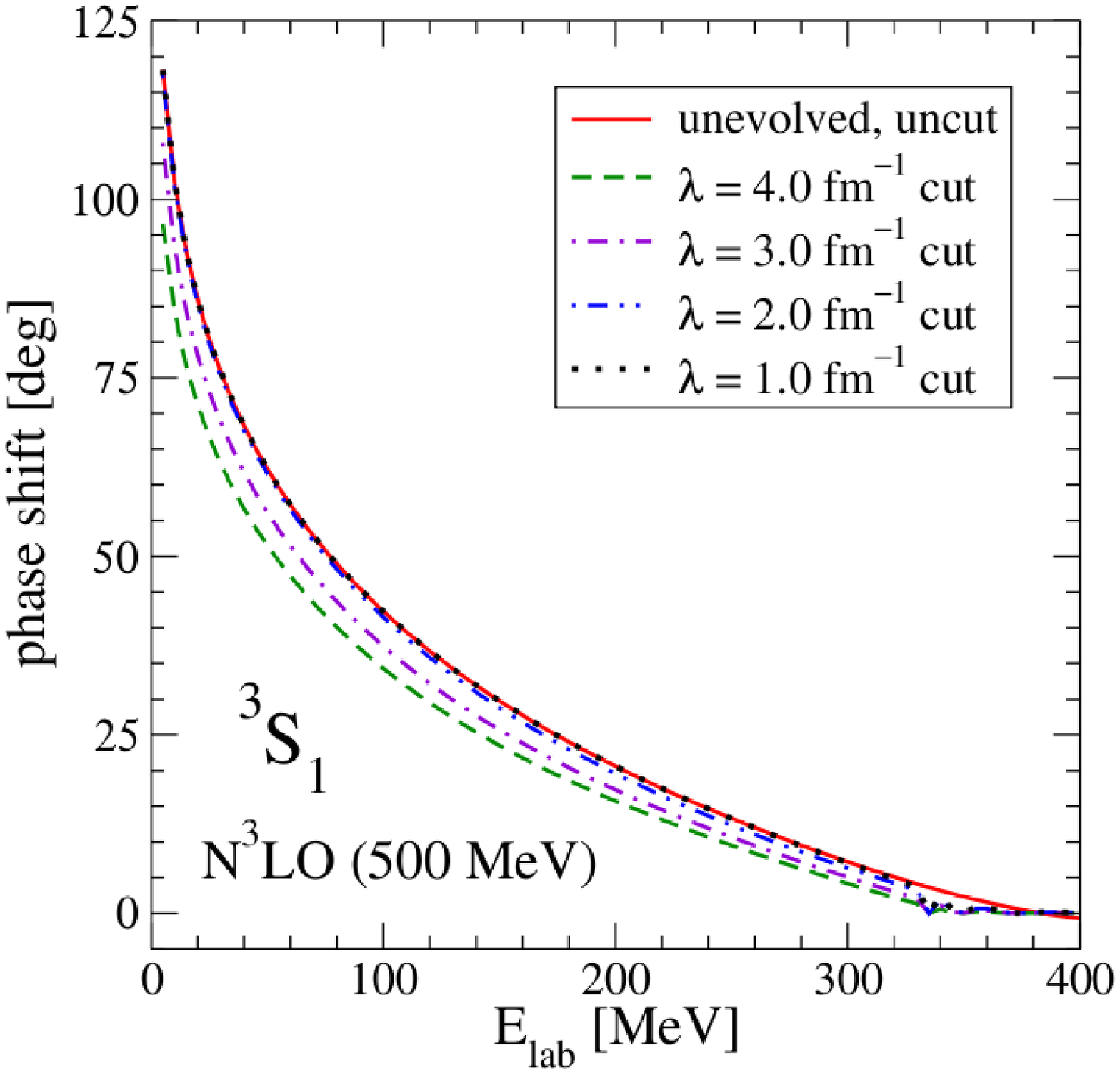}
 \hfill
\dblpic{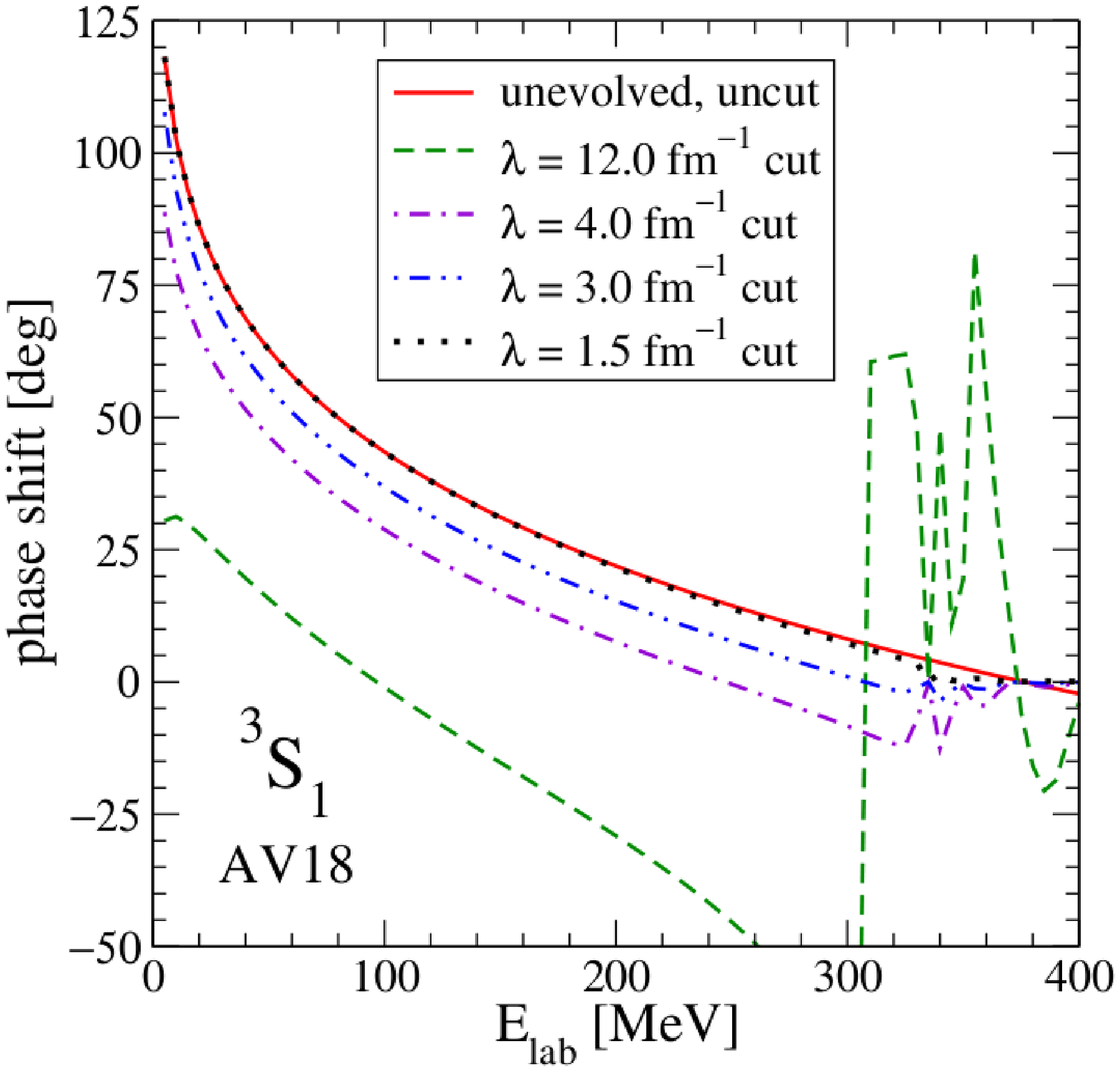}
\ec
\captionspace{Phase shifts for the $^3$S$_1$ partial wave from initial
potentials
N$^3$LO and AV18 and the evolved sharp SRG block-diagonal potential
with $\Lambda_{\rm BD}=2\fmi$ at various $\lambda$, in each case with the
potential set identically to zero above $\Lambda_{\rm BD}$.}
\label{fig:phase_sharp}
\end{figure}

Definitive tests of decoupling for NN observables are now possible for
$\vlowk$ potentials since  the unitary transformation of the SRG
guarantees that no physics is lost.  For example, in
Fig.~\ref{fig:phase_sharp} we show $^3$S$_1$ phase shifts from an SRG
sharp block diagonalization with $\Lambda_{\rm BD}= 2\fmi$ for two
different potentials.  The phase shifts  are calculated with the
potentials cut sharply at $\Lambda_{\rm BD}$. That is, the matrix 
elements of the potential are set to zero above that point. The
improved  decoupling as $\lambda$ decreases is evident in each case.
By $\lambda = 1\fmi$ in Fig.~\ref{fig:phase_sharp}, the unevolved and
evolved curves are indistinguishable  to the width of the line up to
about 300 MeV. 

\begin{figure}[tbh!]
\bc
\dblpic{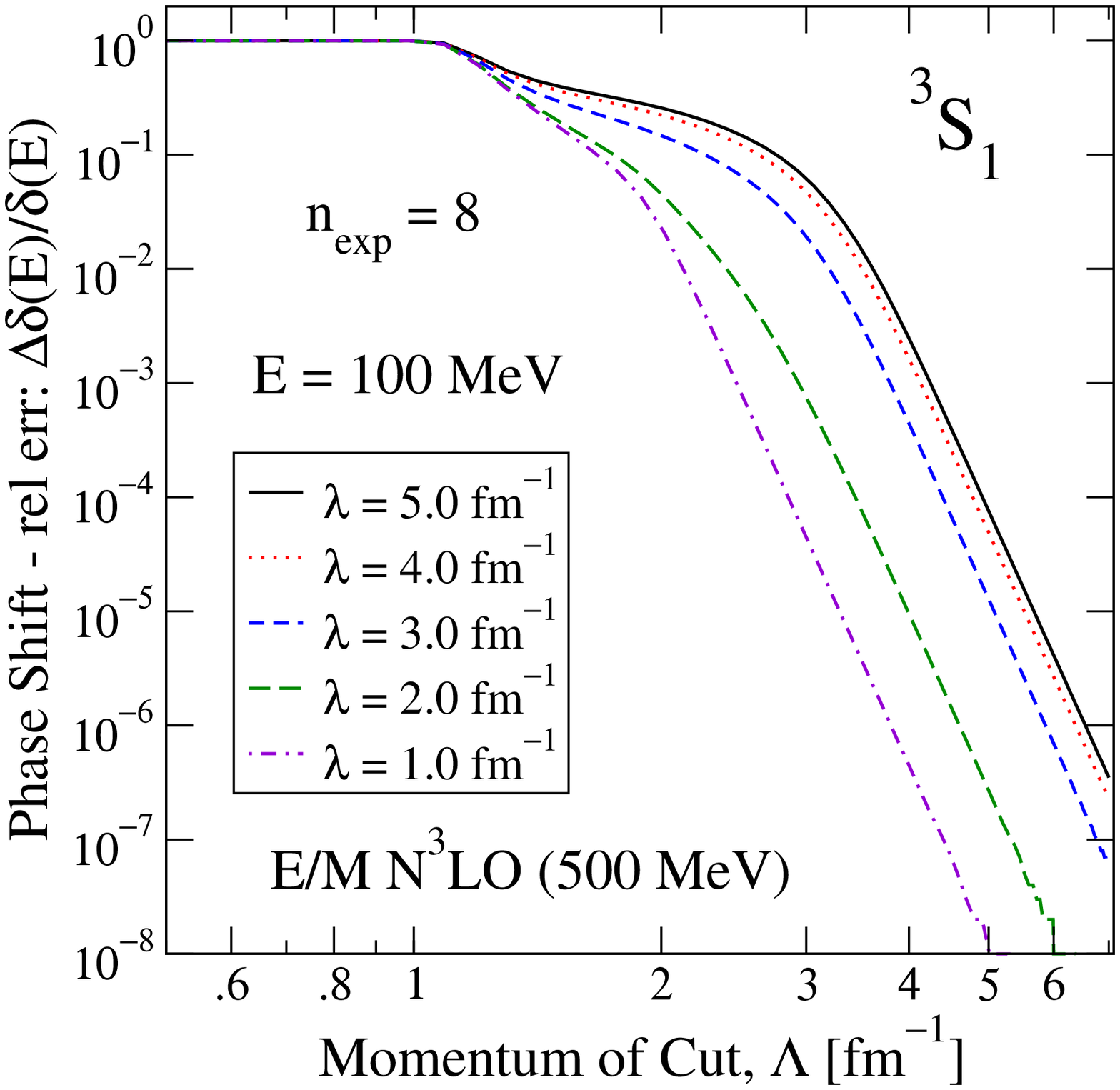}
 \hfill
\dblpic{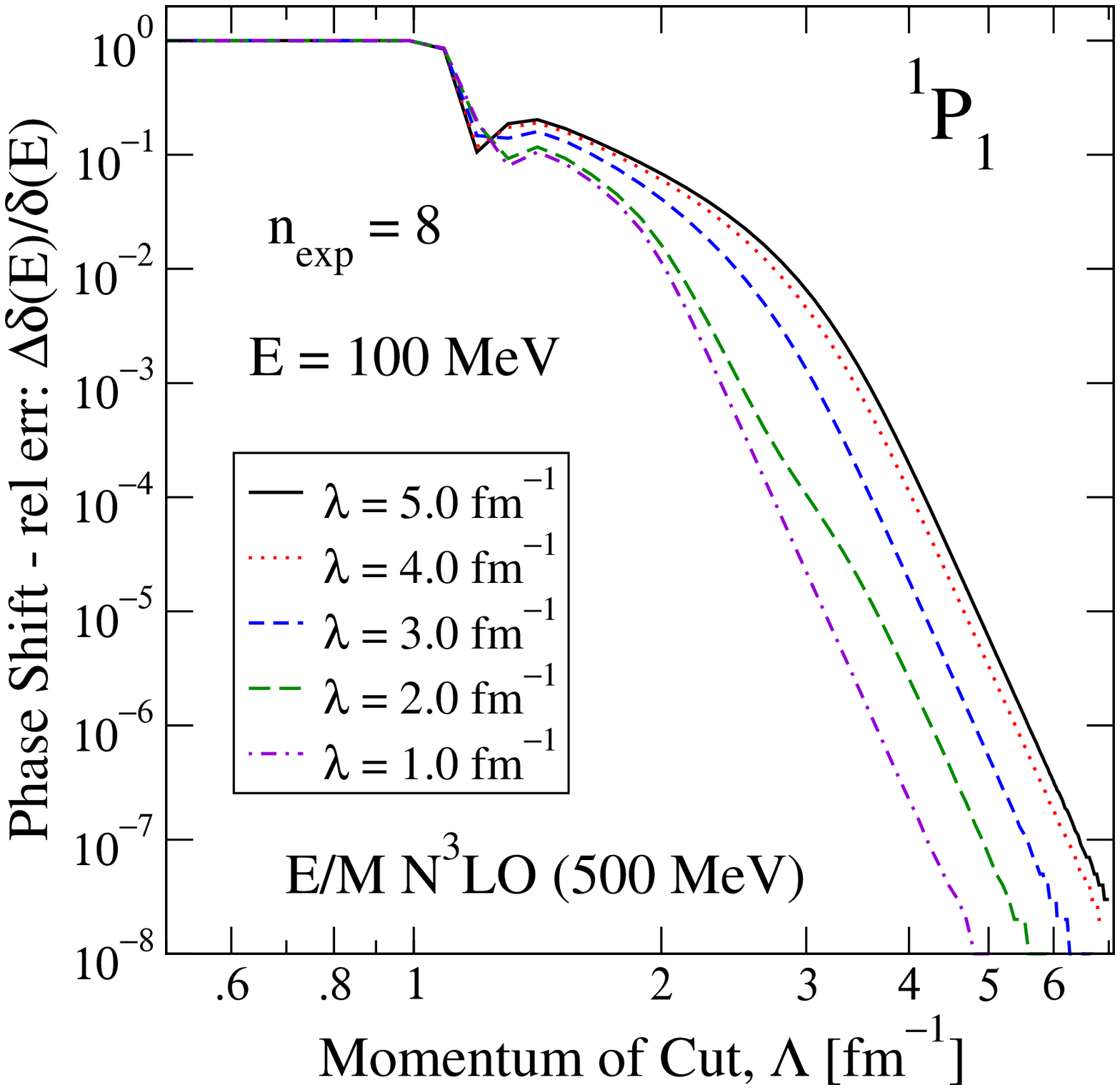}
 \hfill
\dblpic{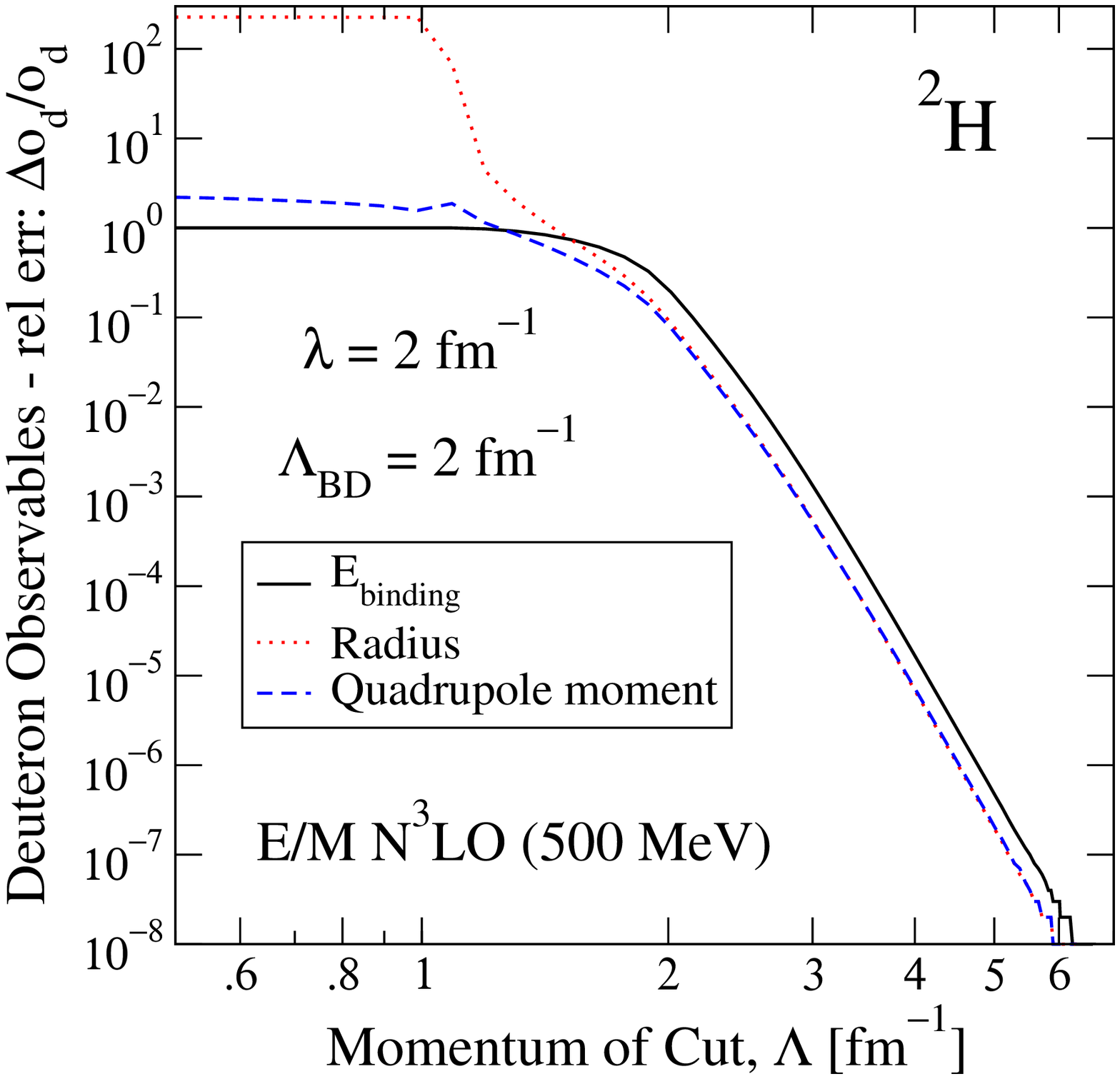}
\ec
\captionspace{ Errors in the phase shift (two partial waves) at $E_{\rm
 lab} = 100\,$MeV and three deuteron observables for the evolved sharp
 SRG block-diagonal  potential with $\Lambda_{\rm BD}=2\fmi$ for a range
 of $\lambda$'s and a regulator with $n=8$. Two partial waves are
 shown.}
 \label{fig:phase_err_sharp}
\end{figure}

In Fig.~\ref{fig:phase_err_sharp} we show a quantitative analysis of
the  decoupling as in section \ref{sec:phase_shifts}. The figure shows
the relative  error of the phase shift at 100\,MeV calculated with a
potential that is cut off as before at a series of values
$\Lambda_{\rm cut}$.  We observe the same universal decoupling
behavior: a shoulder indicating the start of a perturbative decoupling
region, where the slope matches the power $2n$ fixed by the smooth
regulator. 

\begin{figure}[ptbh!]
\bc
\includegraphics*[width=5in]{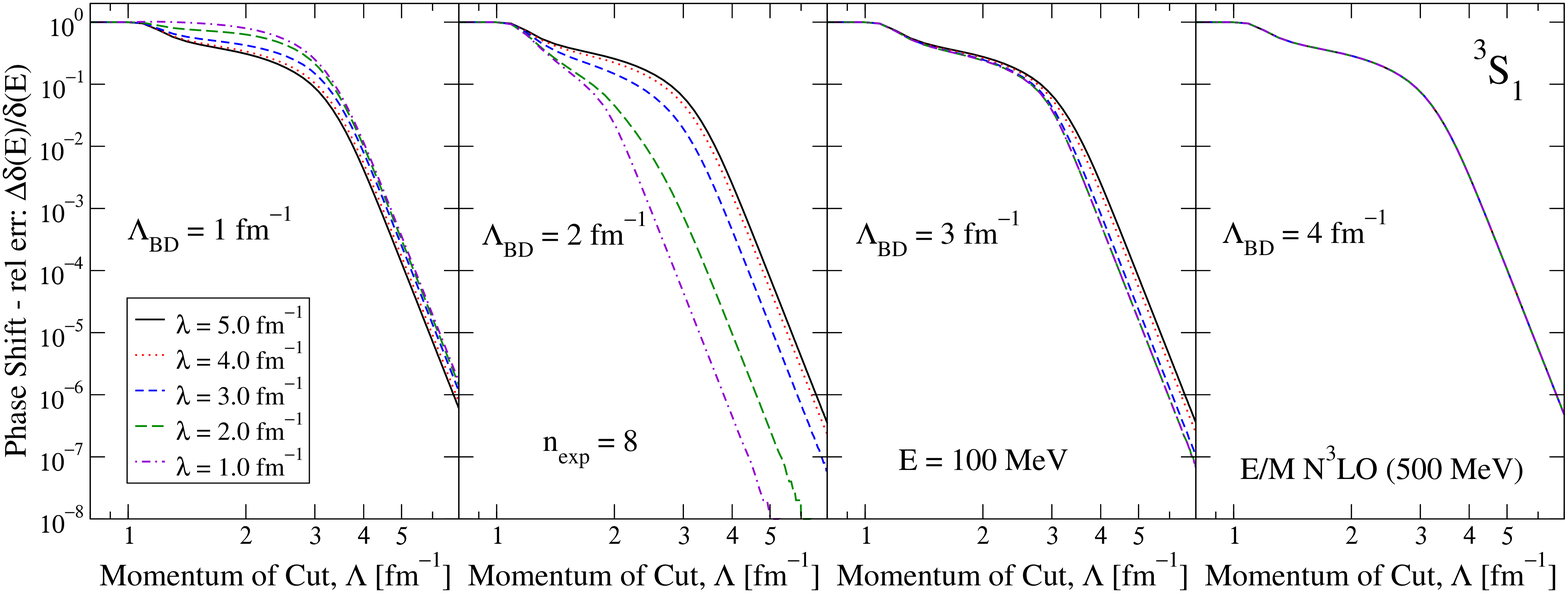}

\includegraphics*[width=5in]{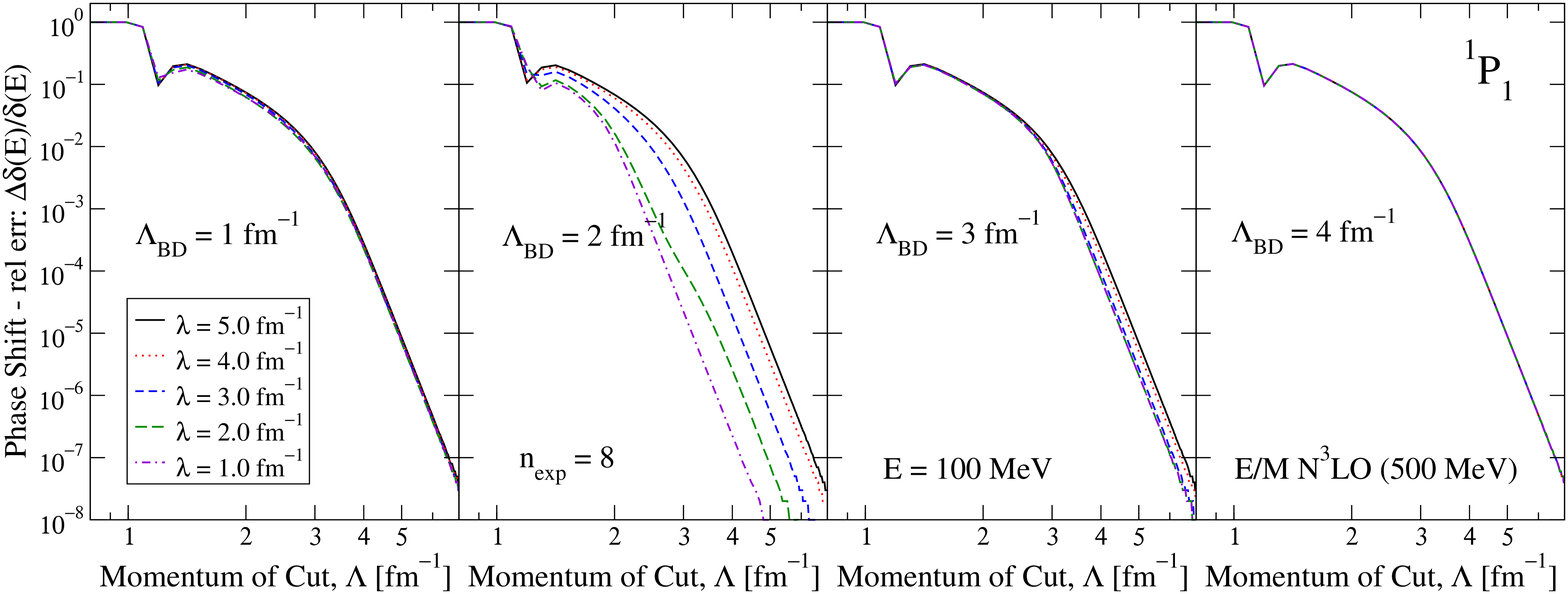}

\includegraphics*[width=5in]{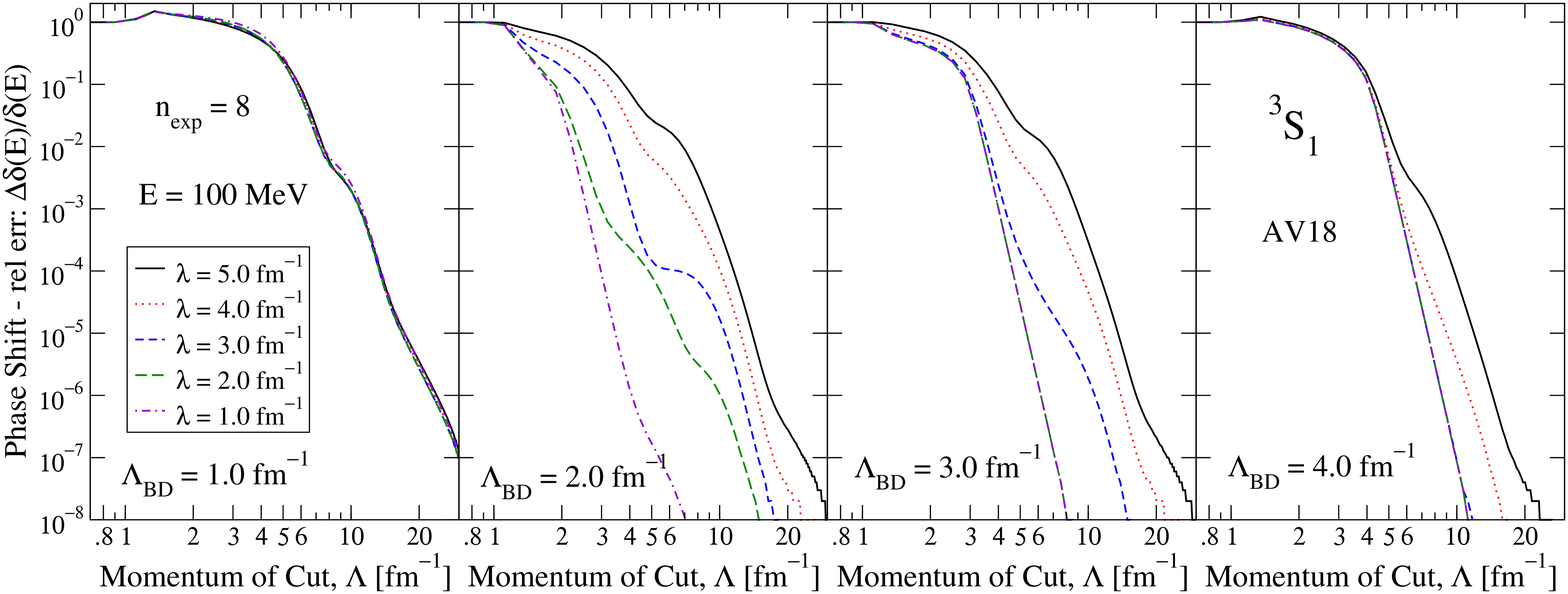}

\includegraphics*[width=5in]{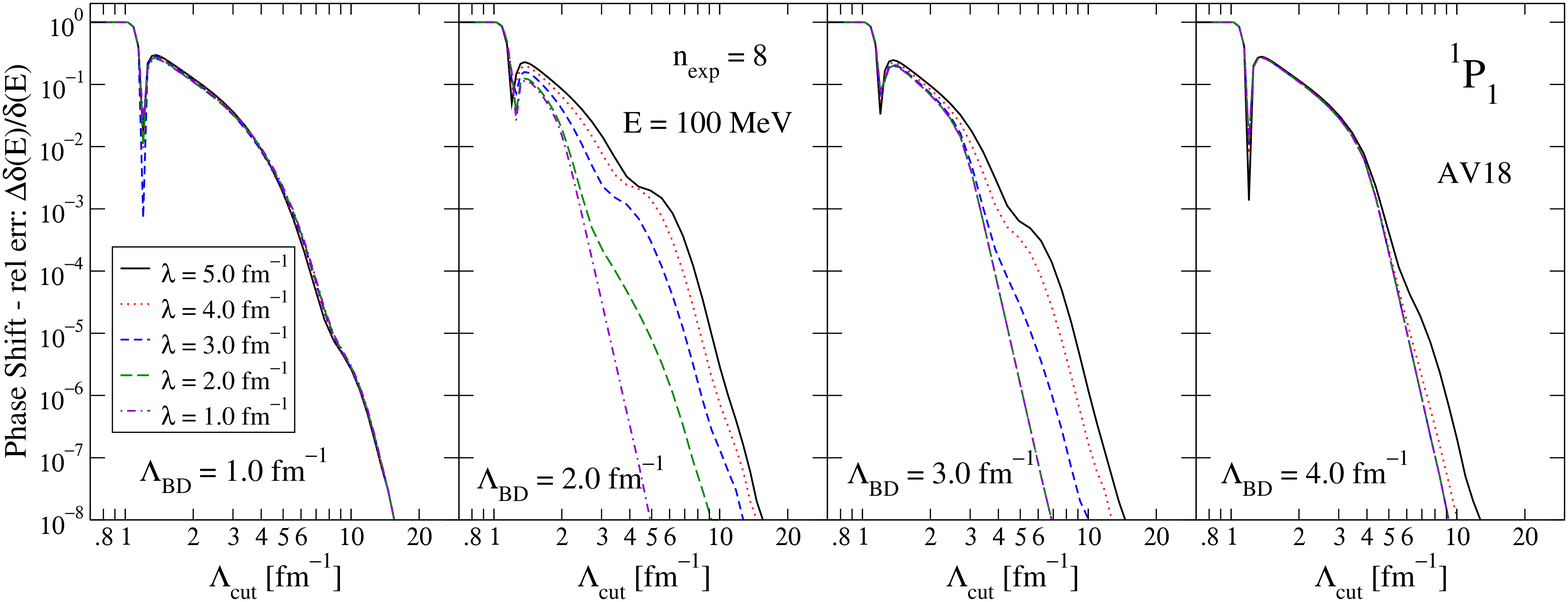}
\ec
\vspace*{-.5in}
\caption{Decoupling error plots using the sharp block-diagonal
generator on two partial waves ($^3S_1$ and $^1P_1$) in two different
initial potentials (AV18 and N$^3$LO (500 MeV)).}
\label{fig:phase_err_var_Lambda_BD}
\end{figure}

Figure \ref{fig:phase_err_var_Lambda_BD} shows a limiting factor with
respect to the decoupling generated by the block-diagonal SRG. The
onset of the shoulder in $\Lambda_{\rm cut}$ decreases with $\lambda$
until it saturates for $\lambda$ near $\Lambda_{\rm BD}$, leaving the
shoulder at $\Lambda_{\rm cut} \approx \Lambda_{\rm BD}$. Thus, as
$\lambda\rightarrow 0$ the decoupling scale is set by the cutoff
$\Lambda_{\rm BD}$. Of course, the $P$ space inside $\Lambda_{\rm BD}$ is not
transformed (diagonally or otherwise) by the SRG and the potential
cannot be decoupled below this scale.

Taking the block diagonal idea further we can apply more general
definitions of $P$ and $Q$. To smooth out the cutoff, we can introduce
a smooth regulator $f_\Lambda$, which we take here to be an
exponential form:
\beqn
   f_\Lambda(k) = e^{-(k^2/\Lambda_{\rm BD}^2)^n}  \;,
   \label{eq:fLambdaa}
\eeqn 
with $n$ an integer. For $\vlowk$ potentials,  typical values used are
$n=4$ and $n=8$ (the latter is considerably sharper but still
numerically robust).
By replacing $\Hbd_s$ with
\beqn
  G_s = f_\Lambda H_s f_\Lambda + (1-f_\Lambda)H_s(1-f_\Lambda)  \;,
  \label{eq:Gs}
\eeqn
we get a smooth block-diagonal potential.

\begin{figure}[tbh!]
\bc
\strip{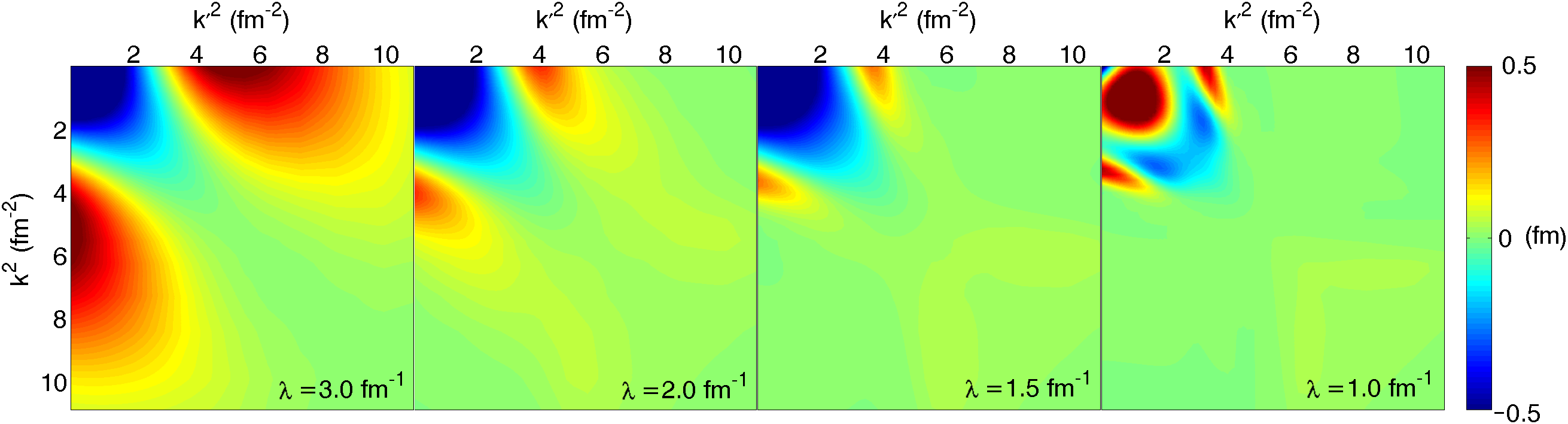}
\ec
\captionspace{Evolution of the $^3$S$_1$ partial wave with a smooth ($n=4$)
block-diagonal flow equation with $\Lambda_{\rm BD} =2.0\fmi$, starting
with the N$^3$LO potential from Ref.~\cite{N3LO}.  The flow parameter
$\lambda$ is  3, 2, 1.5, and $1\fmi$.  The  axes are in units of $k^2$
from 0--11 fm$^{-2}$. The color scale ranges from $-0.5$ to $+0.5\fm$ 
as in Fig.~\ref{fig:vlowk}. }
\label{fig:bd_srg_smooth}
\end{figure}

Fig.~\ref{fig:bd_srg_smooth} shows an example of the smooth
block-diagonal SRG with $\Lambda_{\rm BD}=2\fmi$ and $n=4$. As we
evolve the potential down toward  $\lambda = 1$, the overlap of the
$P$ and $Q$ spaces becomes significant and the potential develops new
structure instead of becoming simpler. This sort of behavior indicates
that there is no further benefit to evolving in $\lambda$ very far
below $\Lambda_{\rm BD}$; in fact the decoupling worsens for $\lambda
< \Lambda_{\rm BD}$ with a smooth regulator.

\begin{figure}[tbh!]
\bc
\dblpic{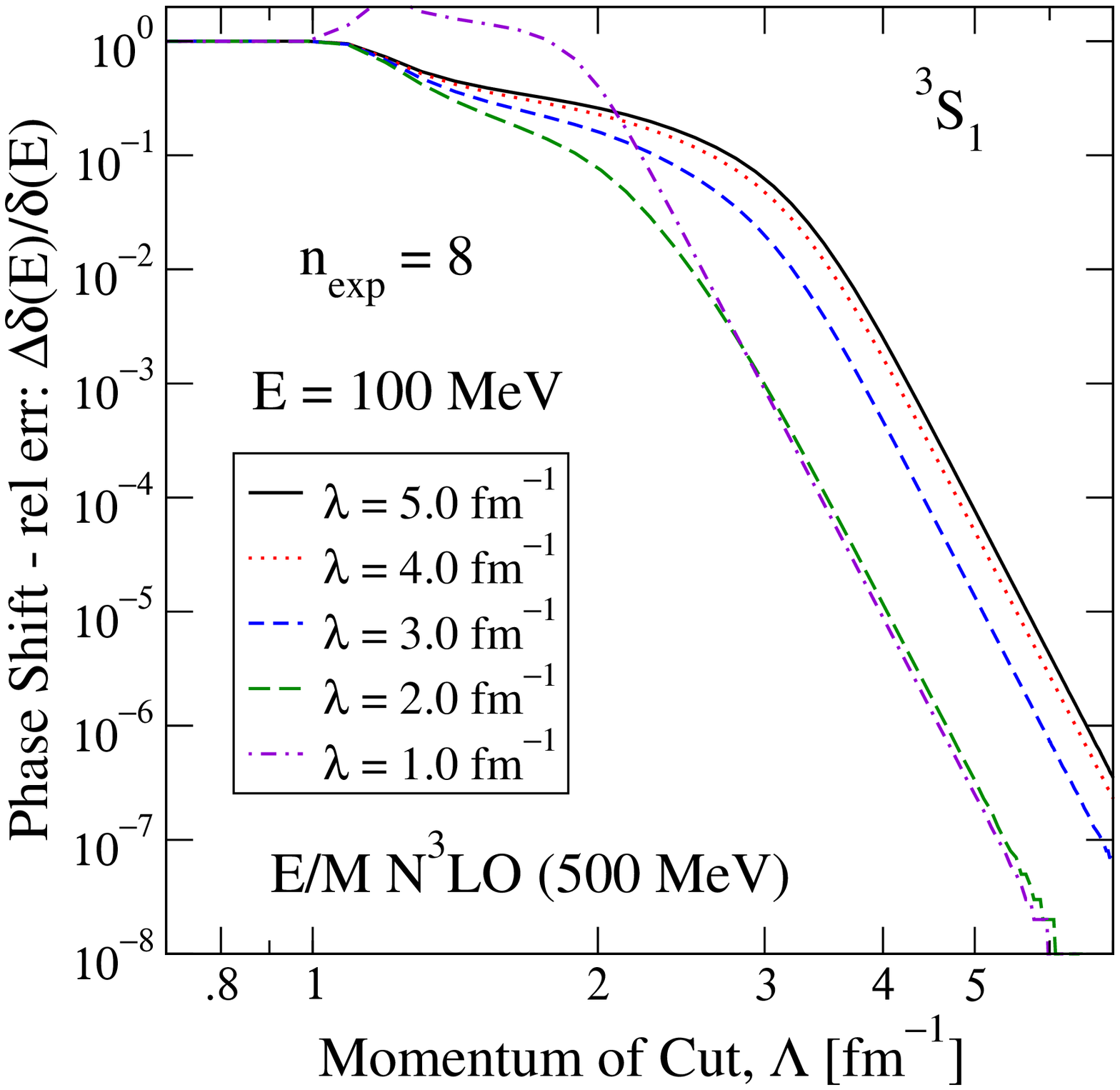}
\hfill
\dblpic{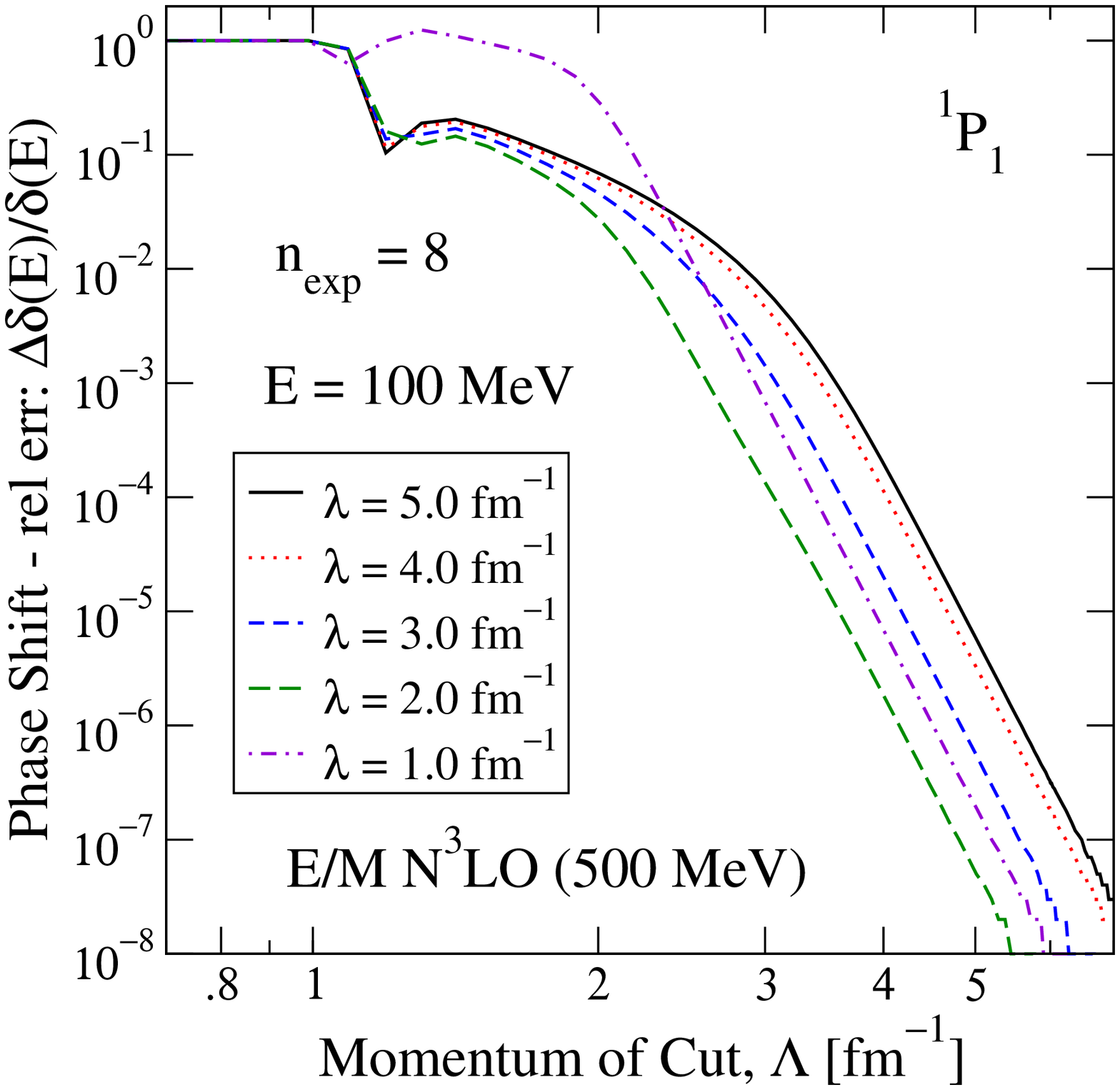}
\ec
\captionspace{Phase shift errors from potentials evolved with the
smooth block-diagonal SRG. The block parameter is $\Lambda_{\rm BD} =
2$. Here the regulator of Eq.~\eqref{eq:fLambdaa} uses $n=4$.}
\label{fig:ps_err_bd_smooth}
\end{figure}

Figure \ref{fig:ps_err_bd_smooth} shows the phase shift relative error
plots for the smooth block-diagonal SRG applied in two partial waves
of the N$^3$LO potential. Note that as $\lambda$ is run down below the
block parameter $\Lambda_{\rm BD}$ the error worsens, signaling a
collision of high- and low-energy degrees of freedom due to the
overlap of $P$ and $Q$ space wavefunctions as defined by the smooth
block transformation.


\begin{figure}[tbh!]
\bc
\strip{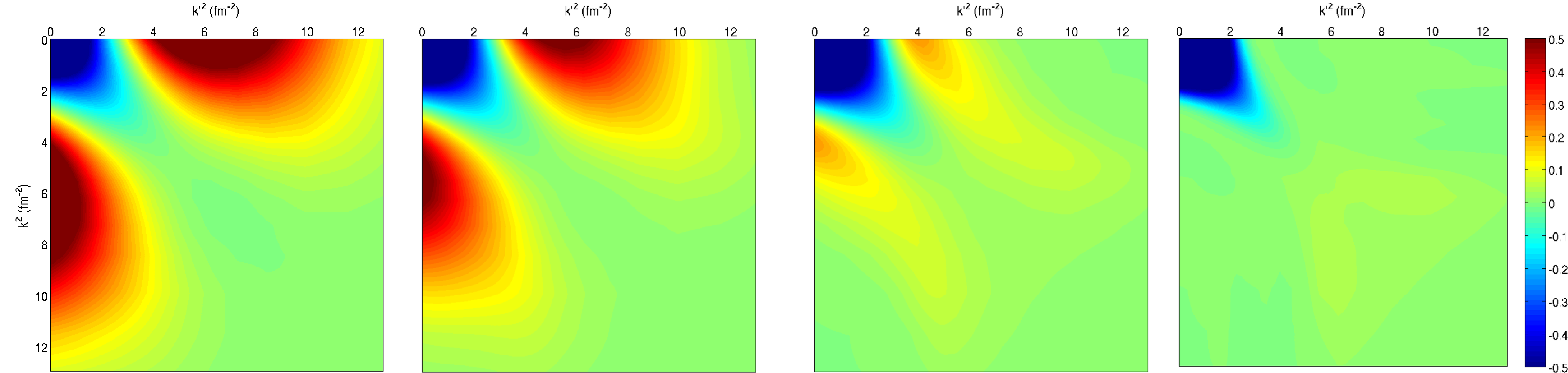}
\ec
\captionspace{Evolution of the $^3$S$_1$ partial  wave with a second-order
exact  block-diagonal flow of equation \ref{eq:second_order} with
$\Lambda_{\rm BD}=2\fmi$ at $\lambda = 4$, 3, 2, and $1\fmi$. The initial
N$^3$LO potential is from Ref.~\cite{N3LO}. The axes are in units of
$k^2$ from 0--11 fm$^{-2}$.  The color scale ranges from  $-0.5$ to
$+0.5\fm$ as in Fig.~\ref{fig:vlowk}.}
\label{fig:vsrg_second_exact}
\end{figure}

Another type of SRG that is second-order exact and yields similar
block diagonalization is defined by
\beqn
  \eta_s = [T, PV_sQ + QV_sP] \;,  
  \label{eq:second_order}
\eeqn
which can be implemented with $P \rightarrow f_\Lambda$ and $Q
\rightarrow (1-f_\Lambda)$, with $f_\Lambda$ either sharp or smooth.
Figure \ref{fig:vsrg_second_exact} shows an example of evolution of
the N$^3$LO potential from Ref.~\cite{N3LO} using this choice.

\begin{figure}[tbh!]
\bc
\dblpic{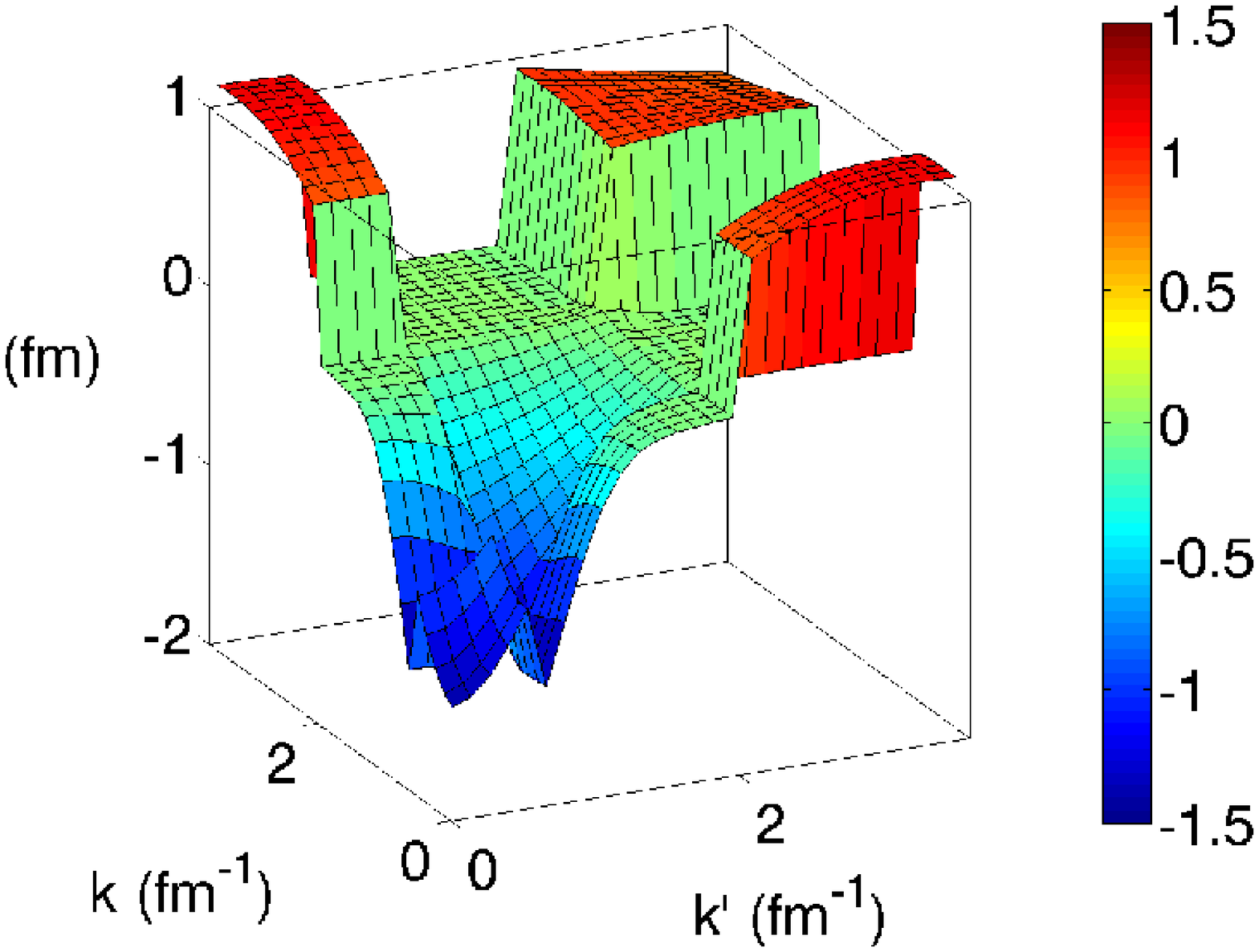}
\hfill
\dblpic{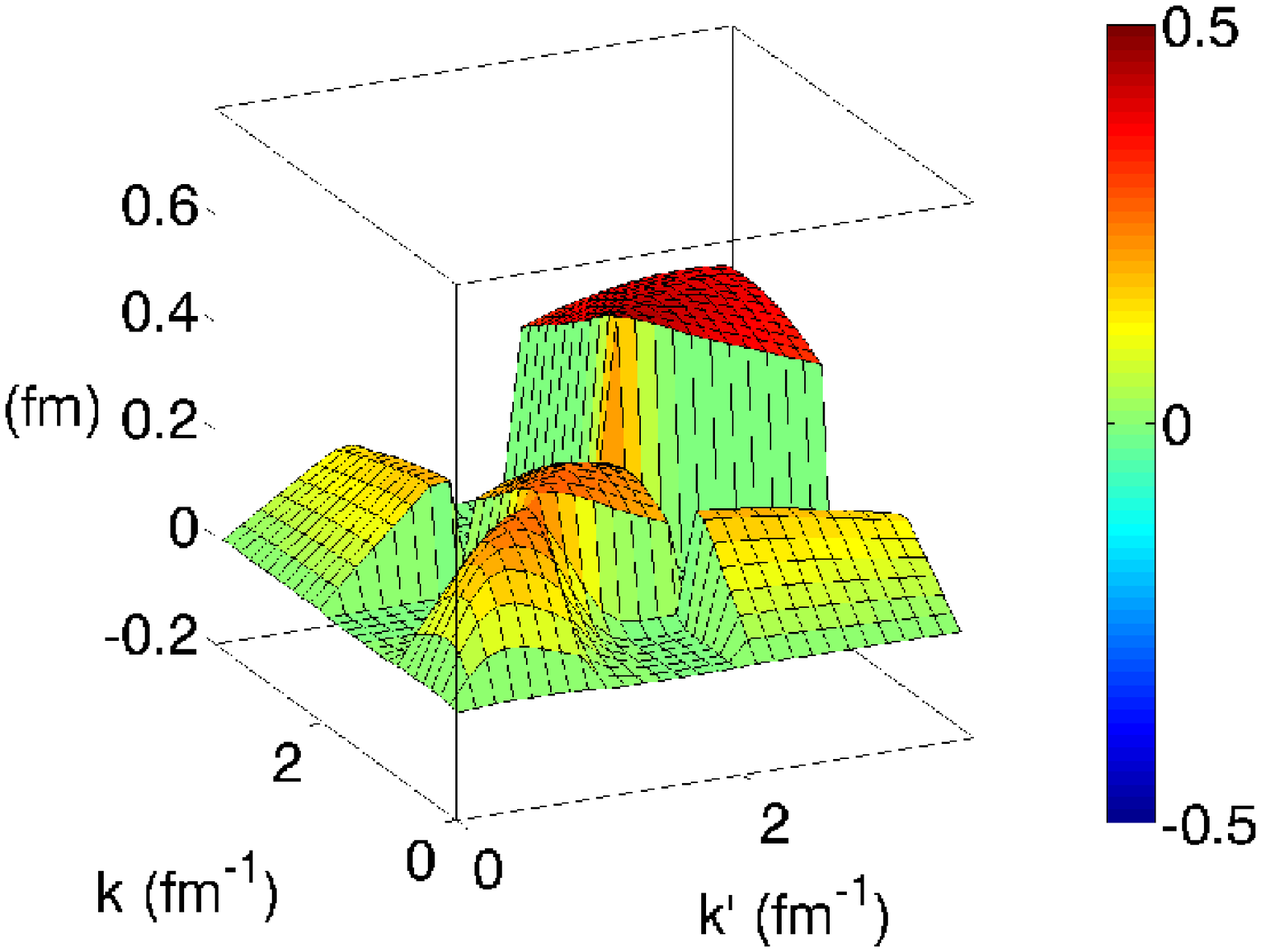}
\ec
\captionspace{Evolved SRG potentials starting from Argonne $v_{18}$ in the
$^1$S$_0$ (left) and $^1$P$_1$ (right) partial waves to $\lambda = 1\fmi$
using a bizarre choice for $G_s$ (see text).}
\label{fig:vsrgweird}
\end{figure}

We can also consider bizarre choices for $f_\Lambda$ in
Eq.~\eqref{eq:Gs}, such as defining it to be zero out to $\Lambda_{\rm
lower}$, then unity out to  $\Lambda_{\rm BD}$, and then zero above
that.  This means that $1 - f_\Lambda$  defines both low and
high-momentum blocks and the region that is driven to zero consists of
several rectangles.  Results for two partial waves starting from the
Argonne $v_{18}$ potential are shown in Fig.~\ref{fig:vsrgweird}.
Despite the strange appearance, these remain unitary transformations
of the original potential, with phase shifts and other NN observables
the same as with the original potential. These choices provide a
proof-of-principle that the decoupled regions can be tailored to the
physics problem at hand.

In the more conventional SRG, where we use $G_s = \Trel$, it is easy
to see that the evolution of the two-body potential in the
two-particle system can be carried over directly to the three-particle
system. In particular, it follows that the three-body potential does
not depend on disconnected two-body
parts~\cite{Bogner:2006srg,Bogner:2007qb}. If we could implement 
$\eta_s$ as proposed here with analogous properties, we would have a
tractable method for generating $\vlowk$ three-body forces. While it
seems possible to define Fock-space operators  with projectors $P$ and
$Q$  that will not have problems with disconnected parts, it is not
yet clear whether full decoupling in the few-body space can be
realized. In fact, if $P$ and $Q$ are $A$-dependent than the two-body
potential will not be completely fixed by evolution in the two-body
space. This is an essential ingredient in the many-body calculations
done in this thesis.

\chapter{One-Dimensional Model \\
in the Harmonic Oscillator Basis}
\label{chapt:OneD}

As shown formally in Section~\ref{sec:second_quantization}, the SRG
induces many-body forces as it evolves the Hamiltonian in an $A$-body
space. For the SRG to be a useful tool, we must develop methods for
calculating these induced many-body interactions and establish the
conditions under which an initial hierarchy of many-body forces is
maintained. In this chapter, we study a one-dimensional system of
bosons as a proof-of-principle of a practical method to evolve and
evaluate such forces, and establish a road map for full
three-dimensional calculations carried out in
chapter~\ref{chapt:ncsm}. Beyond this, a one-dimensional system
provides a straightforward laboratory that is indispensable in gaining
deeper understanding of the SRG's effect on many-body potentials.


Most previous applications of the SRG to nuclear structure have been
in a momentum basis, where decoupling between low-energy and
high-energy matrix elements is naturally achieved by choosing a
momentum-diagonal flow operator such as the kinetic energy $\Trel$.
However, evolution in a three-body space requires proper handling of
the ``dangerous delta functions" associated with the 
spectator particle to the two-body interaction. In a momentum
representation, these delta functions are difficult to resolve in a
practical mesh and require solving separate flow equations for each
combination of three-body matrix elements.


Since we can evaluate the SRG flow equations Eq.~\eqref{eq:srg8}
with $G_s = \Trel$ in any convenient basis, an alternative is a
discrete basis that allows direct application of the SRG flow
equations in each $A$-body sector. We adopt this approach in the
present chapter, using harmonic oscillator wave functions as our basis
and mimicking the formalism used in the no-core shell model (NCSM)
\cite{NCSM1a,NCSM1b,NCSM1c} to create properly symmetrized (for
bosons) matrix elements in relative (Jacobi) coordinates. The
restriction to one dimension makes the construction particularly
straightforward and requires only moderate matrix sizes. We use bosons
in this model for easy comparison with existing model analysis.
However, the boson ground states also coincide with fermion ground
states when the flavor degeneracy is greater than the number of
particles, because the overall anti-symmetrization is realized by the
flavor wavefunction. 

The choice of the harmonic oscillator basis is important because it is
the only orthogonal basis in which the center-of-mass can be
factorized.
With a translationally-invariant basis we have many fewer basis states
and can achieve better convergence in a given $A$-body system. Other
bases can be translationally invariant, like a system of correlated
gaussians, but are non-orthogonal. Others, like Coupled-Cluster
calculations are orthogonal but do not have a straightforward
truncation in $\nmax$.


We use simple flavor-independent potentials that imitate the
short-range repulsion and mid-range attraction characteristic of
realistic local nuclear potentials. Previous studies of the SRG imply
that properties of the transformations are primarily due to the matrix
structure ($G_s$, $H_s$, choice of basis, etc), so we will be able to
directly carry over some of our observations to three dimensions.
Because the NCSM formalism is already developed in three dimensions,
we will see in chapter~\ref{chapt:ncsm} that the generalization to
three-dimensional fermionic calculations with spin-isospin degrees of
freedom and using realistic nuclear interactions is algebraically
straightforward, though far more computationally intensive. 


Another form of the NCSM uses oscillator wavefunctions associated with
lab-frame coordinates and momenta instead of the relative Jacobi
coordinates. This basis uses Slater determinants, making it much easier
to antisymmetrize. This method is often known as
``$m$-scheme"\cite{ncsm_basic} as it is organized according to the $m$
quantum number of the oscillator states. Similarly, a lab-frame NCSM
scheme referred to as $jj$-coupled~\cite{jjcoupled}, takes advantage of
the rotational symmetry of specific nuclei to reduce the size of the basis
used in CC calculations. The major benefit of lab-frame coordinate
based oscillator bases is the ability to generalize to arbitrary $A$
due to its reliance on a Slater determinant to obtain an
antisymmetrized basis as opposed to the recursive procedure used in
Jacobi systems that builds on previous antisymmetric subsystems. On
the downside, this results in poorer scaling of basis sizes for the
$A$-body Hamiltonian matrices. As a check of the one-dimensional Jacobi coordinate
oscillator basis we built a one-dimensional analog of the $m$-scheme
NCSM, though it was unwieldy for our current purposes and not developed
further. The details of this model are discussed in
appendix~\ref{chapt:app_mscheme}.

\section{Symmetrized Jacobi Harmonic Oscillator Basis}

The No-Core Shell Model is a wavefunction-based  method used to make
\abinit calculations of nuclear bound-state properties. The basis it
uses is built from the  eigenfunctions of the harmonic oscillator
Hamiltonian (described in Appendix~\ref{chapt:app_osc_truncation}) up
to a maximum oscillator number, $\nmax$. The functions are products of
a gaussian and polynomials of order less than the value $\nmax$. The
other free parameter in defining this basis is $\omega$, which sets
the scale of the oscillator wavefunctions. This is usually multiplied
by $\hbar$, which is set to one in this unitless model.

Each boson (fermion) wavefunction in an $A$-body system must be fully
symmetric (antisymmetric). Therefore, we must identify and include all
the symmetric states which the system may occupy. A general and
systematic way to build this basis is to list all the possible states
and build a symmetrization operator in this full basis. The
eigenstates of this operator with eigenvalue one will be the exclusive
set of symmetric states relevant for the system up to the stated
$\nmax$; those with eigenvalue zero are irrelevant and can be thrown
out. Also note that bases of smaller $\nmax$ are a subset of larger
bases. The details involving the organization of states for this
method is discussed in Appendix~\ref{chapt:app_osc_basis}.

The initial two-body potential, usually given in the relative momentum
basis, can be converted to the relative oscillator basis
simply by inserting a (semi-\footnote{see
Appendix~\ref{chapt:app_osc_truncation} for a discussion on the
incompleteness of the oscillator basis.}) complete set of states to
expand the momentum states in terms of oscillator wavefunctions. For
three-body calculations, this interaction must be embedded in the
three-body symmetric basis obtained from the symmetrizer
diagonalization. Calculations in higher-body bases require a general
iterative procedure. The $A$-body basis, being symmetric already, is
taken as an explicit set of states to which is added the additional
Jacobi oscillator coordinate. A new symmetrizer is built and the 
$A+1$-body basis is obtained. The potential in the $A$-body basis is
embedded in the new space. Diagonalization or evolution can proceed as
usual. The details of this iterative process, including potential
embedding and symmetrizer construction, are given in
Appendix~\ref{chapt:app_osc_basis}.

\begin{figure}[th]
\begin{center}
\includegraphics[height=2in,clip]{V_srg_lam_40p0_negele_alpha_contour_k}
\includegraphics[height=1.8in,clip]{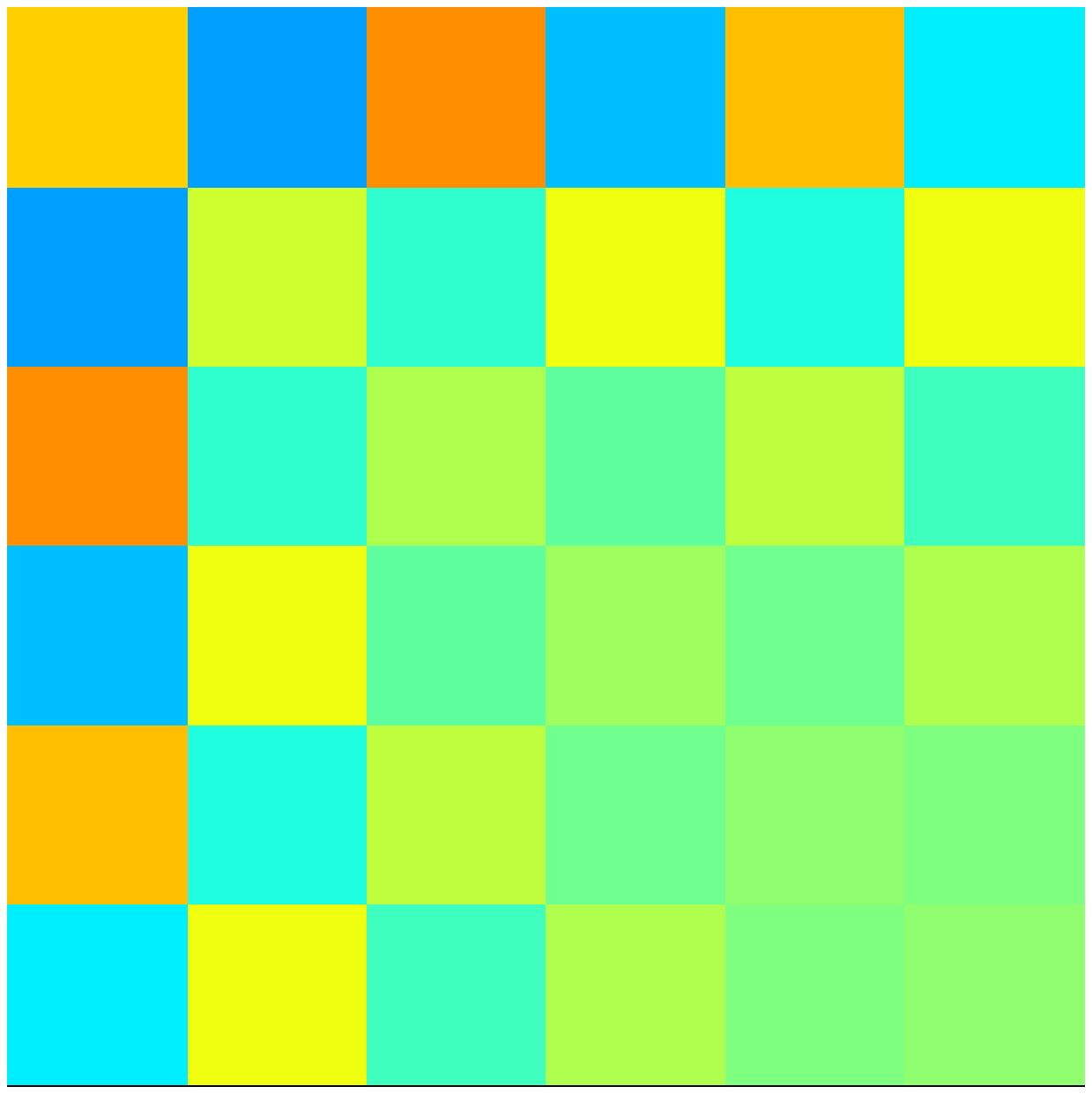}
\includegraphics[height=1.8in,clip]{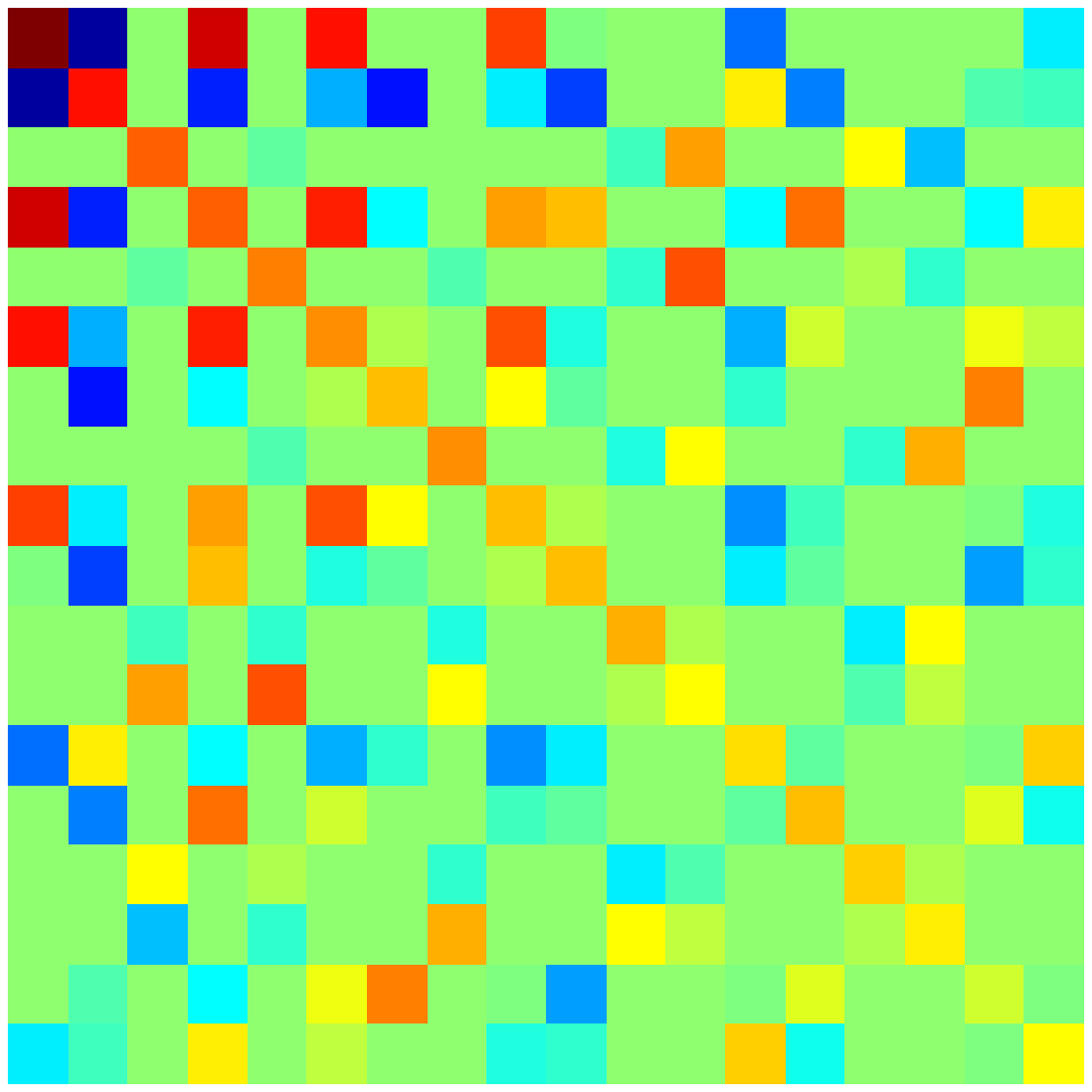}
\includegraphics[height=1.8in,clip]{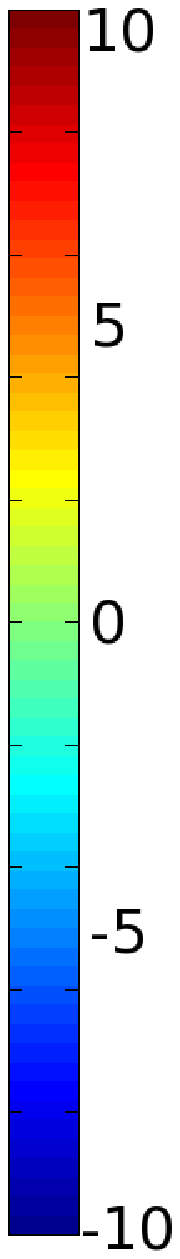}
\end{center}
\captionspace{A snapshot of the embedding process involving the initial
(unevolved) two-body potential. The far left is a two-body,
one-dimensional, potential in momentum space~\cite{negelePot} with axes labelled by
$k^2$, the middle shows that potential converted to the two-body
symmetric oscillator space, and the far right shows it embedded in the
symmetric three-body basis. Each box is a matrix element in the
oscillator basis, and the matrix elements are organized in order of
increasing energy. The oscillator basis axes are unmarked
due to the complicated organization of those bases (see text). For
visibility, a small basis with $\nmax=12$ is shown.} 
\label{fig:embedding_example}
\end{figure}

Figure~\ref{fig:embedding_example} shows an example of the process to
embed a two-body momentum basis potential into the oscillator bases
for $A =$ 2, and 3. The left shows an initial two-body potential in
the momentum basis. The center shows the same potential in the
two-body oscillator basis and is physically identical to the momentum
basis version except for truncation errors induced by oscillator basis conversion
(see appendix~\ref{chapt:app_osc_truncation}). The far right shows the
potential embedded in the three-body basis.  The axes here are
unmarked because the order of the states is partially arbitrary. Here,
as is usual, they are organized in order of increasing energy,
labeled by $N_A$ (for an $A$-body state), up to the maximum $\nmax$.
Any degeneracies in $N_A$ are labeled by an arbitrary index $i_A$.
Note the many matrix elements on the far right plot which are zero
(green in this color scheme) indicating transitions which cannot be
described by only two-body forces and a spectator particle. These
matrix elements represent unique three-body transitions and, while
some of these may remain zero due to symmetry properties, they are now
available to the SRG during evolution. Any change in the value of
these matrix elements represents three-body forces being induced
during evolution in this basis.

Once we have constructed a complete symmetrized basis for $A$
particles (specified by the value of $\nmax$) and evaluated the
Hamiltonian matrix elements in this basis, applying the SRG
transformations is immediate. That is, we have coupled, first-order
differential equations  for each matrix element of the Hamiltonian,
with the right side of each equation given by a series of matrix
multiplications. This is efficiently implemented in any computer
language with matrix operations and differential equation solvers.
Just as in the momentum basis, individual matrix
elements\footnote{The $A$-body states are labeled by the total
energy, $N_A$ and a quantum number, $i_A$, introduced to keep track of
states degenerate in $N$. In three dimensions this list will include
angular momenta and isospin. See appendix~\ref{chapt:app_osc_basis}
for the complete notation.} of the Hamiltonian obey the SRG's
differential equations:
\bea
\frac{d}{ds}\la N_A'i_A'|(V_A)_s|N_Ai_A\ra &=&
\la N_A'i_A'|\big[\big[T,H_s\big],H_s\big]|N_Ai_A\ra \nonumber \\
&=& \la N_A'i_A'|T H_s H_s|N_Ai_A\ra 
+ \la N_A'i_A'|H_s H_s T|N_Ai_A\ra \nonumber \\
 & & \null - 2\la N_A'i_A'|H_s T H_s|N_Ai_A\ra \;.
 \label{eq:SRGmatrices} 
\eea
We have defined ${dT}/{ds}=0$ so that all of the flow occurs in the
matrix representation of the potential, $(V_A)_s$.  Using the matrix
representations of $T$ and $H_s$ in the $|N_A i_A\ra$ basis, the right
side of Eq.~\eqref{eq:SRGmatrices} is simply a series of matrix
multiplications. The initial condition at $s=0$ is the initial
Hamiltonian, $\la N_A'i_A'|T+V_A|N_Ai_A\ra$, which can have few-body
components in $V_A$. We consider here both two-body-only and two-body
plus a three-body component (for $A \geq 3$).

\begin{figure}[th]
\begin{center}
\includegraphics[height=1.8in,clip]{V_srg_lam_2p0_negele_alpha_contour_k}
\includegraphics[height=1.7in,clip]{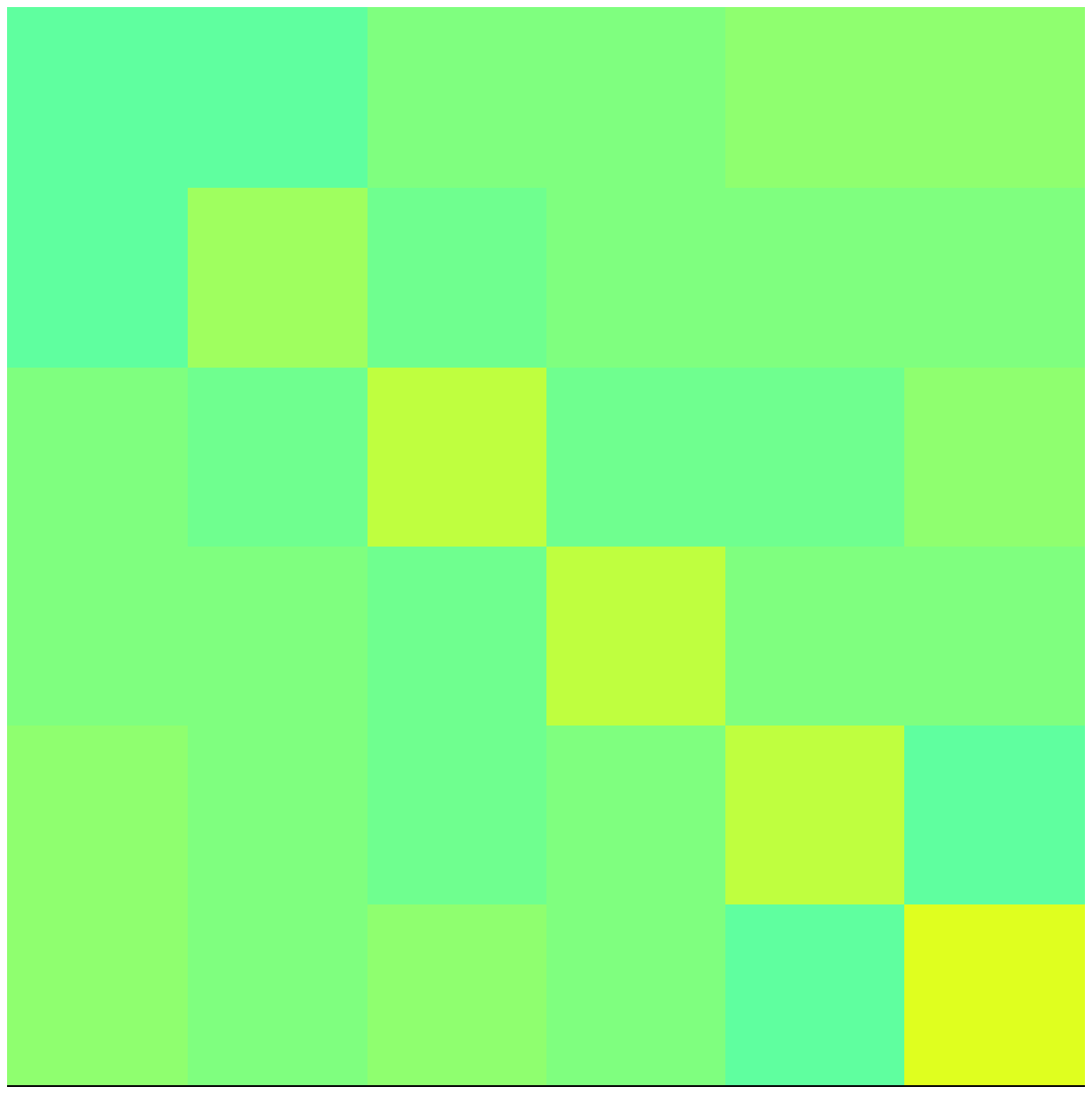}
\includegraphics[height=1.7in,clip]{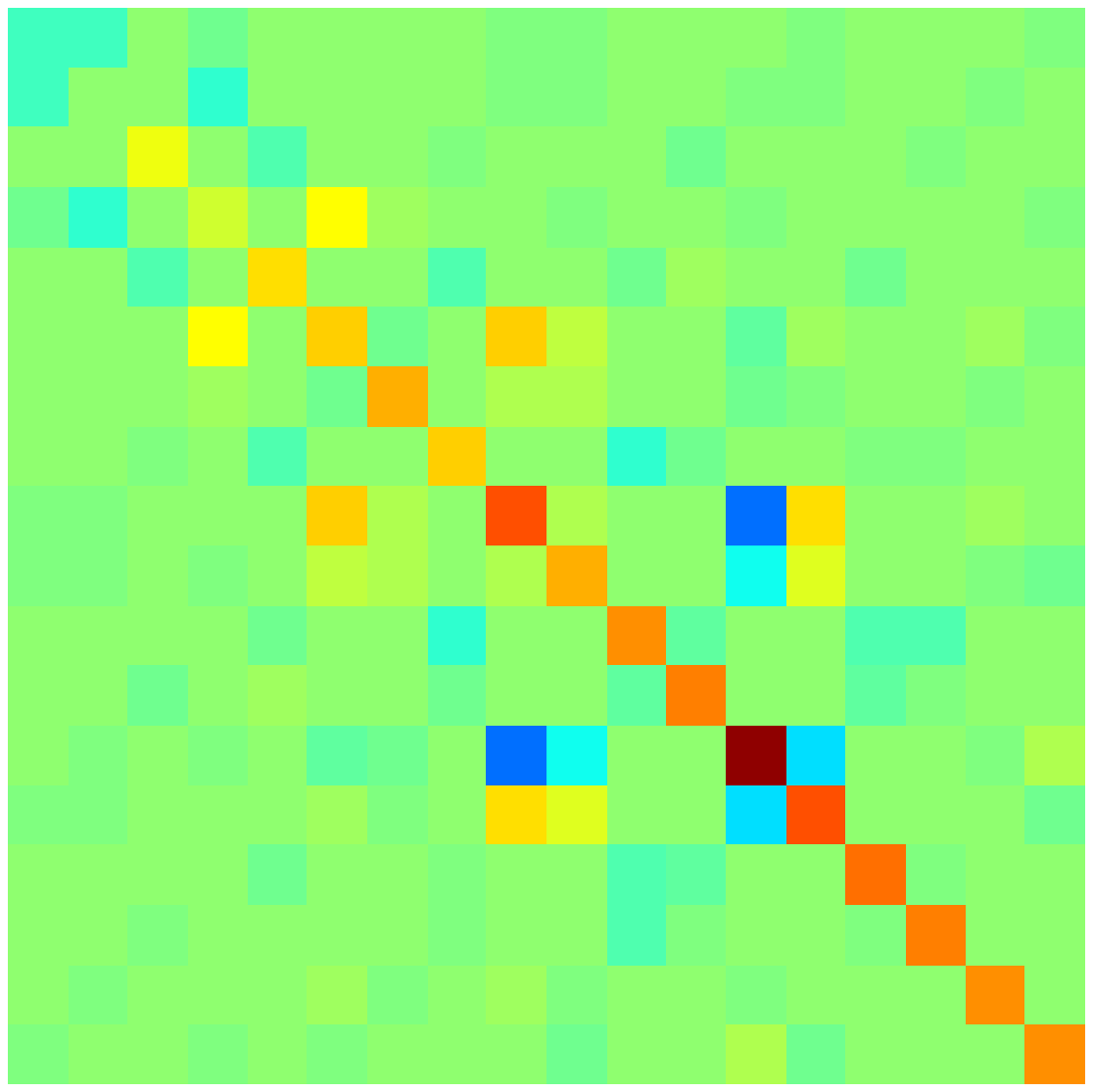}
\includegraphics[height=1.7in,clip]{osc_scale}
\end{center}
 
\captionspace{Same as in \ref{fig:embedding_example} but each matrix is now
evolved to $\lambda = 2$ in each basis shown. Again, $\nmax = 12$.}
\label{fig:embedding_example_evolved}
\end{figure}

Figure~\ref{fig:embedding_example_evolved} shows an example of
evolution in the oscillator bases for $A =$ 2, and 3. The left shows
the evolution, to $\lambda=2$, of a two-body potential in the
momentum basis. The center shows the evolution of the potential in the
two-body oscillator basis and is physically identical to the momentum
basis version again except for errors induced by oscillator basis
conversion. The far right shows the evolution in the three-body basis
and includes the evolution of two body forces but now also includes
induced three-body forces not available in either of the two-body
spaces.

The SRG induces few-body forces as it evolves an initial interaction
in a few-particle space. To study the contributions of different
few-body forces we must isolate these components of the full
interaction. The two-body force evolution keeps the $A=2$ binding
energy invariant under evolution and \emph{defines} the evolved
two-body forces. We can isolate the three-body force from the two-body
matrix elements by embedding the evolved two-body-only force in the
three-particle space and subtracting it from the full
two-plus-three-body evolved interaction. Because we are in a
symmetrized basis, this operation is relatively straightforward.
However, we must be careful to keep track of the correct combinatoric
factors associated with embedding A-body forces in higher spaces. In
our MATLAB implementation these procedures take only a few lines of
code in addition to the fully unitary calculation.

The NCSM, both one- and three-dimensional, is a variational
calculation in the two parameters, $\nmax$ and $\hw$. Largely this can
be understood in terms of the truncation effects discussed in
Appendix~\ref{chapt:app_osc_truncation}. Essentially, conversion to an
oscillator basis imposes both ultraviolet and infrared cutoffs
$\Lambda_{\rm UV}$ and $\Lambda_{\rm IR}$ determined by the parameters
of the basis $\nmax$ and $\hw$. As derived in the appendix, increasing
$\nmax$ improves both cutoffs by raising $\Lambda_{\rm UV}$ and
lowering $\Lambda_{\rm IR}$, but changing $\hw$ shifts both cutoffs in
the same direction. So we will be looking for a sufficiently large
value of $\nmax$ and the optimal value of $\hw$ for a given $\nmax$.


\section{One Dimension Potential}

\begin{figure}[tbh!]
\begin{center}
 \dblpic{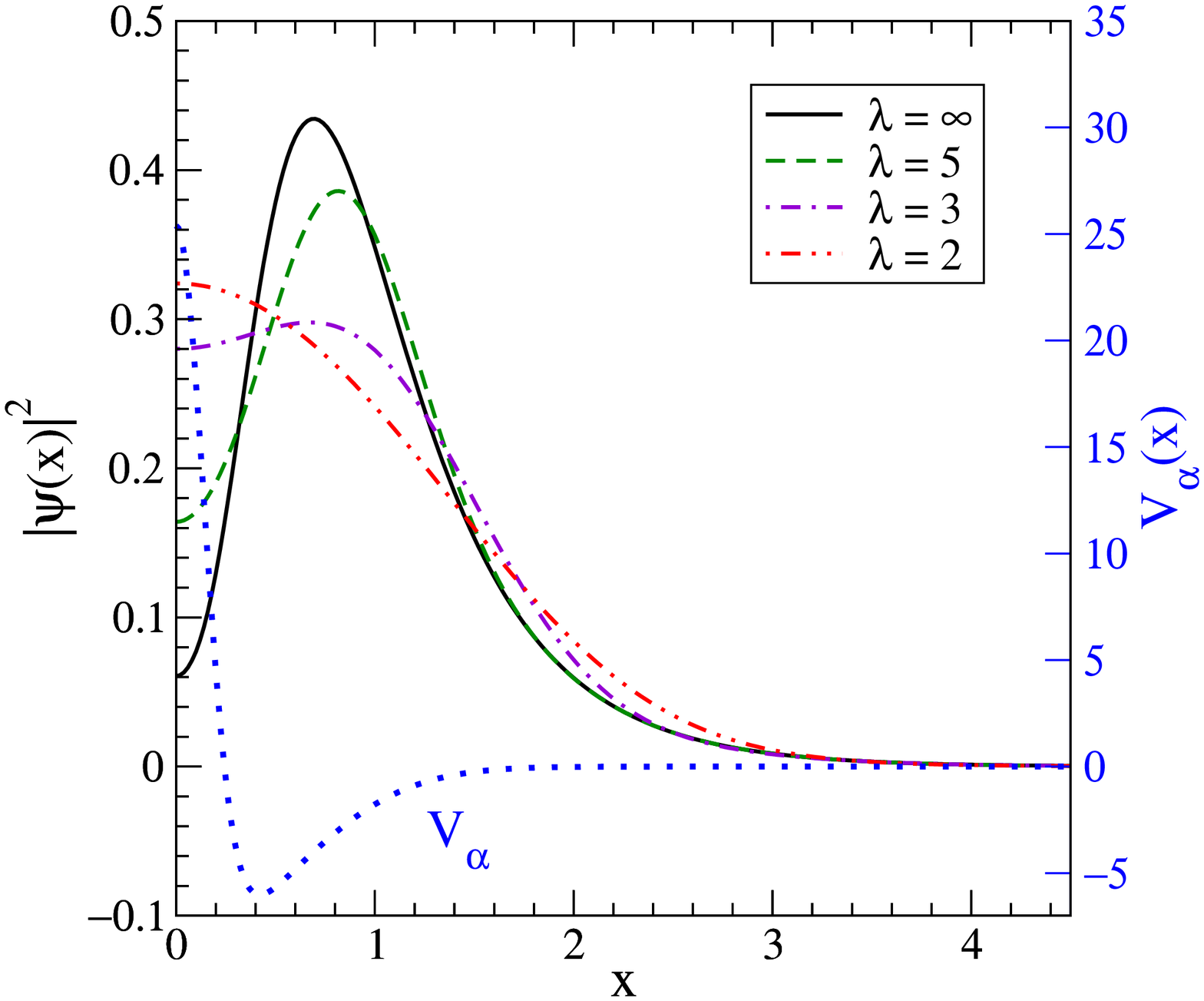}
 \hfill
 \dblpic{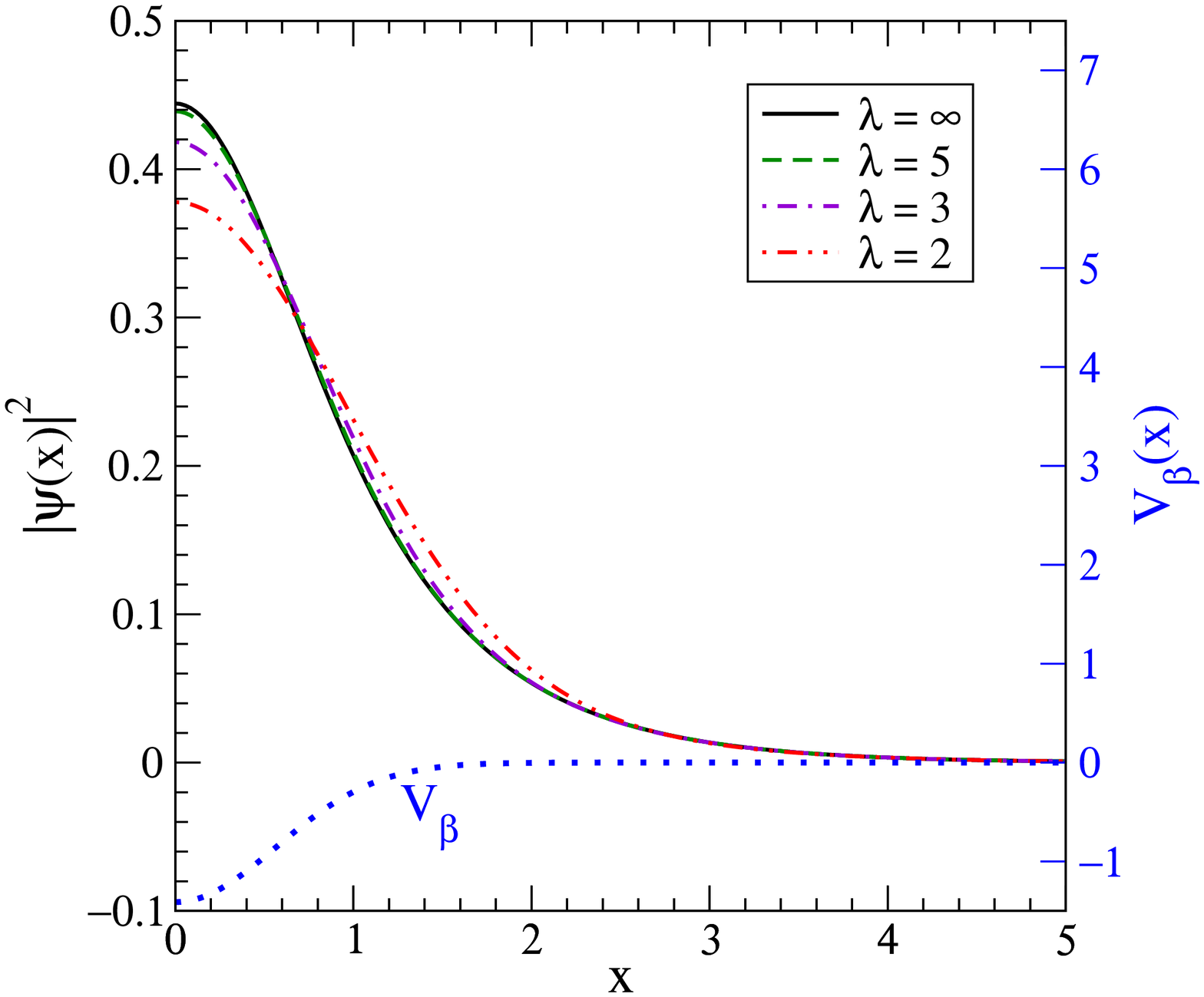}
\end{center}
\captionspace{Potentials (dotted line, with axis on
  right)  and probability distributions (other lines, with axis
  on left) for the lowest two-body bound state as a function of
  $x = |x_1 - x_2|$ at different stages in the SRG evolution
  ($\lambda = 1/s^{1/4}$). The left plot is $V_{\alpha}$ and the
  right plot is  $V_{\beta}$ (see Table~\ref{tab:negelePars}).}
\label{fig:srg_2_body}
\end{figure}

The bulk of the calculations in this chapter adopt a model from
Ref.~\cite{negelePot} that uses a sum of two gaussians to simulate
repulsive short-range and attractive mid-range nucleon-nucleon
two-body potentials:
\beqn
  V^{(2)}(x) = \frac{V_1}{\sigma_1\sqrt{\pi}} e^{-x^2/\sigma_1^2}
    + \frac{V_2}{\sigma_2\sqrt{\pi}} e^{-x^2/\sigma_2^2} 
\eeqn
or
\beqn
   V^{(2)}(p,p') = \frac{V_1}{2 \pi\sqrt{2}}e^{-(p-p')^2 \sigma_1^2/8} 
          + \frac{V_2}{2 \pi\sqrt{2}}e^{-(p-p')^2 \sigma_2^2/8} \;.
  \label{eq:gaussians}
\eeqn
The parameters used in Ref.~\cite{negelePot} were chosen so that the
one-dimensional saturation properties correspond to empirical
three-dimensional properties, but we also want to explore a range of
parameters to test what behavior is general and what relies on
specific features. Otherwise, the units for these model interactions
are straightforward with the mass (same for all particles) and $\hbar$
taken to one, though they are displayed explicitly in some
equations. We start with the parameters listed in
Table~\ref{tab:negelePars}. The potential  $V_{\alpha}$ is from
Ref.~\cite{negelePot} and is plotted in Fig.~\ref{fig:srg_2_body}. We
will fix the range of the attractive part and vary the relative
strength and range of the repulsive parts and vice versa.  We also
vary the purely attractive potential $V_{\beta}$, which was used in
Ref.~\cite{VanNeck:1996} and is also plotted in
Fig.~\ref{fig:srg_2_body}. The eigenvalue problem for the relatively
small matrices considered here can be solved by any conventional
matrix diagonalization program (MATLAB was used here).

\bt
\caption{Parameter sets for the two-body potential of 
Eq.~\eqref{eq:gaussians}.}
\begin{tabular}{c|cccc}
name & $V_1$ & $\sigma_1$ & $V_2$ & $\sigma_2$ \\
\hline
$V_{\alpha}$ & 12. & 0.2 & $-12.$ & 0.8 \\
$V_{\beta}$ & 0. & 0.0 & $-2.0$ & 0.8 \\
\end{tabular}
\label{tab:negelePars}
\et

To test that the symmetrized harmonic oscillator basis was
correctly constructed for $A = 2$, $3$, and $4$ (see below), we
first diagonalized the Hamiltonian using the purely attractive
gaussian  two-body potential $V_{\beta}$.  The normalization is
such that $V_{\beta}(x)$ becomes a delta function with strength
$V_2$ as $\sigma_2 \rightarrow 0$~\cite{VanNeck:1996} (note the
numerical factors from the Fourier transform because of our
normalization of the Jacobi momenta). This limiting case has a
known analytic solution for the (only) bound state of $A$
bosons.  For finite $\sigma_2$, we were able to confirm the
accuracy of the diagonalizations as a function of the basis size
$\nmax$ by comparison to coordinate-space stochastic variational
method (SVM) calculations using a published code~\cite{svmcpc}
adapted to one dimension~\cite{EricPC,SVM_book}.

We also explore the evolution of Hamiltonians with an initial
three-body force. We choose a regulated contact interaction in
the three-particle momentum space,
\beqn
 V^{(3)}(p,q,p',q') =  c_E f_\Lambda(p,q) f_\Lambda(p',q') \; ,
\label{eq:three_body_force}
\eeqn
where $c_E$ is the strength of the interaction and the
form factor $f_\Lambda$ depends on the Jacobi momenta as
\beqn
  f_\Lambda(p,q) \equiv e^{-((p^2+q^2)/\Lambda^2)^n} \;.
  \label{eq:fLambda}
\eeqn
The regulator cutoff $\Lambda$ sets the scale of the fall-off in
momentum and $n$ determines the sharpness of this fall-off. This form
is analogous to the regulated three-body contact interactions used in
chiral effective field theory~\cite{Epelbaum:2002vt} discussed in
Appendix~\ref{chapt:app_eft}. We have not explored in detail the
impact of adjusting $\Lambda$ and $n$ but have focused on how the SRG
handles a varying strength $c_E$. All results here are for $\Lambda =
2$ and $n=4$.

\newcommand{\ph}{\phantom{-}}

\bt
 \caption{Ground-state energies for two-body potentials from
 Table~\ref{tab:negelePars} with various strengths of the
 initial three-body potential
 Eqs.~\eqref{eq:three_body_force}--\eqref{eq:fLambda} 
  with $\Lambda = 2$ and $n=4$ 
 for $A = 2$, 3, and 4.}
 \begin{tabular}{cc|ccc}
   $V^{(2)}$ &  $c_E$  &  $E_2$  & $E_3$  &  $E_4$  \\
   \hline
$V_{\rm \alpha}$ & $-0.10 $ & $ -0.920 $ & $ -3.223 $ & $ -7.125 $ \\
$V_{\rm \alpha}$ & $-0.05 $ & $ -0.920 $ & $ -2.884 $ & $ -5.832 $ \\
$V_{\rm \alpha}$ & $-0.01 $ & $ -0.920 $ & $ -2.628 $ & $ -4.906 $ \\
$V_{\rm \alpha}$ & $ 0.00 $ & $ -0.920 $ & $ -2.567 $ & $ -4.695 $ \\
$V_{\rm \alpha}$ & $ 0.01 $ & $ -0.920 $ & $ -2.507 $ & $ -4.494 $ \\
$V_{\rm \alpha}$ & $ 0.05 $ & $ -0.920 $ & $ -2.278 $ & $ -3.798 $ \\
$V_{\rm \alpha}$ & $ 0.10 $ & $ -0.920 $ & $ -2.027 $ & $ -3.179 $ \\
$V_{\rm \beta}$  & $-0.10 $ & $ -0.474 $ & $ -3.379 $ & $ -8.412 $ \\
$V_{\rm \beta}$  & $-0.05 $ & $ -0.474 $ & $ -2.283 $ & $ -5.727 $ \\
$V_{\rm \beta}$  & $-0.01 $ & $ -0.474 $ & $ -1.792 $ & $ -4.183 $ \\
$V_{\rm \beta}$  & $ 0.00 $ & $ -0.474 $ & $ -1.708 $ & $ -3.846 $ \\
$V_{\rm \beta}$  & $ 0.01 $ & $ -0.474 $ & $ -1.626 $ & $ -3.517 $ \\
$V_{\rm \beta}$  & $ 0.05 $ & $ -0.474 $ & $ -1.370 $ & $ -2.451 $ \\
$V_{\rm \beta}$  & $ 0.10 $ & $ -0.474 $ & $ -1.240 $ & $ -1.874 $
\end{tabular}   
 \label{tab:gs_energies}
\et

A sampling of ground-state energies are given in
Table~\ref{tab:gs_energies}. A few simple features to note are the
fact that the varied three-body force has no effect on the two-body
binding energy and that a negative $c_E$ is added attraction and thus
the 3- and 4-body bound states are deeper. For completeness we also
include in Table~\ref{tab:gs_hws} the optimal $\hw$ for each of these
interaction choices.

\bt
 \caption{Optimal $\hw$s for the potentials in
 Table~\ref{tab:gs_energies} with $A = 2$, 3, and 4.}
 \begin{tabular}{cc|ccc}
   $V^{(2)}$ &  $c_E$  &  $\hw(E_2)$  & $\hw(E_3)$  &  $\hw(E_4)$  \\
   \hline
$V_{\rm \alpha}$ & $-0.10 $ & $ 4.4 $ & $ 5.8 $ & $ 5.5 $ \\
$V_{\rm \alpha}$ & $-0.05 $ & $ 4.4 $ & $ 6.1 $ & $ 5.8 $ \\
$V_{\rm \alpha}$ & $-0.01 $ & $ 4.4 $ & $ 5.1 $ & $ 5.6 $ \\
$V_{\rm \alpha}$ & $ 0.00 $ & $ 4.4 $ & $ 5.0 $ & $ 5.3 $ \\
$V_{\rm \alpha}$ & $ 0.01 $ & $ 4.4 $ & $ 4.9 $ & $ 5.0 $ \\
$V_{\rm \alpha}$ & $ 0.05 $ & $ 4.4 $ & $ 4.7 $ & $ 4.4 $ \\
$V_{\rm \alpha}$ & $ 0.10 $ & $ 4.4 $ & $ 4.5 $ & $ 3.8 $ \\
$V_{\rm \beta}$  & $-0.10 $ & $ 1.1 $ & $ 0.1 $ & $ 0.1 $ \\
$V_{\rm \beta}$  & $-0.05 $ & $ 1.1 $ & $ 0.1 $ & $ 0.1 $ \\
$V_{\rm \beta}$  & $-0.01 $ & $ 1.1 $ & $ 0.5 $ & $ 0.5 $ \\
$V_{\rm \beta}$  & $ 0.00 $ & $ 1.1 $ & $ 1.6 $ & $ 2.1 $ \\
$V_{\rm \beta}$  & $ 0.01 $ & $ 1.1 $ & $ 0.5 $ & $ 0.5 $ \\
$V_{\rm \beta}$  & $ 0.05 $ & $ 1.1 $ & $ 0.2 $ & $ 0.3 $ \\
$V_{\rm \beta}$  & $ 0.10 $ & $ 1.1 $ & $ 0.1 $ & $ 0.1 $
\end{tabular}   
 \label{tab:gs_hws}
\et

Most of the figures in this chapter show calculations with $\nmax =
28$. With this basis size, the ground-state energies   are
generally converged to one part in $10^{4}$, which is more than
sufficient for our purposes.  As usual, increasing $\nmax$ leads
to rapidly increasing matrix sizes and computation times;  times
for $A=3$ with $\nmax = 32$ are a factor of 3 longer than with
$\nmax = 28$ and with $\nmax = 40$ the time increases by another
factor of 10. These timing and scaling properties of the one
dimensional NCSM are explored in depth in Appendix~\ref{chapt:app_scaling}.


\section{Evolution of Many-Body Forces in Bound States}

\subsection{Two-body Results}

We first consider the bound state of two identical bosons using
the potential $V_{\alpha}$. Because the SRG is a series of
unitary transformations, we expect that the binding energy will
not be changed by evolving the two-body interaction in the
two-particle space.  Indeed, we find it to be constant to high
accuracy. The ground-state wave function, however, changes
dramatically, as seen from the probability densities plotted in
Fig.~\ref{fig:srg_2_body}. The initial probability density
exhibits a sizable ``wound'' near the origin that is filled in as
$\lambda$ decreases.  By $\lambda = 2$ there is no signature of a
repulsive core (and the wave function is modified out to larger
$x$). This is the same pattern seen for the S-wave component of
deuteron wave functions starting from three-dimensional
nucleon-nucleon S-wave potentials with strong repulsive cores
such as Argonne $v_{18}$, with the ``uncorrelated'' final wave
function at $\lambda = 2$ roughly comparable to $\lambda =
1.5\,\mbox{fm}^{-1}$ for the deuteron~\cite{Bogner:2007srg}.

\begin{figure}[tbh!]
\begin{center}
 \includegraphics*[height=1.3in]{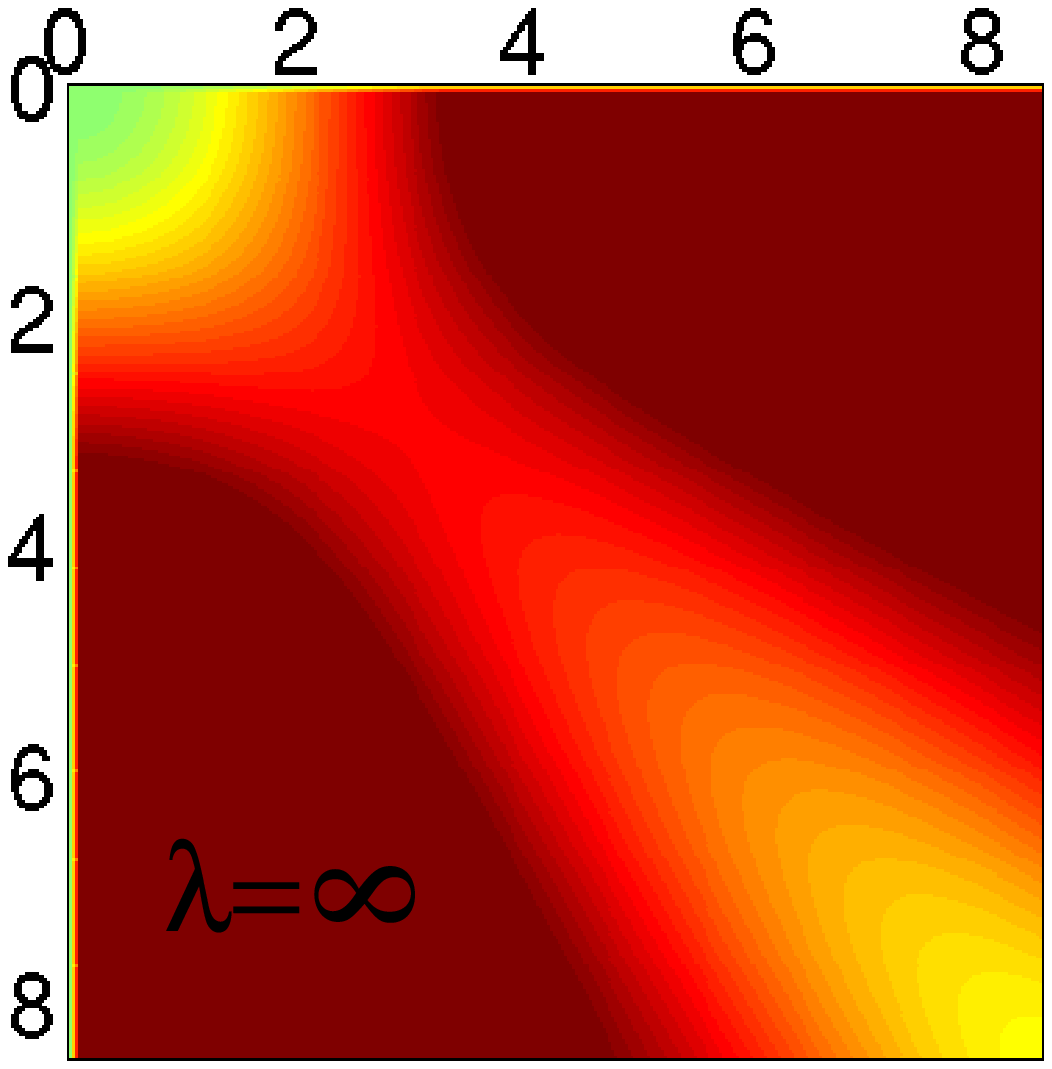}
 \includegraphics*[height=1.3in]{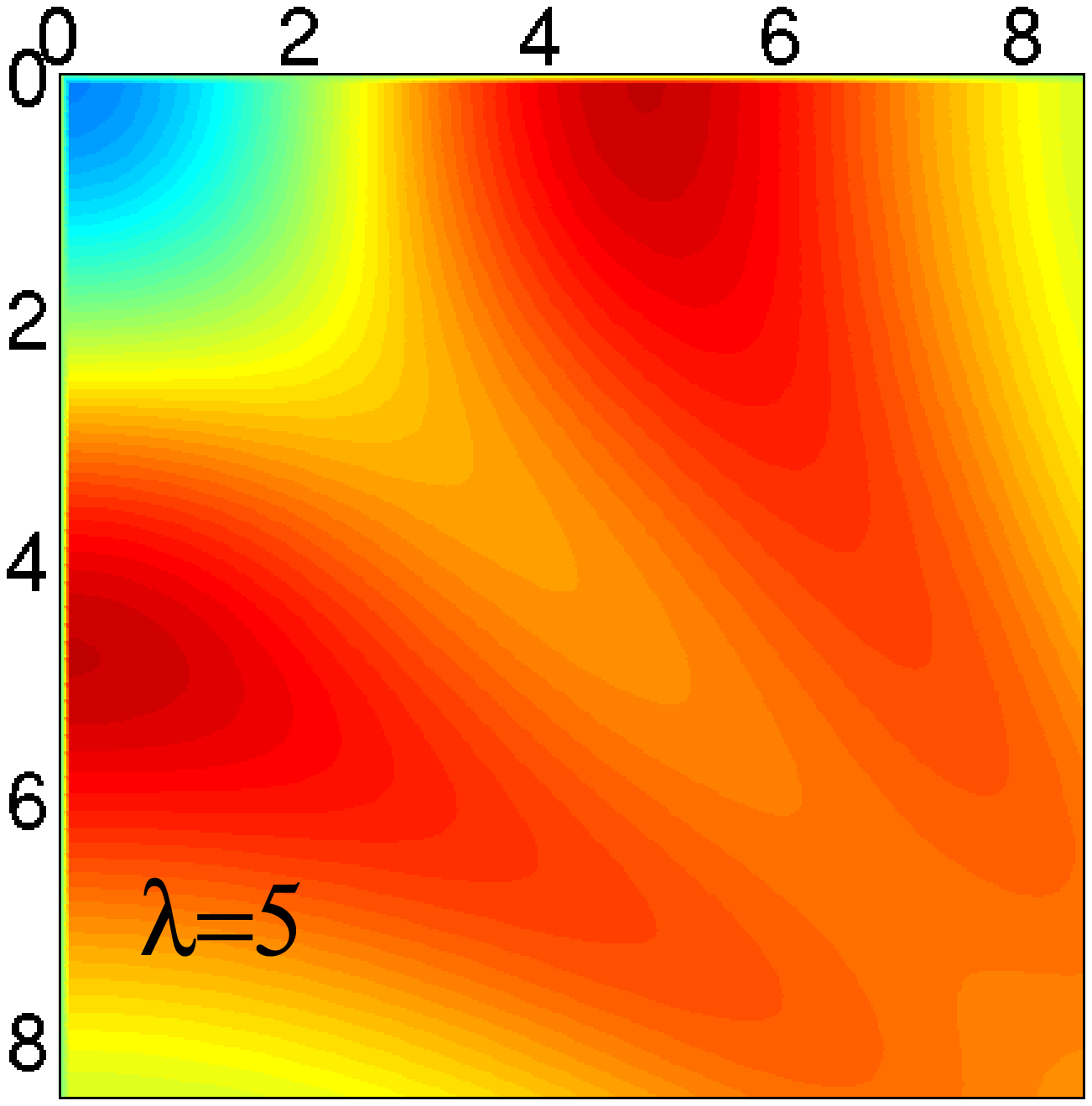}
 \includegraphics*[height=1.3in]{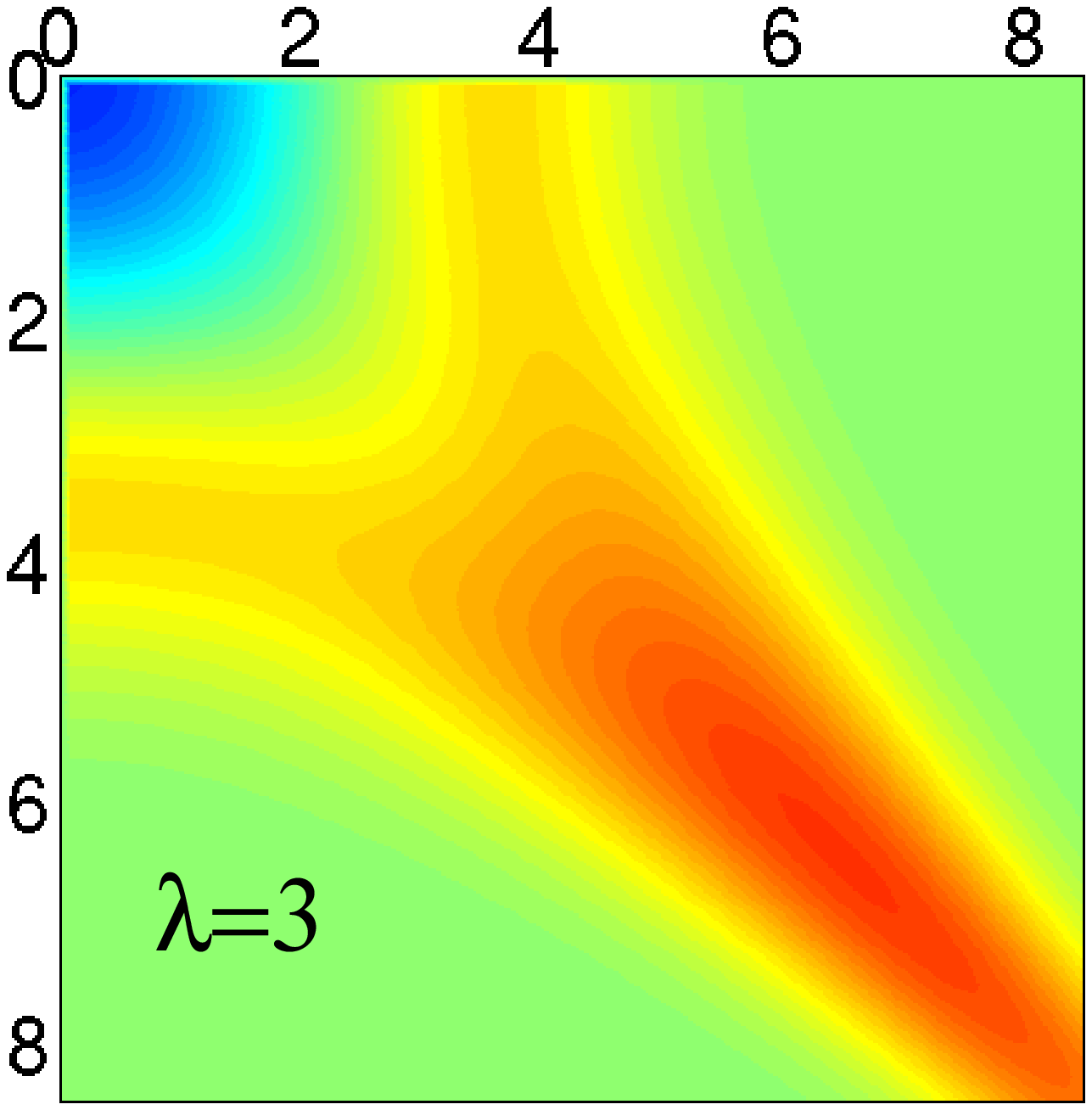}
 \includegraphics*[height=1.3in]{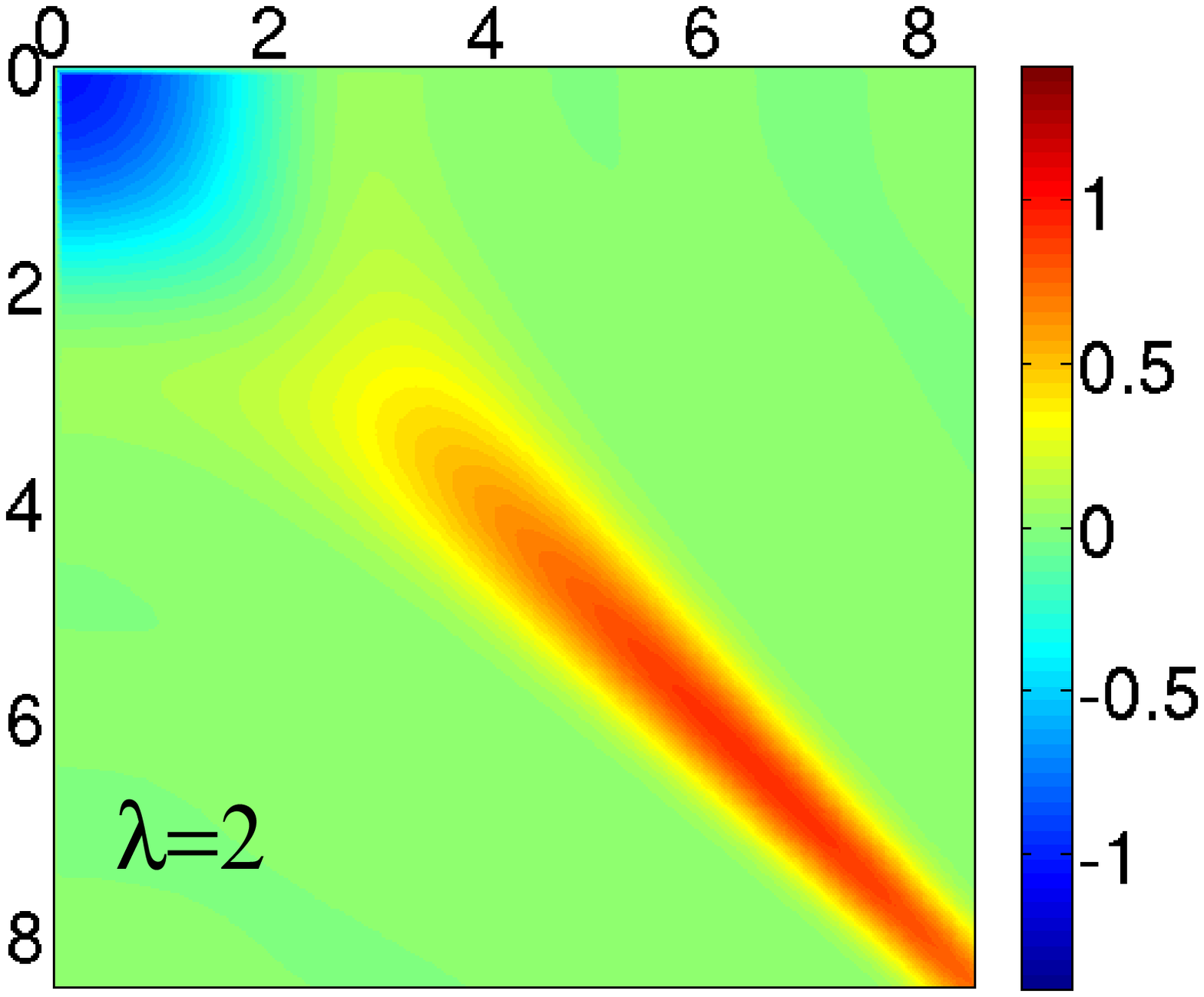}
\end{center}
\captionspace{Even part of the SRG potential  $
[V^{(2)}_{s}(p,p') + V^{(2)}_{s}(p,-p')]$ in dimensionless
units as a function of $p$ and $p'$ for  $\lambda = \infty$, 5,
3, and 2 (where $\lambda = 1/s^{1/4}$).  The initial potential is
$V_{\alpha}$ from Table~\ref{tab:negelePars}.}
\label{fig:srg_2_body_mom}
\end{figure}

The evolution of the potential in the momentum basis, shown as a color
contour plot in Fig.~\ref{fig:srg_2_body_mom}, also demonstrates
this behavior. (The even part of the potential is shown, which is
the analog of the S-wave part.) The initial potential is
dominated by strongly repulsive matrix elements coupling low and
high momenta.  The evolution in $\lambda$ band diagonalizes the
potential to a width in $p^2$ of roughly $\lambda^2$ while a soft
attractive part emerges in the low-momentum region. The pattern
in Fig.~\ref{fig:srg_2_body_mom} reflects increasing non-locality
as $\lambda$ is lowered, which in turn reduces the wound in the
wave function. From the probability density and the momentum
space plots we estimate that evolving to halfway between $\lambda
= 2$ and 3 for $V_{\alpha}$  corresponds roughly to the $\lambda$
scale typically used in nuclear structure calculations (around
$2\,\mbox{fm}^{-1}$).


\subsection{Three-body Results}

To calculate properties of the three-particle system we construct
the Hamiltonian in the basis of symmetric three-particle
eigenstates as described in Sec.~\ref{sec:hamiltonian}. The SRG
evolution of the potential in the three-particle space leaves the
ground state energy invariant if the full Hamiltonian is kept,
because the transformations are unitary. However,  the
Hamiltonian matrix elements in this space  do not follow simply
from the pairwise sum of the two-body potential matrix elements; 
as the SRG evolves a three-body force is induced even if its
initial strength is zero.

\begin{figure}[tbh!]
\begin{center}
 \dblpic{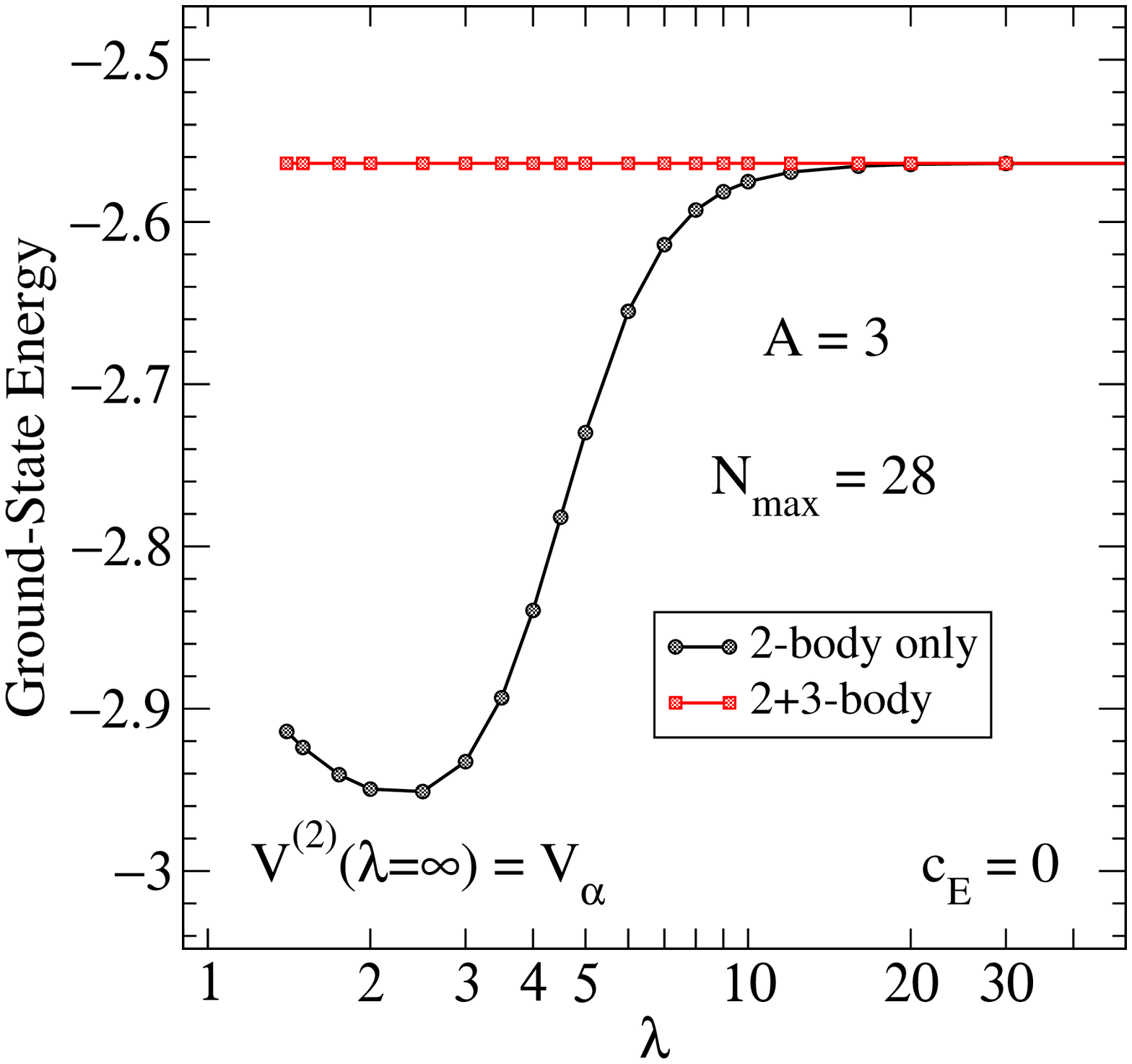}
 \hfill
 \dblpic{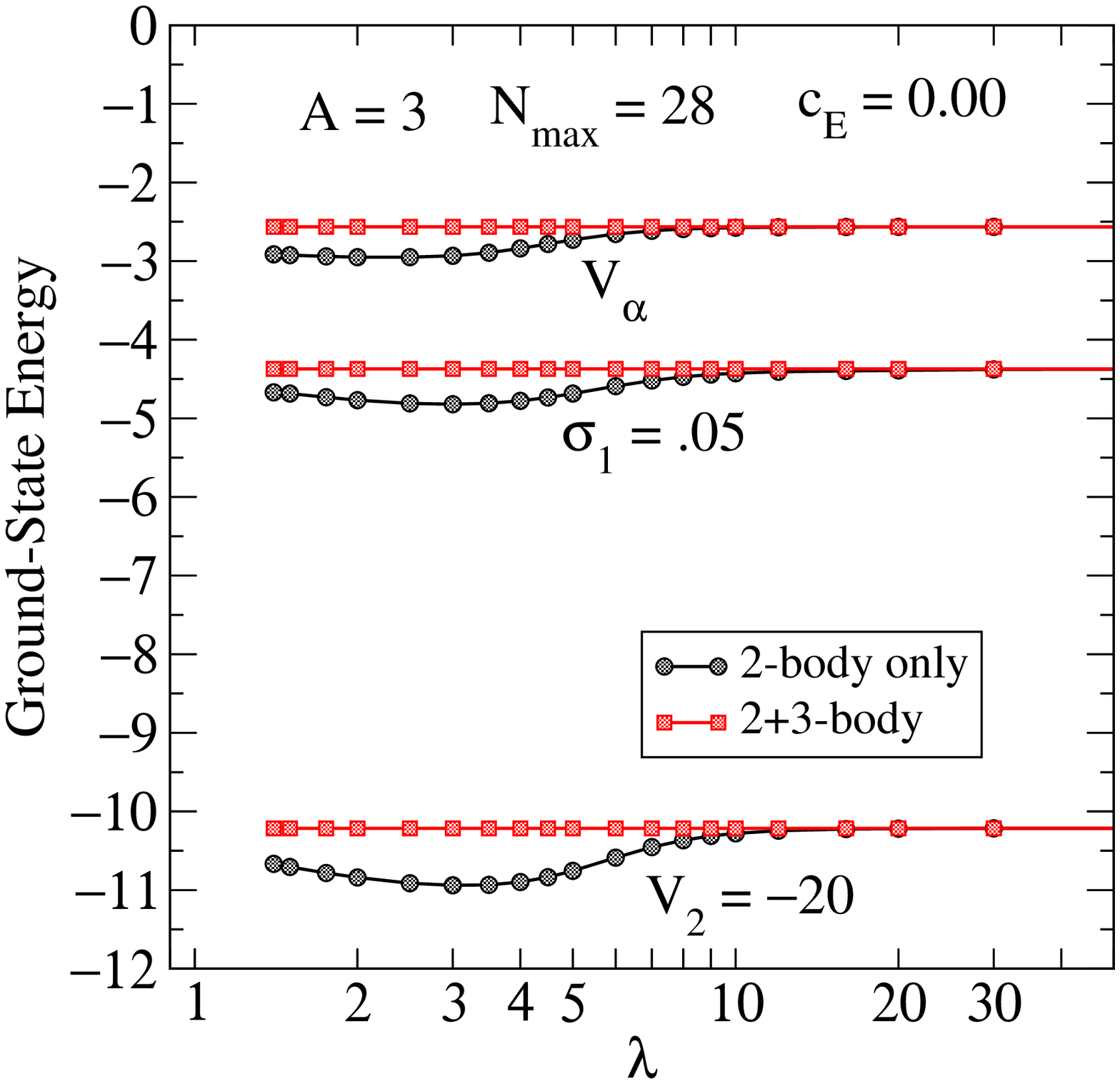}
\end{center}
\captionspace{The lowest bound-state energy $E_3$ for a
three-particle system as a function of $\lambda$ with the initial
two-body-only potential $V_{\alpha}$.   The (red) curves with
squares include the full evolution of the Hamiltonian while the
(black) curves with circles use the two-body potential evolved
in the two-particle system.  The right frame shows two additional
results from varying $\sigma_1$ and $V_2$ from the values in
Table~\ref{tab:negelePars}.}
\label{fig:srg_3_body}
\end{figure}

\begin{figure}[tb]
\begin{center}
  \dblpic{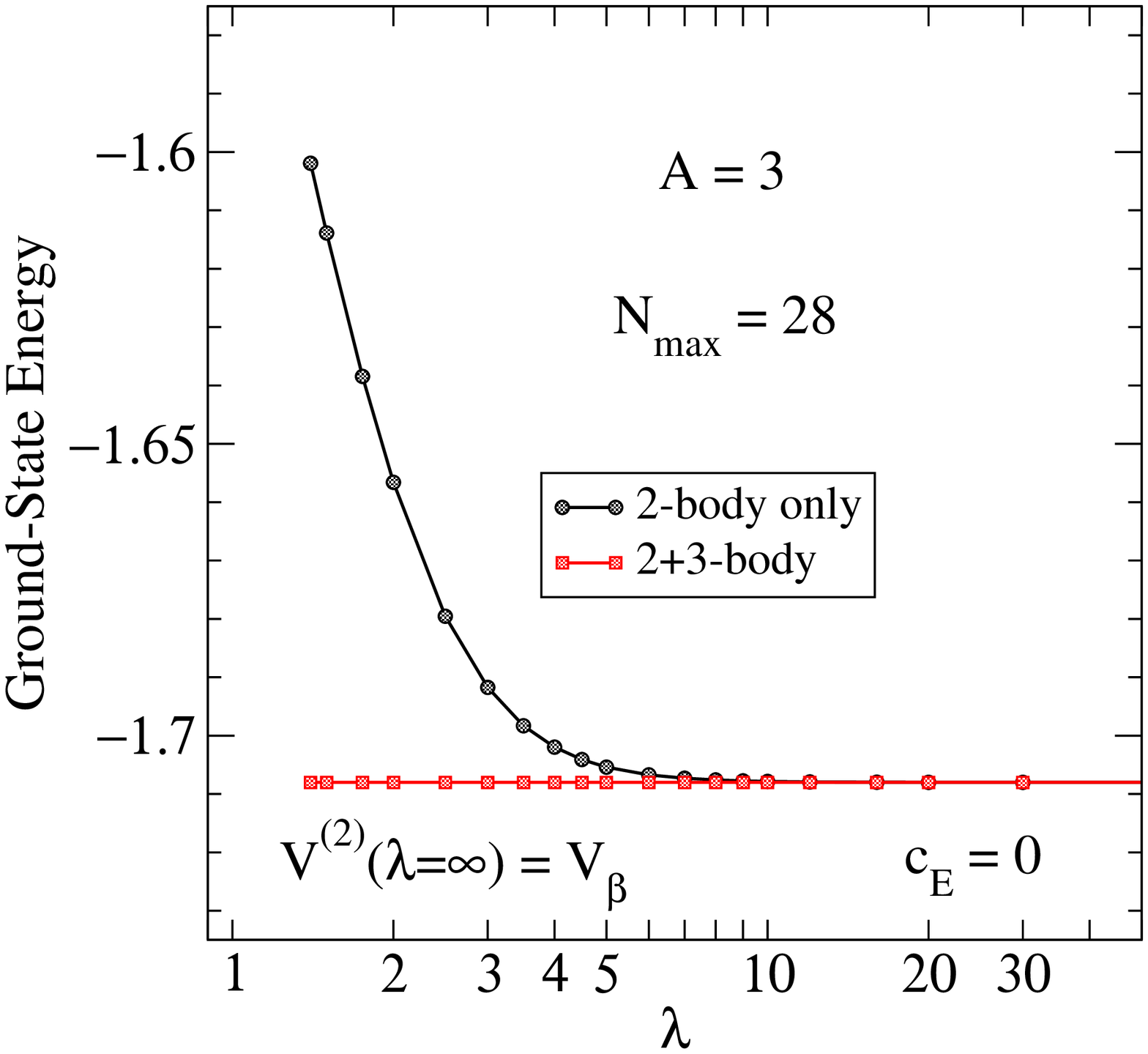}
 \hfill
 \dblpic{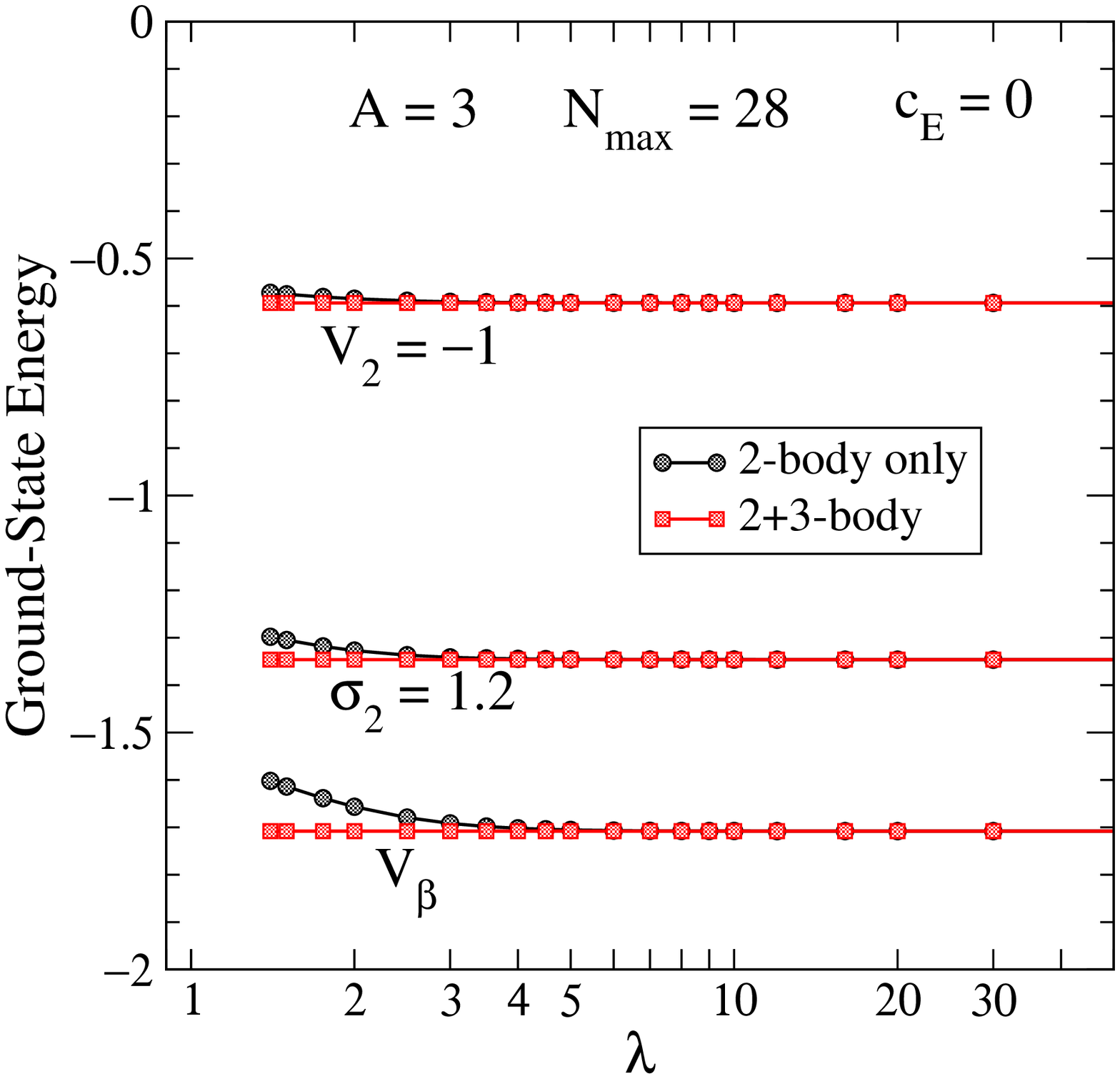}
\end{center}
\captionspace{Same as Fig.~\ref{fig:srg_3_body} but
  with initial potential $V_{\beta}$.  The right frame shows two
  additional results from varying  $\sigma_2$ and $V_2$ from the
  values in Table~\ref{tab:negelePars}.}
\label{fig:3N_other}
\end{figure}

The effect of this full three-particle space SRG evolution is
shown in Fig.~\ref{fig:srg_3_body} for initial two-body potential
$V_{\alpha}$ and with initial $V^{(3)} = 0$ (non-zero values of
$c_E$ are considered in the next section). We plot the
ground-state energy for the three-particle system both with the
initial two-body interaction embedded in the three-particle
symmetric space and then evolved (the red curve with squares) and
also with the two-body interaction evolved in the two-particle
space before embedding in the three-particle space at each
$\lambda$ (the black curve with circles).  We can see that the energy
evaluated with the two-body interaction alone deviates noticeably
as $\lambda$ drops below 10. This variation is the signature that the
two-body transformation is only approximately unitary in the
three-particle sector. The error reaches a peak in $\lambda$
between 2 and 3 and then decreases. The same pattern has been
observed for NN potentials in three
dimensions~\cite{Bogner:2007rx} and remains qualitatively the
same when parameters in the potential are varied (e.g., see the
right plot in Fig.~\ref{fig:srg_3_body}). We made the same
calculation using the purely attractive initial two-body
potential $V_{\beta}$, which is shown in Fig.~\ref{fig:3N_other}.
Here the induced three-body force has the opposite sign and there
is no  maximum, which implies that the qualitative pattern of
evolution is dictated by the interplay between attractive
long-range and repulsive short-range parts of the potential. 
These features are explored further in
Sect.~\ref{sec:oned_vev}.

\subsection{Results for $A = 4$ and $A=5$}

Next we turn to $A=4$ and $A=5$, where we expect to see  the effects
of induced three-, four- and five-body forces. The key issue is the
relative sizes of these contributions; we are looking to test whether
an initial hierarchy of few-body interactions is preserved and
therefore can be truncated with a controlled error. In
Chapter~\ref{chapt:ncsm} we will address the same question in the
realistic NCSM.

\begin{figure}
\begin{center}
\strip{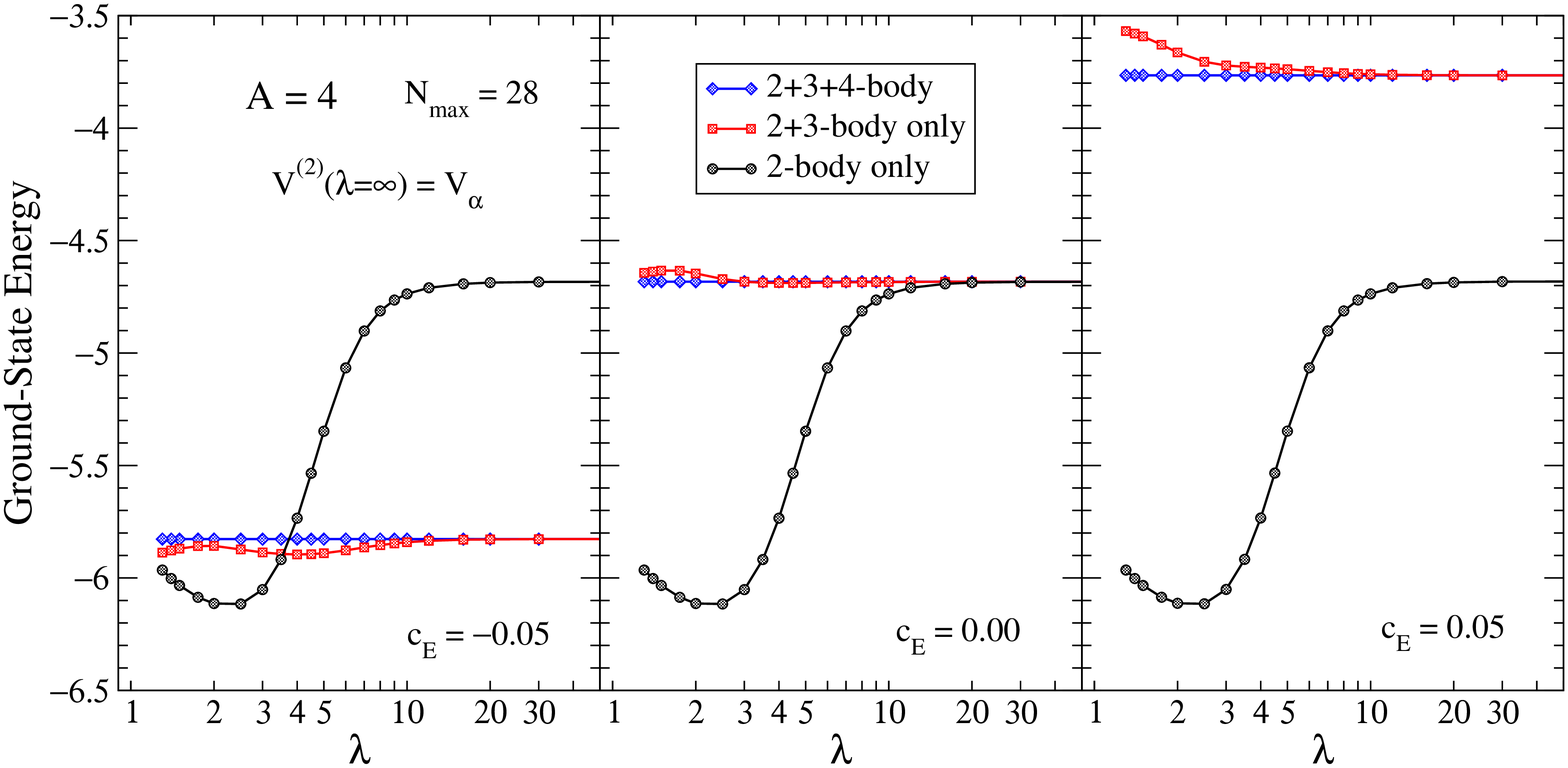}
\end{center}
\captionspace{The lowest bound-state energy $E_4$ for a
four-particle system as a function of $\lambda$ with the initial
two-body potential $V_{\alpha}$ and different initial  three-body
force strengths ($c_E=\pm 0.05$).}
\label{fig:srg_4_body_Va}
\end{figure}

In applying the SRG in the four-particle space we have three different
calculations of the ground-state energy to compare. The first is the
two-body potential embedded successively in the three- and
four-particle spaces and then evolved  in the four-particle space. 
The resulting unitary transformations will induce three- and four-body
interactions that leave the eigenvalues invariant. We can also evolve
in the two-particle space before embedding in the three- and
four-particle spaces and diagonalizing. As we saw before in
Fig.~\ref{fig:srg_3_body} and see now in Fig.~\ref{fig:srg_4_body_Va},
the two-body-only evolution deviates because the Hamiltonian is not
evolved by an exactly unitary transformation. Finally, to find the
relative size of the three and four-body interactions we can evolve in
the three-particle space, thereby inducing only three-body forces.
Note that the two and three-body forces must be embedded differently
in the four particle space because they have different combinatoric
factors associated with them, i.e., there are ${4 \choose 2} = 6$
pairs and ${4 \choose 3} = 4$ triplets. So, the proper mixture of two-
and three-body force contributions to the four-particle system
interaction is $V = 6V^{(2)} + 4V^{(3)}$.

All three of these calculations for $A=4$ are shown in
Fig.~\ref{fig:srg_4_body_Va} for the two-body potential $V_{\alpha}$
and several choices of the initial three-body force.  The magnitude of
$c_E$  was chosen so that the fractions of the $A=3$ and $A=4$
ground-state energies from the three-body interaction are roughly
comparable to the corresponding fractions for nuclei using typical
realistic NN potentials. The qualitative behavior is similar for other
choices of $c_E$ and $V^{(2)}$.    In all plots the curves for the
two-body-only (black line with circles), the two-plus-three (red line
with squares), and the full two-plus-three-plus-four interaction (blue
line with diamonds) show the hierarchy of different few-body
interaction components.  The difference between the square and diamond
lines represents the contribution of the four-body force, and the
difference between the circle and square lines is the contribution of
the three-body force alone. In the left and right panels, where the
initial three-body force has been switched on, we can see that it is
perturbatively small from the balanced attractive/repulsive
contributions.

The four-body contribution is at most ten percent that of the
three-body, which is itself small compared to the two-body
contribution except when the latter gets small  for small $\lambda$
(note the expanded scales on the figures).  Considering calculations
with different $c_E$ values, we see that the $\lambda$ dependence of
the induced four-body part depends on the interplay of initial and
induced forces. In some cases noticeable (but small) evolution starts
at $\lambda =10$ while in other cases it is deferred until much
smaller $\lambda$. Regardless of the details, we stress that there is
no sign that induced many-body forces have rapid growth with $A$ or
exhibit unusual scaling.

We repeated for $A=4$ our test of decoupling that was shown in
Fig.~\ref{fig:srg_3_body_decoupling} for $A=3$. A similar pattern of
decoupling is found, namely an increased degree of decoupling until  a
$\lambda$ corresponding to the minimum of the two-body-only
ground-state energy of the $A=4$ system, after which it deteriorates. 

\begin{figure}
\begin{center}
\strip{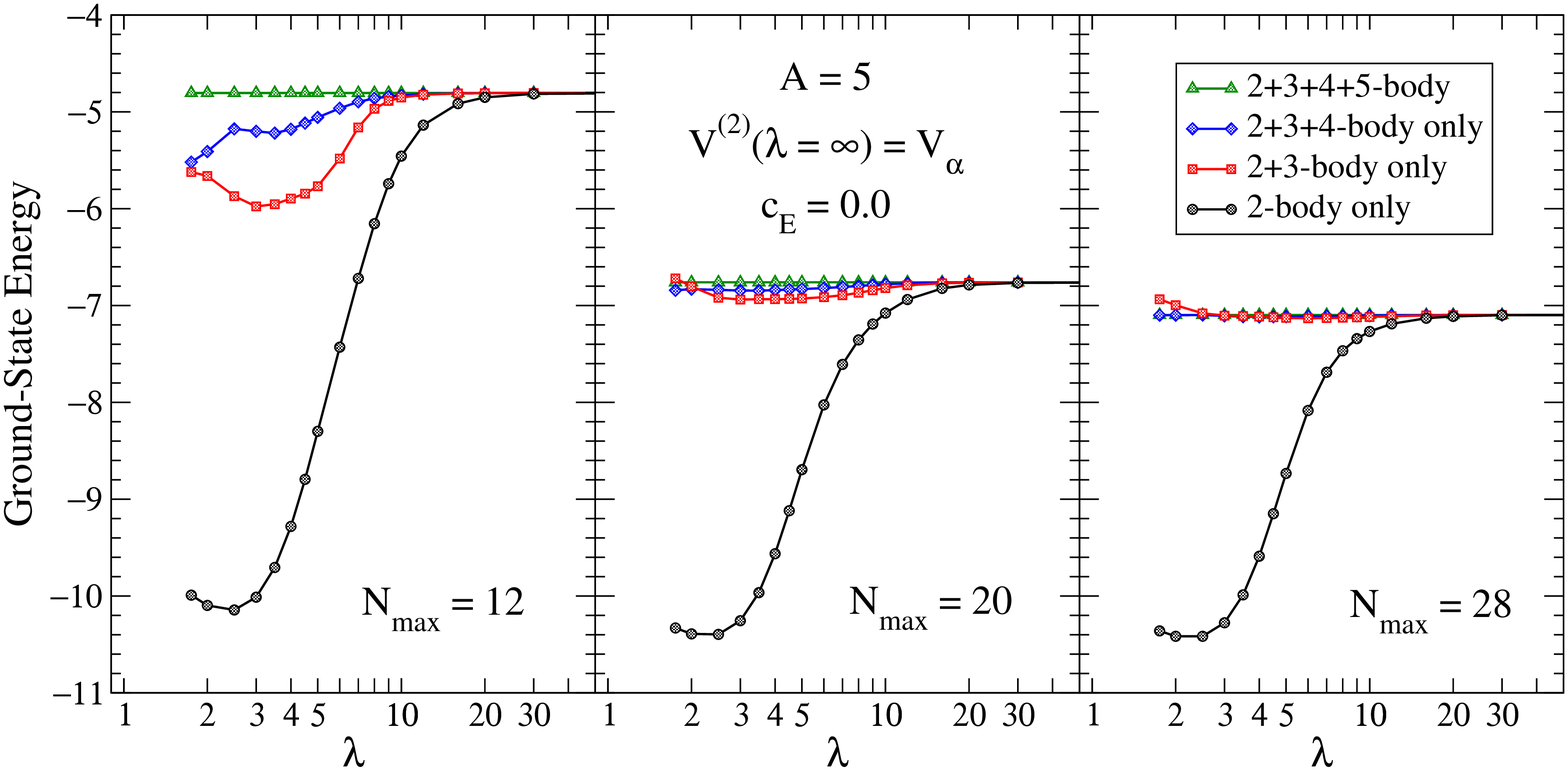}
\end{center}
\captionspace{The lowest bound-state energy $E_5$ for a 5-particle system 
as a function of $\lambda$ with an initial two-body-only $V_{\alpha}$
potential for several values of $\nmax$.}
\label{fig:srg_5_body}
\end{figure}

In Fig.~\ref{fig:srg_5_body} we show results for the SRG evolution,
with initial potential $V_{\alpha}$ and no initial three-body
interactions, in a five-particle system for several values of $\nmax$.
The right panel, with $\nmax = 28$, shows the converged result. We see
a decreasing hierarchy of induced many-body contributions for all
$\lambda$; the five-body contribution is essentially negligible (or
not distinguishable from numerical noise). Differences at the lower
$\nmax$ sized spaces arise both because the space needs to be large
enough for convergence to the exact energy eigenvalues but also
because the initial evolution of the potential needs a sufficiently
large space. Decoupling may improve this feature but is dependent on
the type of SRG used and the basis in which it is implemented. 


\section{Diagrammatic Analysis of Many-Body Force Evolution}
\label{sec:oned_vev}

In Ref.~\cite{Bogner:2007qb}, a diagrammatic approach to the SRG
equation was introduced, which organized the independent evolution of
two- and three-body (and higher-body) potentials. This formalism is
necessary in a momentum basis to avoid ``dangerous'' delta functions
from spectator particles.  In this section we examine how induced
three-body interactions evolve in our one-dimensional laboratory and
make connections to the diagrammatic expansion.

\begin{figure}[ptb]
\dblpic{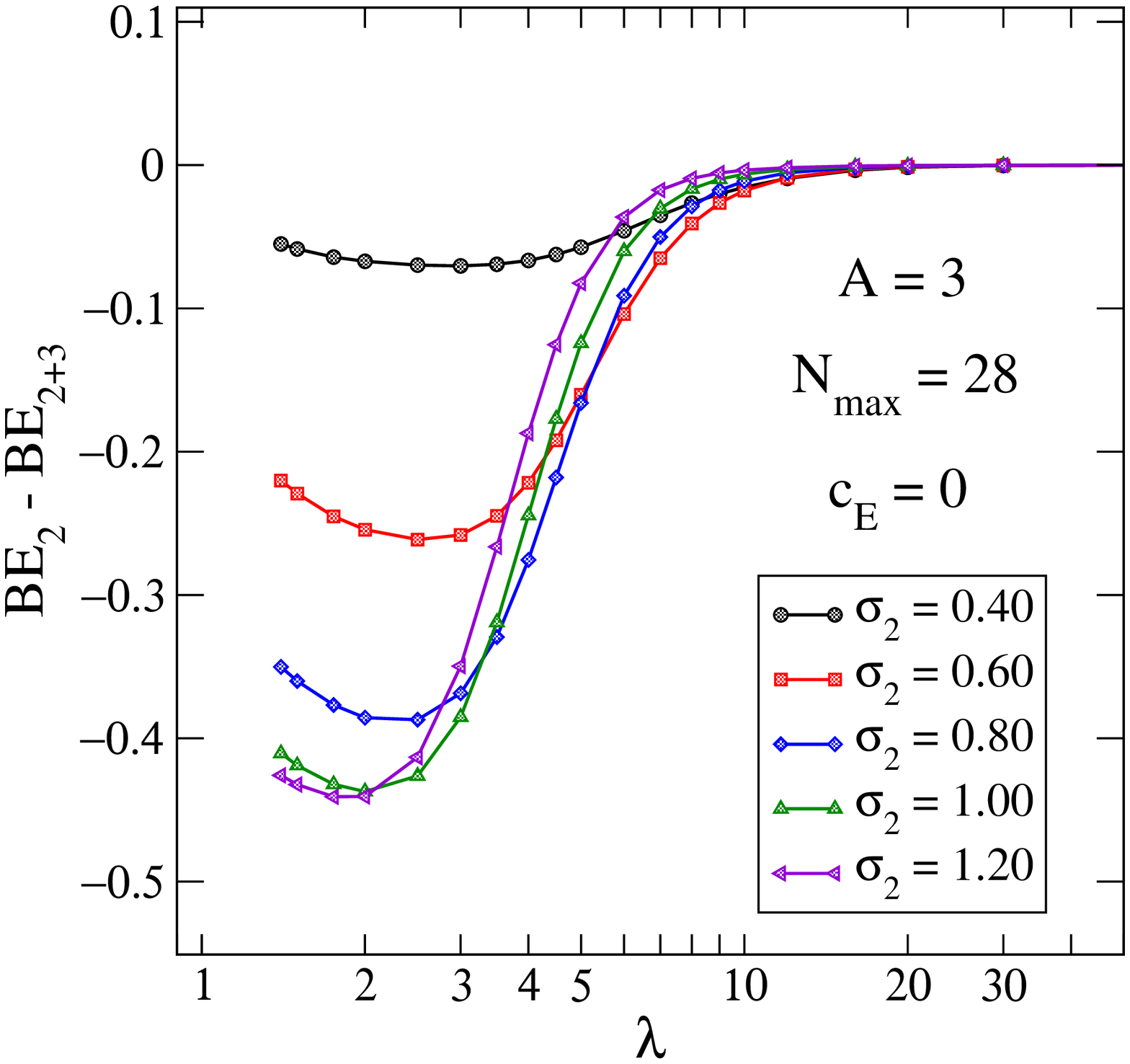}
  \hfill
\dblpic{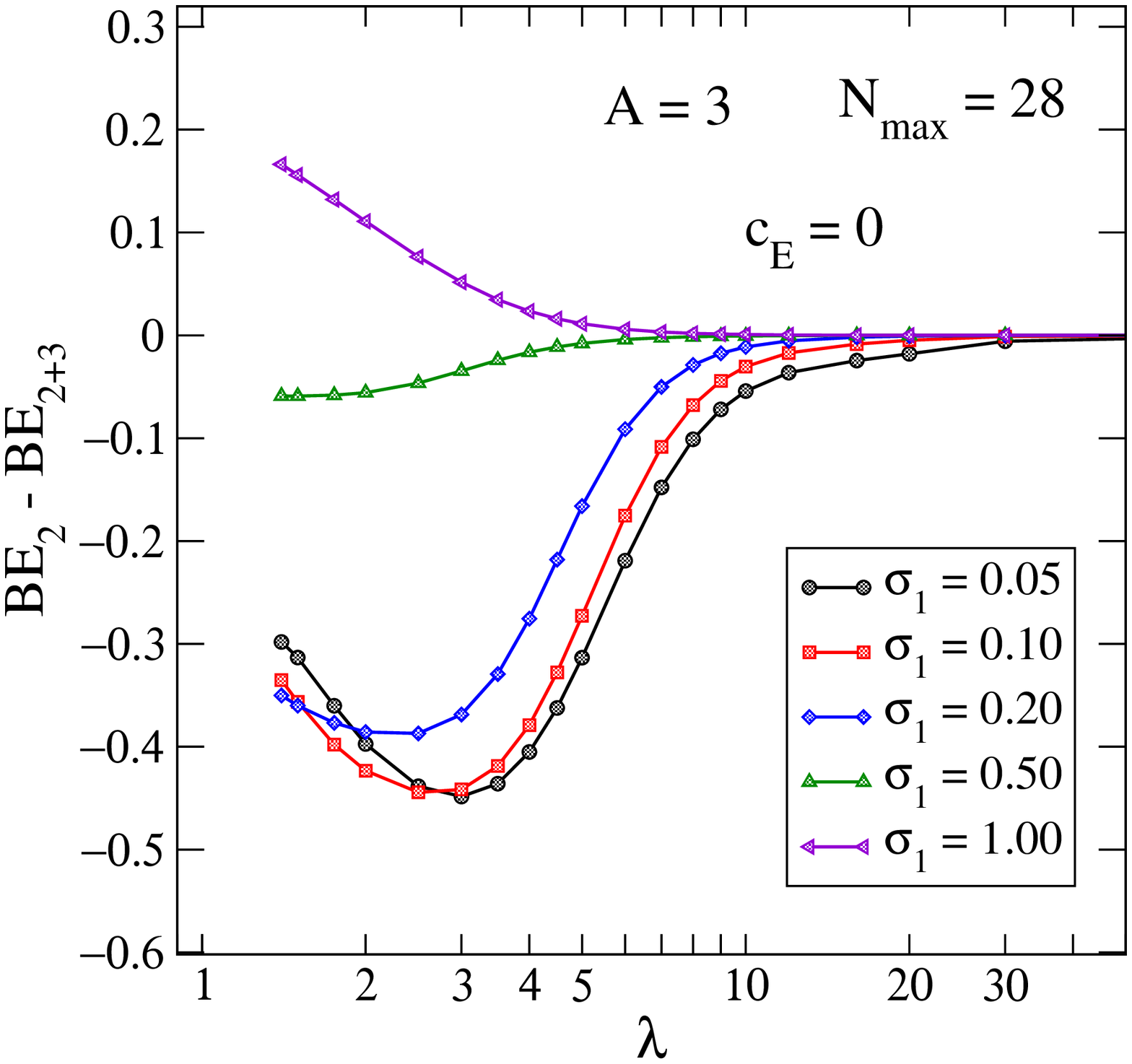}

\dblpic{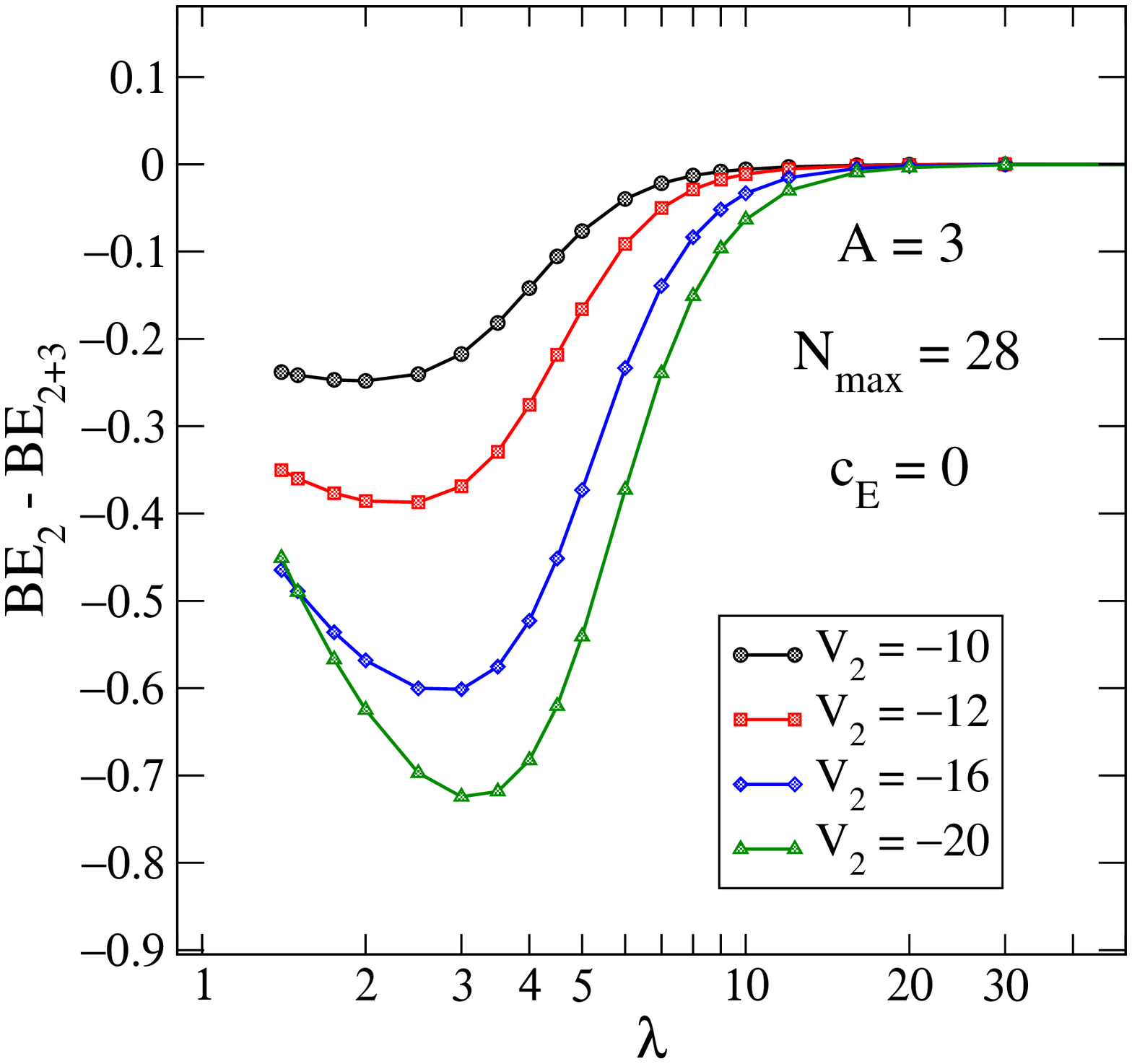}
  \hfill
\dblpic{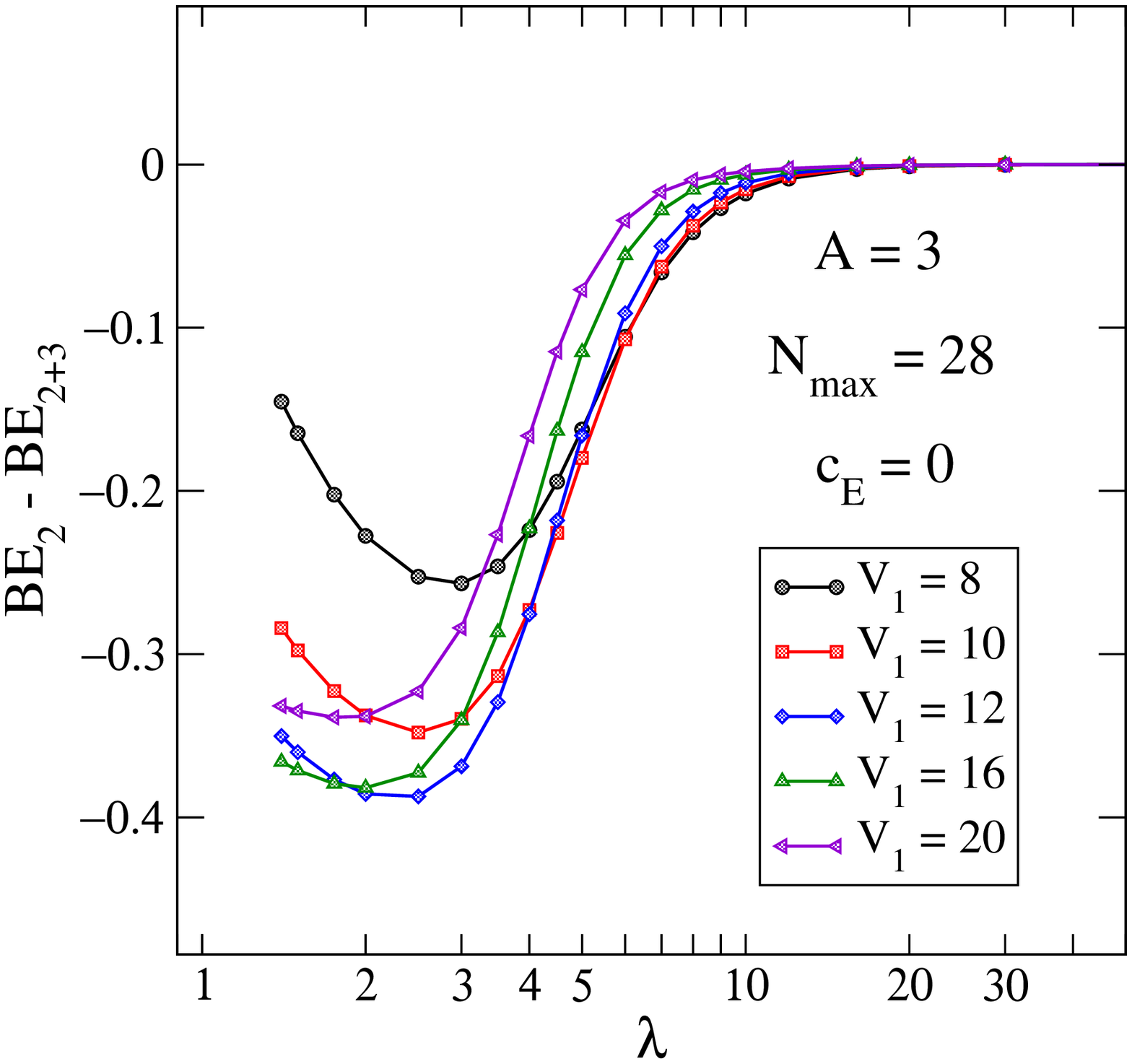}
\captionspace{Differences of two-body-only and 
two-plus-three-body $A=3$ ground-state energies as a function of
$\lambda$. Each of the parameters of the potential $V_{\alpha}$
are varied in each plot as the other parameters are held
constant. The upper panels vary the ranges while the lower vary
the strengths; the left vary the attractive part and the right
vary the repulsive part.} 
\label{fig:3N_rel_err_Va}
\end{figure}

\begin{figure}[ptb]
\dblpic{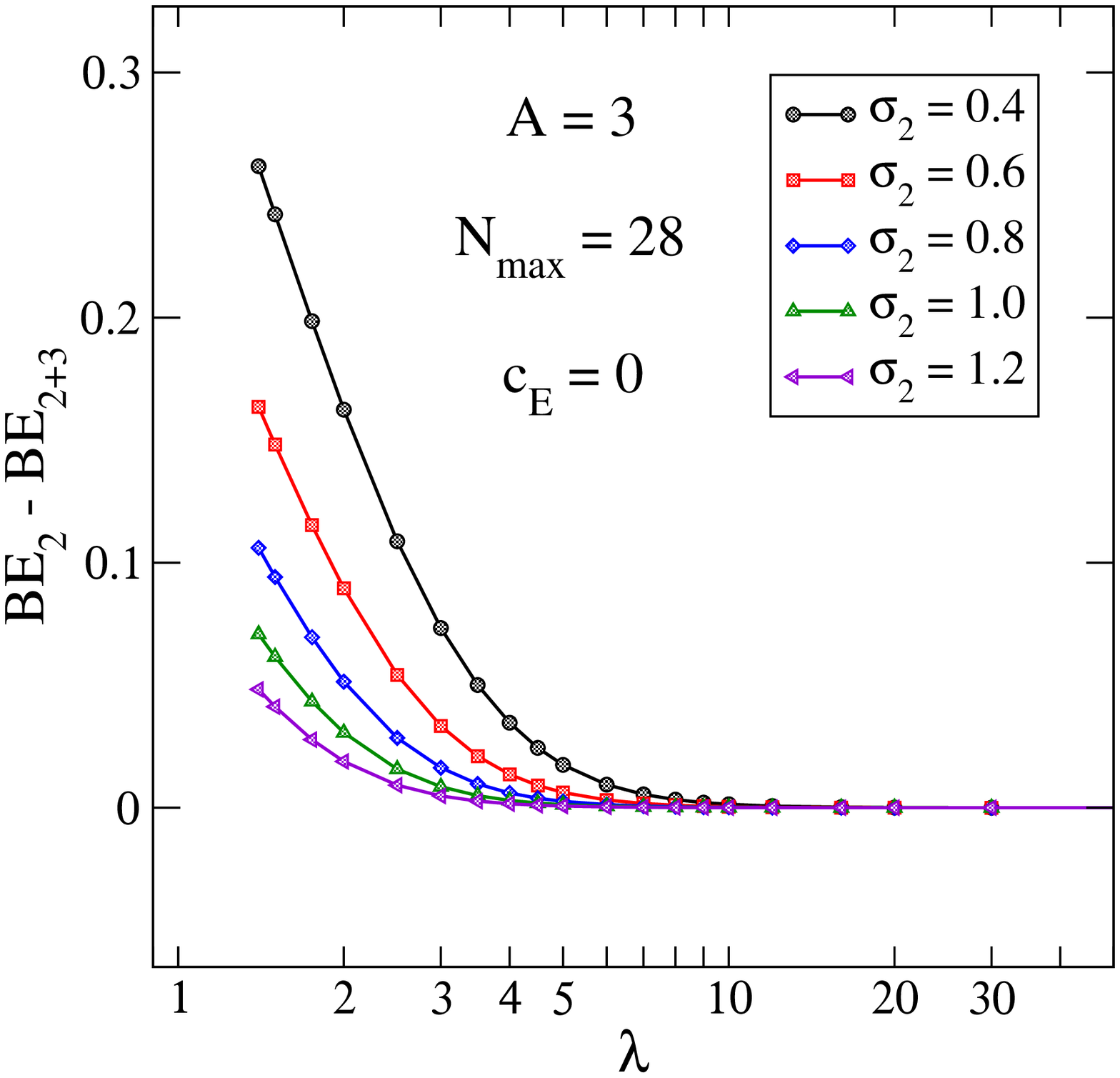}
  \hfill
\dblpic{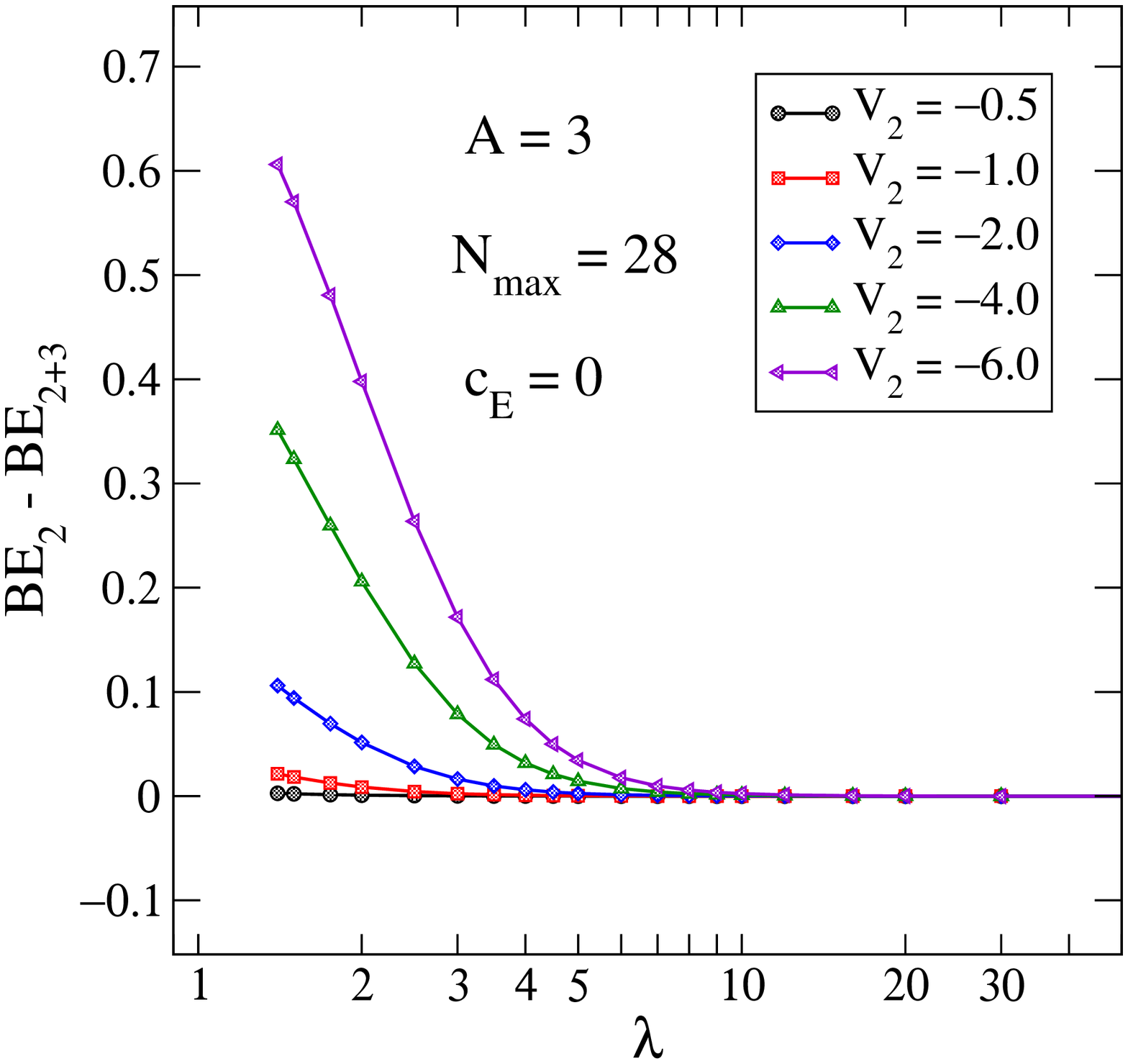}
\captionspace{Same as Fig.~\ref{fig:3N_rel_err_Va} but for $V_{\beta}$.}
\label{fig:3N_rel_err_Vf}
\end{figure}

As raw material for this analysis,  we
plot in Fig.~\ref{fig:3N_rel_err_Va} the error of the evolving
two-body-only binding energy while varying several parameters of
the initial two-body interaction $V_{\alpha}$ for $A=3$
ground-state energies. In the left plots we vary the range (top)
and the strength (bottom) of the attractive part of
$V_{\alpha}$.  In the right plots we vary the range (top) and the
strength (bottom) of the repulsive part. In
Fig.~\ref{fig:3N_rel_err_Vf} we present similar plots for a
simpler system that starts with the initial attraction-only
two-body potential $V_{\beta}$. Note again that the plots here
show the induced error between the two-body-only and
two-plus-three-body binding energies.

Certain qualitative features are found as expected in these
figures. Shorter ranges imply enhanced
coupling from low to high momentum and therefore we anticipate
that the evolution will start sooner (i.e., at higher
$\lambda$).  This is seen clearly on the left in
Fig.~\ref{fig:3N_rel_err_Vf}  and for the variation of the
(shorter-ranged) repulsive potential in
Fig.~\ref{fig:3N_rel_err_Va} (top right plot). There is also an
unsurprising increase in the magnitude of the induced three-body
interaction at each $\lambda$ with increased magnitude of the potential, as
seen on the right in Fig.~\ref{fig:3N_rel_err_Vf} and on the
bottom left in Fig.~\ref{fig:3N_rel_err_Va}. For a more
definitive analysis we need to recall the discussion from
Ref.~\cite{Bogner:2007qb}.

\begin{figure}[bt]
\begin{center}
\strip{figures/feynman_graphs.ps}
\end{center}
\captionspace{A diagrammatic decomposition of the SRG
 Eq.~\eqref{eq:commutator2}. A circle at a vertex denotes a commutator
 with $T_{\rm rel}$. }
\label{fig:srg_diagrams}
\end{figure}

The SRG evolution equation for the three-particle sector in
the notation of 
Ref.~\cite{Bogner:2007qb} is
\bea
  \frac{dV_{\flow}^{(2)}}{d\flow} +\frac{dV_{\flow}^{(3)}}{d\flow} 
  &=&  \mybar{\mybar{V}}_{\flow}^{(2)}+\mybar{\mybar{V}}_{\flow}^{(3)}+   
  [\mybar{V}_{\flow}^{(2)},V_{\flow}^{(2)}]
  +[\mybar{V}_{\flow}^{(2)},V_{\flow}^{(3)}]
  +[\mybar{V}_{\flow}^{(3)},V_{\flow}^{(2)}]
  +[\mybar{V}_{\flow}^{(3)},V_{\flow}^{(3)}]  
  \;,
  \label{eq:commutator2}
\eea 
where each bar denotes a commutator with $T_{\rm rel}$. We remind
the reader that $dT_{\rm rel}/ds = 0$ by construction. A
diagrammatic decomposition of this equation is shown in
Fig.~\ref{fig:srg_diagrams}. In the two-body sector, the
equation  reduces to the first term on the left and the first and
third terms on the right (the first row in 
Fig.~\ref{fig:srg_diagrams}). These terms keep two-particle
energy eigenvalues invariant under evolution. In the
three-particle sector, Eq.~\eqref{eq:commutator2} results in not
only these two-body graphs with a disconnected spectator but
additional graphs involving connected combinations of two and
three-body interactions. The diagrams with two-body interactions
and a disconnected spectator line satisfy the two-body evolution
equations, and so will cancel out of the full three-particle-sector
evolution equation. Thus the evolution of the three-body
interaction is dictated by the connected diagrams (the second row
in Fig.~\ref{fig:srg_diagrams}). In summary, the evolution of the
$A$-body potential in the $A$-particle system is given by
\beqn 
  \frac{dV^{(A)}_s}{ds} = [\eta_s,H_s]_A \;, 
  \label{eq:srg_dVds}   
\eeqn
where the ``$A$'' subscript on the right side means the
fully connected $A$-particle terms.

To make a connection between the individual terms in the
three-body interaction evolution and the running of the ground-state
energy, we need to derive the evolution equations for the
\emph{expectation value} of $V^{(3)}_s$ in the ground state.
Denoting the ground-state wave function for the $A$-particle system
by $|\psi^A_s\ra$, it evolves according to 
\beqn
  |\psi^A_s\ra = U_s |\psi^A_{s=0}\ra \;,
  \quad \mbox \quad 
  \frac{d}{ds} |\psi^A_s\ra = \eta_s |\psi^A_s\ra
  \;,
\eeqn
where $U_s$ is the SRG unitary transformation at $s$ and
\beqn
  \eta_s = \frac{dU_s}{ds} U^\dagger_s = - \eta^\dagger_s
  \;.
\eeqn  
Then the matrix element of an operator $O_s$ evolves according to
\beqn
  \frac{d}{ds} \la \psi^A_s | O_s | \psi^A_s \ra
  = \la \psi^A_s |
  \frac{d O_s}{ds} - [\eta_s,O_s] 
    | \psi^A_s \ra
  \;.
  \label{eq:Oevolve}
\eeqn
If the operator $O_s$ evolves according to $O_s = U_s O_{s=0}
U_s^\dagger$, then the matrix element vanishes, as when $O_s = H_s$.

However, if we wish to see how one part of $H_s$
evolves, such as the expectation value of $V^{(3)}$, we obtain
\beqn
  \frac{d}{ds}\la \psi^A_s |V_s^{(3)}|\psi^A_s\ra = 
  \la \psi^A_s|\frac{dV^{(3)}_s}{ds} - [\eta_s,V_s^{(3)}]|\psi^A_s\ra
  \;,
\label{eq:dds_vev}
\eeqn
which does not give zero in general because $V_s^{(3)} \neq
U_sV_{s=0}^{(3)}U_s^{\dagger}$. 
In the two-particle case, the analog of
Eq.~\eqref{eq:dds_vev} gives $d\la V^{(2)} \ra/ds
= \la [\eta_s,T_{\rm rel}] \ra$. In the three-particle case, 
we can expand Eq.~\eqref{eq:dds_vev} as
\bea
\frac{d}{ds}\la \psi_s |V^{(3)}_s|\psi_s\ra &=& \la \psi_s|[\eta_s,H_s]_3 -
[\eta_s,V^{(3)}_s]|\psi_s\ra  \nonumber \\
&=& \la \psi_s |
   [\vbtr,T_{\rm rel}] + [\vbt,V^{(2)}_s]_c + [\vbt,V^{(3)}_s] + [\vbtr,V^{(2)}_s] +
[\vbtr,V^{(3)}_s] \nonumber \\
&& \qquad - [\vbt,V^{(3)}_s] - [\vbtr,V^{(3)}_s]  | \psi_s \ra  \nonumber \\
&=& \la \psi_s |  [\vbtr,H_s] + [\vbt,V^{(2)}_s]_c -
[\vbtr,V^{(3)}_s]| \psi_s \ra \nonumber \\
&=& \la \psi_s | [\vbt,V^{(2)}_s]_c - [\vbtr,V^{(3)}_s]| \psi_s \ra \;,
\label{eq:A3vevs}
\eea
where $\vbt$ and $\vbtr$ are the commutators $\vbt = [T_{\rm
rel},V^{(2)}_s]$ and $\vbtr = [T_{\rm rel},V^{(3)}_s]$. In the
third line, the expectation value of the commutator,
$[\vbtr,H_s]$, vanishes identically.

The term with the subscript ``$c$'' has had the two-body
disconnected diagrams removed. In our MATLAB implementation, this
subtraction is achieved by first embedding the two-particle-space
evolved version of this commutator in the three-particle space.
Computing $[\vbt,V^{(2)}]$ in the two-particle space alone
involves only the one-loop two-body interactions, so embedding 
in the three-particle sector results in only the disconnected
parts. This disconnected part can then be subtracted from the
total three-particle sector version of the same commutator,
leaving only the three-particle fully connected part.

\begin{figure}[bt!]
\begin{center}
\dblpic{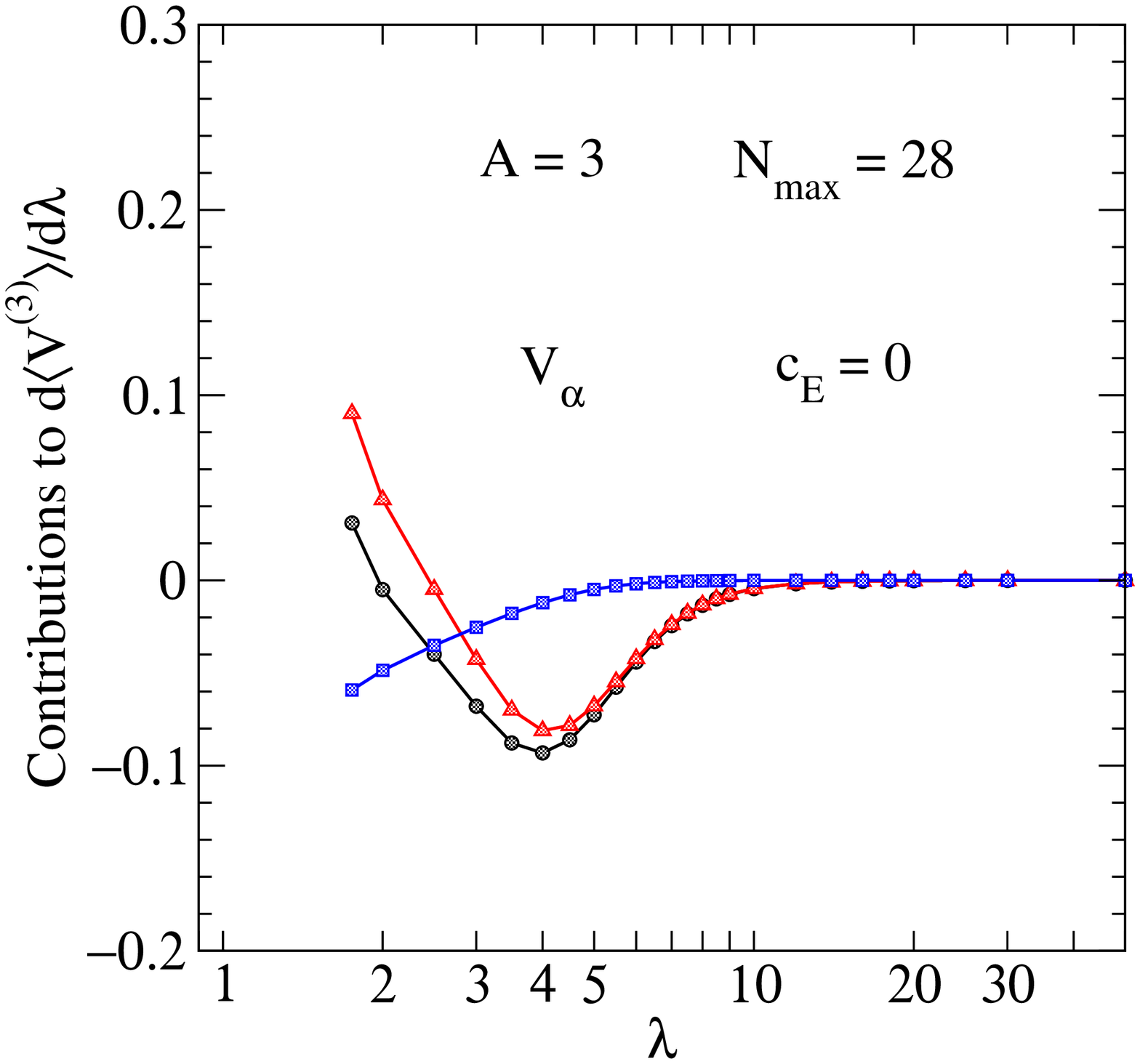}
\hfill
\dblpic{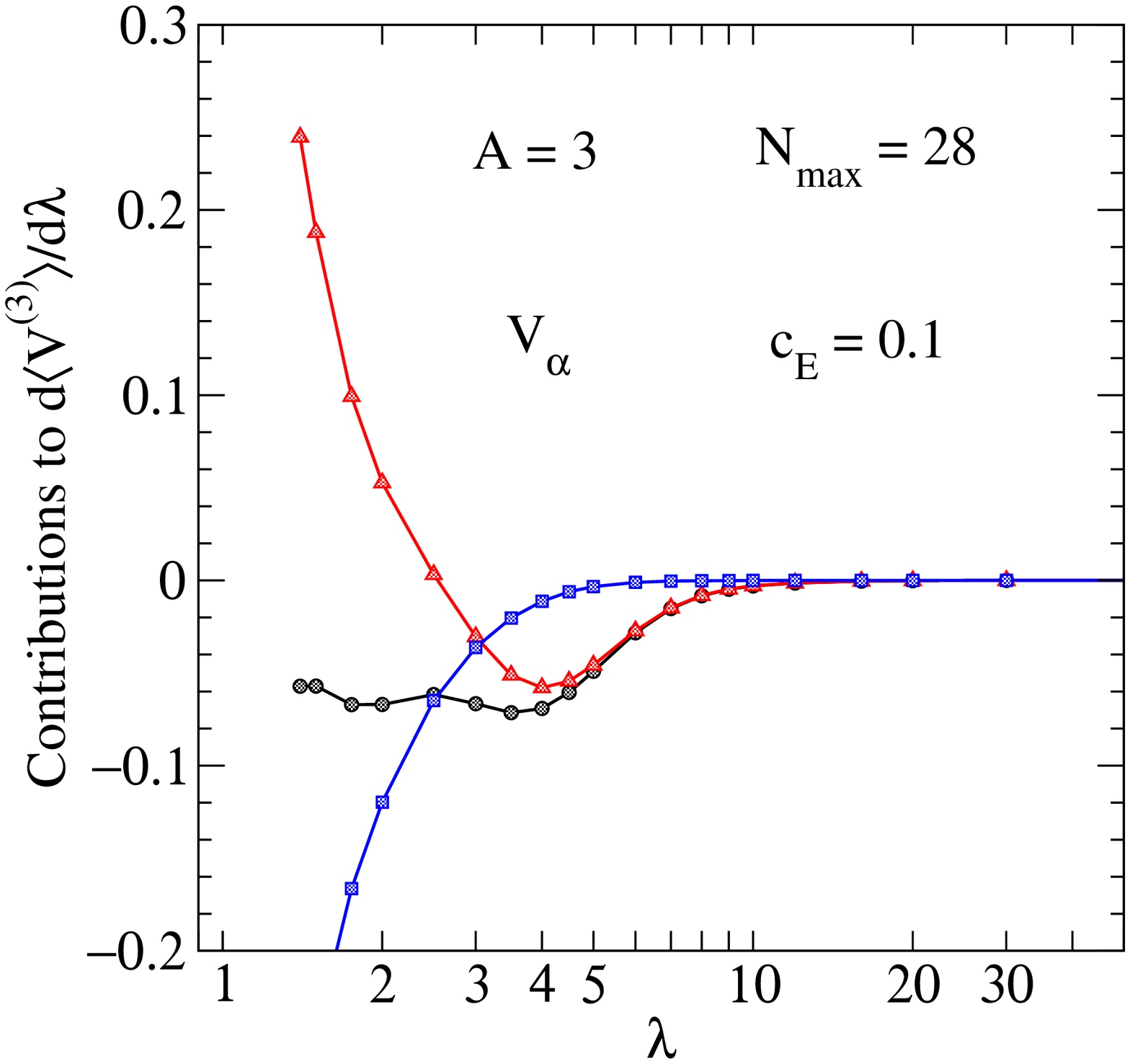}

\dblpic{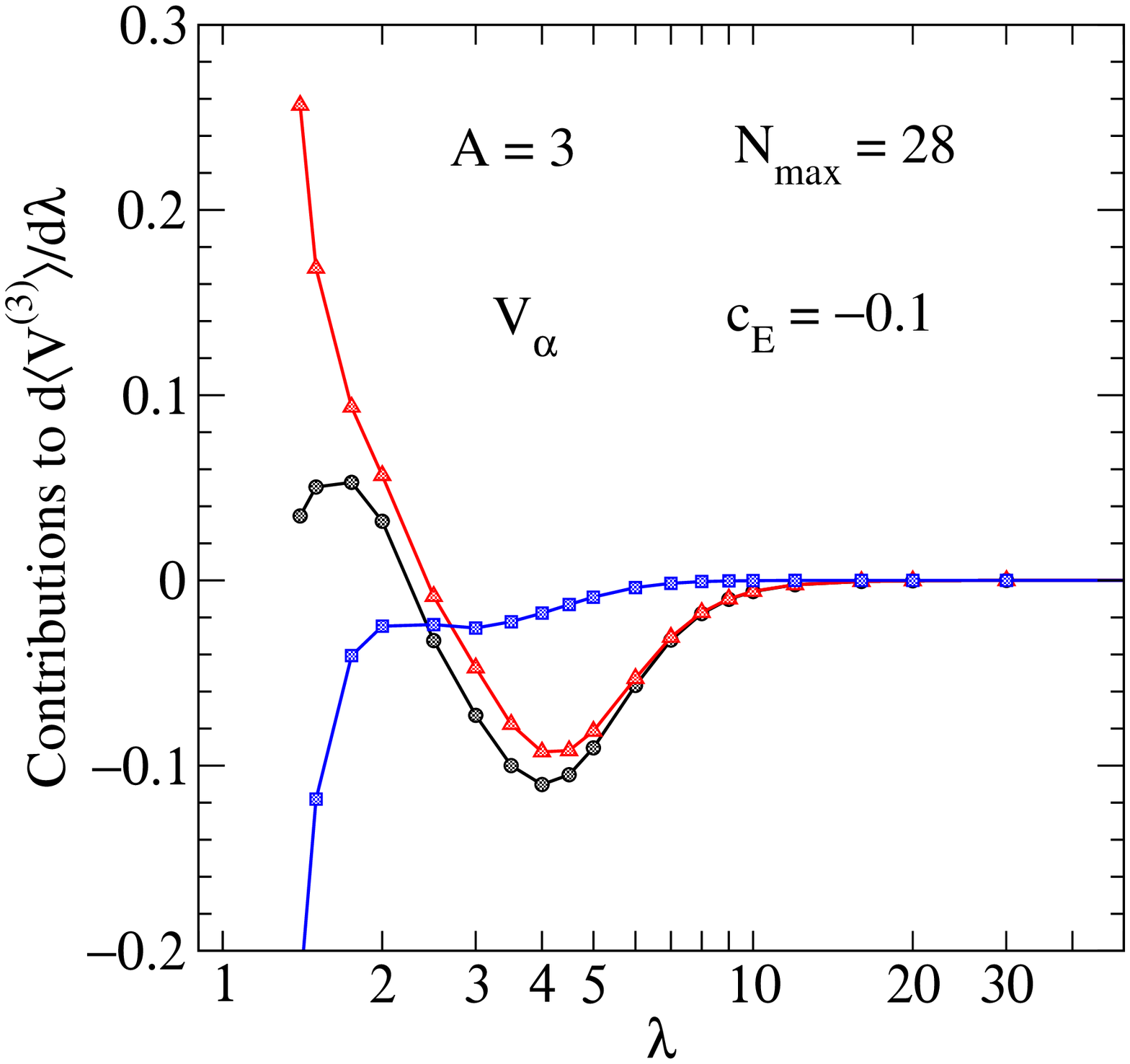}
\hfill
\dblpic{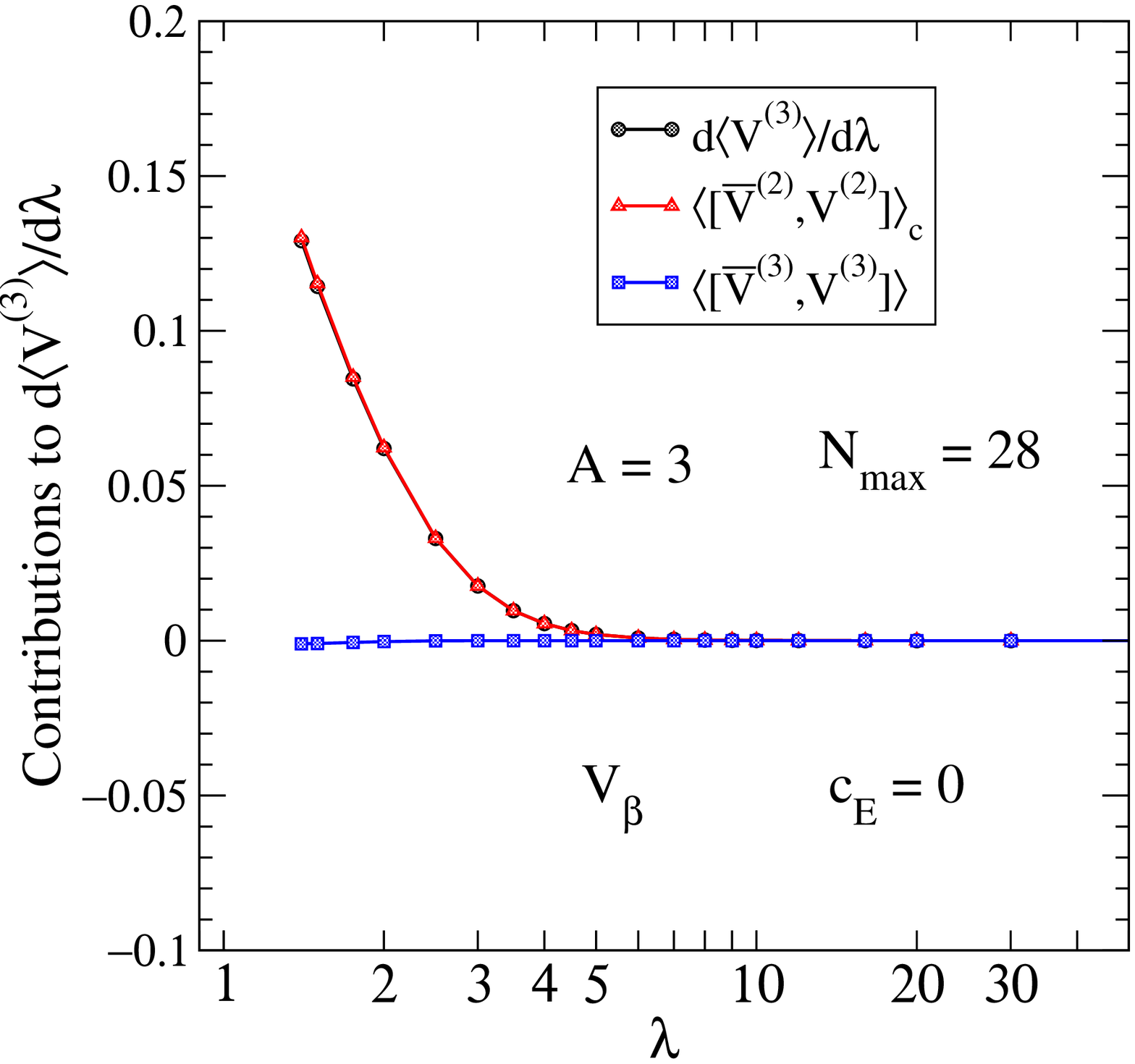}
\end{center}
\captionspace{Contributions from individual terms to
the $A=3$ ground-state expectation value $d \la
V^{(3)}_\lambda\ra / d \lambda$ for several different two- and
three-body potentials, as indicated in the plots. We emphasize
that $\lambda \leq 2$ is very small, comparable to $\lambda \leq
1.5\,{\rm fm}^{-1}$ for NN forces in analogous calculations with the
NCSM~\cite{Bogner:2007rx}.}
\label{fig:srg_VEV3}
\end{figure}

It is most useful for our analysis to convert from derivatives
with respect to $s$ to derivatives with respect to $\lambda$
using $\frac{d}{ds} = -\frac{\lambda^5}{4}\frac{d}{d\lambda}$. In
Fig.~\ref{fig:srg_VEV3} we show the ground-state expectation
values of the right side of Eq.~\eqref{eq:dds_vev}, which are
broken down into the two terms from the right side of
Eq.~\eqref{eq:A3vevs} for $A=3$ and various potentials. It
is apparent that the drivers of three-body matrix element
evolution depend on the interplay between long- and short-range,
attractive and repulsive parts. The lower right panel of
Fig.~\ref{fig:srg_VEV3} shows an increasing attractive strength of
the three-body force when starting from an attractive-only
two-body potential. In this regime the dominant contribution to
the evolution of the three-body potential matrix element is the tree-level
two-body connected part $[\vbt,V^{(2)}_s]_c$. This observation
accounts for the behavior in the right graph of
Fig.~\ref{fig:3N_rel_err_Vf}, where the size of the error scales
(roughly) like $(V^{(2)})^2$. Varying the long-range attraction
strength in Fig.~\ref{fig:3N_rel_err_Va} shows a similar effect.

More generally, the impact on ground-state energies of the
induced three-body interaction depends on the details of the
correlations in the wave function. The other plots in
Fig.~\ref{fig:srg_VEV3} for the more realistic initial two-body
potential, $V_{\alpha}$, show the interplay between two- and
three-body contributions to the three-body matrix element
evolution. The three-body contribution to the three-body
evolution stays small until the longer-range attractive part of
the potential begins to affect the evolution. Most of the change
is from $\lambda = 8$ to $\lambda = 3$, which is dominated by
$\la \overline{V}^{(2)},V^{(2)}\ra$. Thus, the feedback of the
three-body potential depends on the initial conditions, but
in general insures that the binding energy contribution stays
small.

We can repeat the above analysis for $A = 4$
\bea
\frac{d}{ds}\la
\psi^{(4)}_s|V_s^{(4)}|\psi^{(4)}_s\ra  
&=& \la \psi^{(4)}_s|[\overline{V}^{(2)}_s,V^{(3)}_s]_c 
+ [\overline{V}^{(3)}_s,V^{(2)}_s]_c  \nonumber \\
&& \quad + [\overline{V}^{(3)}_s,V^{(3)}_s]_c 
- [\overline{V}^{(4)}_s,V^{(4)}_s]| \psi^{(4)}_s \ra 
\label{eq:A4vevs}
\eea
where we find no fully
connected terms with only two-body forces. Again, disconnected
terms involving two and three body potentials cancel out in the
lower sectors. The leading terms are commutators with one
$V^{(2)}_s$ and one $V^{(3)}_s$, followed by connected terms
quadratic in $V^{(3)}_s$ and one term quadratic in $V^{(4)}_s$.
All terms are small and additional cancellations among them
further suppress the four-body contribution. Thus, the initial
hierarchy of many-body forces implies that induced four-body (and
higher-body) forces will be small.

\begin{figure}[bt!]
\begin{center}
\strip{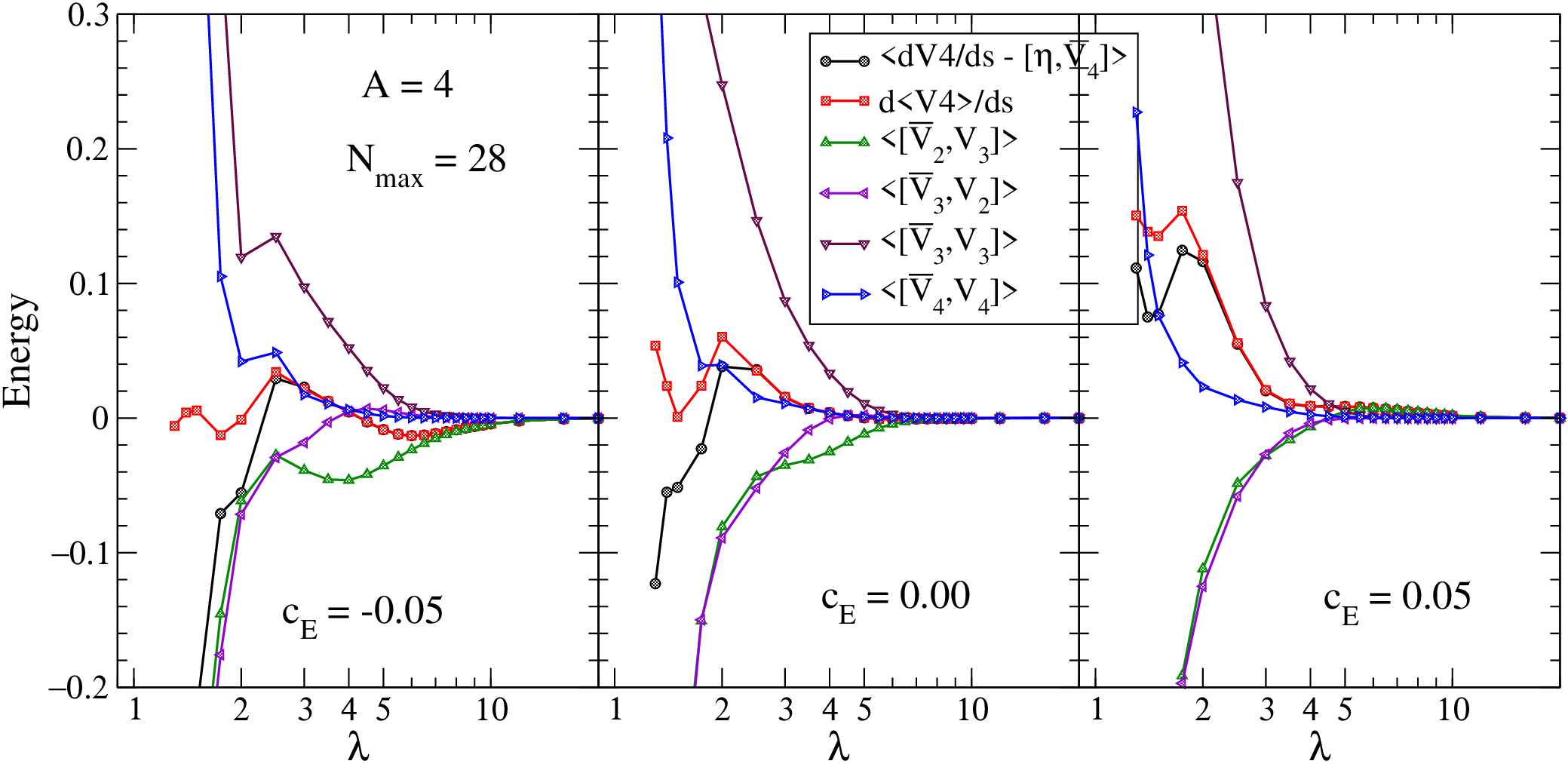}
\end{center}
\captionspace{Contributions from individual terms to
the $A=4$ ground-state expectation value $d \la
V^{(4)}_\lambda\ra / d \lambda$ for several different initial 
three-body potentials, as indicated in the plots. We emphasize
that $\lambda \leq 2$ is very small, comparable to $\lambda \leq
1.5\,{\rm fm}^{-1}$ for NN forces in analogous calculations with the
NCSM~\cite{Bogner:2007rx}.}
\label{fig:srg_VEV4}
\end{figure}

In Fig.~\ref{fig:srg_VEV4} we plot these contributions to the
evolution of the four-body expectation value. Again it is more useful
to convert the derivatives in $s$ to derivatives in $\lambda$. The
interplay of contributions is much more complicated here. We can
see the cancellations between the two terms involving $V^{(2)}_s$ and
$V^{(3)}_s$, and the term quadratic in $V^{(3)}_s$. The total
derivative of $V^{(4)}_s$ is kept small, along with the in term quadratic
in $V^{(4)}_s$, until below $\lambda = 2$. No
terms quadratic in $V^{(2)}_s$ appear as no connected diagrams can be
constructed for the four-particle evolution. This keeps the evolution
of $V^{(4)}_s$ low because they are one more order removed from the
initial interaction.

Also, note the differences in combinatoric factors when comparing this
analysis in different sectors. Between the $A=3$ and 4 sectors we can
only consider the $[\overline{V}^{(3)}_s,V^{(3)}_s]$ contribution.
Diagrammatically speaking, in the $A=3$ sector, this can only include
$\overline{V}^{(3)}_s$ and $V^{(3)}_s$ fully contracted with
each other. However, in the $A=4$ sector the connected diagrams are
only those contractions with one loop. There are $3!=6$ ways to choose
the pair legs composing the loop.\footnote{There are 6 ways to choose
the legs on the receiving term of the diagram but this is true in both
sectors and this 6 cancels out of the comparison.}  So the
$[\overline{V}^{(3)}_s,V^{(3)}_s]$ contribution is 6 times larger in
four-body sector. This does not negate the effects of the absence of
two-body contributions since they would also have large associated
combinatoric factors.


\section{Fitting Three-Body Forces}

\begin{figure}[ht!]
\begin{center}
\strip{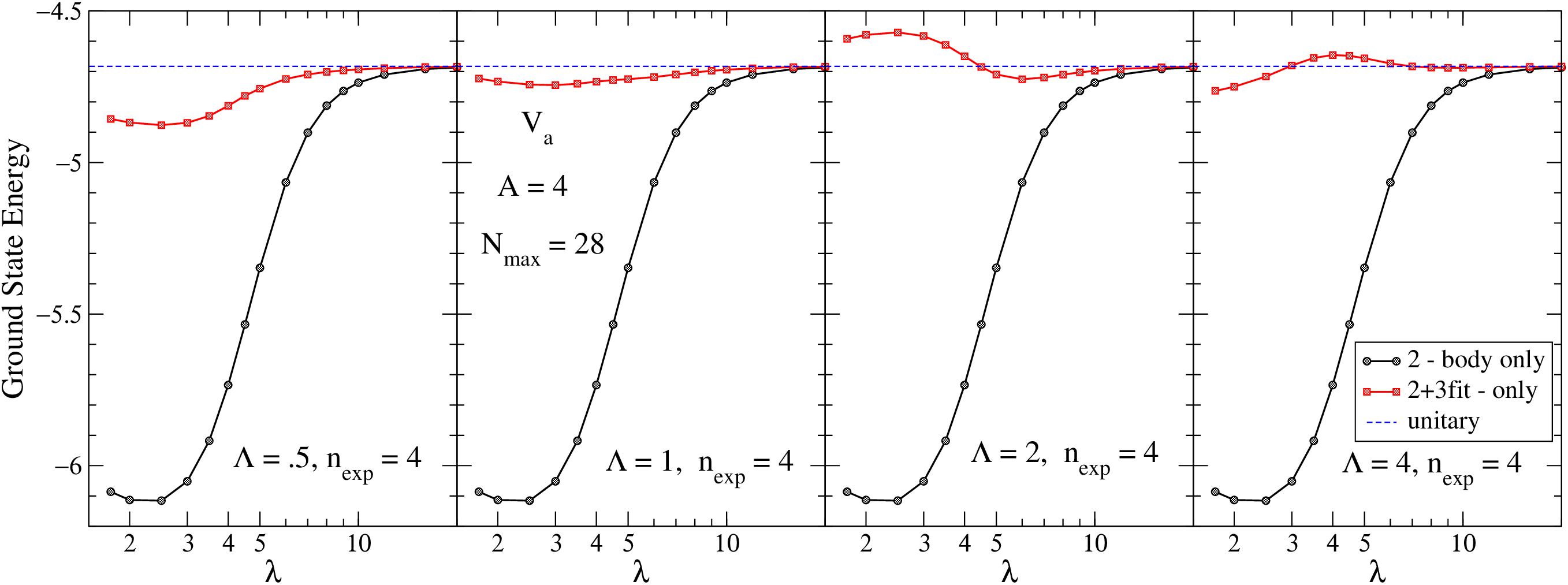}
\end{center}
\captionspace{Ground-state energy of the $A=4$ system with $NN$
interaction evolved and then a three-body interaction term (with an
exponential regulator) fit to the $A=3$ ground-state energy. The
calculation is repeated with a variation of the range parameter,
$\Lambda$}
\label{fig:oned_fit_var_Lam}
\end{figure}

\begin{figure}[ht!]
\begin{center}
\includegraphics*[height=2.5in]{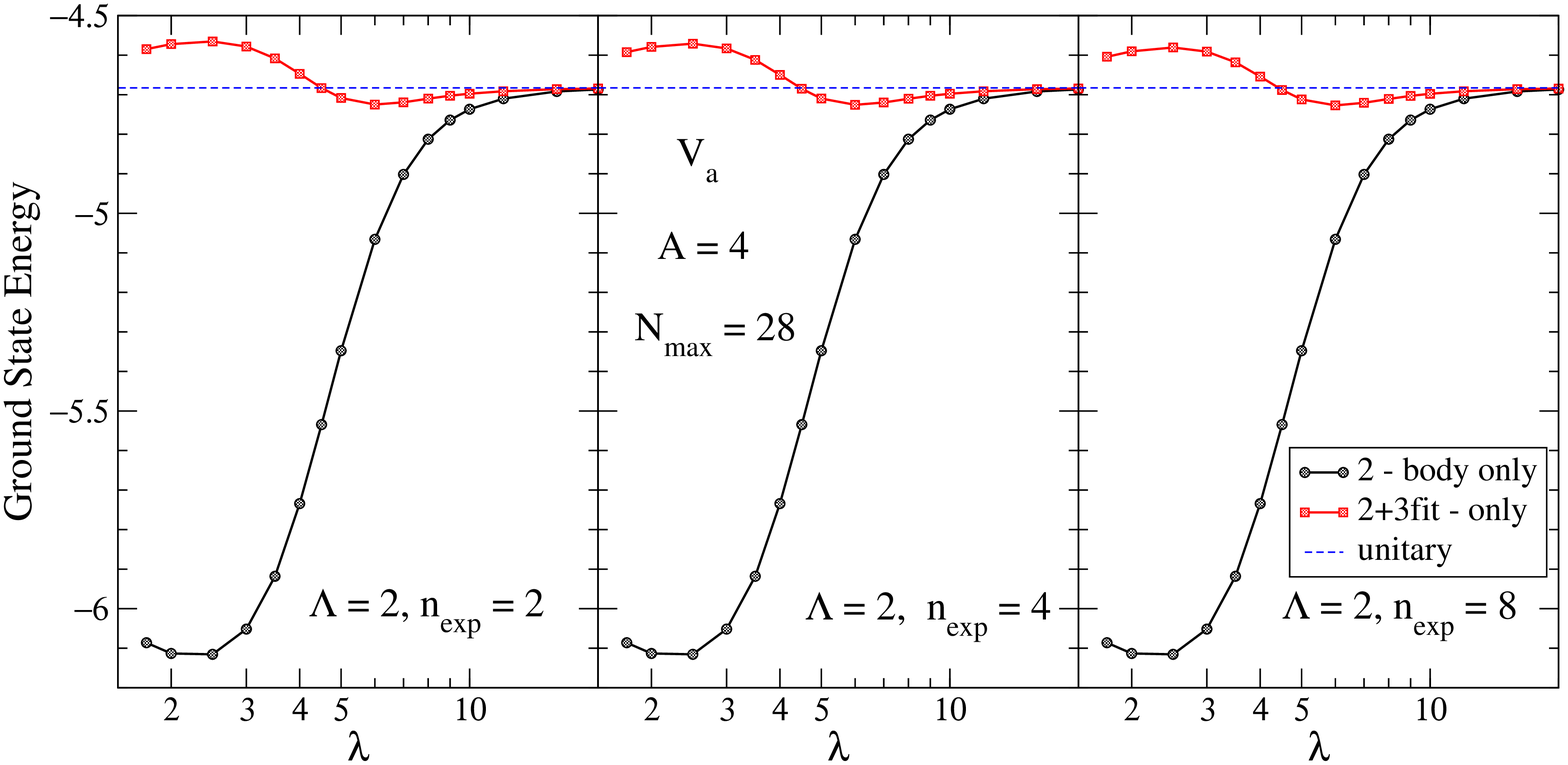}
\end{center}
\captionspace{Same as Fig.~\ref{fig:oned_fit_var_Lam} but at one value
of  $\Lambda$ and a range over the regulator's sharpness parameter,
n$_{\rm exp}$}
\label{fig:oned_fit_var_nexp}
\end{figure}

Figures \ref{fig:oned_fit_var_Lam} and \ref{fig:oned_fit_var_nexp}
show a simple fitting procedure for the three-body force. In this
calculation the initial interaction was evolved only in the $A=2$
basis and then embedded up to the $A=4$ basis for calculations of the
three- and four-particle binding energies. This corresponds to the
black curves (circles) in the figures, which are all identical.
Before, the interaction was also evolved in the $A=3$ basis to obtain
a renormalization that is unitary for $A=3$ and used to compute the
induced four-body forces. Here an independent operator, inspired by a
three-nucleon contact term from \xeft, \footnote{In the \xeft
paradigm, any regularized delta function will work for the fitting
performed here.} is used to fit the missing three-body interaction to
the three-body binding energy:
\beqn
V^{(3)}_{D_0} = D_0 e^{-[(p_1^2+p_2^2)/\Lambda^2]^n}e^{-[(p_1'^2+p_2'^2)/\Lambda^2]^n}
\eeqn
where $p_1$ and $p_2$ are the usual Jacobi momenta, the two regulator
parameters, $\Lambda$ and $n_{\rm exp}$, set the cutoff and sharpness
of the regulator and are kept constant over the range of $\lambda$ in
each panel shown, and the strength, $D_0$, was fit at each value of
$\lambda$. Then this fitted term plus the evolved
two-body only interaction is embedded to $A=4$ to obtain the four-body
binding energy. This corresponds to the red lines (squares).

As can be seen in the figures, this fitting procedure reduced the
error due to missing three-body forces by 80 to 90 percent. In
Fig.~\ref{fig:oned_fit_var_Lam} the regulators sharpness
exponent, $n_{\rm exp}$, is held constant and the range, $\Lambda$,
is varied. A significant dependence on $\Lambda$ is observed
indicating significant momentum dependent structure in the
evolved two-body force. In Fig.~\ref{fig:oned_fit_var_nexp}, now
the range is held fixed but sharpness is varied. Here very little
dependence is found indicating less sensitivity to the exact
shape of the regulator. 

The fits shown in Figs.~\ref{fig:oned_fit_var_Lam} and
\ref{fig:oned_fit_var_nexp} show that this one term can
accommodate much of the missing three-body force. A
natural progression would suggest adding a second term with gradients
as in a Taylor expansion:
\beqn
V^{(3)}_{D_2} = D_2 e^{-[(p_1^2+p_2^2)/\Lambda^2]^n}[p_1^2 + p_2^2 +
p_1'^2 + p_2'^2]e^{-[(p_1'^2+p_2'^2)/\Lambda^2]^n}
\eeqn
where $D_2$ is the new low-energy constant that also must be fit to data. In
this one-dimensional model, what we choose as data is arbitrary. A 
particular fitting strategy used for \xefts\ uses the average of $^3$H
and $^3$He binding energies and the triton $\beta$ decay. In an
attempt to be consistent with that work, we have chosen to fit our two
parameters to the three-body binding energy (as we did with the one
term fitting above) and the first excited state in the three-body
spectrum as another piece of three-body data.

\begin{figure}[ht!]
\begin{center}
\strip{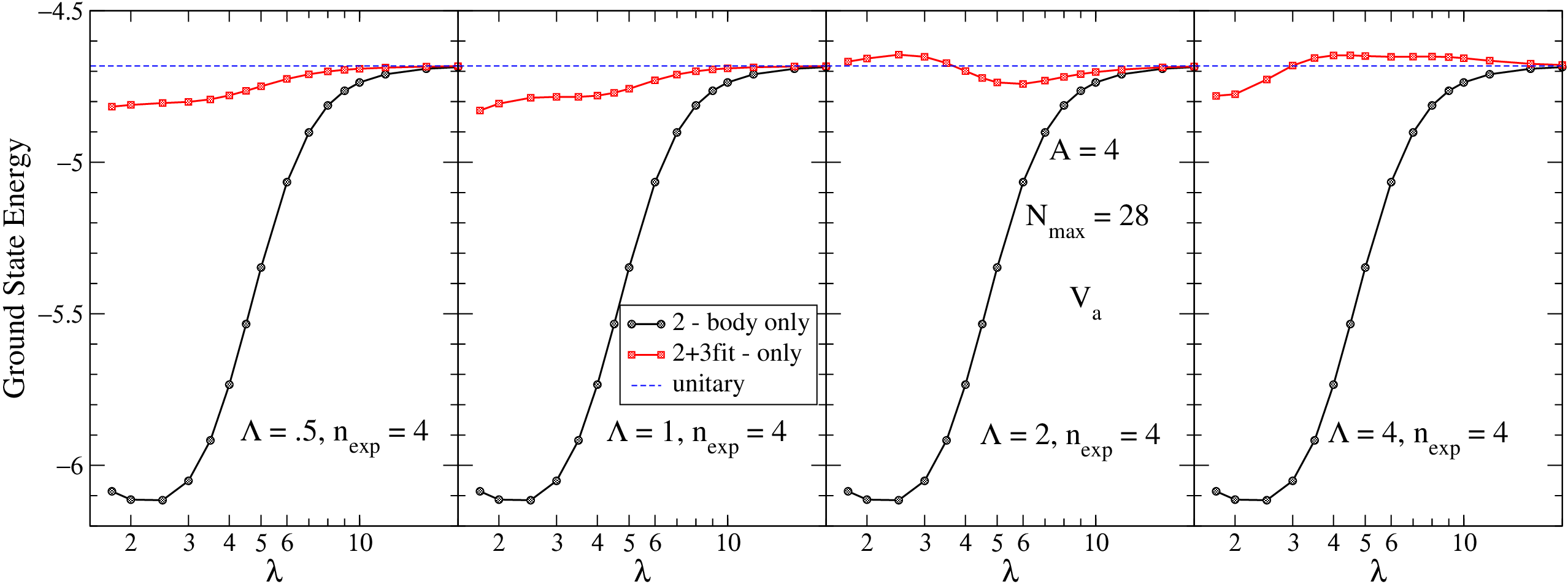}
\end{center}
\captionspace{Same as Fig.~\ref{fig:oned_fit_var_Lam} but now the $A=3$
ground-state energy and a second piece of data, the first
excited state for $A=3$, are fit to two three-body interaction terms. The
first term is the same as in Fig.~\ref{fig:oned_fit_var_Lam} and the second
includes a gradient correction.}
\label{fig:oned_fit_double_var_Lam}
\end{figure}

Figure \ref{fig:oned_fit_double_var_Lam} shows the predictions for the
$A=4$ binding energy after fitting the two terms to two different
three-body quantities. As in Fig.~\ref{fig:oned_fit_var_Lam}, $n_{\rm
exp}$, is held constant and the range, $\Lambda$, is varied. Now we
have an expansion in the strength, $D$, which we fit to three-body
quantities. The resulting fit three-body interaction can then be used
to make predictions of four- and higher-body observables. Here we
display the prediction of the $A=4$ binding energy after fitting both
$D_0$ and $D_2$. The improvement is inconclusive. This is an ongoing
investigation and we must next check several features of the system
and procedure to determine how much correction we should expect. We
may need to choose a different second value for fitting if the one
chosen now is already well fit by the first term. A logical next step
is to reproduce the fitting approach previously employed in realistic
calculations by varying $\Lambda$ as a function of $\lambda$.

Fitting three-body interactions in this manner is another tool to
study the form of induced many-body forces. It is complimentary to the
vacuum expectation value analysis done in
section~\ref{sec:oned_vev}, providing another quantitative tool to
understand the SRG's behavior. In addition, well controlled fitting
procedures will be indispensable in estimating the effects of some
induced $A$-body forces that cannot be induced explicitly due to the
size of a given $A$-body basis.


\section{Evolving Individual Operators: A Study of Cutoff Behavior}

An aspect of the SRG which contributes to its versatility and power is
the ability to evolve individual operators as well as the Hamiltonian.
The majority of work in this area is contained in
Ref.~\cite{Anderson:2009a,Anderson:2009b}, but some preliminary work was done for
few-body systems using a simple operator for the momentum distribution
in the wavefunction of few-body systems \cite{Frankfurt:2008zv}:
\bea
n(k) &=& \sum_{i=1}^A \int \psi_A^2(k_1k_2\ldots k_A)
\delta(k-k_i) \delta(\sum_{j=1}^A k_j) dk_1 dk_2 \ldots dk_A \nonumber \\
&=& A \int \psi_A^2(k_1k_2\ldots k_A)
\delta(k-k_A) \delta(\sum_{j=1}^A k_j) dk_1 dk_2 \ldots dk_A \nonumber \\
&=& A \int \psi_A^2(p_1p_2\ldots p_{A-1})
\delta(\sqrt{\frac{A-1}{A}}(p-p_{A-1})) dp_1 dp_2 \ldots dp_{A-1} \nonumber \\
\eea
where $A$ is the number of particles, $k$'s are lab-frame momenta and
$p$'s are Jacobi momenta. The second line is a simplification due to
the symmetry properties of the present basis. The wavefunctions are
symmetric with respect to all the particles so the total is simply $A$
times the momentum distribution of one particle.  We choose the $A$'th
particle (the last one) due to a special property of the last Jacobi
coordinate that it can be related directly to the last single-particle
coordinate when the center of mass momentum is zero. Using that
property, the third line is converted from single particle coordinates to
the Jacobi coordinates of the basis. 

\begin{figure}[ht!]
\begin{center}
\strip{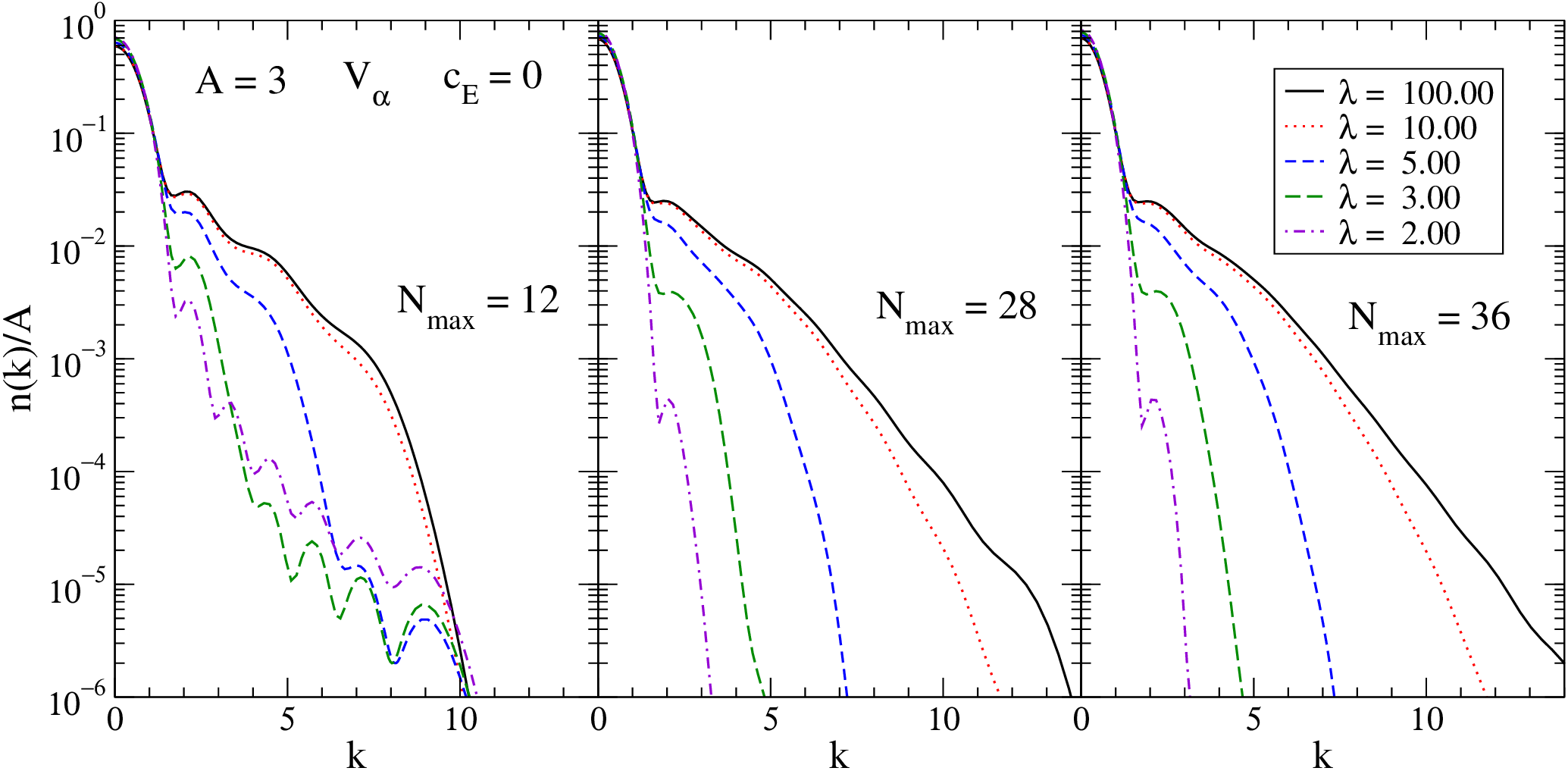}
\end{center}
\captionspace{A sequence of plots showing the evolving components of the
3-body wavefunction for increasing $\nmax$. Each of these plots uses
$\hw=5$.}
\label{fig:oned_adaga_nmax}
\end{figure}

Figure \ref{fig:oned_adaga_nmax} shows the evolution of the momentum
distribution within the three-body wavefunction for a series of
$\nmax$'s. In these plots the operator itself is not actually evolved.
The expectation value of the evolved operator on the evolved
wavefunctions is simply a verification of the unitarity of  $U_s$,
with the curves for all $\lambda$'s on top of the $\lambda = \infty$
curve. Here the interesting part is using the bare operator with the
evolved wavefunctions (or vice versa) to compute the momentum
distribution. The horizontal axis is in single particle momenta while
the cutoffs are in terms of Jacobi momenta. A conversion factor of
$\approx\sqrt{A}$ is needed when analyzing the figures.

The three panels in Fig.~\ref{fig:oned_adaga_nmax} show convergence as
$\nmax$ is increased. The series of curves in each plot shows that the
evolving wavefunction has a considerable decrease in high-momentum
components as $\lambda$ is lowered. At large $\nmax$, the oscillator
basis ultraviolet cutoff ($\Lambda_{\rm UV}$) is large and sufficient
high-momentum information is included. Hence the convergence in the
unevolved wavefunctions. At larger $\nmax$, the evolved wavefunctions
are more converged than the unevolved because the high-momentum
components have been suppressed and therefore the wavefunction is
unaltered by $\Lambda_{\rm UV}$. One can see $\Lambda_{\rm UV}$
explicitly in the unevolved wavefunctions using $\Lambda_{\rm UV}
\approx \sqrt{\hw\nmax}$ with $\hw=5$. At $\nmax=12$ we get
$\Lambda_{\rm UV} \approx 8 \fmi$ which is about the location of the
right-most shoulder of the unevolved wavefunction. Again at $\nmax=28$
the cutoff is $\Lambda_{\rm UV} \approx 12 \fmi$ corresponding to that
shoulder. The shoulder in the momentum distribution is the clearest
high-momentum structure to compare between the panels, hence in this
instance and at constant $A$, we have ignored the Jacobi conversion
factor. 

As $\nmax$ is lowered we can see the infrared cutoff ($\Lambda_{\rm
IR} \approx \sqrt{\frac{\hw}{\nmax}}$) become significant. The
oscillatory behavior in the $\nmax=12$ panel is the ringing from the 
$\Lambda_{\rm IR}$ cutoff inherent in the small oscillator basis.
Given a wavefunction in position space (recall coordinate space plots
of the wavefunction in Fig.~\ref{fig:srg_2_body}) which is cutoff at
some point $x = 2\pi / \Lambda_{\rm IR}$ will show ringing artifacts
from the Fourier transform of the sharp cut. For a square function the
minima should be $\Lambda_{\rm IR}$ apart. Due to the single-to-Jacobi
conversion factor, the minima of the ringing should be
$\sqrt{A}\Lambda_{\rm IR} \approx \sqrt{(A)\hw/\nmax}$ apart from each
other.

\begin{figure}[ht!]
\begin{center}
\strip{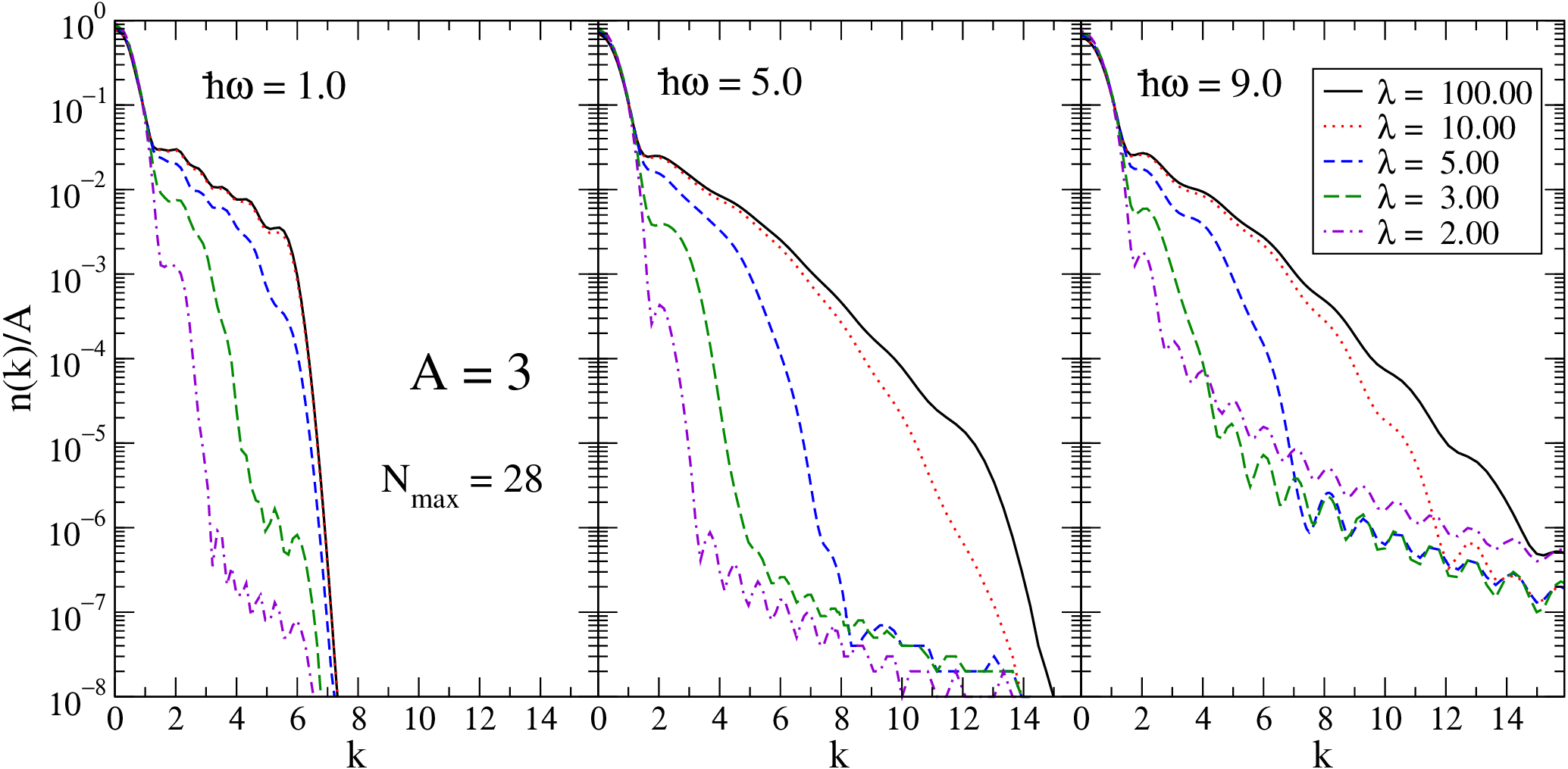}
\end{center}
\captionspace{A sequence of plots showing the evolving components of the
3-body wavefunction for fixed $\nmax$ and varied $\hw$.}
\label{fig:oned_adaga_varhw}
\end{figure}

\begin{figure}[ht!]
\begin{center}
\strip{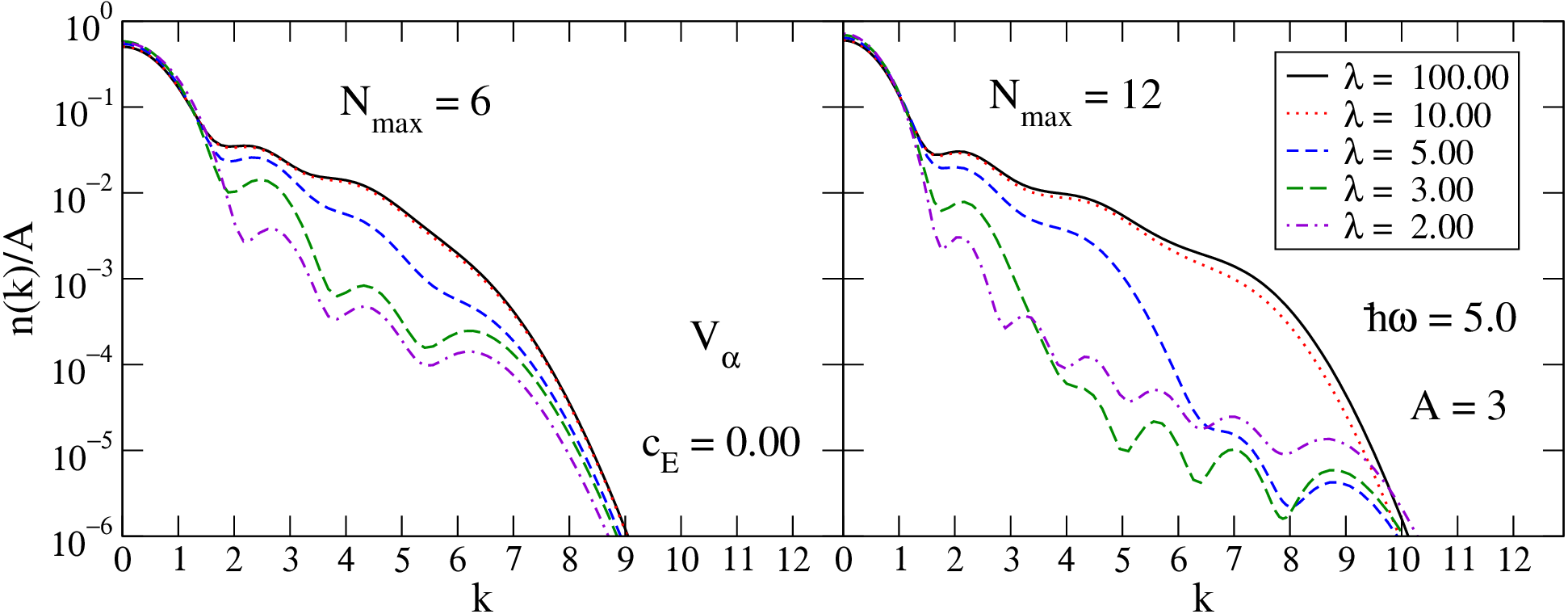}

\strip{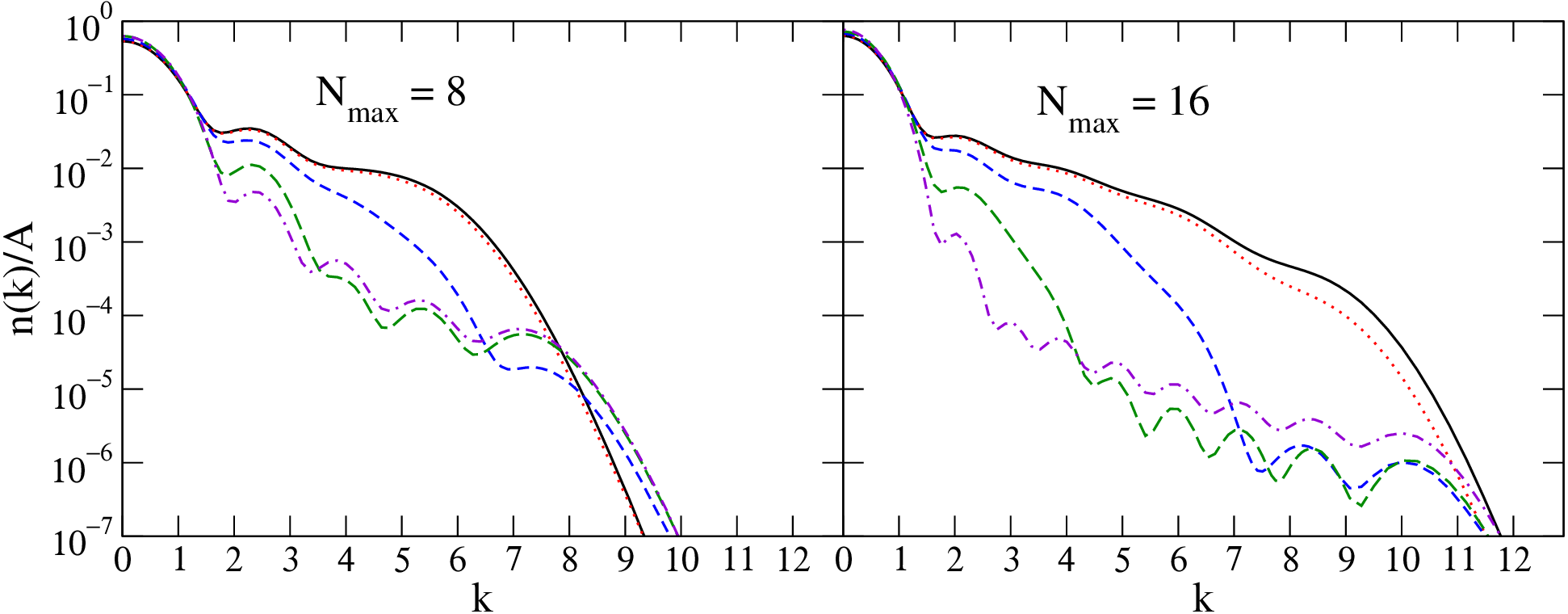}
\end{center}
\captionspace{A sequence of plots showing the evolving components of the
3-body wavefunction for fixed $\hw$ and varied $\nmax$.}
\label{fig:oned_adaga_varnmax}
\end{figure}

Figures \ref{fig:oned_adaga_varhw} and \ref{fig:oned_adaga_varnmax} further
explore the behavior of the cutoffs, $\Lambda_{\rm IR}$ and
$\Lambda_{\rm UV}$. In the former, all the panels are at the same
relatively high $\nmax=28$ with $\hw$ varied. The behavior of the
ultraviolet cutoff is evident with the right-most shoulder given by
$\sqrt{\hw\nmax}$. The infrared is clearer here, with the spacing
between oscillations given by $\approx\sqrt{A\hw/\nmax}$. This is also
evident in Fig.~\ref{fig:oned_adaga_varnmax} where $\nmax$ is varied with
a fixed $\hw$. The spacing decreases by a factor $\sqrt{2}$ from
$\nmax=6 \longrightarrow 12$ or $\nmax=8 \longrightarrow 16$ This is
surprisingly accurate in the regions with relatively uniform
oscillation spacings at midrange momenta. This is indicative of how
much high-momentum details have been transformed by the SRG. We are
left with a smooth gaussian-like wavefunction which is chopped at
large distance by the infrared cutoff, leaving straightforward ringing
effects.


\begin{figure}[ht!]
\begin{center}
\strip{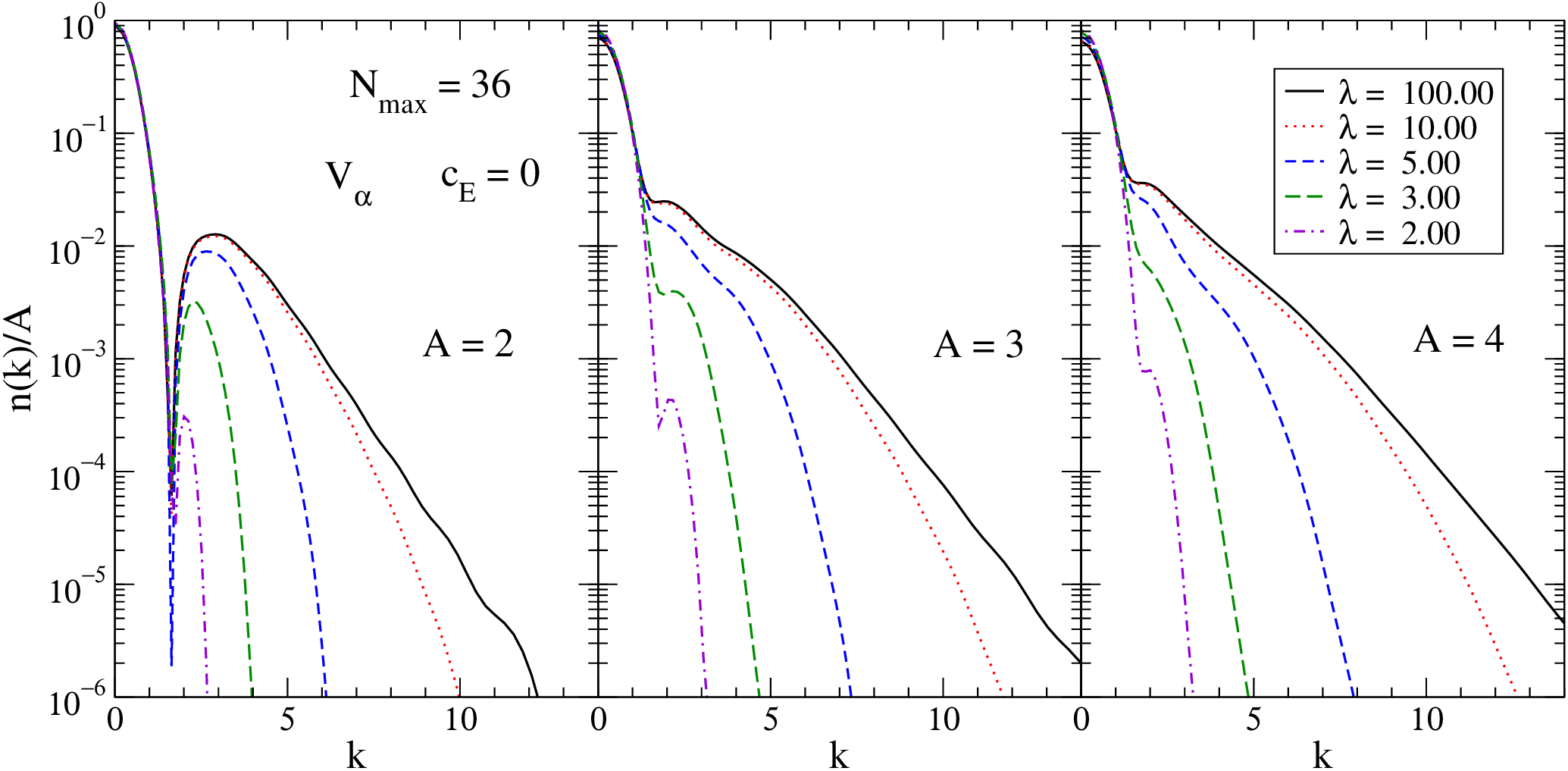}
\end{center}
\captionspace{A sequence of plots at high $\nmax$ showing the evolving
components of the many-body wavefunction for $A = $ 2,3, and 4.}
\label{fig:oned_adaga_AN}
\end{figure}

The momentum distribution plots of Figure \ref{fig:oned_adaga_AN}
shows the same quantity as the previous plots but at a single large
$\nmax$ and for different numbers of particles, $A=2,3,4$. The same
renormalization pattern for high-momentum components is evident
regardless of the considered one-dimensional system.


\section{Convergence in the Oscillator Basis}

\begin{figure}[ht]
\begin{center}
\dblpic{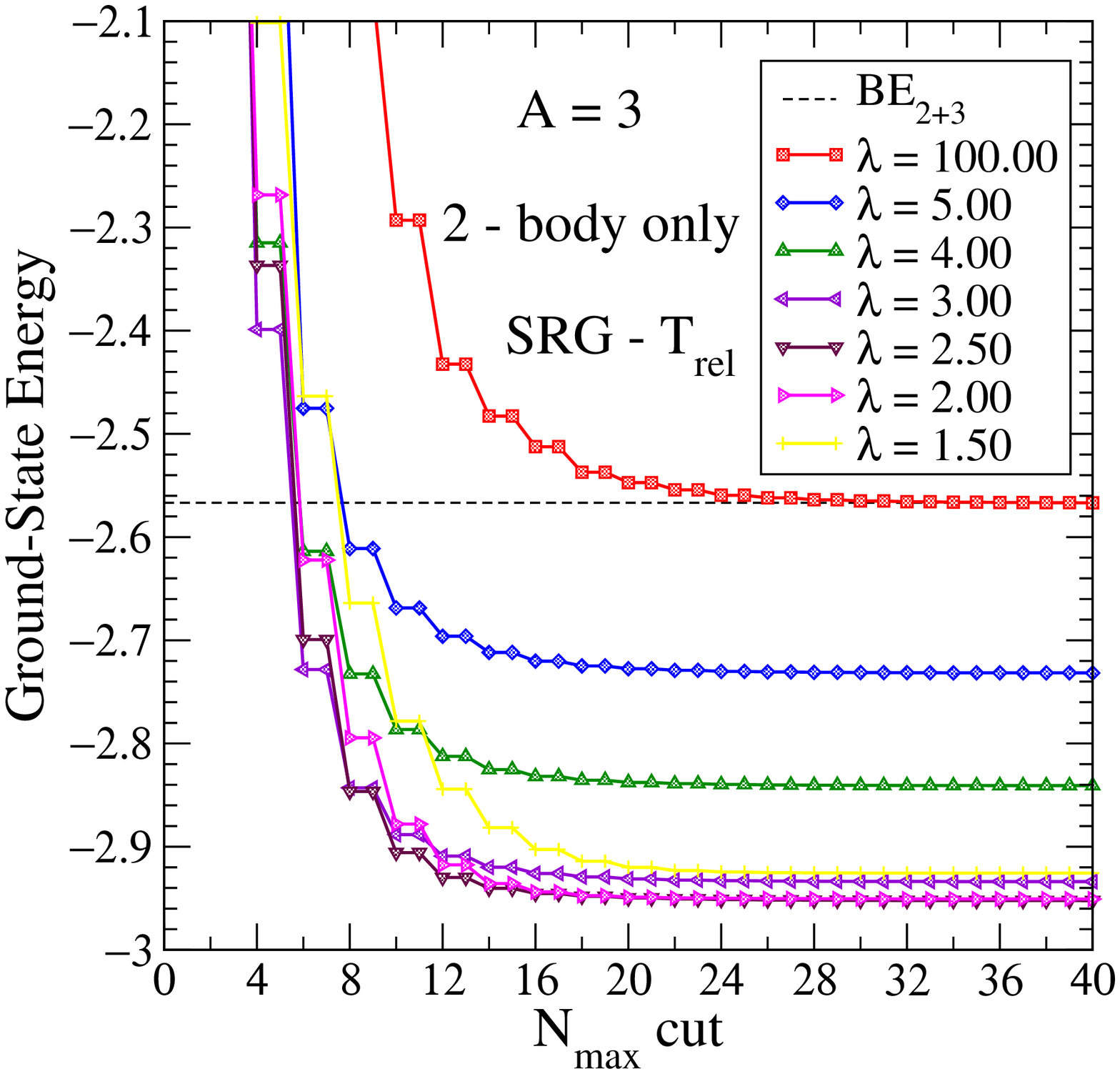}
 \hfill
\dblpic{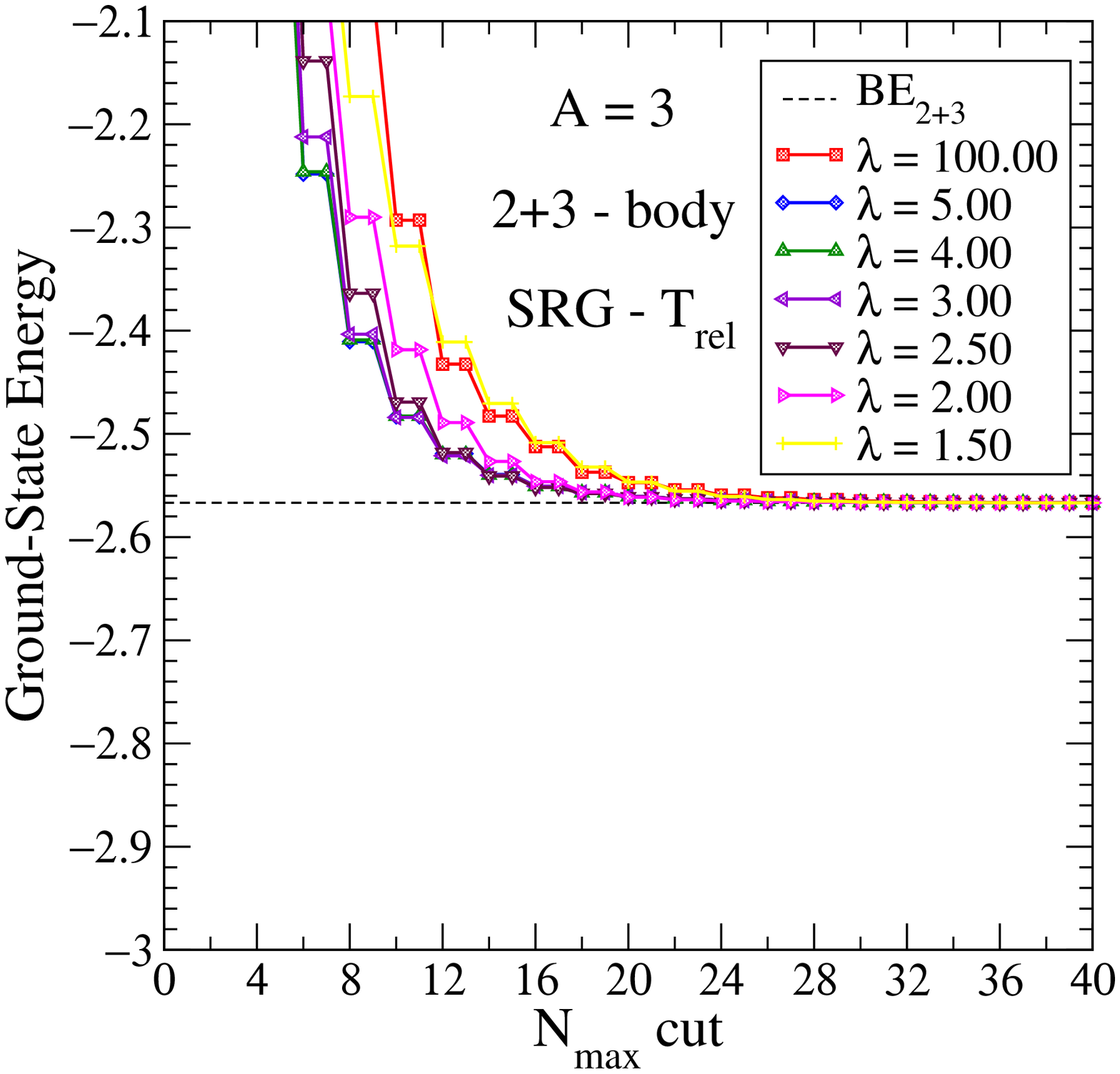}
\end{center}
\captionspace{Decoupling in the three-particle system using the choice $G_s
= \Trel$. The initial $V_{\alpha}$ potential is evolved to each
$\lambda$ shown in a basis with $\nmax = 40$. On the left only the
two-body potential is evolved and embedded while the right involves
the full unitarily transformed potential. Matrix elements of the
potential are set to zero if one or both states have $N >
\mbox{${\rm N}_{\rm cut}$}$ and the resulting Hamiltonian is diagonalized to
obtain the ground-state energies plotted.}
\label{fig:srg_3_body_decoupling}
\end{figure}

In Fig.~\ref{fig:srg_3_body_decoupling} we test SRG decoupling
\cite{Jurgenson:2007td} within the harmonic oscillator basis. The left
panel shows the results when only the two-body evolved potential is
used to calculate $A=3$ binding energies. For the right plot, the
potential was evolved in the full three-body space, allowing
three-body forces to be induced. The Hamiltonians for selected
$\lambda$ values are diagonalized in bases of decreasing size, as
measured by ``${\rm N}_{\rm cut}$'', which is the cut-off  applied to the
potential to study its decoupling properties. The potential is set to
zero for matrix elements for which one or both states has $N > N_{\rm
cut}$. The degree of decoupling is measured by the point of
departure, as ${\rm N}_{\rm cut}$ is lowered, from the asymptotic energy for
$\nmax = 40$. As the potential is evolved from the initial potential
($\lambda = \infty$) down to $\lambda = 2.5$, decoupling is achieved
for smaller spaces, which means convergence is reached for smaller
basis sizes. This is the same pattern as found for realistic NN 
potentials~\cite{Jurgenson:2007td}. In the left panel of
Fig.~\ref{fig:srg_3_body_decoupling}, the ground-state energies
asymptote to different values because the evolution is not completely
unitary. One might imagine that this would affect the decoupling, but
we show this is not the case here. In the right panel, the induced
three-body interaction is kept, so the  curves asymptote to the same
energy at large $\nmax$, while the same pattern of decoupling is
observed.

We repeated for $A=4$ our test of decoupling that was shown in
Fig.~\ref{fig:srg_3_body_decoupling} for $A=3$. A similar pattern of
decoupling is found, namely an increased degree of decoupling until  a
$\lambda$ corresponding to the minimum of the two-body-only
ground-state energy of the $A=4$ system, after which it deteriorates. 

\begin{figure}[ht]
\begin{center}
  \dblpic{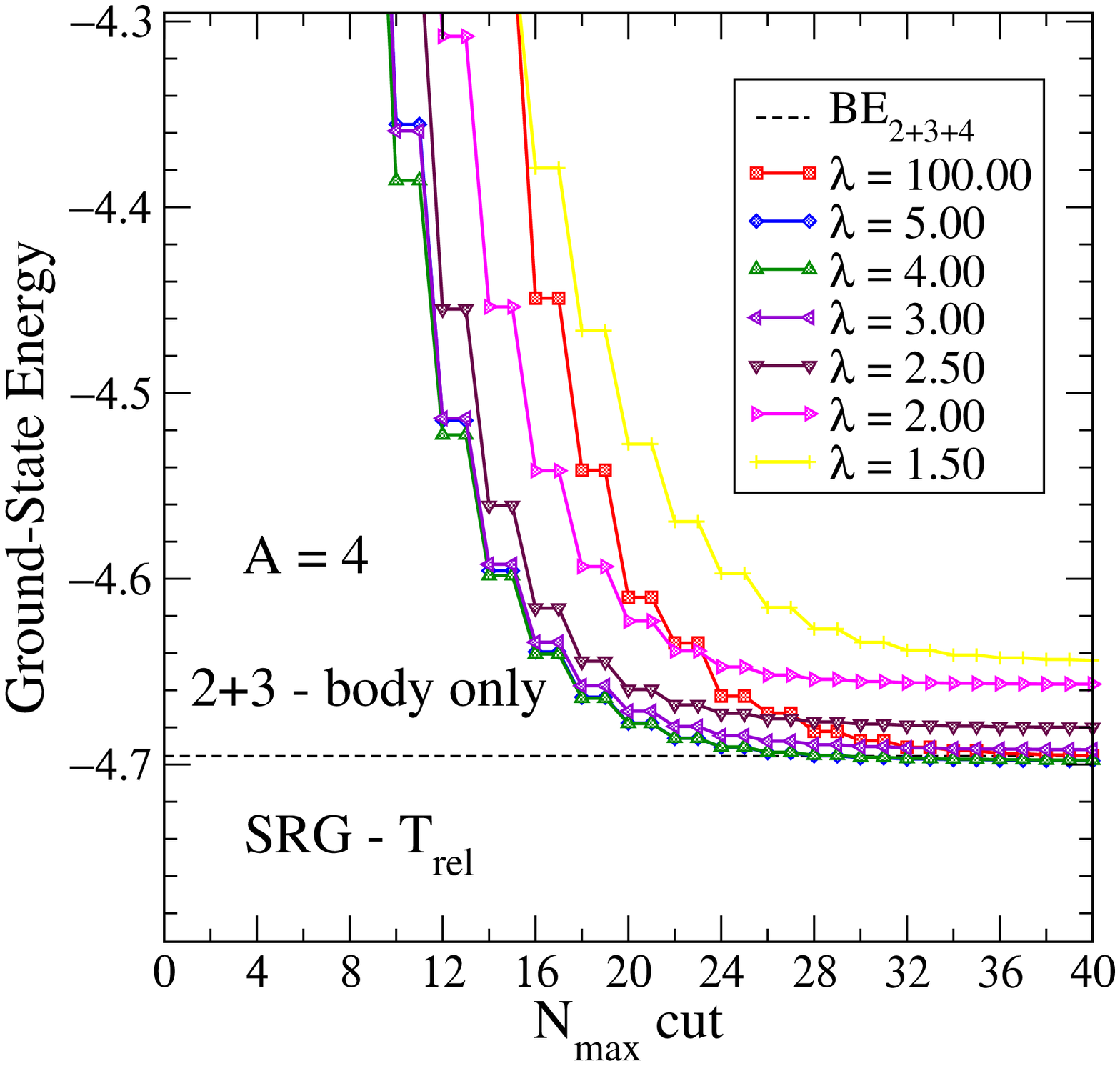}
 \hfill
  \dblpic{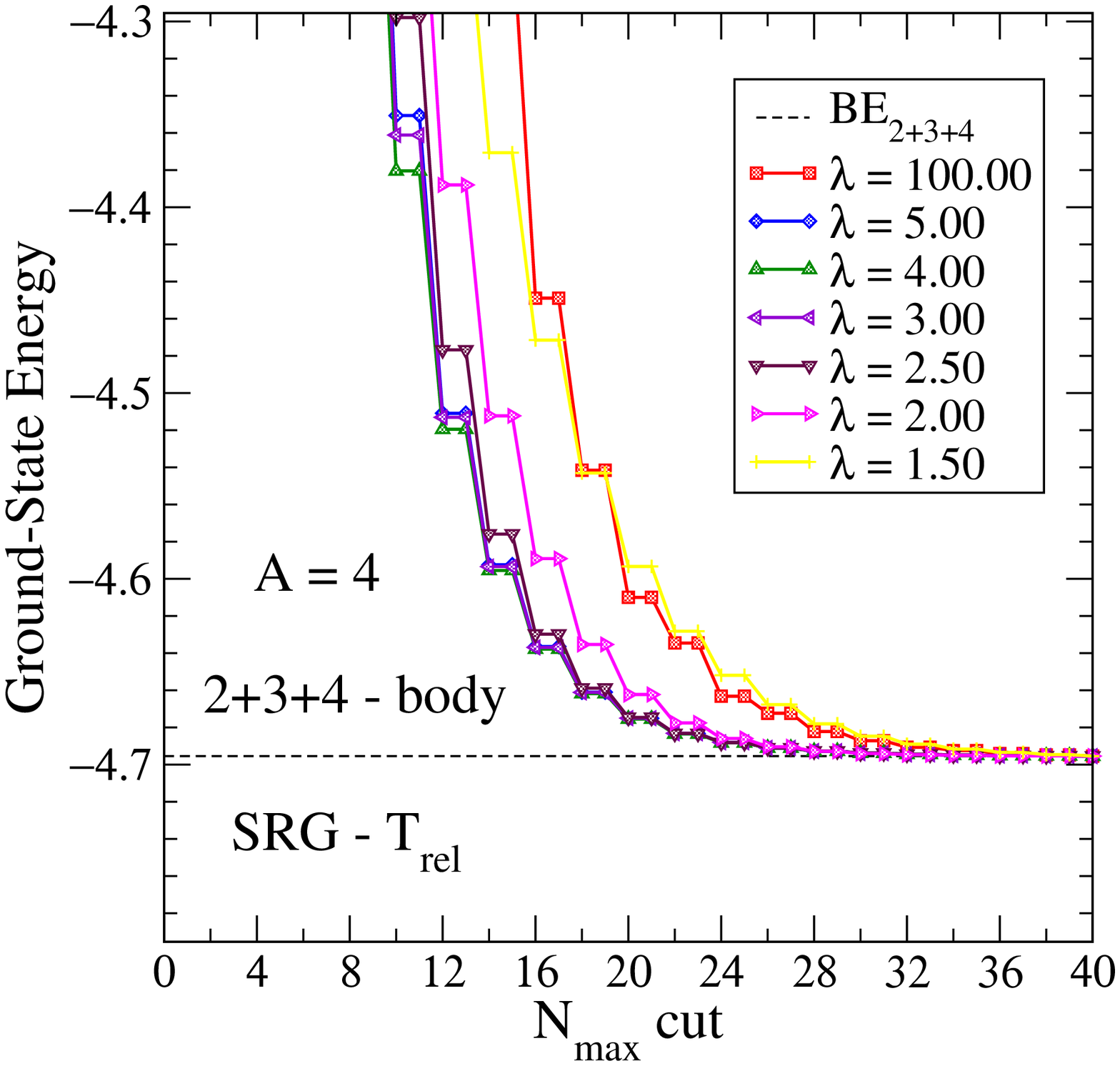}
\end{center}
\captionspace{Decoupling in the four-particle system using $G_s
= \Trel$. The initial $V_{\alpha}$ potential is evolved to each
$\lambda$ shown in a basis with $\nmax = 40$. On the left only the
two-body potential is evolved and embedded while the right involves
the full unitarily transformed potential. Matrix elements of the
potential are set to zero if one or both states have $N >
\mbox{${\rm N}_{\rm cut}$}$ and the resulting Hamiltonian is diagonalized to
obtain the ground-state energies plotted.}
\label{fig:srg_4_body_decoupling}
\end{figure}

We note that the decoupling benefits afforded by evolution in the
oscillator basis are less straightforward than in the two-body
momentum basis studied for NN in \cite{Jurgenson:2007td}. In
particular the cut-off errors  \emph{increase} for $\lambda$
smaller than the point at which the two-body-only binding energy
is at a minimum (i.e., for $\lambda = 1.5$ in
Fig.~\ref{fig:srg_3_body_decoupling}). In the oscillator basis
the Hamiltonian is not being driven towards a diagonal form by
the choice $G = \Trel$, since it is not diagonal in this basis.
In addition, the cuts made in the oscillator basis do not
correspond to analogous cutting done to study decoupling in the
momentum representation. However, control over cutting in the
oscillator basis is the ultimate goal of the improved convergence
in calculations with SRG evolution. Therefore, when working in the
oscillator basis, one should look
for a choice of $G$ which optimizes the decoupling in
that basis.

One possibility is the harmonic oscillator Hamiltonian, $\Hho$,
\beqn
\Hho = \Trel + \frac{1}{2}m\omega^2r^2
\eeqn
which in the oscillator basis is simply a diagonal matrix with
each state's value equal to $N\hw$ for that state. Such a matrix
is easy to build from the lists of state quantum numbers readily
available from the oscillator basis code. This choice of $G$
should drive the Hamiltonian to a diagonal form in the oscillator
basis.

\begin{figure}[ht]
\begin{center}
  \dblpic{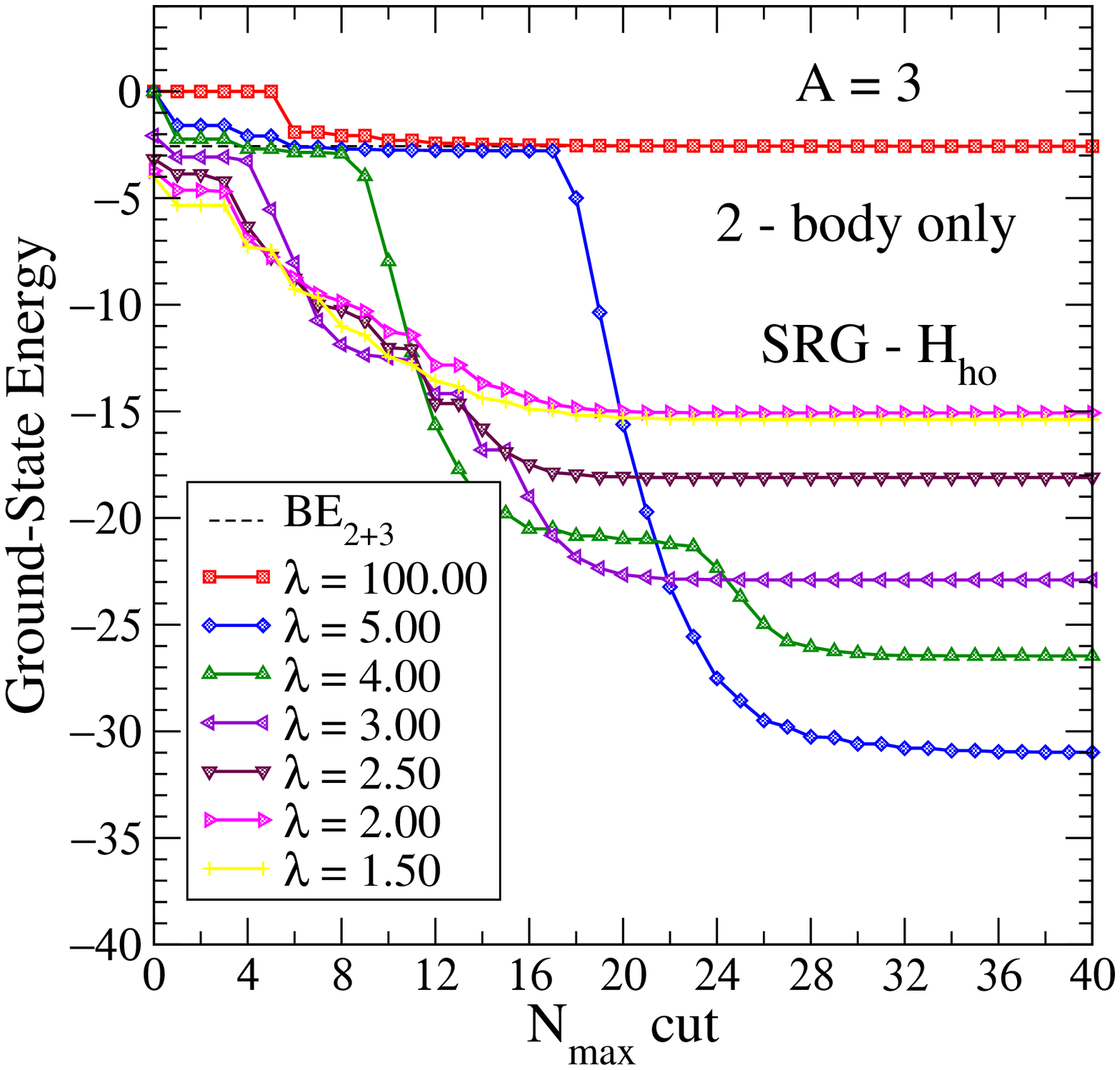}
 \hfill
  \dblpic{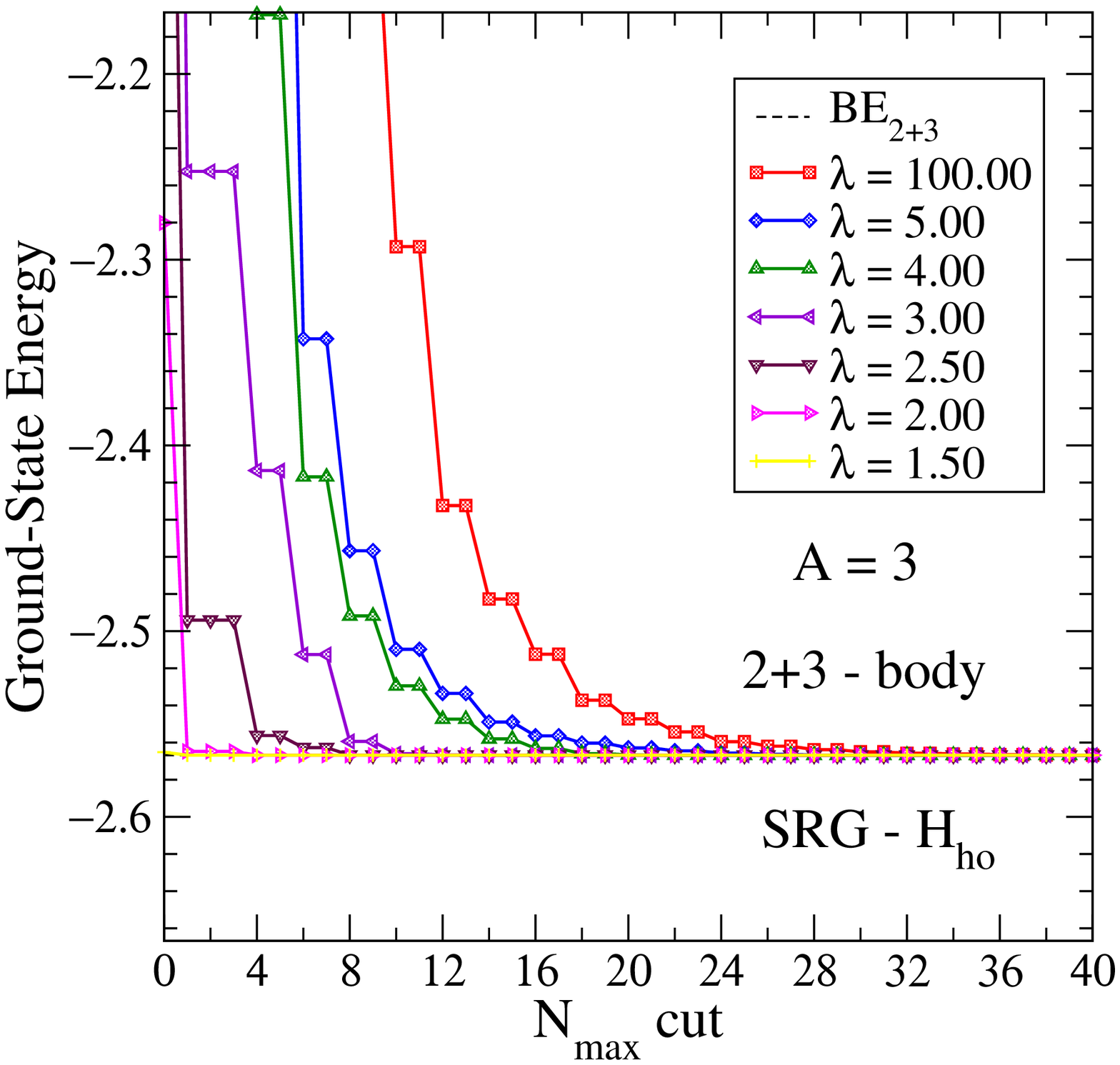}
\end{center}
\captionspace{ Decoupling in the three-particle system with $G_s = \Hho$. 
The initial $V_{\alpha}$ potential is evolved to each $\lambda$ shown
in a basis with $\nmax = 40$. On the left only the two-body potential
is evolved and embedded while the right involves the full unitarily
transformed potential. Matrix elements of the potential are set to
zero if one or both states have $N > \mbox{${\rm N}_{\rm cut}$}$ and the
resulting Hamiltonian is diagonalized to obtain the ground-state
energies plotted.}
\label{fig:srg_3_body_Hho_decoupling}
\end{figure}

\begin{figure}[ht]
\begin{center}
  \dblpic{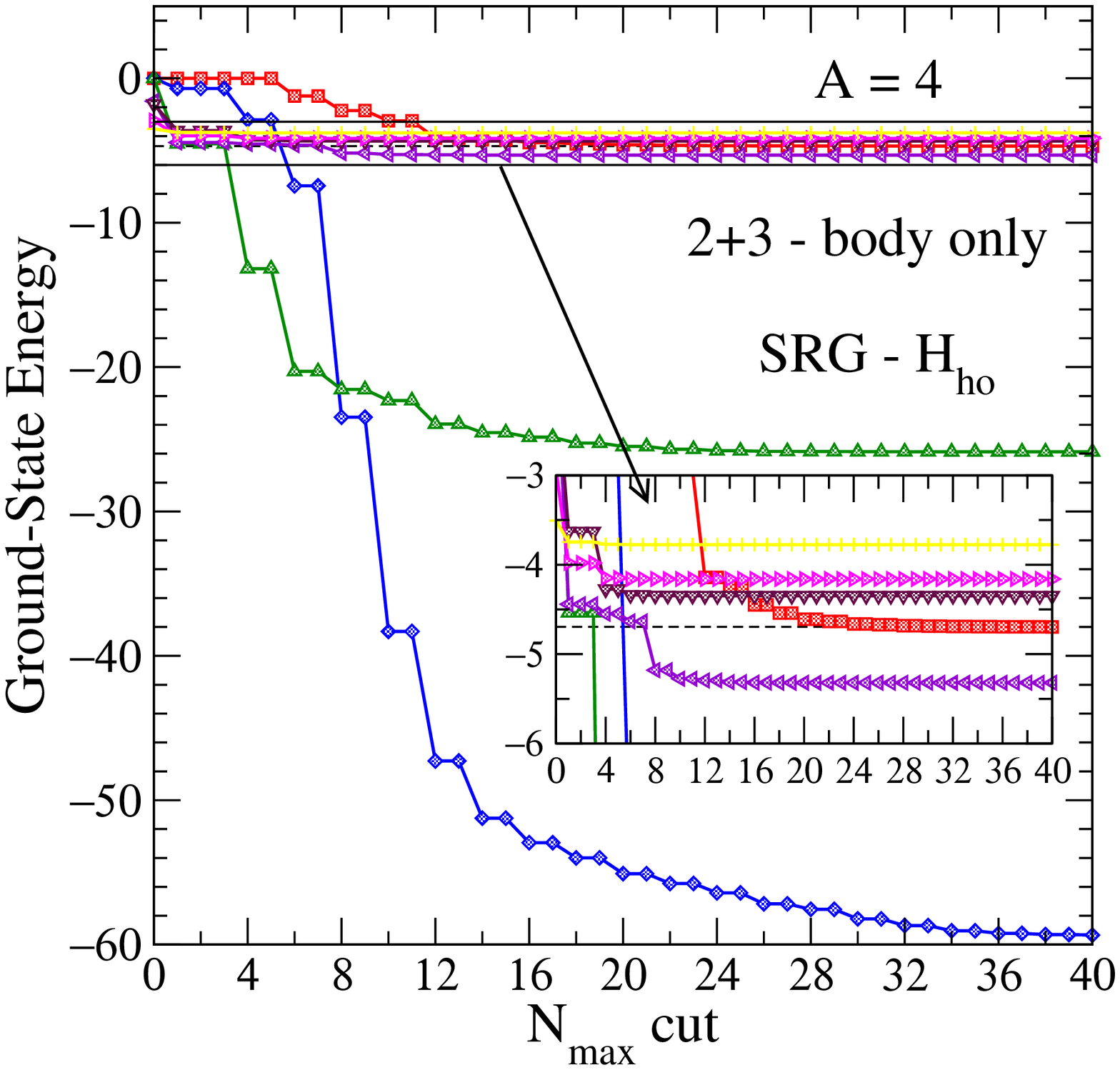}
 \hfill
  \dblpic{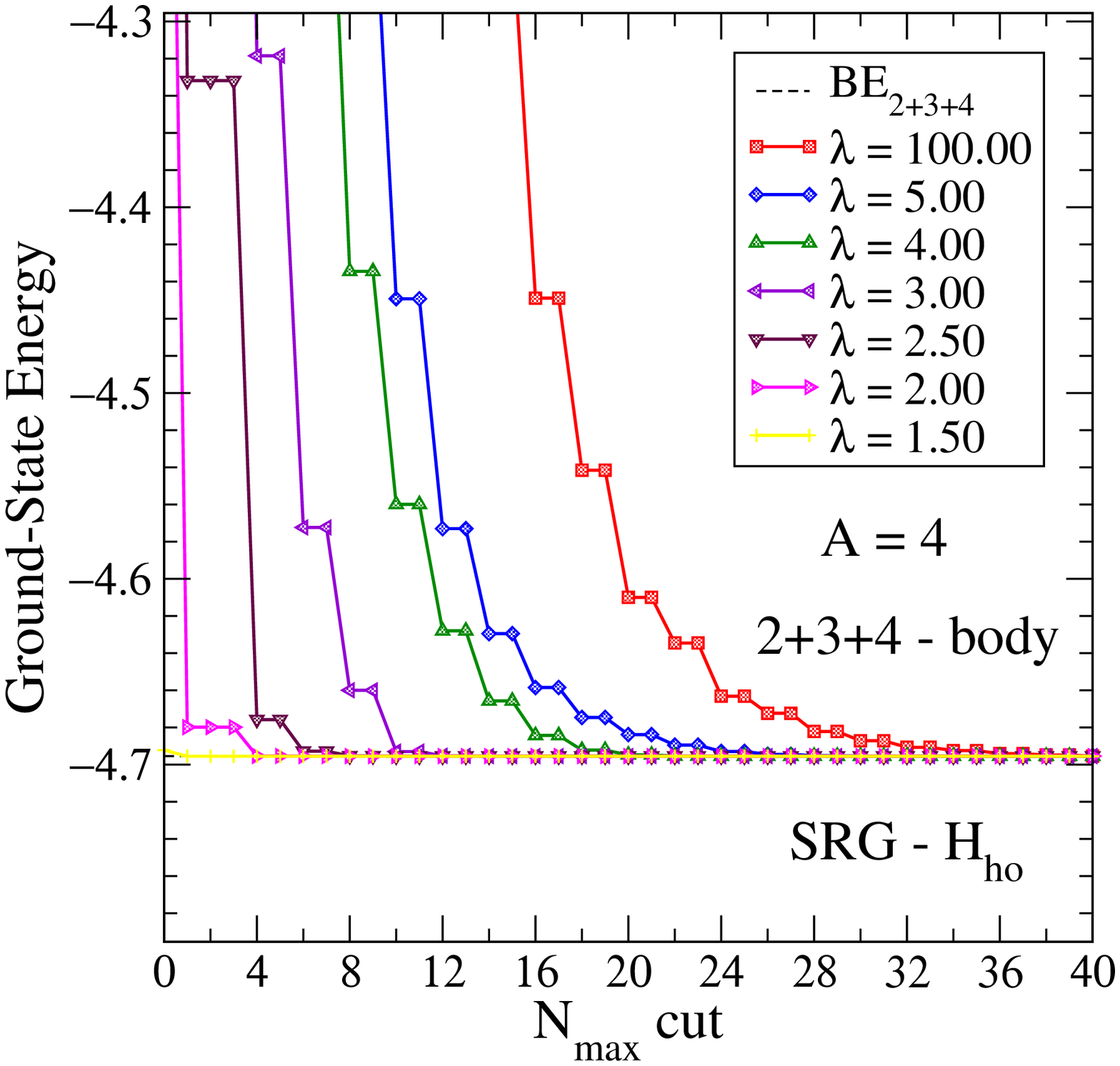}
\end{center}
\captionspace{Same as for Fig.~\ref{fig:srg_3_body_Hho_decoupling} but for
the $A=4$ binding energy. On the left only the two+three-body potential
is evolved and embedded while the right involves the full unitarily
transformed potential.}
\label{fig:srg_4_body_Hho_decoupling}
\end{figure}

Figures \ref{fig:srg_3_body_Hho_decoupling} and
\ref{fig:srg_4_body_Hho_decoupling} shows the decoupling analogous to
Figs.~\ref{fig:srg_3_body_decoupling} and
\ref{fig:srg_4_body_decoupling}, but now using $G_s = \Hho$. Here the
SRG drives the Hamiltonian towards a form diagonal in $\nmax$ (and
hence, energy), which is block diagonal in the oscillator basis. Again
the calculation on the right has the full three-body (or four)
Hamiltonian evolved before being cut to measure the convergence. Here,
with $G = \Hho$ evolution, there is no saturation effect in the
increased convergence. Instead the convergence continues to steadily
improve with decreasing $\lambda$. However, the plot on the left side
of the figure has the two-body (two+three-body) Hamiltonian evolved
before being embedded to the $A=3$ ($A=4$) space and cut for study.
Here, as in the analogous $\Trel$ evolution plot, we expect to see the
asymptotic values of the different $\lambda$ curves vary according to
the missing induced three-body (four-body) forces. But in the $\Hho$
evolution, the variance is dramatically larger, due to spurious bound
states produced during the evolution of two-body (two+three-body) only
matrix elements. These spurious states are a problem for the use of
$\Hho$ as the SRG generator and they are further explored in
Appendix~\ref{chapt:app_spurious}.

In addition to the effects of the generator itself on the Hamiltonian,
the method of cutting used here was applied directly from that used to
study decoupling in the momentum basis. While cutting in $\nmax$ is a
variational technique it is not an exclusive test of high-low energy
decoupling because a cut made in the oscillator basis imposes both IR
and UV cutoffs while a cut in momentum space is simply a UV cutoff.
Therefore the decoupling studies performed here in the oscillator
basis cannot be directly connected to the decoupling studies of
chapter \ref{chapt:decoupling}. However, cutting in $\nmax$ does
establish the convergence with basis size and is a direct measure of
success in the broader program of controlling resource requirements.
The issue of oscillator space decoupling is further informed by the
discussion on general oscillator basis features in
appendix \ref{chapt:app_osc_truncation}.

\chapter{Evolution of Many-Body Forces
in the No-Core Shell Model (NCSM)}
\label{chapt:ncsm}

Our calculations are performed in the Jacobi coordinate harmonic
oscillator (HO) basis of the No-Core Shell Model
(NCSM)~\cite{ncsm_basic}, which is a direct extension of the
one-dimensional oscillator basis built in chapter~\ref{chapt:OneD}.
This is a translationally invariant, anti-symmetric basis for each
$A$, with a complete set of states up to a maximum excitation of
$\nmax\hw$ above the minimum energy configuration, where $\omega$ is
the harmonic oscillator parameter. Further details of the basis
construction are explained in section~\ref{sec:osc_basis_3d}. Note
that the realistic NCSM was built by Navr\'atil \cite{ncsm_basic} and
collaborators long before this thesis work. The work done here
consisted of the code modifications necessary to implement the SRG.
The procedures used here build directly on those of
chapter~\ref{chapt:OneD} where we studied the convergence benefits of
SRG evolution along with a general analysis of the evolving many-body
hierarchy.

Just as in chapter~\ref{chapt:OneD} with the one-dimensional model, we
start by evolving $H_s$ in the $A=2$ space, which completely fixes the
two-body matrix elements. Next, by evolving $H_s$ in the $A=3$ space
we determine the combined two-plus-three-body matrix elements. We can
isolate the three-body matrix elements by subtracting the evolved 
two-body  elements embedded in the $A=3$ basis. Having obtained the
separate NN and NNN matrix elements, we can apply them to any nucleus
with their appropriate combinatoric factors exactly as described in
the one-dimensional case. We are also free to include any initial
three-nucleon force in the initial Hamiltonian without changing the
procedure.  We summarize in Table~\ref{tab:one} the different
calculations to be made for  $^3$H and $^4$He to confront these
questions. If applied to $A \geq 4$, four-body (and higher) forces
will not be included and so the transformations will be only
approximately unitary for those nuclei.  The question to be addressed
is whether the decreasing hierarchy of many-body forces is maintained
as observed in chapter~\ref{chapt:OneD} or the induced four-body
contribution is unnaturally large.

\bt
\label{tab:one}
\caption{Definitions of the various calculations used to study SRG
evolution.}
\vspace{.05in}
\begin{tabular}{rl}
\hline
\hline
 \multirow{2}{*}{NN-only} & No initial NNN interaction and \\
  & do not keep NNN-induced interaction.   \\
  \hline
  \multirow{2}{*}{NN + NNN-induced} & No initial NNN interaction but \\ 
  & keep the SRG-induced NNN interaction.   \\
  \hline
  \multirow{2}{*}{NN + NNN} & Include an initial NNN interaction \emph{and} \\
  & keep the SRG-induced NNN interaction.   \\
\hline
\hline
\end{tabular}
\et

If needed in the future, we could evolve 4-body matrix elements in
$A=4$ and will do so when nuclear structure codes can accommodate them.
The prospects for such work are straightforward. For $A=4$ the Jacobi
NCSM used here can be modified to produce the necessary matrix
elements for evolution by the SRG. These matrices are very large (see
appendix~\ref{chapt:app_scaling} on basis scaling issues) and may
require more sophisticated techniques for computing the many
simultaneous differential equations involved in the evolution. 
Implementing these algorithms should be straightforward, though
computationally intensive. For $A>4$ we would have to implement the
SRG in a different type of NCSM which uses single particle coordinates
and Slater determinants to antisymmetrize the states. This scheme is
also well developed and, due to its basis organization, easy to scale
in $A$ but accordingly computationally intensive.

The initial ($\lambda = \infty$) NN potential used here is the
500\,MeV N$^3$LO interaction from Ref.~\cite{N3LO}.  The initial NNN
potential is the N$^2$LO interaction~\cite{Epelbaum:2002vt}  in the
local form of Ref.~\cite{Navratil:2007} with the additional LEC
constants, $c_D$ and $c_E$, fit to the average of triton and $^3$He
binding energies and to triton beta decay according to
Ref.~\cite{Gazit:2008ma}. The values used here are $c_D = -.205$ and
$c_E = .029$. We expect similar results from other initial
interactions because the SRG drives them toward near universal
form~\cite{Bogner:2006srg}; a survey will be given in
Ref.~\cite{Jurgenson:2009}. NCSM calculations with these initial
interactions yield energies of $-8.473(4)\,$MeV for $^3$H and
$-28.50(2)\,$MeV for $^4$He compared  with $-8.482\,$MeV and
$-28.296\,$MeV from experiment, respectively. So there is a 20\,keV
uncertainty in the calculation of $^4$He from incomplete convergence
(shown in the third decimal place - in parentheses) and a 200\,keV
discrepancy with experiment.  The latter is consistent with the
omission of three- and four-body chiral interactions at
N$^3$LO~\cite{Rozpedzik:2006yi}. These provide a scale for assessing
whether induced four-body contributions are important compared to
other uncertainties.

It is remarkable how well the one-dimensional model predicts behavior
in these three-dimensional calculations. Often serious qualitative
differences appear between one- and three-dimensional systems. In this
case several factors conspire to damp those differences. First, the
three-dimensional potential is usually stated in a partial wave basis,
which breaks up the three-vector information based on angular momentum
states. So, when we work in a given partial wave, we are working with
an effective one-dimensional potential, one which we were able to
successfully mimic in chapter~\ref{chapt:OneD}. Secondly, and more
central to this thesis, the SRG is a simple unitary transformation and
its behavior is governed only by the qualitative structure of the
matrices it is working to evolve, like the Hamiltonian and the
generator. Qualitatively mimicking those, the outcome of SRG evolution
will be predictive. This leaves us with a powerful analytic tool in
the one-dimensional model of the NCSM to quickly explore other methods
to improve many-body calculations beyond just SRG implementation.


\section{Induced Many-Body Forces}

\begin{figure}[thb]
\bc
\singlpic{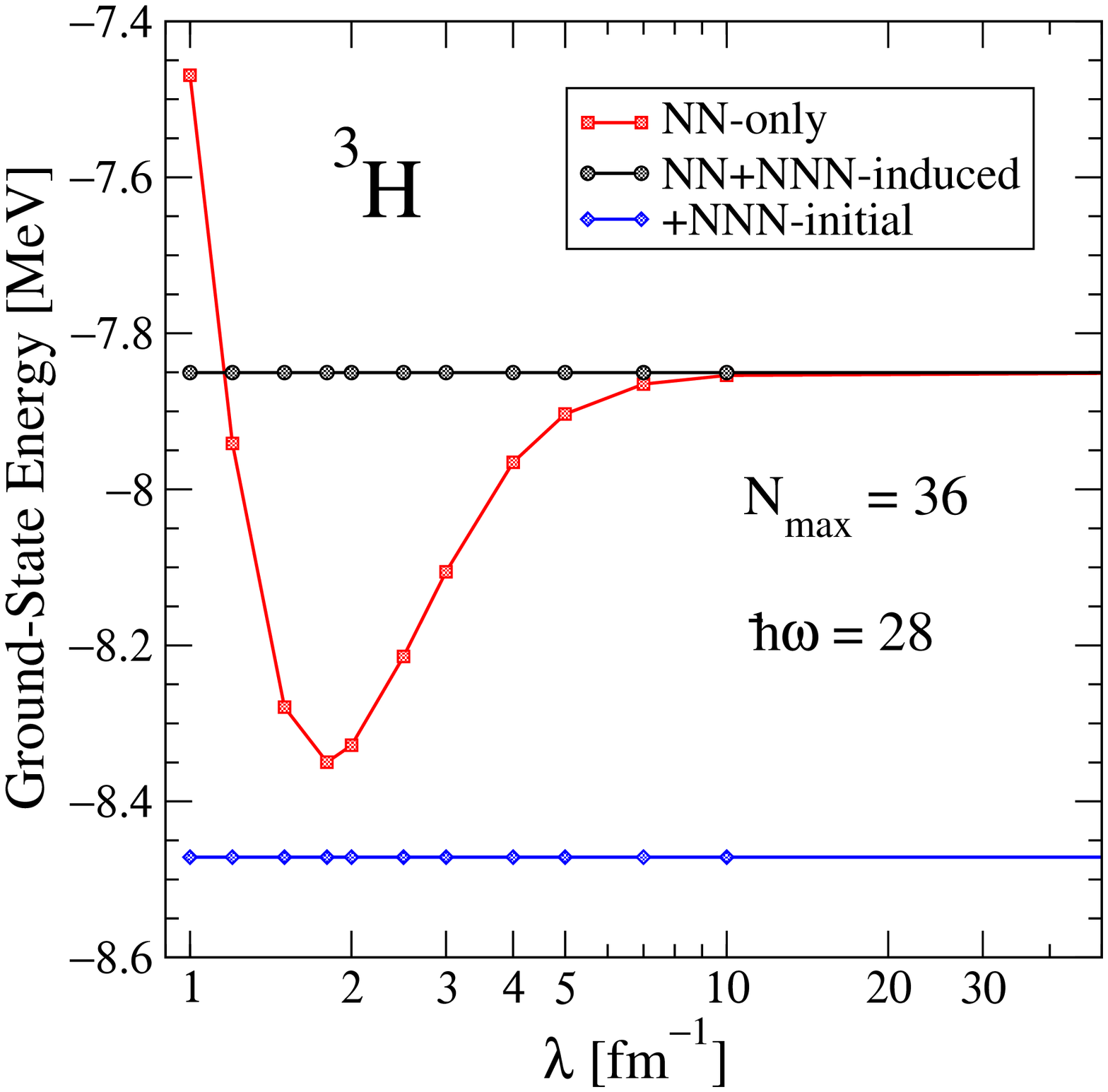}
\ec
\captionspace{Ground-state energy of $^3$H as a function of the SRG
evolution parameter, $\lambda$. See Table~\ref{tab:one} for the
nomenclature of the curves.}
\label{fig:h3_srg}
\end{figure}

In Fig.~\ref{fig:h3_srg}, the ground-state energy of the triton 
is plotted as a function of the flow parameter $\lambda$. 
Evolution is from $\lambda = \infty$, which is the initial (or
``bare'') interaction, toward $\lambda = 0$. We use $N_{\rm max}
= 36$ and $\hw = 28\,$MeV, for which all energies are
converged to better than 10\,keV. We first consider an NN
interaction with no initial NNN (``NN-only'').  If the Hamiltonian is
evolved only in an $A=2$ system, higher-body induced pieces are
lost.  The resulting energy calculations will only be
approximately unitary for $A>2$ and the ground-state energy will
vary with $\lambda$ (squares). Keeping the induced NNN yields a
flat line (circles), which implies an exactly unitary
transformation; the line is equally flat if an initial NNN is
included (diamonds). Note that the net induced three-body is
comparable to the initial NNN contribution and thus is of natural
size.

\begin{figure}[htb]
\bc
\dblpic{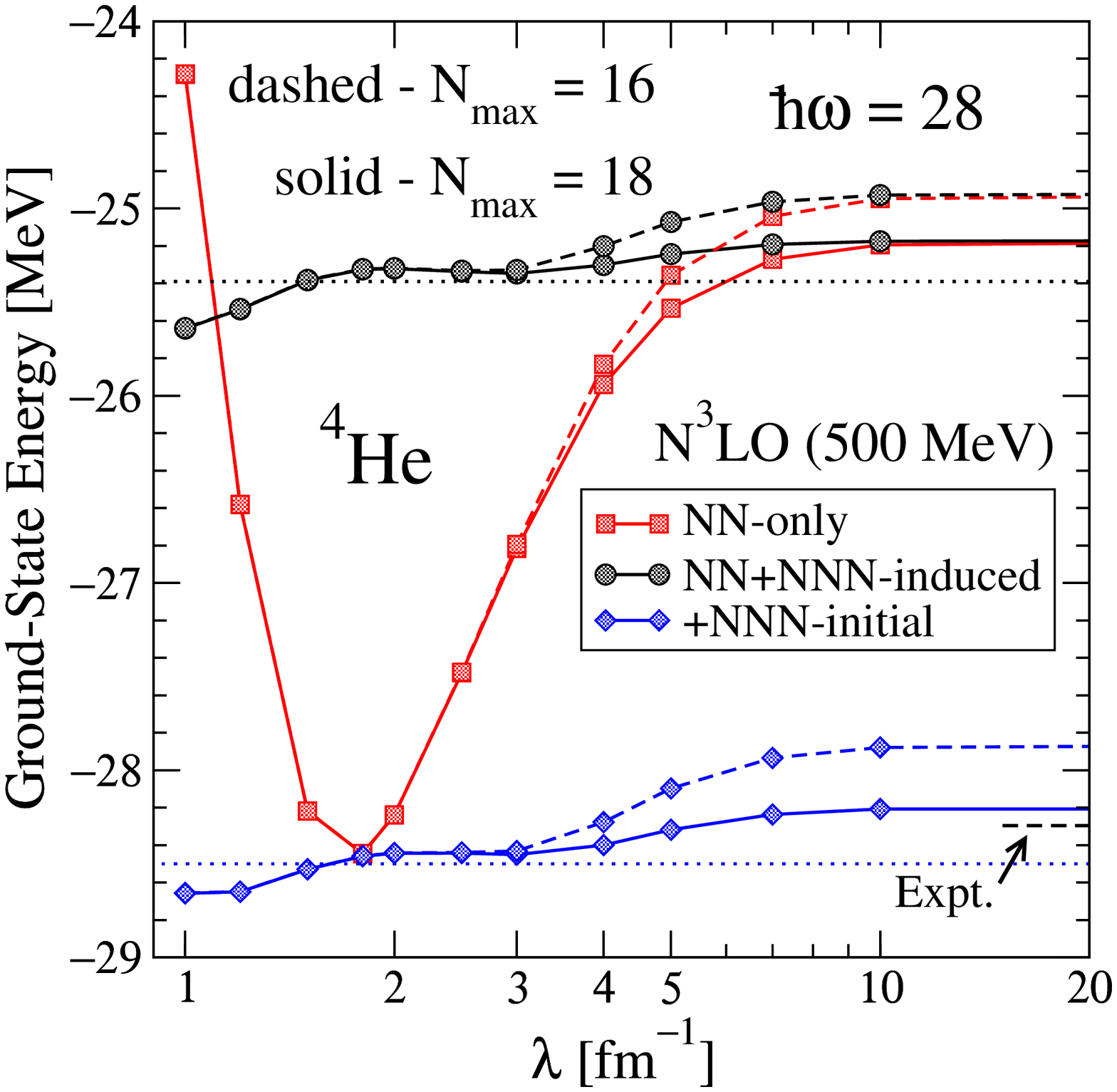}
\dblpic{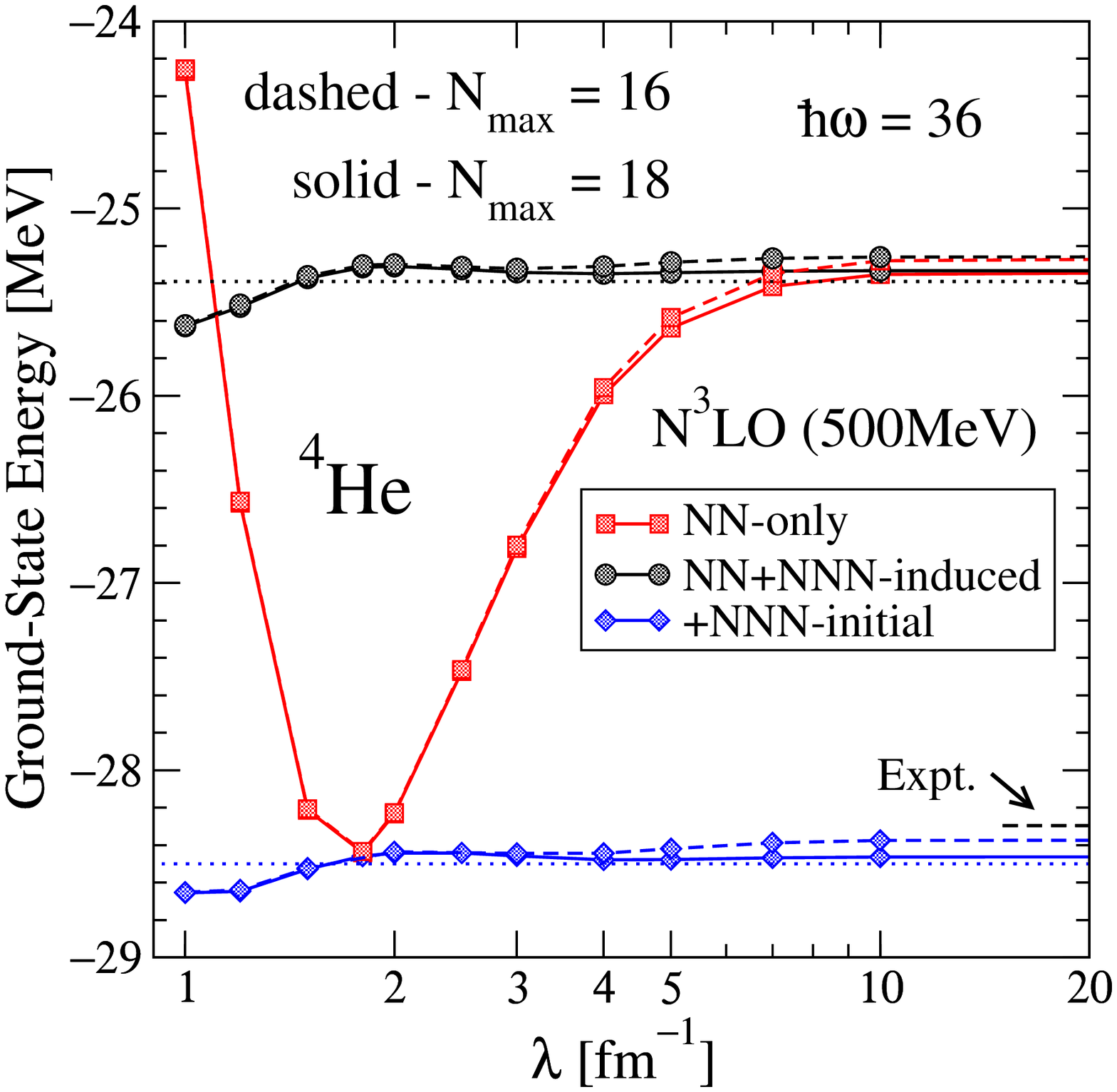}
\ec
\captionspace{Ground-state energy of $^4$He as a function of
$\lambda$. Two $\hw$s (left and right) and two $\nmax$'s (overlayed)
are shown for comparison.}
\label{fig:he4_srg}
\end{figure}

In Fig.~\ref{fig:he4_srg}, we examine the SRG evolution in $\lambda$
for $^4$He with $\hbar\Omega = 36\,$MeV.  The two- and three-body
matrix elements were evolved in $A=2$ and $A=3$ with $N_{\rm max} =
28$ and then truncated to $\nmax = 18$ (solid) and $\nmax = 16$
(dashed) at each $\lambda$ to diagonalize $^4$He. The NN-only curve
has a similar shape as for the triton and when the induced NNN is
included, the evolution is close to unitary, the deviation being an
indication of the induced four-body force.  The pattern only
depends slightly on strength and type of an initial NNN interaction.

In both cases with and without an initial three-nucleon force, the
dotted lines represent the converged values for their respective
initial Hamiltonians. At large $\lambda$, the discrepancy is due to a
lack of convergence at $N_{\rm max} = 18$, (the level of truncation at
this calculation) but by $\lambda < 3\,\mbox{fm}^{-1}$ SRG decoupling
takes over and the discrepancy is due to induced four-body forces.
This is illustrated nicely by the overlay which shows that, for
$\lambda < 3\,\mbox{fm}^{-1}$, cutting from $\nmax = 18$ to $\nmax =
16$ introduces no more basis truncation errors, where above $\lambda =
3\,\mbox{fm}^{-1}$ there is a significant truncation effect just from
18 to 16. This indicates that the SRG has achieved convergence well
inside an $\nmax = 16$ basis for this calculation.

Also, in Fig.~\ref{fig:he4_srg} are shown the same overlay comparison
for two $\hw$ values, that which is optimal for the initial
Hamiltonian to compute $^4$He ($\hw = 36$) and one that is nearer the
optimal for that potential evolved to $\lambda = 2 \,\mbox{fm}^{-1}$
as expected from previous SRG studies. We can see the same converged
results at low $\lambda$ in both $\hw$'s despite the different bare
answer. This is consistent with being well converged and thus flat in
$\hw$.

The induced four-body forces then make up the difference to the
converged value and contribute about 50\,keV net at
$\lambda = 2\,\mbox{fm}^{-1}$. This is small compared to the rough
estimate in Ref.~\cite{Rozpedzik:2006yi} that the contribution from
the long-ranged part of the N$^3$LO four-nucleon force to $^4$He
binding is of order a few hundred keV. If needed, we could evolve
4-body matrix elements in $A=4$ and will do so when nuclear structure
codes can accommodate them.


\begin{figure}[t]
\bc
\singlpic{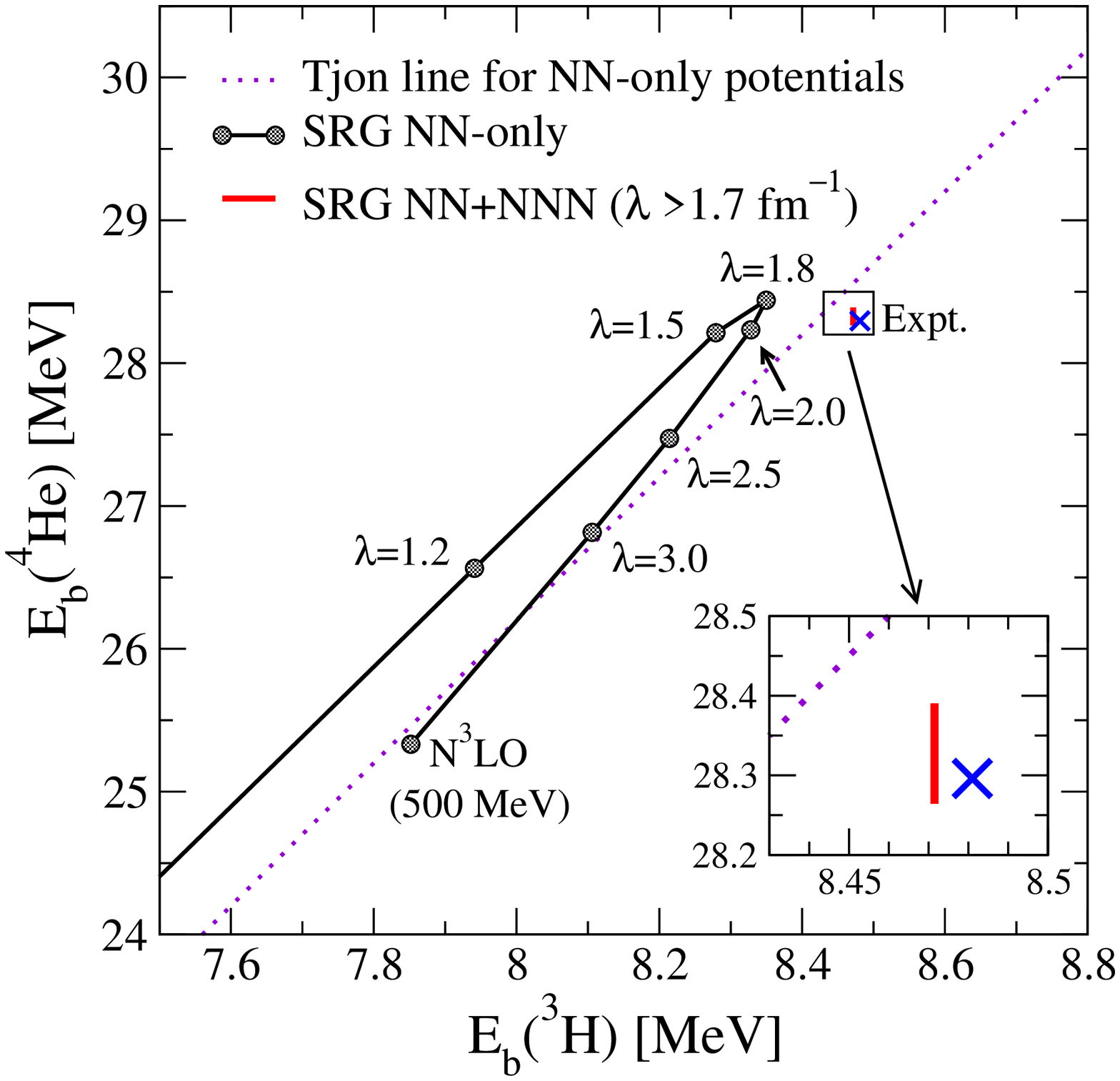}
\ec
\captionspace{Binding energy of the alpha particle vs.\
  the binding energy of the triton. 
  The Tjon line from phenomenological NN potentials
  (dotted) is compared with the trajectory of SRG energies
  when only the NN interaction is kept (circles).  When the
  initial and induced NNN interactions are included, the
  trajectory lies close to experiment for $\lambda >
  1.7\,\mbox{fm}^{-1}$ (see inset). }
\label{fig:tjon_line}
\end{figure}

The impact of evolving the full three-body force is neatly illustrated
as a Tjon plot in Fig.~\ref{fig:tjon_line}, where the binding energy
of $^4$He is plotted against the binding energy of $^3$H.  The
experimental values of these quantities are known to a small fraction
of a keV and define only a point in this plane (at the center of the
X, see inset).  The well-known Tjon line (dotted) is the approximate
locus of points for  phenomenological potentials fit to NN data but
not including NNN~\cite{Vlowk3N}. The SRG NN-only results trace out a
trajectory in the plane that is analogous as it transforms NN forces
unitarily but cannot account for the evolving 3NFs in the $A=2$ space.
In contrast, the short trajectory of the SRG when including the
induced NNN interaction (shown for $\lambda \geq 1.8\,\mbox{fm}^{-1}$)
highlights the small variations from the  omitted four-nucleon force.
Note that a trajectory plotted for NN+NNN-induced calculations  would
be a similarly small line at the N$^3$LO NN-only point where the
initial NNN is zero.


\section{Convergence in $\nmax$: $G = \Trel$}


\begin{figure}[htb]
\bc
\dblpic{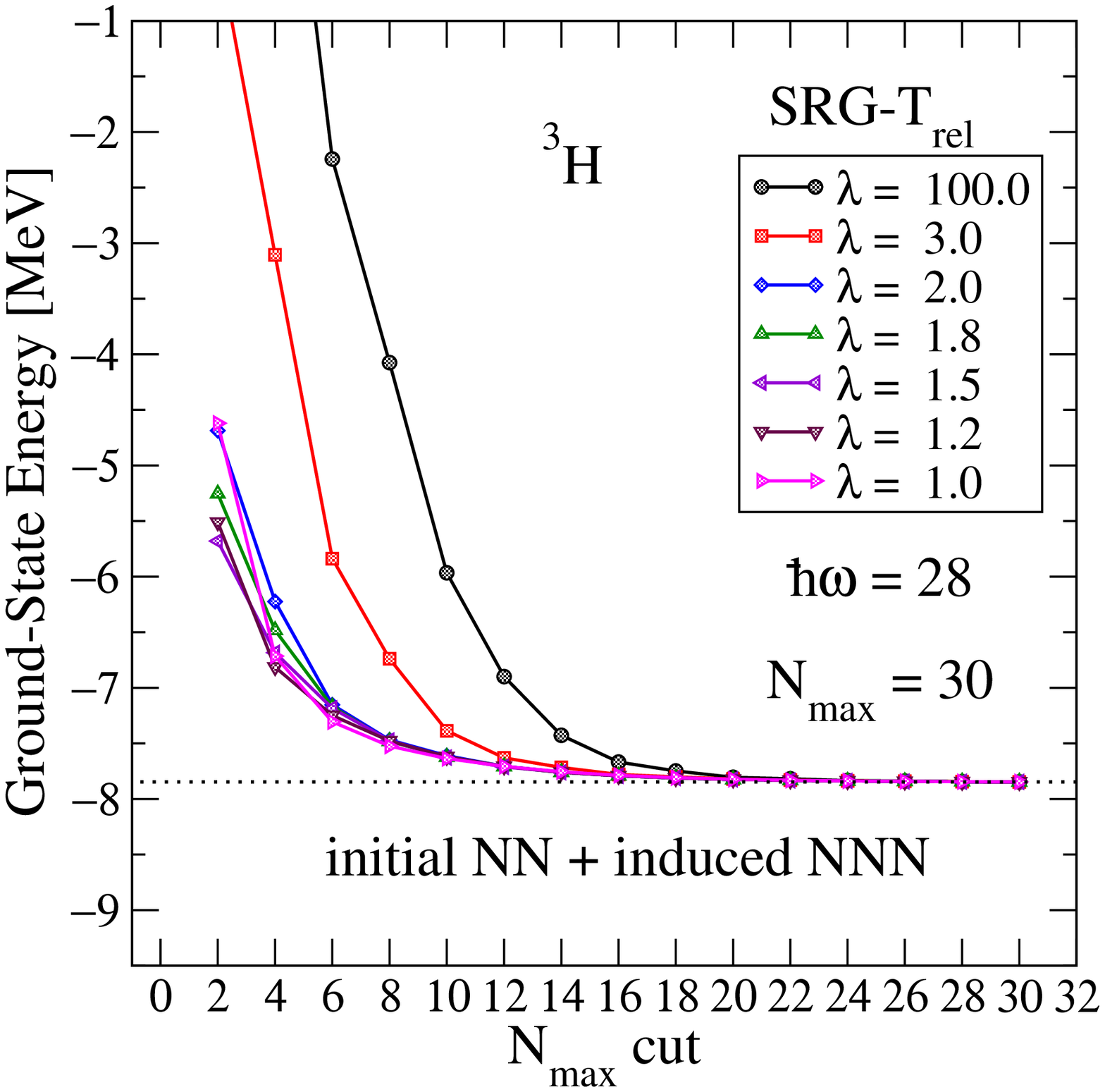}
\dblpic{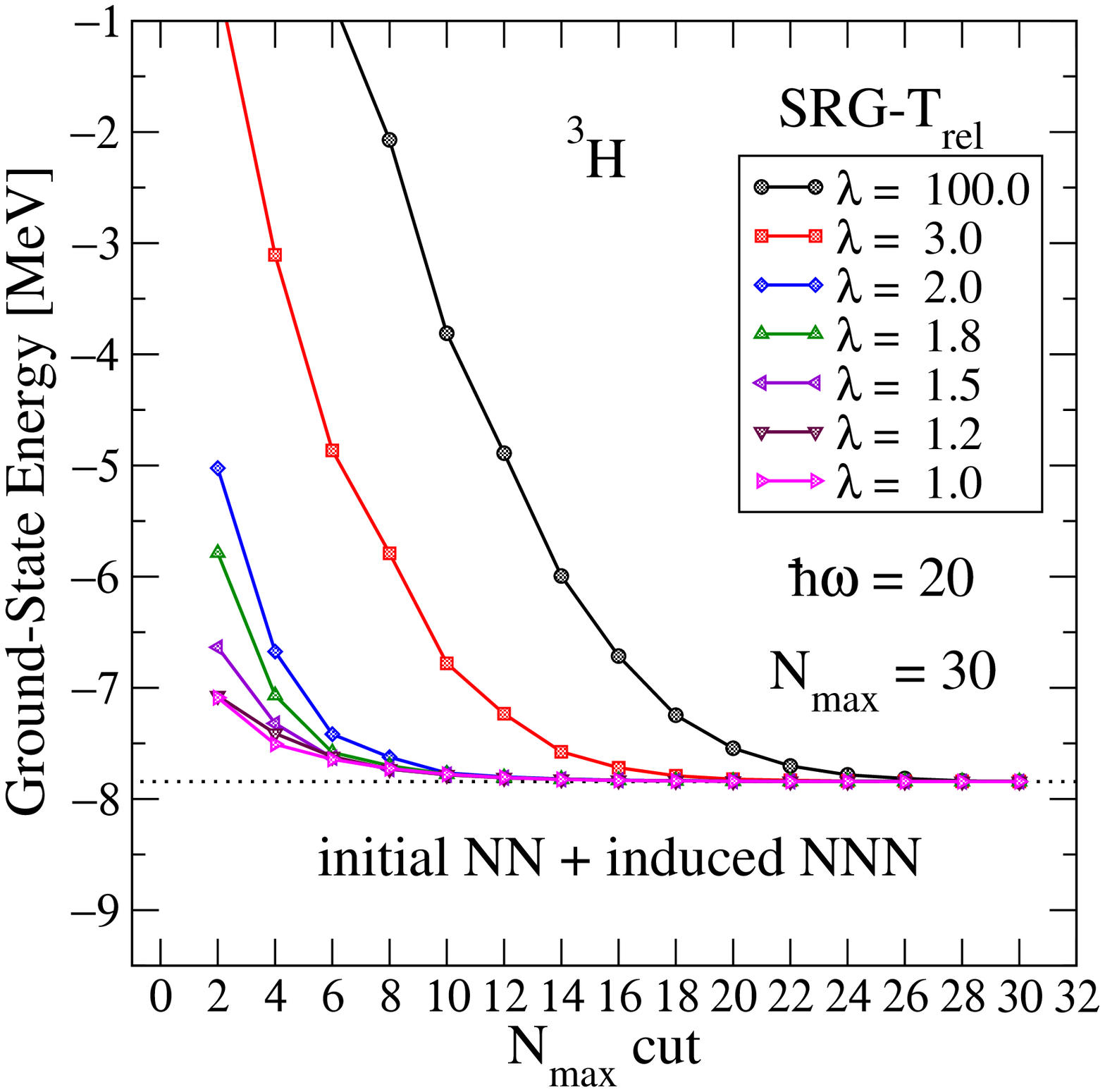}
\ec
\captionspace{Ground-state energy of $^3$H as a function of the basis size
  $\nmax$\ for an N$^3$LO NN interaction~\cite{N3LO} evolved with the
  SRG using $G_s = \Trel$ to selected $\lambda$'s. The same calculation
  is shown at two values of $\hw$.}
\label{fig:h3_convergence}
\end{figure}

Figure~\ref{fig:h3_convergence} shows the convergence properties of
the three-body Hamiltonian as it evolves. The initial interaction used
here is an N$^3$LO NN potential~\cite{N3LO} with no initial NNN. The
figure shows the same plot for two different $\hw$'s. On the left is
$\hw = 28$ MeV which is the optimal value for the initial interaction,
and on the right is the same plot for $\hw = 20$ MeV which is more
optimal for significantly evolved potentials (i.e., those evolved to
$\lambda = 2.0 \fmi$). The calculations are performed by first evolving
the initial Hamiltonian in the three-body space to the $\lambda$
indicated then truncating the potential at each $\nmax$-cut by setting
matrix elements to zero above the extent of that $\nmax$-cut. The
curve for $\lambda = 100 \fmi$ is essentially unevolved and is
approximately equivalent to the initial interaction. 


Notice two major features as the initial Hamiltonian is evolved down
from $\lambda = 100 \fmi$. First, the converged value at each
$\lambda$ is the same - the evolution is unitary. As the SRG evolves
the Hamiltonian in the three-body basis, it evolves the two-body
matrix elements just as in the A=2 basis, but to keep the
transformation unitary it must induce three-body forces to account for
the high-momentum information it has transformed. These induced three
body forces have been kept in the calculation of
Fig.~\ref{fig:h3_convergence} and hence the ground-state energy is
preserved by the transformation. Secondly, the convergence in $\nmax$
improves as $\lambda$ decreases down to some saturation point. The SRG
is suppressing the coupling between states of large and small $\nmax$.
Here we have made the choice of generator $G_s = \Trel$, which was
diagonal in the momentum representation, and should afford the same
decoupling benefits it did there. However, in the oscillator basis,
$\Trel$ has both a diagonal piece and a slightly off-diagonal piece
that couples states at $\nmax$ and $\nmax+2$\footnote{see
Eq.~\eqref{eq:T_exact} for an example of this in the one-dimensional
basis and a sample picture of $\Trel$ can be found in
Fig.~\ref{fig:spurious_Ho3e}.}. So in this basis the convergence
saturates as the Hamiltonian evolves toward the shape of $\Trel$ and
retains a certain amount of coupling. 
However, the evolution is itself basis
independent and this apparent coupling is really due to the structure
of the oscillator basis and the way we are required to cut it.
Converting these Hamiltonians back to the momentum representation (or
working there in the first place) should return to the situation where
the potential asymptotically approaches the diagonal. Work along
these lines is in progress~\cite{lplatter_pc}.


The situation is already improved from the left panel to the right due
to a more optimal value of $\hw$ for the evolved Hamiltonians. The
point of saturation is as far down as $\lambda =  1.0 \fmi$. One would
like to optimize each point on this plot in $\hw$. Since each
combination of $\nmax$ and $\hw$ represents a considerable amount of
computation, a useful tool would be a simple orthogonal transformation of the
Hamiltonian from one value of $\hw$ to another without building up a
separate basis for each parameter set. Until such a tool is coded and
tested, the alternative is to
work at sufficiently large $\nmax$ that the energy is converged and
relatively flat in (insensitive to) $\hw$. This would result in a
curve that may be less smooth and harder to extrapolate but still
variational, converging at the large-$\nmax$ and optimal $\hw$ for the
bare potential.


\begin{figure}[htb]
\bc
\dblpic{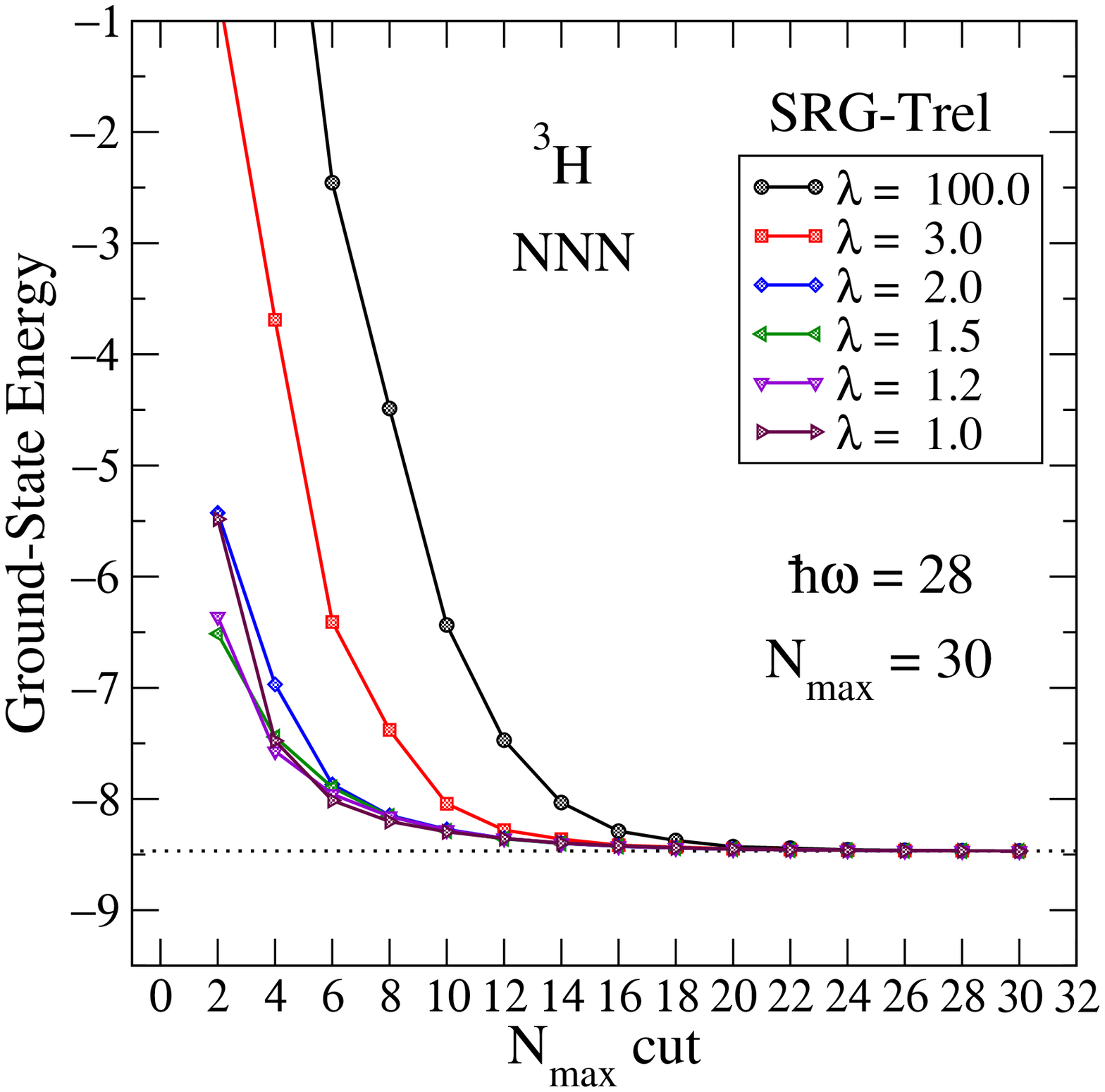}
\dblpic{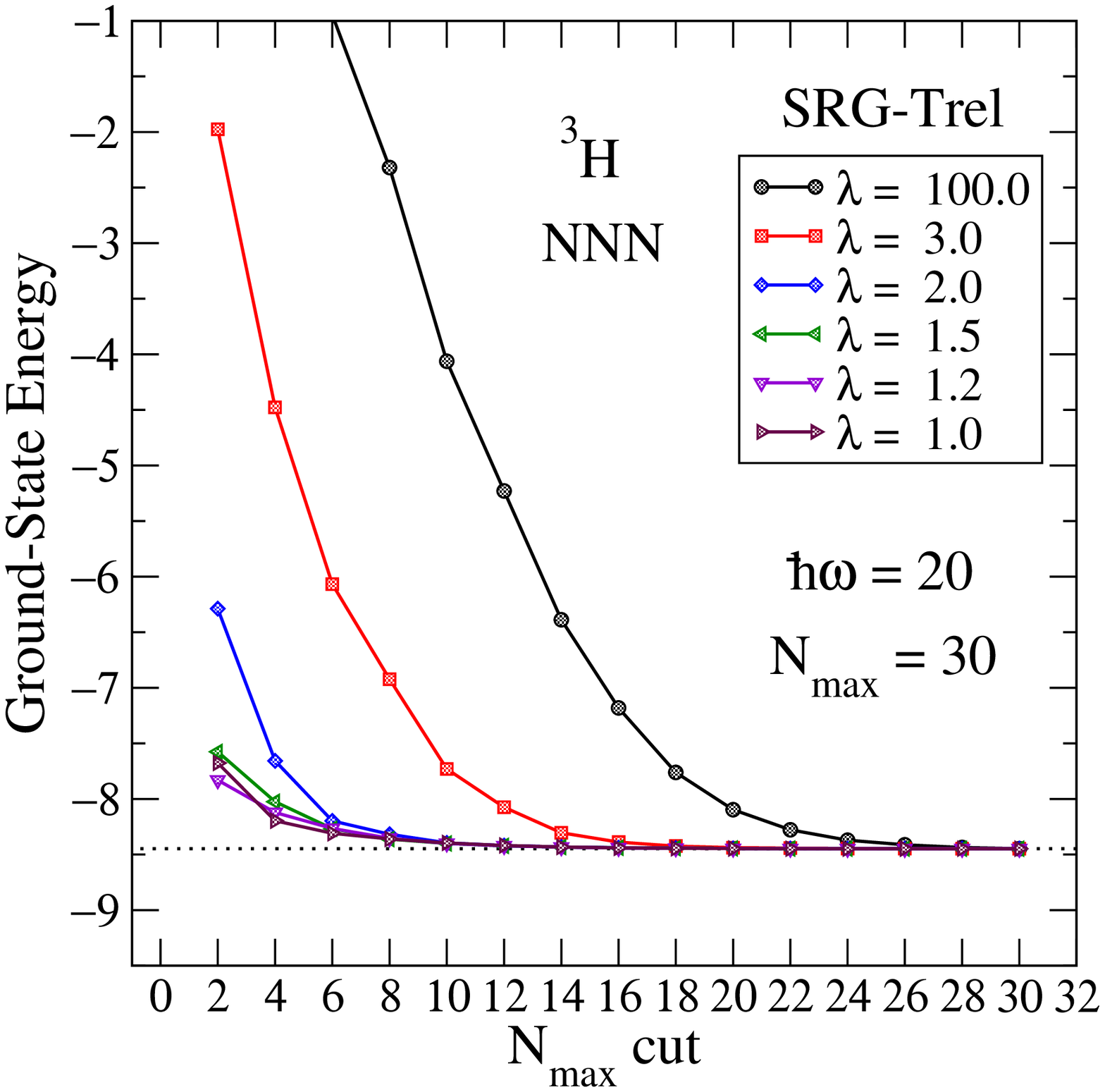}
\ec
\captionspace{Ground-state energy of $^3$H as a function of the basis size
  $\nmax$\ for an N$^3$LO NN  interaction~\cite{N3LO} plus an initial
  NNN interaction~\cite{Gazit:2008ma,Epelbaum:2008ga} evolved with the
  SRG using $G_s = \Trel$ to selected $\lambda$s.}
\label{fig:h3_NNN_convergence}
\end{figure}

The initial Hamiltonians used in Fig.~\ref{fig:h3_convergence} are
NN-only, so to increase the
accuracy in calculating observables we must include initial three-body
forces in our inputs.  One may be concerned that the exact form of any
initial three-body forces will affect the form of the three-body
forces induced to maintain unitarity. This question is tested in
Fig.~\ref{fig:h3_NNN_convergence} which repeats the calculation of
Fig.~\ref{fig:h3_convergence} but now including an initial NNN force
at N$^2$LO from Ref ~\cite{Epelbaum:2002vt} as stated at the beginning
of the chapter. We note that the shapes of the convergence curves in
both figures are identical, indicating that the induced three-body
forces are of natural size and are insensitive to the details of any
initial three-body forces. Therefore
the SRG is insensitive to the particulars of a given interaction,
which is consistent with the observation that different Hamiltonians
evolve to a near universal form.


\begin{figure}[htb]
\bc
\singlpic{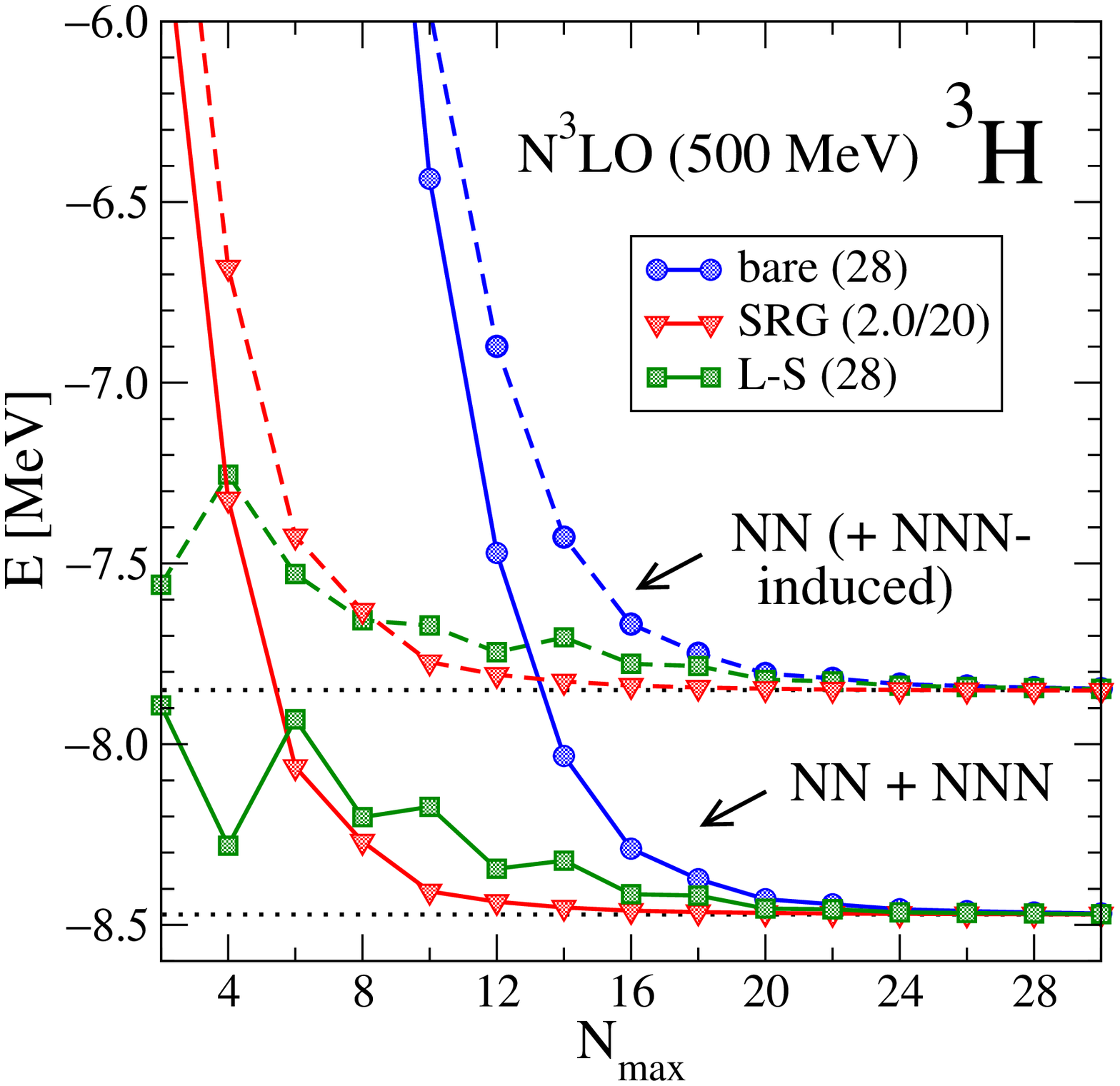}
\ec
\captionspace{Ground-state energy of $^3$H as a function of the basis size
$\nmax$\ for an N$^3$LO NN  interaction~\cite{N3LO} with and without
an initial NNN interaction~\cite{Gazit:2008ma,Epelbaum:2008ga}.
Unevolved (``bare'') and Lee-Suzuki (L-S) results with $\hbar\Omega =
28\,$MeV are compared with SRG at $\hw = 20\,$MeV evolved to
$\lambda = 2.0\,\mbox{fm}^{-1}$.} 
\label{fig:h3_LS_convergence}
\end{figure}

In Fig.~\ref{fig:h3_LS_convergence}, we again show the triton
ground-state energy as a function of the oscillator basis size,
$\nmax$, to make a comparison between various calculations. The lower,
solid (upper, dashed) curves are with (without) an initial
three-body force (see Table~\ref{tab:one}). The convergence of the
bare interaction is shown along with the SRG evolved to $\lambda =
2.0\,\mbox{fm}^{-1}$. The oscillator parameter $\hw$ in each case was
chosen roughly to optimize the convergence of each Hamiltonian. Here,
we compare to a Lee-Suzuki (L-S) effective interaction, which has been
used in the NCSM to greatly improve
convergence~\cite{Nogga:2005hp,Navratil:2007we}. These effective
interactions consist of a block-diagonal type unitary transformation
to achieve an effective interaction in the oscillator basis at low
$\nmax$. The resultant matrix is dependent on the model space of a
given nucleus in contrast to the free-space nature of the SRG. Notice
that the L-S effective interactions are not variational in $\nmax$ due
to their model dependence and therefore it is difficult to extrapolate
their behavior.

The SRG calculations are variational and converge smoothly and
rapidly from above with or without an initial three-body force.
The dramatic improvement in convergence rate is seen even though
the  $\chi$EFT interaction is relatively soft. Thus, once
evolved, a much smaller $\nmax$\ basis is adequate for a desired
accuracy and extrapolating in $\nmax$\ is also feasible.


\begin{figure}[thb]
\bc
\includegraphics*[height=2.5in]{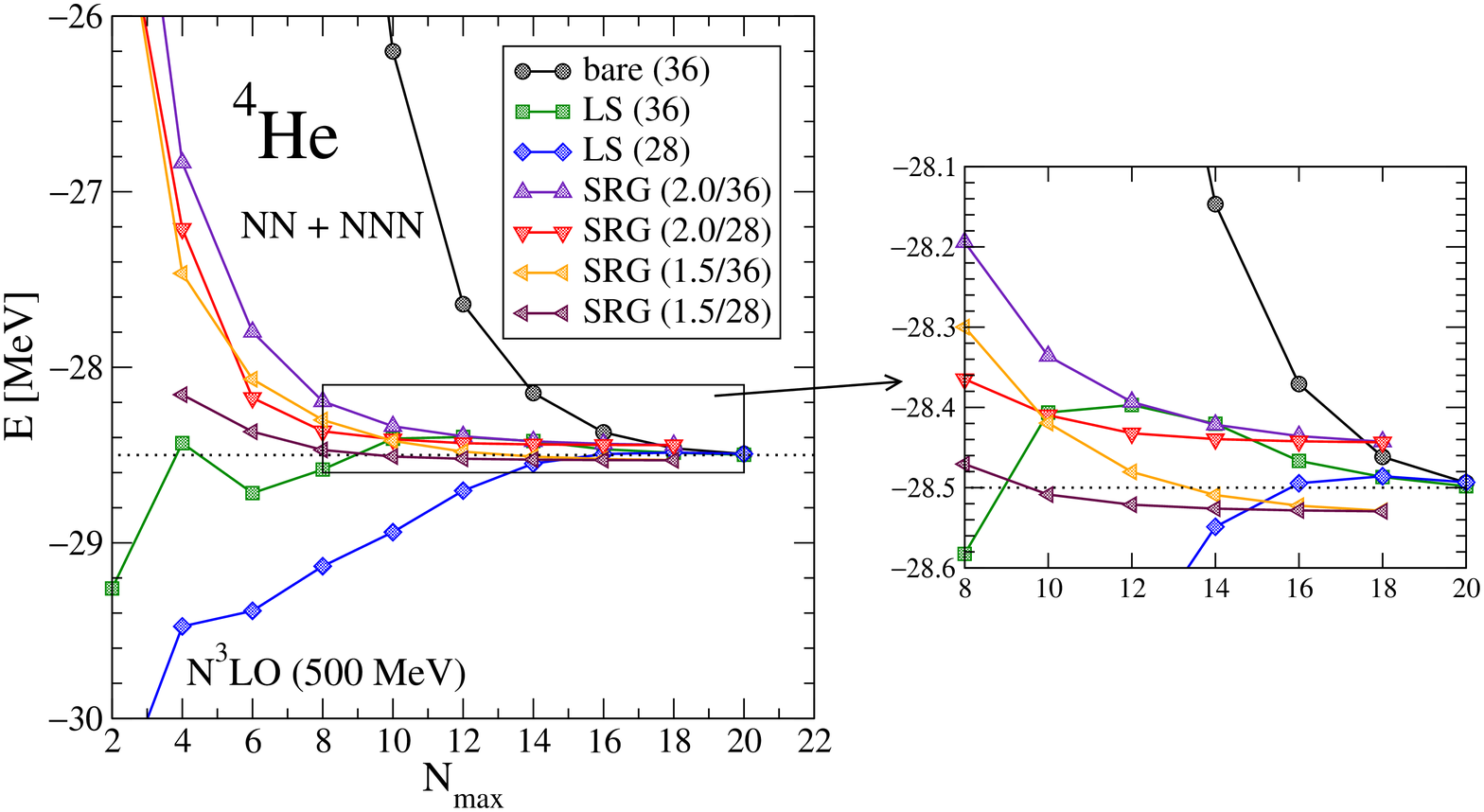}
\ec
\captionspace{Ground-state energy of $^4$He as a
  function of the basis size $\nmax$ for an N$^3$LO NN 
  interaction~\cite{N3LO} with an initial NNN
  interaction~\cite{Gazit:2008ma,Epelbaum:2008ga}.  Unevolved
  (bare) results are compared with Lee-Suzuki (L-S) 
  and SRG evolved to $\lambda = 2.0\,\mbox{fm}^{-1}$ 
  at $\hw = 28$ and $36\,$MeV.}
\label{fig:he4_convergence}
\end{figure}

Figure~\ref{fig:he4_convergence} illustrates for $^4$He the same rapid
convergence with $\nmax$ of an SRG-evolved interaction. Here we show
results in two different $\hw$'s (in parenthesis) for a Lee-Suzuki
calculation, and Hamiltonians evolved to $\lambda = 2.0 \fmi$ and
$\lambda = 1.5 \fmi$. We can see the rapid convergence both in the
main plot and more clearly in the inset. It is again smooth and
variational despite being an approximately unitary calculation for
this sector. In this case the asymptotic value of the energy differs
slightly because of the omitted induced four-body contribution. The
SRG-evolved asymptotic values for different $\hw$  differ by less than
10\,keV, so the gap between the converged bare/L-S results  and the
SRG results is dominated by the induced four-nucleon forces rather
than incomplete convergence. Convergence is even faster for lower
$\lambda$ values~\cite{Jurgenson:2009}, ensuring a useful range for
the analysis of few-body systems. The induced four body forces here
are about 30-60 keV, which is less than the 200 keV discrepancy with
experiment considered to be due to the omission of initial 4NFs from
\xeft. However, because of the strong density dependence of
four-nucleon forces\footnote{This follows from the power-counting of
the \xeft\ lagrangian. The coupling constant must have one more power
of $\Lambda^{-3}$, or density, due to the extra factor of
$N^{\dagger}N$.}, it will be important to monitor the size of the
induced four-body contributions for heavier nuclei and nuclear matter.
It may be necessary to evolve unitarily in the four-body space to
include induced four-body forces, and even initial \xeft\ terms in this
sector. However, it is doubtful that higher body forces ($A > 4$) will
be significant given the strong four particle clustering nature of
larger nuclei. 


\section{Convergence in $\nmax$: $G = \Hho$}

As shown in Figs.~\ref{fig:h3_convergence} and
\ref{fig:h3_NNN_convergence} the choice of SRG generator, $G_s =
\Trel$, is not the most efficient choice in the harmonic oscillator
basis. Due to the organization of the states and their complicated
momentum dependence, $\Trel$ is not diagonal in this basis and
therefore cannot be expected to completely  diagonalize the
Hamiltonian. We can make a different choice of generator, as already
mentioned in chapter~\ref{chapt:OneD}, $G_s = \Hho$, the harmonic
oscillator Hamiltonian. This choice is diagonal in the oscillator basis,
having the eigenvalues $E_n = (\nmax + 3/2)\hw$ of each state along
the diagonal. In
this section we check, in the three-dimensional case, the behavior of
this choice of $G_s$.


\begin{figure}[htb]
\bc
\triplepic{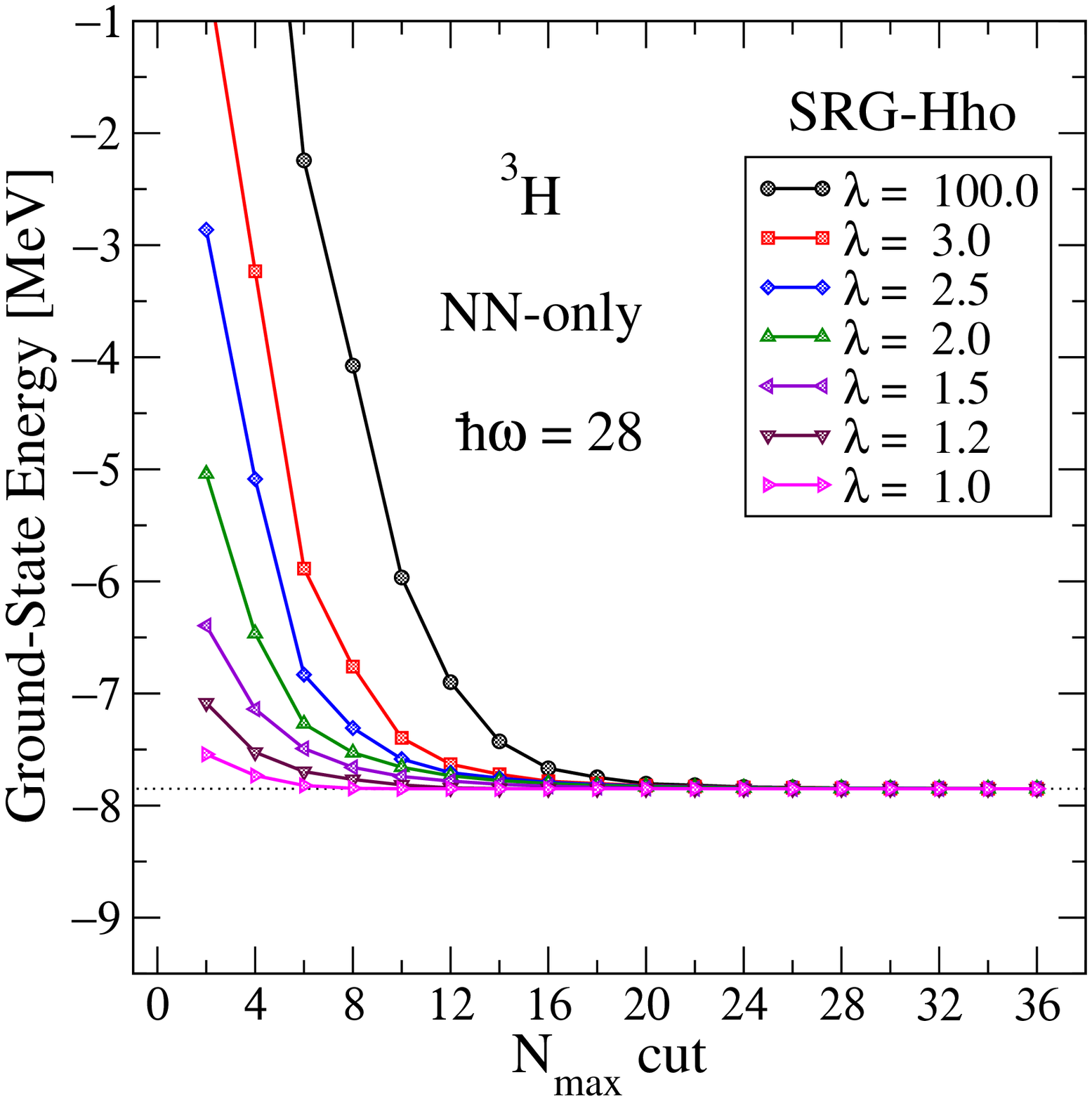}
\hfill
\triplepic{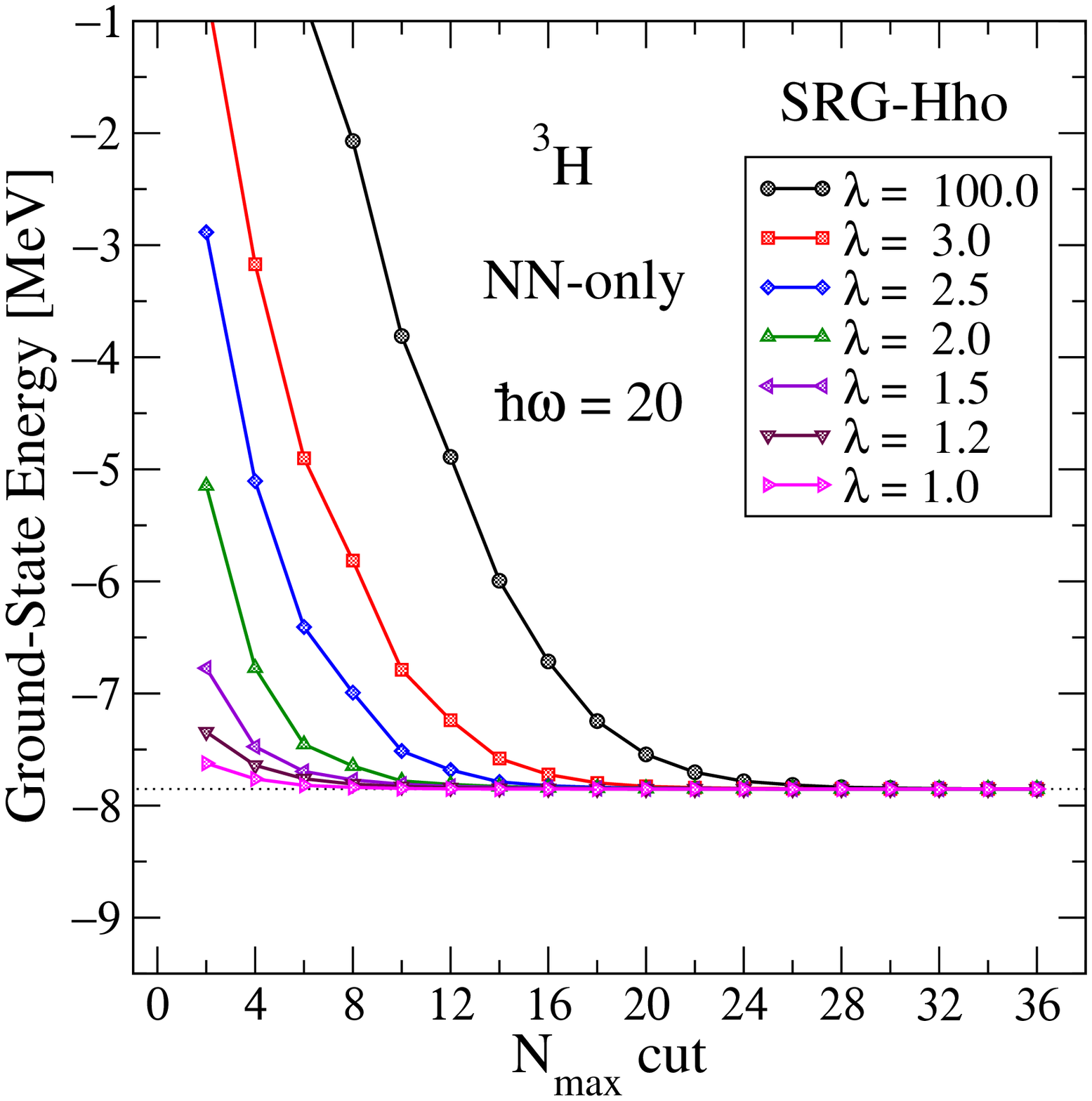}
\hfill
\triplepic{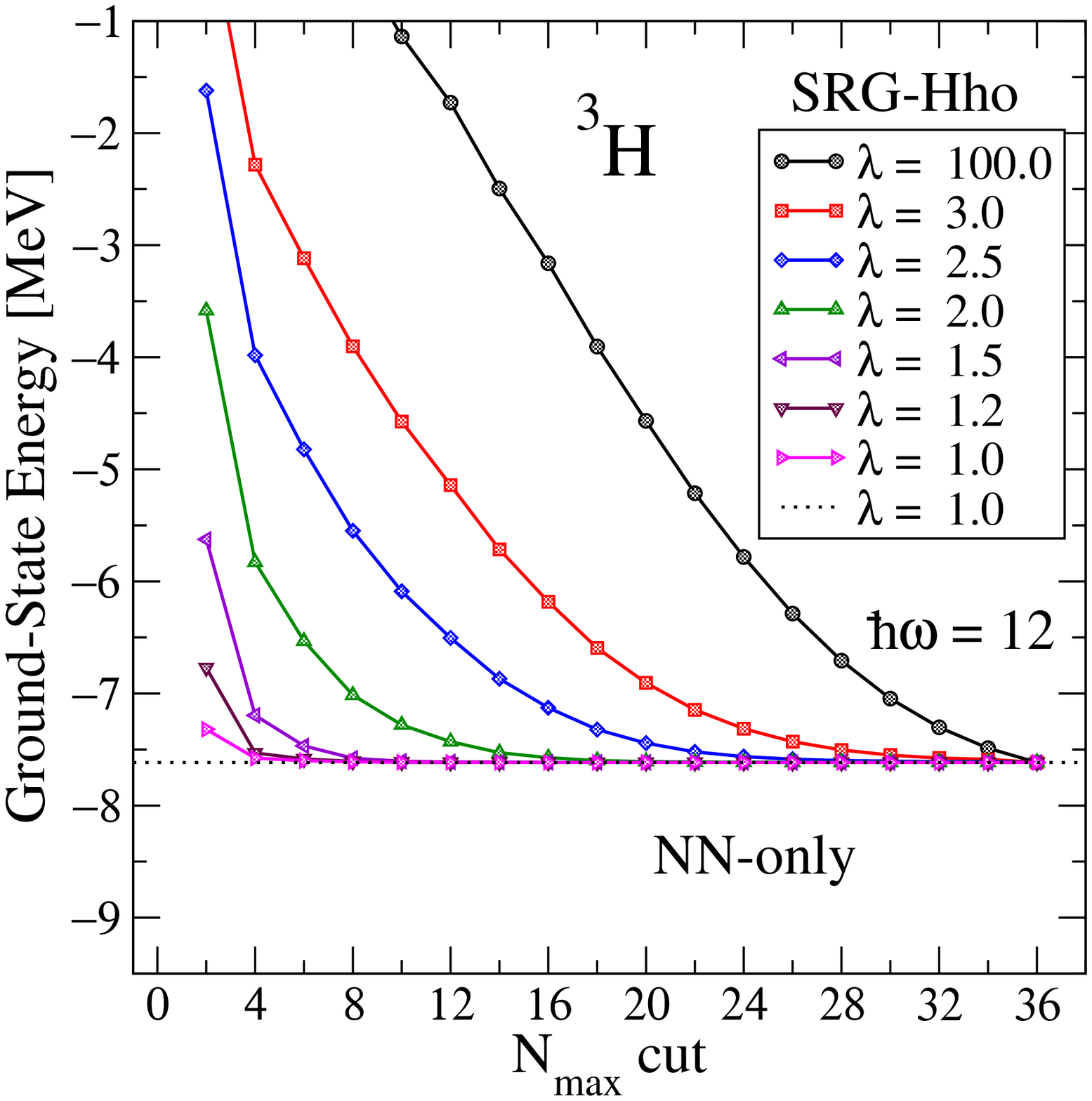}
\ec
\captionspace{Ground-state energy of $^3$H as a function of the basis size
  $\nmax$\ for an evolving N$^3$LO NN interaction~\cite{N3LO} without
  an initial NNN interaction. Three values of oscillator parameter,
  $\hw = 28$, 20, and 12 MeV are shown for comparison.}
\label{fig:h3_Hho_convergence}
\end{figure}

Figure \ref{fig:h3_Hho_convergence} shows the analog of
Fig.~\ref{fig:h3_convergence} but using $\Hho$ as the generator. We
see, just as in the one-dimensional case, that the convergence
improvement does not saturate at a specific point in the evolution but
keeps improving as the SRG evolves it further towards the diagonal in
the oscillator basis. On the left and center are $\hw = 28$ MeV,
optimal for the unevolved Hamiltonian, and $\hw = 20$ MeV for
comparison to the $\Trel$ case. The optimal $\hw$ for $\Hho$-evolved
potentials has not been fully explored for specific $\lambda$'s,
though given the amount of increased convergence it is likely to shift
further. We can see in the center panel with $\hw = 20$ the
convergence is not improved by the same proportions as it was between
the $\Trel$ plots, but on the right we show an even smaller value,
$\hw = 12$, than was used for $\Trel$. Here we see more improvement in
the convergence with a more optimal $\hw$ for evolved potentials.
However, at $\hw = 12$ we start with a less optimal oscillator
representation of the Hamiltonian and therefore converge to an
under-bound value. Again, this is where a $\hw$ switching tool would
come in useful as discussed in the last section. The $\hw$ dependence
would of course have to be tested for each value of $\lambda$ as
discussed in Appendix~\ref{chapt:app_osc_basis}. In
Fig.~\ref{fig:h3_NNN_Hho_convergence} we show again that the specifics
of the initial three-body force do not alter the qualitative pattern
of evolution. 

\begin{figure}[htb]
\bc
\triplepic{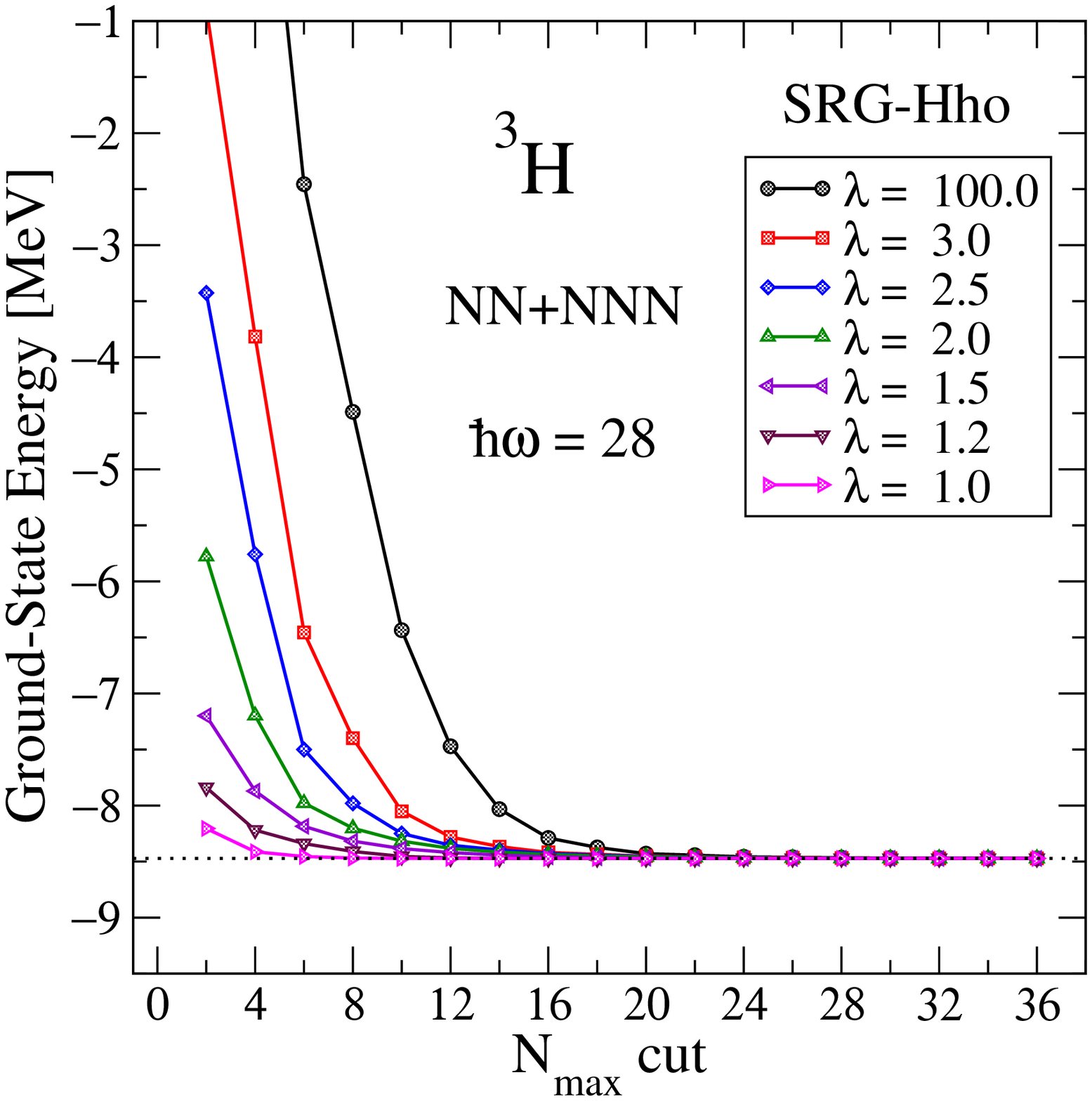}
\hfill
\triplepic{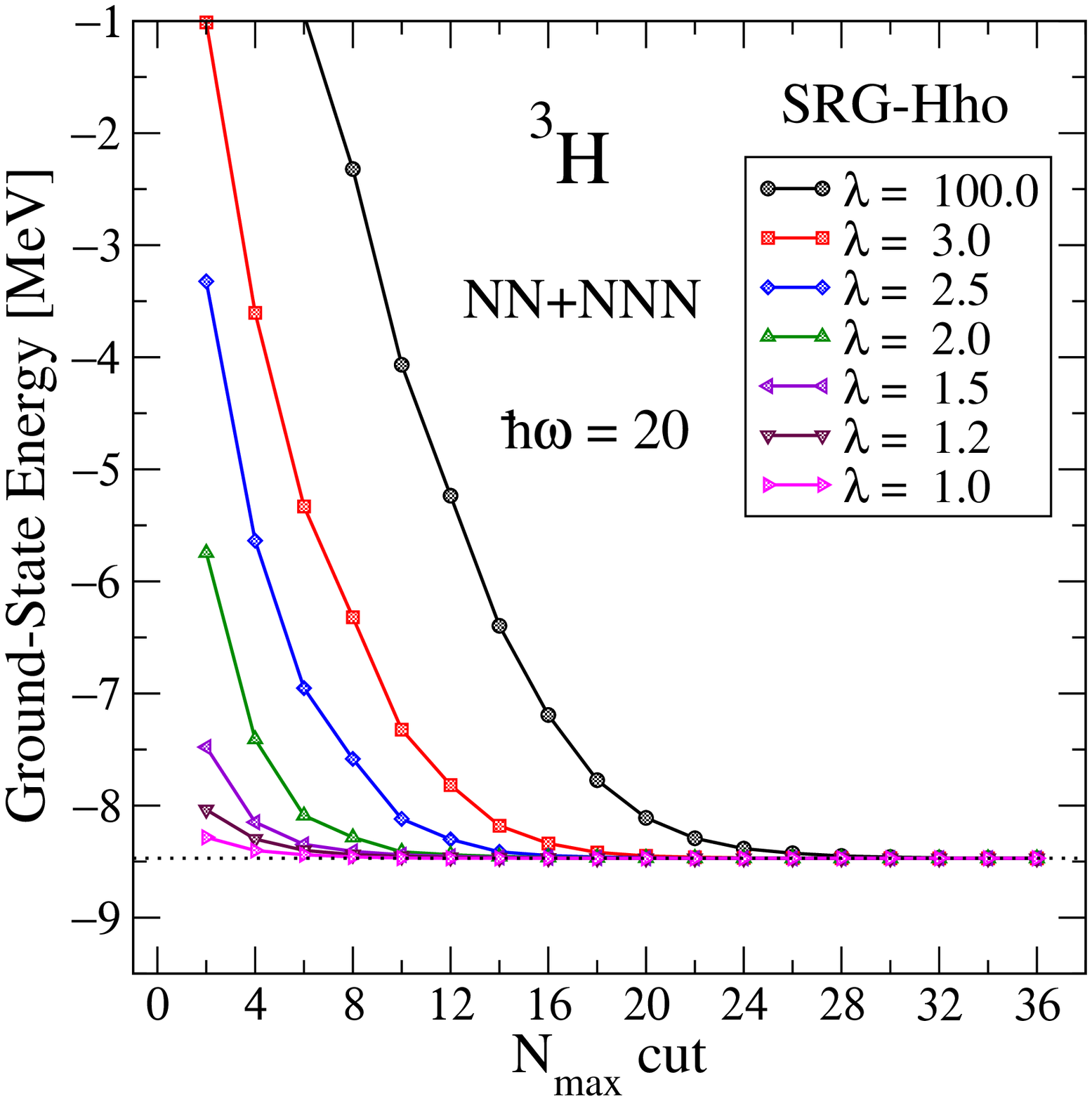}
\hfill
\triplepic{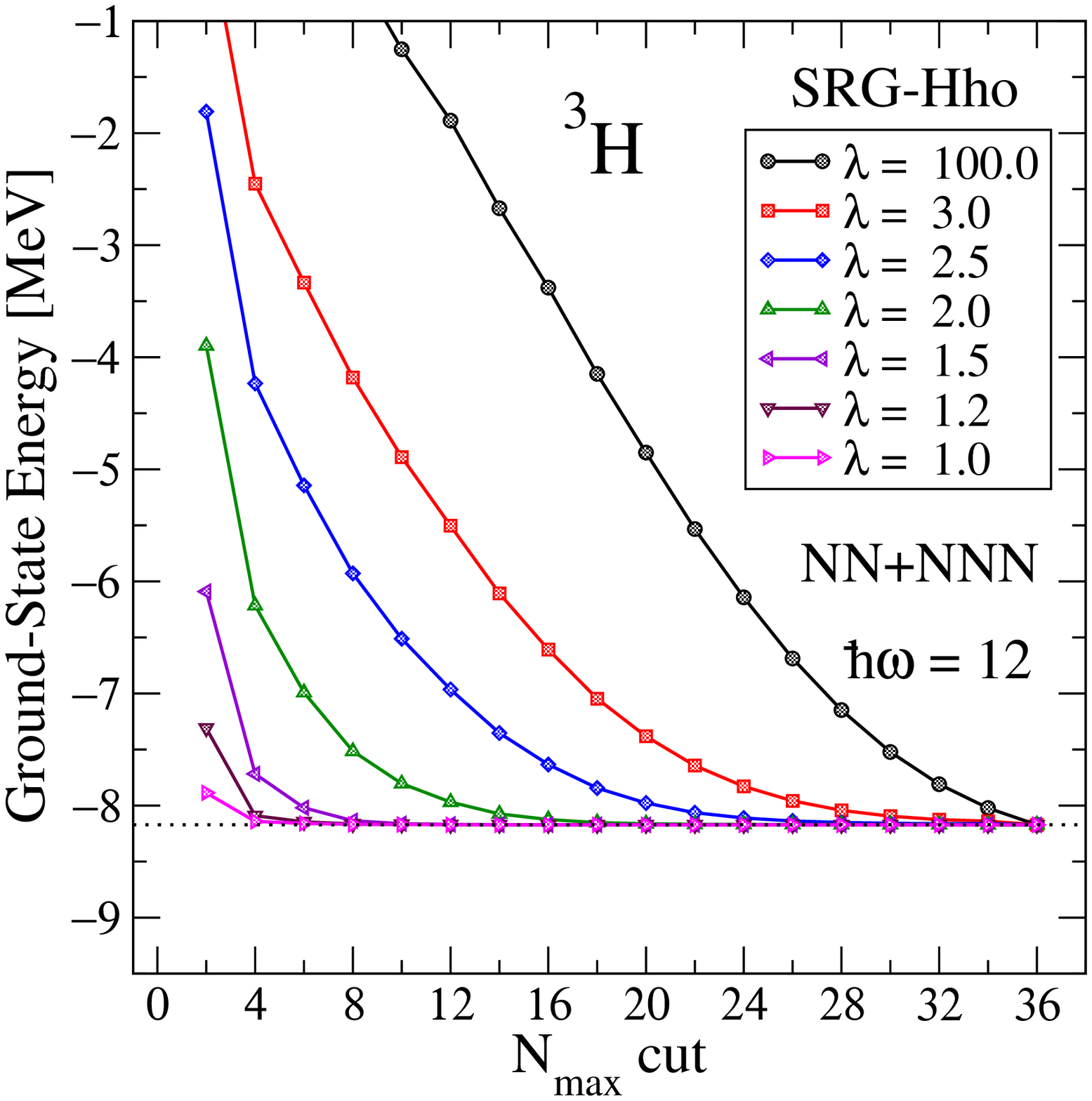}
\ec
\captionspace{Same as in Figure~\ref{fig:h3_Hho_convergence} but now
including the initial NNN force~\cite{Gazit:2008ma,Epelbaum:2008ga}.
Note the qualitative form of convergence is not affected.}
\label{fig:h3_NNN_Hho_convergence}
\end{figure}

Unfortunately, the same spurious states appear here in the
approximately unitary calculations as did in the one-dimensional
version of $G_s = \Hho$, leaving concern about the renormalization by
unnatural operators. Specifically, $\Hho$ contains the long-range
operator $r^2$ which is not present in the Hamiltonian and therefore
one might expect it to induce unnaturally large components in various
parts of the Hamiltonian as it tries to cancel spurious $r^2$ physics. On the
other hand the choice $G_s = \Trel$ does not improve convergence very
well for long ranged operators like, in fact, $r^2$ while the $\Hho$
SRG does. The focus of present investigations~\cite{Anderson:2009b} has
been to consider a hybrid $G_s$ of the form
\beqn
G_s = \Trel + \alpha r^2
\label{eq:hybrid_G}
\eeqn
that is essentially $\Trel$ but incorporates a small admixture of the
$r^2$ operator to renormalize long distance operators without
contaminating the Hamiltonian's flow. This work has progressed well in
the one-dimensional model, and the prospects are promising for such a
hybrid $G_s$ to renormalize both long and short distance operators
without inducing large spurious many-body forces. Calculations to
verify this behavior in three-dimensions have been slowed by coding
complications that will be resolved soon. 


\section{Radius Calculations}

In addition to binding energies, we would like to understand how the
SRG affects other observables, especially long-ranged ones like the
root mean square radius,
\beqn
r_{\rm rms} = \sqrt{\la \phi_0 | \frac{1}{A} \sum_{i=1}^A (x_i - x_{\rm
cm})^2|\phi_0\ra}  \;,
\eeqn
where $|\phi_0\ra$ is the ground-state wavefunction of the $A$-body
system, $x_i$ are the single particle coordinates, and $x_{\rm cm}$ is
the center of mass coordinate. To obtain $ r_{\rm rms}$ in terms of
Jacobi coordinates we can use the identity~\cite{SVM_book}
\beqn
A \sum_{i=1}^A (x_i - x_{\rm cm})^2 \equiv \sum_{j>i=1}^A (x_i -
x_j)^2 = {A \choose 2} (2r_1^2) \;,
\eeqn
where the $x_i$ are single particle coordinates and $r_1$ is the first
Jacobi coordinate. The second equality results from our definitions of the Jacobi
coordinates and the fact that we are working in a symmetrized basis.
This replaces the sum by a multiplicative factor, the total number
of pairs in the $A$-body system.

This calculation is performed directly within the three-dimensional
NCSM code. We can evolve a two-body Hamiltonian and then use the code
to compute $r$ for $^3$H or $^4$He, and we can also evolve the
three-body and embed it to compute $r$ for $^4$He. As of this writing
however, complications with the NCSM code have prevented us from
producing the unitarily evolved version. We cannot yet evolve $A$-body
forces and compute $r$ for the $A$-body nucleus. So, the pictures in
the figures for $^3$H and $^4$He are missing the induced three- and
four-body forces respectively. This is of high priority in the near
future.

\begin{figure}[thb]
\bc
\singlpic{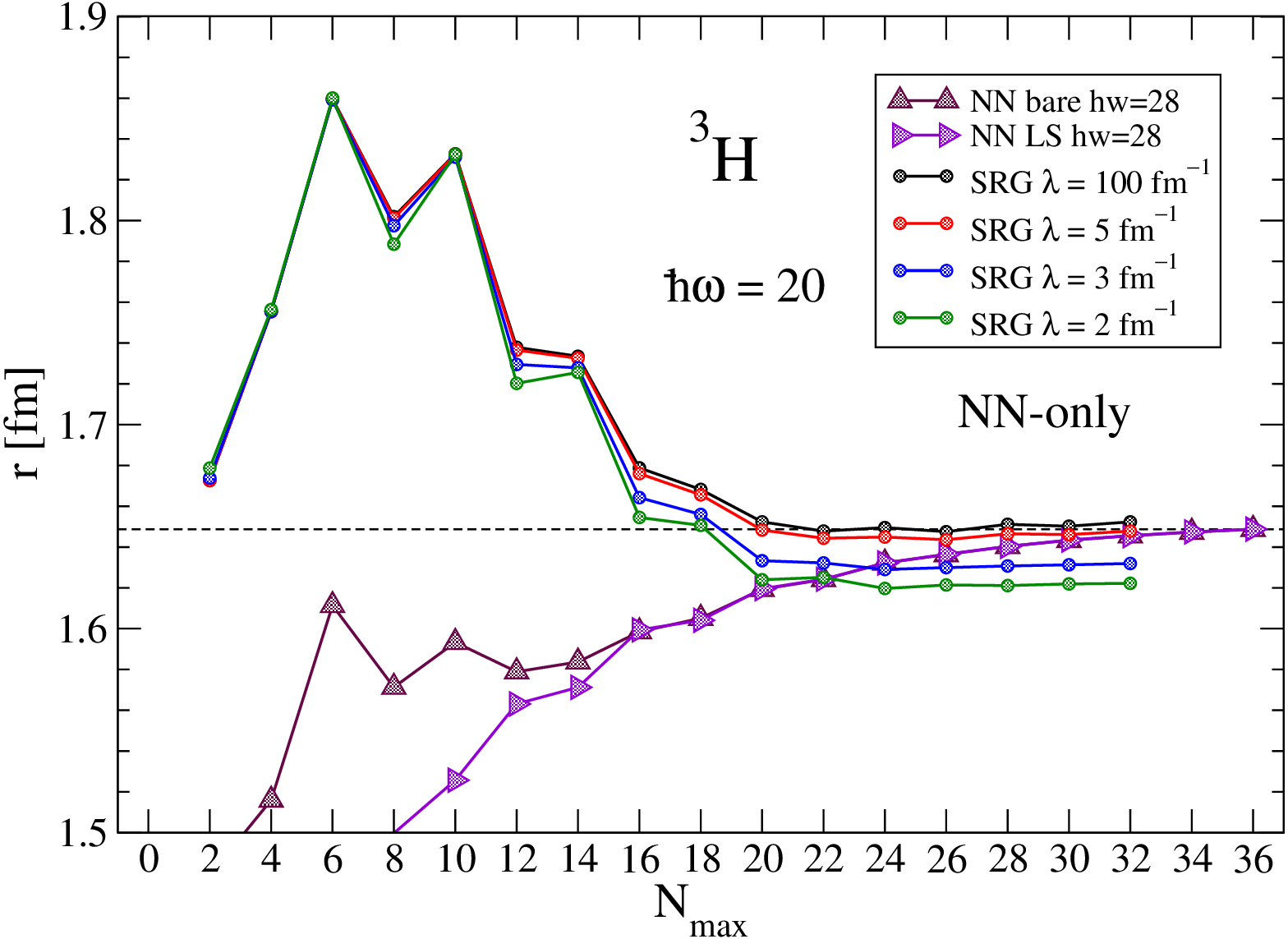}
\singlpic{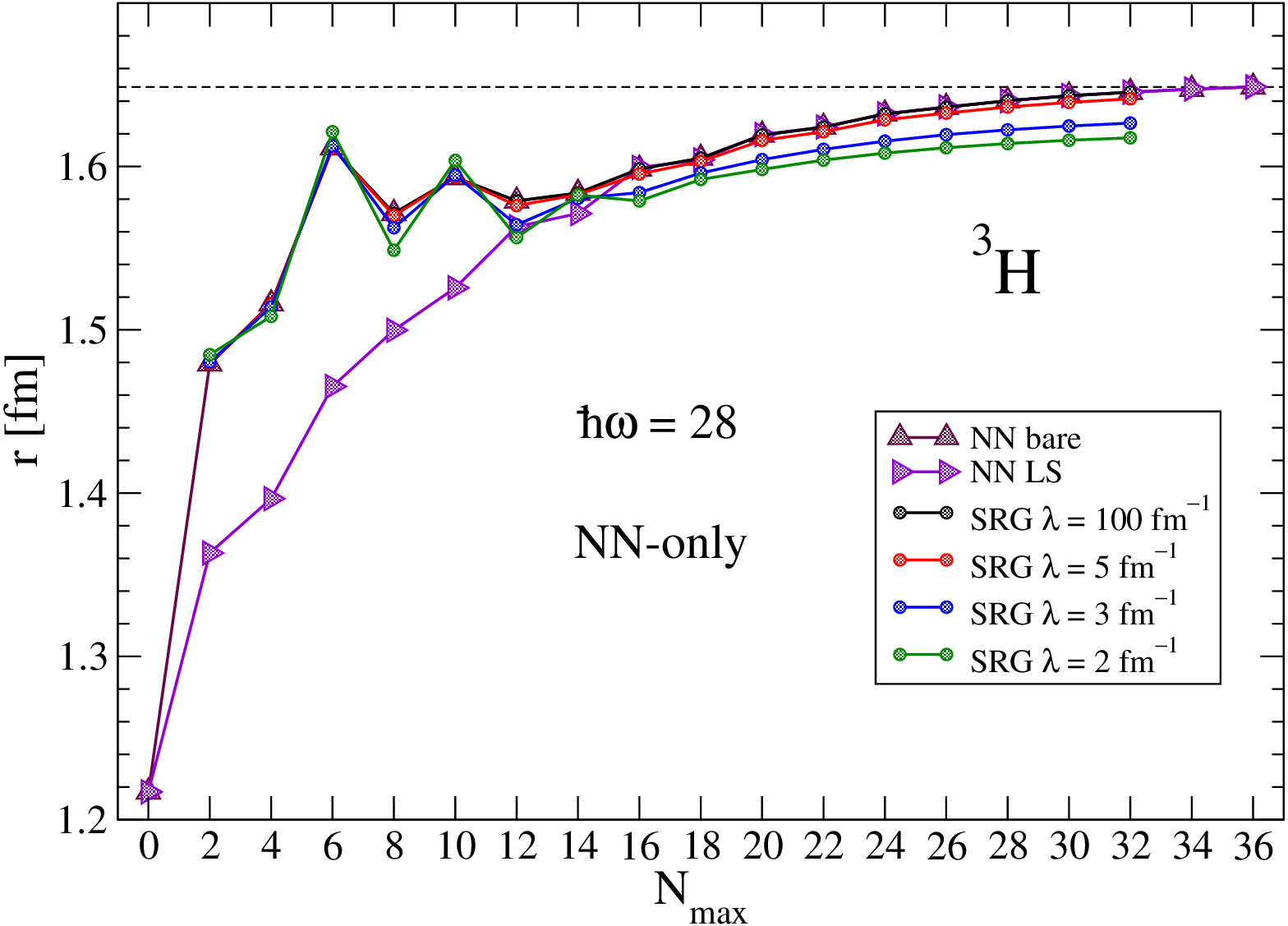}
\ec
\captionspace{The $^3$H radius as a function of basis size $\nmax$ for
two choices of $\hw$, 20, and 28 MeV, corresponding to the optimal
values for the bare and some evolved Hamiltonians. The straight dashed
line indicates the best converged value at the optimal $\hw$ (28 MeV) for the
LS effective potential at $\nmax = 36$.}
\label{fig:h3_radius}
\end{figure}

In Fig.~\ref{fig:h3_radius} we show the proton rms radius of the
triton as a function of $\nmax$ for two values of $\hw$, 20 and 28 MeV
(top and bottom).  We plot several values of $\lambda$ along with the
bare and Lee-Suzuki effective interaction for comparison. Note the
poor convergence in basis size. Unfortunately, evolution with the SRG
does not significantly improve convergence for this long ranged
quantity, though when comparing the two values of $\hw$ the
convergence is improved from 28 to 20 MeV.

\begin{figure}[thb]
\bc
\strip{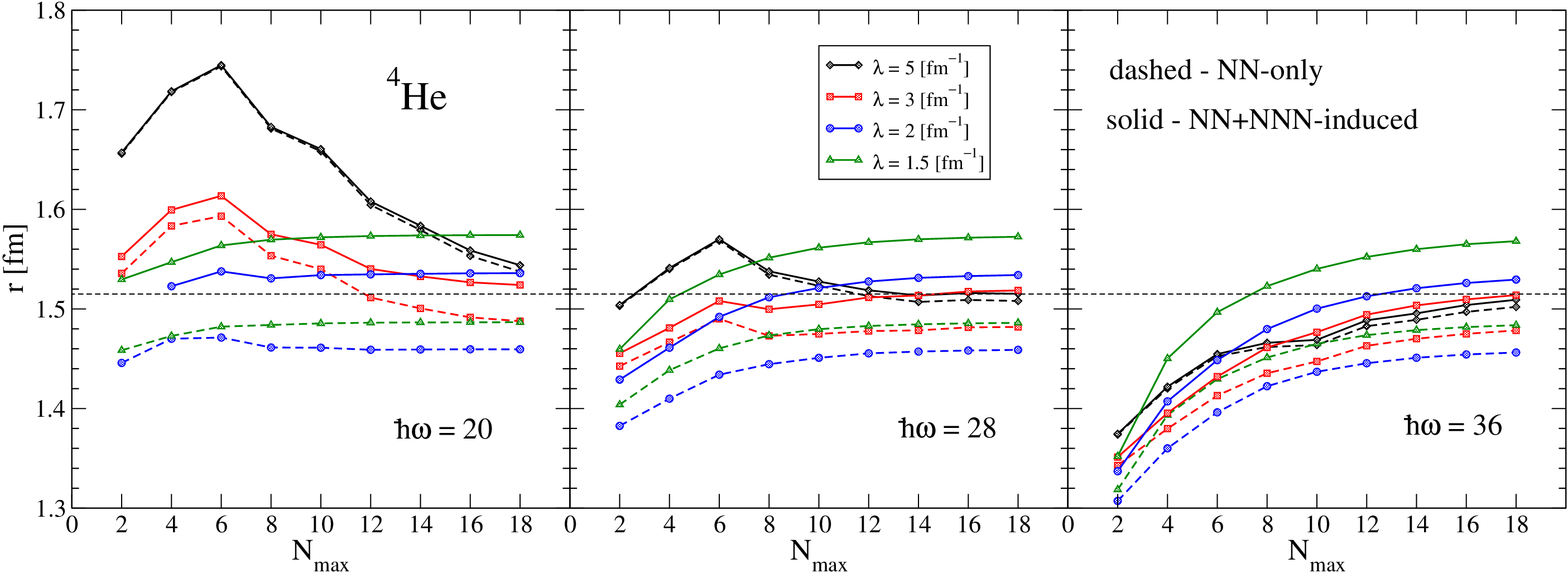}
\ec
\captionspace{The $^4$He radius as a function of basis size $\nmax$
for three choices of $\hw$, 20, 28, and 36 MeV (corresponding to the
optimal values for the bare and some evolved Hamiltonians). The solid
lines are calculations that include induced NNN forces. The dashed
lines are NN-only calculations. The
straight dashed line indicates the best converged value at the optimal
$\hw$ (36 MeV) for the LS effective potential at $\nmax = 20$.}
\label{fig:he4_NN_radius}
\end{figure}

In Fig.~\ref{fig:he4_NN_radius} we show the same radius calculation
for $^4$He but for three $\hw$'s from left to right 20, 28, and 36
MeV, which are optimal values for the evolved and bare Hamiltonians
respectively. Here both NN-only and NN+NNN-induced calculations are
shown, but again the full unitary evolution calculation (including
induced four-body forces) is not yet available. Again there is some
improvement in convergence for lower values of $\hw$. Those bases have
a lower infrared cutoff, $\Lambda_{\rm IR}$, and thus keep more
long-distance information. Decoupling due to the SRG is then more
beneficial to the convergence of the radius calculation.
Unfortunately, the ultraviolet cutoff, $\Lambda_{\rm UV}$, is also
lower and limits the accuracy of the converged binding energy, in turn
limiting the accuracy of the radius. The black and red solid curves at
$\hw = 20$ approach the converged value much slower than at 28 or 36.
At the same time, the green and blue curves are better converged (to
the wrong value) at $\hw = 20$. Of course, a higher $\nmax$ would
help the situation by pushing $\Lambda_{\rm IR}$ down and
$\Lambda_{\rm UV}$ up.

Also, notice in Fig.~\ref{fig:he4_NN_radius} the spread in converged value
(at $\nmax = 18$) as a function of $\lambda$. Here we present the bare
operator acting on the evolved four-body wavefunctions. Not only are
the effects of induced three- and four-body forces on the wavefunction
contributing to the spread, but the evolution of the operator,
necessary to preserve unitarity in this calculation, is not included.
At $\hw = 20$, this seems to be a much bigger effect than that of
the three- and four-body parts the evolved wavefunction,
considering a lack of any clear hierarchy between the intervals from
the NN-only to the NN+NNN-induced and the NN+NNN-induced to the
converged value. At the higher $\hw$'s, the $r^2$ operator is evolved
much less though the converged value is more accurate. Here the three-
and four-body hierarchy in the wavefunction is more evident.

As discussed in the previous section, a possible strategy
around this has focused on using a hybrid generator, $G_s = \Trel +
\alpha r^2$, where $\alpha$ is a constant determining the mixture of
the two terms. With this choice, one hopes to balance the
renormalization of short- and long-range parts of the potential and
increase convergence for all operators~\cite{Anderson:2009b}. Results
of this work in one-dimension are promising with regard to the ability
to tune the SRG to given operators and basis structure. As noted in
several places, progress in the three-dimensional code has been slowed
by coding issues. Most of these relate to the same technical issue of
explicitly building the $r^2$ operator in the NCSM basis. Upon
resolution of this problem, many more calculations will become
available quickly.



It must be stressed that difficulties in finding an optimal SRG
generator stem not from the SRG itself, but from the particular basis
we have been working in. The harmonic oscillator basis is convenient
to work in, both because it is variational in two parameters with
straightforward cutoff implications, and because the center of mass
separates trivially from the calculations. However, it is not the best
choice for describing certain physical properties, especially
long-ranged operators. The oscillator functions simply do not have the
same fall-off as an operator like $r^2$. Other bases that may prove
useful to this work include hyper spherical harmonics, and a basis of
correlated gaussians as used in the Stochastic Variational Method
(SVM)~\cite{svmcpc}. These bases should be more efficient at resolving
long-distance observables, such as nuclear radii, and the choice of
SRG generator may not be as constrained. However, work using these
bases has not been as developed as that of the NCSM, especially work
to implement the SRG, but progress is on the horizon.



\chapter{Concluding Remarks}
\label{chapt:conclusion}

\section{Recapitulation}


For many decades progress in nuclear structure theory has been
hampered by the complexity of the nuclear interaction. A strong
repulsive core, significant tensor force, and poorly understood
many-body forces have all contributed to difficulty in accurately
calculating nuclear observables consistently. These features are
associated with high-momentum/short distance structure of the nuclear
interaction, the details of which are irrelevant to the low-energy
structures of bound nuclear systems. More precisely, infinitely many
different choices of high-energy details can produce the same
low-energy results. The difficulty in many-body calculations stems
from the fact that high- and low-energy states are coupled. These
coupling matrix elements of the potential require an increased basis
size to represent the relevant low-energy physics. Such extra size
increases dramatically with the number of particles in the system
under consideration. This situation is ripe for a method, such as the
SRG, that damps high-energy details without adverse effects to the
physics relevant to low-momentum interactions.

In recent years many inroads have been made into the problem of
expensive computation being spent on irrelevant interaction
components. Foremost is the development of \xefts, a model independent
formulation of pion exchange interaction theory in the mold of Yukawa,
but now informed by the underlying principles of QCD. Among its
successes, these potentials are able to parametrize many-body forces
consistently and robustly with controlled errors at each order.
However, while they are much softer than their phenomenological
predecessors, they are still computationally expensive for computing
all but the lightest nuclei. Also of note is the $\vlowk$
renormalization technique, which has provided a proof of principle that
one can unitarily transform a given potential to decouple the
irrelevant high-momentum degrees of freedom and retain the relevant
physics inside a smaller space. However, this technique has not left a
clear path to the consistent treatment of many-body forces inherent in
the \xeft\ (or any other) potential on which it is used.

The Similarity Renormalization Group has proved to be a powerful tool
for addressing the above general problems. Its flowing unitary
transformations leave the input physics unaffected while transforming
the potential to a softened form. Implementation of the SRG is simple
in any basis and on any given free-space Hamiltonian while allowing
freedom of choice as to the general form of the flow operators.
Furthermore, the SRG provides a self-consistent treatment of many-body
forces easily understood in either first or second quantized forms. It
has been demonstrated to be robust in its scaling of both evolution
times and residual errors. With these features, the SRG is well suited
to address the needs of the nuclear structure community. 


The predominant improvement provided by SRG evolution of a given
potential is the decoupling between low- and high-energy states in the
two-nucleon (NN) interaction as explored in
chapter~\ref{chapt:decoupling}. The degree of decoupling achieved by
evolution is studied by measuring errors induced by an arbitrary
cutoff on evolved potentials. In this way we can see the clean
perturbative scaling of the residual coupling above a given amount of
evolution denoted by the momentum scale $\lambda$ or $s =
1/\lambda^4$. Decoupling in the two-body sector was found to propagate
universally to many-body calculations and was checked explicitly for
few-nucleon systems. In addition, an investigation into the behavior
of other SRG forms, set by the choice of the generator $G_s$, showed
no significant impact on the general decoupling properties expected at
a given $\lambda$.


The formal discussion in section~\ref{sec:decoupling_mechanics} shows
that, in principle, the same mechanism of decoupling also improves the
convergence properties of the three-body (and higher) forces in an
appropriate three-body ($A$-body) basis. In addition the SRG induces
three-body forces to account for the changes in the two-body potential
and keep the evolution fully unitary in the three-body system. In
fact, it induces $A$-body forces in an $A$-particle space and we must
verify that these induced forces can be controlled in order for the
two-, three-, and four-body evolved potentials to be useful to
many-body calculations.

These issues were explored in chapter~\ref{chapt:OneD} through the use
of a one-dimensional model in a symmetrized harmonic oscillator basis
inspired by the highly successful No-Core Shell Model. We demonstrated
that the induced three-body forces were as large as expected from
previous calculations, and that induced four- and higher-body forces
exhibited a hierarchy of decreasing importance. We covered several
other topics relevant to realistic calculations such as the
convergence properties of the three-body evolution, the possibility of
fitting omitted induced forces (thereby avoiding an explicit evolution
in a higher-body space), and evolution of external operators involved
in scattering processes. All of these calculations serve to build
experience and intuition for the realistic three-dimensional case
where angular momentum and other details complicate the calculations.


In chapter~\ref{chapt:ncsm} we achieved the first evolution of
realistic three-body forces using the antisymmetrized Jacobi harmonic
oscillator basis of the No-Core Shell Model. Here we duplicated the
calculations tested in one-dimension, showing induced three- and
four-body forces as a function of the evolution parameter, $\lambda$,
and studied the improved convergence in $^3$H and $^4$He energies due
to the decoupling achieved in the three-body basis. The same hierarchy
of induced forces was exhibited here. We also took a look at a
long-range observable, the rms radius of these nuclei, to extend work
being done by others in one dimension.

Our experience from the 1D calculations guided progress rapidly and
that model continues to be a testing ground for new calculations. All
the features of many-body evolution observed in the one-dimensional
case were born out semi-quantitatively with the realistic
interactions. For instance, all of the induced three- and four-body
forces were in the same proportions to those in one-dimension. Often
one is wary of trusting the results of a model calculation in one
dimension to provide qualitative predictions of three-dimensional
physics. Here, the basis structure is very similar to that of the
three-dimensional problem which is already broken up in a partial-wave
expansion. The SRG is only affected by the general form of the
matrices, Hamiltonians and kinetic energies, involved in its flow.

\section{Plans for Future Investigations}

A broad range of topics should be considered, or reconsidered, in
light of the advantages presented by the Similarity Renormalization
Group. Here we discuss a few of the most pressing issues.


While this thesis has shown the induced many-body forces to follow a
decreasing hierarchy in many-body calculations, they will continue to
be a source of error in all sectors from light nuclei all the way up
to nuclear matter. Interactions evolved in a three- or four-nucleon
basis are unitarily equivalent for their respective nuclei. However,
the use of those potentials, embedded to higher-body spaces, is
complicated by the $A$-body forces that were not induced in the
evolution at a lower space. For example, in chapter~\ref{chapt:ncsm}
results showing increased convergence in $^4$He also show a small
variability, with $\lambda$, in the converged value. This cutoff
dependence is small in this calculation, smaller than the error with
respect to the experimental value, but that space is only one nucleon
higher than that in which the SRG was applied. More calculations are
needed to test how much the cutoff dependence will be reduced by
including just the induced three- and four-body forces.


A first priority to achieve calculations of larger nuclei is to make
the $A=3$ evolved potentials available in a usable form to those
groups who can perform calculations of larger nuclei. The coupled
cluster community uses the single-particle Slater determinant basis to
perform their calculations. They have so far achieved much success
pushing the boundaries of their calculations with NN-only evolved
potentials, reaching as high as $^{56}$Ni. Likewise, NCSM calculations
which use the Slater determinant basis can also reach much higher
nuclei and would be complementary calculations. Codes already exist to
convert the matrix elements in the Jacobi basis used here to the
Slater determinant basis, or ``m-scheme", so progress on this front
should be rapid.


While induced many-body forces in all sectors will be a source of
error, this thesis has shown them to follow a decreasing hierarchy in
many-body calculations. To avoid applying SRG evolution in
increasingly large $A$-body bases, one might control the cutoff
dependence by fitting the omitted induced forces to terms inspired by
the \xeft\ expansion. An implementation of this idea was made in
chapter~\ref{chapt:OneD} with encouraging results and further progress
should come easily. Through such a fitting program the pattern of
induced many-body forces may even lend itself to extrapolation and
increased control on the errors present in approximately unitary
calculations.


Another avenue currently under development to test the viability of
SRG evolved interactions in larger nuclear systems is to calculate
nuclear matter properties using the Hartree-Fock approximation. Here
the SRG is applied in the second quantized form demonstrated in
section~\ref{sec:second_quantization}. Instead of evolving the
free-space interaction, the SRG is applied in the medium where normal
ordering has different consequences with respect to the induced
many-body forces. The induced many-body forces can then be controlled
in a qualitatively different way than the free-space calculations.


As mentioned previously the major goal of the UNEDF collaboration is
the development of a density functional for nuclei which incorporates
microscopic inputs such as those from \xefts. In this ongoing
project, the SRG is playing a vital role by softening initial
interactions~\cite{dft_review}.


Fundamental to the idea of convergence within a particular basis size
is the question of correspondences between various cutoffs in
different bases. While the cutting done in the momentum representation
in chapter~\ref{chapt:decoupling} is a straightforward
ultraviolet cutoff, the same procedure in the oscillator basis
requires a different interpretation as briefly discussed in
appendix~\ref{sec:cutoffs}. Furthermore, many basis choices exist,
each being more suitable to a different physics task. While the NCSM
is a relatively straightforward and familiar basis, other bases like
hyperspherical harmonics and correlated gaussians are promising
choices for describing other nuclear features. In fact, an
implementation of the SRG in three-body momentum representation is
currently under development, and should provide a check on the
evolution done here in the NCSM~\cite{lplatter_pc}. 


One of the major advantages of \xefts\ is their natural framework for
the inclusion of external operators, for example electromagnetic
interactions. The SRG can also be used for consistent, independent,
evolution of these and other operators. Work on this topic is already
proceeding~\cite{Anderson:2009a} and involves, among other things, the
one-dimensional tool developed for chapter~\ref{chapt:OneD} for
preliminary studies of operator evolution in a many-body space.


While reproducing the static properties of nuclear bound states is an
important task for testing our quantitative understanding of the
nuclear interaction, we must address the dynamics associated with
reactions. One current avenue is to couple the NCSM technique with the
Resonating Group Method, which allows access to information on
scattering amplitudes between states within the NCSM basis. These
calculations will contribute greatly to the knowledge of how nuclei
interact with one another and are a vital diagnostic tool for many
experimental setups, especially the inertial confinement fusion
experiments coming online at the National Ignition Facility (NIF) at
Lawrence Livermore National Laboratory. This work is an exciting new
direction for research using the SRG.


The SRG can be applied term by term, as demonstrated by the vacuum
expectation value analysis in section~\ref{sec:oned_vev}, to explore
the interplay between different parts of the flow equations. This
analysis needs to be continued in the realistic NCSM where we have
access to the different contributions to the three-body force in
\xeft. Further down the road, one can envision an analytical
application of the SRG to gain general insight to the SRG's treatment
of \xeft\ potentials (or any other physics one wishes to address). The
diagrammatic decomposition of SRG evolution would inform and
provide checks on the results of such work, but understanding the
effects of SRG evolution on individual coupling constants may have
significant advantages to the formulation of \xefts. Specifically, an
analytical understanding of the SRG's behavior may provide a practical
means to extrapolate its behavior with regard to induced many-body
forces.


The Similarity Renormalization Group has enjoyed a meteoric rise in
its provision of benefits to nuclear few- and many-body problems.
This thesis has provided a review of some of its basic features and
initial applications to address calculations. It appears much more is
to come.

\startsinglespace

\startdoublespace

%
%
\appendix
\chapter{Chiral Effective Field Theories}
\label{chapt:app_eft}


The original efforts to build Chiral Effective Field Theories (\xefts)
with more than one nucleon were based on a suggestion by Weinberg
\cite{weinberg_eft} to use a non-perturbative treatment of the power
counting rules from the perturbative Chiral Perturbation Theory
($\chi$PT). A non-perturbative development is necessary in order to
describe nuclear bound states. Weinberg used a $\chi$PT power counting
scheme to build the nuclear potential and used the Lippmann-Schwinger
equation to iterate it non-perturbatively.

The motivation for this non-perturbative scheme is the reproduction of
the deuteron binding energy, scattering lengths, and other low energy
observables. The above discussion of renormalization scales and
momentum contributions is important to this, but it does not solve the
bound state problem completely. Naive dimensional analysis (NDA) can
only give an estimate of the effective coupling constants, referred to
as Low Energy Constants (LECs). Only a careful calculation and
renormalization procedure can fix their values and even signs. The
power counting rule used here is based on an assumption of
naturalness. That after renormalization, the LEC's will be of order
one. Also, the LECs are assumed to serve as counter terms to loops in
the irreducible diagrams. In practice, in \xeft\ these absorptions are
achieved by a fit to data.

Shortly after Weinberg's proposal there was an attempt to develop a
perturbative treatment of pions in an EFT. This was formally worked
out by Kaplan, Savage, and Wise \cite{KSW}, and is known as KSW power
counting. While still non-perturbative in the contact interactions,
the authors conjectured that the pion dynamics might be weak enough to
treat perturbatively. Unfortunately, it turned out that the breakdown
scale in this approach is too low to be of practical use in nuclear
structure, though new variations have been proposed~\cite{KSW_new}. This
approach will not be discussed here further.


\section{Basics of power counting}

To explain the method of power counting due to Weinberg and used in
Ref.~\cite{Epelbaum_05}, we will start by writing out part of  a
chiraly symmetric Lagrangian for \xeft\ 
\bea
\mathcal{L}^{(0)} &=& \frac{1}{2} \partial_\mu \fet \pi \cdot \partial^\mu \fet \pi  
- \frac{1}{2} M^2 \fet \pi^2 + N^\dagger  \bigg[i \partial_0 +
\frac{g_A}{2 F_\pi} \fet \tau \vec \sigma \cdot \vec \nabla \fet \pi 
- \frac{1}{4 F_\pi^2} \fet \tau \cdot ( \fet \pi \times \dot{\fet \pi } ) \bigg] N  \nonumber \\ 
&&{} - \frac{1}{2} C_S ( N^\dagger  N  )  ( N^\dagger  N  )  - \frac{1}{2} C_T  
( N^\dagger \vec \sigma N )  ( N^\dagger \vec \sigma N )+ \ldots  \;, \nonumber \\  [0.5ex]
\mathcal{L}^{(1)} 
&=& N^\dagger  \bigg[ 4 c_1 M^2  
-\frac{2 c_1}{F_\pi^2} M^2 \fet \pi^2 + \frac{c_2}{F_\pi^2} \dot{\fet \pi}^2  
+\frac{c_3}{F_\pi^2} (\partial_\mu \fet \pi \cdot \partial^\mu \fet \pi) \nonumber \\
&&{} -  \frac{c_4}{2 F_\pi^2} \epsilon_{ijk} \, \epsilon_{abc} \,
 \sigma_i \tau_a (\nabla_j \, \pi_b ) (\nabla_k \, \pi_c ) \bigg] N  \nonumber \\ 
&& {} - \frac{D}{4 F_\pi} (N^\dagger N)  (N^\dagger \vec \sigma \fet \tau N)
 \cdot \vec \nabla \fet \pi - \frac{1}{2} E \, (N^\dagger N) (N^\dagger \fet \tau N) 
 \cdot (N^\dagger \fet \tau N) + \ldots \;. 
\label{eq:lagrangian}
\eea
where the superscripts refer to the chiral dimension, $\Delta_i$, of a
vertex $i$. The operators $\sigma$ and $\tau$ are Pauli spin and
isospin matrices. The epsilons are Levi-Civita symbols necessary to
achieve proper tensor multiplication. The constants $F_\pi$ and $g_A$
are associated with basic pion interactions and are determined
elsewhere\footnote{$F_\pi$ and $g_A$ are parametrizations of weak
nuclear interactions between pions and nucleons. They can be measured
in processes such as pion decay and nuclear $\beta$ decay.}. The LEC's
here are $c_i$, $C_{S,T}$, D, and E. Notice E is the LEC corresponding
to the first 3NF contact term. This Lagrangian is complete up to
$N^2LO$ including only isospin symmetric terms.

In this power counting approach we calculate the T-matrix directly via
the Lippmann-Schwinger equation (given here schematically),
\beqn
T = V + VG_0T \;,
\label{eq:LS}
\eeqn   
where we will define the potential, $V$, to be the sum of all
irreducible diagrams up to a certain order in $\nu$ as defined below.
An irreducible graph is defined as one which has no purely
nucleonic intermediate states. The need for this definition will
become apparent momentarily. The intermediate state Green's function,
$G_0$ will play a crucial role in the power counting and will be
defined also.

All irreducible graphs can be categorized by their contribution $\sim
O(Q/\Lambda)^\nu$ governed by the power counting rule
\beqn
\nu = 2(N + L - C - 1) + \sum_i V_i \Delta_i \hspace{.5 cm} 
    \mathrm{and} \hspace{.5 cm} \Delta_i = d_i + n_i/2 - 2 \;.
\label{eq:nu}
\eeqn
The variables N, L, C, $V_i$, $\Delta_i$, $d_i$, and $n_i$ are the
numbers of nucleons, loops, connected pieces, vertices of type $i$,
chiral dimension of vertex type $i$, derivatives or factors of $m_\pi$
at vertex $i$, and nucleon lines at vertex $i$ respectively. At any
given order only a finite number of time-ordered irreducible graphs
are needed to compute the exact potential at that order. Chiral
Symmetry sets a lower bound at $\Delta_i \geq 0$. The simplest graph we
can have is one with no nucleons and two pions (pion propagator $\sim
m_\pi^2$) or two nucleons and one derivative provided by a pion
interaction or four nucleons and no pions (contact interaction). Those
are the lowest-order options in the Lagrangian in
Eq.~\eqref{eq:lagrangian}. Given the bounds on $\Delta_i$, the lower
bound on $\nu$ is 0, thus we have a lowest order in the expansion and
a valid power counting scheme.

Let us apply Eq.~\eqref{eq:nu} to a few example graphs from
Fig.~\ref{fig:EFT_hierarchy} in section~\ref{sec:intro_eft}. Take the
simple 1PE graph at LO. The graph has two vertices, each with two nucleon
lines and one derivative (from pion interaction) giving $\Delta_i=0$.
There are 2 nucleons in this graph, no loops, and 1 connected piece.
The grand total is $\nu=0$ (LO) which matches with the table. For a
different example, consider the 1PE-contact graph for a 3NF at
$\nu=3$ ($N^2LO$). There are two vertices, one with two nucleon lines and one
pion based derivative gives $\Delta_i=0$, and the other with four
nucleon lines and the pion interaction derivative (the term with the
LEC ``D'') gives $\Delta_i=1$. So, with the sum over vertices giving
one, three nucleons, no loops, and one connected piece we have a total
of $\nu=3$. So, all graphs are organized according to their
contribution $Q^\nu$ given by Eq.~\eqref{eq:nu} and
Eq.~\eqref{eq:lagrangian}. Note that no contributions survive of order
$\nu = 1$.

The derivation of Eq.~\eqref{eq:nu} appears in many forms throughout the
literature \cite{epelbaum_98,epelbaum_04,Bedaque_vanKolck}. Start by
considering a general graph and sum up momentum contributions via
naive dimensional analysis.
The momentum contribution, $Q$, for that graph is 4 for each
loop integration ($L$) , $-2$ for each internal pion
($P$), $-1$ for each internal nucleon ($I$), $+1$ for each vertex derivative($D$). We obtain the
expression
\beqn
\nu = 4L - I - 2P + \sum_i V_id_i \;.
\eeqn
We can change the form of this using the well-established topological
identities
\beqn
L=P+I-\sum_iV_i+1 \;,
\eeqn
and
\beqn
2I+E=\sum_iV_in_i \;,
\eeqn
where $n_i$ is the number of nucleon lines at vertex of type $i$ and E
is the number of external nucleon legs ($E=2N$). The particular form
of Eq.~\eqref{eq:nu} presented here also requires some non-trivial
transformations associated with the formulation of graphs in an energy
independent way \cite{epelbaum_98} such as the method of unitary
transformations that separates the purely nucleonic part of the Fock
space from the rest~\cite{method_unitary_trans}. Finally one arrives
at the convenient expression of Eq.~\eqref{eq:nu} for $\nu$. The
variables N, L, and C are easy to count and $\Delta_i$ provides us
with a way of organizing vertex types as discussed above.

\section{Many-body forces}

One of the beautiful aspects of using field theoretic techniques to
describe the nuclear interactions is the consistent and
straightforward generalization to few and many body forces. In
chapter~\ref{chapt:introduction}, we briefly mentioned why few-body
forces are guaranteed to exist in the \xeft\ expansion. We argued that
the renormalization intrinsic to an EFT is responsible for integrating
out the high-energy effects, so that in a few-body problem, certain
components of the interaction will be hidden from the experiment (i.e.
they will occur at distance scales much smaller than we can resolve).
These effects can be divided roughly into three types. First, the
$\Delta$ resonance can play a role in the interactions and is
discussed in the next section. Second, simple high-momentum nucleonic
intermediate states like a box diagram with a large momentum running
around the loop. Third, relativistic corrections which modify the
kinetic energy of the intermediate nucleon states (i.e. through pair
production) allow for irreducible diagrams with high-energy
intermediate states that are (at low lab energy) highly virtual and
will occur over very small distances. As stated above, these are the
major contributions that cannot be resolved at the relevant lab
energies and therefore must be renormalized into a 3N contact force.
Thus a 3NF appearing in the EFT Lagrangian is inevitable due to any of
the above three effects.

A simple NDA consideration leads us to expect that the basic 3NF terms
should become relevant at $\nu = 3$. To see this, consider the 3NF
contact term compared to the NN contact term at LO ($\nu = 0$). All
other things being equal, the only difference should be that the 3NF
has an extra pair of nucleon legs ($NN^\dagger$) each of order 3/2 for
a total of 3 extra orders of the momentum scale Q. Thus we should
expect 3NFs to become relevant at $\nu = 3$.

\begin{figure}
\bc
\strip{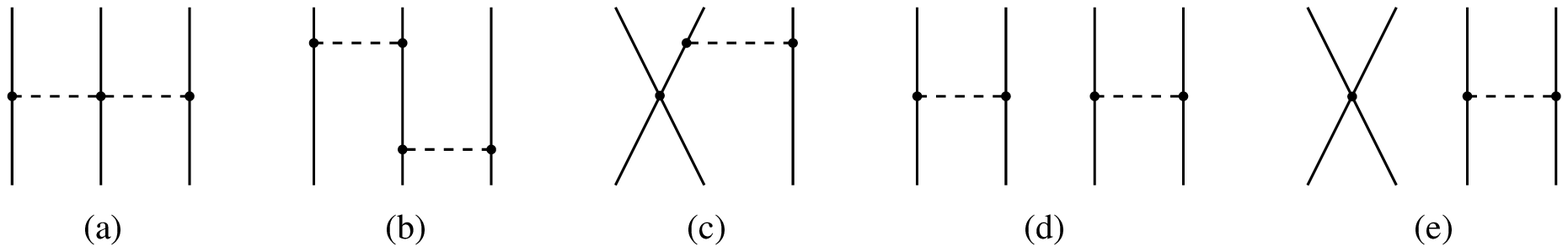}
\ec
\captionspace{3 and 4 nucleon amplitudes that occur at $\nu=2$. These
terms will cancel each other out in an energy-independent
formulation. Figure courtesy of Ref.~\cite{Epelbaum_05}.}
\label{fig:34NFs_nu2}
\end{figure}

If we use the power counting of Eq.~\eqref{eq:nu} to build the
simplest three- and four- body diagrams we find the first
contributions appearing at $\nu = 2$ which seems to contradict the
above NDA argument. However, these first 3NF and 4NF terms (shown in
Fig.~\ref{fig:34NFs_nu2}) will not actually contribute. If we draw out
all possible graphs from the possible topologies at this order we will
find that they cancel each other out. This vanishing is evident when
we formulate the nuclear force in an energy-independent
way~\cite{tamm_dancoff,okubo}.

\begin{figure} 
\begin{center}
\strip{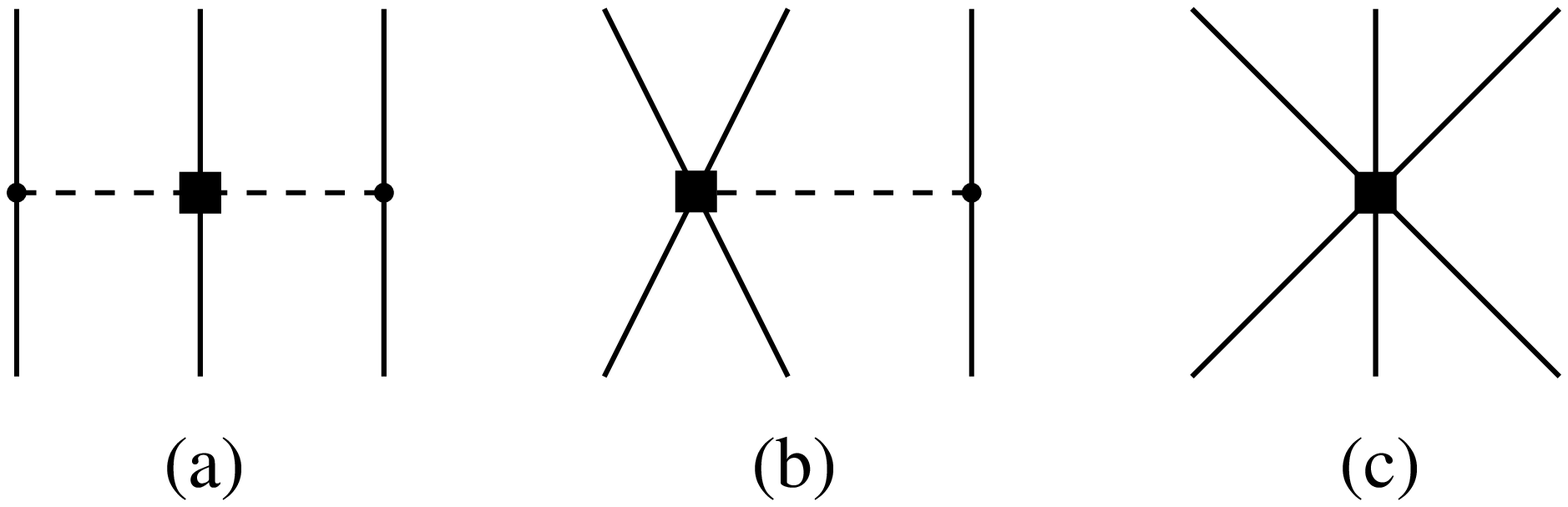} 
\end{center}
\captionspace{The leading order 3NFs at $\nu=3$. Figure courtesy
of Ref.~\cite{Epelbaum_05}}
\label{fig:34NFs_nu3}
\end{figure}

Finally, at $\nu = 3$ we will find the leading 3NF terms.  With $E=3$,
$L=0$, $C=1$, and one $\Delta=1$ vertex we can draw all such diagrams
as shown in Fig.~\ref{fig:34NFs_nu3}.  The contributions to the
potential from these three graphs are given by \cite{Epelbaum_05}
\beqn
\label{3nftpe}
V^{\rm (3)}_{\rm 2 \pi}=\sum_{i \not= j \not= k} \frac{1}{2}\left(
  \frac{g_A}{2 F_\pi} \right)^2 \frac{( \vec \sigma_i \cdot \vec q_{i}
  ) 
(\vec \sigma_j \cdot \vec q_j  )}{(\vec q_i\, ^2 + M_\pi^2 ) ( \vec
q_j\, ^2 + M_\pi^2)}  F^{\alpha \beta}_{ijk} \tau_i^\alpha 
\tau_j^\beta ,
\eeqn
with
\beqn
F^{\alpha \beta}_{ijk} = \delta^{\alpha \beta} \left[ - \frac{4 c_1
    M_\pi^2}{F_\pi^2}  + \frac{2 c_3}{F_\pi^2}  
\vec q_i \cdot \vec q_j \right] + \sum_{\gamma} \frac{c_4}{F_\pi^2} \epsilon^{\alpha
\beta \gamma} \tau_k^\gamma  
\vec \sigma_k \cdot [ \vec q_i \times \vec q_j  ],
\eeqn
and
\beqn
\label{3nfrest}
V^{\rm (3)}_{1 \pi, \; \rm cont} = - \sum_{i \not= j \not= k} \frac{g_A}{8
  F_\pi^2} \, D \, \frac{\vec \sigma_j \cdot \vec q_j }{\vec q_j\, ^2
  + M_\pi^2}  
\, \left( \fet \tau_i \cdot \fet \tau_j \right) 
(\vec \sigma_i \cdot \vec q_j ),
\eeqn
and
\beqn
V^{\rm (3)}_{\rm cont} = \frac{1}{2} \sum_{j \not= k}  E \, ( \fet \tau_j \cdot \fet \tau_k ) \;,
\eeqn
where $i,j,k$ refer to nucleons, and $\vec q_i \equiv \vec p_i \, ' -
\vec p_i$; $\vec p_i$ ($\vec p_i \, '$) is the initial (final)
momentum of the nucleon $i$. These will involve vertices with
$\Delta_i = 1$. These expressions for the potential look complicated,
but they must reproduce the structure of the pion interactions that
they are parametrizing. Thus the intricate structures of spin and
isospin matrices.

\begin{figure} 
\begin{center}
\strip{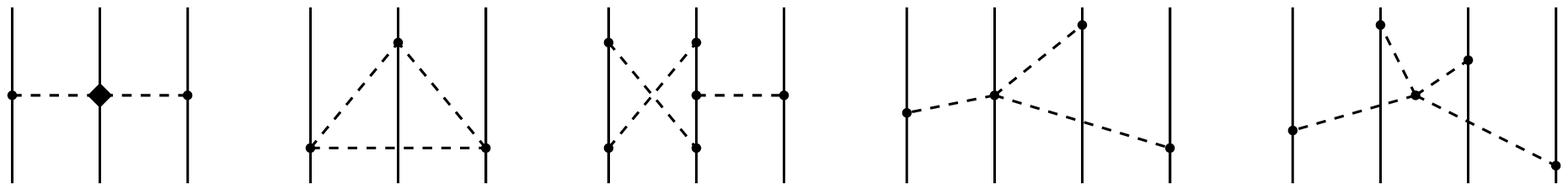}
\end{center}
\captionspace{Sample 3NF corrections and leading 4NFs at $\nu=4$. A more
complete list of 4NFs is in Fig.~\ref{fig:4NFs_nu4}. Figure courtesy
of Ref.~\cite{Epelbaum_05}}
\label{fig:34NFs_nu4}
\end{figure}

\begin{figure} 
\begin{center}
\strip{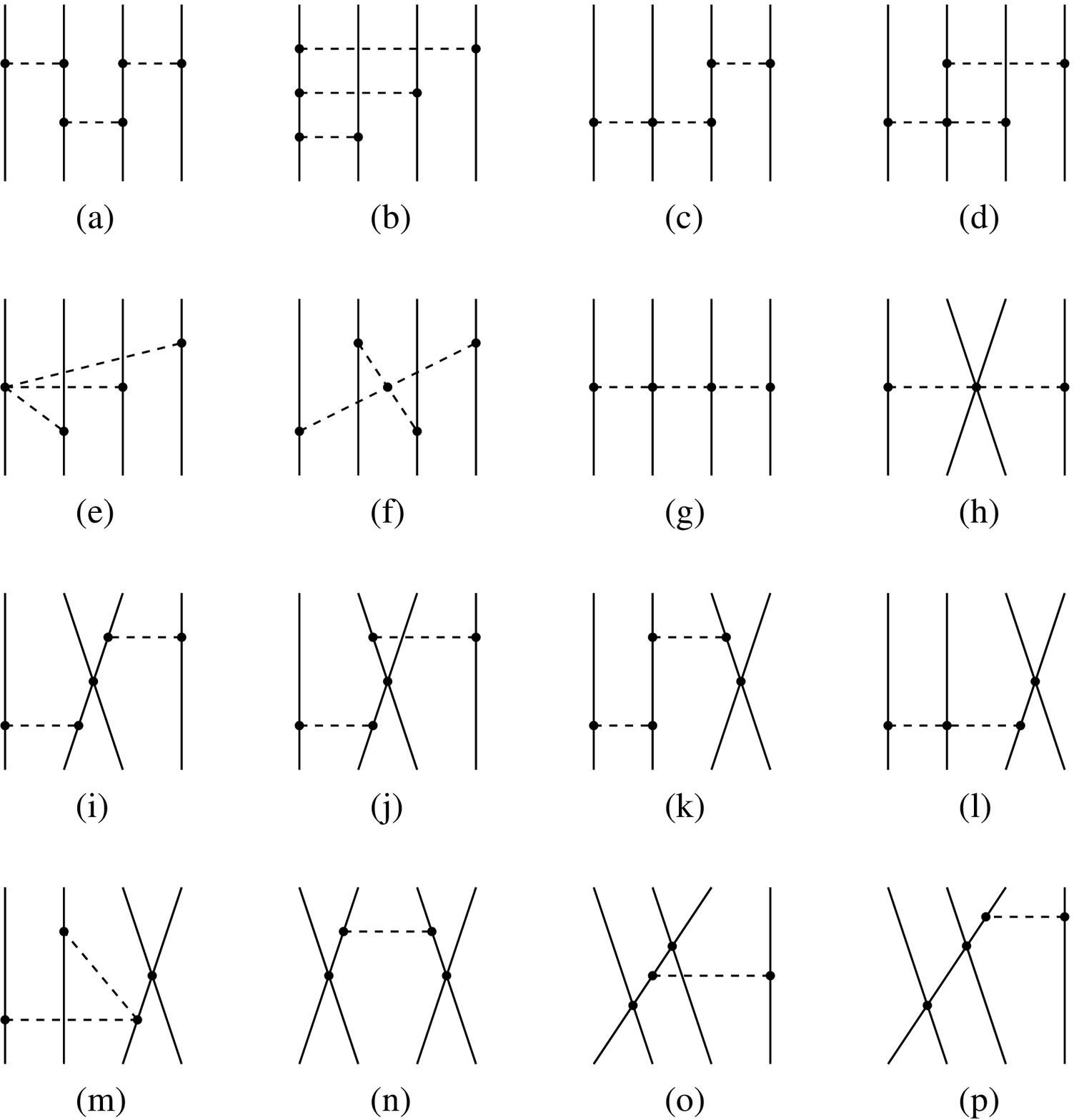}
\end{center}
\captionspace{Leading 4NFs at $\nu=4$. Permutations of vertex ordering and
nucleon lines are implicit. Figure courtesy
of Ref.~\cite{epelbaum_05_4nf}}
\label{fig:4NFs_nu4}
\end{figure}

At $\nu = 4$ we find the first corrections to the 3NF terms and the
leading order 4NF's. Some of these diagrams are shown in
Fig.~\ref{fig:34NFs_nu4} while a more complete list of 4NF diagrams
can be found in Fig.~\ref{fig:4NFs_nu4} from ref.
\cite{epelbaum_05_4nf}, which also presents the analytic expressions
for the potential contribution from these graphs. It has also been
shown that all disconnected diagrams at this order vanish completely
in an energy independent formulation~\cite{method_unitary_trans}.

In many-body systems we must also consider disconnected graphs that
involve fewer-body interactions and spectators. This is accounted for
in Eq.~\eqref{eq:nu} by the value of $C$. For instance a contribution
to a three-body system might include a diagram with a 2N interaction
and one spectator nucleon. Another might be a four nucleon system with
a 3NF interaction and one spectator or two disconnected NN
interactions (such as the $\nu=2$ interactions in
Fig.~\ref{fig:34NFs_nu2}). Such disconnected diagrams are expected to
be two orders of momentum more important than their corresponding tree
graphs, though many formally vanish and must be carefully worked
out~\cite{Epelbaum_05}.

It should be clear that the complexity and sheer number of graphs
increases quickly with the order in momenta ($\nu$) under
consideration. However, numerical calculations indicate
\cite{epelbaum_05_4nf} that graphs with $\nu \geq 4$ have a negligible
contribution, though these results are very preliminary and not
conclusive.

\section{The role of the $\Delta$ resonance}

The $\Delta$-isobar is a baryon with mass $m_\Delta = 1232$ which is
only 300 MeV above the mass of the nucleon. The $\Delta$ is composed
of 3 non-strange quarks and being in the quartet of isospin $3/2$ and
spin $3/2$ can have an integral charge from $-1$ to +2. For historical
reasons, one might also hear the $\Delta$ referred to as an excited
state of the nucleon especially given the strong resonance in $\pi N$
scattering. This view comes out of the simple quark model where the
$\Delta$ is a very short lived (decays via strong interactions)
particle whose main decay channel is to a pion and nucleon.

So far we have ignored the question of the effects of an explicit
$\Delta$ degree of freedom in the Lagrangian. Whatever effects there
might be would be treated as short ranged and encoded in the LEC's of
our existing theory. If they are not sufficiently short the LEC's will
become unnaturally large. But some work has been done on a
Weinberg-like power counting scheme with the assumption $m_\Delta -
m_N \sim 2m_\pi$. This is a simplification however since  $m_\Delta -
m_N$ is yet another small momentum parameter in which to make a
possible expansion.

An example Lagrangian for a \xeft\ involving an explicit $\Delta$
is~\cite{Epelbaum_05} 
\beqn
\mathcal{L} = \Delta^\dagger \left( i \partial_0 - \Delta m \right) \Delta + \frac{h_A}{2 F_\pi} 
\left( N^\dagger \vec S \fet T \Delta +  \mbox{h.~c.} \right) \cdot \vec \nabla \fet \pi - D_T N^\dagger \vec \sigma \fet \tau N
\cdot \left( N^\dagger \vec S \fet T \Delta +  \mbox{h.~c.} \right) \,,
\eeqn
where $h_A$ and $D_T$ are LECs and $S_i$ and $T_a$ are spin and
isospin matrices. These matrices are $2 \times 4$ because they must
describe the interactions between nucleons and deltas, the doublet and
quartet channels respectively, in spin and isospin space.

\begin{figure}[ht]
\begin{center}
\strip{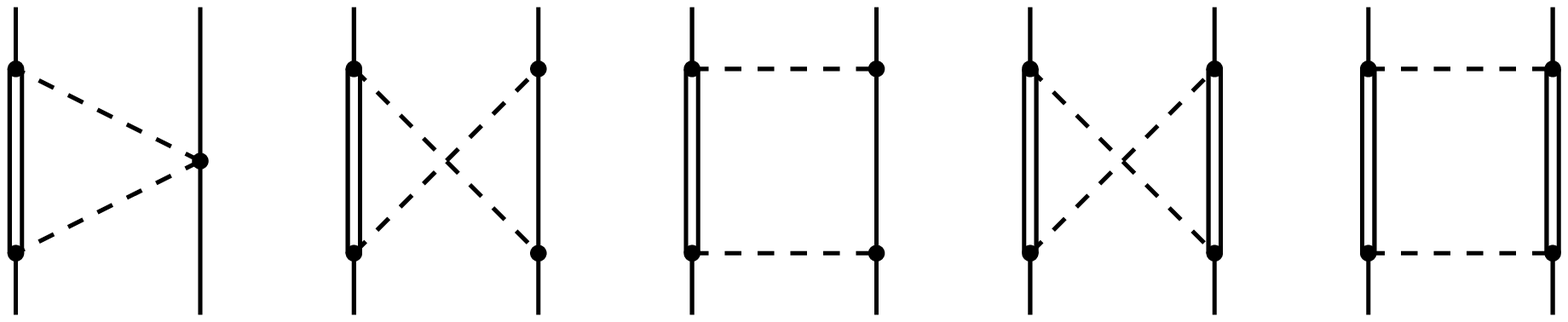} 
\captionspace{Leading NN two pion exchange graphs with explicit
$\Delta$ (double solid line) arising at $\nu=2$. Figure courtesy
of Ref.~\cite{Epelbaum_05}}
\label{fig:NN_delta}
\end{center}
\end{figure}

The first contributions involving explicit $\Delta$'s are NN graphs
involving two pion exchange and shown in Fig.~\ref{fig:NN_delta}. One
can see the $\Delta$'s dominance in the LECs at this order. The
potentials corresponding to the one-$\Delta$ 2PE box graphs are
identical in form to their analogous graphs without $\Delta$s.
Matching up the LECs, we can see $c_3 = 2c_4 = g_A^2/2(m_\Delta -
m_N)$ and as above the mass scale appearing is $(m_\Delta -
m_N)$ rather than $m_\pi$.

\begin{figure}[ht]
\begin{center}
\strip{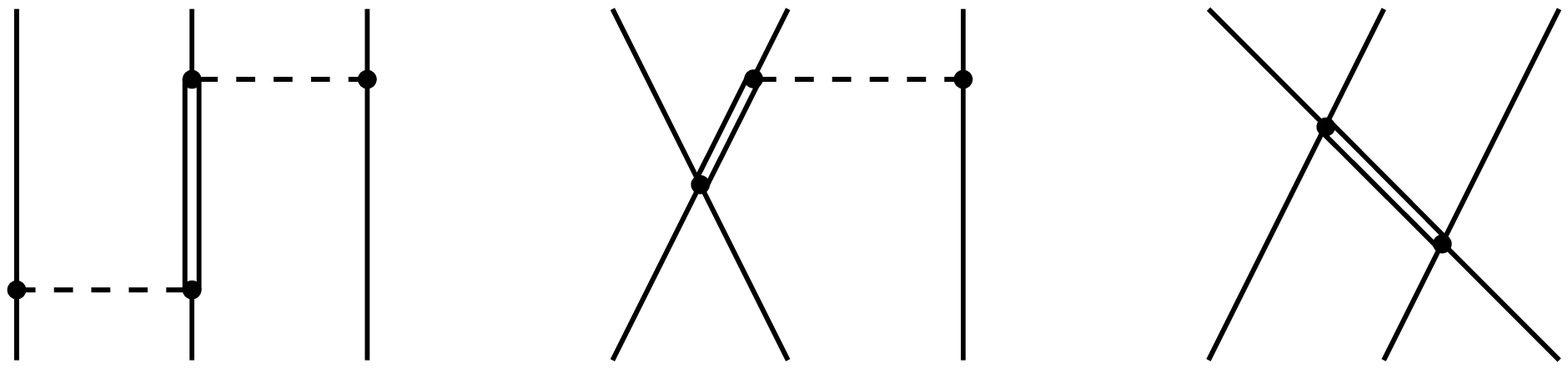} 
\captionspace{Leading order 3NFs with explicit $\Delta$ arising at
$\nu=3$, an enhancement over the deltaless theory. Figure
courtesy of Ref.~\cite{Epelbaum_05}}
\label{fig:3N_delta}
\end{center}
\end{figure}

The 3NF terms with explicit $\Delta$ resonances show up at $\nu =2$,
instead of non-$\Delta$ 3NF terms starting at $\nu = 3$. There is
basically one major contribution at this order; the first graph in
Fig.~\ref{fig:3N_delta} is analogous to the leading 3NF graphs that
had reducibility cancellations. Now because of the existence of a
$\Delta$ in the intermediate state, these cancellations don't occur.
The other two 3NF graphs at this order formally cancel out due to
Pauli exclusion \cite{Epelbaum:2002vt}.

In summary, the $\Delta$ can be included explicitly to achieve higher
accuracy or it can be excluded for simplicity leaving those higher
energy details to be encoded in the LECs of some $\Delta$-less \xeft.

\chapter{General Features of the Harmonic Oscillator Basis}
\label{chapt:app_osc_truncation}

\section{One-Dimensional Harmonic Oscillator Wavefunctions}

The one-dimensional harmonic oscillator functions in momentum space are given
by~\cite{osc_funcs_1D}
\beqn
\psi_n(p) = \left(\frac{1}{\pi m \hw 2^{2n}n!^2}\right)^{1/4}
\exp\left[-\frac{p^2}{2m\hw}\right] H_n\left[\frac{p}{\sqrt{m\hw}}\right]
\label{eq:ho_func_momentum}
\eeqn
where $H_n$ are the Hermite polynomials
\bea
H_0(y) &=& 1 \nonumber \\
H_1(y) &=& 2y \nonumber \\
H_2(y) &=& 4y^2 - 2 \nonumber \\
H_3(y) &=& 8y^3 - 12y \nonumber \\
H_4(y) &=& 16y^4 - 48y^2 + 12 \nonumber \\
H_5(y) &=& 32y^5 - 160y^3 + 120y 
\label{eq:hermite}
\eea
or more formally
\beqn
H_n(y) = (-1)^n \exp^{y^2} \frac{d^n}{dy^n} \exp^{-y^2}.
\eeqn
We will often make the approximation
\beqn
H_n(y) \approx (2y)^n,
\eeqn
since this is the dominant term in the polynomial for large $y$.

These oscillator functions are displayed in
Fig.~\ref{fig:oscillators} on the left, in a standard MATLAB
color scheme, with reds positive, blues negative, and green zero.
The vertical axis is increasing oscillator number, $n$, downward.
The horizontal axis is momentum, $p$, from negative to positive
$\kmax$. On the right, the truncated momentum delta function,
\beqn
\sum_{n=1}^{\nmax} \psi_n(p)\psi_n(p') \approx
\delta(p-p')\Theta(\pmax-p)
\label{eq:delta}
\eeqn
is a demonstration of the incompleteness of the finite oscillator
basis. In the limit $\nmax \longrightarrow \infty$ the sum should
go to a delta function, $\delta(p-p')$. Here, because of the
finite nature of the oscillator basis, the delta function is
truncated at a point we will refer to as $\pmax$ (referred to in other
contexts as the ultraviolet cutoff, $\Lambda_{\rm UV}$). The $\pmax$ of
the truncation and the extent of the last oscillator function in
the basis correspond to one another.

The coordinate space version of \ref{eq:ho_func_momentum} is
\beqn
\psi_n(r) = \left(\frac{m\hw}{\pi \hbar^2 2^{2n}n!^2}\right)^{1/4}
\exp\left[-\frac{m\hw r^2}{2\hbar^2}\right]
H_n\left[\sqrt{\frac{m\hw}{\hbar^2}}r\right].
\label{eq:ho_func_coord}
\eeqn
Note that the only difference in form is that the factor of $m\hw$ is
flipped in relation to the variable $r$ as it was to $p$ in
\ref{eq:ho_func_momentum}. This will be important later in an analytical
discussion of infrared and ultraviolet cutoffs due to basis
conversions.

\bfig[h]
\bc
\dblpic{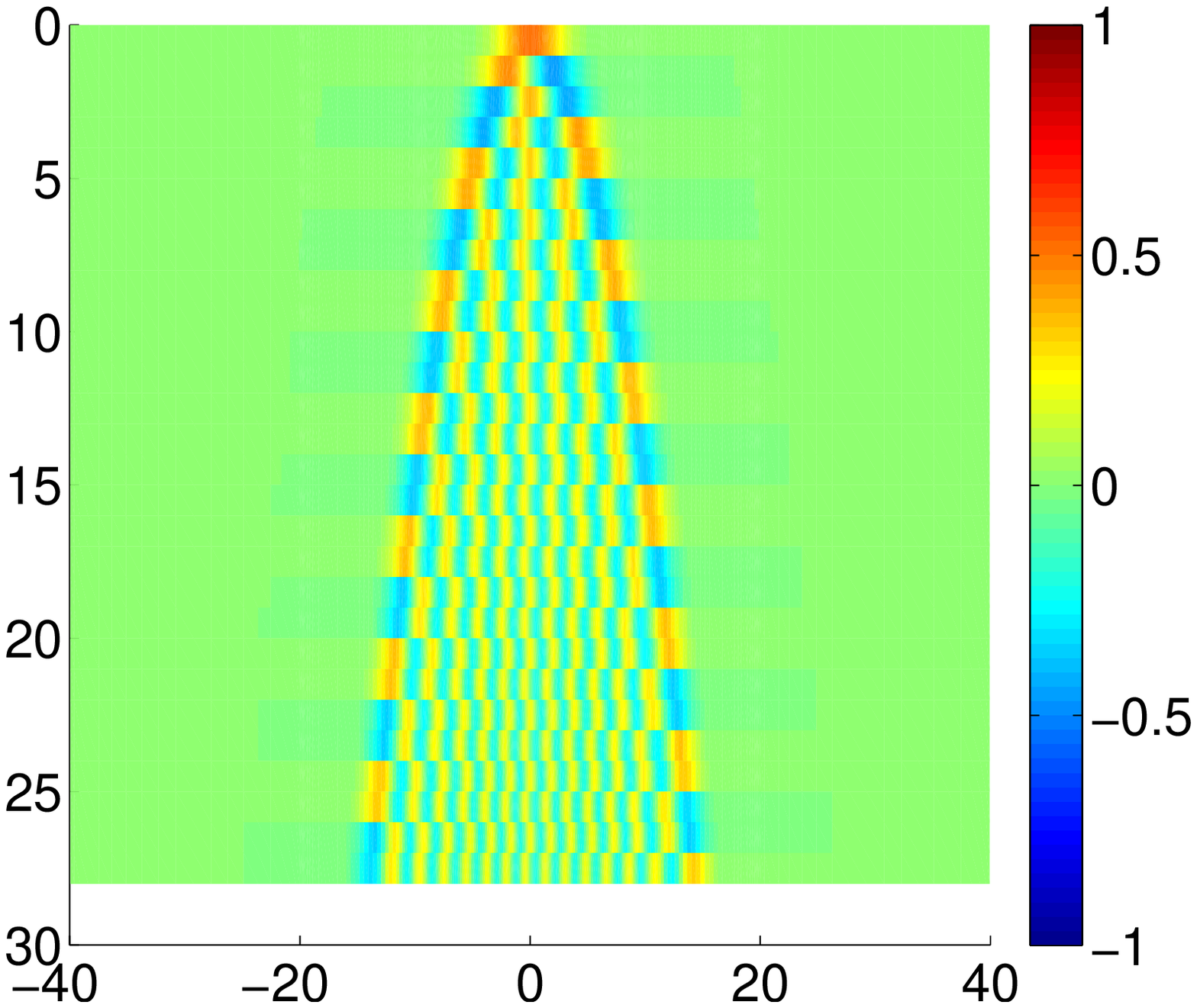}
\dblpic{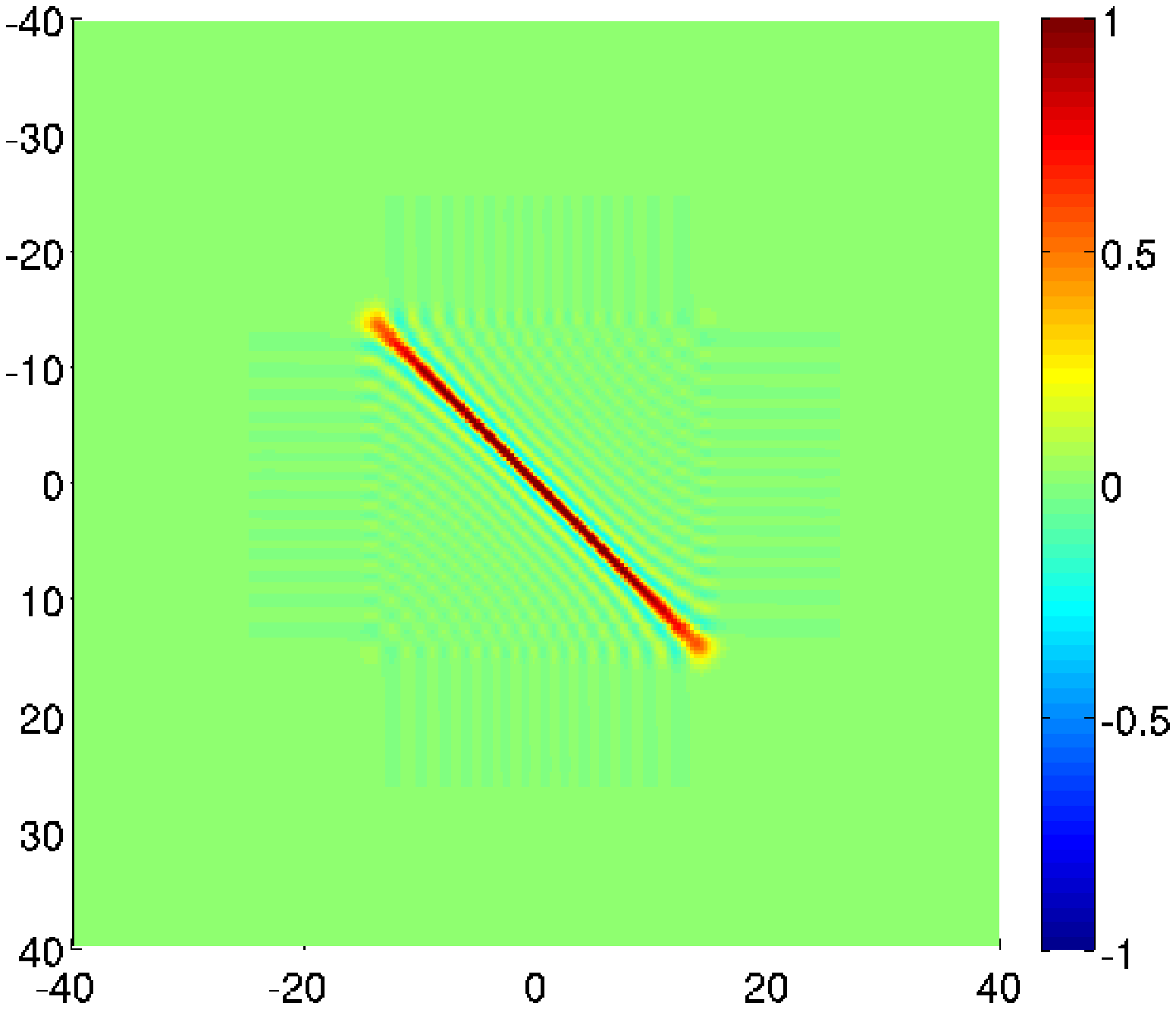}
\ec
\captionspace{On the left are shown the harmonic oscillator functions.
On the right is the truncated momentum delta function
$\delta(p-p')\Theta(\pmax-p)$. In this MATLAB color scheme the red is
positive and blue is negative.}
\label{fig:oscillators}
\efig


\section{Truncation}
\label{sec:trunc}

When converting a momentum space potential into the oscillator
basis as in 
\beqn
V(n,n') = \int_{-\kmax}^{\kmax} \psi_n(p)\psi_{n'}(p') V^{(2)}(p,p')
dp dp' 
\eeqn
or
\beqn
\la n|V|n'\ra = \int_{-\kmax}^{\kmax} \la n|p\ra \la p|V^{(2)}|p' \ra
\la p'|n'\ra dp dp' 
\eeqn
the potential becomes truncated due to the finite nature of the
oscillator basis. When we try to convert back to the momentum
basis we find that we have effectively multiplied by the
truncated delta function of Eq.~\ref{eq:delta}. 

This truncation is shown in Fig.~\ref{fig:truncation} for
different sized oscillator bases, each using the same value
for the oscillator parameter, $\hw$. Notice that as
$\nmax$ increases we retain more of the original potential, and
parts of it smooth out to their original form. There are also
residual oscillatory effects left at the edges of the converted
potential.

\bfig[ht!]
\bc
\dblpic{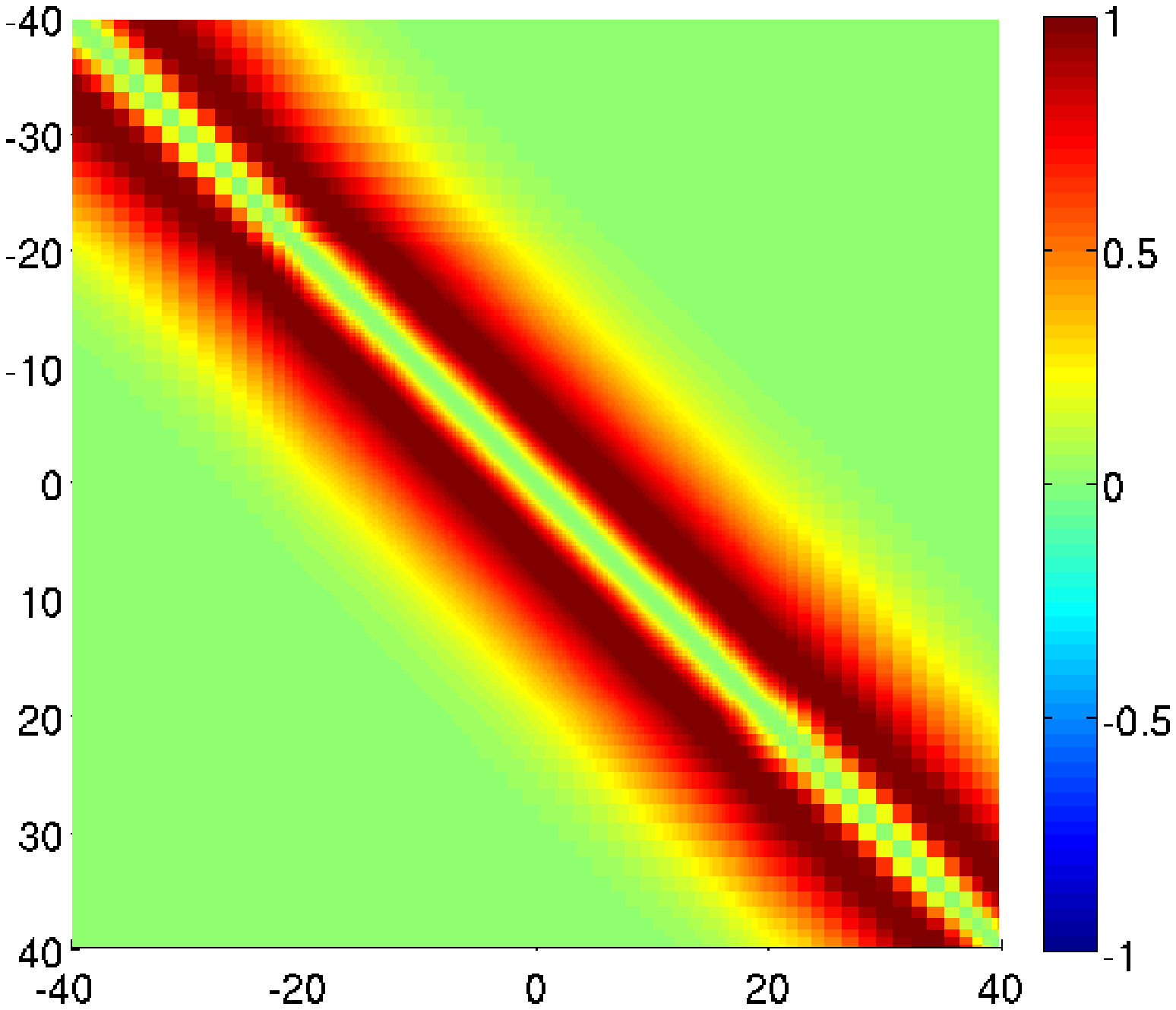}
\hfill
\dblpic{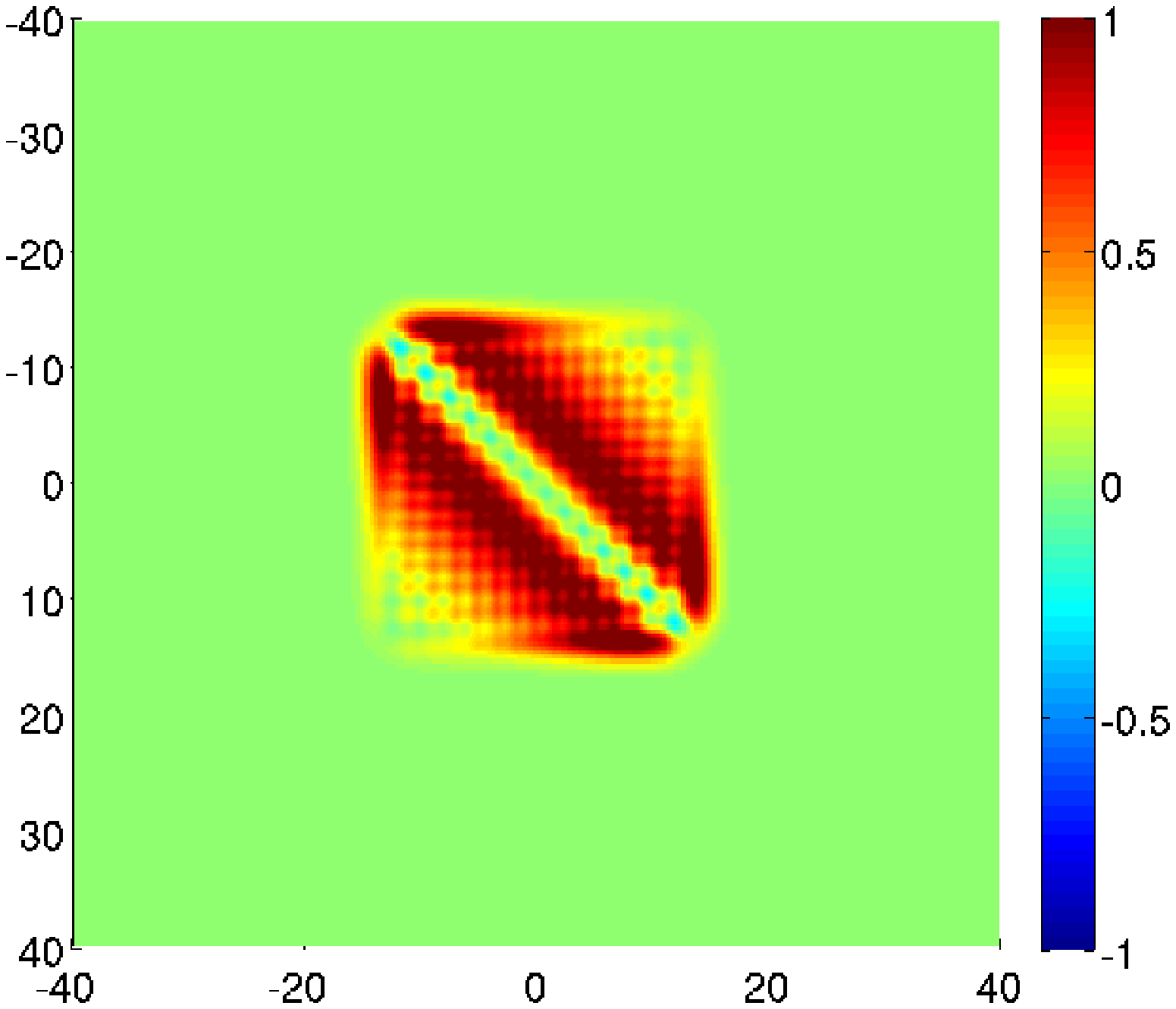}

\dblpic{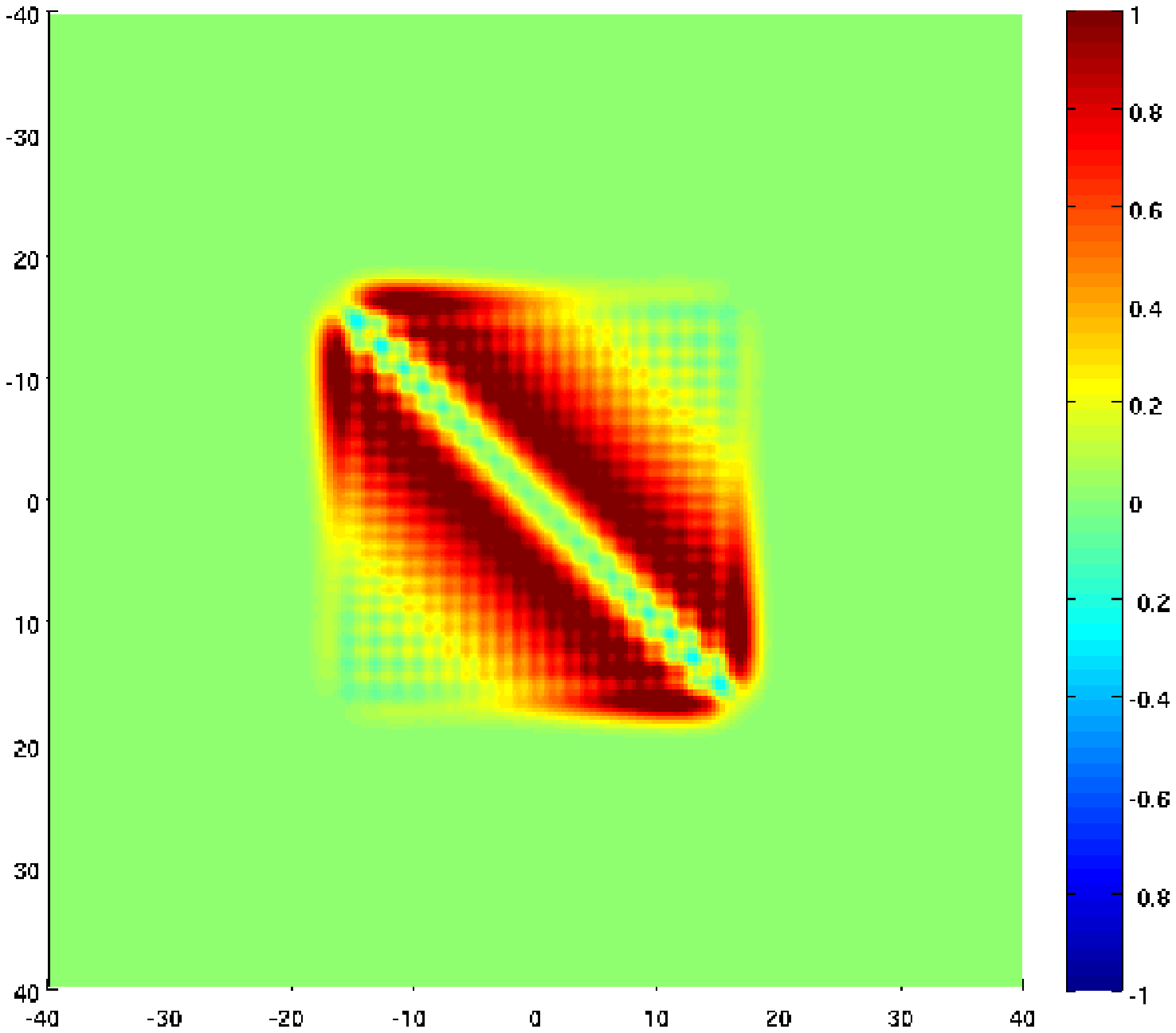}
\hfill
\dblpic{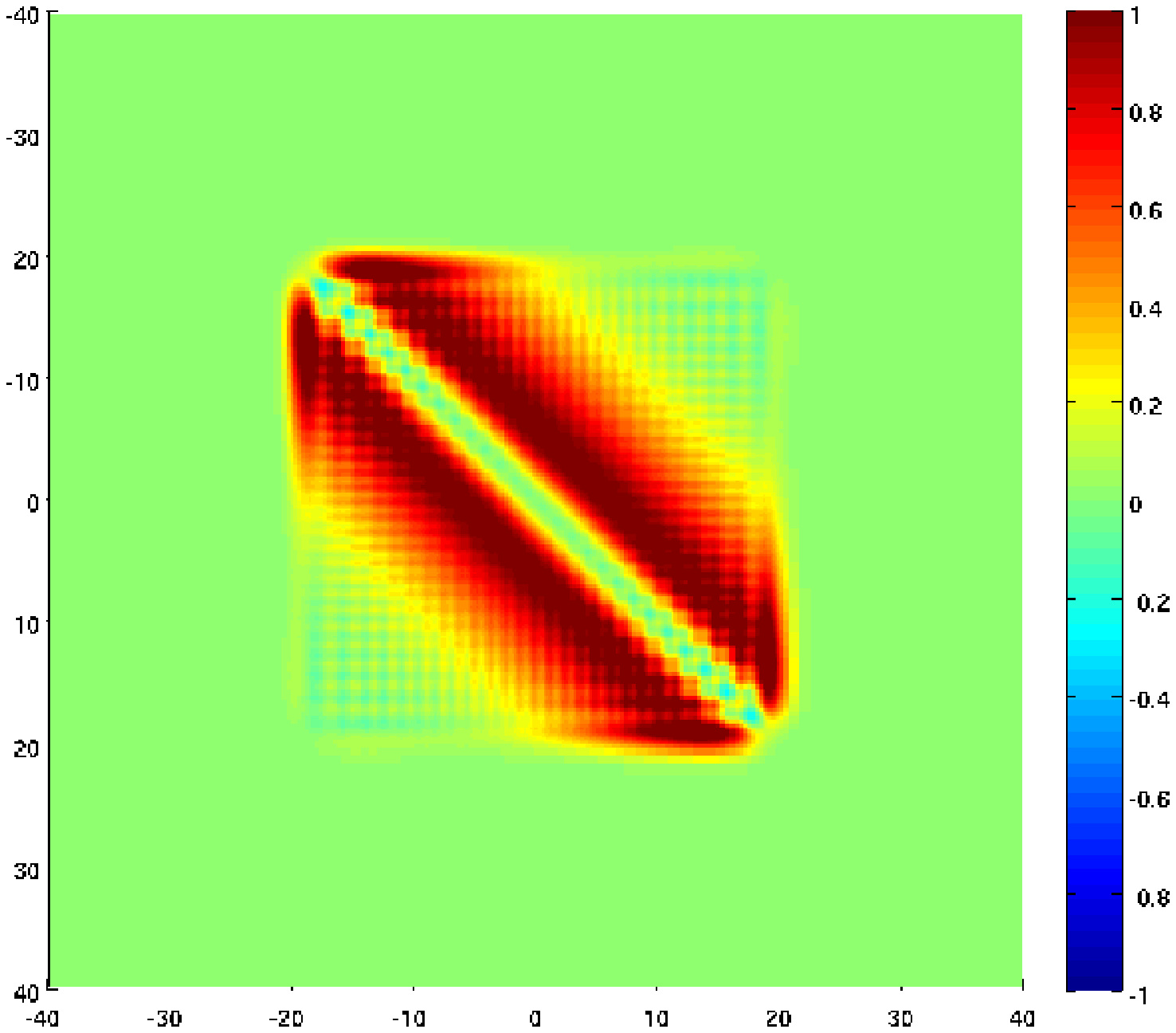}
\ec
\captionspace{A two-body potential in the momentum representation. The axes
are initial and final momenta, $k$ and $k'$, from $-\kmax$ to $\kmax$.
On the upper-left is the original momentum basis two-body potential.
The other three plots are the same potential after being converted to
an oscillator basis and back. Notice the truncation of matrix elements
above each $\pmax$ of the oscillator basis.}
\label{fig:truncation}
\efig

As stated above the truncation at $\pmax$ corresponds to the
extent in momentum space to the highest harmonic oscillator
wavefunction in the oscillator basis, $\psi_{\nmax}(p)$. Two
variables affect the value of $\pmax$: the oscillator parameter 
$\hw$ and the basis size $\nmax$. We will consider the
dependence on each of these variables in order.

A quick look at the harmonic oscillator wavefunctions in
Eq.~\ref{eq:ho_func_momentum} shows a simple proportionality between
$p$ and $\hw$. The change in $\pmax$ should scale like the square root
of the change in $\hw$. Numerically this can be verified by looking at
the furthest extent of oscillator functions for different $\hw$'s.
Shown in Fig.~\ref{fig:hw_variation} are the oscillator bases for
$\nmax$=28 and $\hw$=2,4,8 and 16. An increase in $\hw$ by a factor of
four doubles $\pmax$. In Fig.~\ref{fig:hw_variation}, looking down the
vertical axis in the plots of shows a similar dependence on $\nmax$.
The curve formed by the edge of the oscillator wavefunctions looks
like a square root function implying $\pmax \approx n^{1/2}$.

\bfig[ht!]
\bc
\dblpic{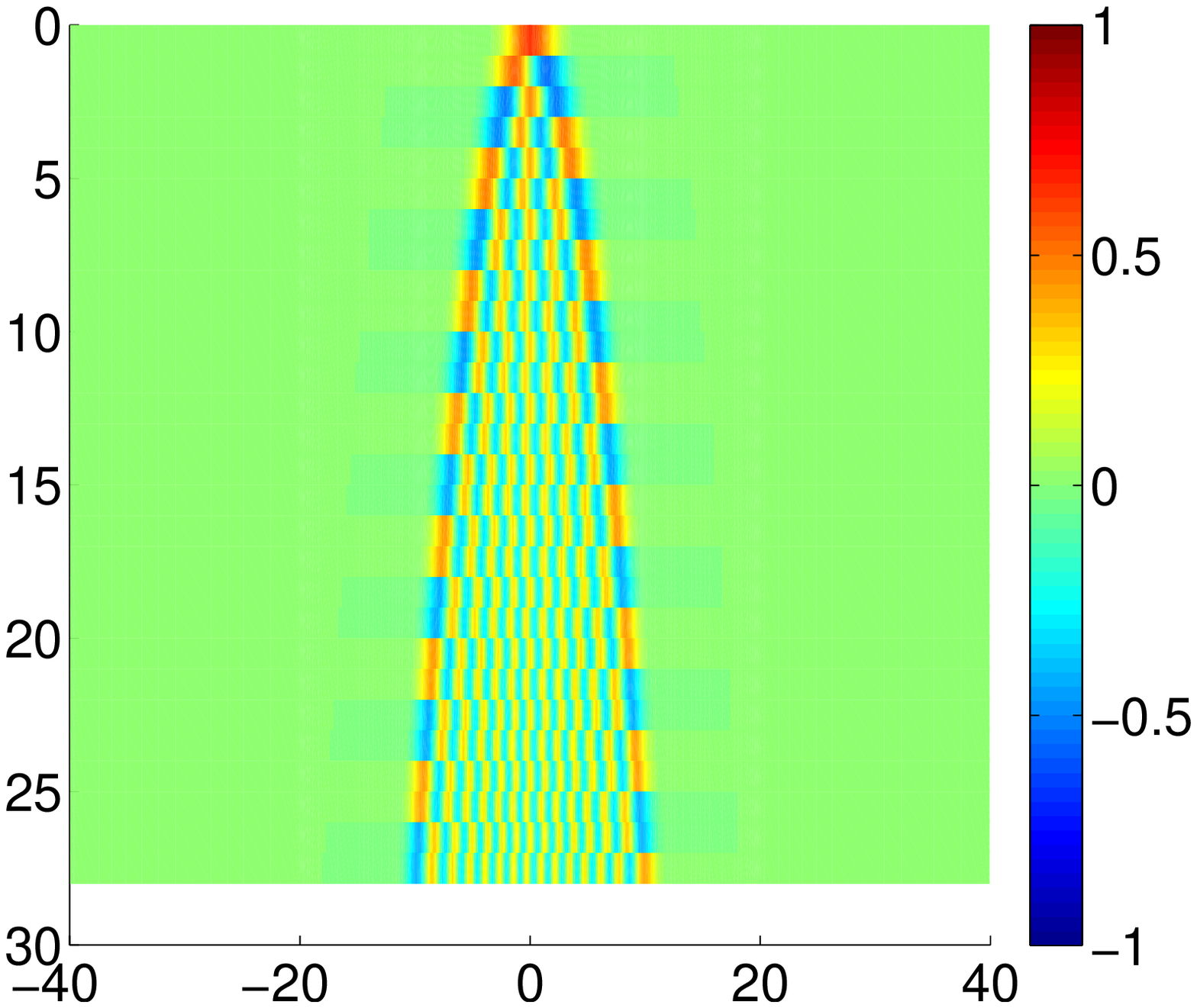}
\hfill
\dblpic{figures/osc_nmax28_hw4}

\dblpic{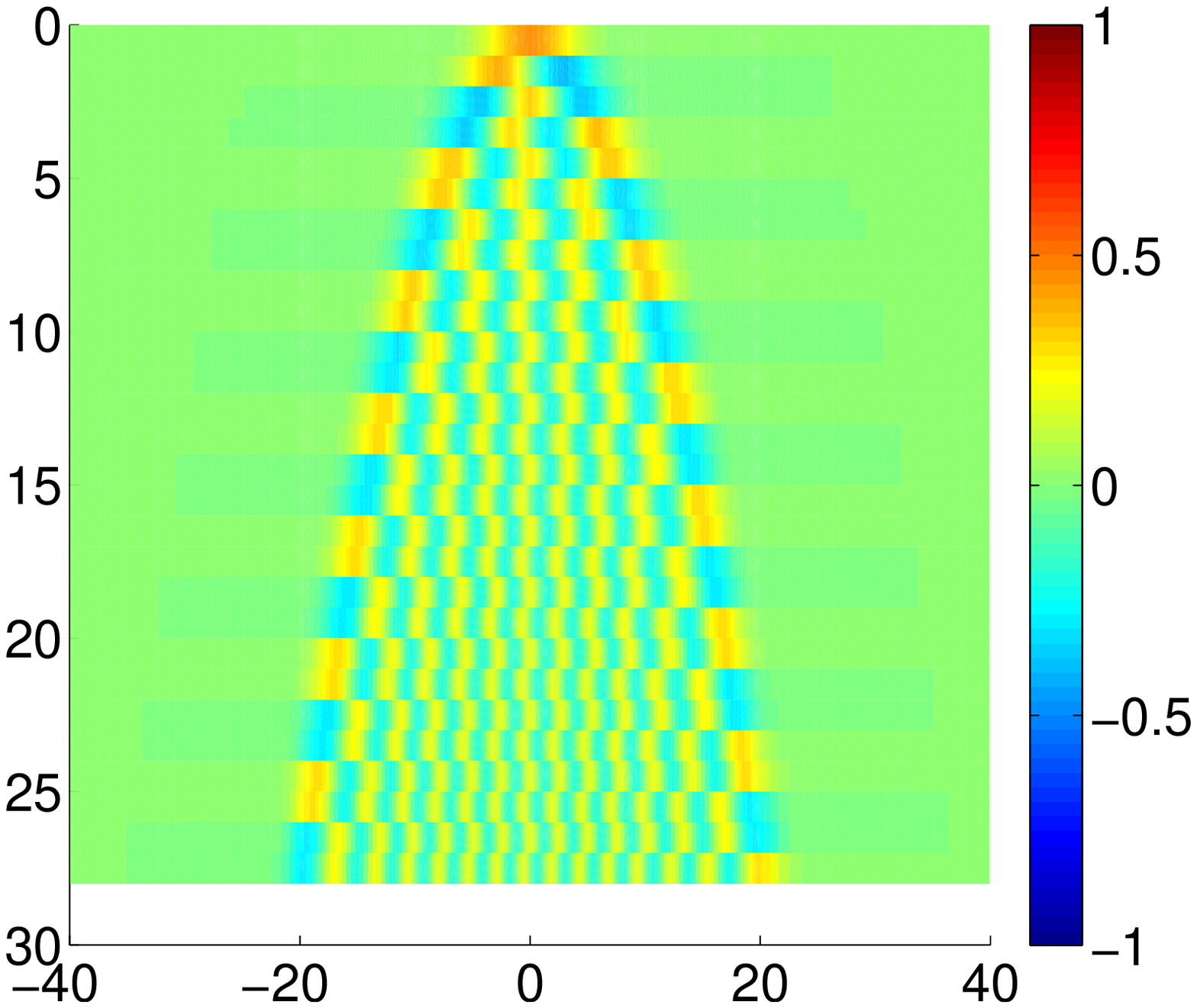}
\hfill
\dblpic{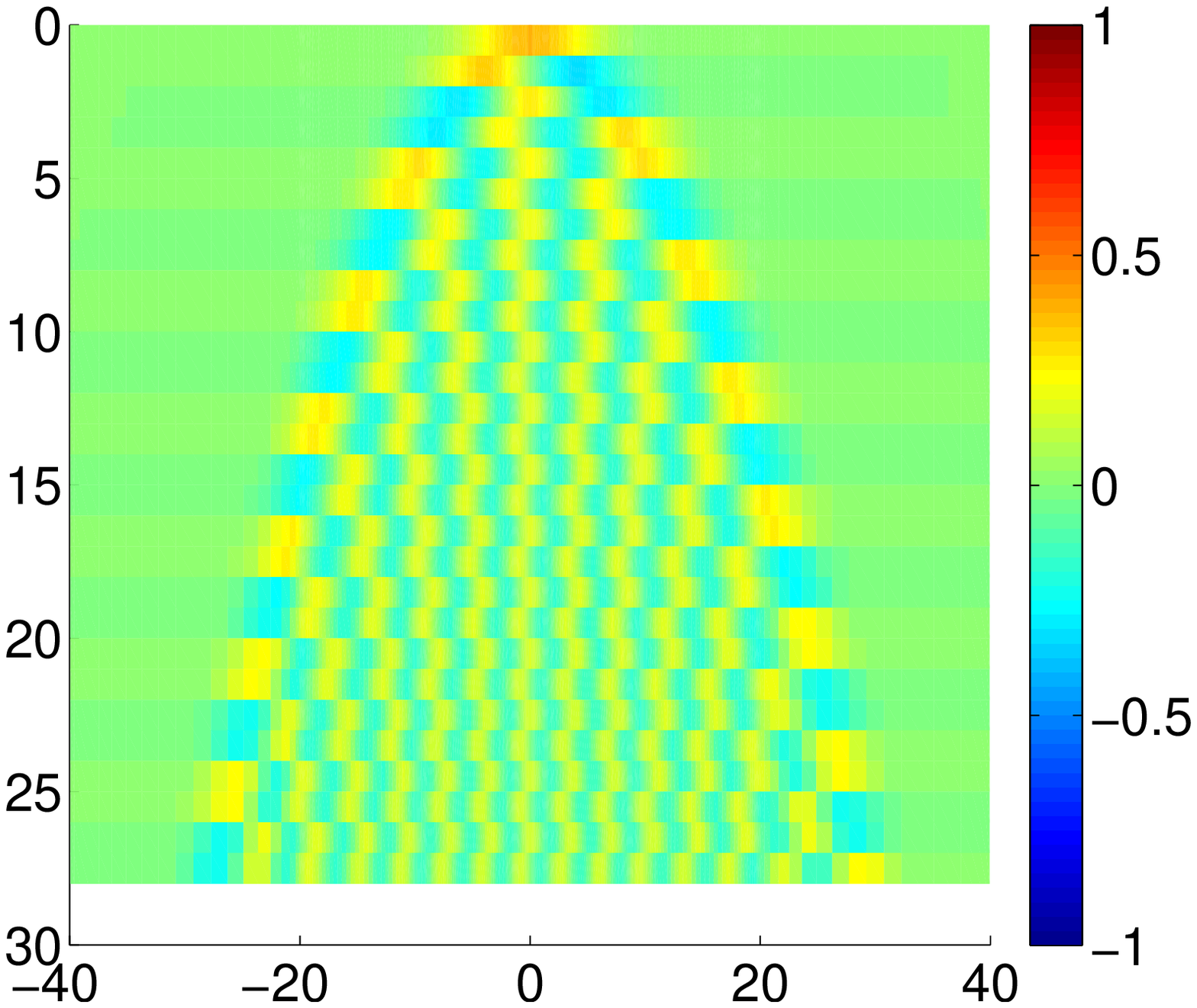}
\ec
\captionspace{Harmonic oscillator functions up to $\nmax$=28 for $\hw$
= 2, 4, 8, and 16.}
\label{fig:hw_variation}
\efig

We can measure these dependencies more quantitatively by picking
out the mesh point at which each oscillator function passes a
certain minimum threshold:
\beqn
|\psi_n(\pmax)| = \psi_{min}\;.
\eeqn
In Fig.~\ref{fig:pmax_plots} we show an example of such a measurement.
The value $\psi_{\rm min}=.01$ is used
to plot the variation of $\pmax$ with $\hw$ on the left and with
$n$ on the right. In this log-log plot the dependence on $\hw$ is
a clean power law with a slope of $.5$ as expected from the oscillator
wavefunctions dependence on $p/\sqrt{\hw}$.

\bfig[ht!]
\bc
\triplepic{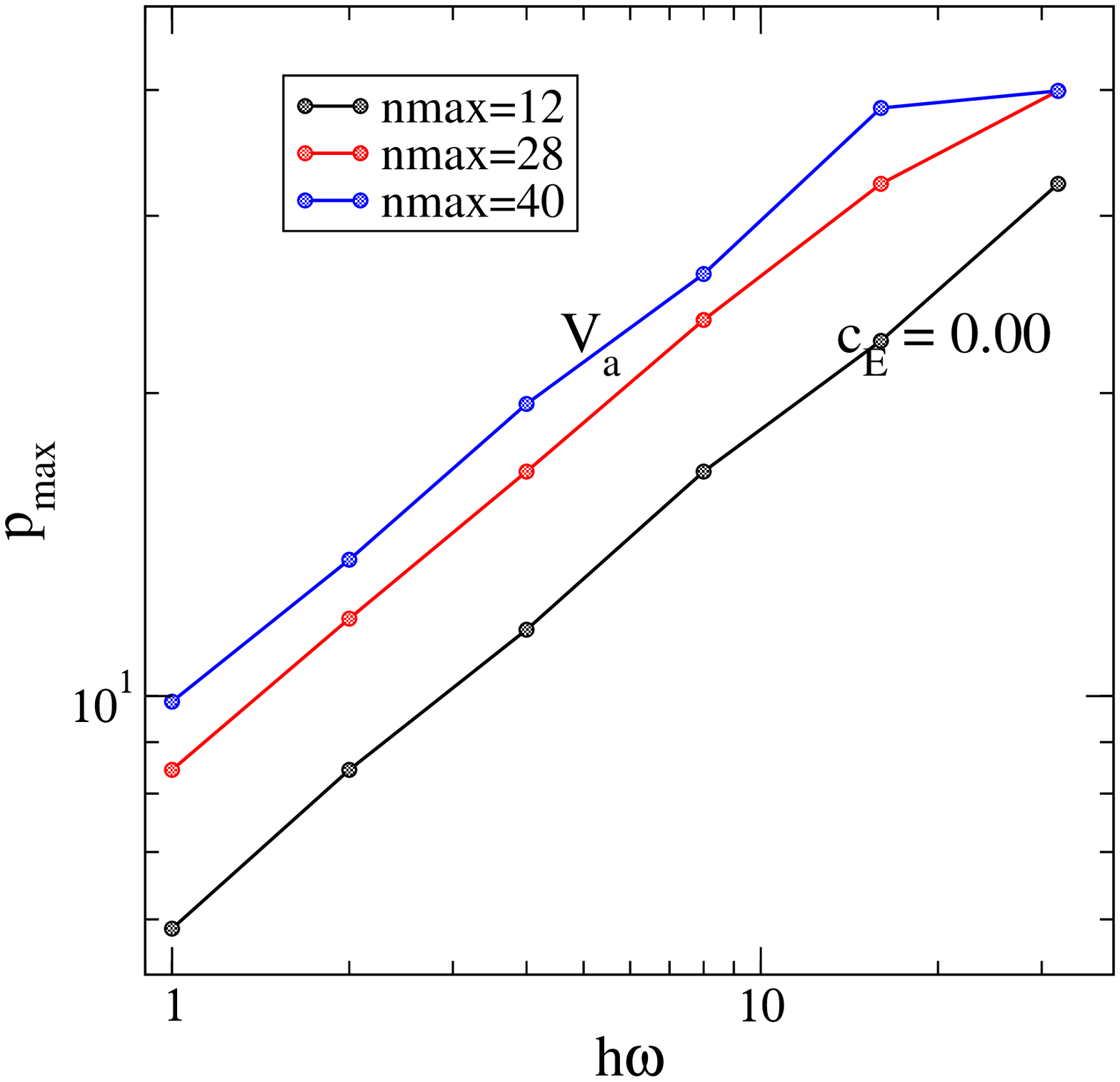}
\triplepic{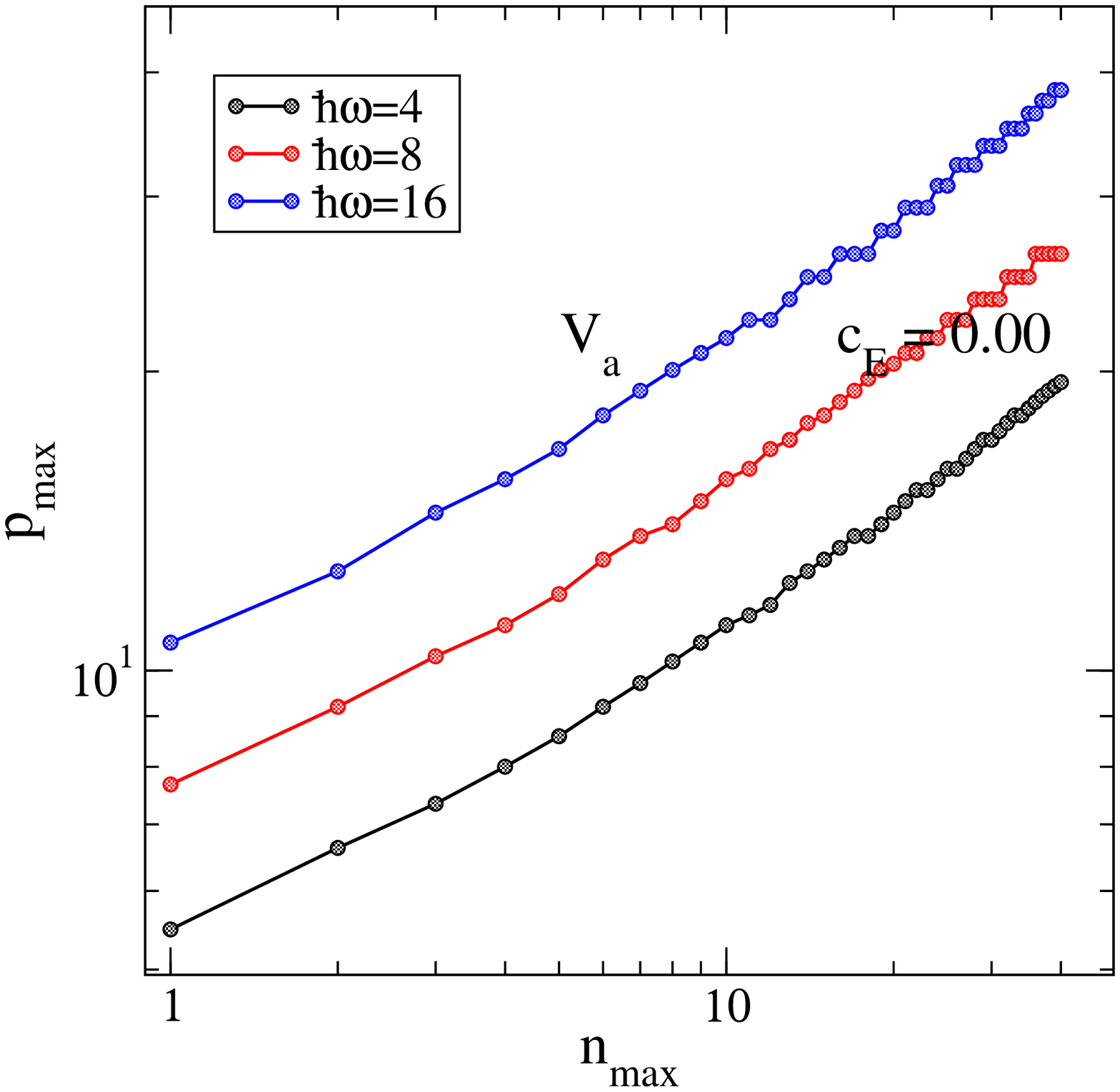}
\triplepic{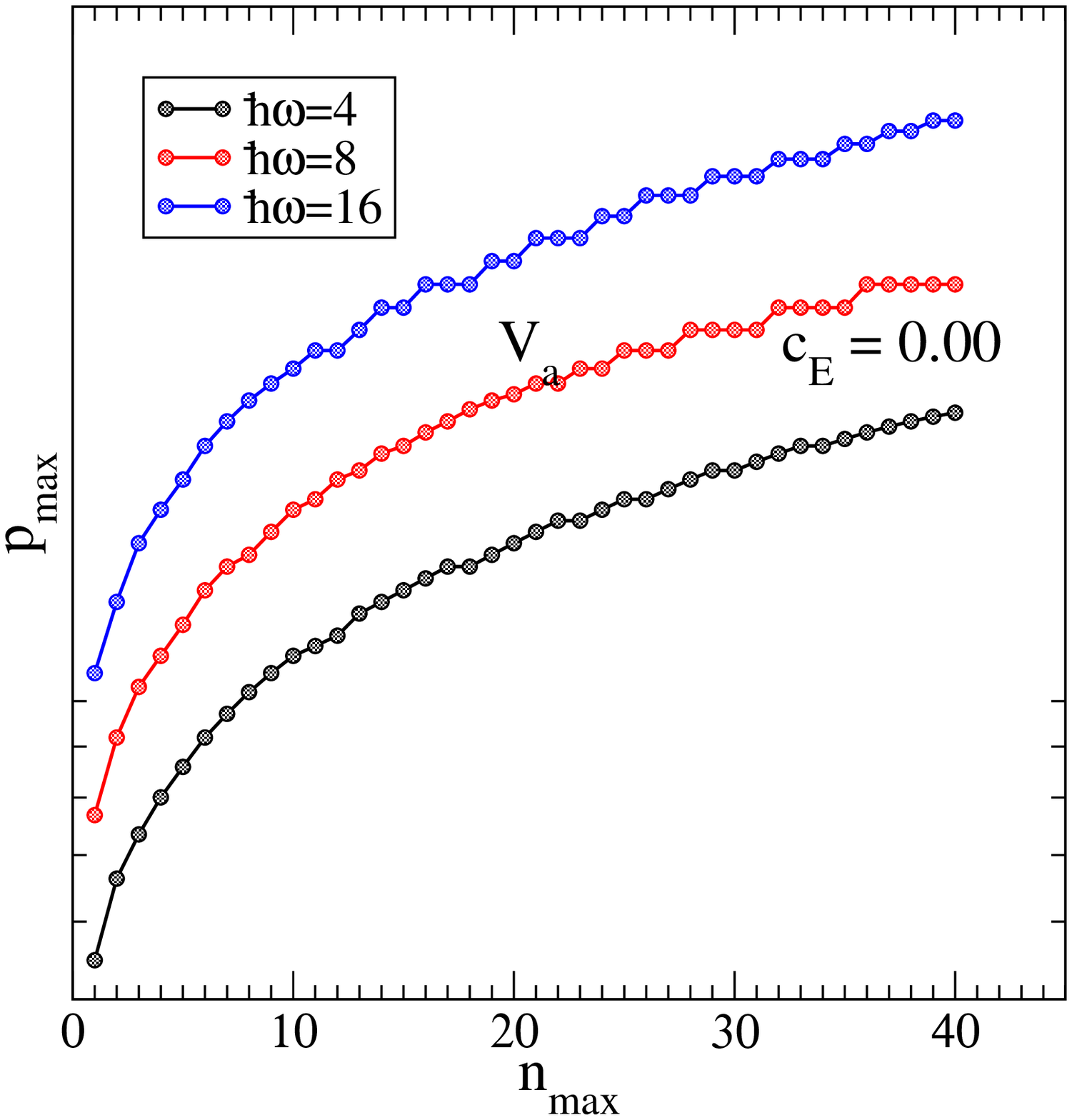}
\ec
\captionspace{Variation of $\pmax$ with $\hw$ on the left and
 with $n$ on the right.}
\label{fig:pmax_plots}
\efig


\section{Cutoff Derivation}
\label{sec:cutoffs}

Now, taking a more analytical approach to understanding the cutoffs
inherent in the finite oscillator basis, let us first consider the
momentum space harmonic oscillator function of
Eq.~\ref{eq:ho_func_momentum}. This function is essentially a gaussian
shifted to the right by the largest $\nmax$ polynomial term. The local
maximum will be correlated with the high-momentum cutoff,
$\Lambda_{\rm UV}$, or $\pmax$. Making the substitution, $U = \pmax /
\sqrt{m\hw}$ we can write the large-$n$ approximation by
\beqn
\psi (U) = A(2U)^n \exp^{-U^2/2}\;.
\eeqn
Differentiating with respect to U, we get
\beqn
\frac{d\psi}{dU} =  (2U)^{n-1}(n - (2U)U) \exp^{-U^2/2} \;.
\eeqn
The maximum, at $\frac{d\psi}{dU} = 0$, yields
\beqn
U\sim\sqrt{n} \longrightarrow 
\Lambda_{\rm UV} = \pmax \sim \sqrt{m\nmax\hw}\;.
\eeqn
which is the large $\nmax$ behavior for the ultraviolet cutoff caused by
converting to the harmonic oscillator basis. As expected, when either
$\nmax$ or $\hw$ is increased, then the scope of the basis also
increases and we keep more of the momentum space potential.

For the infrared cutoff, $\Lambda_{\rm IR}$, we can make the same
analysis, but now using the coordinate space oscillator wavefunctions
where $p^2/m\hw \longrightarrow m\hw r^2$. We again ask about the
$\nmax$ dependence of the local maximum of $\psi(U)$. Here we get a
maximum extent in coordinate space,
\beqn
r_{max} = \sqrt{\frac{n\hbar^2}{m\hw}}
\eeqn
which implies an infrared cutoff 
\beqn
\Lambda_{\rm IR} \sim \sqrt{\frac{m\hw}{\nmax}}
\eeqn
setting a bound on the size of objects that can be accurately
described in a given $\nmax\hw$ basis. When $\hw$ grows large,
individual oscillations are large and lose resolution on the small
details in the momentum basis potential corresponding to large $r$
structures. However, high $\nmax$ polynomials have many small
oscillations at low momenta compensating for the large $\hw$ value.
Thus, both $\Lambda_{\rm IR}$ and $\Lambda_{\rm UV}$ are pushed back
(down and up respectively) by increasing $\nmax$ which one expects as
the basis is extended towards completeness.


\section{Three-Dimensional Harmonic Oscillators}
\label{sec:osc_function_3D}


To build the NCSM in three-dimensions, we must use the isotropic
Harmonic oscillator eigenfunctions~\cite{osc_funcs_3D},
\beqn
\psi_{klm}(r,\theta,\phi) = N_{kl} r^{l}e^{-\nu r^2}
{L_k}^{(l+{1/2})}(2\nu r^2)Y_{lm}(\theta,\phi)\;,
\eeqn
where $\nu$ is defined as
\beqn
\nu \equiv \frac{\mu\omega}{2\hbar}\;,
\eeqn
with $\mu$ the reduced mass ($m$ is now reserved for the magnetic
moment quantum number) and $\omega$ is again the oscillator parameter.
The normalization constant, $N_{kl}$, is%
\beqn
N_{kl}=\sqrt{\sqrt{\frac{2\nu ^{3}}{\pi}}
\frac{2^{k+2l+3}\;k!\;\nu ^{l}}{(2k+2l+1)!!}}\;.
\eeqn
The
\beqn
{L_k}^{(l+{1\over 2})}(2\nu r^2)
\eeqn
are generalized Laguerre polynomials are given by
\beqn
L_n^{(\alpha)}(x)= \frac{x^{-\alpha} e^x}{n!}\frac{d^n}{dx^n}
\left(e^{-x} x^{n+\alpha}\right)\;.
\eeqn
The first few generalized Laguerre polynomials are:
\bea
L_0^{(\alpha)}(x) &=& 1  \nonumber \\
L_1^{(\alpha)}(x) &=& -x + \alpha +1 \nonumber \\
L_2^{(\alpha)}(x) &=& \frac{x^2}{2} - (\alpha + 2)x +
    \frac{(\alpha+2)(\alpha+1)}{2}  \nonumber \\
L_3^{(\alpha)}(x) &=& \frac{-x^3}{6} + \frac{(\alpha+3)x^2}{2} -
    \frac{(\alpha+2)(\alpha+3)x}{2} +
    \frac{(\alpha+1)(\alpha+2)(\alpha+3)}{6}
\label{eq:lagauerre} 
\eea

The first several spherical harmonic
functions
\beqn
Y_\ell^m(\theta,\phi) = \sqrt{\frac{(2\ell+1)}{4\pi}
\frac{(\ell-m)!}{(\ell+m)!}} P_\ell^m(\cos{\theta}) e^{im \phi } \;, 
\eeqn
are given by
\beqn
\begin{array}{ll}
 Y_{0}^{0}(\theta,\phi) 
 = {1\over 2}\sqrt{1 \over \pi} 
&  Y_{2}^{-2}(\theta,\phi) 
 = {1\over 4}\sqrt{15\over 2\pi} 
\sin^{2}\theta e^{-2i\phi}   \\
 Y_{1}^{-1}(\theta,\phi) 
 = {1\over 2}\sqrt{3 \over 2\pi} \sin\theta e^{-i\phi} 
&  Y_{2}^{-1}(\theta,\phi) 
 = {1\over 2}\sqrt{15\over 2\pi} \sin\theta \cos\theta e^{-i\phi}  \\
 Y_{1}^{0}(\theta,\phi) 
 = {1\over 2}\sqrt{3 \over \pi} \cos\theta 
&  Y_{2}^{0}(\theta,\phi) 
 = {1\over 4}\sqrt{5\over\pi} (3\cos^{2}\theta-1)  \\
 Y_{1}^{1}(\theta,\phi) 
  = {-1\over 2}\sqrt{3 \over 2\pi} \sin\theta e^{i\phi} 
&  Y_{2}^{1}(\theta,\phi) 
 = {-1\over2}\sqrt{15\over 2\pi} \sin\theta\cos\theta e^{i\phi}  \\
 &  Y_{2}^{2}(\theta,\phi) 
 = {1\over 4}\sqrt{15\over 2\pi}\sin^{2}\theta  e^{2i\phi} 
\end{array}
\label{eq:list_spherical_harmonics}
\eeqn

At large $r$, we can simplify the expression for $\psi_{k\ell m}$ as
\bea
\psi_{k \ell m}(r,\theta,\phi) &\approx& 
\sqrt{\left(\frac{\mu\omega}{2\hbar}\right)^{\ell + 3/2}} 
r^{\ell} e^{-\nu r^2} (\nu r^2)^k Y_\ell^m(\theta,\phi) \nonumber \\
&=& \left(\frac{\mu\omega}{2\hbar} \right)^{3/4} r^{2k+\ell}
\left(\sqrt{\frac{\mu\omega}{2\hbar}} \right)^{2k+\ell}
e^{-\frac{\mu\omega}{2\hbar} r^2}Y_\ell^m(\theta,\phi)\;.
\eea
If we make the substitution $2k+\ell \longrightarrow n$ this has
exactly the form of Eq.~\eqref{eq:ho_func_coord} and we can make the
same analysis of the cutoff as done in section~\ref{sec:cutoffs} for
the one-dimensional oscillator functions. Indeed, the quantity
$n=2k+\ell$ is the quantum number for the energy of a state in the
three-dimensional harmonic oscillator trap. The eigenvalues are
\beqn
E_n = (2k+\ell+3/2)\hw
\eeqn
with a degeneracy of $(n+1)(n+2)/2$.

So, the three-dimensional oscillator basis has the same variational
properties as the one-dimensional discussed above. The exact values of
$\hw$ and convergence in $\nmax$ (maximum allowed $n=2k+\ell$) will
vary due to different potentials and dimensional factors such as the
cube in the normalization factor. But the cutoff behavior is the same
between the one- and three-dimensional cases.


\section{Variational Properties}

Any harmonic oscillator basis such as the NCSM, both one- and
three-dimensional, is a variational calculation in the two parameters,
$\nmax$ and $\hw$. Larger $\nmax$ allows inclusion of more of the
momentum basis potential, increasing $\Lambda_{\rm UV}$. Larger $\hw$
extends the reach of a given polynomial of order $\nmax$ but damps the
resolution of the polynomials oscillations, raising both
$\Lambda_{\rm UV}$ and $\Lambda_{\rm IR}$. Smaller $\hw$ increases the
resolution but hampers the reach of the $\nmax$th polynomial,
resulting in lowering both $\Lambda_{\rm UV}$ and $\Lambda_{\rm IR}$. So
bigger $\nmax$ is always better (except for increasing computational
requirements) by raising $\Lambda_{\rm UV}$ and lowering $\Lambda_{\rm IR}$,
but the optimal $\hw$ must result from a tuning of  $\Lambda_{\rm IR}$ and
$\Lambda_{\rm UV}$ at a particular $\nmax$.

\begin{figure}[th!]
\begin{center}
\triplepic{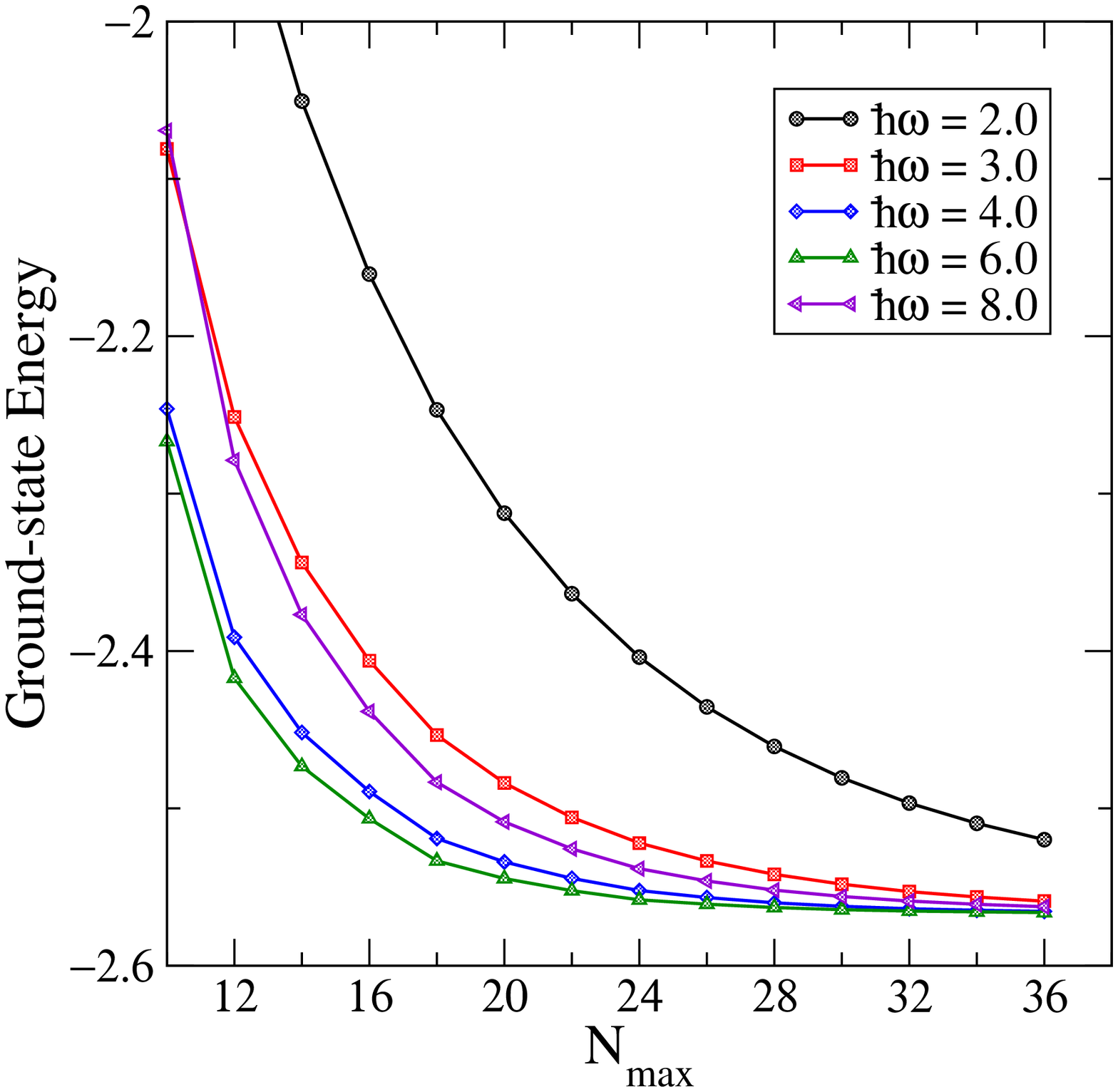}
\triplepic{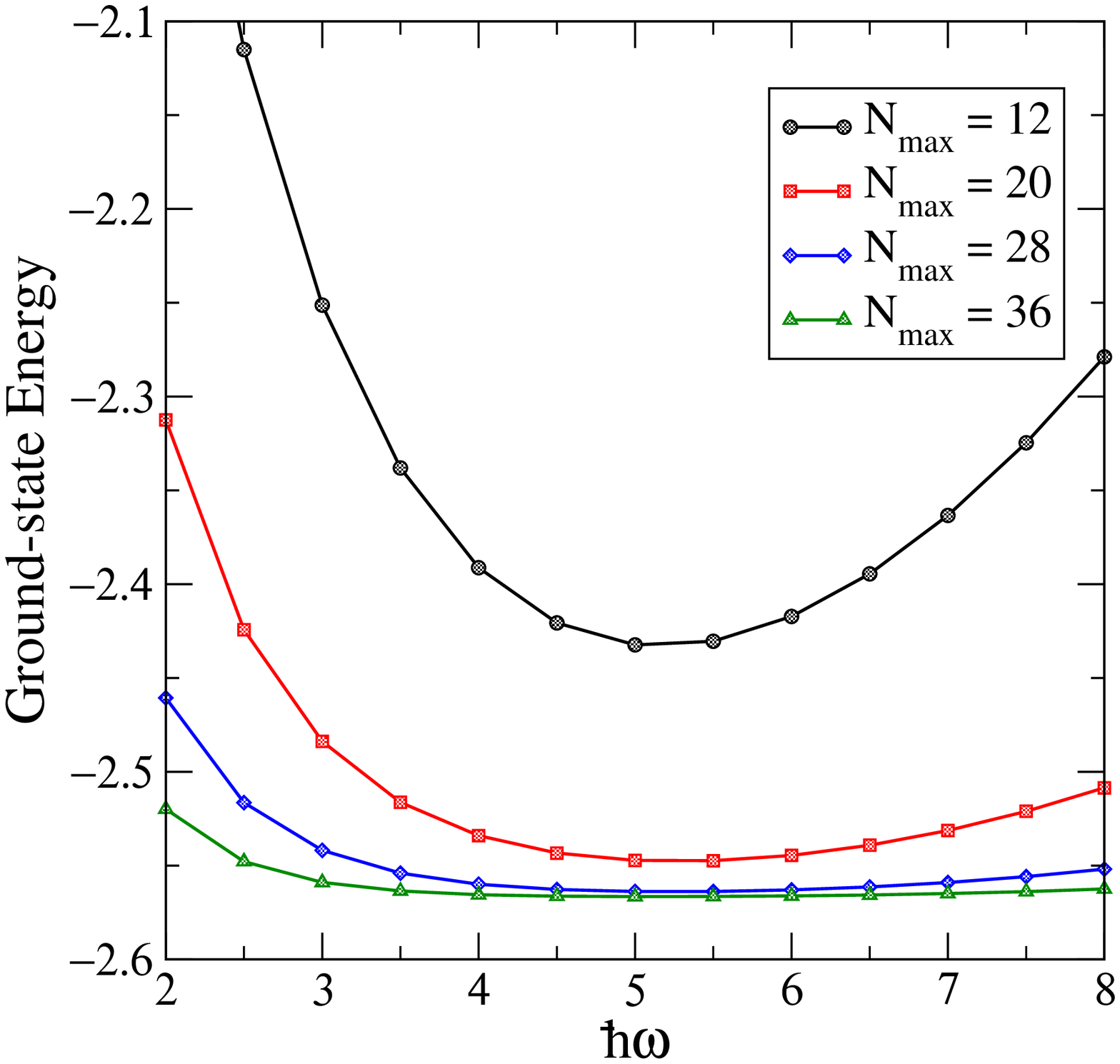}
\triplepic{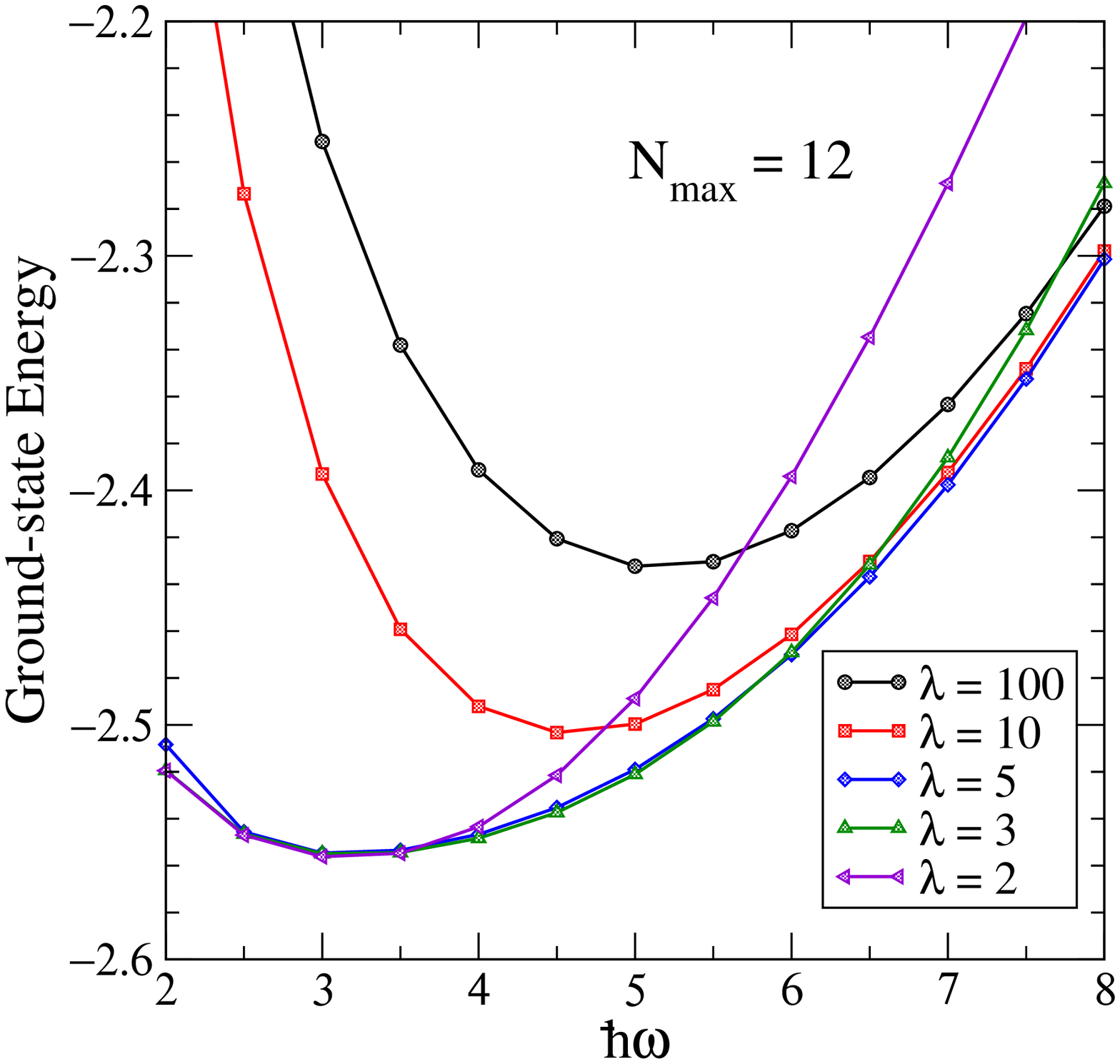}
\end{center}
\captionspace{Plots showing the variational nature of the harmonic
oscillator basis and the effect of SRG transformations within it.}
\label{fig:ncsm_variational}
\end{figure}

These variational properties are shown nicely in
Fig.~\ref{fig:ncsm_variational}. The left plot shows the increasing
convergence with the size of the basis, $\nmax$. Here several select
values of $\hw$ show that one curve is optimal for most of the range
in $\nmax$ considered. As $\nmax$ grows larger, more polynomials are
available for the expansion of momentum basis wavefunctions and the
dependence on $\hw$ flattens out. The center panel also shows this by
plotting instead verses $\hw$ where the minimum is clear and the
increase in $\nmax$ is apparent but less quantitative. Again the
dependence on $\hw$ flattens as $\nmax$ grows. The overall goal in
these calculations is to balance convergence in $\nmax$ (bringing a
reduced sensitivity to $\hw$) with the increasing computational
requirements the larger basis demands. This is further explored in
Appendix~\ref{chapt:app_scaling}.

The far right plot of Fig.~\ref{fig:ncsm_variational} shows the
effects the SRG has on convergence properties of a potential in the
oscillator basis. The curves are plotted by taking a potential at a
large $\nmax$ ($\sim 40$) and evolving it in that basis, then cutting
it to a specified $\nmax$ ($\sim 28$) before computing the binding
energy\footnote{This calculation is analogous to the situation in
Ref.~\cite{Bogner:2007rx} where the NN potential was evolved in
momentum space ($\nmax \longrightarrow \infty$) and then converted to
some finite oscillator basis.}. This procedure was repeated at
multiple values of $\hw$ for comparisons and we can see several
effects. First, the SRG evolves the initial potential to a new and
different Hamiltonian and therefore the optimal $\hw$ for that
Hamiltonian's expansion will be different. In fact, it will be lower
due to the way in which the Hamiltonian has been altered; high
momentum matrix elements have been reduced and simplified so that the
Hamiltonian needs less basis size to achieve the same convergence as
before. Therefore the basis can accommodate information with a smaller
$\hw$ and gets the benefit of increased resolution due to the smaller
$\hw$. In other words, a lower $\Lambda_{\rm UV}$ is sufficient so a
lower $\Lambda_{\rm IR}$ is affordable. Thus the optimal $\hw$ shifts
downward. Furthermore, each step in the evolution increases the
convergence at a given basis size and so the overall dependence on
$\hw$ decreases and the curves flatten out as $\lambda$ decreases.

\chapter{Symmetric Jacobi Oscillator Basis
in One and Three Dimensions}
\label{chapt:app_osc_basis}

\section{Jacobi Coordinates}

The initial (i.e., unevolved) one-dimensional  Hamiltonian for
$A$ bosons of equal mass $m$ with a local two-body potential has
the first-quantized form (in units with $\hbar=1$)
\beqn
  H = \frac{1}{2m}\sum_{i=1}^Ak_i^2 + \sum_{i<j=1}^AV(x_i-x_j) \;,
  \label{eq:hamiltonian}
\eeqn
where the $x_i$ are single-particle coordinates and the $k_i$ are
single-particle momenta. To connect to the nuclear problem of interest
that uses potentials given in a momentum basis (e.g., chiral effective
field theory potentials), we calculate matrix elements using harmonic
oscillator basis states in Jacobi momentum coordinates. This
representation also provides a clean visual interpretation of the SRG
evolution of potentials.

With equal-mass particles, a convenient set of relative momentum
Jacobi coordinates is defined by (for $j = 1 $ to $A-1$)
\beqn
 p_j = \sqrt{\frac{j}{j+1}}\left(\frac{1}{j}\sum_{i=1}^j k_i -k_{j+1}\right) 
 \;,
 \label{eq:Jacobi_coords}
\eeqn
where the $k_i$ are the single-particle momenta of the $A$ particles.
This is one particular choice of normalization for the Jacobi
coordinates. This choice is especially useful here because it sets the
reduced masses for each coordinate equal to each other, simplifying
rotation operations performed later. We define the Fourier transform
$V(x_1-x_2)$ to ${V}(p_1,p_1')$ as
\beqn
  {V}(p_1,p_1') 
    = \int V(\sqrt{2}\ell_1)e^{-i(p_1-p_1')\ell_1} d\ell_1 \;,
  \label{eq:fourier_transform}
\eeqn
where $\ell_1 = (x_1-x_2)/\sqrt{2}$ is the coordinate conjugate
to the Jacobi momentum $p_1$. We introduce a set of harmonic
oscillator states $| n_j \ra$ corresponding to each of the
coordinates of Eq.~(\ref{eq:Jacobi_coords}), so a general product
basis state has the form
\beqn
 \prod_{j=1}^{A-1} \la p_j|n_j\ra \;,
 \label{eq:ho_basis}
\eeqn
with $n_j = 0,1,2,\ldots, \nmax$ for each $j$. In the next section
we discuss how to build linear combinations of these states that
have the appropriate symmetry.

\section{Transformation Brackets}
\label{sec:trans_brack}

Transformation brackets are the expansion coefficients in the
oscillator basis of one system of coordinates in terms of
another~\cite{moshinsky,shlomo}. We apply them to relate two different
choices of Jacobi coordinates. Here, we show the relevant
transformation using the three-particle harmonic oscillator states
defined in Eq.~\eqref{eq:ho_basis} and then generalize at the end.

The single particle momenta are $k_1$, $k_2$, and $k_3$. The
unprimed Jacobi momenta [see Eq.(~\ref{eq:Jacobi_coords})] are 
\bea
p_1 &=&  \frac{1}{\sqrt{2}} (k_1 - k_2) \;, \nonumber \\
p_2 &=& \sqrt{\frac{2}{3}} ((k_1+k_2)/2 - k_3) \;,
\eea
and the primed coordinates are obtained from exchanging $k_2$ and
$k_3$:
\bea
p_1' &=&  \frac{1}{\sqrt{2}} (k_1 - k_3) \;, \nonumber \\
p_2' &=& \sqrt{\frac{2}{3}} ((k_1+k_3)/2 - k_2) \;.
\eea
After some algebra, the transformation that exchanges the last
two particles (i.e., $k_2$ and $k_3$) can be written as
\bea
  \left( \begin{array}{c} p_1' \\ p_2' \end{array}\right) 
   &=& \left( \begin{array}{cc} \frac{1}{2} & \frac{\sqrt{3}}{2} 
   \\[4pt]
  \frac{\sqrt{3}}{2} & -\frac{1}{2}  \end{array}\right) 
  \left( \begin{array}{c} p_1 \\ p_2 \end{array}\right) \;.
  \label{eq:p_transformation}
\eea
which enables us to express the primed oscillator states in terms
of the unprimed ones. 

We denote the three-particle oscillator basis by $|n_1n_2\ra =
\eta_1^{\dagger}\eta_2^{\dagger}|0\ra$ where we have set
$\hw = 1$ for simplicity in this appendix. Note that the
transformation brackets will not depend on the value of
$\hw$. The
transformation that exchanges the last two single-particle
coordinates can again be written as
\bea
  \left( \begin{array}{c} \eta_1' \\ \eta_2' \end{array}\right) 
   &=& \left( \begin{array}{cc} \frac{1}{2} & \frac{\sqrt{3}}{2}  
   \\[4pt] 
  \frac{\sqrt{3}}{2} & -\frac{1}{2}  \end{array}\right) 
  \left( \begin{array}{c} \eta_1 \\ \eta_2 \end{array}\right) \;,
  \label{eq:eta_transformation}
\eea
The derivation of the
harmonic oscillator transformation bracket follows directly as
\newcommand{\Anorm}{\frac{1}{\sqrt{n_1!n_2!n_1'!n_2'!}}}
\bea
  \la n_1'n_2' | n_1n_2 \ra_3 &=& \la 0| \Anorm
   \eta_1'^{n_1'}\eta_2'^{n_2'} \eta_1^{\dagger n_1} 
   \eta_2^{\dagger n_2}|0\ra \nonumber \\ 
&=& \la 0|\Anorm \Bigl[\frac{1}{2}\eta_1 
   + \frac{\sqrt{3}}{2}\eta_2\Bigr]^{n_1'} \nonumber \\
&& \hspace{2cm}  \null\times \Bigl[\frac{\sqrt{3}}{2}\eta_1 -
  \frac{1}{2}\eta_2\Bigr]^{n_2'} 
  \eta_1^{\dagger n_1} \eta_2^{\dagger n_2}  |0\ra \nonumber \\
&=& \la 0|\Anorm  \sum_{k=0}^{n_1'} {n_1' \choose k}
  \Bigl[\frac{1}{2}\eta_1\Bigr]^{n_1'-k}
  \Bigl[\frac{\sqrt{3}}{2}\eta_2\Bigr]^k 
     \nonumber \\
&& \hspace{2cm}  \null\times\sum_{j=0}^{n_2'} {n_2' \choose j}
 \Bigl[\frac{\sqrt{3}}{2}\eta_1\Bigr]^{n_2'-j}
 \Bigl[-\frac{1}{2}\eta_2\Bigr]^j 
\eta_1^{\dagger n_1}\eta_2^{\dagger n_2} |0\ra \nonumber \\
&=& \Anorm \sum_{k=0}^{n_1'} \sum_{j=0}^{n_2'} 
{n_1' \choose k}{n_2' \choose j} 
\biggl[\frac{1}{2}\biggr]^{n_1'-k+j}
\biggl[\frac{\sqrt{3}}{2}\biggr]^{n_2'-j+k}(-1)^j \nonumber \\
&& \hspace{2cm}  \null\times n_1!n_2!\delta_{n_1'-k+n_2'-j,n_1}
\delta_{k+j,n_2} \nonumber \\
&=& \sqrt{\frac{n_1!n_2!}{n_1'!n_2'!}} 
\sum_{k=0}^{n_1} {n_1' \choose k}{n_2' \choose n_2-k}
\biggl[\frac{1}{2}\biggr]^{n_1'+n_2-2k}
\biggl[\frac{\sqrt{3}}{2}\biggr]^{n_2'-n_2+2k}(-1)^{n_2-k} \;.
\label{eq:trans_bracket}
\eea
The second line is obtained from operating the transformation
on the creation operators  $\eta_s^\dagger$. The third line is
the application of the binomial theorem. The fourth  balances the
oscillator creation and annihilation, and the fifth is just some
simplification. 

In the general A-particle system the transformation to exchange
the last two particles, $k_{A-1}$ and $k_A$, can be written as
\bea
\left( \begin{array}{c} \eta_{A-2}' \\ \eta_{A-1}' \end{array}\right) 
 &=& \left( \begin{array}{cc} 
  \sqrt{\frac{1}{d+1}} & \sqrt{\frac{d}{d+1}}  \\[4pt] 
  \sqrt{\frac{d}{d+1}} & -\sqrt{\frac{1}{d+1}}  \end{array}\right) 
\left( \begin{array}{c} \eta_{A-2} \\ \eta_{A-1} \end{array}\right) \;,
\label{eq:etap_trans}
\eea
where $d = (A-1)^2-1$ is the number of generators of the rotation
group, $U(A-1)$, or the group $U(A)$ with the center of mass
coordinate held fixed.  An expression for the bracket $\la
n_{A-2}n_{A-1} | n_{A-2}'n_{A-1}' \ra_{A(A-2)}$, which appears in
Eq.~(\ref{eq:symmetrizer_A}), is obtained from
Eq.~(\ref{eq:trans_bracket}) by substituting the general
coordinate transformation Eq.~(\ref{eq:etap_trans}) for the
three-particle transformation Eq.~(\ref{eq:eta_transformation}),
or $\sqrt{1/(d+1)}$ and $\sqrt{d/(1+d)}$ for $1/2$ and
$\sqrt{3}/2$.


\section{Symmetrization}
\label{sec:symmetrization}

We carry out the SRG evolution for each $A$-particle subsystem in a
complete basis of properly symmetrized states, which will be linear
combinations of the basis states of Eq.~(\ref{eq:ho_basis}). The
symmetrization procedure is adapted from the procedure developed for
NCSM calculations~\cite{NCSM1a,NCSM1b,NCSM1c}. This entails
symmetrizing first the two-particle system and then using a recursive
procedure to go from the $(A-1)$-particle  basis to an $A$-particle
basis. At each stage we keep only symmetric states, identified as
eigenstates of the symmetrizer with eigenvalue unity.

The two-particle system is specified by the oscillator number $n_1$. 
The symmetrizer is $(1+P_{12})/2$, where $P_{ij}$ is the exchange
operator between particles $i$ and $j$.  Because $P_{12} |n_1\ra =
(-1)^{n_1} |n_1 \ra$, the symmetrizer in the two-particle case has
eigenvalue one acting on states with $n_1$ even and zero when acting
on states with $n_1$ odd. Thus the symmetric basis states have $n_1$
even and we simply omit the odd states. Following conventions from
Ref.~\cite{NCSM1a}, we label these eigenstates as $|N_2i_2\ra$, where
$N_2$ is  the total oscillator number of the symmetric state and $i_2$
is an arbitrary label that distinguishes states degenerate in $N_2$.
In the two-particle case the notation is trivial, with $N_2 = n_1$
even  and $i_2 = 1$. We write eigenstate projection coefficients as 
$\la N_2 i_2 \| n_1\ra = \delta_{N_2,n_1}(1+(-1)^{n_1})/2$. These are
referred to in the literature as the ``coefficients of fractional
parentage".

A three-particle basis is specified by product states of  the two-body
symmetric eigenstates, $|N_2i_2\ra$, and single-particle states with
the oscillator number corresponding  to the third particle, $|n_2\ra$.
The symmetrizer for this system is governed  by the permutation group,
$S_3$, which can be defined by just two of its generators. Here we
choose the permutation operators $P_{12}$ and $P_{23}$. The
symmetrization operator can be written as
\beqn
  S = \frac{1}{6}(1 + P_{12} + P_{23} + P_{12}P_{23} 
          + P_{23}P_{12} + P_{12}P_{23}P_{12}) \; .
\label{eq:symmetrizer_3N}
\eeqn
We build this symmetrizer in the basis $|N_2i_2;n_2\ra \equiv | N_2
i_2 \ra | n_2 \ra$ where the states $|N_2i_2\ra$ are already
eigenstates of $P_{12}$ with eigenvalue one, so
Eq.~(\ref{eq:symmetrizer_3N}) reduces to $S = (1+2P_{23})/3$.

In this basis, the matrix elements of $P_{23}$ can be expressed as 
\beqn
  \la N_2'i_2';n_2'|P_{23}|N_2i_2;n_2\ra 
    = \delta_{N',N}\la n_1'n_2'|n_1n_2\ra_3 \;,
  \label{eq:trans_bracket_3N}
\eeqn
where $N \equiv N_2+n_2 = N_2'+n_2' = N'$ and $\la
n_1'n_2'|n_1n_2\ra_3$ is the one-dimensional harmonic oscillator
transformation bracket for particles with mass ratio
3~\cite{Trlifaj:1972zz}. We construct
these transformation brackets and generalize to mass ratio $d$ in
section~\ref{sec:trans_brack}. By diagonalizing this symmetrizer we
identify the symmetric eigenstates of the system as the ones with
eigenvalue unity.  We keep only those states and discard the others. 
This set of eigenvectors gives us the coefficients of fractional
parentage, $\la N_2 i_2; n_2 \| N_3 i_3 \ra$, of the three-boson
symmetric eigenstates, $| N_3 i_3 \ra$, in terms of the original
partially symmetrized three-particle space, $| N_2 i_2; n_2 \ra $. 
Note that $i_3$ is not trivial like $i_2$, because in the three-body
system there are eigenstates degenerate in the total oscillator
number, $N_2+n_2$. The label $i_3$ keeps track of those degeneracies.
We find in the one-dimensional system of bosons that the fraction of
symmetric basis states for $A=3$ is about one-fifth. For $A=4$ the
reduction in number of states is above 90\%.

\begin{figure}
\begin{center}
\strip{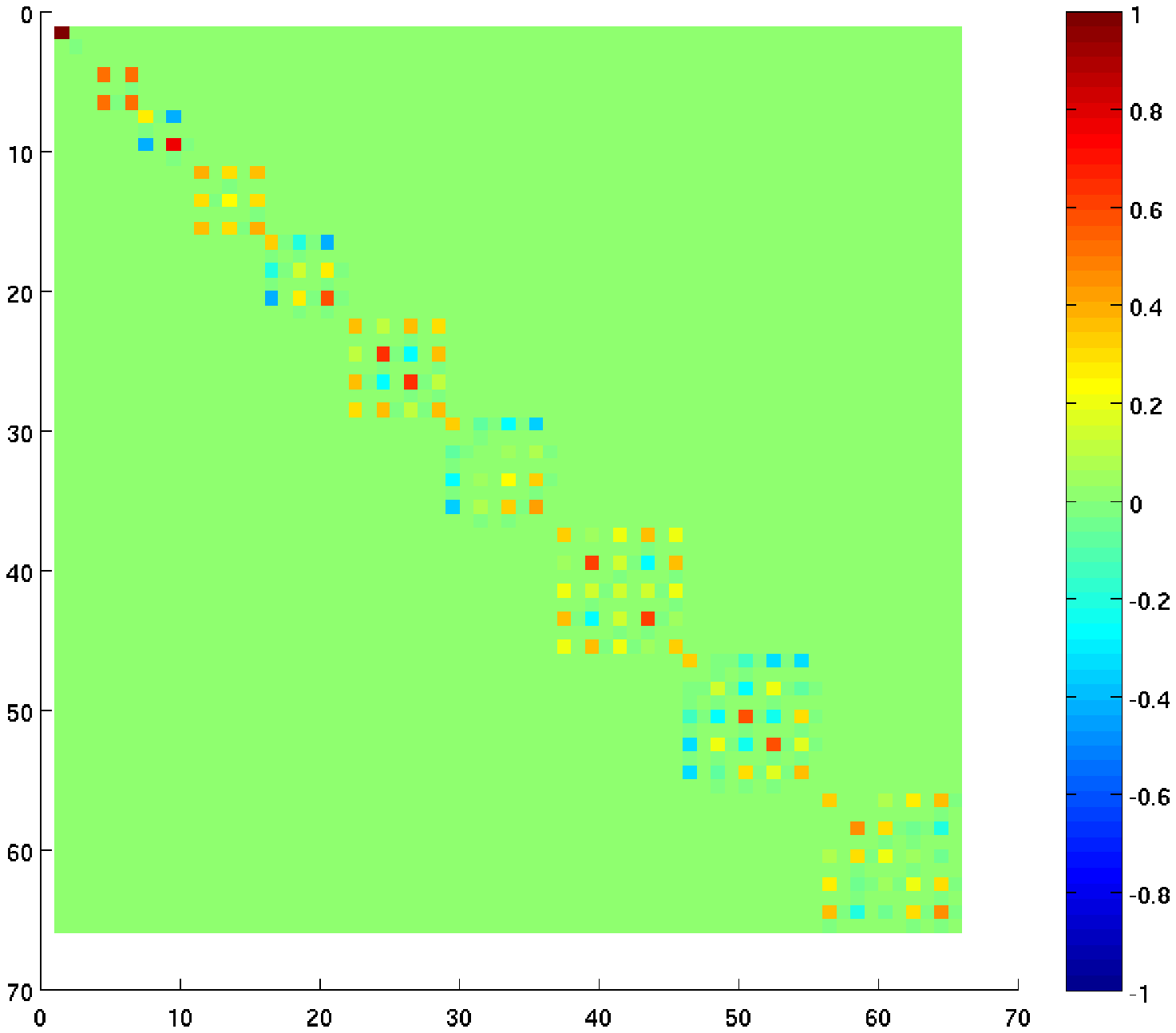}
\end{center}
\captionspace{The full symmetrizer for the $|N_2i_2;n_2\ra$ space, with
both physical and spurious states. A small, $\nmax=10$, basis is shown
for clarity. Note also that Matlab has chopped off the last column 
when making the figure; this is a symptom of my very simple plotting
routine.}
\label{fig:S_ho}
\end{figure}

The symmetrizer can now be coded directly in the three-particle
oscillator space. A picture of the resulting matrix is shown in
Fig.~\ref{fig:S_ho}. This matrix has the three-particle
oscillator basis states organized in a block diagonal form
because the total oscillator number, $N=n_1+n_2$, is a conserved
quantum number proportional to the total energy of a state.

\begin{figure}
\begin{center}
\strip{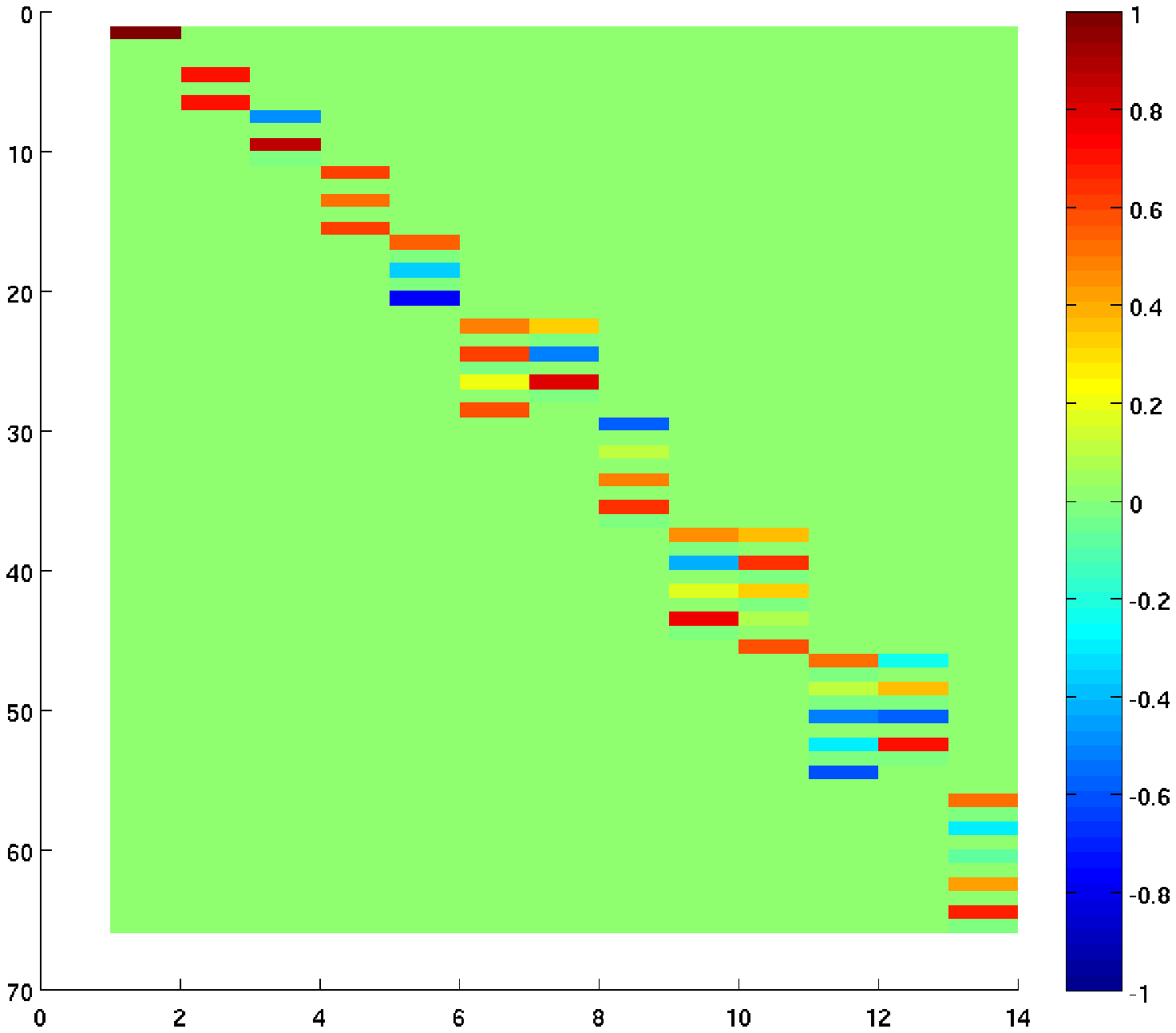}
\end{center}
\captionspace{The physical eigenvectors of the symmetrizer. Note vector
components corresponding to $\nmax$ blocks in Fig.~\ref{fig:S_ho}. The
same size basis as in Fig.~\ref{fig:S_ho}, $\nmax=10$, is shown.
Again, Matlab has chopped off the last column.}
\label{fig:U_Nn1n2}
\end{figure}

Remembering that we need only the physical eigenstates of the
symmetrizer, we diagonalize the matrix and obtain the eigenvectors. We
keep only those vectors which have nonzero eigenvalue (the physical
states), thus resulting in a symmetrizing matrix about one-fifth the
size  of the original oscillator basis. A picture of the sorted
eigenvectors is shown in  Fig.~\ref{fig:U_Nn1n2}. These are all the
symmetric states as expressed in the  three-particle oscillator basis.
The symmetrizer in this basis is, of course, diagonal, and all the
eigenvalues are unity.

The color scheme in Figs.~\ref{fig:S_ho} and \ref{fig:U_Nn1n2} is
displayed to the right  of each graph; green is zero, red is
positive, and blue is negative. Notice that the eigenvectors are
normalized to one. The easiest non-trivial case is the second
vector where each of the two components is $1/\sqrt{2}\approx
.7$, which is shown in a bright red.

To construct the basis states for higher $A$, we generalize this
procedure. To go from $A-1$ to $A$ we need only to symmetrize
between the last two particles, so we construct the symmetrizer
\beqn
  S_A = \frac{1}{A}\bigl(1+(A-1)P_{(A-1)A} \bigr)
  \label{eq:symmetrizer_AN}
\eeqn
in the space of $(A-1)$-particle symmetric eigenstates and the
additional Jacobi state, $n_{A-1}$. We label the basis of
this space as $|N_{A-1} i_{A-1}; n_{A-1}\ra$.  The matrix element
of the exchange operator in this space is
\bea
 && \la N_{A-1}' i'_{A-1}; n_{A-1}'| P_{(A-1)A} | N_{A-1} i_{A-1}; n_{A-1}\ra 
    \nonumber \\
    && \qquad\qquad =   
   \delta_{{N'_{A-1}+n'_{A-1}},{N_{A-1}+n_{A-1}}}\sum
   \la N_{A-1}' i'_{A-1} \| N_{A-2} i_{A-2}; n_{A-2}'\ra 
    \nonumber \\
    &&  \qquad\qquad\quad \null\times
    \la N_{A-2} i_{A-2}; n_{A-2} \| N_{A-1} i_{A-1} \ra
   \la n_{A-2}' n_{A-1}' | n_{A-2} n_{A-1}\ra_{A(A-2)} 
   \;,
  \label{eq:symmetrizer_A}
\eea
where the sum is over $N_{A-2}$, $i_{A-2}$, $n_{A-2}$ and
$n_{A-2}'$. The only significant difference from the
three-particle case is that we must sum over the components of
the $A-1$ subcluster symmetric states to get all the
contributions to the exchange of the last two bosons, $n_{A-2}$
and $n_{A-1}$. The parameter $d=A(A-2)$ can be derived by taking
the last two Jacobi coordinates $p_{A-2}$ and $p_{A-1}$, as
defined in Eq.~(\ref{eq:Jacobi_coords}), and finding the
transformation that exchanges particles labeled by $k_{A-1}$ and
$k_A$. This procedure was explained in section~\ref{sec:trans_brack}

For fermions, we need a complete basis of fully anti-symmetrized
states. If we consider the one-flavor case, the procedure for our
one-dimensional model is a trivial modification of
Eqs.(\ref{eq:symmetrizer_3N}) and (\ref{eq:symmetrizer_AN}), namely
all odd permutations come with a minus sign. Thus, for $A=3$ the
anti-symmetrizer can be written
\beqn
A = \frac{1}{3}(1-2P_{23}) \;,
\eeqn
where $P_{23}$ acts on the flavor space as well. If there are more
flavors than particles and the interaction is flavor independent, the
spatial wavefunction for the ground state will be symmetric and
correspond to our boson ground state wavefunctions. For realistic
three-dimensional nuclei, the required construction of an
anti-symmetric Jacobi basis with full angular momentum coupling has
been worked out for the NCSM by Navratil et
al.~\cite{NCSM1a,NCSM1b,NCSM1c} and is briefly discussed in
section~\ref{sec:osc_basis_3d}.


\section{Hamiltonian Matrix Elements}
\label{sec:hamiltonian}

To obtain the Hamiltonian in the symmetric eigenbasis of the
general $A$-particle system, we employ a recursive embedding
procedure that utilizes the partially symmetric bases developed
for the symmetrization operator. First we treat the kinetic
energy and then the potential.

The relative kinetic energy in the three-particle system is the total 
minus the center-of-mass kinetic energies:
\bea
  T_{\rm rel} &\equiv& T_{\rm tot} - T_{\rm cm} \nonumber \\ 
  &=& \frac{k_1^2}{2m} + \frac{k_2^2}{2m} + \frac{k_3^2}{2m} 
  - \frac{(k_1+k_2+k_3)^2}{2(3m)} \nonumber \\ 
  &=& \frac{p_1^2 + p_2^2}{2m} \;,
  \label{eq:Trel}
\eea
where the $p_i$'s are defined in  Eq.~(\ref{eq:Jacobi_coords}).
Momentum basis states are organized by  increasing kinetic
energy. We can project $T_{\rm rel}$ directly onto the
three-particle oscillator basis by using the ladder operator
definitions of the Jacobi momenta. The projection of $T_{\rm
rel}$ into the $\noscr$ basis is
\bea
\la n_1' n_2' | T_{\rm rel} |n_1n_2\ra
  &=&  \la n_1' n_2'|\frac{p_1^2 + p_2^2}{2m} |n_1n_2\ra
    \nonumber \\ 
  &=&  \frac{1}{2m} \frac{-m\omega}{2} 
  \la n_1' n_2'|(\eta_1^\dagger - \eta_1)^2 + (\eta_2^\dagger - \eta_2)^2 |n_1n_2\ra 
 \;,
\eea
where the $\eta_1$ and $\eta_2$ operators act on the $n_1$ and $n_2$ spaces, 
respectively.  Continuing, we get
\bea
\la n_1' n_2' | T_{\rm rel} |n_1n_2\ra  &=& \frac{1}{2m} \frac{-m\omega}{2} 
    [\la n'_1 | (\eta_1^\dagger - \eta_1)^2 | n_1 \ra \delta_{n_2,n'_2} 
    + \la n'_2 | (\eta_2^\dagger - \eta_2)^2 | n_2 \ra \delta_{n_1,n'_1}] 
   \nonumber \\ 
&=& \frac{-\omega}{4} 
          \Bigl[ \bigl( \sqrt{(n_1+1)(n_1+2)}\,\delta_{n'_1,n_1+2} 
   + \sqrt{n_1(n_1-1)}\,\delta_{n'_1,n_1-2} 
   \nonumber \\ 
   & & \qquad \null 
      - (2n_1+1)\delta_{n'_1,n_1}
   \bigr)\, \delta_{n'_2,n_2} 
   \nonumber \\ 
 && \qquad \null + \bigl( \sqrt{(n_2+1)(n_2+2)}\,\delta_{n'_2,n_2+2} 
   + \sqrt{n_2(n_2-1)}\,\delta_{n'_2,n_2-2}
   \nonumber \\ 
   & & \qquad \null 
    - (2n_2+1)\delta_{n'_2,n_2} 
    \bigr)\, \delta_{n'_1,n_1}\Bigr] \;.
  \label{eq:T_exact}
\eea
As noted, we keep only the $n_1$-even states using the projector
$\la N_2 i_2 \| n_1\ra$, and we can symmetrize the full
three-particle system with the symmetric eigenstates, $|N_3i_3\ra$
 whose components are given by $\la N_3i_3\|N_2i_2;n_2\ra$. 

To derive the $A$-body kinetic energy in the symmetrized basis,
$(T_A)_{\rm sym}$, we use a recursive procedure on the
$(A-1)$-body result to find the $A$-particle space operator
matrix elements: 
\bea
  (T_{A})_{\rm sym} 
         &=& \la N_A' i'_A|T_A|N_A i_A\ra 
	        \equiv \la N_A' i'_A| \sum_{i=1}^{A-1} p_i^2/2m |N_A i_A\ra 
            \nonumber \\
         &=& \la N_A'\|N_{A-1}'n_{A-1}'\ra
             \la N_{A-1}'n_{A-1}'|T_A|N_{A-1}n_{A-1}\ra
             \la N_{A-1}n_{A-1}\|N_A\ra 
             \nonumber \\
         &=& \la N_A'\|N_{A-1}'n_{A-1}'\ra
             \bigl[ 
             \la N_{A-1}'n_{A-1}'|(T_{A-1})_{\rm sym} 
             \nonumber \\
         &&  \null + p_{A-1}^2/2m|N_{A-1}n_{A-1}\ra
             \bigr] \la N_{A-1}n_{A-1}\|N_A\ra 
             \nonumber \\
         &=& \la N_A'\|N_{A-1}'n_{A-1}'\ra
             \bigl[
             \la N_{A-1}'|(T_{A-1})_{\rm sym}|N_{A-1}\ra \delta_{n_{A-1}',n_{A-1}} 
             \nonumber \\
         &&  \null + \delta_{N_{A-1}',N_{A-1}}
             \la n_{A-1}'|p_{A-1}^2/2m|n_{A-1}\ra \bigr]
             \la N_{A-1}n_{A-1}\|N_A\ra 
             \nonumber \\
         &=& \la N_A'\|N_{A-1}'n_{A-1}'\ra
             \bigl[(T_{A-1})_{\rm sym}\delta_{n_{A-1}',n_{A-1}} 
             \nonumber \\
         &&  \null  - \frac{\omega}{4}\delta_{N_{A-1}',N_{A-1}}
             \bigl( \delta_{n_{A-1}',n_{A-1}}(2n_{A-1}+1) 
             - \delta_{n_{A-1}'+2,n_{A-1}}\sqrt{n_{A-1}^2 - n_{A-1}} 
             \nonumber \\
         &&  \null - \delta_{n_{A-1}'-2,n_{A-1}}\sqrt{(n_{A-1}+1)(n_{A-1}+2)} 
             \bigr)\bigr] \la N_{A-1}n_{A-1}\|N_A\ra  \; ,
	     \label{eq:T_AS}
\eea
where we have suppressed the $i_A$'s and $i_{A-1}$'s for
simplicity after the first line.  Intermediate summations over
$N_{A-1}$, $n_{A-1}$, $i_{A-1}$, $N_{A-1}'$, $n_{A-1}'$, and
$i_{A-1}'$ are implicit.

In the same manner as the kinetic energy, we can recursively
embed the potential in the $A$-particle space, starting with the
two-body interaction between the first two particles. Because we
are working in fully symmetrized few-body spaces we do not need
to consider all pair-wise interactions, but only one such pair
and scale by the number of interactions. For instance, in the
three-particle system the full two-body interaction is $V^{(2)} =
V_{12} + V_{23} + V_{13} = 3 V_{12}$. In a general $A$-particle
space, this becomes $V^{(2)} = {A \choose 2} V_{12}$. The matrix
element of a two-body potential, $V_{12}$, in the relative
coordinate harmonic oscillator basis, $|n_1\ra$, is $\la n_1'|
V_{12} | n_1\ra = \int \la n_1'| p_1' \ra \la p'| V_{12} | p \ra
\la p |n_1\ra dp\, dp'$ where the matrix elements of $\la
p'|V_{12}|p\ra$ are given by  Eq.~\eqref{eq:fourier_transform}. 

Once in the oscillator basis, embedding in a larger particle
space is a straightforward process. Starting with the two-body
interaction, $V_{12}$, the two-body oscillator symmetric states
are isolated using the projector $\la N_2 i_2|n_1 \ra$ which picks
out just the $n_1$-even states. Embedding this interaction in the
three-particle space involves adding a new Jacobi coordinate,
$|n_2\ra$, to the existing system. With respect to the two-body
interaction, $V_{12}$, this additional coordinate is associated
with a delta function, $\delta_{n_2,n_2'}$. Finally we obtain the
symmetric three-particle states by using  the projector,  $\la
N_3i_3||N_2i_2;n_2\ra$. Multiplying by ${3 \choose 2} = 3$ gives
us the full strength of the two-body interaction.

In general we can write this procedure as an expansion of the final
$A$-particle symmetric space matrix elements of $V_{12}$:
\bea
  (V^{(2)}_{A})_{\rm sym} &=& \la N_A' i'_A|V^{(2)}_A|N_A i_A\ra 
  	\equiv  {A \choose 2} \la N_A' i'_A|V_{12}|N_A i_A\ra \nonumber \\
         &=& \la N_A'\|N_{A-1}'n_{A-1}'\ra
             \la N_{A-1}'n_{A-1}'|V_A|N_{A-1}n_{A-1}\ra
             \la N_{A-1}n_{A-1}\|N_A\ra 
             \nonumber \\
         &=& \la N_A'\|N_{A-1}'n_{A-1}'\ra \nonumber \\
	 && \null \times \la N_{A-1}'n_{A-1}'|(V_{A-1})_{\rm sym}
	     \delta_{n_{A-1}',n_{A-1}}|N_{A-1}n_{A-1}\ra
             \la N_{A-1}n_{A-1}\|N_A\ra  
\label{eq:v_osc_A}
\eea
where again we have dropped the $i_A$'s after the first line for
simplicity and intermediate sums are implicit. We remind the
reader that $n_{A-1}$ can only take values from $0$ to $N -
N_{A-1}$, where $N$ is the total oscillator quantum number used
to organize the states. We start with the two-particle space and
work our way up to the $A$-body space, embedding the interactions
successively in each sector using Eq.~(\ref{eq:v_osc_A}). 

When symmetrizing $V_A$ we must embed the symmetrized $V_{A-1}$
with the appropriate combinatoric factor included as explained
above. This factor derives from the fact that we had embedded a
2-body force in the $A-1$ space that is now to be extracted and
embedded in the $A$-particle space. Thus we must remove the old
factor $A-1 \choose{2}$ and multiply by the new $A \choose{2}$
factor,  which has the net effect of  multiplying by $A/(A-2)$. 

Any initial three-body force (discussed below) is embedded in the
same manner as above except that it originates in the
three-particle space. The initial three-body force is a function
of two Jacobi momenta, which we transform directly into the
partially symmetrized three-particle oscillator space and then
use all of the same embedding procedures developed above. Note
that two- and three-body forces must be embedded in higher spaces
with different symmetry factors, ${A \choose 2}$ and ${A \choose
3}$ respectively. 

In previous formulations of this recursive approach~\cite{NCSM1a},
subsequent potential embeddings are achieved by making a change
of coordinates for the last two Jacobi momenta. For systems with
$A>5$, the three-body force requires a similar change of
coordinates for the last three Jacobi momenta. Such a scheme 
is unnecessary here.


\section{Three-Dimensional NCSM}
\label{sec:osc_basis_3d}

In this thesis, only the one-dimensional NCSM was built for SRG
studies. However, the three-dimensional codes were modified and used
for the realistic calculations in chapter~\ref{chapt:ncsm}. In three
dimensions the Jacobi harmonic oscillator basis differs in several
significant ways that complicate the expressions involved
considerably, though in a straightforward way. However, the basic form
of the basis remains the same as in the one-dimensional case. 


The three-dimensional NCSM basis is, like the one-dimensional case,
organized in $\nmax$ blocks, so scaling up the basis size or making
cuts to study decoupling is a simple matter. It is variational in
$\nmax$ and $\hw$ as noted in App.~\ref{chapt:app_osc_truncation}. It
is built recursively from one $A$-body sector to the next, and the
Hamiltonian is embedded in subsequent sectors with combinatoric
factors, taking advantage of the symmetry properties. 


The three-dimensional NCSM was built to calculate systems of nucleons,
which are fermions, while the one-dimensional model was built based on
a system of bosons with no spin or angular momentum. So in
three-dimensions, we have to build an antisymmetrizer to find fully
antisymmetric states in which fermions would reside. This still
employs the same strategy of building on the $A-1$ subcluster. So a
new Jacobi coordinate is added to the $A-1$ space and the matrix
elements of the antisymmetrizer,
\beqn
\chi = \frac{1}{A}(1-(A-1)P_{(A-1)A}) \;,
\eeqn
are computed. As before, $P_{(A-1)A}$ is the permutation operator
between the $A$th and $(A-1)$st nucleons. Here the only difference is
the minus sign. And the fact that $P$ now also transforms angular
momenta. 


Now in three dimensions, we use the three-dimensional isotropic
harmonic oscillator wavefunctions as displayed in
Sec.~\ref{sec:osc_function_3D}. The quantum numbers of the nuclear
system are now $n,\ell,s,j,$ and $t$, the radial, orbital angular
momentum, spin, total angular momentum, and isospin of the state. In
the three-dimensional oscillator, the conserved energy is $N=2n+\ell$
which will impact embedding procedures from one $\nmax$ sized basis to
another. Now the anti-symmetrized states are labeled as
\bea
|N_Ai_AJ_AT_A\ra &=& \sum \la N_{A-1}i_{A-1}J_{A-1}T_{A-1};
n_{A-1}\ell_{A-1}j_{A-1}|| N_Ai_AJ_AT_A\ra \nonumber \\
&& \quad \times |N_{A-1}i_{A-1}J_{A-1}T_{A-1};
n_{A-1}\ell_{A-1}j_{A-1} \ra \;,
\eea
where the $\la N_{A-1}i_{A-1}J_{A-1}T_{A-1};
n_{A-1}\ell_{A-1}j_{A-1}|| N_Ai_AJ_AT_A\ra$ are the three-dimensional
coefficients of fractional parentage. These are given by the
eigenstates of the antisymmetrizer in the non-antisymmetrized basis just
as their one-dimensional counterparts were obtained from the
eigenstates of the symmetrizer.

To build the antisymmetrizer, $\chi$, the matrix elements of
$P_{(A-1)A}$ in the non-antisymmetrized basis can be written as
\bea
&& \la N'_{A-1}i'_{A-1}J'_{A-1}T'_{A-1}; 
n'_{A-1}\ell'_{A-1}j'_{A-1}JT|P_{(A-1)A}|
N_{A-1}i_{A-1}J_{A-1}T_{A-1};n_{A-1}\ell_{A-1}j_{A-1}JT\ra \nonumber \\
&& \quad = \delta_{N',N} \sum \la N_{A-2}i_{A-2}J_{A-2}T_{A-2};
n'_{A-2}\ell'_{A-2}j'_{A-2}||N'_{A-1}i'_{A-1}J'_{A-1}T'_{A-1}\ra \nonumber \\
&& \quad\quad\times \la N_{A-2}i_{A-2}J_{A-2}T_{A-2};
n_{A-2}\ell_{A-2}j_{A-2}||N_{A-1}i_{A-1}J_{A-1}T_{A-1}\ra \nonumber \\
&& \quad\quad\times T_{A-1}T'_{A-1}(-1)^{T_{A-1}+T'_{A-1}+j_{A-2}+j'_{A-2}}
\left\{ \begin{array}{ccc}
1/2 & T_{A-2} & T_{A-1} \\
1/2 & T & T'_{A-1}
\end{array}\right\} \nonumber \\
&& \quad\quad\times j'_{A-2}j_{A-2}j'_{A-1}j_{A-1}J'_{A-1}J_{A-1}K^2
\left\{ \begin{array}{ccc}
J_{A-2} & j'_{A-2} & J'_{A-1} \\
j_{A-2} & K & j'_{A-1} \\
J_{A-1} & j_{A-1} & J
\end{array}\right\} \nonumber \\
&& \quad\quad\times
\left\{ \begin{array}{ccc}
\ell'_{A-2} & \ell_{A-1} & K \\
j_{A-1} & j'_{A-2} & 1/2
\end{array}\right\}
\left\{ \begin{array}{ccc}
\ell_{A-2} & \ell'_{A-1} & K \\
j'_{A-1} & j_{A-2} & 1/2
\end{array}\right\}
\left\{ \begin{array}{ccc}
\ell'_{A-1} & \ell_{A-2} & K \\
\ell_{A-1} & \ell'_{A-2} & L
\end{array}\right\} \nonumber \\
&& \quad\quad\times
L^2(-1)^{\ell_{A-2}+\ell'_{A-1}+L}
 \la n'_{A-1}\ell'_{A-1}n'_{A-2}\ell'_{A-2}L|
n'_{A-2}\ell'_{A-2}n'_{A-1}\ell'_{A-1}L\ra_{A(A-2)} \;,
\eea
where the quantities in curly brackets are the Wigner 6$j$ and 9$j$ symbols
used to compute the couplings between the angular momenta and isospin
quantum numbers. The last line is the three-dimensional oscillator
transformation bracket as derived in various sources with various
methods~\cite{moshinsky,shlomo,Trlifaj:1972zz}.



Given the scaling properties of this basis discussed in
appendix~\ref{chapt:app_scaling}, several techniques have been
developed to improve the efficiency of the three-dimensional
technique. In Ref.~\cite{Deveikis:1998mu} a method is introduced that
allows computation of only some antisymmetrizer matrix elements
instead of requiring the full antisymmetrizer. Another technique deals
with the cumbersome increase of angular momentum channels with $A$.
Some authors~\cite{barnea:1997} have developed techniques to better
organize these channels and make the coding more manageable. Another
approach to this problem is to convert to the single-particle, or
Slater determinant, basis where the all the angular momentum embedding
procedures are more straightforward. However, those calculations
require large clusters and sophisticated memory handling techniques to
calculate even the lightest nuclei at small $\nmax$. The details of
the single-particle basis scaling are discussed in
appendix~\ref{chapt:app_scaling}. A one-dimensional single-particle
basis was constructed and is presented in
appendix~\ref{chapt:app_mscheme}.

\chapter{Scaling properties}
\label{chapt:app_scaling}

In this appendix we document the computational scaling features of the
various calculations in this thesis. An introduction to the scaling
issues involved in schemes such as the NCSM is necessary to appreciate
the computational improvements brought about by the SRG.

\section{Momentum Space NN calculations}
\label{sec:timing_momentum}

\begin{figure*}
\begin{center}
\dblpic{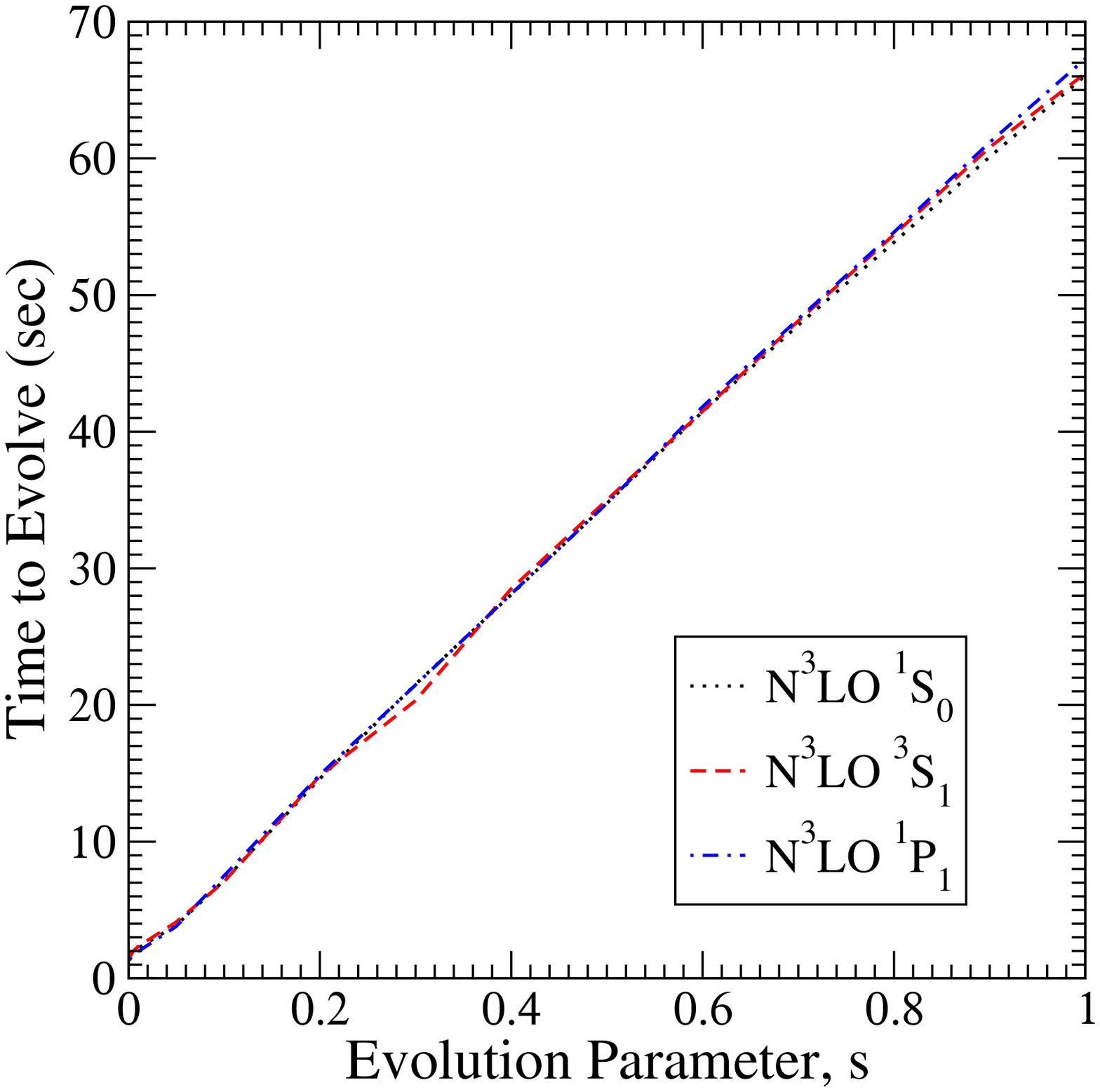}
\dblpic{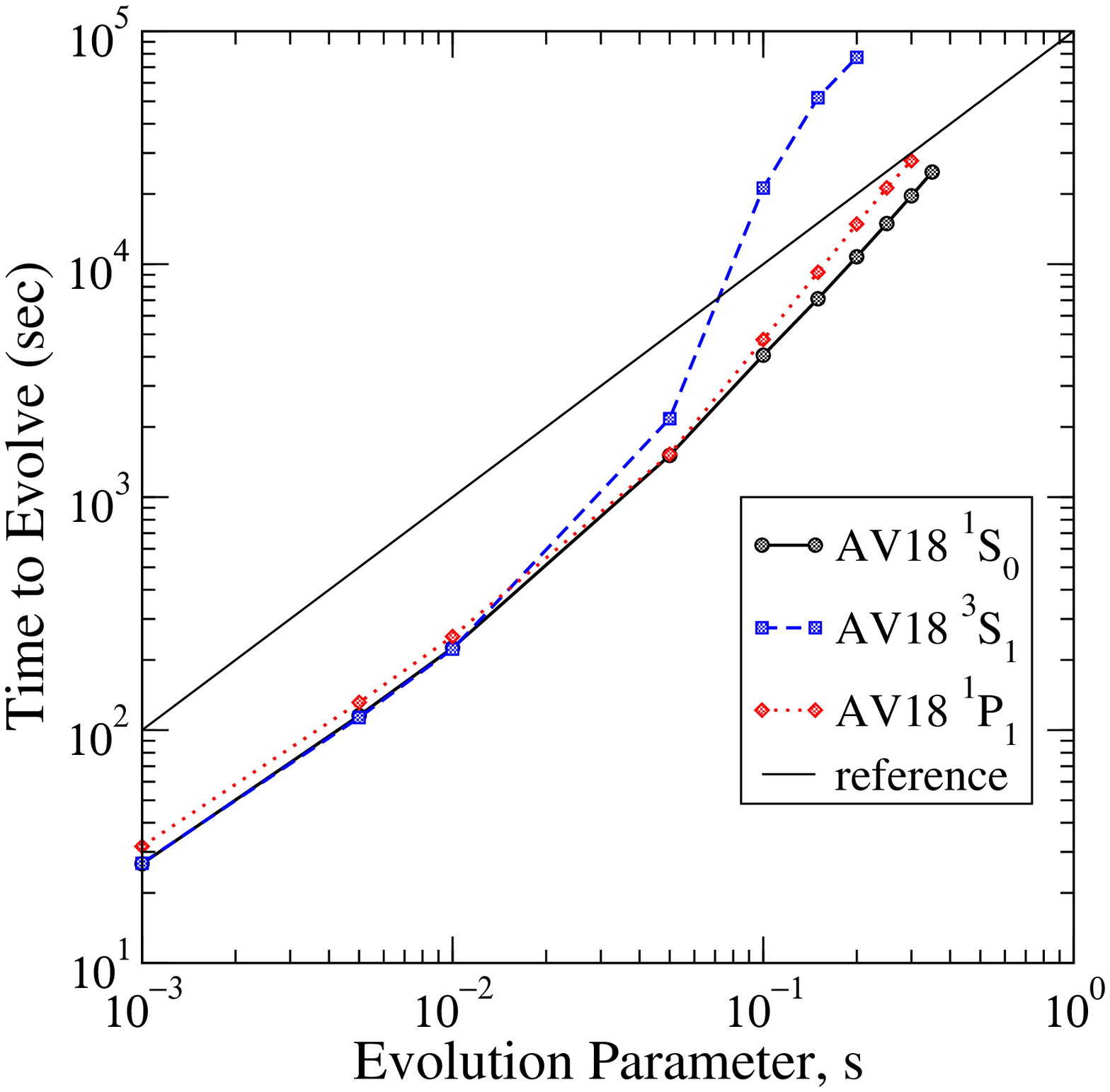}
\end{center}
\captionspace{SRG timings vs the parameter $s$ show no stiffening for
large $s$ (small $\lambda$). The left shows timing results for N$^3$LO
(500 MeV) potentials in several partial waves and the right shows the
same for Argonne $v_{18}$ but in a log-log plot.}
\label{fig:srg_timing_momentum}
\end{figure*}

The first problems addressed in chapter~\ref{chapt:decoupling}
involved applying the SRG to interactions projected onto a momentum
representation. The mesh used here is most often gaussian quadrature
with the grid dimension at about 100. Figure
\ref{fig:srg_timing_momentum} shows the time as a function of the
evolution parameter, $s$ to evolve two different potentials, on the
left a \xeft\ and on the right AV18, in several partial waves. We
expect the evolution to be linear in $s$ and indeed find such
behavior. On the left plot, as $s$ increases, we find no hint of
stiffening of the differential equations. The dependence is strictly
linear all the way to $s=1$ ($\lambda = 1$). 

The right, log-log, plot of Argonne partial waves shows something a
little more interesting. At smaller $s$ the evolution is also linear,
but as $s$ increases, the dependence in some partial waves stiffens
somewhat. Some investigations suggest this may happen when the
potential has been driven to the desired decoupling and due to the
larger number of matrix elements the flow is working hard to maintain
a degree of accuracy that is unnecessary. It seems that freezing the
high energy parts of the potential, when they have reached a
sufficient evolution, works extremely well to improve the time
scaling.

\section{One Dimensional Jacobi Model}
\label{sec:timing}

\begin{figure*}
\begin{center}
\triplepic{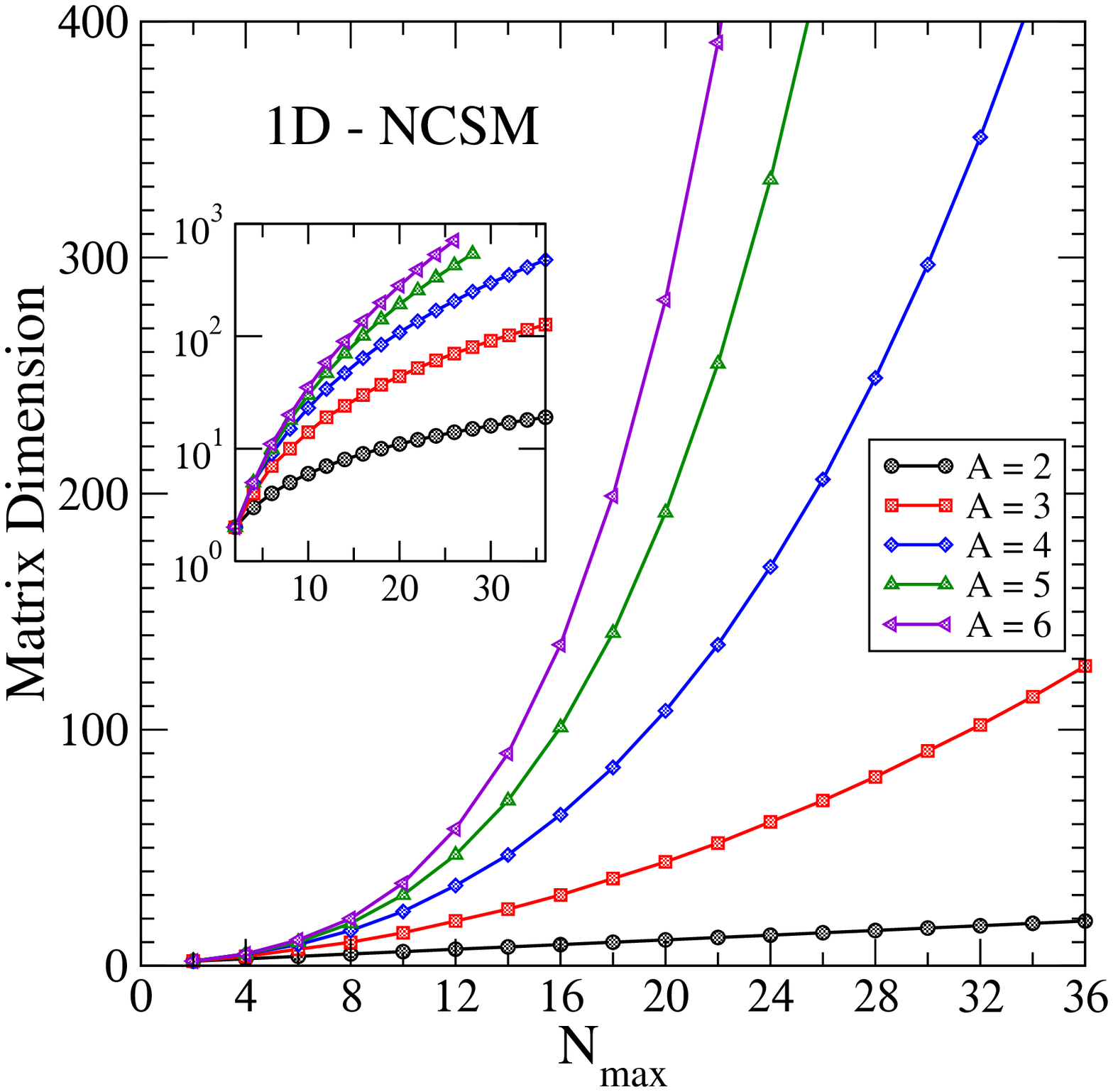}
\triplepic{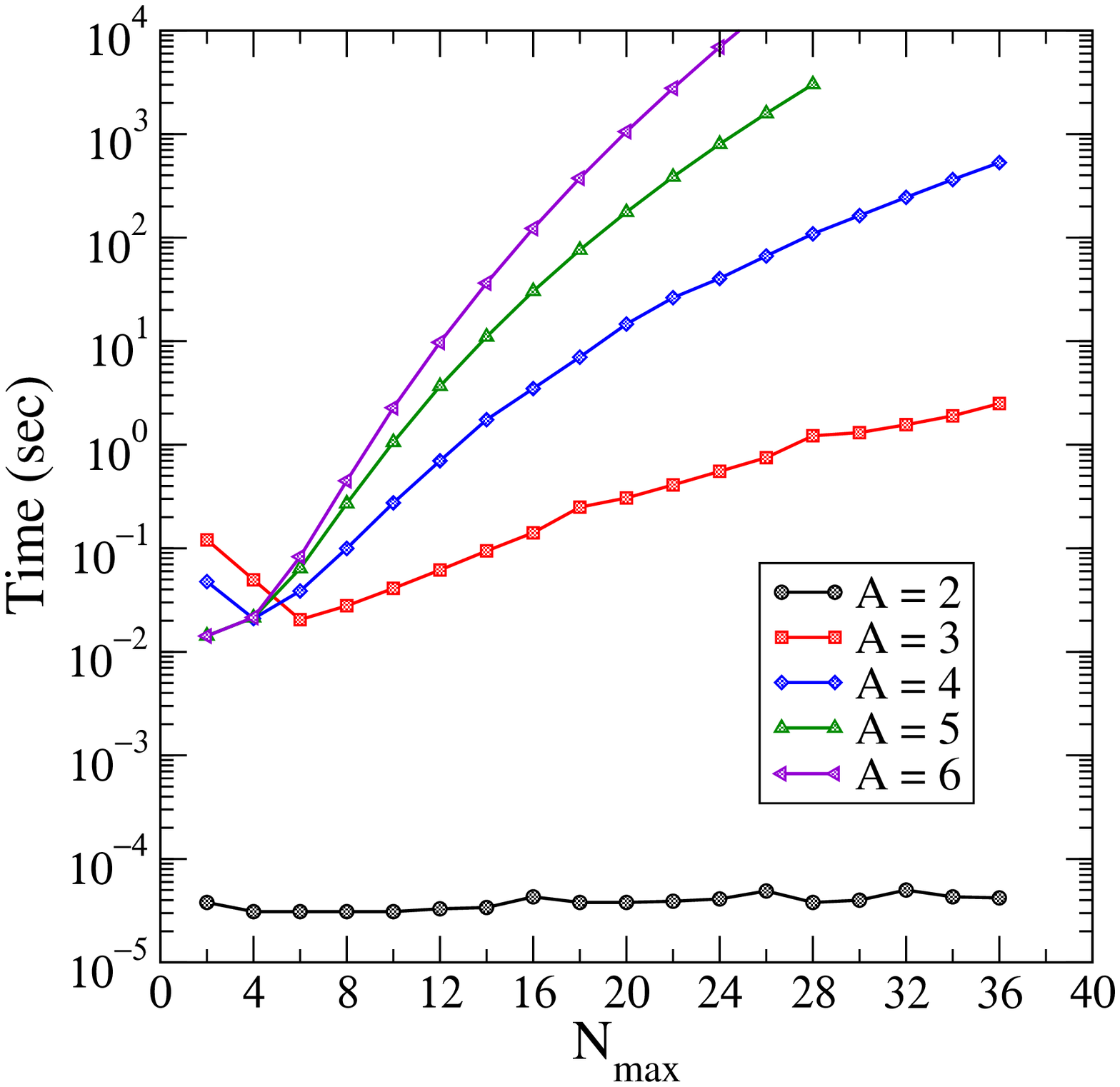}
\triplepic{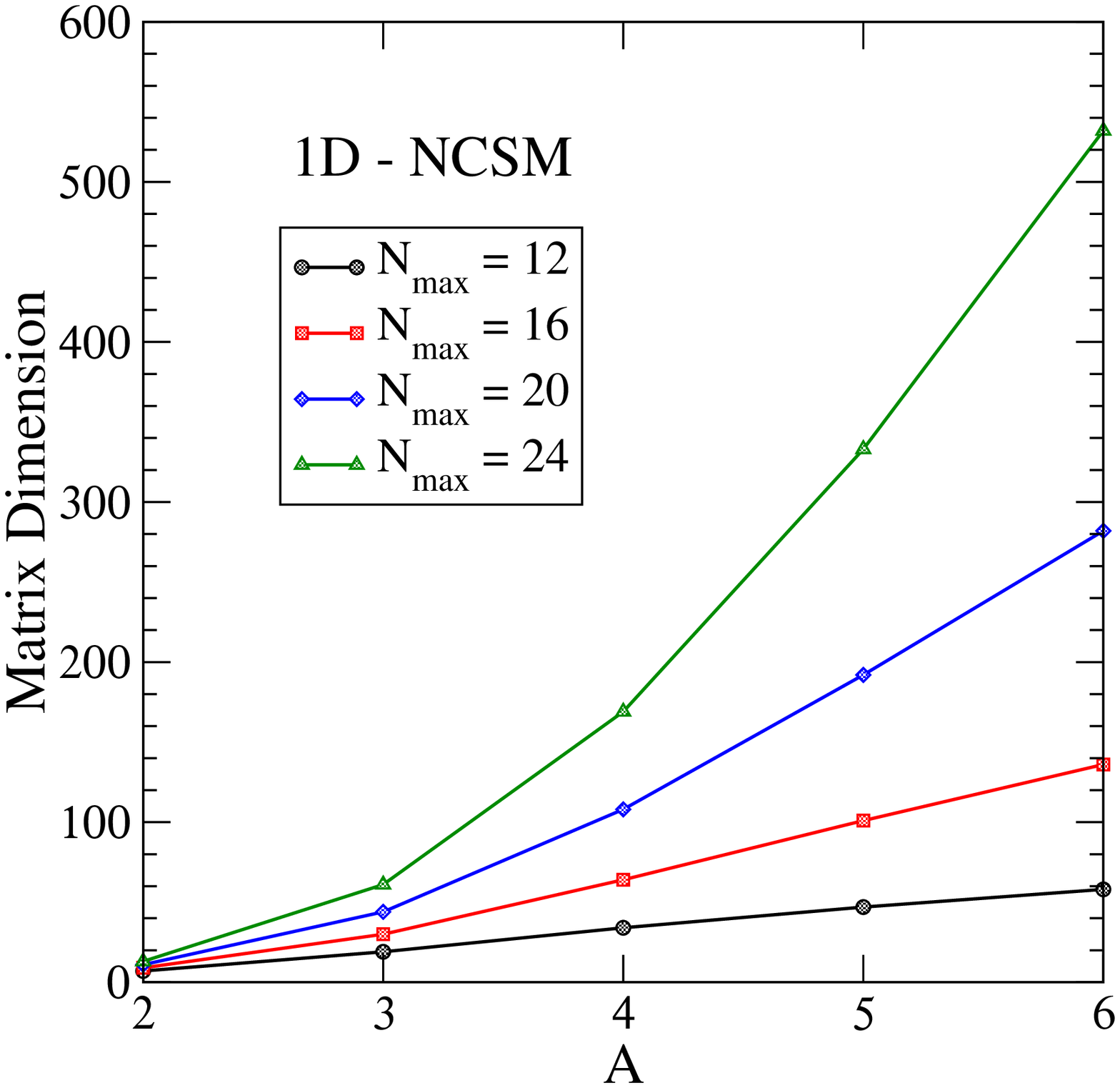}
\end{center}
\captionspace{Plots showing the scaling of One-D NCSM basis size and time to build
those bases as a function of $\nmax$ for $A=2,3,4,5,$ and 6 bosons.}
\label{fig:OneD_basis_scaling}
\end{figure*}

In the one-dimensional version of the no-core shell model, we must
build a basis of harmonic oscillator states (see
Chap.~\ref{chapt:app_osc_basis}). This basis provides for a
variational calculation with the accuracy directly connected to the
the basis size, $\nmax$. Therefore, we need to understand how the
basis scales and the resources that are consumed to build larger
bases. Figure \ref{fig:OneD_basis_scaling} has on the left the size of
matrices for different $A$-body spaces as a function of $\nmax$ and in
the center a log-log plot showing the time to build the list of states
for those bases. Note the log-log inset on the left plot shows the
same dependence in $\nmax$ as the timing plot indicating the build
time is linear with the number of matrix elements. Here we can see
dramatic increase in size and time as A increases from 2 to 6. The
far right plot recasts this data as basis dimension versus $A$ for
several choices of $\nmax$. Again we can see the importance of
bringing down the $\nmax$ required for convergence in larger $A$
systems. At small $\nmax$ the size dependence on $A$ is roughly
linear. As $\nmax$ increases we get back into the exponential region.

\begin{figure*}
\begin{center}
\dblpic{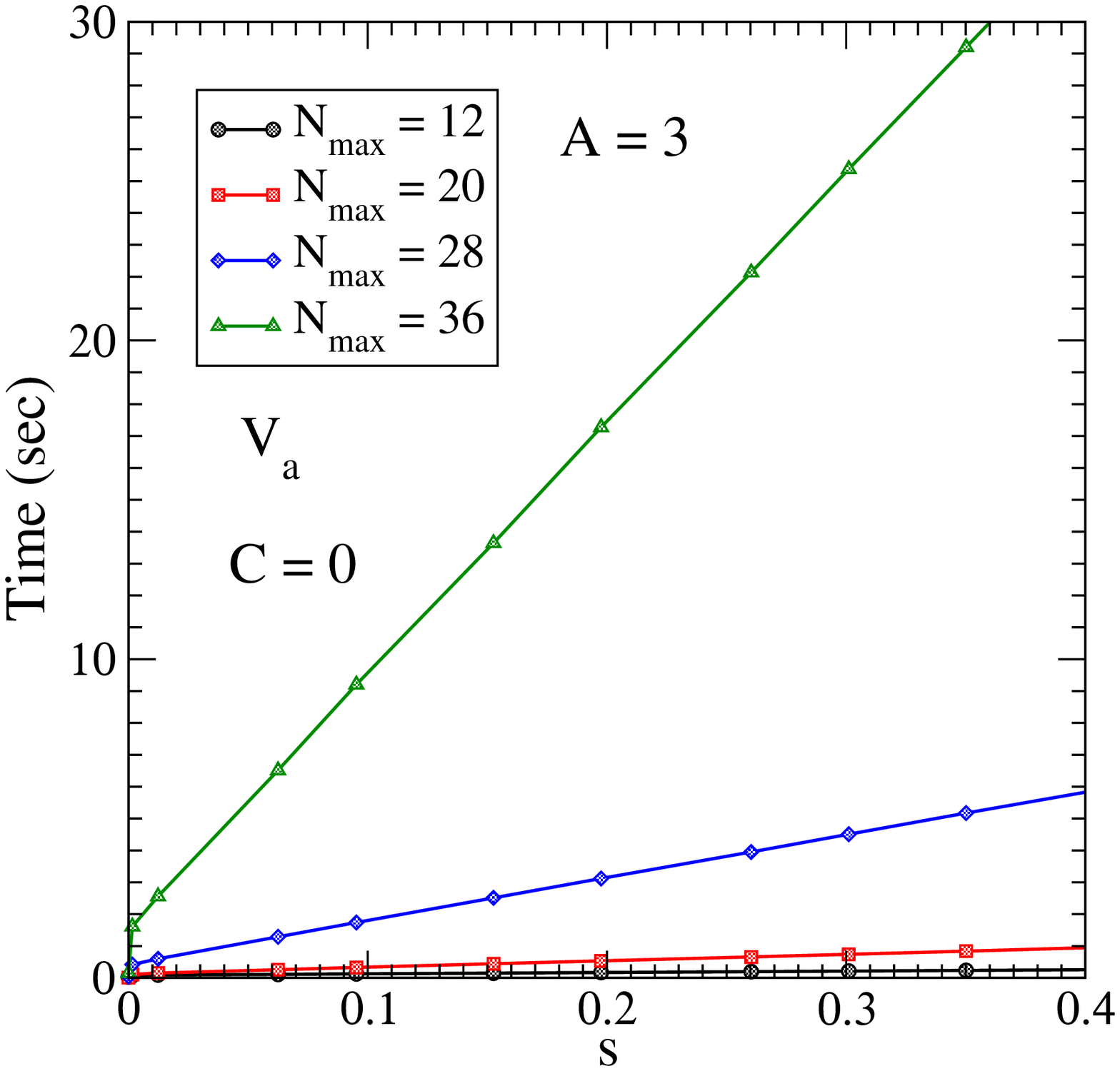}
\dblpic{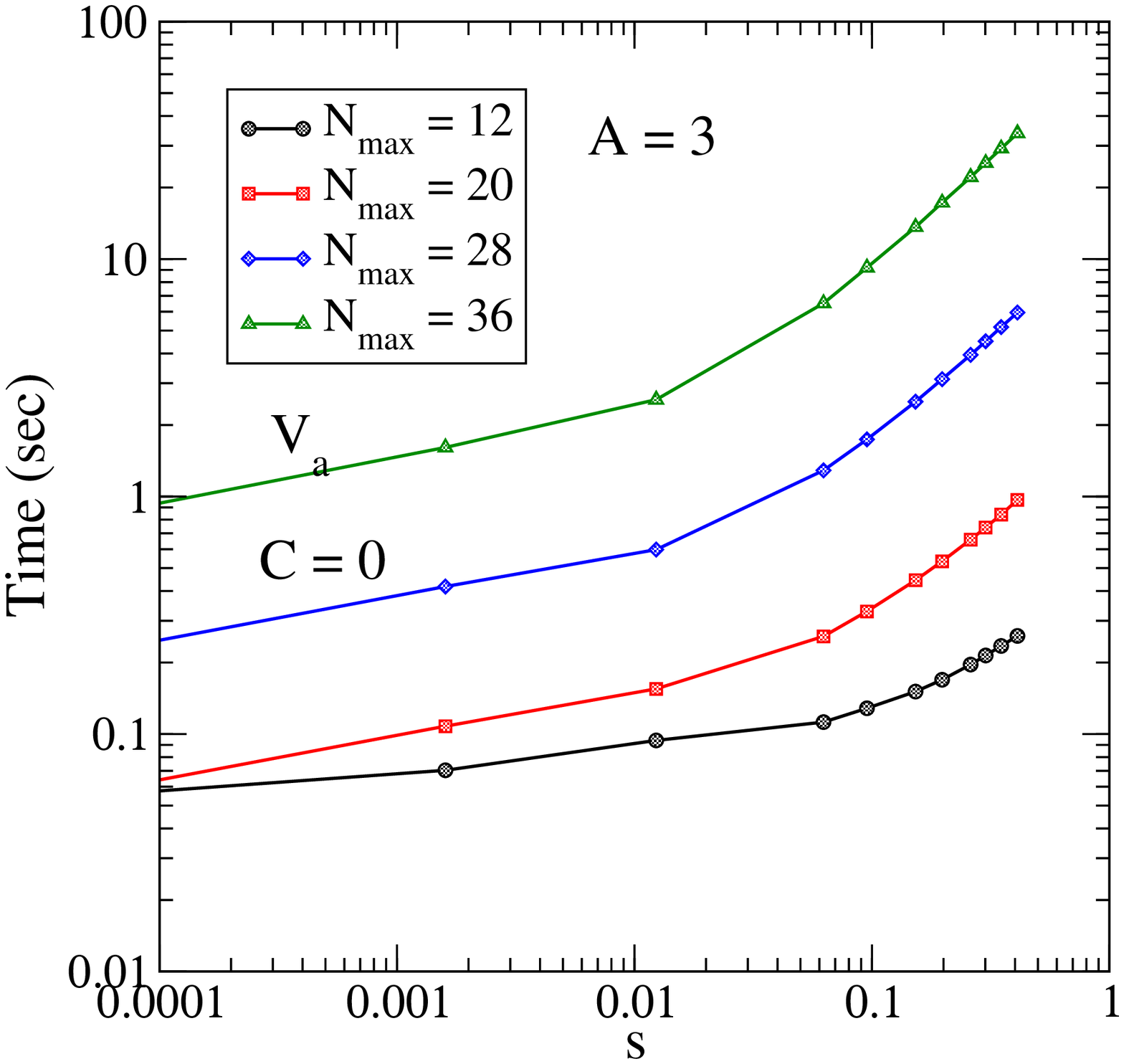}
\end{center}
\captionspace{SRG timings vs the parameter $s$ show no stiffening for
large $s$
(small $\lambda$). The log-log plot on the right gives a better view
of the low $\nmax$ samples.}
\label{fig:OneD_3N_srg_timing}
\end{figure*}

Again we must check on the scaling behavior of the SRG evolution with
increasing basis size. The left panel of
Fig.~\ref{fig:OneD_3N_srg_timing} shows the evolution time in seconds
as a function of the linear flow parameter, $s$, for several sample
$\nmax$ values. All are linear in $s$, again indicating no stiffening
of the problem. On the right panel we display the same quantities in a
log-log plot for better visibility. The slope of a straight line in
this type of plot indicates the exponent of a power-law dependence,
here the slope is unity and the limits are set to make this apparent.
The vertical shift indicates the difference in the slopes on the
linear scale.

\section{Three Dimensional Jacobi NCSM}
\label{sec:osc_scaling}

\begin{figure*}
\begin{center}
\dblpic{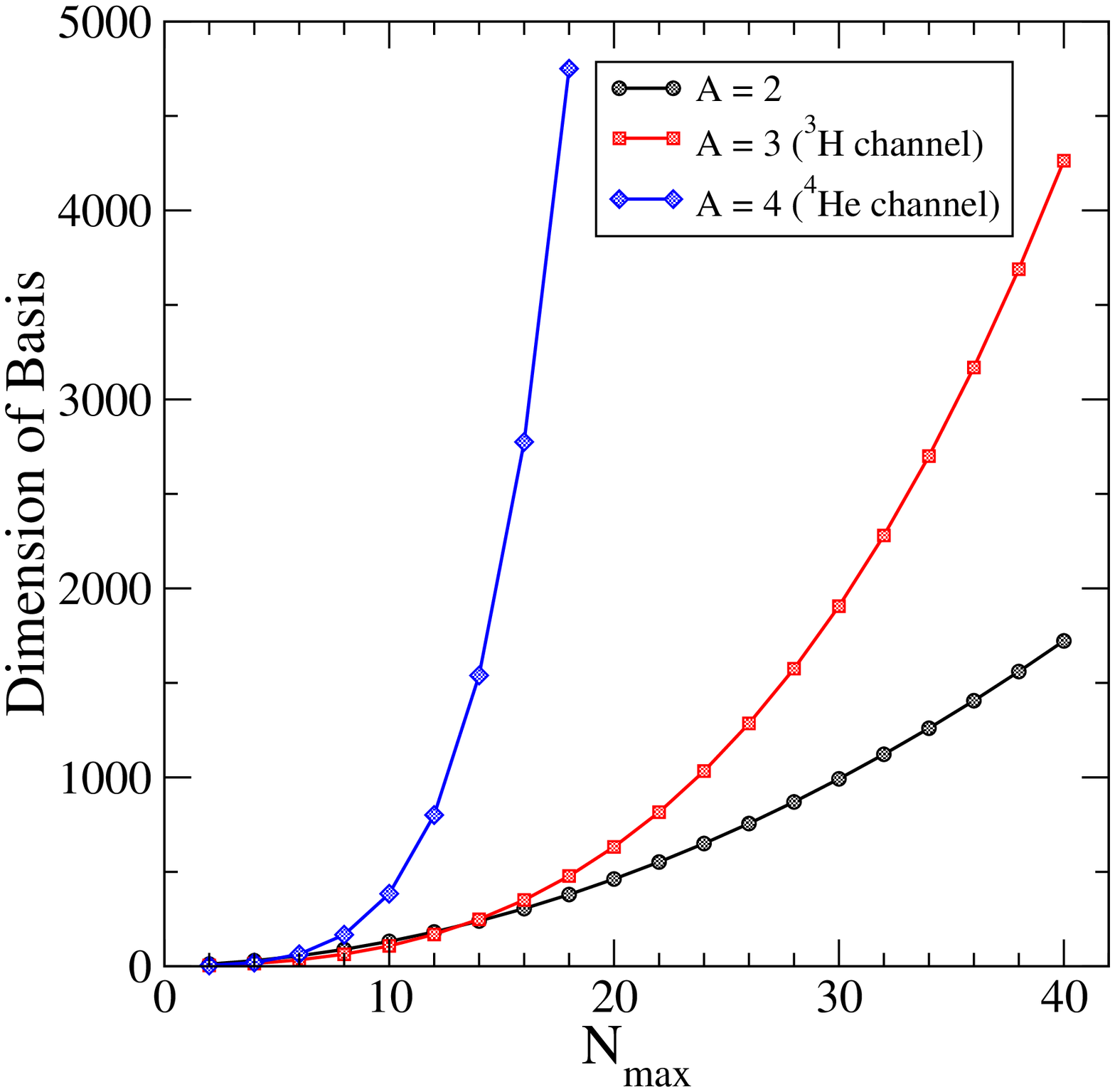}
\dblpic{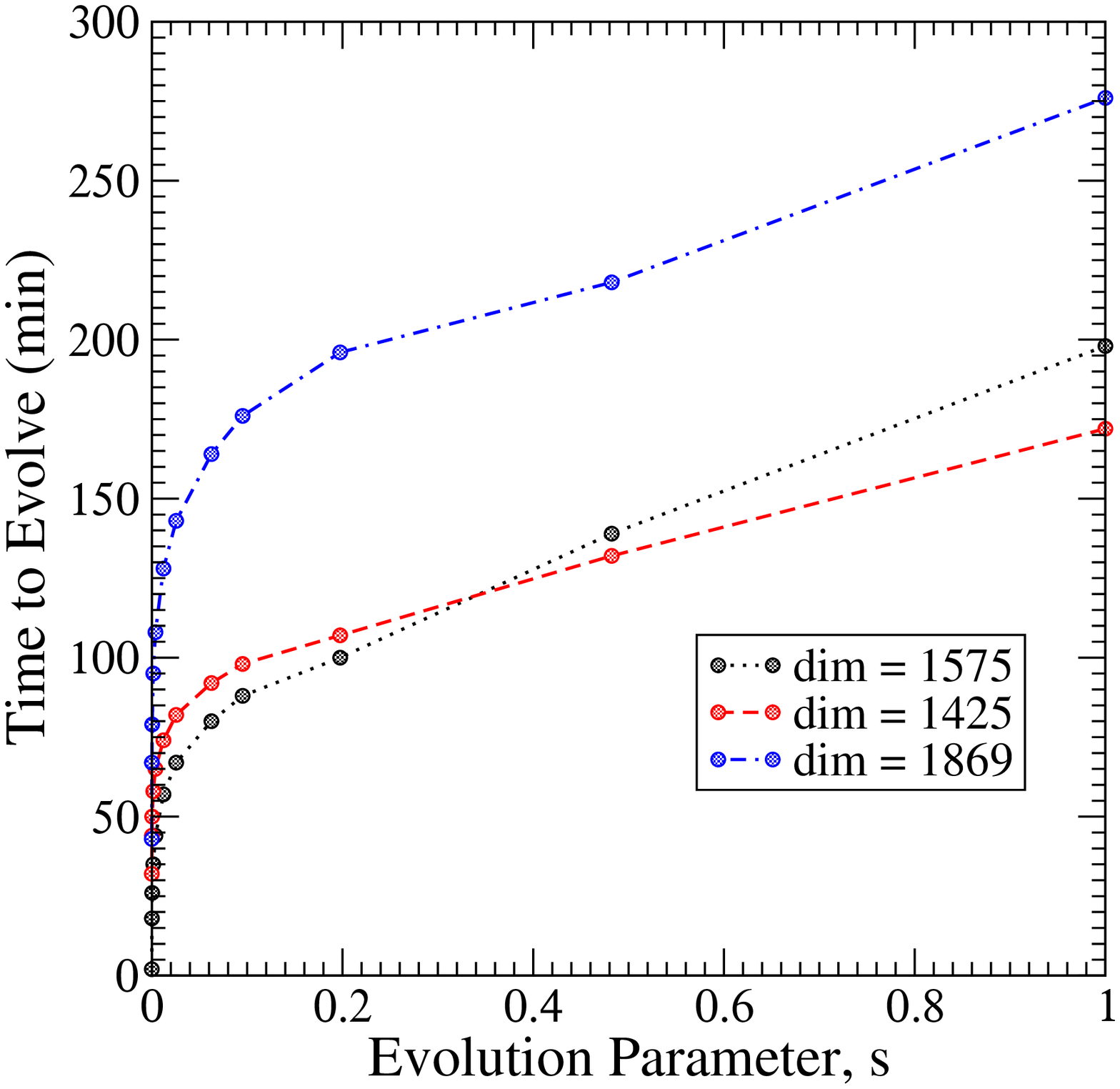}
\end{center}
\captionspace{The left plot shows scaling of NCSM basis size with $\nmax$
for the two-, three-, and four-body systems. The right plot shows evolution
timings vs. $s$ in realistic NCSM calculations for selected
$A = 3$ partial waves. The legend indicates the dimension of the
matrices being evolved.}
\label{fig:srg_ncsm_scaling}
\end{figure*}

In the left plot in Fig.~\ref{fig:srg_ncsm_scaling} we can see a
similar scaling of basis size with $\nmax$ as in the one-dimensional
version. This is yet another area where the one-dimensional model
mirrors the behavior of realistic calculations and lends credence to
its predictive power. The right panel of
Fig.~\ref{fig:srg_ncsm_scaling} shows the SRG evolution is again
robustly linear in $s$ despite the different and complicated basis and
embedding procedures of the realistic NCSM. This data had to be
obtained from file write times so the curves are shifted
vertically by an indeterminate amount of disk writing time.

Practical coding considerations have so far prevented realistic
calculations in the relative Jacobi NCSM for nuclei larger than $A=4$.
As covered in appendix~\ref{chapt:app_osc_basis} the calculation of
larger systems is accomplished by converting the Jacobi matrix
elements into the single-particle, or $m$-scheme, basis. This scaling is
discussed in the next section with Fig.~\ref{fig:mscheme_scaling}.
These problems have nothing to do with SRG implementation, but they
have an incidental impact on how quickly progress can be made.
Implementing the SRG independently in the $m$-scheme basis is a
different body of work and will be an important check though the
return may not be high for evolving extremely large sparse matrices.
The initial potential would have to be in a basis close to
convergence; previous work using the SRG within $m$-scheme
calculations~\cite{Bogner:2007rx,Jurgenson:2007td} did the evolutions
in the momentum basis, which is effectively $\nmax=\infty$.

\section{Lab Frame in One and Three Dimensions}
\label{sec:labframe_scaling}

\begin{figure*}
\begin{center}
\strip{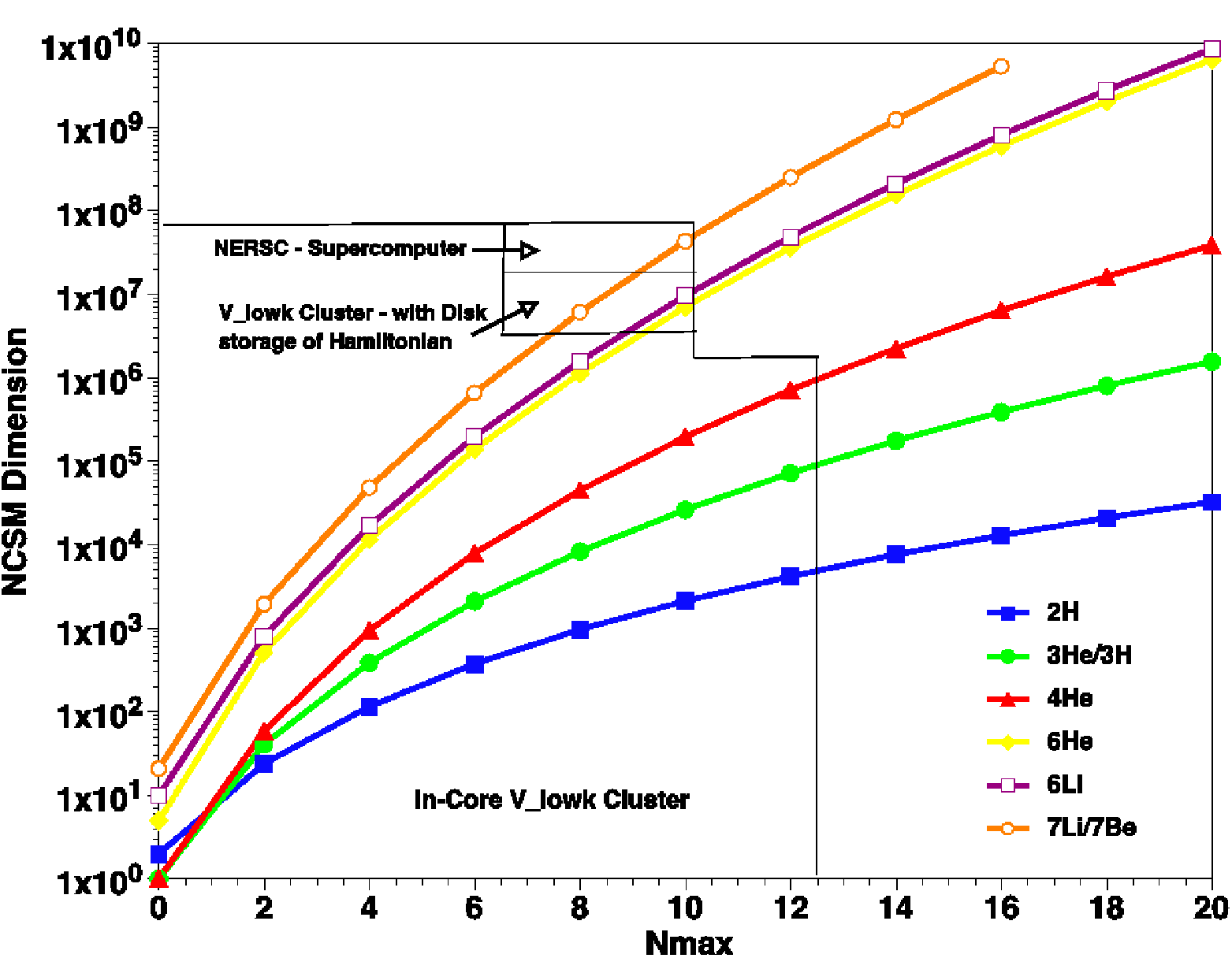}
\end{center}
\captionspace{Size of $m$-scheme basis as a function of $\nmax$ for several
light nuclei. Note the logarithmic scale in basis size.}
\label{fig:mscheme_scaling}
\end{figure*}

In the decoupling studies of chapter~\ref{chapt:decoupling} we used
evolved NN forces embedded in the single particle NCSM, referred to as
$m$-scheme, to compute light nuclei. It is important to note how much
benefit the SRG provides to these particular calculations. Figure
\ref{fig:mscheme_scaling} shows a plot on a log scale of the basis
size verses $\nmax$ for several lighter nuclei up to $A=7$. The boxes
indicate the extent of computational abilities for various systems.
The box marked ``V\_lowk Cluster - with Disk storage of Hamiltonian"
indicates the limits of our abilities with our local cluster of 32
processors with the limitation of writing the large files to disk,
which is a considerable slowdown. Also the basis dimension is not
quite as daunting as it seems since these matrices are fairly sparse;
the number of actual nonzero elements is in the range of the basis
dimension itself. Still this is a very large number of matrix elements
to compute and store. Clearly we would like to have converged results
for the $\nmax$ values within the larger box marked ``In-Core V\_lowk"
so that we can perform the calculations without expensive
supercomputer resources. As shown in section~\ref{sec:ncsm_calcs} the
decoupling afforded by the SRG enables these calculations to achieve
just such a convergence.

\begin{figure*}
\begin{center}
\dblpic{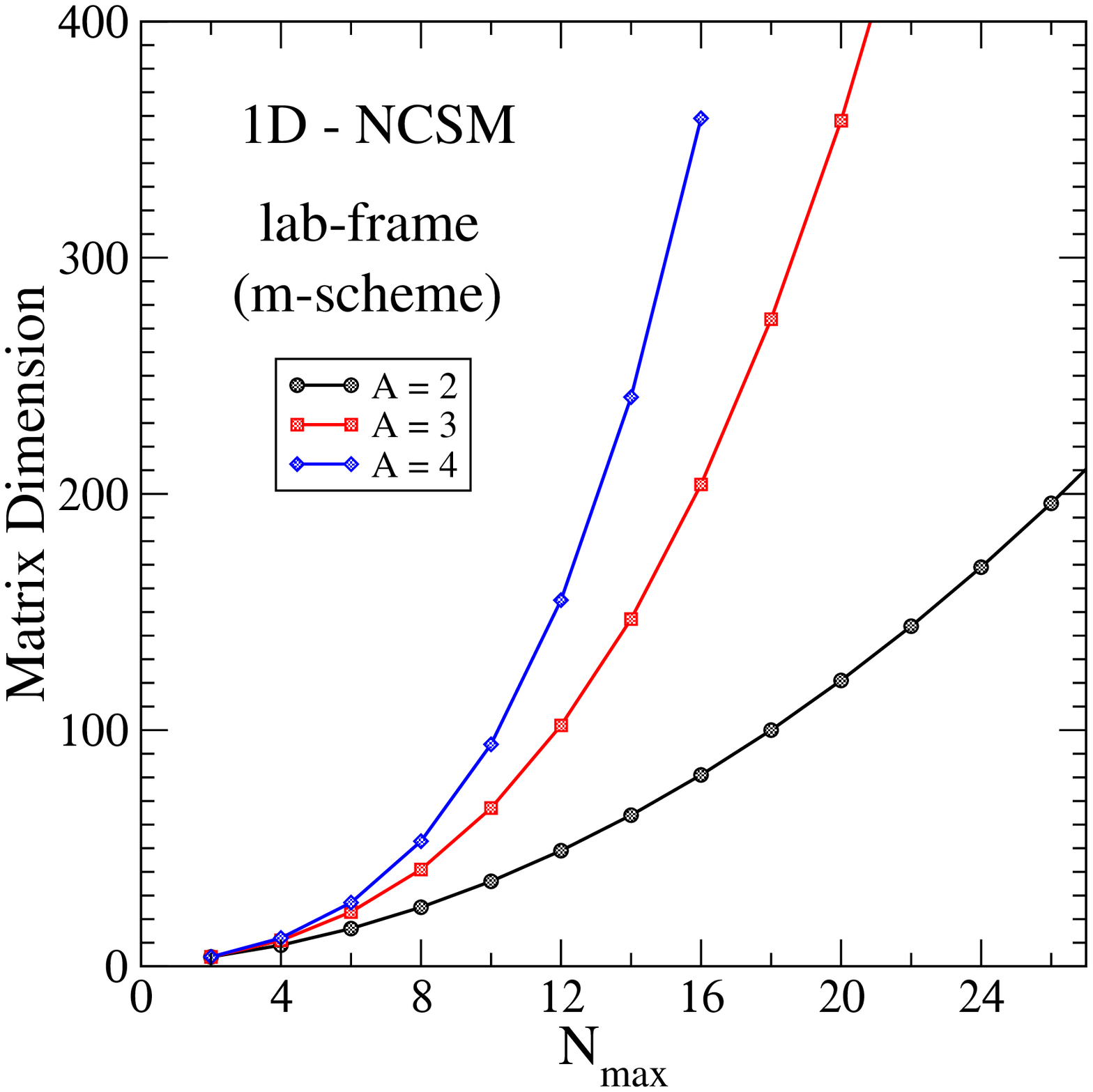}
\end{center}
\captionspace{A plot showing the scaling of One-D NCSM lab-frame basis or
``$m$-scheme" size as a function of $\nmax$ for $A=2,3,4$ bosons.
Compare to Jacobi plots of Jacobi basis size in Fig.~\ref{fig:OneD_basis_scaling}.}
\label{fig:OneD_mscheme_scaling}
\end{figure*}

Figure \ref{fig:OneD_mscheme_scaling} shows the scaling of basis sizes
in the one-dimensional lab-frame basis. Note the dramatic increase in
basis size over the one-dimensional Jacobi basis. This calculation
has not been developed further due to its poor scaling properties as
compared to the Jacobi version. However, it is an important check on
the accuracy of calculations performed in the Jacobi basis, and
further developments for this purpose will be straightforward as
discussed in section~\ref{sec:lab_evolution}.

In one dimension, the lab-frame basis is not as efficient as the
Jacobi basis in achieving convergence with basis size and therefore
the Jacobi basis is preferred. However, in three-dimensions, angular
momentum considerations change the situation. In the Jacobi basis for
a given $A$-body calculation, one must include multiple angular
momentum channels, or partial waves, to obtain accurate results. The
coding involved in correctly accounting for all the necessary channels
is apparently very involved due to the iterative procedure that builds
on successive symmetrized $A$-body clusters. In contrast, the $m$-scheme
strategy builds an $A$-body basis  and symmetrizer from scratch and
embeds the two- and three-body interactions directly into that space.
Therefore the balance is tipped in favor of using large supercomputing
to handle the large bases of the $m$-scheme, instead of continuing to
develop code in the Jacobi basis.

\chapter{Spurious States from $G_s = \Hho$}
\label{chapt:app_spurious}

In this appendix we will consider the spurious bound states which
appear during the evolution with certain choices of the SRG generator,
$G_s$. In the one dimensional model, we explored various features of
these spurious states qualitatively through observing the evolution of
potential matrix elements in the 1D basis. Watching the flow of
individual elements can be very instructive when trying to understand
the qualitative behavior of the SRG. Here we plot various potentials
in the oscillator basis at $\nmax = 12$, which is convenient for looking
directly at the potentials and seeing the interplay among matrix
elements.

\begin{figure}
\begin{center}
\spuriousscale{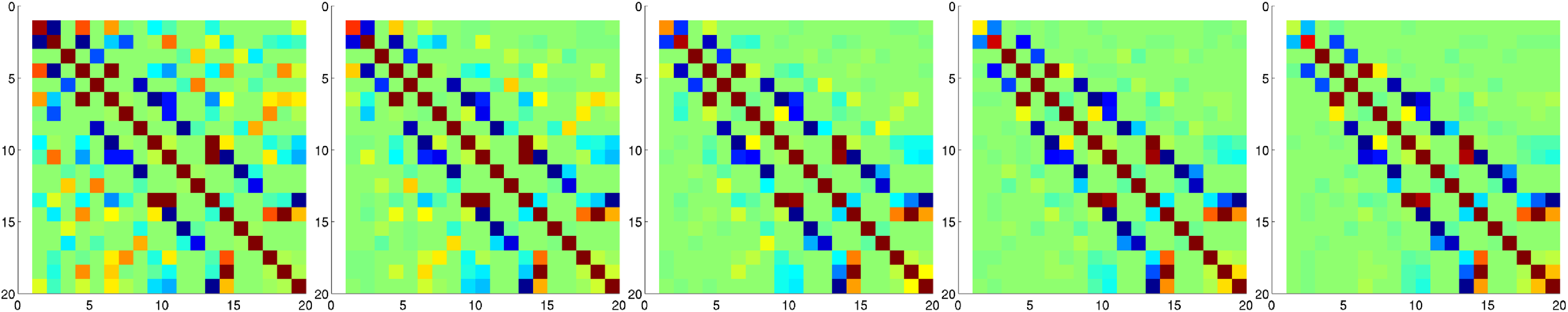}
\spuriousscale{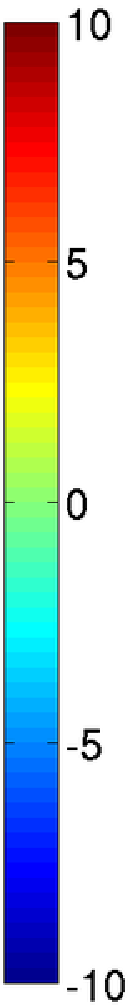}
   
\spuriousscale{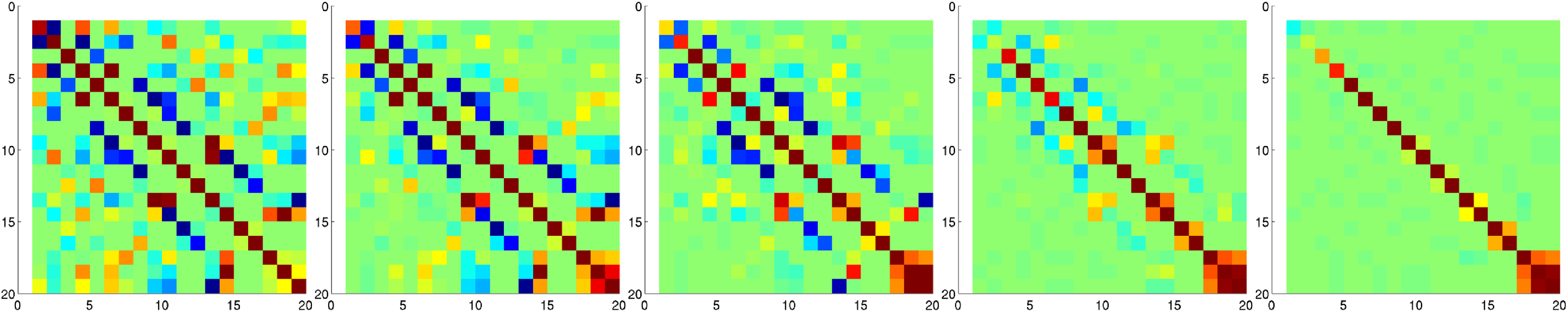}
\spuriousscale{figures/spurious_color_scale}
\end{center}
\captionspace{A series of plots of the oscillator basis Hamiltonian in a
reduced $\nmax$ basis as it is evolved using $G_s = \Trel$ (top) and
$G_s = \Hho$ (bottom) to $\lambda = 7,5,4,3,2$. The size of the basis
shown is only $\nmax = 12$ for visibility. Larger bases look
qualitatively similar.}
\label{fig:spurious_Ho3e}
\end{figure}

Figure \ref{fig:spurious_Ho3e} shows a film strip of evolving
potentials in the three-body basis of the one-dimensional model. The
top shows evolution using the choice $G_s = \Trel$ and the bottom uses
$G_s = \Hho$. The advantage of $\Hho$ over $\Trel$ with respect to the
amount of diagonalization achieved in this basis is obvious as noted
in chapter~\ref{chapt:OneD}. However, problems arise when we try to
isolate the evolving two- and three-body forces.

\begin{figure}
\begin{center}
\triplepic{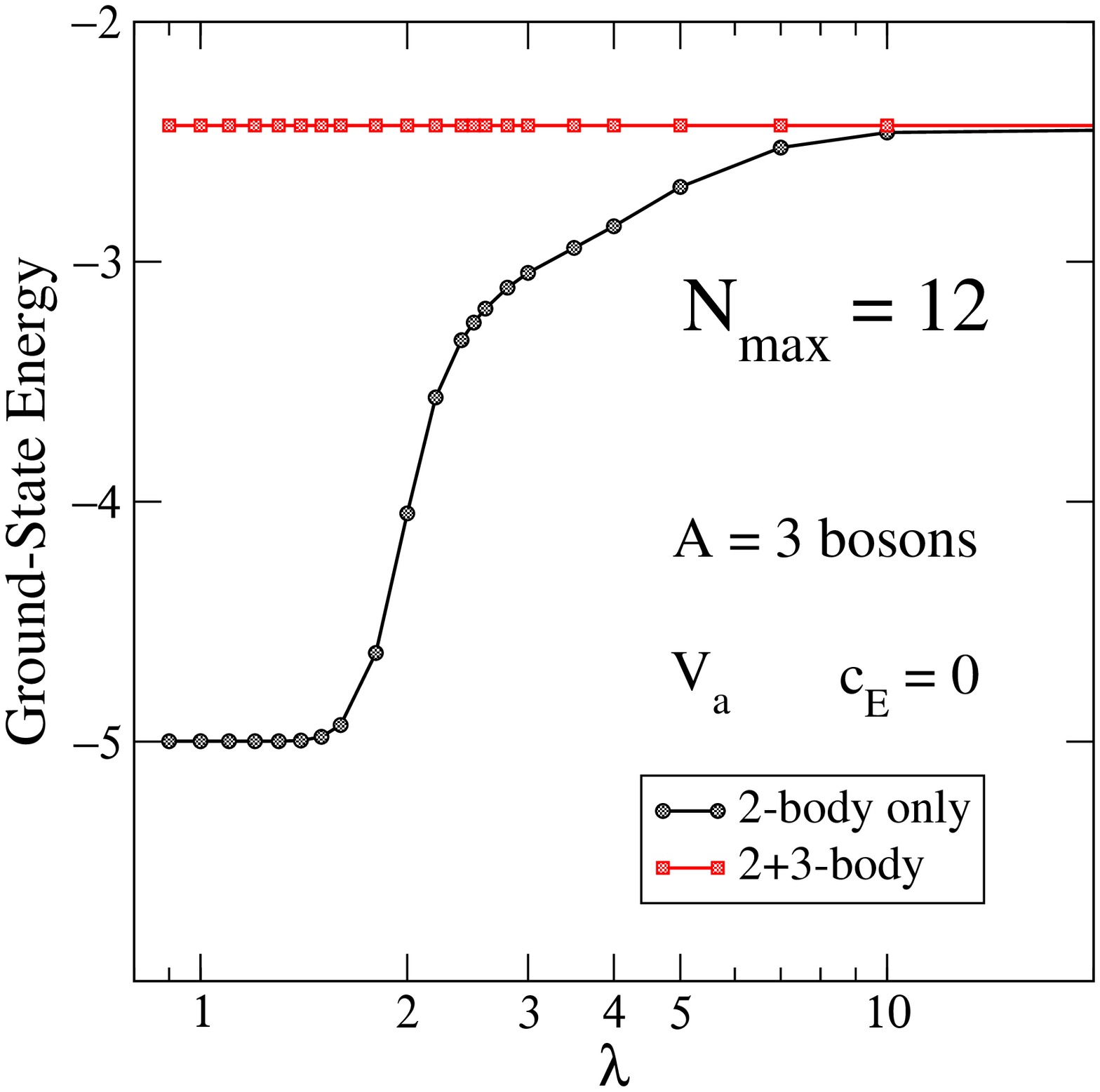}
\triplepic{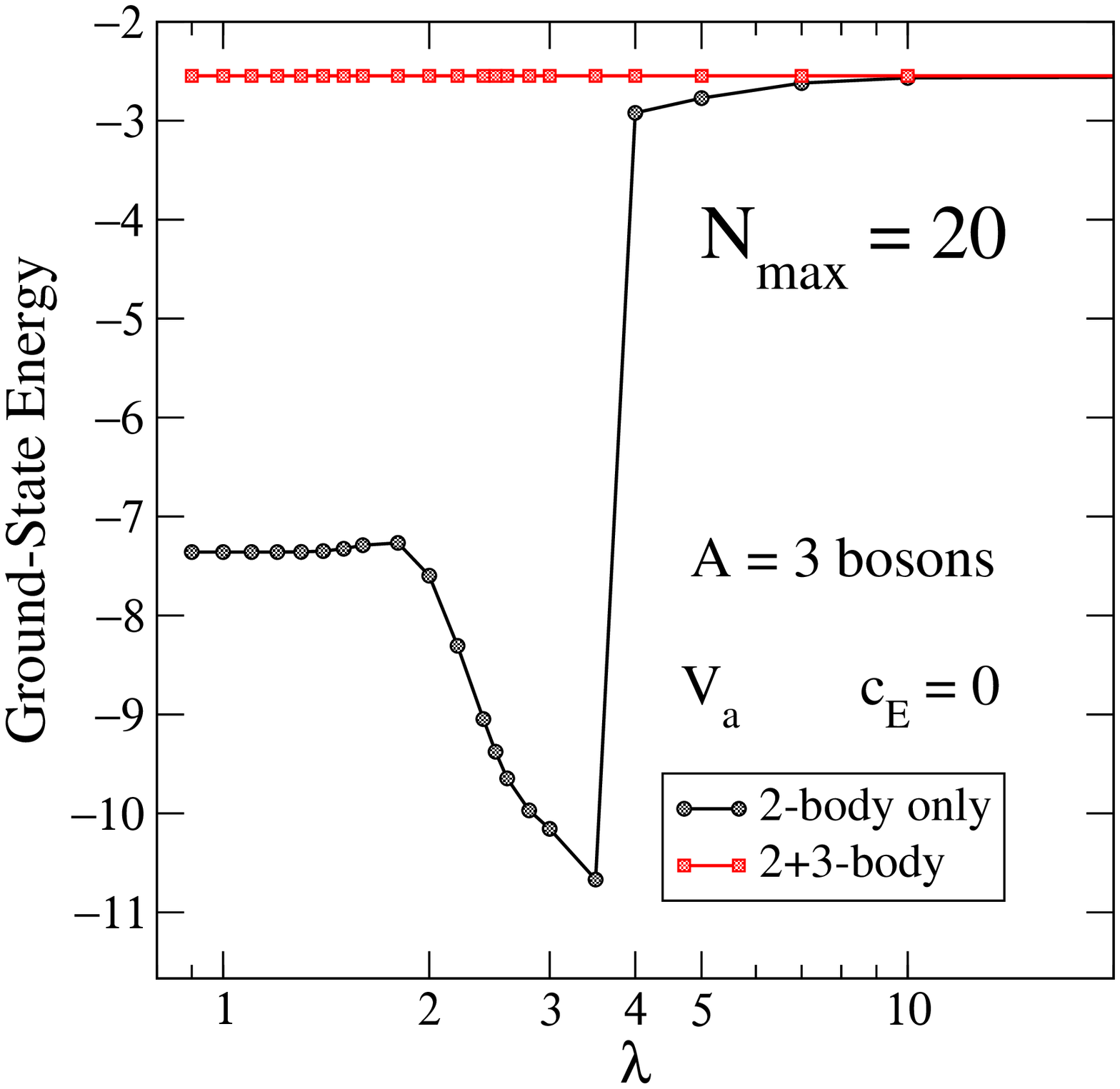}
\triplepic{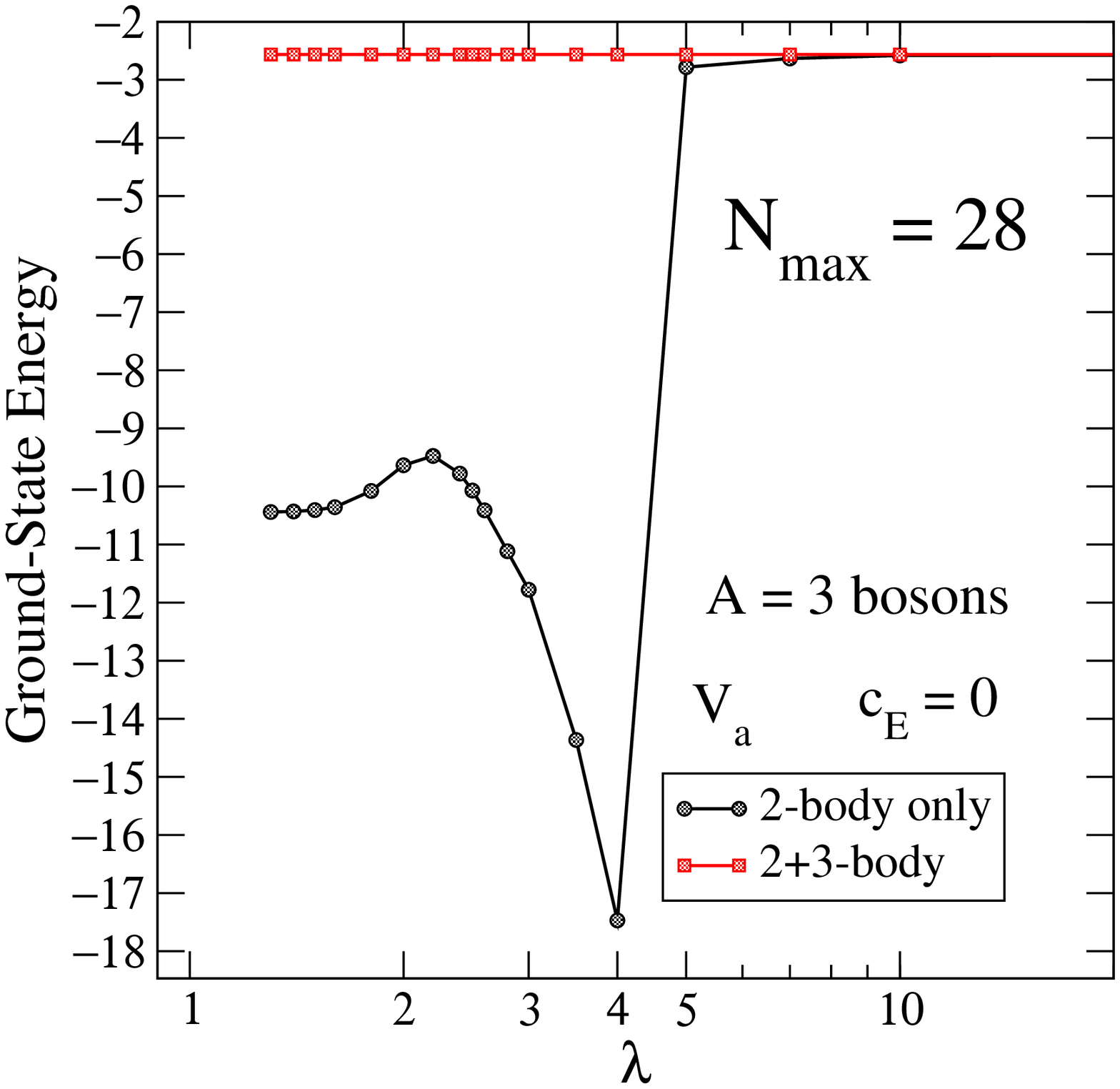}
\end{center}
\captionspace{An example of the discontinuous curves of two-body-only
calculations characteristic of the $G_s = \Hho$ SRG. The two-body-only
curves computed by evolving the $A=2$ potential using $G_s = \Hho$ and
embedding the result in the $A=3$ basis to obtain the three-boson
binding energy. The straight line labeled ``2+3-body" is a check on
the unitary evolution in the $A=3$ space.}
\label{fig:spurious_Hho_evolution}
\end{figure}

Figure~\ref{fig:spurious_Hho_evolution} shows the result of evolving
the two-body potential using $G_s = \Hho$ in that basis before
embedding in the $A=3$ space and computing the two-body-only
contribution to the $A=3$ bound state. The resulting evolution curve
is analogous to that of Fig.~\ref{fig:srg_3_body} where we chose $G_s
= \Trel$ and the result was a nice smooth curve indicating the effect
of induced three-body forces, in the lone bound state, as a function
of $\lambda$. Now, using $G_s = \Hho$, the curves are
discontinuous\footnote{Note that the lines in
Fig.~\ref{fig:spurious_Hho_evolution} are drawn to guide the eye from
one point to the next. More sample $\lambda$'s may reveal this drop to
be simply very steep.}
and the magnitude of induced three-body forces is much larger than
expected. This plot turns out to be contaminated by the appearance of
multiple three-body bound states due to the evolution of the two-body
matrices with this $G_s$.

\bt
\caption{A sampling of evolving $A=3$ spectra under the choice $G_s =
\Hho$. The table uses values at $\nmax = 20$, which is large enough to
display the effect, but small enough to display the entire
spectrum.}
\vspace*{.1in}
\begin{tabular}{c|ccccc}
    state &  $\lambda = 100$  &  $\lambda = 5$  & $\lambda = 4$  & 
    $\lambda = 3$   & $\lambda = 2$  \\
    \hline
    $1$ & $ -2.547 $ & $ -2.772 $ & $ -2.920 $ & $-10.156  $ & $ -7.597 $ \\
    $2$ & $  0.049 $ & $ -0.295 $ & $ -0.456 $ & $-10.090  $ & $ -7.497 $ \\
    $3$ & $  1.713 $ & $  1.192 $ & $  0.370 $ & $ -3.377  $ & $ -5.129 $ \\
    $4$ & $  3.103 $ & $  2.479 $ & $  0.760 $ & $ -2.994  $ & $ -4.199 $ \\
    $5$ & $  5.709 $ & $  4.847 $ & $  2.126 $ & $ -0.895  $ & $ -3.649 $ \\
    $6$ & $  6.615 $ & $  5.595 $ & $  2.701 $ & $ -0.571  $ & $ -3.544 $ \\ 
    $7$ & $  7.525 $ & $  6.830 $ & $  3.771 $ & $  0.672  $ & $ -1.171 $ \\ 
    $8$ & $ 10.828 $ & $  9.221 $ & $  5.090 $ & $  1.113  $ & $ -0.878 $ \\ 
    $9$ & $ 12.276 $ & $ 10.627 $ & $  7.192 $ & $  1.634  $ & $  0.114 $ \\ 
    $10$ &$ 13.401 $ & $ 11.291 $ & $  7.670 $ & $  2.614  $ & $  0.190 $
\end{tabular}   
\label{tab:spurious_spectrum_3N}
\et

Table~\ref{tab:spurious_spectrum_3N} shows a sample of the first few
states in the spectrum as the initial Hamiltonian is evolved with
$G_s = \Hho$. The evolution proceeds as expected up to a $\lambda$
value where
new states begin to appear. As $\lambda$ is increased further all the
bound states deepen dramatically, though the precise behavior is
dependent on the choice of $G_s$.

\begin{figure}
\begin{center}
\spuriousscale{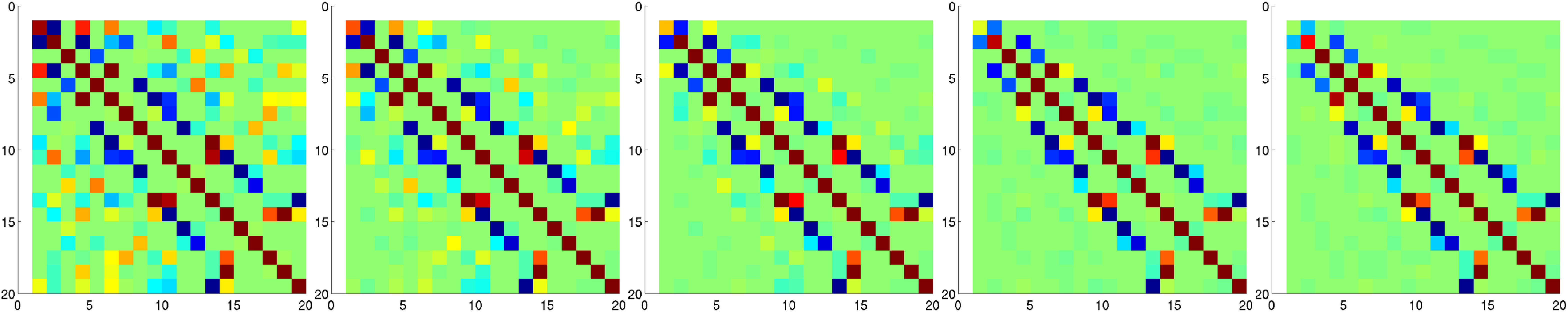}
\spuriousscale{figures/spurious_color_scale}

\spuriousscale{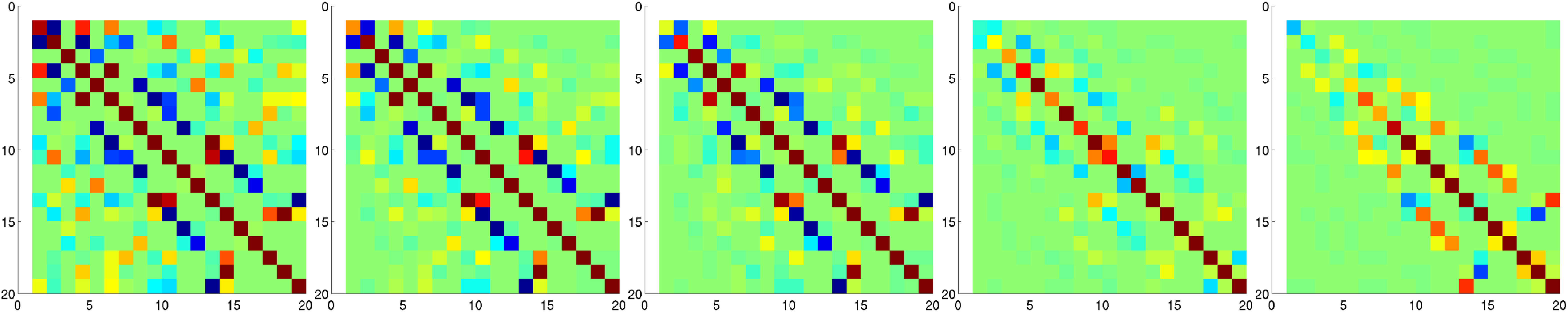}
\spuriousscale{figures/spurious_color_scale}
\end{center}
\captionspace{A series of plots of the $A=3$ oscillator basis Hamiltonian in
a reduced $\nmax$ basis as it is evolved using $G_s = \Hho$.}
\label{fig:spurious_Hoe3}
\end{figure}

To further explore the origins of these states, we looked at the
NN-only $A=3$ matrices as they are evolved and embedded.
Figure~\ref{fig:spurious_Hoe3} compares the choices of $\Trel$ and
$\Hho$ with film strips of these. Notice that while the $\Trel$
version converges on a shape similar to $\Trel$ itself, the $\Hho$
version moves toward the diagonal and then recedes from it, ending up
at a shape similar to the $\Trel$ component but of opposite sign. This
suggests that the fully unitary transformations are kept constant by
means of large three-body forces being induced to compensate for some
unknown effect in transformation of the two-body forces.

\begin{figure}
\begin{center} 
\spuriousscale{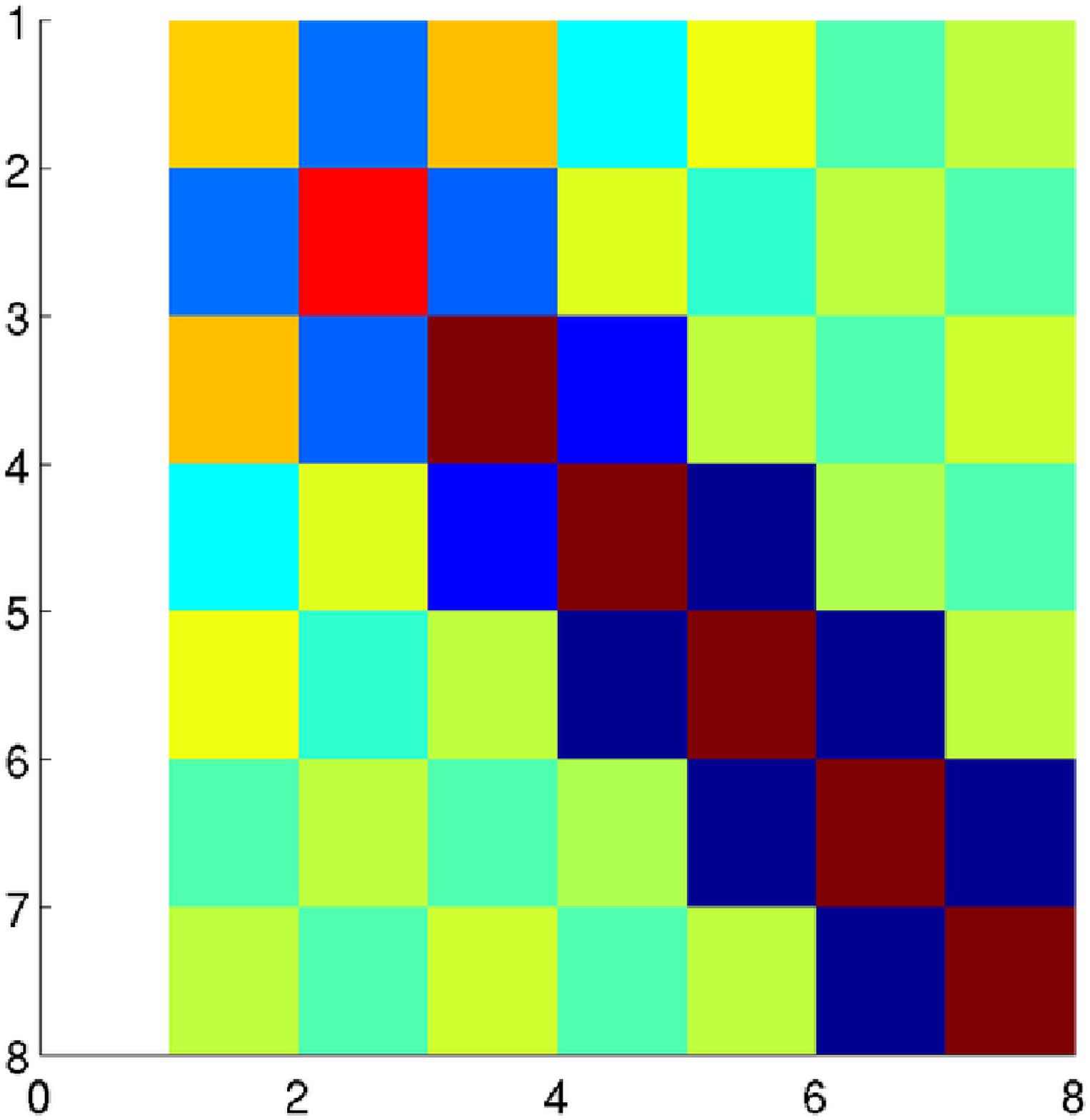}
\spuriousscale{figures/spurious_color_scale}
\end{center}
\captionspace{A series of plots of the $A=2$ oscillator basis Hamiltonian in
a reduced $\nmax$ basis as it is evolved using $G_s = \Hho$.}
\label{fig:spurious_Hoe}
\end{figure}

In figure~\ref{fig:spurious_Hoe} a film strip is shown of evolving
$A=2$, matrices corresponding to the $A=3$ evolutions shown above.
Notice how some matrix elements along the diagonal mysteriously
decrease so that they are not being driven to a monotonically
increasing pattern as set by $G_s = \Hho$. As seen in
chapter~\ref{chapt:decoupling}, the SRG tries to drive the matrix
toward a shape dictated by the choice of $G_s$. With the matrix
elements out of order according to the SRG form, it appears to induce
strength in off-diagonal matrix elements coupling these smaller
states. This in turn demands large off-diagonal three-body forces to
be induced to compensate for this reordering. The exact cause of the
disrupted ordering of states is not well understood presently but is a
subject of ongoing investigation.

\bt
\caption{A sampling of evolving spectra for $A=4$ under the choice
$G_s = \Hho$. Again the table uses values at $\nmax = 20$. The upper
block displays the two-body-only calculation while the lower block
includes the two- and induced three-body forces.}
\vspace*{.1in}
\begin{tabular}{c|cccccc}
    state &  $\lambda = 100$  &  $\lambda = 5$  & $\lambda = 4$  & 
    $\lambda = 3.5$   & $\lambda = 3$   & $\lambda = 2$  \\
    \hline
    $1$ & $ -4.609 $ & $ -5.496 $ & $ -9.697 $ & $ -42.757 $ & $-10.156  $ & $ -33.059 $ \\
    $2$ & $ -1.558 $ & $ -2.752 $ & $ -8.415 $ & $ -34.195 $ & $-10.090  $ & $ -27.979 $ \\
    $3$ & $  0.812 $ & $ -0.792 $ & $ -6.021 $ & $ -34.112 $ & $ -3.377  $ & $ -27.549 $ \\
    $4$ & $  2.254 $ & $  0.446 $ & $ -3.215 $ & $ -22.265 $ & $ -2.994  $ & $ -27.152 $ \\
    $5$ & $  2.733 $ & $  0.990 $ & $ -2.301 $ & $ -21.183 $ & $ -0.895  $ & $ -26.934 $ \\
    $6$ & $  5.483 $ & $  2.999 $ & $ -1.974 $ & $ -20.878 $ & $ -0.571  $ & $ -25.851 $ \\ 
    $7$ & $  6.742 $ & $  4.252 $ & $ -1.912 $ & $ -17.088 $ & $  0.672  $ & $ -25.692 $ \\ 
    $8$ & $  7.620 $ & $  5.152 $ & $ -0.665 $ & $ -16.396 $ & $  1.113  $ & $ -23.746 $ \\ 
    $9$ & $  8.077 $ & $  5.741 $ & $  0.147 $ & $ -15.059 $ & $  1.634  $ & $ -22.920 $ \\ 
    $10$ &$  9.372 $ & $  6.465 $ & $  0.951 $ & $ -14.269 $ & $  2.614  $ & $ -21.483 $ \\
    \hline
    $1$ & $ -4.609 $ & $ -4.674 $ & $-29.265 $ & $ -11.822 $ & $ -6.029  $ & $  -4.039 $ \\
    $2$ & $ -1.558 $ & $ -2.969 $ & $-11.135 $ & $  -5.366 $ & $ -4.016  $ & $   0.745 $ \\
    $3$ & $  0.812 $ & $ -0.235 $ & $ -8.177 $ & $  -5.155 $ & $ -3.004  $ & $   1.856 $ \\
    $4$ & $  2.254 $ & $  0.735 $ & $ -7.319 $ & $  -4.190 $ & $ -1.838  $ & $   1.983 $ \\
    $5$ & $  2.733 $ & $  2.475 $ & $ -4.210 $ & $   0.154 $ & $  0.398  $ & $   5.157 $ \\
    $6$ & $  5.483 $ & $  5.361 $ & $  0.083 $ & $   2.472 $ & $  2.378  $ & $   8.707 $ \\ 
    $7$ & $  6.742 $ & $  5.529 $ & $  1.500 $ & $   3.532 $ & $  4.328  $ & $  11.501 $ \\ 
    $8$ & $  7.621 $ & $  6.095 $ & $  2.001 $ & $   4.142 $ & $  6.327  $ & $  11.619 $ \\ 
    $9$ & $  8.077 $ & $  7.414 $ & $  6.263 $ & $   5.964 $ & $  8.342  $ & $  13.510 $ \\ 
    $10$ &$  9.372 $ & $  7.473 $ & $  6.382 $ & $   7.188 $ & $  9.963  $ & $  16.678 $
\end{tabular}   
\label{tab:spurious_spectrum_4N}
\et

Moving up to the next many-body sector, we can look at the effect of
the $\Hho$ choice in a four-body system. In
Table~\ref{tab:spurious_spectrum_4N} we show the evolving spectra for
the two-body-only and two-plus-three calculations of the $A=4$ system.
The former is on the top block of the table and the latter is on the
bottom. In the top we can see the effect is magnified in the larger
system. In the bottom the onset of the spurious states appears to be
earlier, which is consistent with the different shape of
two-plus-three body calculations of the four-body system. That is, the
induced four-body force may be compensating for additional spurious
contributions occurring in the $A=3$ sector.

In order to study the scale at which these spurious states appear, we
also tried setting the potential to zero, leaving the initial
Hamiltonian to be $H_{s=0} = \Trel$. We find that the spurious states
appear suddenly at a specific value of the evolution parameter,
$\lambda = \sqrt{\hw}$. This is expected since with $V=0$, and most
physical constants set to one, $\hw$ is the only scale left in the
problem, having units of energy or $[\lambda]^2$. By increasing the
sampling of $\lambda$ around the value, $\sqrt{\hw}$, we find that the
onset of spurious states is indeed discontinuous, at least in the case
of zero initial potential.
%

\begin{figure}
\begin{center}
\spuriousscale{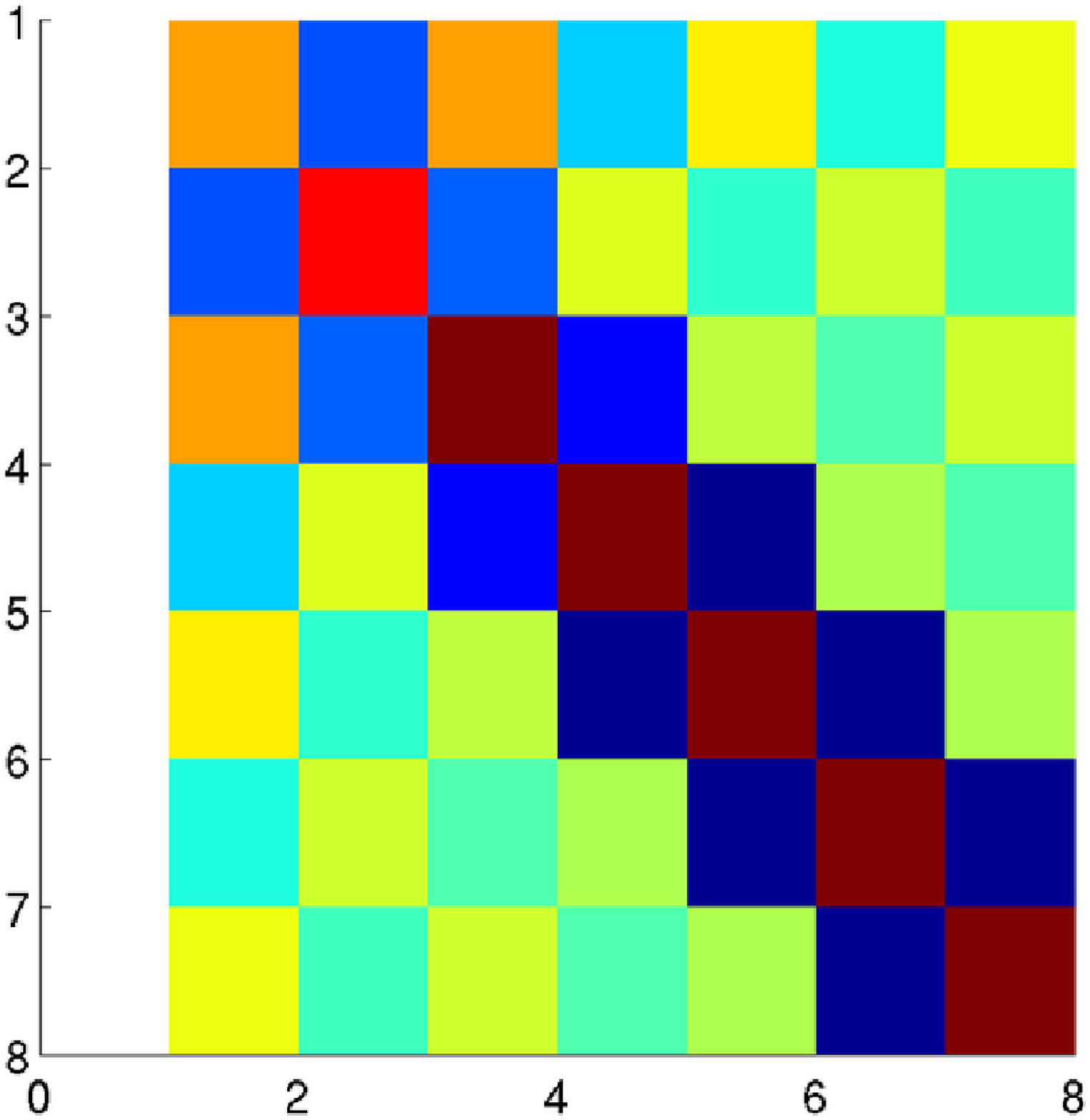}
\spuriousscale{figures/spurious_color_scale}
\end{center}
\captionspace{A series of plots of the $A=2$ oscillator basis Hamiltonian in
a reduced $\nmax$ basis as it is evolved using $G_s = H_{\rm d,osc}$.}
\label{fig:spurious_Hoe_Hd}
\end{figure}

We also tried other choices of $G_s$ that are diagonal in the
oscillator basis. In Fig.~\ref{fig:spurious_Hoe_Hd} we show the
evolution the $A=2$ Hamiltonian using $G_s = H_{\rm d,osc}$, where
$H_{\rm d,osc}$ is the diagonal part of the evolving Hamiltonian in the
oscillator basis. Here we can still see the matrix elements evolving out of
order and subsequent development of large off-diagonal elements in
the evolution to push strength to low momentum. Here the effect is
more dramatic than in the $\Hho$ case, since the moving strength is
jumping several states. This again forces large induced three-body
forces to compensate for these evolved short range interactions.

\bt
\caption{A sampling of evolving $^3$H spectra under the choice $G_s = \Hho$
in the realistic three-dimensional NCSM. The table uses values at
$\nmax = 20$ which is large enough to display the effect, but small
enough to display the entire spectrum.}
\vspace*{.1in}
\begin{tabular}{c|cccccc}
state &  $\lambda = 100$  &  $\lambda = 5$  & $\lambda = 2 $  & $\lambda = 1.5$  & 
$\lambda = 1.2$   & $\lambda = 1$  \\
\hline
$1$ & $ -7.4758 $ & $ -7.5540 $ & $ -8.7177 $ & $ -10.0938$ & $ -145.6345 $ & $ -118.5970 $ \\
$2$ & $ 2.1186  $ & $ 2.0412  $ & $ 0.9895  $ & $ -0.8918 $ & $ -130.2285 $ & $ -116.3076 $ \\
$3$ & $ 6.5599  $ & $ 6.4646  $ & $ 5.1126  $ & $ 3.1332  $ & $ -123.0114 $ & $ -109.9408 $ \\
$4$ & $ 6.8592  $ & $ 6.8040  $ & $ 5.8956  $ & $ 4.5028  $ & $ -106.4009 $ & $ -91.1749  $ \\
$5$ & $ 12.7642 $ & $ 12.7365 $ & $ 11.0206 $ & $ 6.5762  $ & $ -105.4125 $ & $ -88.3281  $ \\
$6$ & $ 13.5303 $ & $ 13.3832 $ & $ 11.7719 $ & $ 10.0159 $ & $ -97.4654  $ & $ -87.8343  $ \\ 
$7$ & $ 15.7025 $ & $ 15.6187 $ & $ 13.9796 $ & $ 10.6716 $ & $ -94.8004  $ & $ -86.9702  $ \\ 
$8$ & $ 16.4911 $ & $ 16.4027 $ & $ 14.7041 $ & $ 11.5136 $ & $ -88.8669  $ & $ -84.5732  $ 
\end{tabular}
\label{tab:spurious_spectrum_3D}
\et

We repeated this experiment in the three-dimensional NCSM code for
calculations of the triton. The scale of the spurious states, relative
to the initial bound states in that calculation, was similarly large.
The spurious states appeared in the NN-only calculations in
qualitatively similar ways. Table~\ref{tab:spurious_spectrum_3D}
recreates the results of table~\ref{tab:spurious_spectrum_4N} using
the three-dimensional NCSM. The first few states in the spectrum are
shown at selected $\lambda$'s. 

As can be seen in Fig.~\ref{fig:spurious_Hoe_Hd} the
SRG evolution on the two-body Hamiltonian seems to be driving
information from high to low $\nmax$ in an off-diagonal way. This
pattern is indicative of spurious states being developed as the
SRG builds large strength matrix elements at high-momenta and
compensates for that by inducing large three-body forces.

So, while the lone constraint on $\eta_s$ of anti-hermiticity grants a
freedom of choice in choosing the form, $G_s$, of evolution, we must
be careful of the effects that choice may have on the physics during
renormalization of different parts of the interaction. Work is
currently underway to investigate various choices of $G_s$ which
achieve convergence not only for binding energies but for other
long-range observables such as the RMS radius.~\cite{Anderson:2009b}

\chapter{Single Particle Coordinate Oscillator Basis}
\label{chapt:app_mscheme}


An important check on the Jacobi oscillator basis calculations done in
this thesis is to repeat them in a different basis. An oscillator
basis in single particle, or lab-frame coordinates has already been
developed in three-dimensions and is used to compute nuclei above the
practical range of the Jacobi basis. Here we develop a one-dimensional
analog to the single-particle basis in the spirit of
Ref.~\cite{mscheme_ref} and compare the results with our Jacobi basis
calculations.


The general strategy used here for the lab-frame oscillator basis
approach to the many-body problem is very similar to the Jacobi basis
version. The allowed basis states are explicitly constructed and the
states' quantum numbers listed. The nuclear interaction is embedded in
the constructed basis. A symmetrization operator is built to isolate
the physical states of the system. The symmetric $A$-body Hamiltonian
is then diagonalized to obtain the energy spectrum. However there are
several important differences in each of these steps; we will treat
them in turn.

\section{Embedding}
\label{sec:lab_embedding}


Here we are working exclusively in single particle coordinates, $k_1$
and $k_2$, but we have the potential in terms of Jacobi momenta
because we already had a function for it from the previous
one-dimensional work. We must first convert our interaction from the
Jacobi basis to the single particle basis. We will use the
Talmi-Moshinsky transformation brackets that we are already familiar
with from the Jacobi basis in appendix~\ref{chapt:app_osc_basis}.
Formally, we can express this transformation of the potential as:
\bea
\label{eq:potential}
\la n_1n_2|V|n_1'n_2'\ra &=& \la n_1n_2|Nn\ra\la Nn|V|N'n'\ra\la
N'n'|n_1'n_2'\ra \nonumber \\
&=& \la n_1n_2|Nn\ra\delta_{N,N'}\la n|p\ra V(p,p')\la p'|n'\ra\la N'n'|n_1'n_2'\ra
\eea
In our MATLAB implementation, Eq.~\eqref{eq:potential} is accomplished
by matrix multiplications. The matrix representing the transformation
from momenta to oscillators, $\la n|p\ra$, is the same set of
single-coordinate oscillator wavefunctions as used previously. The
matrix representation of $V(p,p')$ is read in directly from the
expression of Eq.~\eqref{eq:gaussians} as before. The delta function,
$\delta_{N,N'}$, is embedded so that we are working in a full two
particle basis organized by the total oscillator number, $N_{tot} =
N+n$ (i.e., $|Nn\ra =
|00\ra,|01\ra,|10\ra,|02\ra,|11\ra,|20\ra,|03\ra,...$). The
transformation brackets used here can be obtained from
Eq.~\eqref{eq:trans_bracket} by using $d=1$ instead of $d=3$ used
there. This gives the rotation from the Jacobi center of mass, and
first relative coordinate to the two single-particle coordinates.


As in the Jacobi case, the relative kinetic energy can be built from
creation and annihilation operators:
\bea
\la n_1 n_2 |\Trel|n_1'n_2'\ra
  &=&  \la n_1 n_2|\frac{(k_1 - k_2)^2}{2m}|n_1'n_2'\ra \nonumber \\ 
  &=&  \frac{1}{2m} \frac{-m\omega}{2} 
  \la n_1 n_2|(a_1^\dagger - a_1  - a_2^\dagger + a_2)^2
  |n_1'n_2'\ra\;.
\eea
Or, we can take the kinetic energy as built in the Jacobi
basis and convert using the transformation brackets as for the
potential above. These two approaches are identical and important
checks that our construction is consistent. However, for $A>2$ the only
practical option is the first option if we do not want to build
transformation brackets for each $A$-body space conversion.


To calculate properties of an $A$-body system, we must be able to
embed the two-, three-, and higher-body forces into the $A$-body
space. First we build a list of the possible states in that space,
$|n_1n_2n_3...n_A\ra$, and organize them into groups of total
$N\equiv\sum_{i=1}^A n_i$ up to the maximum, $\nmax$. For example, for
$A=3$ the list starts out as: 0,0,0; 0,0,1; 0,1,0; 1,0,0; 0,0,2;
0,1,1; 0,2,0; 1,0,1; 1,1,0; 2,0,0; etc up to $N=\nmax$.

It is a relatively simple matter to look through the lab-frame states
and place matrix elements of the two-body lab-frame interaction in the
appropriate matrix elements of the $A$-body lab-frame basis. For a
two-body interaction, we look through every pair of incoming and
outgoing particles in every matrix element of the $A$-body basis and
require a delta function in the initial and final energy of all other
particles in the system. If these conditions are met, the appropriate
two-body interaction matrix element is added to the $A$-body
representation of $V^{(2)}$. A similar routine must be coded
separately for embedding the three-body forces. This routine would
likewise consider all triplets of particles and require a delta
function in the others. This method of embedding is
not the most efficient especially with regard to obtaining matrix
elements between symmetric states as discussed below. 

\section{Symmetrization}
\label{sec:lab_symmetrizer}


In lab-frame coordinates, to get a symmetric state of bosons
(or antisymmetric for fermions), we must simply take the permanent
(determinant) of matrices in the full oscillator representation. For
example, the symmetric state with one particle in an oscillator state
$n_i=2$ and the other with oscillator 0 would be: 
\beqn
(\la 02| + \la 20|)V^{(2)}(|02\ra + |20\ra)/2 = (V(02;02) + V(02;20)
+ V(20;02) + V(20;20))/2; \; ,
\eeqn
where the 2 is the product of both normalization factors, $1/\sqrt{2}$.

Here we have used existing machinery to accomplish the permanent by
building a symmetrizer matrix, $S$, and requiring that it be a projector,
$S^2 \equiv S$. Instead of the Jacobi version, where we needed some
complicated combination of transformation brackets to build $S$,
simple factors of one over the number of permutations of a given state
ensure that $S$ is a projector for the single particle basis. By
keeping the eigenstates of this matrix with eigenvalue one (and
discarding those with eigenvalue zero), we obtain a non-square matrix
analogous to the ``coefficients of fractional parentage" used in the
Jacobi case. The Hamiltonian and other operators expressed in the full
single particle basis can be symmetrized using the physical eigenstate
matrix just as done in the Jacobi case to obtain an operator in that
basis.

\begin{figure}
\begin{center}
\includegraphics*[width=4in]{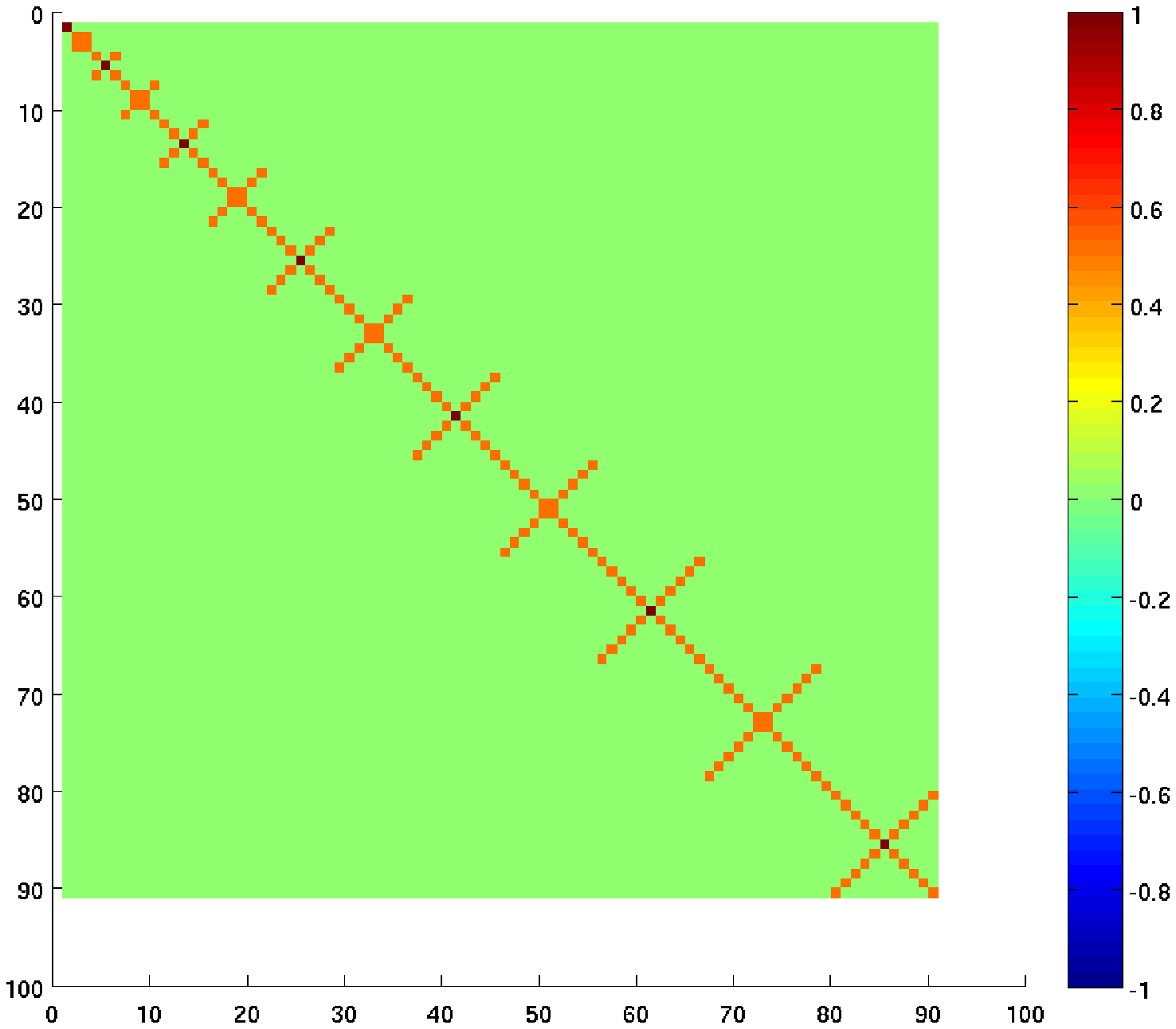}
\vspace*{.1in}
\includegraphics*[width=4in]{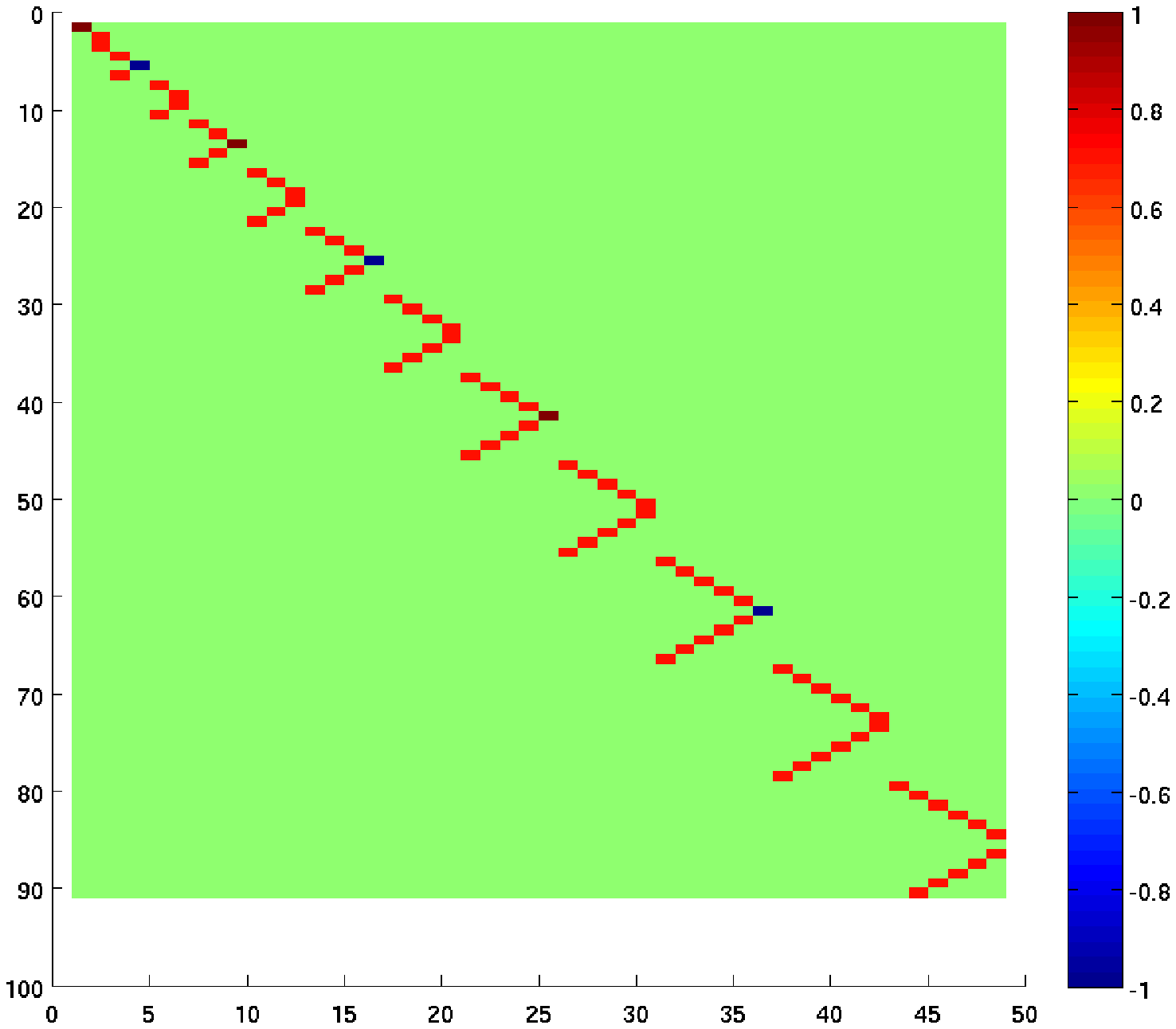}
\captionspace{A sample of the symmetrizer for the 2-body lab-frame
problem on the top, and the corresponding non-zero eigenstates of the
symmetrizer on the bottom. Here $\nmax=12$ as a more realistic $\nmax$
would not be visible on the page.}
\label{fig:symmetrizer_2N}
\end{center}
\end{figure}

For $A=2$, both of these matrices can be seen in
Fig.~\ref{fig:symmetrizer_2N}. The symmetrizer shown on top has a
simple structure with factors of $1/2$ for those states with 2
permutations (i.e.: $|02\ra$ and $|20\ra$) and 1 for those which are
already symmetric like (i.e.: $|11\ra$). Note that the eigenvectors of
displayed in the bottom panel, have the correct normalization factors,
$1/\sqrt{2}$ for a symmetrized two-particle state.

For $A>2$, we can build a symmetrizer operator from the list of total
possible states discussed above. Now we have to be more careful of the
permutations of states involved in symmetrizing. The number of
permutations of a given state is computed and recorded as the list is
built. By counting the number of particles with a certain oscillator
number, the number of permutations of the state is given by
$(A!/\prod_i^{\nmax} k_i)$ where $k_i$ is the number of particles in
the oscillator state $i$. To build the symmetrizer, we simply loop
through the matrix elements of the full single-particle basis and
compare the initial and final states. If one state is a permutation of
the other and their number of permutations are recorded as the same,
then one over that number is recorded for that matrix element of the
symmetrizer.

\begin{figure}
\begin{center}
\includegraphics*[width=4in]{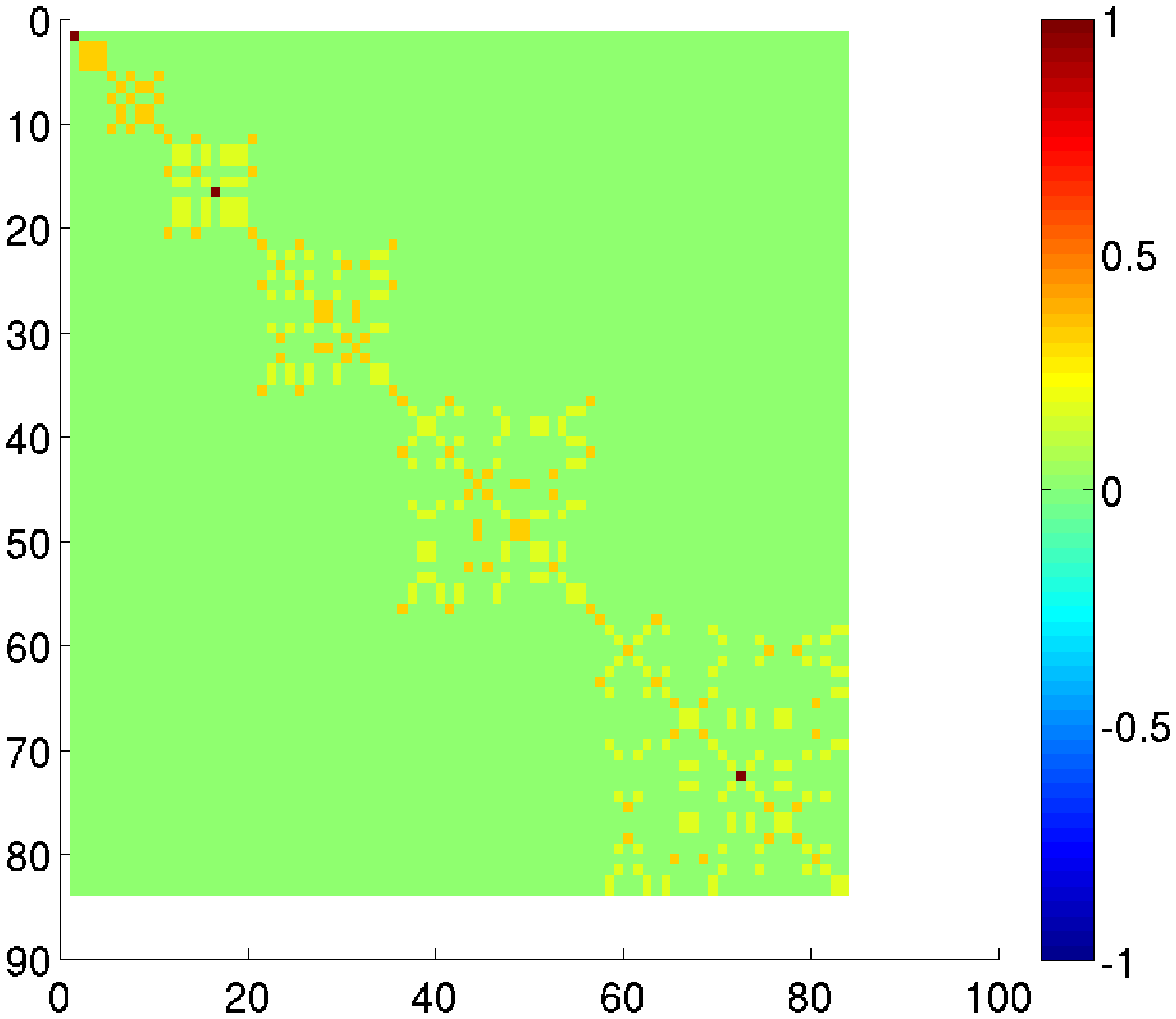}
\vspace*{.1in}
\includegraphics*[width=4in]{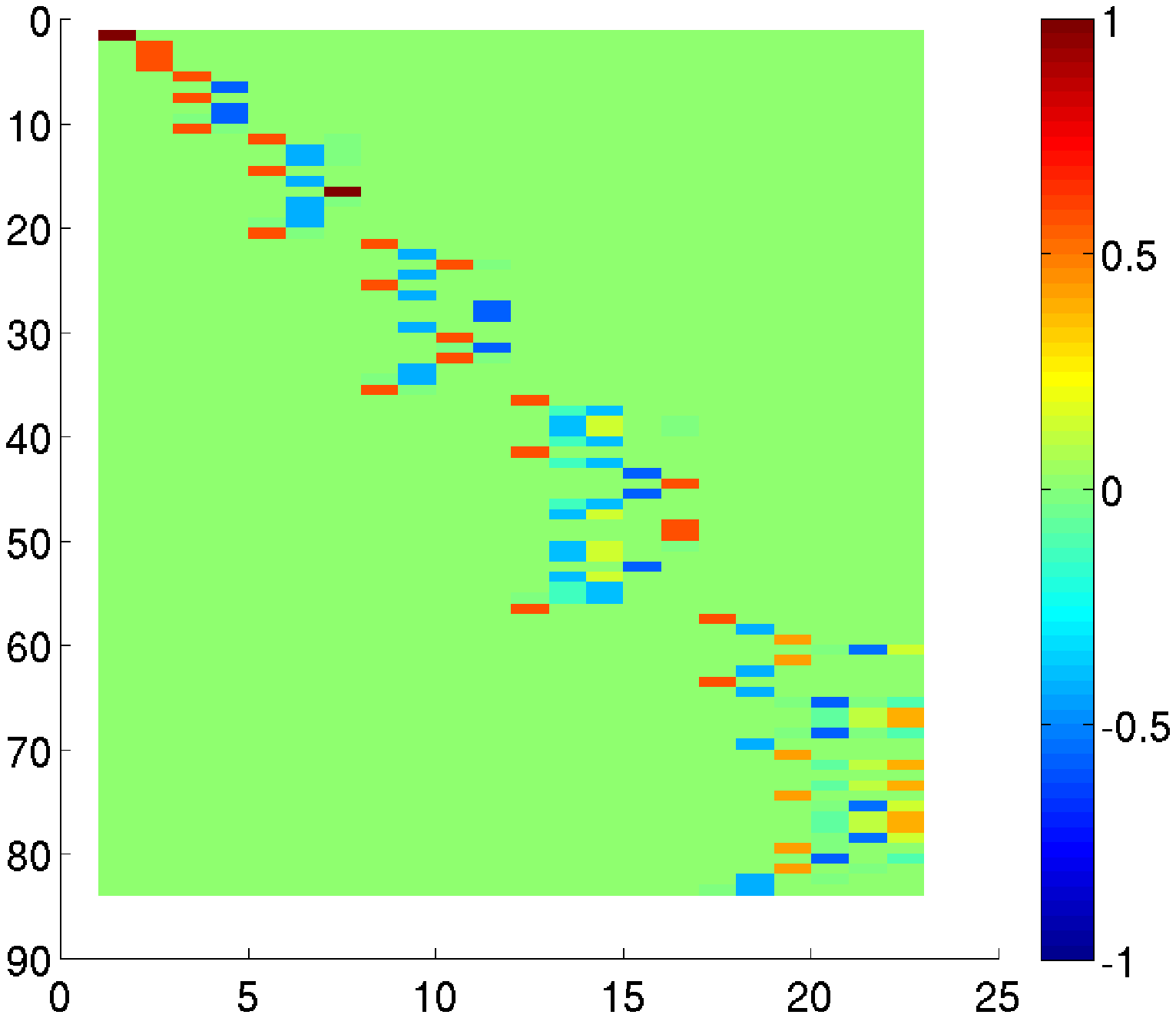}
\captionspace{The same as for Fig.~\ref{fig:symmetrizer_2N} but for
$A=3$. Here $\nmax=6$ as a more realistic $\nmax$ would not be visible
on the page.}
\label{fig:symmetrizer_3N}
\end{center}
\end{figure}

Figure \ref{fig:symmetrizer_3N} shows an example of the symmetrizer
and the corresponding symmetric eigenstates for the $A=3$ system up to
$\nmax=6$. Note the smaller colors compared to the $A=2$ case because
with more particles, there are more permutations of a given energy
content. For instance, the faintest yellow corresponds to a value of
$1/6$ for states with 6 permutations (i.e.: $|021\ra$). The darker
yellow is $1/3$ for states like $|112\ra$.

The matrices displayed in Figs.~\ref{fig:symmetrizer_2N} and
\ref{fig:symmetrizer_3N} for the physical symmetric eigenstates are
the lab-frame analog to the coefficients of fractional parentage used
in the Jacobi basis. Now, however, the construction of an $A$-body
basis does not rely on an iterative procedure on the $A-1$ symmetrized
basis. We can build the $A$-body basis from scratch, embed the two-
and three- (and higher) body forces in that basis, and symmetrize them
with the $A$-body symmetrizer built as just described.

This approach to building a symmetrized basis for lab-frame momenta is
not the same as employed in three-dimensions. The overhead of building
a large set of states and distilling out the symmetric states is very
large in three-dimensions. Convergence is only feasible in
one-dimensional calculations because of an overall reduction in basis
size do to the omission of angular momentum quantum numbers. However,
this is a useful check on the general procedure used in the Jacobi
case, and can still be a useful tool for scaling up in $A$ in one
dimension. The method used here of building a relatively simple
symmetrizer operator can also be viewed as a first step to check
future algorithms which may be less straightforward to code.

\section{Center of Mass Separation}
\label{sec:cm_separation}


A major difference from the Jacobi basis approach is the presence
of the center of mass energy in the single-particle basis. For each
value of the total center of mass energy, the $A$-body system has the
same intrinsic spectrum. However, in our basis these copies of the
spectrum are mixed up together, so we must separate them to obtain
meaningful results. We can add some quantity which will boost the
unwanted copies up in energy and isolate them. The harmonic oscillator
Hamiltonian, $H_{\rm cm}$, has the straightforward behavior in the
oscillator basis that it doesn't mix up the intrinsic physics.
Therefore, we can subtract off its effects to regain the
desired spectrum.

So, in order to isolate the different center of mass states we will
have to add
\beqn
\beta (H_{\rm cm} - (n+1/2)\hw)\;
\eeqn
to the total interaction. The constant $\beta$ is a large positive
number which will provide large separation between the spectra at each
of the center-of-mass eigenstates denoted by $n$ ($n=0$ is the
center-of-mass ground state). The value of $n$ determines which
center-of-mass
solution we will isolate exactly, free of factors of $\beta\hw$.
Spectra at center of masses less than $n$ will be kicked downward and
those greater than $n$ will be kicked upward.


Writing out the center-of-mass Hamiltonian using creation and
annihilation operators is a simple matter in the general $A$-body
space:
\newcommand{\ad}{a^{\dagger}}
\bea
&& \la n_1n_2...n_A|H_{\rm cm}|n_1'n_2'...n_A'\ra  \nonumber \\
&& \quad\quad = \frac{\hbar\omega}{8mA}
\la n_1n_2...n_A|(\sum_{i=1}^A \ad_i + a_i)^2 
- (\sum_{i=1}^A \ad_i - a_i)^2)|n_1'n_2'...n_A'\ra \nonumber \\
&& \quad\quad = \frac{\hbar\omega}{8mA}
\la n_1n_2...n_A|2\sum_{i,j=1}^A (\ad_ia_j + a_i\ad_j)
|n_1'n_2'...n_A'\ra \nonumber \\
&& \quad\quad = \frac{\hbar\omega}{8mA}
\la n_1n_2...n_A|2\sum_{i=1}^A (\ad_ia_i + a_i\ad_i) +
4\sum_{i=1}^A\sum_{j=i+1}^A (\ad_ia_j)|n_1'n_2'...n_A'\ra \nonumber \\
&& \quad\quad = \frac{\hbar\omega}{8mA}
\la n_1n_2...n_A|2\sum_{i=1}^A (2n_i'+1)\delta_{n_i,n_i'}
\nonumber \\
&& \quad\quad \quad + 4\sum_{i=1}^A\sum_{j=i+1}^A\sqrt{n_j'}\sqrt{n_i'+1})
\delta_{n_i'+1,n_i}\delta_{n_j'-1,n_j}|n_1'n_2'...n_A'\ra \;.
\eea
This is written in the same basis as the potential and kinetic energy
and can be symmetrized by the same process as described in
section~\ref{sec:lab_symmetrizer}.

\begin{table}[ht!]
\begin{center}
\caption{A sample of the spectrum resulting from the lab-frame basis
at a small $\nmax$ before and after separation by a $H_{\rm cm}$ term.
The columns right of each list of energies show the center of mass and
intrinsic energy level numbers, $n$ and $l$. The last column shows the
effective $\nmax$ for the right-most spectra.  The value of $\beta\hw$
here is 5000. 
Note the ``separated" values should be scaled by $10^4$.}
\vspace*{.1in}
\begin{tabular}{c|c|c|cc}
   mixed & $(n,l) $ & separated (x$10^4$) & $(n,l)$ & eff. $\nmax$ \\
\hline
-0.812294429336271 & $(0,0)$ & -0.000081229442933 & $(0,0)$ & 6 \\
-0.487362126673450 & $(1,0)$ &  0.000409822038397 & $(0,1)$ & 6 \\
-0.487362126673446 & $(2,0)$ &  0.001080597080727 & $(0,2)$ & 6 \\
-0.482179966274186 & $(3,0)$ &  0.002687021583751 & $(0,3)$ & 6 \\
-0.482179966274182 & $(4,0)$ &  0.499951263787332 & $(1,0)$ & 5 \\
 4.098220383969666 & $(0,1)$ &  0.500489798221359 & $(1,1)$ & 5 \\
 4.817772759274547 & $(5,0)$ &  0.501990571277867 & $(1,2)$ & 5 \\
 4.817772759274551 & $(6,0)$ &  0.999951263787332 & $(2,0)$ & 4 \\
 4.897982213589284 & $(1,1)$ &  1.000489798221359 & $(2,1)$ & 4 \\
 4.897982213589287 & $(2,1)$ &  1.001990571277867 & $(2,2)$ & 4 \\
10.805970807270517 & $(0,2)$ &  1.499951782003372 & $(3,0)$ & 3 \\
12.846759311682062 & $(4,1)$ &  1.501284675931167 & $(3,1)$ & 3 \\
12.846759311682076 & $(3,1)$ &  1.999951782003373 & $(4,0)$ & 2 \\
19.905712778674406 & $(1,2)$ &  2.001284675931168 & $(4,1)$ & 2 \\
19.905712778674410 & $(2,2)$ &  2.500481777275925 & $(5,0)$ & 1 \\
26.870215837516941 & $(0,3)$ &  3.000481777275927 & $(6,0)$ & 0 
  \end{tabular}   
\label{tab:lab_spectrum_2N}
\end{center} 
\end{table}

Table \ref{tab:lab_spectrum_2N} shows an example of the separation of
different center of mass spectra due to $H_{\rm cm}$. Columns showing
the labels of the center-of-mass and intrinsic states are meant to aid
the reader's eye when viewing these spectra. The left shows the
spectra from different center-of-mass energies entangled. For
instance, all the ground states are collected together as the lowest
energies and next the first excited states. Separating these by hand
is not a trivial matter, especially in larger systems. In contrast,
the right side shows the same spectrum but now separated by a strong
$\beta$ term. The spectrum from the lowest center of mass appears at
the head of the list. The next spectrum is multiplied by $\beta\hw$
(here $\beta = 1000$ and $\hw=5$), the third by  $2\beta\hw$, etc.

Each of the spectra obtained in this way are identical in a complete
$N\hw$ space, but we must work in a truncated space. The most accurate
spectrum is that of $n=0$ because the other spectra begin at larger
and larger center of mass energies and have fewer basis states
available to contain their excited states. In this way we can see the
effect of truncation in $\nmax$ as we look up in center-of-mass
energy. There are fewer and fewer states in the spectrum until the
center-of-mass energy is $\nmax$ and there is no more ``space" for
excitations in the basis. Note that this truncation is identical to
that done explicitly in the Jacobi calculations. On the right side of
table~\ref{tab:lab_spectrum_2N} we show the effective $\nmax$ for each
center-of-mass spectrum. The ground states for each of these spectra
are identical to the ground states from the Jacobi calculations
truncated to the $\nmax$ shown in the last column. 

\section{Evolution Results}
\label{sec:lab_evolution}


Just as in the Jacobi basis, implementation of the SRG's flow
equations is a straightforward matter in the lab-frame coordinates. 
The Hamiltonian and generator are given to a differential equation
solver and the resulting evolved Hamiltonian is diagonalized to
produce a spectrum of states for that $A$-body system. Also as in the
Jacobi basis, evolved $n$-body forces must be isolated and embedded
with the appropriate symmetry factor into the $A$-body space. At
present, existing algorithms are only sufficient for embedding
two-body operators into an $A$-body space. This limits us to NN-only
and fully unitary calculations of $A$-body spaces. Additional codes
are needed before the $m$-scheme version can be used to check the
hierarchy of induced many-body forces in systems with $A>3$. However,
for $A=3$ we can perform the calculation analogous to that of
Fig.~\ref{fig:srg_3_body} and find identical results to within
expected numerical errors. Further coding to check the hierarchy for
larger $A$ is a straightforward next step for this tool.

%
%

\end{document}